\documentclass[traditabstract]{aa}

\usepackage{graphicx}
\usepackage{txfonts}
\usepackage{natbib}
\usepackage{subfigure}
\usepackage{url}
\usepackage{longtable}
\usepackage{supertabular}
\usepackage{rotating}
\usepackage{multirow}
\usepackage{gensymb} 
\usepackage{lscape}
\usepackage{color}

\def\mum{\,$\mu$m}
\def\msun{\,M$_{\odot}$}
\def\msunyr{\,M$_{\odot}$yr$^{-1}$}
\def\lsun{\,L$_{\odot}$}
\def\nodata{...}

\def\herge{HeRG\'E}
\def\mbh{M$_{\rm BH}$}
\def\mbhedd{M$_{\rm BH}^{\rm Edd}$}

\def\spitzer{{\it Spitzer}}
\def\herschel{{\it Herschel}}
\def\hst{{\it HST}}

\def\mips1{MIPS (24\,\mum)}
\def\pacsb{PACS (70\,\mum)}
\def\pacsg{PACS (100\,\mum)}
\def\pacsr{PACS (160\,\mum)}
\def\spires{SPIRE (250\,\mum)}
\def\spirem{SPIRE (350\,\mum)}
\def\spirel{SPIRE (500\,\mum)}

\def\liragn{L$^{\rm IR}_{\rm AGN}$}
\def\lbolagn{L$^{\rm Bol}_{\rm AGN}$}
\def\lirsb{L$^{\rm IR}_{\rm SB}$}
\def\lirtot{L$^{\rm IR}_{\rm tot}$}

\def\lirobs{L$^{\rm IR}_{\rm obs}$}
\def\ledd{L$^{\rm Edd}$}

\def\mbh{M$_{\rm BH}$}
\def\mstel{M$_{\rm stel}$}

\def\mdotbh{$\dot{\rm M}_{\rm BH}$}
\def\dsini{$D$sin($i$)}
\def\smdot{s$\dot{\rm M}$$_{\rm BH}$}
\def\mbhbulge{M$_{\rm BH}$-M$_{\rm Bulge}$}
\def\mbhgal{M$_{\rm BH}$-M$_{\rm Gal}$}
\def\fagnten{$f^{10\mu \rm m}_{\rm AGN}$}
\def\fagnfifty{$f^{50\mu \rm m}_{\rm AGN}$}
\def\fagnhundred{$f^{100\mu \rm m}_{\rm AGN}$}

\def\l500{L$^{\rm 500 MHz}_{\rm ext}$}
\def\weakdetect{2$\sigma$$<$$F^{\rm gal}$$<$3$\sigma$}

\begin{document}

\title{Rapidly growing black holes and host galaxies in the distant Universe from 
 the {\it Herschel} Radio Galaxy Evolution Project \thanks{{\it Herschel} is an ESA space observatory with
    science instruments provided by European-led Principal
    Investigator consortia and with important participation from
    NASA.}}

\author{
G. Drouart
\inst{1,2,3,4},
C. De Breuck
\inst{1},
J. Vernet 
\inst{1},
N. Seymour
\inst{3},
M. Lehnert
\inst{2},
P. Barthel
\inst{5},
F. E. Bauer
\inst{6,7},
E. Ibar
\inst{6,8},
A. Galametz
\inst{9},
M. Haas
\inst{10},
N. Hatch
\inst{11},
J. R. Mullaney
\inst{12},
N. Nesvadba
\inst{13},
B. Rocca-Volmerange
\inst{2},
H. J. A. R\"ottgering
\inst{14},
D. Stern
\inst{15},
D. Wylezalek
\inst{1}}

\institute{\inst{1}European Southern Observatory, Karl Schwarzschild Stra\ss e 2, 85748 Garching bei M\"unchen, Germany \\
           \inst{2}Institut d'Astrophysique de Paris, 98bis boulevard Arago, 75014 Paris, France \\
           \inst{3}CSIRO Astronomy \& Space Science, PO Box 76, Epping, NSW 1710, Australia \\
           \inst{4}Department of Earth and Space Science, Chalmers University of Technology, Onsala Space Observatory, 43992 Onsala, Sweden \\
           \inst{5}Kapteyn Astronomical Institute, Univ. of Groningen, Netherlands \\
           \inst{6}Instituto de Astrof\'{\i}sica, Facultad de F\'{i}sica, Pontificia Universidad Catlica de Chile, 306, Santiago 22, Chile \\
           \inst{7}Space Science Institute, 4750 Walnut Street, Suite 205, Boulder, Colorado 80301 \\
           \inst{8}Instituto de F\'isica y Astronom\'ia. Universidad de Valpara\'iso. Avda. Gran Breta\~na 1111. Valpara\'iso. Chile \\
           \inst{9}INAF - Osservatorio di Roma, Via Frascati 33, I-00040, Monteporzio, Italy \\
           \inst{10}Astronomisches Institut, Ruhr-Universit\"at Bochum, Universit\"atstr. 150, 44801 Bochum, Germany \\
           \inst{11}School of Physics and Astronomy, University of Nottingham, University Park, Nottingham, NG7 2RD, UK \\
           \inst{12}Department of Physics, Durham University, South Road, Durham, DH1 3LE, UK \\
           \inst{13}Institut d'Astrophysique Spatiale, CNRS, Universit\'e Paris-Sud, 91405, Orsay, France \\
           \inst{14}Leiden Observatory, University of Leiden, P.O. Box 9513, 2300 RA Leiden, Netherlands \\
           \inst{15}Jet Propulsion Laboratory, California Institute of Technology, Mail Stop 169-221, Pasadena, CA 91109, USA}

\date{accepted for publication, \aap, 28 March 2014}

\abstract{ {We present results from a comprehensive survey of 70 radio galaxies
at redshifts 1$<$$z$$<$5.2 using the PACS and SPIRE instruments on-board
the \herschel\ {\it Space Observatory}. Combined with existing mid-IR
photometry from the \spitzer\ {\it Space Telescope}, published
870\mum\ photometry and new observations obtained with
LABOCA on the APEX telescope, the spectral energy distributions (SEDs)
of galaxies in our sample are continuously covered across 3.6--870\mum.
The total 8-1000\mum\ restframe infrared luminosities of these radio galaxies are such that
they almost all are either ultra-(\lirtot$>$10$^{12}$ \lsun) or hyper-luminous
(\lirtot$>$10$^{13}$ \lsun) infrared galaxies.
We fit the infrared SEDs with a set of empirical templates which represent
dust heated (1) by a variety of starbursts (SB) and (2) by an
active galactic nucleus (AGN). We find that the SEDs of radio galaxies require the
dust to be heated by both AGN and SB, but the luminosities of these two
components are not strongly correlated. Assuming empirical relations and
simple physical assumptions, we calculate the star formation rate (SFR), the
black hole mass accretion rate (\mdotbh), and the black hole mass (\mbh)
for each radio galaxy.  We find that the host galaxies and their black
holes are growing extremely rapidly, having SFR$\approx$100-5000\msunyr
and \mdotbh$\approx$1-100\msunyr.  The mean specific star formation rates
(sSFR) of radio galaxies at $z$$\ga$2.5 are higher than the sSFR of typical
star-forming galaxies over the same redshift range but are similar or
perhaps lower than the galaxy population for radio galaxies at $z$$\la$2.5.
By comparing the sSFR and the specific black hole mass accretion rate,
we conclude that black holes in radio loud AGN are already, or soon
will be, overly massive compared to their host galaxies in terms of
expectations from the local \mbhgal\ relation.  In order to ``catch up''
with the black hole, the galaxies require about an order-of magnitude
more time to grow in mass, at the observed SFRs, compared to the time the
black hole is actively accreting. However, during the current cycle of
activity, we argue that this catching-up is likely to be difficult due 
to the short gas depletion times. Finally, we speculate on how the host
galaxies might grow sufficiently in stellar mass to ultimately fall
onto the local \mbhgal\ relation.} {}{}{}{} }

\titlerunning{Projet HeRG\'E}
\authorrunning{G. Drouart et al.}

\maketitle

\section{Introduction}


At high redshifts, deep sub-mm observations suggest that massive
galaxies have high flux densities and vigorous, on-going star formation
\citep[e.g.][]{Hughes1998, Barger1998, Stevens2003, Chapman2005,
Wardlow2011, Swinbank2014}.  The sensitivity of wide-field bolometer
arrays limits these studies to only the brightest sub-mm emitters
\citep[e.g.][]{Weiss2009}. Such bright sub-mm galaxies (SMGs) frequently
appear to be highly disturbed, which favours gas inflows driven by mergers
as the chief instigator for generating the high observed sub-mm fluxes 
\citep[e.g.][]{Somerville2001, Engel2010}. Whether these intense starbursts
are indeed driven by mergers or by high rates of cold gas accretion
is a question that is still actively debated \citep[e.g.][]{Noeske2007, Daddi2007a,
Tacconi2008}.

Often, vigorous star formation is accompanied by powerful active
galactic nuclei \citep[AGN; e.g.][]{Hopkins2010,Wang2011,Seymour2012,Rosario2012}. 
The presence of AGN is revealed throughout the electromagnetic spectrum, 
from X-rays to radio, and in both continuum and line emission \citep[e.g.][]{Carilli1997,
Hardcastle1999, Vernet2001, Alexander2005, Ogle2006, Nesvadba2008, Ivison2012,Wang2013}.
How AGN are triggered remains one of the most challenging questions
 of contemporary extragalactic astrophysics \citep[see ][for a
recent review]{Alexander2012}. Even if current solutions and simulations are
not completely satisfying \citep[e.g.][]{Hopkins2010}, it is evident that the same material, 
cold molecular gas, is the reservoir out of which stars are formed and
the AGN is fuelled \citep[e.g.][]{Hicks2009}.

Interestingly, the expected correlation between AGN activity
and star-formation rate is not obvious in observations, both locally and at high redshift
\citep[e.g.][]{Netzer2009,Hatziminaglou2010,Asmus2011,Dicken2012,Bongiorno2012,Harrison2012,Rosario2012,Rosario2013,Feltre2013,Videla2013,Esquej2014,Leipski2014}.
This may be due to high variability of AGN
\citep{Hickox2011} or the differences in timescales it takes for gas
to become unstable, collapse to form stars over kpc scales compared to
the time it takes for gas to lose sufficient angular momentum to reach
the inner central parsec of the galaxy \citep{Jogee2005}. Despite our
difficulties in understanding how relationships between the host galaxy
and super massive black holes come about, we observe a tight correlation
between the black hole and the physical properties of their host galaxies in the local universe
\citep[e.g.][]{Magorrian1998, Gebhardt2000, Ferrarese2000, Neumayer2004}.
These relations suggest that both components of galaxies grew
simultaneously \citep[e.g.][]{Hopkins2006p}. Nevertheless, some discrepancies
have been observed from the local relation implying either 
an observational bias or a possible evolution of this relation
 with redshift \citep[e.g.][]{Lauer2007,Zhang2012}. Currently,
there are no complete answers that reconcile all the observations 
\citep[][for a recent review]{Kormendy2013}.

Observations of infrared emission plays a key role in disentangling
the relative importance of star formation and AGN to the bolometric
emission from galaxies.  As the IR emission is a mixture of dust
heated by both the stars and the AGN, the nature of the IR spectral energy distribution (SED) can be
used to probe the relative growth of galaxies and supermassive black
holes and how their growth rates are related (the ``AGN-starburst
connection''). The short cooling time of the dust provides us with a
snapshot of the heating rate of a galaxy due to the re-emission of absorbed UV and optical photons
\citep[e.g.][]{Draine2003}. However, the peak of the IR SED, where both heating of dust grains by AGN and star formation
make important contributions, was not
completely covered with good sensitivity by \spitzer\ or by ground-based
sub-mm photometry for distant galaxies \citep[e.g.][]{Archibald2001,
Reuland2004, Cleary2007, DeBreuck2010, Rawlings2013}. \herschel\ now
provides the first opportunity to explore the complete IR SED of high
redshift AGN, and thus to examine the relative contribution of the AGN
and star formation to the bolometric luminosity of galaxies over a wide
range of redshift.

Powerful radio galaxies are crucial objects in understanding
the evolution of massive galaxies.  They present all phenomenology
undergoing both active star formation and rapidly accreting supermassive
black holes.  Powerful radio jets, strong and highly ionized
optical and near-IR emission lines, and luminous mid-IR continuum,
for example, betray the presence of an accreting supermassive
black hole \citep[e.g][]{Carilli1997, Vernet2001, Nesvadba2008,
DeBreuck2010, Drouart2012, Rawlings2013}. They also have luminous submm
emission, which is directly related to their vigorous star formation.
Moreover, they have elliptical light profiles \citep{Matthews1964,
vanBreugel1998, Pentericci1999,Zirm2003}, are extremely massive \citep{Rocca2004,
Seymour2007} and are often associated with high density environments
\citep[e.g.][]{Venemans2007,Falder2010,Hatch2011a, Kuiper2011, Galametz2012, Wylezalek2013b}.
In other words, they have many hallmarks of a massive (perhaps cluster)
galaxy in formation \citep{Miley2008}.

By their fortuitous edge-on orientation, the radio galaxies present a dusty 
torus occulting the light from the hot accretion disk (type 2 AGN), enabling the simultaneous 
study of the host galaxy and the AGN, more easily than in the case of quasars 
(i.e. type 1 AGN, for a recent review, see \citealp{Antonucci2011}). Therefore, observing 
and characterising the different constituents of high redshift radio galaxies appears 
as our best chance to gain insights on the connection of the galaxy and black hole 
growth at much earlier stage in their history, more especially during the peak of the 
cosmic AGN and star formation activity \citep{Hopkins2006a,Aird2010}. 
Since characterising the host galaxy/BH through dynamic properties at 
high redshift is observationally expensive \citep{Nesvadba2011}, and beyond the reach of most of the 
current facilities, one has to rely on energetic diagnostics (such as SED decomposition) 
and empirical relations \citep[e.g. M$_{\rm BH}$-$\sigma$, M$_{\rm BH}$-M$_{\rm bulge}$, 
M$_{\rm BH}$-M$_{K}$]{Ferrarese2000,Neumayer2004,Merloni2010} to investigate
this (non-)relation during the first half of the history of the Universe in larger samples.

In this paper, we analyse the characteristics of the IR SEDs of a sample
of 70 powerful radio galaxies spanning the redshift range from 1 to
5.2. This large sample allows us to compare the properties of the IR SED with their
other characteristics (e.g.  radio luminosities and sizes).  The paper
is organised as follows: \S~2 outlines the \herschel\ and sub-mm observations and data
reduction; \S~3 demonstrates how the photometry was calculated in
cases of isolated and blended sources in the \herschel\ images; \S~4
discusses the IR luminosities and the SED fitting procedure which was
used to estimate the bolometric, AGN and starburst luminosities; \S~5
compares the IR emission with other properties of the radio galaxies; \S~6
discusses the interpretation of these luminosities in terms of physical
parameters allowing us to put new constraints on the evolution of radio
galaxies. Throughout this paper, we adopt the concordance cosmological
model ($H_0= 70$\,kms$^{-1}$\,Mpc$^{-1}$, $\Omega_{\Lambda}=0.7$,
$\Omega_{M}=0.3$).

\section{Observations and Data reduction}

This paper aims to disentangle the IR SED of a sample of 70 powerful radio
galaxies spanning the redshift range 1--5.2. This HErschel Radio Galaxies
Evolution (\herge) sample is identical to the \spitzer\ High-$z$ Radio Galaxies (SHzRG) sample described by
\citet{Seymour2007} and \citet{DeBreuck2010}. We  briefly summarize here 
the selection criteria to build this sample. The radio galaxies have been selected 
to cover homogeneously the radio luminosity-redshift plane, applying the criteria 
$L^{\rm{3\,GHz}}>10^{26}$\,WHz$^{-1}$, where $L^{\rm{3\,GHz}}$ is
the total luminosity at a rest-frame frequency of 3\,GHz \citep[Table 1;][]{Seymour2007}. 

We first describe the new \herschel\ data of our entire sample, followed by a presentation of
sub-mm data which were obtained with the LArge Bolometer
CAmera (LABOCA) on the APEX telescope to complete the submm observations of
our sample\footnote{For our sample of radio galaxies, synchrotron contamination 
at submm wavelength is negligible as all sources present steep radio spectral indices.}.

\subsection{Far-IR data, \herschel}

The far-IR data for all 70 sources were obtained with the \herschel\
{\it Space Observatory} \citep{Pilbratt2010} in five broad bands: in two bands with PACS 
\citep[Photodetector Array Camera and Spectrometer;][at 160\mum\ and either 70\mum\ or 100\mum\ depending on the redshift of the radio galaxy]{Poglitsch2010} 
and in three bands with SPIRE \citep[Spectral and Photometric Imaging REceiver;][at 250, 350 and 500\mum]{Griffin2010}. 
Our program was observed between 2011 February
and 2012 March. Several sources were already observed as part of
guaranteed time observations, and those data were obtained from the
\textit{Herschel} Science Archive (see Table \ref{tab:herschel_info} for program and ObsID).

\begin{table*} \centering
\caption{Main parameters of the \herschel\ bands and photometry for isolated sources. The 3$\sigma$ sensitivity limit is the average sensitivity calculated over our entire sample for each band. Absolute calibration uncertainties and aperture correction from \protect\url{http://herschel.esac.esa.int/Documentation.shtml}. Note that this correction is only applied for the aperture photometry (see \S~\ref{sec:phot}).}
\label{tab:main_param}
\begin{tabular}{ccc ccc ccc}
\hline
Bands & beam size & absolute cal. & final pixel size & Av. 3$\sigma$ sensitivity & aperture & inner sky & outer sky & Ap. corr. \\
      & [arcsec] &  & [arcsec] & [mJy] & [arcsec] & [arcsec] & [arsec] & \\
\hline \hline
\pacsb & 5.6 & 5\% & 2 & 8.7 & 7 & 15 & 25 & 1.33 \\
\pacsg & 6.7 & 5\% & 2 & 10.8 & 7 & 15 & 25 & 1.39 \\
\pacsr  & 11 & 5\% & 3 & 24.6 & 11 & 15 & 30 & 1.37 \\
\spires & 18 & 7\% & 6 & 15.9 & 22 & 60 & 90 & 1.28 \\
\spirem & 25 & 7\% & 10 & 17.7 & 30 & 60 & 90 & 1.19 \\
\spirel & 36 & 7\% & 14 & 18.9 & 42 & 60 & 90 & 1.26\\
\hline
\end{tabular}
\end{table*}

\subsubsection{PACS reduction}
\label{sec:pacs_reduc}

PACS covers the spectral region 60 to 210\mum.
The ``mini-scan map mode" was used on each
science target, using the \pacsb/\pacsr\ and \pacsg/\pacsr\ configurations
for sources at $z$$<$2 and $z$$>$2, respectively. Each observation consisted of two
cross-scan images centred on the source. The final map covers about 2x4
arcmin, with homogeneous coverage of 50 arcsec diameter around the target. The
observations of PKS~1138-262 cover a larger field but were reduced using
the same procedure.  As each PACS observation consists of a simultaneous
scan in two bands at medium scan speed (20 arcsec.s$^{-1}$), two sub-images are
produced for each band and co-added to obtain the final maps. Each data
set was reduced from level 0 using the \herschel\ Interactive Processing
Environment, version 8 \citep[HIPE;][]{Ott2010}
using the standard \textit{deep miniscan} pipeline. As we are looking
for faint sources, the \textit{MMTDeglitching} task was applied on
level 0.5 of the data and we checked on the coverage map that no flux was 
potentially removed from the source. The PACS data is dominated by 1/$f$ noise, so we applied a \textit{high-pass filter} on level 1 data
with a \textit{high-pass filtering radius (hpfradius)} value of 15 readouts for the
blue/green channel and 25 readouts for the red channel with a circular mask of 15 arcsec
 radius centred on the galaxy coordinates (best strategy available, see \citealp{Popesso2012}). Finally, each map was
projected onto a user-defined world coordinate system (WCS) grid centred
on the source. As our observations have a high redundancy, we chose a
small \textit{pixfraction} value (0.01) and set the \textit{pixsize} to the recommended 
values: 2 arcsec for the blue/green channel and 3 arcsec for the red channel (Table
\ref{tab:main_param})\footnote{We also test the reduction with smaller pixel size 
(1.2 and 2.1 arcsec for the blue/green and red channels) and find the differences 
on the final flux to be negligible ($<$5\%). See also PICC-ME-TN-033 (April 4, 2012 v2) at 
\protect\url{http://herschel.esac.esa.int/twiki/pub/Public/PacsCalibrationWeb/bolopsf_20.pdf} for further
information on the pixelisation effect on the PSF.}. Finally, the 2 sub-maps were co-added into a
final map with the \textit{MosaicTask}.

\subsubsection{SPIRE reduction}

SPIRE covers the spectral region 200 to 700\mum.  Each observation with SPIRE
consists of three successive scans centred on the source with all three bands
(250, 350 and 500\mum) at 30 arcsec.s$^{-1}$ scan speed. The only exception
is again PKS~1138-262 which had four scans over a wider area. 
The final map for each source covers 8x10 arcmin, with
a homogeneous exposure level throughout the entire field. We reduced
the data with the {\it Photometer small map pipeline} within version 8
of HIPE.  As glitches are present in the SPIRE timeline, several
deglitching procedures were applied to the level 1 data. We choose the
{\it linearadaptive20} option for the {\it wavelet deglitcher} with all
the other options at their default values. We used the {\it
naivemapmaker} to create the final map with pixel sizes of 6, 10, and 14 arcsec
for the maps at 250, 350, and 500\mum\ maps, respectively
(Table \ref{tab:main_param}).

\subsection{\herschel\ Photometry}
\label{sec:phot}

Thanks to high resolution radio observations \citep{Carilli1997,Pentericci2000,DeBreuck2010}, radio galaxy positions are 
known to sub-arcsec accuracy. As the average \herschel\ pointing uncertainties 
are $\sim$1 arcsec, we performed fixed aperture photometry directly on the known
position of the each radio galaxy. Due to the depth of the images and the large beam of \herschel,
the aperture photometry is often contaminated by nearby companions,
which may contribute significantly to the estimated flux of the radio
galaxy. In order to mitigate against this contamination, we visually
checked each galaxy in six bands, from \mips1\ to \spirel. As the 24\mum\ image provides the
best spatial resolution with which to investigate the dust emission, it was
used to isolate potential companions contributing to the total flux in the
\herschel\ data. The \spires\ image, since it is taken through one of
the most sensitive channels and has reasonable resolution, was used to
provide long wavelength information about possible contaminated sources.
We first mark all the positions of detected sources in the 24\mum\ image
onto the \herschel\ maps. When a possible contaminating source was found
within 60 arcsec, it was deblended to remove its contribution from
the radio galaxy flux (see \S~\ref{sec:blended}). Otherwise, a single aperture 
was used to estimate the flux (Table \ref{tab:herschel_flux}). 

\subsubsection{Isolated sources}

When the image does not show a contamination in the \mips1\
and \spires\ images, aperture photometry was performed using
the \textit{AnnularSkyAperturePhotomery} task within HIPE.
For a comparison between the different strategies, we refer to \cite{Popesso2012,Pearson2013}, 
for PACS and SPIRE, respectively. We summarise here briefly for our sample. 
For PACS, the optimal strategy of masking is applied (see \S~\ref{sec:pacs_reduc}),
and aperture and PSF photometry gives similar results. For SPIRE, 
as our sample contains mainly faint sources ($F^{gal}$$<$30mJy), 
automatic procedures would normally be preferred (SUSSEXtractor or DAOphot). 
However, as our sample is subject to blending effects (see next section),  
PSF-photometry is mainly performed to measure source flux making use of Starfinder (similar to SUSSEXtractor).
 See Table \ref{tab:main_param} for a summary of the parameters. Table
\ref{tab:herschel_flux} reports the final flux, obtained after aperture
correction in the case of aperture photometry.

\subsubsection{Blended sources}
\label{sec:blended}

Blending becomes more important, particularly for the SPIRE bandpasses 
where the large beams encompass a large area around the radio galaxy (for 
example, the SPIRE 500\mum\ beam corresponds to $\sim$300\,kpc at $z$$=$1). 
While this is particularly problematic for blind source extraction at a single wavelength 
\citep{Nguyen2010}, we can use here the prior information given by higher resolution 
observations such as \mips1\ images. 

We use Starfinder, software optimised for crowded fields, performing Point
Spread Function (PSF) photometry to estimate the fluxes of sources which
are blended \citep{Diolaiti2000}. StarFinder requires both the estimated
position of each source and the characteristics of the PSF. We defined a 2D
Gaussian PSF with the FWHM equal to the beam size. Even if the PACS and SPIRE beams
are slightly different from Gaussian, the energy in the secondary
lobes is only a small fraction of the total integrated energy, and the Gaussian approximation is still valid. 
We checked this difference in the SPIRE images where several sources can be used to estimate the PSF. 
We found the differences to be negligible. 

Even with input positions on possible sources, sometimes Starfinder
did not converge on a solution, especially in the case where two sources
are separated by less than the FWHM of the PSF
for SPIRE. The SPIRE 500\mum\ band, which has the largest beam, is the most affected
by this effect.  For sources that could not be accurately deblended, we
assume the total flux to be the upper limit for the radio galaxy. 

The main caveat to this technique is the assumption that a source detected in SPIRE has a counterpart in the MIPS images. Using the average sensitivity of our \mips1\ and \spires\ images, we calculate the corresponding colour limit, log ($F_{250\,\mu m}/F_{24\,\mu m}$)=2.12. Making use of templates from DecompIR (see \S~\ref{sec:decompir}), 
we indeed find that this approach can miss some sources at $z$$>$3 given our achieved sensitivities. 
Nevertheless, such contamination is estimated to be only a few percent \citep{Roseboom2010,Magdis2011} 
and is therefore not taken into account for the remainder of this paper. 

\subsubsection{Uncertainties}
\label{sec:noise}

The design of the PACS detectors makes those data prone to correlated noise
 \citep{Popesso2012}. While a formal estimate of such noise is almost 
impossible, it is possible to estimate the average total noise from the images. 
Given the observing strategy, we focused on the most homogeneous,
central part of the images to estimate the noise. We drew identical,
non-overlapping apertures around the source in a hexagonal pattern,
and performed the same aperture photometry as used to estimate the flux of the central source. 
We considered the total noise on the map to be the standard deviation of these distributed apertures around the source. 

For the SPIRE images, the uncertainties are calculated either in the sky annulus
for the aperture photometry in the case of an isolated source, or by
the standard deviation of the pixel value distribution of the map for
the PSF-photometry. 

\subsection{Final uncertainty and \herschel\ flux}

As the observations are centred at the position of the radio galaxies, 
which are well detected at shorter wavelength, we have a strong prior on the detection 
of a source at a given position. We define ``strict non-detection", ``tentative detection"
and ``strong detection" as sources detected at the  $F^{\rm gal}$$<$2$\sigma$, \weakdetect\ and 
$F^{\rm gal}$$>$$3\sigma$ levels, respectively. 

In the case of a non-detection ($F^{\rm gal}$$<$2$\sigma$), we took the upper limit as
three times the sky standard deviation (we discussed the estimation
of the uncertainties in previous sub-sections). In the case of tentative
detection (\weakdetect), we provide the value of the flux between square
brackets (Table~\ref{tab:herschel_flux}) and display these as
open diamonds on the SED plots (Fig. \ref{fig:sed}). In addition, we
add the calibration uncertainty in the formal errors for detected
sources. Table \ref{tab:herschel_flux} presents the final flux estimates
and their associated total uncertainties (photon, instrumental and confusion noise). They are calculated by adding
quadratically the absolute calibration uncertainty (see Table \ref{tab:herschel_info}) with the uncertainty
estimated directly from the noise characteristics of the images (Section
\ref{sec:noise}).  Because of the additional flux calibration uncertainty,
the \textit{Signal-to-noise} ratio does not correspond to the flux 
uncertainties given in Table \ref{tab:herschel_flux}, as those are calculated \textit{before}
applying the calibration correction.

\begin{table*}[th] \centering
\caption{\herschel observations. All are part of the OT1 seymour 1
  program except when specified.}
\label{tab:herschel_info}
\begin{tabular}{l ccc ccc l}
\hline
Name & RA (J2000) & Dec (J2000) & $z$ & ObsID PACS 1 & ObsID PACS 2 & ObsID SPIRE & Notes \\
\hline \hline
    6C~0032+412  &       00:34:53.1  &      +41:31:31.5  &   3.670  &    1342237860   &    1342237861   &    1342238254  &  \\
   MRC~0037-258  &       00:39:56.4  &      -25:34:31.0  &   1.100  &    1342235414   &    1342235415   &    1342221924  &  \\
   6CE~0058+495  &       01:01:18.8  &      +49:50:12.3  &   1.173  &    1342237864   &    1342237865   &    1342238256  &  \\
   MRC~0114-211  &       01:16:51.4  &      -20:52:06.7  &   1.410  &    1342224381   &    1342224382   &    1342234710  &  \\
  TN~J0121+1320  &       01:21:42.7  &      +13:20:58.0  &   3.516  &    1342238029   &    1342238030   &    1342223211  &  \\
   6CE~0132+330  &       01:35:30.4  &      +33:16:59.6  &   1.710  &    1342237844   &    1342237845   &    1342237505  &  \\
    6C~0140+326  &       01:43:43.8  &      +32:53:49.3  &   4.413  &    1342214047   &    1342214048   &    1342213490  &  KPGT kmeisenh 1 \\
   MRC~0152-209  &       01:54:55.8  &      -20:40:26.3  &   1.920  &    1342238786   &    1342238787   &    1342234712  &  \\
   MRC~0156-252  &       01:58:33.4  &      -24:59:31.7  &   2.016  &    1342238739   &    1342238740   &    1342234716  &  \\
  TN~J0205+2242  &       02:05:10.7  &      +22:42:50.4  &   3.506  &    1342237400   &    1342237401   &    1342237501  &  \\
   MRC~0211-256  &       02:13:30.5  &      -25:25:20.6  &   1.300  &    1342239459   &    1342239460   &    1342234717  &  \\
   TXS~0211-122  &       02:14:17.4  &      -11:58:46.0  &   2.340  &    1342238111   &    1342238112   &    1342237532  &  \\
          3C~65  &       02:23:43.5  &      +40:00:52.7  &   1.176  &    1342238005   &    1342238006   &    1342239821  &  \\
   MRC~0251-273  &       02:53:16.7  &      -27:09:11.6  &   3.160  &    1342237410   &    1342237411   &    1342214558  &  \\
   MRC~0316-257  &       03:18:12.1  &      -25:35:09.7  &   3.130  &    1342239422   &    1342239423   &    1342214555  &  \\
   MRC~0324-228  &       03:27:04.5  &      -22:39:42.1  &   1.894  &    1342239424   &    1342239425   &    1342238288  &  \\
   MRC~0350-279  &       03:52:51.6  &      -27:49:22.6  &   1.900  &    1342239418   &    1342239419   &    1342227718  &  \\
   MRC~0406-244  &       04:08:51.5  &      -24:18:16.4  &   2.427  &    1342225214   &    1342225215   &    1342239859  &  \\
       4C~60.07  &       05:12:55.1  &      +60:30:51.0  &   3.788  &    1342206050   &    1342206051   &    1342203606  &  KPGT kmeisenh 1 \\
   PKS~0529-549  &       05:30:25.4  &      -54:54:23.2  &   2.575  &    1342236654   &    1342236655   &    1342226641  &  \\
  WN~J0617+5012  &       06:17:39.3  &      +50:12:54.2  &   3.153  &    1342242754   &    1342242755   &    1342229114  &  \\
       4C~41.17  &       06:50:52.2  &      +41:30:30.1  &   3.792  &    1342206336   &    1342206337   &    1342204958  &  KPGT kmeisenh 1 \\
  WN~J0747+3654  &       07:47:29.4  &      +36:54:38.1  &   2.992  &    1342229038   &    1342229039   &    1342229478  &  \\
  6CE~0820+3642  &       08:23:48.1  &      +36:32:46.4  &   1.860  &    1342243302   &    1342243303   &    1342230755  &  \\
       5C~7.269  &       08:28:38.8  &      +25:28:27.1  &   2.218  &    1342232224   &    1342232225   &    1342230774  &  \\
   USS~0828+193  &       08:30:53.4  &      +19:13:16.2  &   2.572  &    1342232222   &    1342232223   &    1342230772  &  \\
  6CE~0901+3551  &       09:04:32.3  &      +35:39:04.1  &   1.910  &    1342232232   &    1342232233   &    1342230750  &  \\
     B2~0902+34  &       09:05:30.1  &      +34:07:56.0  &   3.395  &    1342232230   &    1342232231   &    1342230737  &  \\
  6CE~0905+3955  &       09:08:16.9  &      +39:43:26.0  &   1.883  &    1342232236   &    1342232237   &    1342230748  &  \\
  TN~J0924-2201  &       09:24:19.9  &      -22:01:42.3  &   5.195  &    1342198543   &    1342198544   &    1342198865  &  KPGT kmeisenh 1 \\
    6C~0930+389  &       09:33:06.9  &      +38:41:50.1  &   2.395  &    1342232036   &    1342232037   &    1342230738  &  \\
   USS~0943-242  &       09:45:32.7  &      -24:28:49.7  &   2.923  &    1342233248   &    1342233249   &    1342234835  &  \\
         3C~239  &       10:11:45.4  &      +46:28:19.8  &   1.781  &    1342231241   &    1342231242   &    1342230739  &  \\
   MG~1019+0534  &       10:19:33.4  &      +05:34:34.8  &   2.765  &    1342233228   &    1342233229   &    1342222672  &  \\
   MRC~1017-220  &       10:19:49.0  &      -22:19:59.6  &   1.768  &    1342233103   &    1342233104   &    1342234838  &  \\
  WN~J1115+5016  &       11:15:06.9  &      +50:16:23.9  &   2.540  &    1342231395   &    1342231396   &    1342222662  &  \\
         3C~257  &       11:23:09.4  &      +05:30:17.1  &   2.474  &    1342221966   &    1342221967   &    1342210514  &  GT1 pbarthel 1 \\
  WN~J1123+3141  &       11:23:55.7  &      +31:41:26.7  &   3.217  &    1342222681   &    1342222682   &    1342222669  &  \\
   PKS~1138-262  &       11:40:48.4  &      -26:29:08.8  &   2.156  &    1342222456   &    1342222457   &    1342210877  &  GT1 baltieri 3 \\
         3C~266  &       11:45:43.6  &      +49:46:05.2  &   1.275  &    1342222695   &    1342222696   &    1342222663  &  \\
     6C~1232+39  &       12:35:04.7  &      +39:25:38.9  &   3.220  &    1342234365   &    1342234366   &    1342232703  &  \\
   USS~1243+036  &       12:45:38.4  &      +03:23:20.7  &   3.570  &    1342223828   &    1342223829   &    1342224982  &  \\
  TN~J1338-1942  &       13:38:26.1  &      -19:42:30.7  &   4.110  &    1342237906   &    1342237907   &    1342236186  &  \\
       4C~24.28  &       13:48:14.9  &      +24:15:50.5  &   2.879  &    1342233533   &    1342233534   &    1342234787  &  \\
         3C~294  &       14:06:53.2  &      +34:11:21.1  &   1.786  &    1342233525   &    1342233526   &    1342236145  &  \\
   USS~1410-001  &       14:13:15.1  &      -00:22:59.7  &   2.363  &    1342237900   &    1342237901   &    1342236162  &  \\
    8C~1435+635  &       14:36:37.2  &      +63:19:14.4  &   4.250  &    1342209329   &    1342209330   &    1342199362  &  KPGT kmeisenh 1 \\
   USS~1558-003  &       16:01:17.3  &      -00:28:46.2  &   2.527  &    1342238057   &    1342238058   &    1342238311  &  \\
   USS~1707+105  &       17:10:06.9  &      +10:31:10.2  &   2.349  &    1342230116   &    1342230117   &    1342229578  &  \\
    LBDS~53W002  &       17:14:14.8  &      +50:15:30.6  &   2.393  &    1342234084   &    1342234085   &    1342229153  &  \\
    LBDS~53W069  &       17:20:02.5  &      +49:44:51.0  &   1.432  &    1342231668   &    1342231669   &    1342229155  &  \\
    LBDS~53W091  &       17:22:32.9  &      +50:06:01.3  &   1.552  &    1342234086   &    1342234087   &    1342229156  &  \\
       3C~356.0  &       17:24:19.3  &      +50:57:36.2  &   1.079  &    1342219036   &    1342219037   &    1342206197  &  GT1 pbarthel 1 \\
   7C~1751+6809  &       17:50:50.0  &      +68:08:26.4  &   1.540  &    1342233557   &    1342233558   &    1342223224  &  \\
   7C~1756+6520  &       17:57:05.4  &      +65:19:53.1  &   1.416  &    1342233561   &    1342233562   &    1342229141  &  \\
         3C~368  &       18:05:06.4  &      +11:01:33.1  &   1.132  &    1342216599   &    1342216600   &    1342216954  &  GT1 pbarthel 1 \\
   7C~1805+6332  &       18:05:56.8  &      +63:33:13.1  &   1.840  &    1342233563   &    1342233564   &    1342229140  &  \\
       4C~40.36  &       18:10:55.7  &      +40:45:24.0  &   2.265  &    1342225262   &    1342225263   &    1342229165  &  \\
 TXS~J1908+7220  &       19:08:23.7  &      +72:20:11.8  &   3.530  &    1342232248   &    1342232249   &    1342220624  &  \\
  WN~J1911+6342  &       19:11:49.6  &      +63:42:09.6  &   3.590  &    1342233575   &    1342233576   &    1342220864  &  \\
  TN~J2007-1316  &       20:07:53.3  &      -13:16:43.6  &   3.840  &    1342217425   &    1342217426   &    1342230833  &  \\
   MRC~2025-218  &       20:27:59.5  &      -21:40:56.9  &   2.630  &    1342217421   &    1342217422   &    1342230830  &  \\
   MRC~2048-272  &       20:51:03.6  &      -27:03:02.5  &   2.060  &    1342218548   &    1342218549   &    1342218982  &  \\
   MRC~2104-242  &       21:06:58.3  &      -24:05:09.1  &   2.491  &    1342232504   &    1342232505   &    1342218979  &  \\
       4C~23.56  &       21:07:14.8  &      +23:31:45.0  &   2.483  &    1342222551   &    1342222552   &    1342233325  &  \\
   MG~2144+1928  &       21:44:07.6  &      +19:29:14.6  &   3.592  &    1342235313   &    1342235314   &    1342220527  &  \\
   USS~2202+128  &       22:05:14.2  &      +13:05:33.0  &   2.706  &    1342235311   &    1342235312   &    1342220528  &  \\
   MRC~2224-273  &       22:27:43.3  &      -27:05:01.7  &   1.679  &    1342234103   &    1342234104   &    1342234742  &  \\
  B3~J2330+3927  &       23:30:24.8  &      +39:27:12.5  &   3.086  &    1342225457   &    1342225458   &    1342234918  &  \\
       4C~28.58  &       23:51:59.2  &      +29:10:29.0  &   2.891  &    1342225467   &    1342225468   &    1342234922  &  \\
         3C~470  &       23:58:36.0  &      +44:04:46.0  &   1.653  &    1342237858   &    1342237859   &    1342236248  &  \\
\hline
\end{tabular}
\end{table*}

\begin{table*}[th] \centering
\caption{\herschel\ photometry. Values between parenthesis are the
  {\it signal-to-noise} estimates from the images {\it before}
  including the uncertainties in the flux calibration. Values between
  square brackets are weak detections (\weakdetect). Upper limits are given at the
  3$\sigma$ level of the noise (see \S~\ref{sec:noise} for how these
  were determined). Flux given in italic were deconvolued using {\it Starfinder} (See
  \S~\ref{sec:blended}).}
\label{tab:herschel_flux}
\begin{tabular}{l ccc ccc}
\hline
Name & \pacsb & \pacsg & \pacsr & \spires & \spirem & \spirel \\
     & [mJy] & [mJy] & [mJy] & [mJy] & [mJy] & [mJy] \\
\hline \hline
         6C~0032+412 &                   \nodata &                  $<$ 11.2 &                  $<$ 21.3 &                  $<$ 14.2 &                  $<$ 15.4 &                  $<$ 20.2 \\
        MRC~0037-258 &                  $<$  7.5 &                   \nodata &                  $<$ 28.0 &                  $<$ 14.6 &                  $<$ 16.6 &                  $<$ 18.9 \\
        6CE~0058+495 &                  $<$  9.2 &                   \nodata &                  $<$ 35.5 &                  $<$ 19.2 &                  $<$ 19.8 &                  $<$ 29.1 \\
        MRC~0114-211 &       [5.5$\pm$  2.7( 2.0)] &                   \nodata &      32.2$\pm$  9.4( 3.5) & {\it 24.3$\pm$  7.5( 3.3)}&                  $<$ 36.2 &                  $<$ 30.8 \\
       TN~J0121+1320 &                   \nodata &                  $<$  7.9 &                  $<$ 24.2 & {\it 15.9$\pm$ 5.7( 2.8)]}& {\it [18.0$\pm$6.6( 2.8)]}&                  $<$ 18.4 \\
        6CE~0132+330 &                  $<$ 10.2 &                   \nodata &                  $<$ 26.0 &                  $<$ 14.3 &                  $<$ 16.7 &                  $<$ 19.5 \\
         6C~0140+326 &                  $<$  6.9 &                   \nodata &                  $<$ 19.8 &                  $<$ 14.7 &                  $<$ 14.5 &                  $<$ 16.5 \\
        MRC~0152-209 &      22.6$\pm$  3.5( 6.8) &                   \nodata &       119.2$\pm$9.8(15.4) & {\it 105.0$\pm$ 8.6(23.0)}&  {\it 81.3$\pm$ 7.3(17.6)}& {\it 64.4$\pm$  6.8(12.6)}\\
        MRC~0156-252 &                   \nodata &      13.8$\pm$  3.7( 3.8) &                  $<$ 23.3 &                  $<$ 15.0 &                  $<$ 18.0 &                  $<$ 20.9 \\
       TN~J0205+2242 &                   \nodata &                  $<$  7.3 &                  $<$ 30.2 &                  $<$ 14.6 &                  $<$ 14.7 &                  $<$ 17.7 \\
        MRC~0211-256 &                  $<$  9.5 &                   \nodata &   [ 17.4$\pm$  6.6( 2.6)] & {\it 25.0$\pm$  3.8( 7.3)}&  {\it 25.9$\pm$ 5.8( 4.7)}& {\it [15.7$\pm$5.9( 2.7)]}\\
        TXS~0211-122 &                   \nodata &  {\it [7.4$\pm$3.4( 2.2)]}& {\it [11.7$\pm$  5.9( 2.0)]}&                  $<$ 15.9 &                  $<$ 19.2 &                  $<$ 24.5 \\
               3C~65 &                  $<$ 10.5 &                   \nodata &                  $<$ 16.1 &                  $<$ 15.9 &                  $<$ 17.7 &                  $<$ 21.9 \\
        MRC~0251-273 &                   \nodata &                  $<$ 10.0 &                  $<$ 18.7 &                  $<$ 15.7 &                  $<$ 14.1 &                  $<$ 19.5 \\
        MRC~0316-257 &                   \nodata &                  $<$ 11.1 &   [ 17.7$\pm$  7.2( 2.5)] &  {\it 22.7$\pm$ 5.1( 4.7)}& {\it 20.2$\pm$  5.4( 3.9)}&                  $<$ 19.3 \\
        MRC~0324-228 &                  $<$  9.1 &                   \nodata & {\it27.9$\pm$  5.4( 5.4)} & {\it61.8$\pm$  6.7(12.1)} & {\it 35.5$\pm$  5.9( 6.7)}& {\it[ 17.5$\pm$7.4( 2.4)]}\\
        MRC~0350-279 &                  $<$ 11.3 &                   \nodata &                  $<$ 25.5 &                  $<$ 14.2 &                  $<$ 14.0 &                  $<$ 15.9 \\
        MRC~0406-244 &                   \nodata &                  $<$ 12.3 & {\it[ 21.5$\pm$7.9( 2.7)]}& {\it47.6$\pm$  5.6(10.6)} & {\it 38.7$\pm$  5.3( 8.4)}& {\it 22.8$\pm$  5.9( 4.0)}\\
            4C~60.07 &                  $<$  4.9 &                   \nodata &                  $<$ 29.1 & {\it 46.4$\pm$  6.5( 8.2)}& {\it 49.5$\pm$  8.4( 6.5)}& {\it 48.0$\pm$  8.3( 6.3)}\\
        PKS~0529-549 &                   \nodata & {\it [8.3$\pm$ 4.0( 2.1)]}&      31.9$\pm$  9.0( 3.6) & {\it 35.1$\pm$  7.3( 5.1)}& {\it 43.8$\pm$  8.3( 5.6)}& {\it 40.0$\pm$  8.9( 4.8)}\\
       WN~J0617+5012 &                   \nodata &                  $<$  7.9 &                  $<$ 23.2 &                  $<$ 19.3 &                  $<$ 21.4 &                  $<$ 22.4 \\
            4C~41.17 &                  $<$  4.2 &                   \nodata &   [ 17.9$\pm$  6.8( 2.6)] & {\it 28.2$\pm$  4.8( 6.5)}& {\it 35.7$\pm$  5.8( 6.8)}& {\it 31.1$\pm$  6.1( 5.5)}\\
       WN~J0747+3654 &                   \nodata &                  $<$  8.8 &                  $<$ 26.7 &                  $<$ 14.9 &                  $<$ 18.1 &                  $<$ 17.1 \\
       6CE~0820+3642 &                  $<$ 11.3 &                   \nodata &                  $<$ 22.5 & {\it 32.2$\pm$  5.1( 7.0)}& {\it [11.3$\pm$4.7( 2.5)]}&                  $<$ 15.4 \\
            5C~7.269 &                   \nodata &                  $<$  7.8 &                  $<$ 25.0 &                  $<$ 13.4 &                  $<$ 18.0 &                  $<$ 14.5 \\
        USS~0828+193 &                   \nodata &      18.5$\pm$  3.5( 5.4) &   [ 24.0$\pm$  9.6( 2.5)] & {\it 20.2$\pm$  4.5( 4.7)}& {\it 17.5$\pm$  4.7( 3.8)}&                  $<$ 17.2 \\
       6CE~0901+3551 &                  $<$  8.6 &                   \nodata &                  $<$ 26.8 &                  $<$ 13.8 &                  $<$ 16.2 &                  $<$ 19.3 \\
          B2~0902+34 &                   \nodata &                  $<$  9.7 &                  $<$ 20.9 & {\it [12.4$\pm$4.6( 2.7)]}&                  $<$ 15.3 &                  $<$ 15.4 \\
       6CE~0905+3955 &      34.2$\pm$  2.8(15.4) &                   \nodata &      59.8$\pm$ 11.2( 5.5) & {\it 38.8$\pm$  4.9( 9.5)}& {\it 30.9$\pm$  5.4( 6.2)}&                  $<$ 16.0 \\
       TN~J0924-2201 &                  $<$  4.6 &                   \nodata &                  $<$ 16.3 &                  $<$ 11.4 &                  $<$ 16.1 &                  $<$ 14.3 \\
         6C~0930+389 &                   \nodata &                  $<$  9.4 &                  $<$ 18.6 &                  $<$ 16.7 &                  $<$ 15.5 &                  $<$ 16.8 \\
        USS~0943-242 &                   \nodata &                  $<$ 27.6 &      23.6$\pm$  7.7( 3.1) & {\it 25.7$\pm$  5.2( 5.2)}& {\it 31.7$\pm$  5.5( 6.3)}& {\it 35.2$\pm$  7.3( 5.1)}\\
              3C~239 &                  $<$ 12.7 &                   \nodata &                  $<$ 33.3 &                  $<$ 15.5 &                  $<$ 15.2 &                  $<$ 18.4 \\
        MG~1019+0534 &                   \nodata &      15.4$\pm$  2.9( 5.5) &      23.5$\pm$  5.8( 4.2) & {\it 28.6$\pm$  5.4( 5.7)}& {\it 29.9$\pm$  5.3( 6.2)}& {\it 33.2$\pm$  5.3( 7.1)}\\
        MRC~1017-220 &                  $<$  7.7 &                   \nodata &                  $<$ 25.1 &                  $<$ 17.4 &                  $<$ 23.6 &                  $<$ 22.4 \\
       WN~J1115+5016 &                   \nodata &                  $<$  9.3 &                  $<$ 20.7 &                  $<$ 17.4 &                  $<$ 18.7 &                  $<$ 21.4 \\
              3C~257 &       7.7$\pm$  1.5( 5.4) &                   \nodata &   [ 14.6$\pm$  6.1( 2.4)] & {\it 29.8$\pm$  4.8( 6.9)}& {\it 25.6$\pm$  4.6( 6.1)}& {\it 17.4$\pm$  5.3( 3.3)}\\
       WN~J1123+3141 &                   \nodata &   [ 15.6$\pm$  6.2( 2.5)] &                  $<$ 27.2 & {\it 21.0$\pm$  4.6( 4.9)}& {\it 15.8$\pm$  4.6( 3.5)}&                  $<$ 19.1 \\
        PKS~1138-262 &                   \nodata &      25.2$\pm$  2.2(13.9) &      40.2$\pm$ 10.2( 4.0) & {\it 40.4$\pm$  5.9( 7.8)}& {\it 33.0$\pm$  6.1( 5.8)}& {\it 28.9$\pm$  6.7( 4.5)}\\
              3C~266 &                  $<$  9.4 &                   \nodata &      28.7$\pm$  7.6( 3.8) & {\it 14.7$\pm$  4.3( 3.5)}&                  $<$ 17.8 &                  $<$ 20.2 \\
          6C~1232+39 &                   \nodata &                  $<$  6.7 &                  $<$ 26.4 &                  $<$ 13.3 &                  $<$ 14.0 &                  $<$ 17.1 \\
        USS~1243+036 &                   \nodata &   [  8.3$\pm$  3.1( 2.7)] &   [ 14.1$\pm$  6.3( 2.2)] & {\it 16.6$\pm$  5.6( 3.0)}&                  $<$ 19.6 &                  $<$ 22.2 \\
       TN~J1338-1942 &                   \nodata &                  $<$  5.9 &                  $<$ 27.1 &                  $<$ 16.6 &                  $<$ 17.5 &                  $<$ 18.0 \\
            4C~24.28 &                   \nodata &      14.2$\pm$  3.3( 4.4) &      23.3$\pm$  7.9( 3.0) &                  $<$ 15.9 &                  $<$ 14.6 &                  $<$ 14.6 \\
              3C~294 &                  $<$  7.1 &                   \nodata &                  $<$ 27.0 &                  $<$ 15.6 &                  $<$ 17.9 &                  $<$ 20.4 \\
        USS~1410-001 &                   \nodata &                  $<$ 10.1 &                  $<$ 19.2 &                  $<$ 15.0 &                  $<$ 17.3 &                  $<$ 21.8 \\
         8C~1435+635 &                  $<$  5.2 &                   \nodata &                  $<$ 16.2 &                  $<$ 10.4 &                  $<$ 11.4 &                  $<$ 13.5 \\
        USS~1558-003 &                   \nodata &                  $<$  7.4 &                  $<$ 22.4 &                  $<$ 16.4 &                  $<$ 18.8 &                  $<$ 21.0 \\
        USS~1707+105 &                   \nodata &                  $<$  8.6 &                  $<$ 27.7 &                  $<$ 16.8 &                  $<$ 14.9 &                  $<$ 19.5 \\
         LBDS~53W002 &                   \nodata &                  $<$  9.4 &                  $<$ 24.9 &                  $<$ 15.0 &                  $<$ 14.3 &                  $<$ 16.5 \\
         LBDS~53W069 &                  $<$  7.9 &                   \nodata &                  $<$ 24.3 &                  $<$ 14.8 &                  $<$ 12.8 &                  $<$ 17.2 \\
         LBDS~53W091 &                  $<$ 10.5 &                   \nodata &                  $<$ 36.7 &                  $<$ 14.7 &                  $<$ 15.2 &                  $<$ 17.9 \\
            3C~356.0 &                  $<$ 11.8 &                   \nodata &                  $<$ 23.8 &                  $<$ 14.1 &                  $<$ 14.3 &                  $<$ 15.6 \\
        7C~1751+6809 &                  $<$  9.9 &                   \nodata &                  $<$ 26.1 &                  $<$ 15.2 &                  $<$ 13.9 &                  $<$ 15.4 \\
        7C~1756+6520 &                  $<$  8.2 &                   \nodata &                  $<$ 29.8 &                  $<$ 14.6 &                  $<$ 19.3 &                  $<$ 19.5 \\
              3C~368 &      32.3$\pm$  3.5(10.4) &                   \nodata &      52.9$\pm$  9.1( 6.1) &  {\it 35.6$\pm$ 6.2( 6.3)}&  {\it 19.6$\pm$ 6.5( 3.1)}&                  $<$ 19.0 \\
        7C~1805+6332 &                  $<$  7.4 &                   \nodata &                  $<$ 28.4 &                  $<$ 14.6 &                  $<$ 17.6 &                  $<$ 18.3 \\
            4C~40.36 &                   \nodata &                  $<$  8.6 &                  $<$ 29.3 &                  $<$ 16.0 &                  $<$ 18.5 &                  $<$ 15.6 \\
      TXS~J1908+7220 &                   \nodata &      19.4$\pm$  3.1( 6.5) &      36.7$\pm$  8.1( 4.7) & {\it 38.9$\pm$  8.0( 5.2)}& {\it 52.9$\pm$  7.7( 7.9)}& {\it 49.5$\pm$  8.6( 6.3)}\\
       WN~J1911+6342 &                   \nodata &       9.2$\pm$  3.0( 3.1) &                  $<$ 20.4 &                  $<$  9.4 &                  $<$ 15.2 &                  $<$ 11.7 \\
       TN~J2007-1316 &                   \nodata &                  $<$  7.7 &                  $<$ 20.8 & {\it 16.7$\pm$  5.2( 3.3)}& {\it 16.8$\pm$  5.1( 3.4)}&                  $<$ 18.9 \\
        MRC~2025-218 &                   \nodata &                  $<$ 11.1 &                  $<$ 30.7 &                  $<$ 18.5 &                  $<$ 28.4 &                  $<$ 19.5 \\
        MRC~2048-272 &                   \nodata &                  $<$  8.5 &                  $<$ 25.3 &                  $<$ 15.3 &                  $<$ 16.0 &                  $<$ 18.6 \\
        MRC~2104-242 &                   \nodata &      14.4$\pm$  3.5( 4.3) &   [ 22.0$\pm$  8.4( 2.6)] & {\it [14.2$\pm$5.1( 2.8)]}&  {\it 21.1$\pm$ 6.6( 3.3)}&                  $<$ 15.8 \\
            4C~23.56 &                   \nodata &      29.2$\pm$  3.2(10.3) &   [ 17.2$\pm$  6.8( 2.6)] &                  $<$ 18.5 &                  $<$ 24.2 &                  $<$ 25.9 \\
        MG~2144+1928 &                   \nodata &                  $<$ 46.1 &                  $<$ 40.4 &                  $<$ 14.8 &                  $<$ 18.1 &                  $<$ 17.5 \\
        USS~2202+128 &                   \nodata &   [  7.2$\pm$  2.9( 2.5)] &      18.2$\pm$  5.7( 3.2) &                  $<$ 13.7 &                  $<$ 12.6 &                  $<$ 17.1 \\
        MRC~2224-273 &                  $<$ 10.5 &                   \nodata &                  $<$ 28.2 & {\it 14.6$\pm$  4.8( 3.1)}&                  $<$ 17.7 &                  $<$ 18.3 \\
       B3~J2330+3927 &                   \nodata &       9.5$\pm$  2.6( 3.8) &                  $<$ 32.2 & {\it 39.0$\pm$  5.8( 7.7)}& {\it 48.0$\pm$  6.4( 8.8)}& {\it 50.3$\pm$  7.5( 7.6)}\\
            4C~28.58 &                   \nodata &      22.8$\pm$  2.9( 8.6) &   [ 23.6$\pm$  8.7( 2.7)] & {\it 42.5$\pm$  4.9(11.0)}& {\it 29.7$\pm$  5.4( 6.0)}& {\it 15.5$\pm$  4.5( 3.5)}\\
              3C~470 &      21.6$\pm$  3.7( 6.0) &                   \nodata &      22.5$\pm$  7.2( 3.2) & {\it 39.2$\pm$  5.7( 7.9)}& {\it 33.5$\pm$  6.0( 6.1)}& {\it 24.9$\pm$  6.3( 4.1)}\\

\hline
\end{tabular}
\end{table*}

\subsection{Sub-mm data, LABOCA}
\label{sec:laboca}

In addition to the fluxes already available in the literature
\citep{Archibald2001,Reuland2003}, we obtained new sub-mm data for some
of those sources lacking it\footnote{Based on observations made with APEX
telescope obtained during ESO, Chile and 
Sweden time under programme IDs E-090.A-0730. APEX is a collaboration 
between the Max-Planck-Institut f\"ur Radioastronomie, the European Southern Observatory, and the Onsala
Space Observatory.}. We observed 18 sources in the southern hemisphere
spanning 1$<$$z$$<$3. The observations were
done in service mode between 2012 July and December, with precipitable
water vapour generally below 1\,mm. To save observing time, most sources
were observed using the LABOCA wobbler on-off (WOO) photometry mode. As
this WOO mode does not provide any spatial information, it should
only be used on isolated sources. If the \herschel\ maps showed either
multiple point-like sources within a radius of 20\arcsec\ (one LABOCA beam size),
or a spatial offset more than 5\arcsec\ from the radio core position,
we used the raster spiral mapping mode instead. The integration times
per source were adapted to obtain an approximately uniform rms for
all 18 sources. To reduce the data, we used the doOO script inside BoA
\citep{Schuller2012} for the WOO data, and the reduction macro in CRUSH2
\citep{Kovacs2008} for the mapping data. Table \ref{tab:submm_flux}
summarises the observing modes, fluxes, uncertainties and the references
for data taken from the literature.

\begin{table*}[ht] \centering
\caption{Sub-mm flux at 870\mum\ with LABOCA in this paper, at
  850\mum\ for all the rest. Flux in square brackets symbolise the weak
  detection (\weakdetect). For a description of the observing mode, {\it WOO}
  or {\it mapping} see S~\ref{sec:laboca}.}
\label{tab:submm_flux}
\begin{tabular}{l ccc}
\hline
Name & Flux [mJy] & Obs. mode & References \\
\hline \hline
    6C~0032+412  &  [2.6$\pm$1.2]  &  -  & \citet{Archibald2001}  \\
   MRC~0037-258  &  $<$12.9        & WOO  & this paper  \\
   6CE~0058+495  &  \nodata  &  -  &  - \\
   MRC~0114-211  &  $<$16.8   &  mapping  & this paper  \\
  TN~J0121+1320  &  7.5$\pm$1.0  &  -  & \citet{Reuland2004}    \\
   6CE~0132+330  &  \nodata  & -   & -  \\
    6C~0140+326  &  [3.3$\pm$1.5]  &  -  & \citet{Archibald2001}  \\
   MRC~0152-209  &  14.5$\pm$3.3  & WOO   & this paper  \\
   MRC~0156-252  &  $<$21.0  & mapping   & this paper  \\
  TN~J0205+2242  &  $<$5.2  &  -  & \citet{Reuland2004}    \\
   MRC~0211-256  &  $<$26.1  & mapping   & this paper  \\
   TXS~0211-122  &  $<$24.6  & mapping   & this paper  \\
          3C~65  &  \nodata  &  -  & -  \\
   MRC~0251-273  &  $<$8.9  &  -  & \citet{Reuland2004}    \\
   MRC~0316-257  &  $<$8.8  &  -  & \citet{Reuland2004}    \\
   MRC~0324-228  &  $<$9.0  & WOO   & this paper  \\
   MRC~0350-279  &  $<$23.1  & mapping  & this paper  \\
   MRC~0406-244  &  $<$17.8  & WOO   & this paper  \\
       4C~60.07  &  17.1$\pm$1.3  & -   & \citet{Archibald2001}  \\
   PKS~0529-549  &  \nodata  & -   &  - \\
  WN~J0617+5012  &  $<$3.2  &  -  & \citet{Reuland2004}    \\
       4C~41.17  &  12.1$\pm$0.9  &  -  & \citet{Archibald2001}  \\
  WN~J0747+3654  &  4.8$\pm$1.1  &  -  & \citet{Reuland2004}    \\
  6CE~0820+3642  &  [2.1$\pm$1.0]  & -   & \citet{Archibald2001}  \\
       5C~7.269  &  $<$4.7  &  -  & \citet{Archibald2001}  \\
   USS~0828+193  &  \nodata  & -   & -  \\
  6CE~0901+3551  &  $<$3.45  &  -  & \citet{Archibald2001}  \\
     B2~0902+34  &  [2.8$\pm$1.0]  &  -  & \citet{Archibald2001}  \\
  6CE~0905+3955  &  3.6$\pm$0.9  &  -  & \citet{Archibald2001}  \\
  TN~J0924-2201  &  $<$3.2  &  -  & \citet{Reuland2004}    \\
    6C~0930+389  &  $<$3.4  &  -  & \citet{Archibald2001}  \\
   USS~0943-242  &  $<$24.0  &  mapping  & this paper  \\
         3C~239  &  $<$3.8  &  -  & \citet{Archibald2001}  \\
   MG~1019+0534  &  [2.4$\pm$0.9]  & -   & \citet{Archibald2001}  \\
   MRC~1017-220  &  $<$18.6  &  WOO  & this paper  \\
  WN~J1115+5016  &  [3.0$\pm$1.3]  & -   & \citet{Reuland2004}    \\
         3C~257  &  5.4$\pm$1.0  &  -  & \citet{Archibald2001}  \\
  WN~J1123+3141  &  4.9$\pm$1.2  &  -  & \citet{Reuland2004}    \\
   PKS~1138-262  &  12.8$\pm$3.3  & -   & \citet{Reuland2004}    \\
         3C~266  &  $<$4.4  & -   & \citet{Archibald2001}  \\
     6C~1232+39  &  3.9$\pm$0.7  & -   & \citet{Archibald2001}  \\
   USS~1243+036  &  [2.3$\pm$1.1]  & -   & \citet{Archibald2001}  \\
  TN~J1338-1942  &  6.9$\pm$1.1  &  -  & \citet{Reuland2004}    \\
       4C~24.28  &  [2.6$\pm$1.2]  & -   & \citet{Archibald2001}  \\
         3C~294  &  $<$2.5  &  -  & \citet{Archibald2001}  \\
   USS~1410-001  &  $<$10.8  &  WOO  & this paper  \\
    8C~1435+635  &  7.8$\pm$0.8  &  -  & \citet{Archibald2001} \\
   USS~1558-003  &  $<$9.6  &  WOO  & this paper  \\
   USS~1707+105  &  $<$9.3  &  WOO  & this paper  \\
    LBDS~53W002  &  $<$4.3  & -   & \citet{Archibald2001}  \\
    LBDS~53W069  &  $<$3.1  &  -  & \citet{Archibald2001}  \\
    LBDS~53W091  &  \nodata  & -   & -  \\
       3C~356.0  &  $<$4.8  &  -  & \citet{Archibald2001}  \\
   7C~1751+6809  &  \nodata  & -   &  - \\
   7C~1756+6520  &  \nodata  & -   & -  \\
         3C~368  &  4.1$\pm$1.1  &  -  & \citet{Archibald2001}  \\
   7C~1805+6332  &  \nodata  & -   & -  \\
       4C~40.36  &  $<$3.9  &  -  & \citet{Archibald2001}  \\
 TXS~J1908+7220  &  10.8$\pm$1.2  &   - & \citet{Reuland2004}    \\
  WN~J1911+6342  &  $<$11.9  &  -  & \citet{Reuland2004}    \\
  TN~J2007-1316  &  5.8$\pm$1.5  &  -  & \citet{Reuland2004}    \\
   MRC~2025-218  &  $<$10.5  &  WOO  & this paper  \\
   MRC~2048-272  &  $<$21.0  &  mapping  & this paper  \\
   MRC~2104-242  &  \nodata  & -   & -  \\
       4C~23.56  &  $<$4.7  &  -  & \citet{Archibald2001}  \\
   MG~2144+1928  &  [2.3$\pm$0.9]  &  -  & \citet{Reuland2004}    \\
   USS~2202+128  &  $<$11.1  &  mapping  & this paper  \\
   MRC~2224-273  &  $<$12.3  &  mapping  & this paper  \\
  B3~J2330+3927  &  14.1$\pm$1.7   &  -  & \citet{Reuland2004}    \\
       4C~28.58  &  3.9$\pm$1.2   &   - & \citet{Archibald2001}  \\
         3C~470  &  5.6$\pm$1.1   & -   & \citet{Archibald2001}  \\
\hline
\end{tabular}
\end{table*}

\section{The \herge\ Infrared Spectral Energy Distributions}

Combining the \spitzer, \herschel\ and sub-mm data, we continuously cover 
the wavelength range 16--870\mum. The panels of Fig.~\ref{fig:sed} show
the resulting SEDs for our 70 radio galaxies. As our focus is on the warm
to cold dust emission, we do not use \spitzer\ IRAC photometry in our
SEDs because those data are generally dominated by stellar photospheric
emission
\citep[e.g.][]{Seymour2007}. \cite{DeBreuck2010} show that hot dust
emission can also contribute significantly to the IRAC fluxes of
some sources. This hot dust component ($>$500\,K), however, only represents a small
fraction of energy of the total IR SED ($<$5\%), and is influenced by
orientation-dependent effects \citep{Drouart2012}. We therefore do
not include this hot dust contribution in our SED fitting. Moreover,
we add 20\% uncertainties to the MIPS data to account for cross-calibration
uncertainties between \spitzer\ and \herschel. We overplot the \spitzer\ 
spectrum available for a sub-sample of our sources
\citep{Seymour2008,Rawlings2013}.  These spectra are not used to constrain
our fits, but provide a consistency check on our decomposition of the
IR/submm SEDs.

\subsection{Total IR luminosities}
\label{sec:tot_lir}

Since our IR/submm SEDs are well-sampled, we can estimate robustly the
total IR luminosity (\lirtot). We use its most common definition, integrating the flux density in the 8-1000\mum\ restframe range. 
To interpolate between our photometric data points, we assume the models
described in \S~\ref{sec:decompir}. From these estimates of the total IR luminosity,
it appears that almost all radio galaxies from our sample are ultra-luminous infrared galaxies
(ULIRG; \lirobs$>$10$^{12}$L$_{\odot}$).

\begin{figure}[t] \centering
  \includegraphics[width=0.5\textwidth]{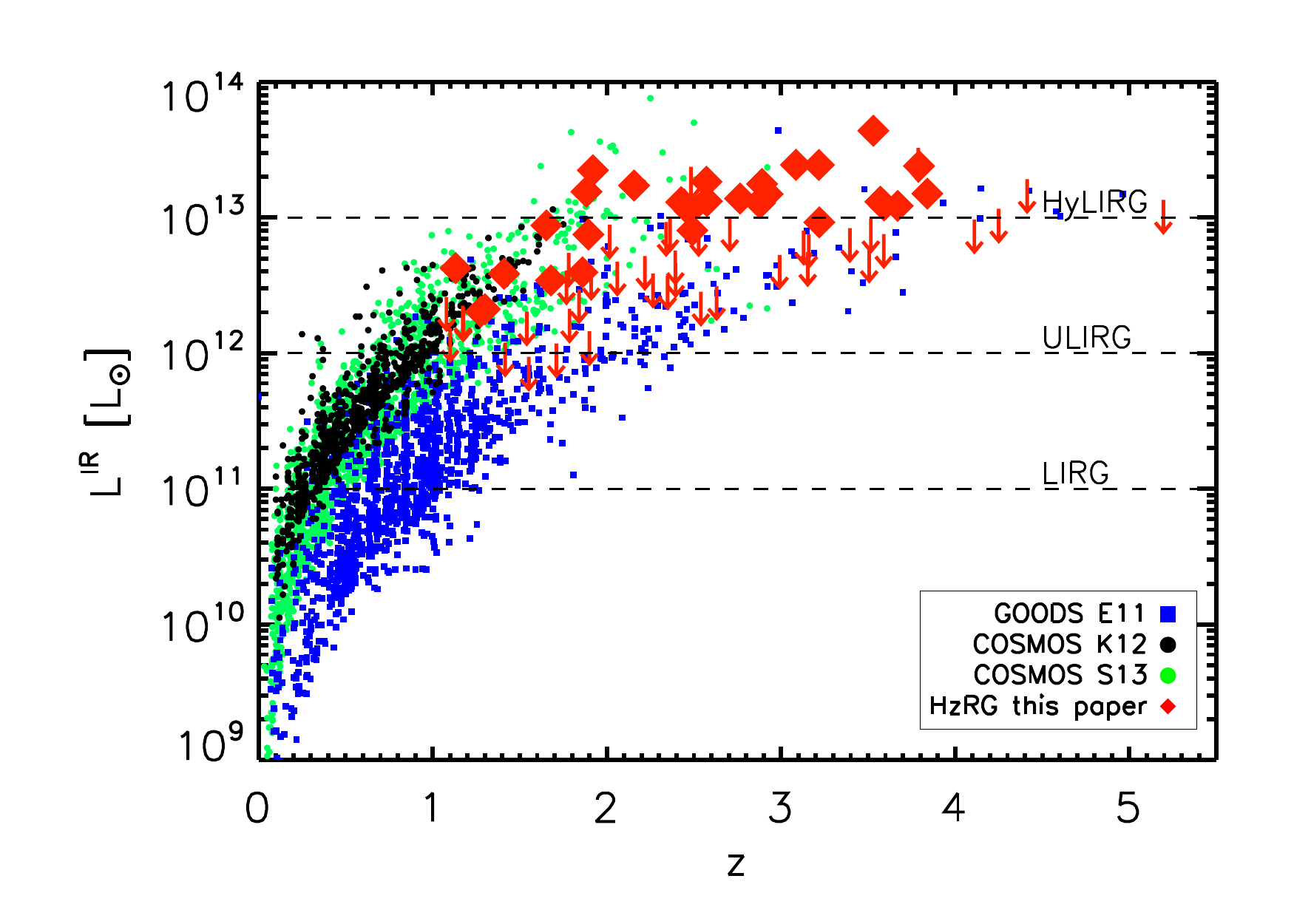}
  \caption{Total IR luminosity (\lirtot) versus redshift. The green dots are the 
    COSMOS sample from \citet[][]{Kartaltepe2010}, using \spitzer\ data. 
The blue squares are the GOODS samples from \citet{Elbaz2011}. 
    The black dots are the selection from \citet{Symeonidis2013}.     
    We also indicate the LIRG, ULIRG and HyLIRG limits.}
  \label{fig:z_vs_lir}
\end{figure}

Fig.~\ref{fig:z_vs_lir} plots our sample along with other samples available in 
literature with \herschel\ observations. Our galaxies are among the 
brightest emitters in the IR in their redshift range. About half of our $z$$>$2 
sample belongs to the HyLIRG regime (\lirtot$>$10$^{13}$L$_{\odot}$). From this diagram, 
it is interesting to note that the high redshift radio galaxies are indistinguishable from the most extreme
IR emitters. 

Interestingly, such luminous objects imply strong AGN and/or star formation activities. 
This should be compared with previous results about the mass of these objects, located at the high-mass 
end of the galaxy mass distribution \citep[e.g.][]{Rocca2004,Seymour2007}. Even if these galaxies 
are identified as the progenitors of the ``red-and-dead" local ellipticals \citep[e.g.][]{Matthews1964,Rocca2004,Thomas2005,Labbe2005}, 
they appear to be pretty active in the past.

\subsection{The warm and cold dust contributions}
\label{sec:class}

While both AGN and SB can heat dust, their input SEDs are significantly different. 
AGN heating tends to contribute at shorter wavelengths
($\sim$10\mum, T$_{\rm dust}^{\rm AGN}$$\sim$300K) while star formation heating 
tends to dominate the emission at longer wavelengths ($\sim$100\mum, T$^{\rm SB}_{\rm dust}$$\sim$30K). 
Given the large variety of the data quality, we want to define a set of criteria to disentangle  the AGN and SB contributions. We therefore classify our 
galaxies into classes depending on the number of detections on either side of 50\mum\ 
restframe \citep[e.g.][]{Leipski2013}. This value is preferred for several reasons. First, in 
the case of an object with both contributions (AGN and SB), the change in regime is expected 
to occur around this wavelength. Second, our sample spans a large range in redshift 
(1$<$$z$$<$5) and therefore a simple colour selection would be severely affected by the $k$-correction.
Third, this wavelength equally splits the number of channels available for each source 
(four bands on either side). We note  that by changing this limit to 30\mum\ or 70\mum\  
only changes the fraction of sources in each class by a small amount ($<$5\%). 

The classes are defined as follows, with their
respective fractions in our sample:
\begin{enumerate}
\item {\bf Warm and cold dust (WCD, 45\%)}: corresponds to detections on both sides of $\lambda_{\rm rest}$$=$50\mum.
\item {\bf Warm dust (WD, 33\%)}: corresponds to detections only in the mid-IR ($\lambda_{\rm rest}$$<$50\mum).
\item {\bf Cold dust (CD, 11\%)}: corresponds to detections only in the far-IR ($\lambda_{\rm rest}$$>$50\mum).
\item {\bf Upper Limit (UL, 11\%)}: corresponds to no detections in either the mid-IR or the far-IR. 
\end{enumerate}

We detect warm, \textit{preferentially} AGN-heated dust emission in the
majority (78\%) of our sample, while the cooler, \textit{preferentially} starburst-heated dust
emission is detected in half (54\%) of our sample (Table
\ref{tab:lir_list}). This difference between the possible constraints on
the two components can be interpreted in two ways: either our \herschel\
(and in particular SPIRE) data are comparatively less sensitive than
\spitzer, or the AGN contributes more significantly to the IR SED while
the strength of the associated SB varies by a larger amount. We note
that only 11\% of our sample does not have any constraints on the relative
contributions of either AGN or starbursts to the IR SED.

To further examine correlations involving IR luminosities
in our sample, we next separate the AGN and SB
components. 


\section{IR SED Decomposition Method}
\label{sec:decompir}

In order to decompose the two main contributions to the IR SED, AGN and
SB emission, we need models for each component. The AGN
dust emission, which contributes mainly in mid-IR emission, comes mainly
from the far-UV through optical light that has been reprocessed by dust
in close proximity to the AGN.  The far-UV through optical emission from
any young stellar population that may exist is largely reprocessed into the far-IR
via dust grains.

One of the most important goals of this analysis is to
determine the relative emission from the AGN and starburst components.
Disentangling this relative emission allows us to investigate the principal physical processes responsible
for the luminous IR emission in distant radio galaxies, since the
dust reprocessed emission is the largest contributor in active galaxies. 
Such analysis provides the best measure of the bolometric luminosity. 


We use the SED fitting procedure DecompIR \citep[][\url{https://sites.google.com/site/decompir/home}]{Mullaney2011}, with
some minor modifications to add the information and constraints provided
by the \herschel\ and sub-mm data. Briefly, DecompIR allows the fitting of one or two templates
thanks to $\chi^2$ minimisation. It considers an empirical library estimated from local starburst
and an empirical unique AGN template consisting of a composite spectrum of broken power-laws and a black body. 
All templates cover the 3-1000\mum\ restframe range\footnote{Due to $k$-correction effects, a part
of the IRS filter falls outside the template for $z$$>$2.5. We therefore
extrapolate the templates to $\lambda$=2\mum\ using a power-law function. This modification does
not impact our results as the energy contribution from these wavelengths
(2\mum$<\lambda<$3\mum) is negligible compared to the total IR luminosity.}, 
and an extinction can be applied independently for each component. This 
procedure has been extensively tested on higher redshift sources and described in \cite{DelMoro2013}.
In order to keep our approach as homogeneous as possible over the whole sample, we minimize the number
of free parameters. We remind the reader that we did not include the IRAC 
data because it contains a significant component of stellar photospheric emission.

\subsection{Additional starburst template}
\label{sec:sb_template}

DecompIR includes five different SB templates. Briefly, they 
represent SB with different peaking temperatures and PAH strength, with the coldest corresponding to SB1.
We refer to \citet[][ their Fig. 4]{Mullaney2011} for presentation of the SEDs. 
For two galaxies ({\it 4C~41.17} and {\it 4C~28.58}), these five available
templates do not converge to an acceptable solution.  The best
fitting SED suggests that either a hotter starburst component or a colder
AGN contribution (Figure \ref{fig:sed}) is required to reproduce the observed SED. However, {\it
4C~41.17} is well fitted by a synthetic SED from the galaxy synthesis
and evolution code, PEGASE.3 \citep{Rocca2013}. Fortuitously,
this galaxy appears to have a relatively small AGN contribution
\citep{Dey1997}. \cite{Rocca2013} show that the IR part of the SED
is clearly dominated by a young stellar population. We have therefore
included the IR part of the best fitting SED of {\it 4C~41.17} from PEGASE.3  as a new
template to the DecompIR library (the ``SB6'' template). This template
has the highest relative dust temperature of any of the SB templates in the
library -- its dust emission peaks at $\sim$60\mum($\sim$50K). This template
does not represent a local SB as the other templates, it is a solution for a 30~Myr old 
starburst \citep{Rocca2013}. However, we seek here only to reproduce the general shape of the IR
SED to estimate IR luminosities, and will not make any further considerations about the age/mass of this template. A
further analysis of the SB/host properties similar to the approach \citealp{Rocca2013} 
is the object of a forthcoming paper (Drouart et al., in prep).

\subsection{AGN template}
\label{sec:agn_template}

The AGN template used in this analysis is calculated using a sample
of AGN that has had the starburst contribution removed from their
mid-IR spectral energy distribution.  The template is an average of the
residual mid-IR SEDs \citep[see ][for details]{Mullaney2011}. Due to the 
empirical nature of this subtraction and the variety of
possible AGN-dominated mid-IR SEDs, this average template is expected
to show discrepancies from object to object, however it represents well the average
 AGN spectrum in mid-IR \citep{Dale2014}. Ideally, we would like to use 
different AGN templates, similar to the SB analysis. In particular, the hottest part of the 
AGN is subject to inclination-dependent effects \citep[e.g.][]{Leipski2010,Drouart2012}. 

In order to test this, we modified DecompIR to include the average type 1 
AGN template from \cite{Richards2006} onto which we apply an extinction from 
 \cite{Fitzpatrick1999}. Even if this AGN template significantly improves our fitting, we 
 decided to discard this template from our library for the following reasons. First, the 
 inclusion of an extra parameter (the extinction) decreases the number of sources 
 on which our fitting can be applied. Second, the Richards template presents an odd, 
 and unrealistic tail in the far-IR (probably due to the poor far-IR coverage of the dataset 
 used to build this template). Third, while increasing the scatter in \liragn, we find no 
 drastically different results from using the built-in AGN template. Finally, being an average
 template, star formation can still contribute at long wavelength and we would therefore 
 overestimate the AGN luminosity.

\subsection{Transition regime: hot starburst or cold AGN ?}

The transition between the two components is the most critical parameter as it 
has an important influence on the calculated IR luminosities. 
On one hand, we want to be sure that the templates we are using are effectively 
representative of the AGN and/or the SB components. On the other hand, 
we want to keep the decomposition as simple as possible to apply it over the entire 
sample. As previously mentioned, we use only one AGN template, deemed 
representative of the general AGN SED. How can we be certain that this only empirical 
template is representative for all our sources? This question is difficult to answer as 
the data quality varies from object to object and we only have broad band photometry 
for our sample. Nevertheless, one can argue that this template is valid considering the 
following assumptions. The cold dust emission for the AGN ($\lambda_{\rm rest}$$>$30\mum) 
can come from (i) an extended torus \citep[e.g.][]{Fritz2006,Nenkova2008a}, or (ii) reprocessed 
light from the NLRs \citep[e.g.][]{Dicken2009,Dicken2010}. While (i) will require the inclusion 
of a large number of new free parameters, (ii) can only be assessed by the [OIII] luminosities 
that are not available for our entire sample. The quantification of these effects is beyond the 
scope of this paper and will therefore be ignored for the remainder of the analysis. Moreover, 
one notes that the empirical AGN template does present a significant contribution at long wavelength. 
It is also interesting to note that the extreme object \textit{4C~23.56} represents the prototype 
of a pure AGN contribution, and is remarkably well reproduced by the built-in template. Part 
of our sample (7 objects) have IRS spectra (12-24\mum) available \citep{Seymour2008,Rawlings2013}. 
Overplotting the spectrum \textit{after} the fitting shows a good agreement between the IRS 
and the results. Finally, the overall results do not seem to bias our results towards one or 
another SB templates (see Table \ref{tab:lir_list}).

\begin{figure*}[th] \centering
\begin{minipage}[c]{0.46\linewidth}
  \includegraphics[width=1.1\textwidth]{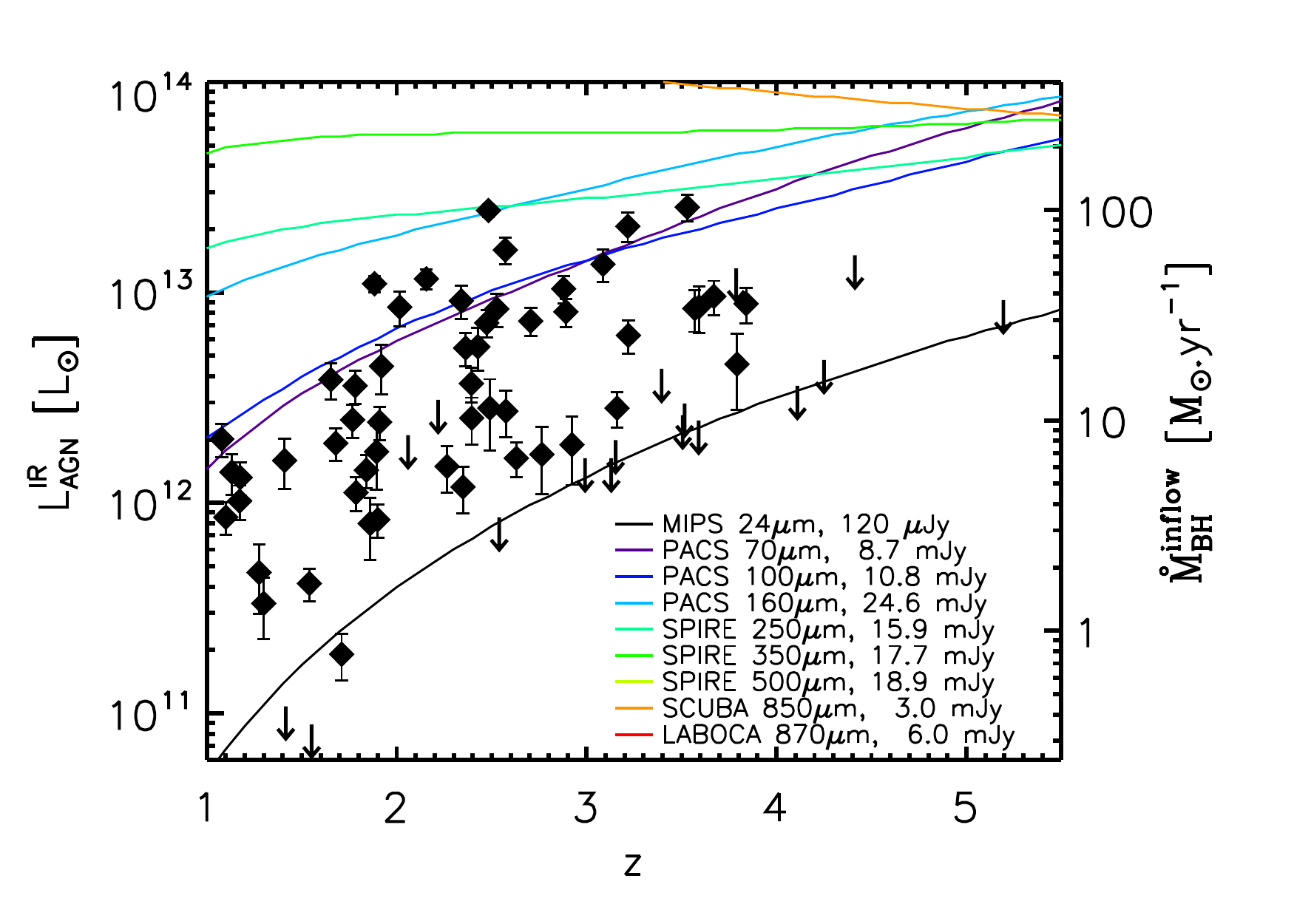}
\end{minipage} \hfill
\begin{minipage}[c]{0.46\linewidth}
  \includegraphics[width=1.1\textwidth]{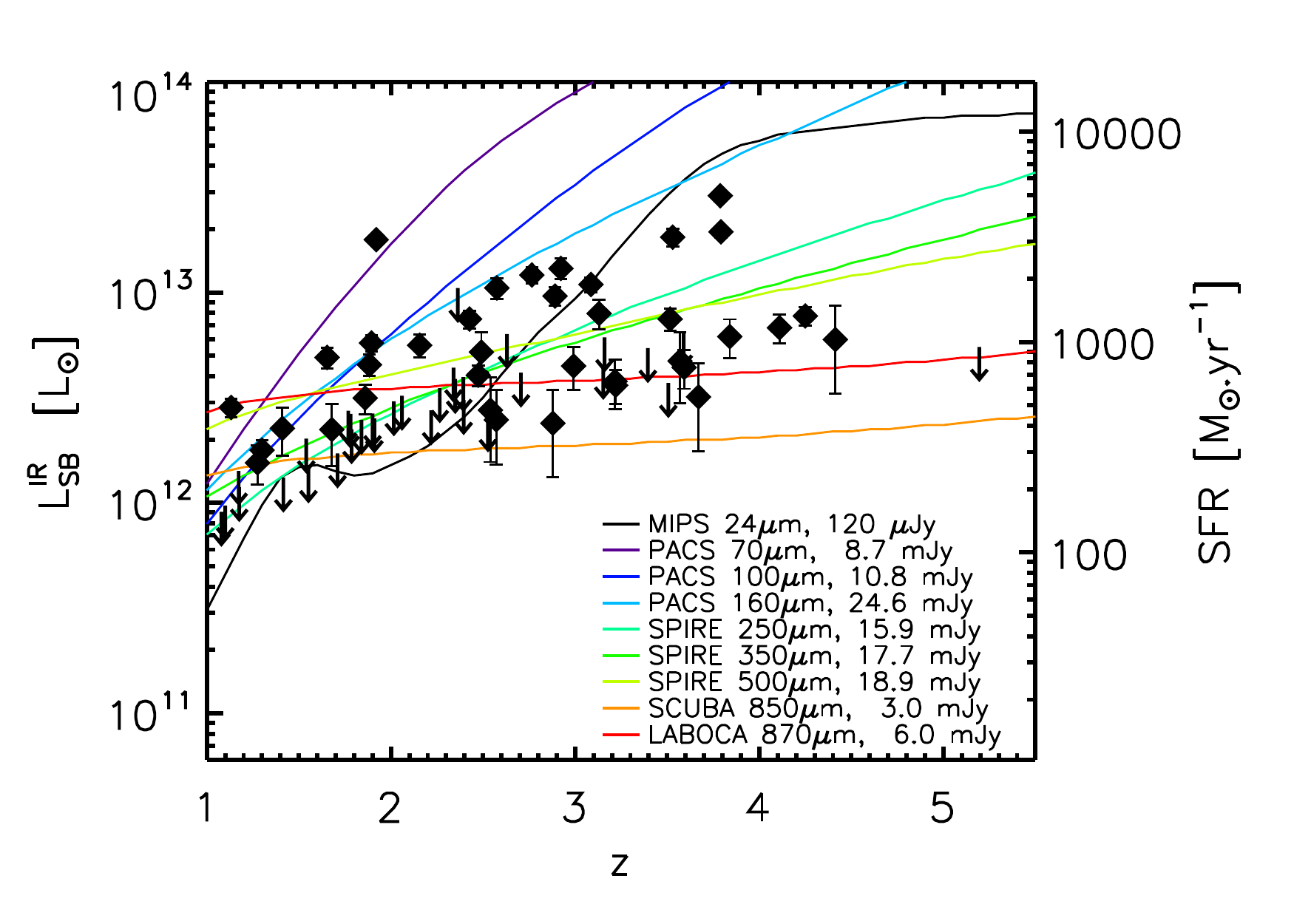}
\end{minipage}
\caption{{\it Left:} \liragn\ against redshift. The ordinate on the right
hand side of the plot is calculated using Eq.~\ref{eq:mdot} assuming a
radiative conversion efficiency $\epsilon$=0.1 and $\kappa_{\rm Bol}^{\rm AGN}$=6 (\S~\ref{sec:mdot}).  {\it Right:} \lirsb\
against redshift.  The ordinate on the right hand side of the plot is
calculated using Eq. \ref{eq:sfr}, assuming the \cite{Kennicutt1998} law. We
indicate our average 3$\sigma$-sensitivity limit for the AGN and the SB3
templates for each filter (see \S~\ref{sec:agnsb_limits} for details, figure inspired from \cite{Elbaz2011}). 
One notes that in a case of pure AGN emission, 
\mips1\ is the most sensitive band over the entire redshift range. On the other hand, in the case of a pure 
starburst emission, the most sensitive band at $z$$>$2 is in the submm (APEX or SCUBA).}
\label{fig:z_vs_liragn_lirsb}
\end{figure*}

\subsection{AGN/SB relative contribution}
\label{sec:fagn}

The relative contribution of the AGN and the SB can vary a lot, depending 
of the physical condition of the SB or the configuration of the dust close to the 
AGN (see previous section). In order to check this potential impact on the 
previously defined classification (UL, CD, WD, WCD) we define three values of
the relative AGN and SB contribution to the flux at 10, 50 and 100\mum\ restframe
($f_{\rm AGN}^{10, 50, 100\mu m}$=$F_{\rm AGN}^{10, 50, 100\mu m}/F_{\rm SB}^{10, 50, 100\mu m}$). 
This relative contribution may vary, depending
on the SB template used for the fit. In the mid-IR (10\mum), this
effect can be strong due to the emission from Polycyclic Aromatic Hydrocarbon (PAH) molecules. For the two extreme
starburst templates, this relative contribution can vary up to a factor
of 4. Nevertheless, in most cases, we can discriminate between the most
extreme templates (SB1 and SB5) which have a factor of $\sim$2 difference
in their relative contributions for the same total luminosity. In the             
far-IR (100\mum) this relative difference is smaller,
the SB dominates the SED except for a few cases (e.g. {\it 4C~23.56}).
The Appendix B shows these fractions as a function of the total infrared 
luminosity \lirtot.

\subsection{Procedure on the sample}
\begin{table}[t] \centering
\caption{Distribution of the sample as a function of their class and their number of detections in the IR.}
\label{tab:compo_sample}
\begin{tabular}{ccc ccc ccc c} 
\hline
 & \multicolumn{9}{c}{N. of Detections} \\
 \cline{2-10}
Class & 0 & 1 & 2 & 3 & 4 & 5 & 6 & 7 & 8 \\
\hline \hline
UL  & 8 &  &  &  & 1 &  &  &  &   \\
WD  &  & 4 & 17 & 0 & 2 & 1 & 0 & 0 & 0  \\
CD  &  & 5 & 1 & 1 & 1 & 0 & 0 & 0 & 0  \\
WCD &  &  & 1 & 3 & 2 & 6 & 6 & 6 & 5 \\
\hline
\end{tabular}
\end{table}

The large difference in data quality prevents us from
 blindly applying the same fitting procedure on all galaxies in the full sample. 
 In order to take full advantage of our data, we apply
different procedures on each source, depending on the number and
the quality of detections, and their previously defined classes (see
\S~\ref{sec:class} and Tab.~\ref{tab:compo_sample}). We also
highlight some special cases which need a specific treatment. We stress
that independently from their designated class, each acceptable solution must
respect the``3-sigma rule": if a solution present a template brighter
that any of our 3$\sigma$ detection limits, this solution was discarded
as not physically acceptable. Note that all calculated infrared 
luminosities are integrated over the 8-1000\mum\ range.



\paragraph{UL sources:}
For sources without any firm detections, only upper limits on
\liragn\ and \lirsb\ can be calculated. We normalise separately the
AGN template and a SB template on the most constraining upper
limit. The upper limit on \lirtot\ is calculated by fitting simultaneously 
both templates on the most constraining upper limits. 

\paragraph{WD sources:}
For sources with detections only in the mid-IR, we fit only an AGN template.
 In a second step, we normalise a SB template to the most
constraining upper limit in the far-IR. The upper limit on \lirtot\ is calculated
fitting simultaneously both templates on the detected points in mid-IR and the
most constraining upper limit in far-IR.

\paragraph{CD sources:}
For sources with detections only in the far-IR, we fit only an SB template. If the number of
detections allows it ($n$$\geq$2), we leave the type (SB1 to SB6) as a
free parameter. In a second step we normalise the AGN template on the
most constraining upper limit in the mid-IR. The upper limit on \lirtot\ is
calculated fitting simultaneously both templates on the detected points in the 
far-IR and the most constraining upper limit in the mid-IR.

\paragraph{WCD sources}
For sources with two or three detections in mid- and far-IR, we fit
both AGN and SB components but choose the SB template which maximises
\lirtot\ with respect to the most constraining upper limits in the far-IR.
For sources with four or more detections, we fit both the AGN and SB
components, with the SB type template as an additional free parameter
in the fitting. In both cases, \lirtot\ is the sum of both templates. 
As mentioned previously, extinction could have a strong impact on the fitting. We tested 
for this effect on this subsample adding the extinction as a free parameter on both components. 
The \lirsb\ and \liragn\ are changed within a factor $<$3. We therefore do not consider extinction
in our fitting procedure for the reminder of this paper.

Table \ref{tab:lir_list} provides the measured \lirtot, \liragn, \lirsb\ and SB template given 
by the best solution and the AGN fraction at 10, 50 and 100 \mum\ restframe. The AGN
fractions are described in \S~\ref{sec:fagn} and Appendix B. 

\section{Results}

\begin{table*}[t] \centering
\caption{Main results from the SED fitting. (*)See Appendix A. SB template refers to the best-fit template (\S~\ref{sec:sb_template}). \lirtot is the total integrated luminosity (\S~\ref{sec:tot_lir}). \liragn,\lirsb are the integrated AGN and SB luminosity in the 8-1000 \mum\ restframe  (\S~\ref{sec:lir_results}). \fagnten, \fagnfifty and \fagnhundred are the ratio S$_{AGN}$/S${_{SB}}$ at 10, 50 and 100\mum\ restframe, respectively  (\S~\ref{sec:fagn} and Appendix B).}
\label{tab:lir_list}
\begin{tabular}{l ccc ccc ccc}
\hline
Name & class & Detect. & SB template & \lirtot [$10^{12}$L$_{\odot}$] & \liragn [$10^{12}$L$_{\odot}$] & \lirsb [$10^{12}$L$_{\odot}$] & \fagnten & \fagnfifty & \fagnhundred \\
\hline \hline
6C~0032+412     &  WCD & 3 & SB2 &            12.2 &       9.6$\pm$   1.8 &       3.2$\pm$   1.4 &     17.5 &     1.56 &    0.350 \\
MRC~0037-258    &   WD & 2 & SB2 &       $<$   1.5 &       0.9$\pm$   0.2 &            $<$   1.0 &  $>$ 4.8 & $>$  0.4 & $>$ 0.09 \\
6CE~0058+495    &   WD & 2 & SB2 &       $<$   2.2 &       1.0$\pm$   0.2 &            $<$   1.4 &  $>$ 4.0 & $>$  0.3 & $>$ 0.07 \\
MRC~0114-211    &  WCD & 5 & SB6 &             3.8 &       1.6$\pm$   0.4 &       2.3$\pm$   0.6 &      3.8 &     0.25 &    0.080 \\
TN~J0121+1320   &   CD & 2 & SB2 &       $<$   9.8 &            $<$   3.0 &       7.5$\pm$   0.9 &  $<$ 2.2 & $<$  0.2 & $<$ 0.04 \\
6CE~0132+330    &   WD & 1 & SB3 &       $<$   1.2 &       0.2$\pm$   0.0 &            $<$   1.7 &  $>$ 1.0 & $>$  0.0 & $>$ 0.01 \\
6C~0140+326     &   CD & 1 & SB6 &       $<$  19.2 &            $<$  15.0 &       6.0$\pm$   2.7 &  $<$13.4 & $<$  0.8 & $<$ 0.28 \\
MRC~0152-209    &  WCD & 8 & SB2 &            22.3 &       4.5$\pm$   1.2 &      17.8$\pm$   0.9 &      1.4 &     0.12 &    0.028 \\
MRC~0156-252    &   WD & 2 & SB2 &       $<$   8.9 &       8.5$\pm$   1.6 &            $<$   3.0 &  $>$15.4 & $>$  1.3 & $>$ 0.30 \\
TN~J0205+2242   &   UL & 0 & SB2 &       $<$   5.9 &            $<$   2.6 &            $<$   3.7 &  \nodata &  \nodata &  \nodata \\
MRC~0211-256    &  WCD & 6 & SB3 &             2.1 &       0.3$\pm$   0.1 &       1.8$\pm$   0.2 &      1.7 &     0.09 &    0.017 \\
TXS~0211-122    &   WD & 4 & SB2 &       $<$   9.3 &       9.1$\pm$   1.7 &            $<$   4.4 &  $>$11.4 & $>$  1.0 & $>$ 0.22 \\
3C~65           &   WD & 2 & SB2 &       $<$   2.2 &       1.3$\pm$   0.2 &            $<$   1.2 &  $>$ 6.1 & $>$  0.5 & $>$ 0.12 \\
MRC~0251-273    &   WD & 2 & SB2 &       $<$   7.6 &       2.8$\pm$   0.5 &            $<$   6.1 &  $>$ 2.5 & $>$  0.2 & $>$ 0.05 \\
MRC~0316-257    &   CD & 3 & SB6 &       $<$   8.0 &            $<$   1.6 &       7.9$\pm$   1.3 &  $<$ 1.1 & $<$  0.0 & $<$ 0.02 \\
MRC~0324-228    &  WCD & 6 & SB4 &             7.5 &       1.7$\pm$   0.6 &       5.8$\pm$   0.5 &      1.1 &     0.18 &    0.023 \\
MRC~0350-279    &   WD & 2 & SB2 &       $<$   1.4 &       0.8$\pm$   0.1 &            $<$   2.7 &  $>$ 1.7 & $>$  0.1 & $>$ 0.03 \\
MRC~0406-244    &  WCD & 6 & SB4 &            13.0 &       5.5$\pm$   1.3 &       7.5$\pm$   0.7 &      2.7 &     0.45 &    0.057 \\
4C~60.07        &   CD & 4 & SB6 &       $<$  32.6 &            $<$  13.0 &      28.8$\pm$   1.7 &  $<$ 2.4 & $<$  0.1 & $<$ 0.05 \\
PKS~0529-549    &  WCD & 7 & SB3 &            13.2 &       2.7$\pm$   0.7 &      10.5$\pm$   1.2 &      2.4 &     0.13 &    0.023 \\
WN~J0617+5012   &   UL & 0 & SB6 &       $<$   5.6 &            $<$   2.0 &            $<$   4.5 &  \nodata &  \nodata &  \nodata \\
4C~41.17$^{*}$       &  WCD & 7 & SB6 &            24.0 &       4.6$\pm$   1.8 &      19.4$\pm$   1.2 &      1.3 &     0.08 &    0.027 \\
WN~J0747+3654   &   CD & 1 & SB2 &       $<$   5.3 &            $<$   1.6 &       4.5$\pm$   1.0 &  $<$ 2.0 & $<$  0.1 & $<$ 0.04 \\
6CE~0820+3642   &  WCD & 5 & SB2 &             4.0 &       0.8$\pm$   0.3 &       3.1$\pm$   0.5 &      1.4 &     0.12 &    0.028 \\
5C~7.269        &   UL & 0 & SB3 &       $<$   5.2 &            $<$   3.1 &            $<$   2.7 &  \nodata &  \nodata &  \nodata \\
USS~0828+193    &  WCD & 6 & SB3 &            18.4 &      15.9$\pm$   2.3 &       2.5$\pm$   1.0 &     59.0 &     3.19 &    0.575 \\
6CE~0901+3551   &   WD & 2 & SB2 &       $<$   4.4 &       2.4$\pm$   0.4 &            $<$   2.5 &  $>$ 5.3 & $>$  0.4 & $>$ 0.10 \\
B2~0902+34$^{*}$      &   UL & 4 & SB6 &       $<$   8.4 &            $<$   4.3 &            $<$   5.4 &  \nodata &  \nodata &  \nodata \\
6CE~0905+3955   &  WCD & 7 & SB2 &            15.5 &      11.0$\pm$   1.0 &       4.5$\pm$   0.5 &     13.4 &     1.19 &    0.268 \\
TN~J0924-2201   &   UL & 0 & SB6 &       $<$  13.5 &            $<$   9.2 &            $<$   5.5 &  \nodata &  \nodata &  \nodata \\
6C~0930+389     &   WD & 1 & SB2 &       $<$   4.6 &       2.5$\pm$   0.6 &            $<$   3.1 &  $>$ 4.5 & $>$  0.4 & $>$ 0.09 \\
USS~0943-242    &  WCD & 6 & SB1 &            14.9 &       1.9$\pm$   0.7 &      13.0$\pm$   1.5 &      0.6 &     0.08 &    0.017 \\
3C~239          &   WD & 2 & SB2 &       $<$   5.5 &       3.6$\pm$   0.6 &            $<$   2.7 &  $>$ 7.5 & $>$  0.6 & $>$ 0.15 \\
MG~1019+0534    &  WCD & 8 & SB1 &            13.8 &       1.7$\pm$   0.6 &      12.1$\pm$   1.1 &      0.6 &     0.08 &    0.016 \\
MRC~1017-220    &   WD & 2 & SB2 &       $<$   4.1 &       2.5$\pm$   0.4 &            $<$   2.7 &  $>$ 5.0 & $>$  0.4 & $>$ 0.10 \\
WN~J1115+5016   &   CD & 1 & SB2 &       $<$   2.8 &            $<$   0.9 &       2.8$\pm$   1.2 &  $<$ 1.7 & $<$  0.1 & $<$ 0.03 \\
3C~257          &  WCD & 8 & SB4 &            11.2 &       7.2$\pm$   1.1 &       4.0$\pm$   0.5 &      6.4 &     1.08 &    0.136 \\
WN~J1123+3141   &  WCD & 5 & SB2 &            24.4 &      20.6$\pm$   3.3 &       3.8$\pm$   1.0 &     30.0 &     2.68 &    0.600 \\
PKS~1138-262    &  WCD & 8 & SB3 &            17.2 &      11.6$\pm$   1.3 &       5.6$\pm$   0.7 &     19.0 &     1.03 &    0.185 \\
3C~266          &  WCD & 4 & SB6 &             2.0 &       0.5$\pm$   0.2 &       1.6$\pm$   0.3 &      1.6 &     0.11 &    0.034 \\
6C~1232+39      &  WCD & 3 & SB2 &             9.2 &       6.2$\pm$   1.1 &       3.6$\pm$   0.7 &      9.3 &     0.83 &    0.186 \\
USS~1243+036    &  WCD & 5 & SB6 &            13.1 &       8.4$\pm$   1.9 &       4.7$\pm$   1.7 &      9.5 &     0.63 &    0.203 \\
TN~J1338-1942   &   CD & 1 & SB2 &       $<$   9.7 &            $<$   3.6 &       6.8$\pm$   1.1 &  $<$ 2.9 & $<$  0.2 & $<$ 0.05 \\
4C~24.28        &  WCD & 5 & SB2 &            12.8 &      10.4$\pm$   1.5 &       2.4$\pm$   1.1 &     23.9 &     2.13 &    0.478 \\
3C~294          &   WD & 2 & SB2 &       $<$   2.1 &       1.1$\pm$   0.2 &            $<$   2.2 &  $>$ 2.8 & $>$  0.2 & $>$ 0.05 \\
USS~1410-001    &   WD & 2 & SB2 &       $<$  10.0 &       5.4$\pm$   1.0 &            $<$  10.5 &  $>$ 2.9 & $>$  0.2 & $>$ 0.05 \\
8C~1435+635     &   CD & 1 & SB2 &       $<$  11.6 &            $<$   4.8 &       7.7$\pm$   0.8 &  $<$ 3.4 & $<$  0.3 & $<$ 0.06 \\
USS~1558-003    &   WD & 2 & SB5 &       $<$   9.2 &       8.4$\pm$   1.5 &            $<$   2.5 &  $>$10.4 & $>$  2.8 & $>$ 0.24 \\
USS~1707+105    &   WD & 1 & SB2 &       $<$   3.7 &       1.2$\pm$   0.3 &            $<$   3.9 &  $>$ 1.7 & $>$  0.1 & $>$ 0.03 \\
LBDS~53W002     &   WD & 2 & SB2 &       $<$   5.7 &       3.7$\pm$   0.7 &            $<$   4.0 &  $>$ 5.1 & $>$  0.4 & $>$ 0.10 \\
LBDS~53W091     &   UL & 0 & SB3 &       $<$   0.9 &            $<$   0.1 &            $<$   1.5 &  \nodata &  \nodata &  \nodata \\
3C~356.0        &   WD & 2 & SB2 &       $<$   2.6 &       2.0$\pm$   0.4 &            $<$   0.9 &  $>$12.2 & $>$  1.0 & $>$ 0.24 \\
7C~1751+6809    &   WD & 2 & SB2 &       $<$   2.0 &       0.4$\pm$   0.1 &            $<$   2.0 &  $>$ 1.1 & $>$  0.1 & $>$ 0.02 \\
7C~1756+6520    &   UL & 0 & SB3 &       $<$   1.2 &            $<$   0.1 &            $<$   1.3 &  \nodata &  \nodata &  \nodata \\
3C~368          &  WCD & 7 & SB2 &             4.2 &       1.4$\pm$   0.3 &       2.8$\pm$   0.3 &      2.7 &     0.24 &    0.055 \\
7C~1805+6332    &   WD & 2 & SB2 &       $<$   3.0 &       1.4$\pm$   0.3 &            $<$   2.5 &  $>$ 3.2 & $>$  0.2 & $>$ 0.06 \\
4C~40.36        &   WD & 1 & SB2 &       $<$   3.9 &       1.5$\pm$   0.4 &            $<$   3.5 &  $>$ 2.3 & $>$  0.2 & $>$ 0.04 \\
TXS~J1908+7220  &  WCD & 7 & SB6 &            43.7 &      25.4$\pm$   3.6 &      18.3$\pm$   1.8 &      7.4 &     0.49 &    0.158 \\
WN~J1911+6342   &   UL & 0 & SB2 &       $<$   7.6 &            $<$   2.5 &            $<$   6.4 &  \nodata &  \nodata &  \nodata \\
TN~J2007-1316   &  WCD & 4 & SB2 &            15.0 &       8.8$\pm$   1.7 &       6.2$\pm$   1.3 &      7.9 &     0.70 &    0.158 \\
MRC~2025-218    &   WD & 2 & SB2 &       $<$   3.1 &       1.6$\pm$   0.3 &            $<$   6.3 &  $>$ 1.4 & $>$  0.1 & $>$ 0.02 \\
MRC~2048-272    &   UL & 0 & SB2 &       $<$   4.7 &            $<$   2.1 &            $<$   3.2 &  \nodata &  \nodata &  \nodata \\
MRC~2104-242    &  WCD & 5 & SB2 &             8.0 &       2.8$\pm$   1.0 &       5.2$\pm$   1.2 &      3.0 &     0.27 &    0.060 \\
4C~23.56$^{*}$        &   WD & 5 & SB2 &       $<$  23.7 &      24.5$\pm$   1.7 &            $<$   5.4 &  $>$24.9 & $>$  2.2 & $>$ 0.49 \\
MG~2144+1928    &  WCD & 2 & SB2 &            12.3 &       8.5$\pm$   2.1 &       4.4$\pm$   0.9 &     11.0 &     0.99 &    0.221 \\
USS~2202+128    &   WD & 4 & SB2 &       $<$  10.0 &       7.3$\pm$   1.1 &            $<$   4.2 &  $>$ 9.7 & $>$  0.8 & $>$ 0.19 \\
MRC~2224-273    &  WCD & 3 & SB1 &             3.4 &       1.9$\pm$   0.3 &       2.2$\pm$   0.7 &      3.3 &     0.45 &    0.092 \\
B3~J2330+3927   &  WCD & 6 & SB4 &            24.5 &      13.6$\pm$   2.4 &      10.9$\pm$   0.9 &      4.5 &     0.76 &    0.096 \\
4C~28.58        &  WCD & 8 & SB6 &            17.7 &       8.1$\pm$   1.2 &       9.6$\pm$   1.0 &      4.5 &     0.30 &    0.096 \\
3C~470          &  WCD & 7 & SB1 &             8.8 &       3.8$\pm$   0.8 &       4.9$\pm$   0.6 &      3.2 &     0.45 &    0.092 \\
\hline
\end{tabular}
\end{table*}

\subsection{AGN/SB detection limits}
\label{sec:agnsb_limits}

Fig. \ref{fig:z_vs_liragn_lirsb} shows both the infrared AGN and SB
luminosities as a function of redshift. In order to verify whether the upper
limits on the AGN and the SB components are mainly due to
physical processes or purely from an observational bias, we calculate the
minimum luminosity for each component related to the band sensitivity. Our
3$\sigma$ sensitivity limits are calculated averaging the uncertainties
over the entire sample. We normalise the AGN and SB3\footnote{We note that using another SB template
would introduce some variation but the general shape of the threshold remains
unchanged.} templates in each observed band and calculate the corresponding
\liragn\ or \lirsb\ at any redshift.

One notes that our upper limits do not
always follow the most sensitive detection limit (for instance the black line on left plot). This can be explained as: (i) the IR emission
is a mixture of AGN- and SB-heated dust; (ii) especially in the IR,
the background emission varies locally and as a function of Galactic longitude,
impacting the final sensitivity; (iii) the depth of the \mips1\ imaging is not uniform throughout 
the sample \citep{DeBreuck2010}.  

From these diagrams, MIPS 24\mum\
data appear to be the most sensitive to the AGN contribution at any
redshift. It is important to remember that a pure AGN contribution
is very unlikely to be detected in the sub-mm as they require AGN of
\liragn$>$10$^{14}$\lsun\ at any redshift (see orange, SCUBA line at the top of 
Fig. \ref{fig:z_vs_liragn_lirsb}, left). Nevertheless, one should be careful 
to associate the 24\mum\ flux to the AGN because PAH contributions from star formation can
be important in this band (see plateau in Fig. \ref{fig:z_vs_liragn_lirsb}, right).
In the case of a pure SB component, the situation is completely
different. Up to $z$$\sim$2, MIPS 24\mum\ is again our most sensitive
band to detect both SB and AGN. However, in the case of a pure SB 
component, the SB will be detected only at \lirsb$>$10$^{12}$\lsun\ where for the same sensitivity 
a pure AGN will be detected at the \liragn$>3\times$10$^{11}$\lsun\ level.
This implies that the MIPS 24\mum\ band is likely to be dominated by AGN 
emission if any hints of AGN activity is detected in a source (which is the case for radio galaxies).

At $z$$>$2, SCUBA (and LABOCA) become our most
sensitive bands for detecting SB components. Moreover, due to $k$-correction effects \citep{Blain1999},
 this limit is roughly constant with redshift. 
Our 3$\sigma$ sensitivity allows us to detect starburst activity of at least
400\msun yr$^{-1}$ (800\msun yr$^{-1}$ for LABOCA) assuming the standard
\lirtot-SFR conversion law \citep{Kennicutt1998}.

\subsection{Infrared AGN and Starburst luminosities}
\label{sec:lir_results}

\begin{figure*}[t] \centering
  \includegraphics[width=1.0\textwidth]{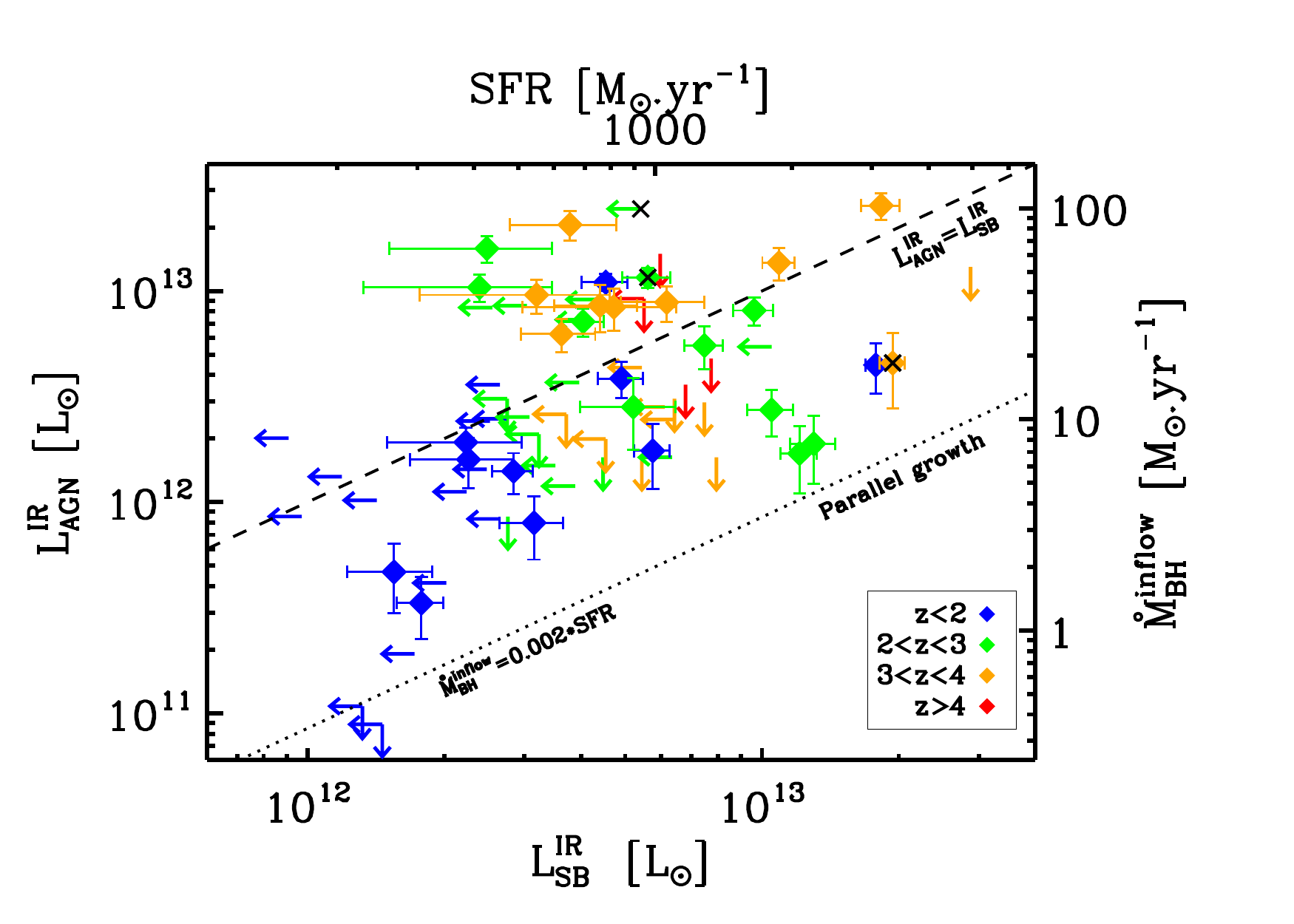}
  \caption{\liragn\ versus \lirsb. The top axis converts \lirsb\ to SFR using the \citet{Kennicutt1998} relation (Eq. \ref{eq:sfr}). The right axis converts \liragn\ to \mdotbh\ assuming $\epsilon$=0.1 and $\kappa_{\rm AGN}^{\rm Bol}$=6 (Eq. \ref{eq:mdot}). The dashed line marks \liragn=\lirsb. This dashed line indicates the relation corresponding to \mdotbh=0.024 $\times$ SFR, using the right and top axis. The dotted line represents the parallel growth mode, where black holes and galaxies are growing simultaneously, following the \mbhgal\ relation (see \S~\ref{sec:parallelgrowth} for details). Colours code redshift bins. The three crosses indicate the three sources of peculiar interest, {\it 4C~23.56}, {\it PKS~1138-262} and {\it 4C~41.17}}.
  \label{fig:liragn_vs_lirsb}
\end{figure*}

Fig. \ref{fig:z_vs_liragn_lirsb} plots \liragn\ and \lirsb\ versus redshift. Both
plots suggest an increasing trend with redshift ($\rho$=0.374, $p$=0.0019 and $\rho$=0.613, $p$$<$0.0001, 
respectively\footnote{Making use of the IRAF survival analysis package, $p$ is the probability of no correlation.}). 
While this trend seems real for \liragn (given the few upper limits), it appears stronger with \lirsb. 
Indeed, even if \lirsb\ is affected by numerous upper limits, especially at $z$$<$2.5
(due to the limited sensitivity of the 250\mum\ SPIRE band), these upper limits are fully consistent with an increasing trend.
 An improvement of one order of magnitude
in our far-IR sensitivities should certainly be enough to detect the missing sources and
therefore confirm the increasing trend with redshift for \lirsb. 

We also estimate the black hole accretion rate 
(\mdotbh) and the star formation rates (SFR) based on the fits to the SED. 
We discuss these parameters in greater detail in \S~\ref{sec:discussion}. First,
we note that our sample spans a large range in both SFR and \mdotbh, almost 
two orders-of-magnitude for both of them, 1\msunyr$<$\mdotbh$<$100\msunyr\ and 
100\msunyr$<$SFR$<$5000\msunyr. Moreover, as they scale linearly with \liragn\ and \lirsb, 
the suggested increasing trend with redshift applies to both the SFR and the black hole accretion rate.
In particular, all our $z$$<$2 sources have a \mdotbh$<$5\msunyr, while at $z$$>$2, \mdotbh\ can reach 100\msunyr. 
The same behaviour, though weaker, can be observed with SFR where all sources (except one) have 
SFR$<$1000\msunyr\ at $z$$<$2.5, while SFR can reach 4000 \msunyr\ for the sources with the highest redshifts.

Fig. \ref{fig:liragn_vs_lirsb} plots \liragn\ versus \lirsb (we discuss this plot more 
extensively in \S~\ref{sec:discussion}). Radio
galaxies cover a wide range of relative contributions: from almost
pure star forming galaxies (e.g. {\it 4C~41.17}), to almost pure AGN
contribution (e.g. {\it 4C~23.56}\footnote{classified 
as a WD, this extreme object appears to be the prototype 
of the pure AGN contribution (c.f. Appendix A and Fig. \ref{fig:sed}).}) but with the majority having SEDs which
are composites of star formation and AGN heating (e.g. {\it PKS~1138-262}). 
We point these three specific sources in Fig. \ref{fig:liragn_vs_lirsb} as black crosses. 

Taking into account only the objects with good constraints on both their
AGN and SB contributions, we find no significant correlation. This provides
confidence about the decomposition as we do not expect, {\it a priori},
to have a correlation between the AGN and SB luminosities 
\citep[as also find in other studies, e.g.][]{Bongiorno2012,Dicken2012,Feltre2013,Leipski2014}. 
Nevertheless, it is interesting to note that each component (AGN and SB) has an integrated 
luminosity of $L^{IR}$$>$10$^{12}$\lsun. This indicates that a high IR luminosity 
does not necessarily imply a high star formation rate or a strong AGN activity.

\subsection{Comparing AGN and SB IR luminosities with radio properties}
\label{sec:ir_radio_cor}

\subsubsection{Radio luminosities}

\cite{DeBreuck2010} calculated the 500\,MHz restframe luminosity for
the entire sample.  In the case of powerful radio galaxies, the radio emission is
dominated by the AGN. The 500\,MHz luminosity (\l500) is
an excellent proxy for estimating the energy injected by the AGN into
the lobes of the radio galaxy\footnote{At 500\,MHz, the radio emission is dominated by the lobes. 
At such frequencies, relativistic beaming effects are not playing a significant role \citep{Blundell1998}.}.

\begin{figure}[t] \centering
  \includegraphics[width=0.5\textwidth]{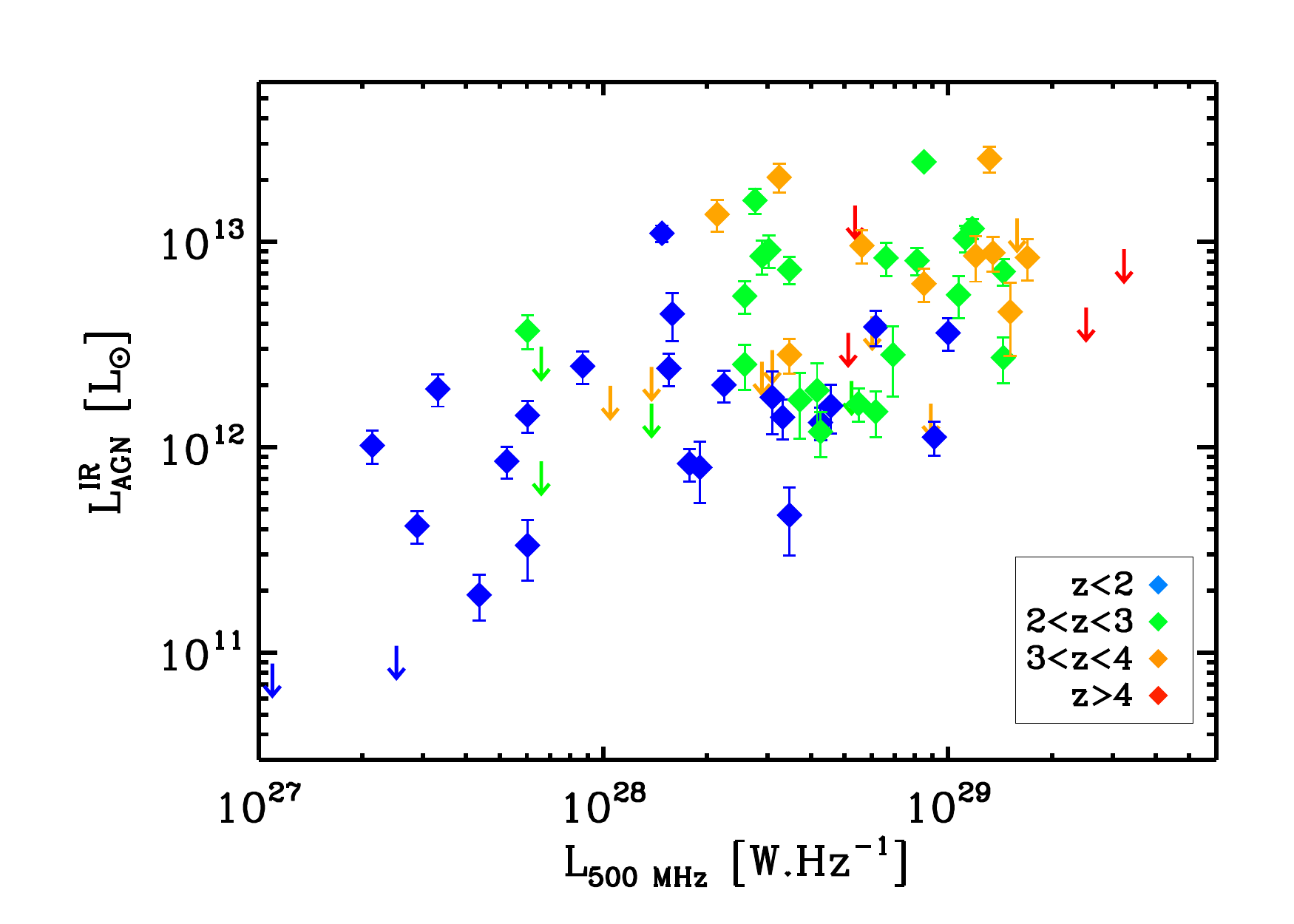}
  \caption{\liragn\ against \l500. The colours code the redshift bins.}
  \label{fig:l500_vs_liragn}
\end{figure}

We see a weak correlation over 2 orders of
magnitude in \liragn\ and \l500 (Fig. \ref{fig:l500_vs_liragn}, $\rho$=0.475,$p$=0.0001). 
However, both \l500\ and \liragn\ present
a correlation with redshift. As we constrained \liragn\ for most of our
sample and \l500\ is well determined, we apply a partial correlation
test\footnote{We make use of the IDL function {\it p\_correlate} and only consider sources with detected \liragn\ and \l500.} 
in order to take this mutual dependence on redshift into account. Indeed, this partial test severely degrades the correlation ($R$=0.10) indicating
that redshift is the determinant factor of this correlation.
It is therefore impossible to conclude much about the correlation between
the radio and the IR in radio galaxies (at least with this sample which
spans a wide redshift range but $<$2 orders of magnitude in radio luminosity). 

This apparent lack of correlation can be easily explained by
comparing the timescales in the IR and radio to respond to changes in
the energy output of the AGN.  The dust heated by the AGN is likely to be
circumnuclear given its emission temperature. Dust cools quickly and the
timescale for the photons to stream through the nebula is relatively short.
The radio emission, on the other hand, has a much longer response time to
changes in the AGN output and the aging time of electrons is of the order of tens of Myr \citep{Blundell1999},
especially at low frequencies and considering shock re-energization in
the lobes themselves. Also, it is not clear if the relative fraction of
energy and emission in the radio and IR should be similar anyway.

\lirsb\ and \l500 tend to also present a positive weak correlation ($\rho$=0.536
and $p$=0.001 applying a survival analysis), similarly to \liragn\ and \l500. 
The numerous upper limits and poor statistics make it even more difficult 
to conclude anything about this correlation. Moreover, as for \liragn, this correlation seems 
also mostly driven by redshifts effects.


\subsubsection{Radio sizes}
\label{sec:ir_size}

Spatially resolved radio observations can measure the distance between
the core and the lobe (or lobe-lobe), providing useful information on the age of the
radio phase assuming a simple ballistic trajectory for the ejected
particules \citep{Blundell1999}. The radio size, $las$ \citep[Table
  2 in][]{DeBreuck2010}, corresponds to the largest extension in 1.4GHz
radio maps. As all our radio galaxies have a spectroscopically
determined redshift, we calculate the projected size \dsini\ in kpc of the radio galaxy where $D$ is
the physical size of the galaxy and the $sin(i)$ term refers to its
projection onto the sky plane. A degeneracy appears here due to the
inclination $i$ of the radio galaxy. Nonetheless, this latter quantity is not
expected to be important, as we are dealing with type 2 AGN,
i.e. mostly oriented in the plane of the sky \citep{Drouart2012}.
The real size $D$ will likely be at most 30\% larger due to
projection effects. 

\begin{figure}[t] \centering
  \includegraphics[width=0.5\textwidth]{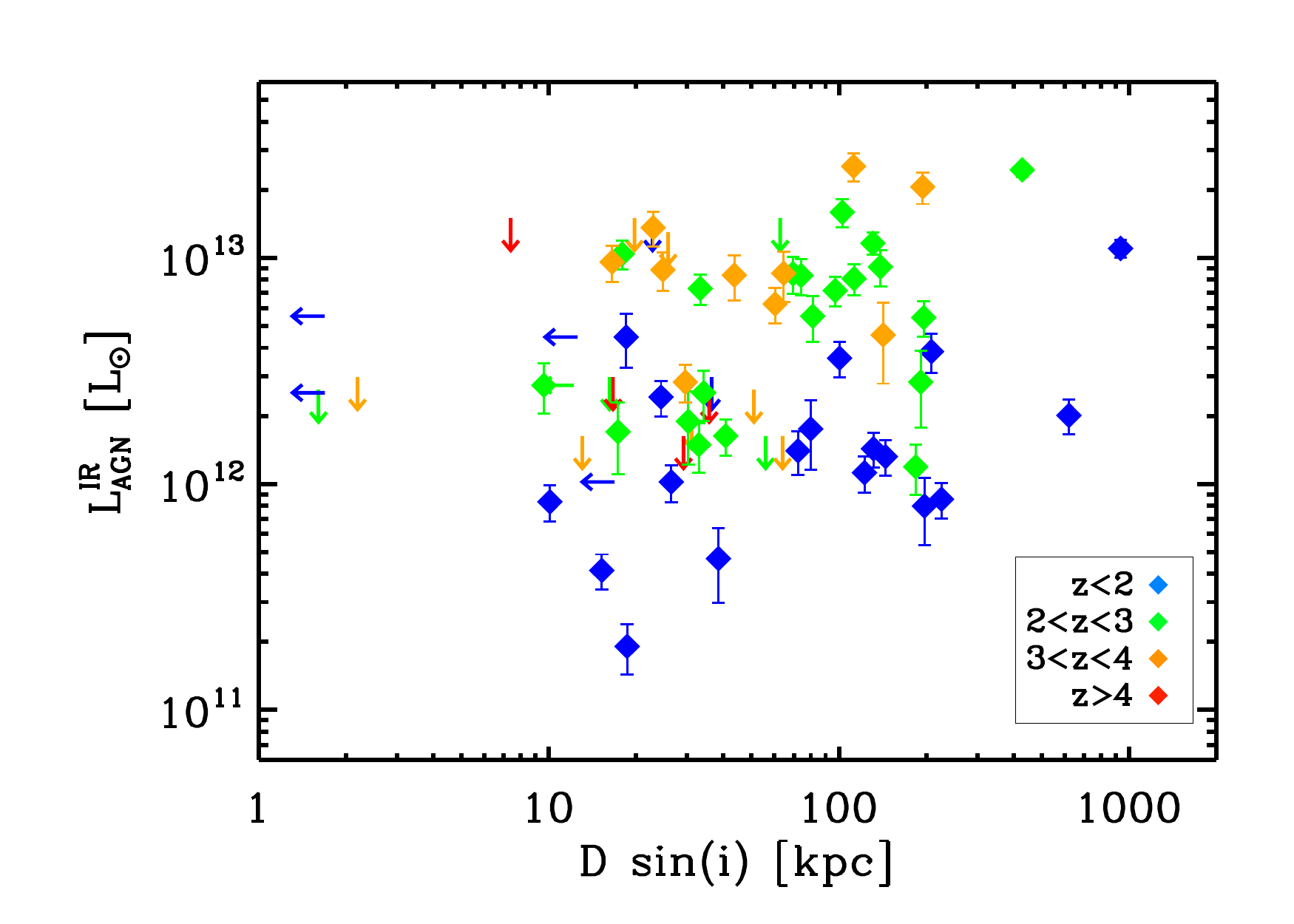}
  \caption{\liragn\ versus projected size \dsini. See \S~\ref{sec:ir_size} regarding the details of the calculation of the projected size. The colours code redshift bins.}
  \label{fig:d_vs_liragn}
\end{figure}

Fig. \ref{fig:d_vs_liragn} plots \liragn\ against the
projected size \dsini. Similar to \l500, coloured points show a redshift effect in our data;
the most compact AGN are at higher redshift. Radio sizes could be also 
affected by two effects: (i) it can depend on environment \citep[e.g.][]{Kaiser1997,Klamer2006, Bornancini2010,Ker2012}; and (ii)
our sample presents a weak selection bias in size, with larger objects located at 
lower redshift (see Fig. \ref{fig:d_vs_liragn}).

\subsection{Comparing IR luminosity with stellar mass}


Our sample benefits from stellar mass estimates thanks to 
\spitzer\ data. By fitting the 3.6-24\mum\ range with a sum of 
an elliptical template from PEGASE.2 \citep{Fioc1997} and 
blackbodies at various temperatures, \cite{DeBreuck2010} estimated
reliable stellar masses for our sample, finding values of $\sim$10$^{11-12}$\msun. Although massive, our low dynamic 
range in mass prevents us from drawing any conclusion about possible correlations. However, it
is interesting to note the large scatter in the AGN luminosity, over two
orders of magnitude, but the relatively small scatter in \mstel \citep[see Fig. 5 in ][]{DeBreuck2010}. 
Plotting \mstel\ against \lirsb\ exhibits the same behaviour, with the range of \lirsb\ 
being smaller given the larger number of upper limits. It is interesting to compare these masses
and \lirsb\ with SMGs at similar redshift. While the SFRs are
similar \citep[e.g.][\S~\ref{sec:sfr}]{Archibald2001,Reuland2004,Engel2010}, 
our sample of radio galaxies appears more massive by a factor of 2-5 
\citep{DeBreuck2010,Hainline2011,Michalowski2012,Simpson2013}.

\section{Discussion}
\label{sec:discussion}

Having disentangled the total IR luminosities of the star formation and AGN
components in the SEDs, we can now estimate star
formation rates (SFR), because we have stellar mass estimates (\mstel), the specific
star formation rates (sSFR), the black hole accretion rates (\mdotbh), Eddington ratios 
($\lambda$) and the specific black hole growth rates (\smdot).  Indeed,
these estimates will allow us to characterize the evolutionary state of
powerful radio galaxies, since we have a sample which spans a wide range
of redshifts. 

Are the host galaxies and their black holes co-evolving or is one of them 
outgrowing the other? Because
it is difficult to provide reliable uncertainties for individual sources and parameters
and undoubtedly our estimates suffer from systematic uncertainties,
we will have to interpret these estimates as ensemble averages instead
of focusing on individual measurements.  With these insights, we will
attempt to characterize the place of radio galaxies in the population
of distant galaxies and what their future evolution might be within the
context of their place in the ensemble population of galaxies.

\subsection{How rapidly are radio galaxies and their black holes growing?}

To put the radio galaxies into the context of the evolution of galaxies
and into the broad range of black hole demographics (i.e. growth rate
and masses), we need to convert estimates of the bolometric luminosities
of both the recent star formation and the black hole accretion, \lirsb
and \liragn\ , into SFR and \mdotbh. These estimates
 depend on the rate at which short wavelength emission (e.g. blue
optical, UV) from young stars is reprocessed into the IR and submm and
the rate at which the accreted mass on to the black hole is converted
into radiated energy and the ratio of the IR luminosity to the total
bolometric luminosity.

\subsubsection{Star formation rate}
\label{sec:sfr}

We have found that powerful radio galaxies are extremely bright in the IR 
(L$^{IR}_{tot}>$10$^{12}$ L$_{\sun}$), which may indicate that they have very
high SFRs. We have seen in \S~\ref{sec:lir_results} that we 
can disentangle SB from AGN emission. We can thus provide much more 
reliable determinations of the SFR than previous submm only determinations 
\citep[e.g.][]{Archibald2001,Reuland2004}. Given the high IR luminosities and the fact
that we are concerned here with the ensemble properties (averages,
ranges, changes with redshift) not the details of individual sources,
we will use the simple relation between the SFR and IR
luminosity given for local galaxies \citep{Kennicutt1998}:

\begin{equation}
  \textrm{SFR} =  1.72 \, 10^{-10} \times L^{\rm IR}_{\rm SB},
  \label{eq:sfr}
\end{equation}
where \lirsb is in units of L$_{\sun}$ and SFR in \msunyr.  Our galaxies span a large range
of SFR, from 100 to $\sim$5000\msunyr. These results are similar to SFRs
estimated for SMGs over the same redshift range \citep[e.g.][]{Engel2010,
Wardlow2011,Swinbank2014} and radio galaxies from the 3C catalogue \citep[e.g.][]{Barthel2012}\footnote{We 
note that the \lirsb\ and SFR from our SED decomposition are 
also compatible with the previous estimates based on IRS spectra \cite{Seymour2008,Rawlings2013}.}. 
This wide range of SFRs is somewhat
surprising.  Radio galaxies obviously have very active and luminous
AGN which emit across the electromagnetic spectrum and as such, the AGN
must have a significant impact on the host galaxy.  However, despite the
evidence for the impact of the AGN, these galaxies exhibit a
very wide range of SFR which is not correlated with the AGN luminosity (see Fig. \ref{fig:liragn_vs_lirsb}).
One must be very careful about
both correlation or lack of correlation being causal, the fact that
global star formation and AGN activity occur over different timescales,
and that estimates of the instantaneous power output of an AGN may not
be closely related to its longterm average \citep{Hickox2013, Chen2013}.
Such variations might mask any underlying relationship.

None of the radio galaxies at $z$$<$2.5 (except one) show a high \lirsb\ and
hence a high SFR compared to radio galaxies at higher redshifts
(Fig. \ref{fig:z_vs_liragn_lirsb}). The brightest IR sources,
have SFRs up to 5000\msun yr$^{-1}$. 
Whether or not this is a physical limit \citep[e.g.][]{Lehnert1996}, we caution that this large value may
be partially a result of the low angular resolution of our submm data
($\sim$20 arcsec at 850\mum). At $z$$>$1, 20 arcsec corresponds to $\sim$160 kpc
and so our observations may include contributions from many nearby star
forming galaxies \citep[e.g.][]{Hatch2008,Ivison2008,Ivison2012}. The SFRs we have estimated are therefore in some cases
best thought of as an upper limit to the SFR of the radio galaxy itself.
If indeed several sources are in the same beam, the low resolution means
that we are measuring an upper envelope of the SFR for the whole system
\citep[e.g.][]{Karim2013}. The multi-object nature of some IR sources is
evident in recent ALMA high resolution observations of submm galaxies
\citep{Hodge2013}.  The overall similarity in the star formation
rate estimates for our radio galaxies and the SMG population suggests
that perhaps the most luminous radio galaxies are affected in the same
manner. However, this is unlikely to be more than a
factor of a few \citep{Karim2013}.

\subsubsection{Black hole accretion rate}
\label{sec:mdot}

Assuming a fraction of the rest-mass energy of the material accreting
onto the black hole is converted into radiation over the whole of the
electromagnetic spectrum, one can estimate the accretion rate from an
estimate of the bolometric luminosity. The accretion rate (\mdotbh)
can be defined as:

\begin{equation}
\kappa^{\rm Bol}_{\rm AGN} \times L^{\rm IR}_{\rm AGN}=\epsilon \dot{\rm M}_{\rm BH}^{\rm acc} c^{2},
\label{eq:mdot}
\end{equation}

where $\epsilon$ is the efficiency factor for converted accreted
mass into bolometric luminosity and $\kappa^{\rm Bol}_{\rm
AGN}$ is a bolometric correction factor to convert \liragn\ into \lbolagn. There are only a small number of
empirical constraints on $\epsilon$.  Results on quasar clustering
suggest $\epsilon$$>$0.2 \citep{Shankar2010c,Shen2007}, other
studies suggest a mass-dependent factor ranging from 0.06 to 0.4
\citep{Davis2011,Cao2008,Volonteri2007,Merloni2004b}. We adopt here a conservative value
of $\epsilon$=0.1. If $\epsilon$ is actually higher than our preferred
value, all the relations will move by the necessary factor. 
The $\kappa^{\rm Bol}_{\rm AGN}$ correction is uncertain,
as it depends mostly on how much of the radiative energy is reprocessed
by dust, the wavelength of the observations that must be converted to the
bolometric luminosity, and the AGN type and their selection (X-ray AGN,
quasars, etc.).  This conversion factor to the bolometric luminosity
can vary from 1.4 to 15 for the IR (see Appendix C). Assuming the full 
unobscured AGN SED is similar to the \cite{Elvis1994} or \cite{Richards2006} templates, 
we find L$_{AGN}^{Bol}$$\approx$$6\times$L$_{AGN}^{IR}$ (we note other 
unobscured AGN templates produce similar numbers). We therefore 
decided to fix $\kappa^{\rm Bol}_{\rm AGN}$=6. We mark the 
influence of this choice with a vector in the relevant figures. 

Similar to the star formation rates, the black holes in powerful
radio galaxies appear to have a wide range of accretion rates,
1-100 M$_{\odot}$yr$^{-1}$ and similarly cover about two orders-of-magnitude
(Fig.~\ref{fig:z_vs_liragn_lirsb}).  To put this in perspective,
powerful radio galaxies have accretion rates similar to those of
high redshift quasars \citep{Hao2008}. Moreover, the accretion rates also
appear to increase with redshift as do the star formation rates
(Fig.~\ref{fig:z_vs_liragn_lirsb}).  

Assuming $\epsilon$ and $\kappa^{\rm Bol}_{\rm AGN}$ are constant for the ensemble of radio galaxies,
\mdotbh\ also appears as an upper limit of accretion rates in these
radio loud AGN. A simple order-of-magnitude calculation suggests that
$\sim$10$^{7-9}$M$_{\odot}$ of gas is needed to continuously support 
such AGN activity over a $~$10\,Myr timescale. This
quantity of gas is similar to the gas mass observed at $<$1\,kpc scale in some early-type
gas-rich galaxies at low redshift \citep[e.g.][]{Young2011,Crocker2011}. 
At higher redshift, where more molecular gas is expected to be present to fuel both 
the AGN and the star formation activity, only a few percent of the available gas mass observed in radio galaxy systems
\citep[$\sim10^{10-11}$\msun,][]{Ivison2012,Emonts2011, Emonts2013} is
necessary to fuel the central black hole. This transport of the gas to the inner part of 
the galaxy needs a process to efficiently remove the angular momentum of the gas
to fall within the sphere of influence of the central black hole \citep[e.g.][]{Jogee2005}. 
Even if some hypotheses are proposed, the dominant process is still unclear \cite[e.g.][ for a review]{Alexander2012}. 

\subsubsection{Co-eval stellar population and black hole growth?}
\label{sec:parallelgrowth}

How do the growth rates of the stellar population compare to that of the
AGN in these powerful radio galaxies?  If the galaxies and super massive
black holes were growing sufficiently rapidly to remain on the spheroid
mass-black hole mass relation, we would expect the growth rate of the
BH (i.e. \mdotbh) to be about 0.2\% of the growth rate of the stellar population
(i.e. SFR).  However, with the parameters given in the previous two
subsections, we find high redshift powerful radio galaxies are found 
to lie around the relation represented by \mdotbh$=0.024\times$SFR (i.e. 
offset by one order-of-magnitude, Fig.~\ref{fig:liragn_vs_lirsb}).  Although obviously these
estimates are very uncertain for individual sources, we see that overall, 
radio galaxies represent a phase in the evolution of both
the galaxy and the black hole where relatively
speaking, it appears as a more important growth of the black hole. In fact, it appears
that the black hole is out-growing its host galaxy, in spite of the high
observed SFR \citep[similar to SMGs at similar redshift][]{Alexander2005}, 
by about a factor of 10 relative to what would be expected if they were growing in lock step. 
It is important to keep in mind that we set $\kappa^{\rm Bol}_{\rm AGN}$=6 and
the exact value of the offset between the relative rate of black hole to
galaxy growth is dependent on this choice.  However, even if we choose
a lower but still reasonable value, say $\kappa^{\rm Bol}_{\rm AGN}$=2 (see Appendix C),
the general population of powerful radio galaxies would still have a
significant offset toward more rapid black hole growth.

We also stress that this result is completely mass-independant,
as neither the mass of the black hole nor of the galaxy are needed, only the local
spheroid mass-black hole mass to draw the parallel growth mode (dotted line in Fig. 
\ref{fig:liragn_vs_lirsb}). This behaviour is similar to moderate redshift quasars
\citep[$z$=1,][]{Urrutia2012} and high redshift quasars
\citep[$z$=6,][]{Willott2013}. This similarity suggests that high accretion rates are more directly related
to the fact that the AGN are bolometrically luminous with copious output
rates of ionizing photons but are not directly related to the production
of the powerful radio emission in the extended radio lobes (see \S~\ref{sec:ir_radio_cor}). 
Notably, the presence of strong emission lines in our sample of radio galaxies \citep[e.g.][]{Vernet2001,DeBreuck2002}
suggests that these powerful radio galaxies have indeed high relative accretion rates 
\citep[e.g.][]{Janssen2012,Hardcastle2007} as expected if the black holes are growing rapidly as we purport.
We remind the reader, as discussed earlier, the calculated \mdotbh\ represents the instantaneous accretion
rate of the BH, not the long term average accretion rate. Variability in
the bolometric luminosity and hence the accretion rate may be important
\citep{Hickox2013}.  However, such an effect is not likely to be important
for the ensemble since the mean will remain the same, variability will only 
introduce more scatter. 

\subsection{Black hole mass and Eddington ratio}

\begin{figure*}[th] \centering
\begin{minipage}[c]{0.46\linewidth}
  \includegraphics[width=1.1\textwidth]{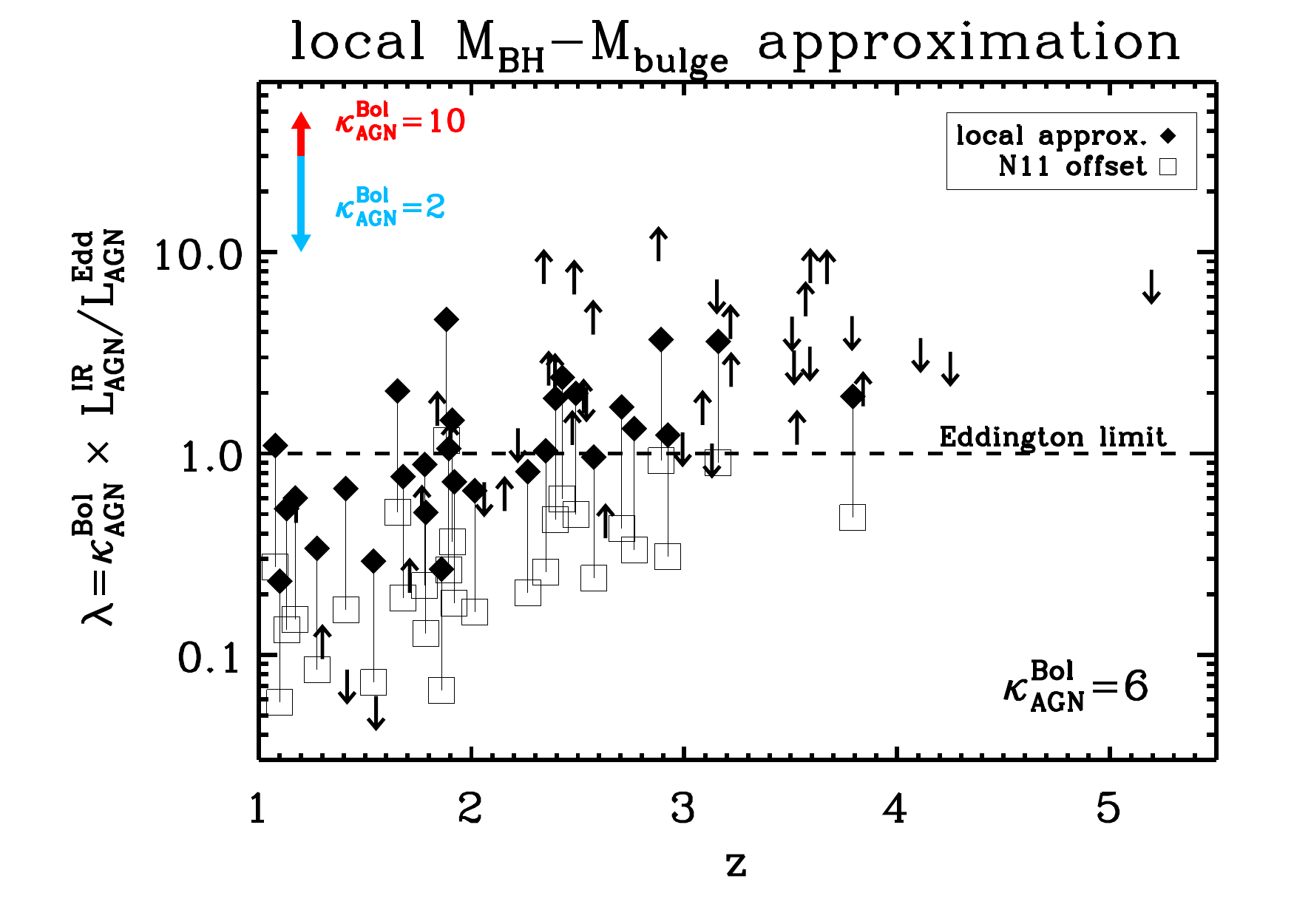}
\end{minipage} \hfill
\begin{minipage}[c]{0.46\linewidth}
  \includegraphics[width=1.1\textwidth]{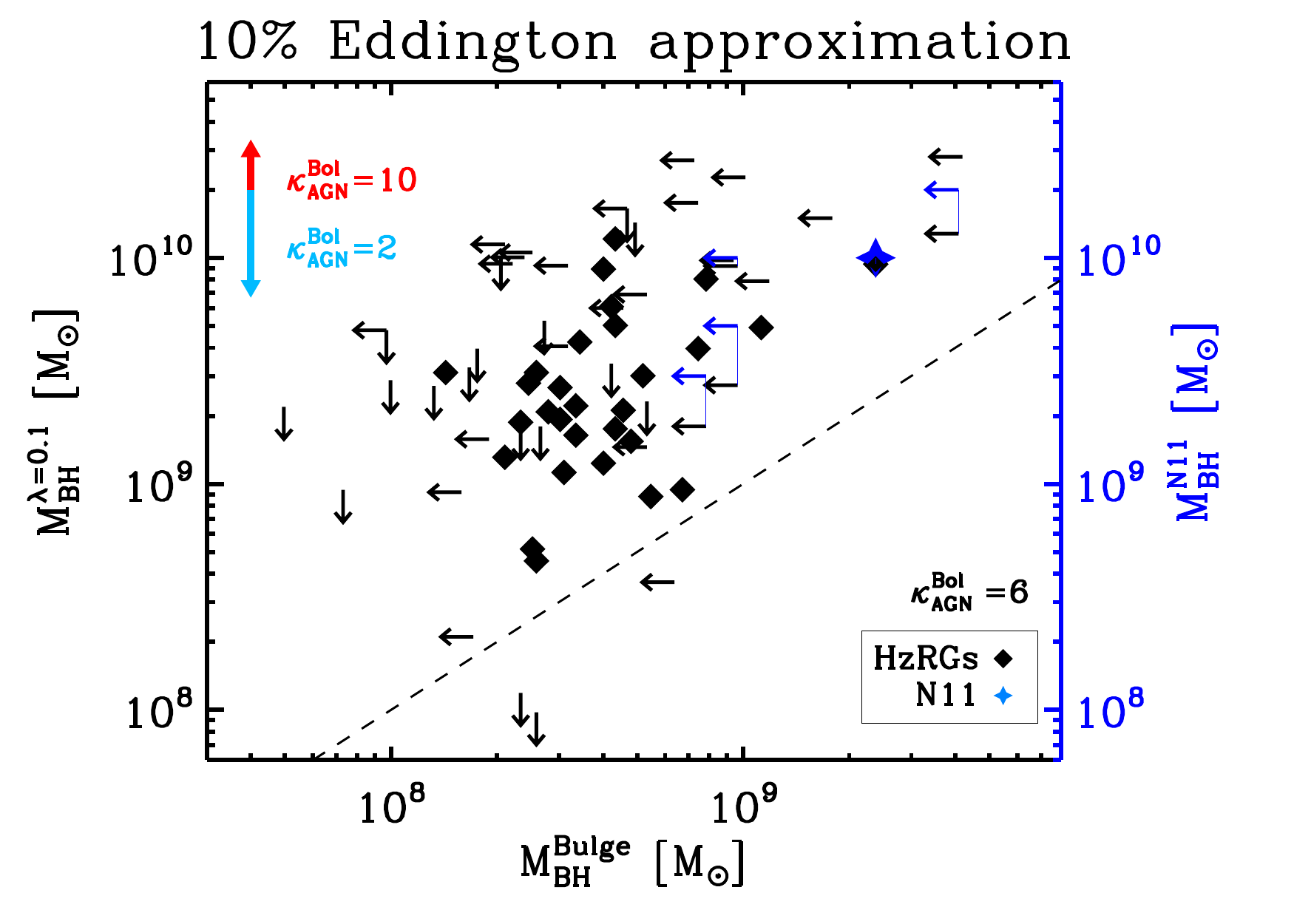}
\end{minipage}
\caption{Illustration of the two different, mutually exclusive assumptions (\S~\ref{sec:bh_hypo}). 
On both figures, we assume  $\kappa^{\rm Bol}_{\rm AGN}=6$. We also plot the 
influence of  $\kappa^{\rm Bol}_{\rm AGN}$ with a vector of the most likely values (see Appendix C).
{\it Left:} Eddington ratio $\lambda$ (Eq. \ref{eq:lambda}) versus
redshift. We draw the Eddington limit at $\lambda$=1.0 (dashed line). We
assume $\kappa^{\rm Bol}_{\rm AGN}=6$. Black diamonds are the calculated Eddington ratios assuming the 
\mbhgal\ relation while the open squares linked by a thin line represent the 
same black hole assuming the offset from the local relation from 
\cite[][see text for details]{Nesvadba2011}. {\it Right:} \mbh\ from the two 
different hypothesis, assuming the local \mbhgal\ (bottom axis) and 
assuming $\lambda$=0.1 (left axis). The dashed line represent the 
one-to-one relation. The blue points (arrows and star) are the black hole 
masses measured from H$\alpha$ lines \citep{Nesvadba2011} and 
compared to the black hole masses from the stellar masses from 
\citep{DeBreuck2010}. These points therefore do not assume the $\lambda$=0.1.}
\label{fig:BH_hypo}
\end{figure*}

There are black hole mass estimates for five of our objects from broad 
components of the H$\alpha$ emission \citep{Nesvadba2011}. They show
that the black holes in powerful radio galaxies are extremely massive, \mbh$>$10$^9$\msun. 
Only based on this small number, a characterisation of the black holes properties 
over our entire sample is difficult. However, these measurement are especially useful
for comparison with the assumptions in the following sections. 

To increase the number of \mbh\ estimates, we will use empirical relationships based on both \mstel\
and \liragn\ to estimate the black hole mass in our sample. These two
approaches are somewhat degenerate, as they are using the same data,
with two different sets of assumptions. We first define the assumptions
made in using the total stellar mass and the local \mbhbulge\ relation
to estimate black hole masses. We then use the infrared luminosity of the
AGN combined with this mass to calculate the Eddington ratio.
We also present a second approach where we fix the Eddington ratio ($\lambda$=0.1)
and then use the infrared luminosity of the AGN to estimate the black hole mass.
We took these two approaches to constrain the possible ranges for
the Eddington ratios and/or black hole masses and to isolate the impact of the various assumptions
that go into these sorts of estimates. We also mention the systematics that would
affect our results when relevant, and summarise them in Appendix D. 

\subsubsection{Black hole mass}
\label{sec:mbh}

As previously noted, all the galaxies in our sample have estimated
stellar masses \citep{Seymour2007,DeBreuck2010}. All are very massive,
with M$_{\rm stel}$$>$10$^{11}$M$_{\odot}$. When \hst\ imaging is available, the best fit light profiles
are consistent with $n$=4 profile \citep[][Appendix D]{Pentericci2001}, suggesting
that the luminosity weighted mass distribution has a spheroidal
morphology (even if some discrepancies are observed). Since the 
mass of the black hole is related to the spheroidal mass\footnote{Even 
in the case of a non-spheroidal geometry, the most important factor 
appears to be observed the mass of the galaxy, with little evolution 
with redshift \citep[][ Appendix D]{Jahnke2009}}, we can use the local \mbhbulge\ relationship 
to estimate the black hole mass, \mbh \citep{Neumayer2004}:

\begin{equation}
  {\rm log}_{10}(M_{\rm BH}/M_{\odot}) = 8.2 + 1.12 {\rm log}_{10}(M_{\rm bulge}/10^{11} M_{\odot}),
  \label{eq:mbh}
\end{equation}
where \mbh\ and $M_{\rm bulge}$ are in \msun.  We therefore refer to this approximation as the local approximation (see Fig. \ref{fig:BH_hypo}, left).

Out of the five sources in our sample with independent black hole mass estimates, only one \mbh\ is directly
comparable given the upper limits on the stellar mass (due to AGN torus contribution 
in the other 4 sources in the near-IR). For this source,
{\it MRC\, 0156-252}, the derived \mbh\ from stellar mass lies at a factor $\sim$4
below that estimated using the broad H$\alpha$ emission. The four remaining
sources suggest also a significant offset in respect to the \mbhbulge\ 
relation \citep[see Fig. 4 in][]{Nesvadba2011}. We will therefore refer
to this offset as the N11 offset. 

\subsubsection{Eddington ratio}
\label{sec:edd_rate}

The Eddington ratio represents 
the rate at which a black hole is accreting compared to the maximal accretion
rate considering a spherical accretion (i.e. Eddington limit). This Eddington ratio 
($\lambda$) is defined as follows:

\begin{equation}
\lambda= \frac{\kappa^{\rm Bol}_{\rm AGN} L_{\rm AGN}^{\rm IR}}{L^{\rm Edd}},
\label{eq:lambda}
\end{equation}
where \liragn\ is in \lsun, $\kappa^{\rm Bol}_{\rm AGN}$ is the bolometric correction from
IR (set to 6 here, see \S~\ref{sec:mdot}, Appendix C and Appendix D) and the Eddington 
luminosity (the maximal luminosity radiated at given black hole mass), \ledd, is defined as:

\begin{equation}
L^{\rm Edd}=\frac{4\pi {\rm GM}_{\rm BH}m_{\rm p}c}{\sigma_{\rm T}}=3.29\times10^{4}\times M_{\rm BH}^{\rm Edd},
\label{eq:ledd}
\end{equation}
where $m_{\rm p}$ is the mass of the proton, $G$ the gravitational constant, $c$ the speed of light, $\sigma_{\rm T}$ the Thomson cross
section of the electron, \ledd\ is in \lsun\ and \mbhedd\ in \msun. Rearranging 
Eq. \ref{eq:lambda} and Eq. \ref{eq:ledd}, one can obtain an estimate of the black hole
mass for a given Eddington ratio and IR luminosity. Observations on quasars 
show a typical Eddington ratio $\lambda$$\sim$0.1 \citep{Kollmeier2006,Vestergaard2009,Ballo2012}. We therefore
consider an alternate black hole mass defined through Eq. \ref{eq:lambda} and Eq \ref{eq:ledd}, 
making use of \liragn\ and setting $\lambda$=0.1. We refer to this approximation as the 10\% Eddington approximation (see Fig. \ref{fig:BH_hypo} right, Appendix C and Appendix D for a discussion of the systematic effects).

\subsubsection{Two hypotheses on Eddington ratio and black hole mass}
\label{sec:bh_hypo}

Figure \ref{fig:BH_hypo} summarises the two previously introduced methods to estimate \mbh\
(with stellar mass, \S~\ref{sec:mbh} the local approximation and 
Eddington ratio, \S~\ref{sec:edd_rate}, the 10\% Eddington approximation). 

The left panel presents the Eddington ratios calculated assuming the local \mbhbulge\ 
relation with black diamonds and with the N11 offset from the same relation 
from \cite{Nesvadba2011} as empty squares. The latter implies a lower $\lambda$ 
as they have a larger black hole mass. We also illustrate the Eddington limit ($\lambda$=1). 
While $\lambda$ suggests an increasing trend with redshift (factor of $\sim$10 
between $z$=1 and $z$=3, with or without the N11 offset), the main difference holds in the 
range of  Eddington ratios. We stress that our uncertainties on \mbh\ are still 
consistent with black holes close to the Eddington limit in both approximations without any 
need to invoke super Eddington accretion.  Anyway, this is interesting as it suggests that 
to grow rapidly, the SMBH need to accrete close to the Eddington limit to produce their high
bolometric luminosities. Moreover, this result seems consistent 
with quasars, where an increase in $\lambda$ between $z$=2 and $z$=6 has 
been observed \citep[e.g.][]{Willott2010, Urrutia2012}. 

The right panel presents the two different inferred black hole masses 
(local and 10\% Eddington) plotted against each other. It is clear from this 
plot that in the case of the 10\% Eddington approximation, all black holes 
appear more massive than suggested by the local \mbhbulge\ relation, i.e. 
above the dashed line (1:1 relation). As a comparison, we overplot the black 
hole mass measurements for \cite{Nesvadba2011} scaled from the right axis 
(in blue). It seems that the 10\% Eddington approximation reproduces  
well the measured black hole masses from \cite{Nesvadba2011} (within a factor 2).

Independently from these assumptions, we observe here \mbh$>$10$^9$\msun\ at $z$$>$1. 
Optical studies of SDSS quasars \citep[e.g.][]{Vestergaard2009}
show that, though rare, $>$10$^9$\msun\ \mbh\ are not exceptional 
at high redshift. This implies that such extremely massive black holes have 
acquired most of their mass by $z$=2-4 as no significantly 
more massive black holes are found in the local Universe 
\citep[][for a recent review]{Kormendy2013}. We would therefore be observing 
the progenitors of the most massive and quiescent black holes at $z$=0. 

We warn that the last results are degenerate. The only
way to overcome this deficiency is through independent measurements of
the black hole masses or better constraints on the Eddington ratio ($\lambda$).
Constraining the former allows us to bypass the \mbhbulge\ relation,
while constraining the latter allows us to estimate the black hole mass 
without the 10\% Eddington approximation ($\lambda$=0.1). 

\subsection{Specific growth properties}

Two of the most challenging questions in modern astrophysics are determining
the relative growth rate of galaxies and how this growth is related
to the growth and activity of their central supermassive black holes.
The relative growth of galaxies and their black holes can be specified
as the specific star formation rate (sSFR) and the specific black
hole accretion rate (\smdot). Galaxies at high and low redshift follow
 a reasonably tight ``main sequence'' of star formation in the
SFR-M$_{\star}$ plane \citep[e.g.][]{Noeske2007, Elbaz2007, Elbaz2011}. How
is the relative growth rate of the stellar mass of radio galaxies
related to the general population of star forming galaxies?  We have
already shown that powerful high redshift radio galaxies are forming
stars at very high rates but they are also massive.  Are their relative
growth rates, their sSFR, higher than normal star forming galaxies?
Being very luminous AGN, we know their black hole accretion rates are
high, but is the supermassive black hole growing at a relative rate
that is consistent with maintaining the relationship of spheroid mass
and black hole mass similar to what is observed at low redshift?

\subsubsection{sSFR of high redshift radio galaxies}

\begin{figure}[t]
\includegraphics[width=0.5\textwidth]{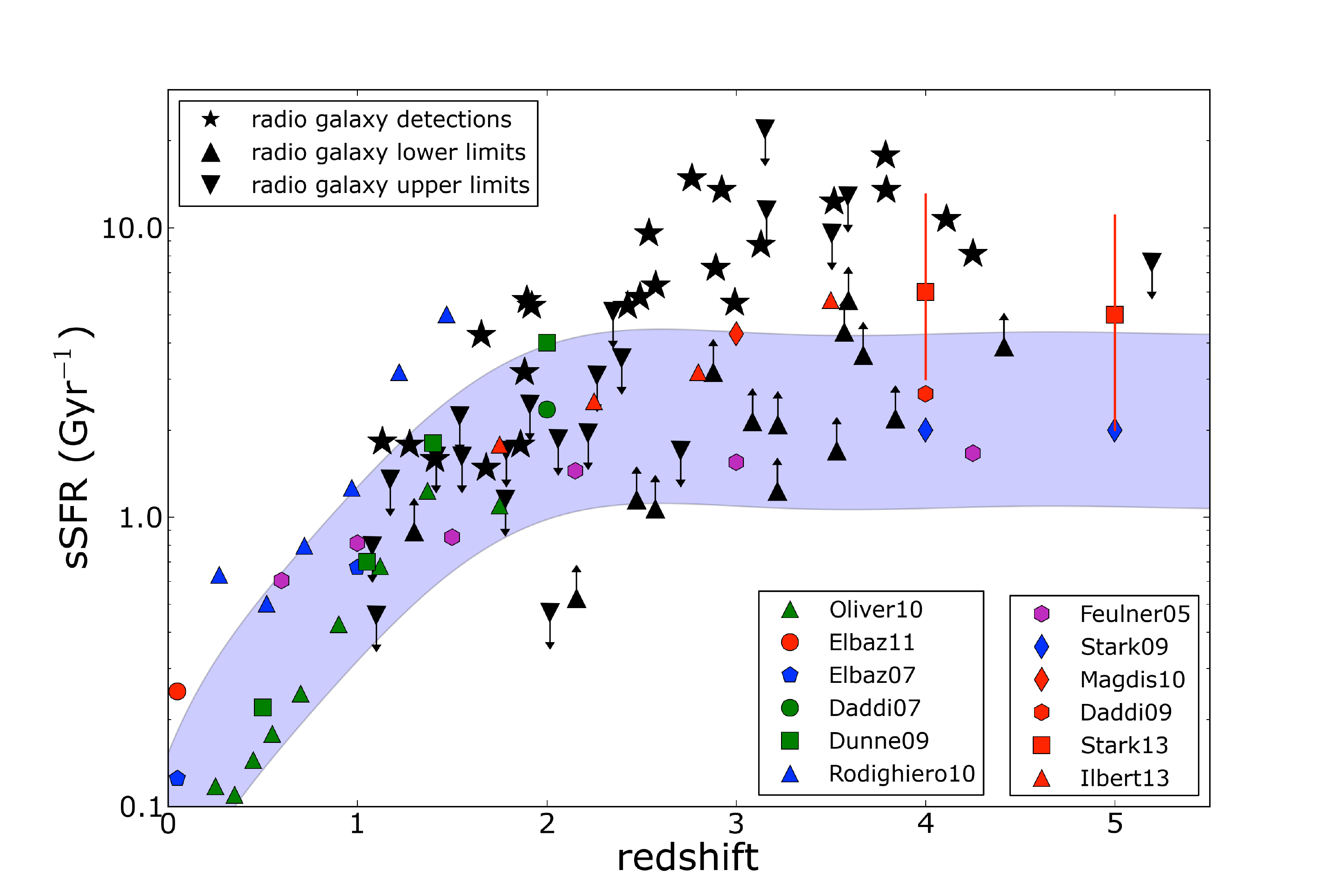}
\caption{The specific star-formation rate (sSFR, in Gyr$^{-1}$) as a
function of redshift. The various colored points represent measurements
from the literature at M$_{\star}$$\sim$10$^{10}$ M$_{\sun}$; see the
references in the legend at the bottom right. Since the slope of the
sSFR-M$_{\star}$ relation is approximately zero, the rate at which
the sSFR evolves is largely independent of M$_{\star}$.  Thus this is
an appropriate comparison, even for galaxies as massive as the radio galaxies
studied here.
The black stars, triangles, and upside-down triangles
represent the radio galaxy detections, lower limits, and upper limits to
the sSFR, respectively.  The blue shaded region represents the scatter in
the observed sSFR values ($\pm$0.3 dex). This rendition of the evolution
of the sSFR is inspired by a similar plot in \citet{Weinmann2011}. See also 
Lenhert et al., in prep. }
  \label{fig:z_vs_ssfr}
\end{figure}

The specific star formation rate provides an estimate of the instantaneous
relative stellar mass growth rates of galaxies.  If galaxies were able to sustain
their star formation over a significant time and at the rate observed, the
inverse of the sSFR is the mass doubling time scale. Interestingly, the
mass doubling time scale is shorter than a Hubble time at all redshifts,
becoming comparable to the Hubble time at $z$=0.  This suggests that if
the galaxies have long duty cycles, they can grow their masses relatively
quickly at high redshift. Over the redshift range spanned by our radio
galaxy sample, the sSFR of the population of star forming galaxies is
approximately constant ($\sim$2 Gyr$^{-1}$) or only slowly increasing with redshift
\citep[e.g.][]{Feulner2005, Rodighiero2010, Stark2013, Ilbert2013}.

Compared to the ensemble of
distant star forming galaxies, we find that generally the radio galaxies
are growing relatively more rapidly (Fig.~\ref{fig:z_vs_ssfr}). The
combination of upper limits in the \herschel\ photometry and the stellar
masses, the number of galaxies with constrained sSFR is only a fraction
of the entire sample ($\sim$30\%).  Nevertheless, a significant trend for the sSFRs
of the radio galaxies to increase with increasing redshift can be seen.
This increase can be characterized as at z$\la$2.5, the sSFR of the
radio galaxies is comparable to that of the normal star-forming galaxy
population (i.e., no AGN) while at $z$$\ga$2.5, the radio galaxies
lie significantly above (about a factor of 3) the galaxy population.
The scatter for the radio galaxies is a bit higher than in the galaxy
population, $\sim\pm$0.5 dex compared to $\sim\pm$0.3 dex for the non-AGN
galaxies.  However, given the more limited number of radio galaxies,
this difference is not significant. The offset to higher values by 
a factor $\sim$3 for the radio galaxies at $z$$\ga$2.5 is significant.
This offset is to be compared to the systematic uncertainties associated 
with the calculation of the sSFR (see Apendix D). Notably, the uncertainties 
on the IMF are already taken into account in the blue area \citep[][]{Weinmann2011}.
Overall, the results might shift due to systematic uncertainties, but differentially, it is unlikely to wipe out any 
differences between our sample and the ensemble of galaxies. We also note 
that radio galaxies are at the bright-end of the K-$z$ diagram \citep{Rocca2004}, 
and present comparable \lirsb\ than SMGs (see \S~\ref{sec:sfr}), so there are naturally expected to lie in a
different area of the SFR-M$_{*}$ diagram than normal, more quiescent galaxies \citep[e.g.][]{Elbaz2011}.

The cause of the offset at higher redshifts in the sSFR of radio galaxies
compared to the normal population of star forming galaxies is not known.
Morphological evidence for galaxies lying above the main sequence of
star formation suggests that mergers may play a significant role in
increasing the sSFR \citep[e.g.][]{DeBreuck2005,Elbaz2011, Sargent2013}.  This picture seems
consistent with the trend for radio galaxies to appear in disturbed
systems \citep[e.g.][]{Ivison2008,Ivison2012,Seymour2012,Wylezalek2013a}.  However,
whether or not merging is the only cause of perturbed systems is
still an open question for the sample of radio galaxies studied here,
especially in light of the fact that radio galaxies generally lie
in galaxy over-densities \citep{Pascarelle1996,Venemans2007, Overzier2008,
Kuiper2010, Wylezalek2013b}.  Galaxies in over densities at high redshift
may preferentially have higher sSFRs \citep[][Cooke et al., in prep.]{Elbaz2007}.

Whatever the cause of their elevated sSFRs,
what is clear is that the mass doubling time of the powerful radio galaxy
population is short, only about 100 Myr at z$\ga$2.5.  If such a
relative growth rate could be sustained for 1 Gyr, the typical radio
galaxy would have grown by a factor of 1000.  Thus despite their high
masses, the current star formation rate and relative growth rate do not
need to be sustained for a significant fraction of the local Hubble time
(1-2 Gyrs over the redshift range spanned in our sample). Notably, the
mass depletion time scales are generally very short, of order 100 Myr
or less \citep{Ivison2012, Emonts2011, Emonts2013}.  This either suggests
that powerful radio galaxies generally represent the last phases of
their rapid growth or that, given their relatively rich environments,
they are being continuously (re-)fuelled.  Their large stellar masses,
significantly greater than the fiducial stellar mass turnover in the galaxy
co-moving volume density and their overall consistency with the old ages
derived for local early type galaxies, suggest these are the almost fully
formed progenitors of local early type galaxies \citep{Bernardi2010}.  So it
may well be that these galaxies are at the end of their formation epoch.
They likely formed the bulk of their stars at much higher redshifts,
consistent with the stellar synthesis fitting to the broad-band SEDs of
a few of these galaxies \citep{Rocca2013}.

\subsubsection{Specific black hole accretion rate}
\label{sec:smdotbh}

\begin{figure}[t] \centering
\includegraphics[width=0.5\textwidth]{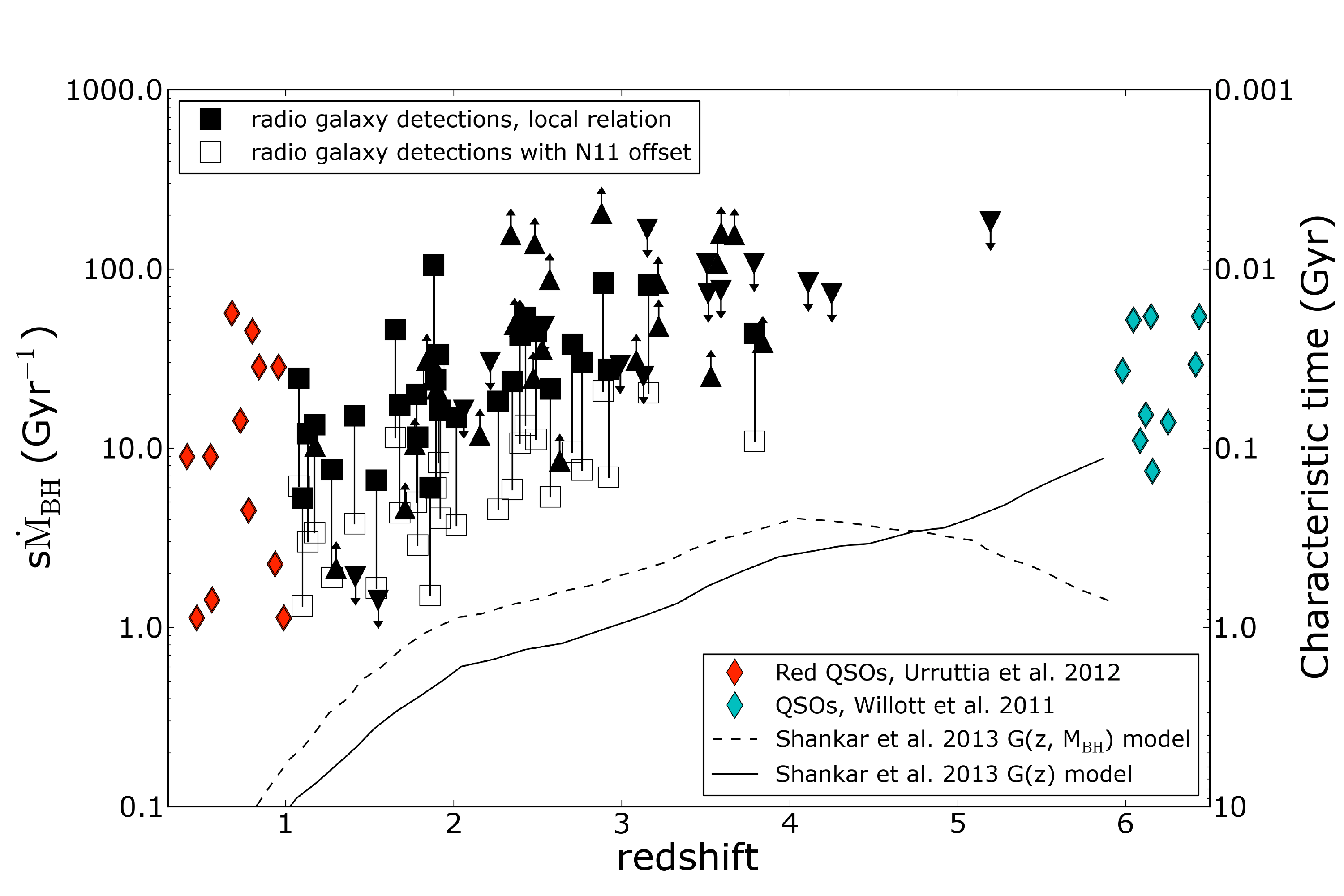}
\caption{\smdot\ versus redshift. \smdot\ is reported here with the two assumptions
on black hole masses discussed on Fig. \ref{fig:BH_hypo}. The black filled squares are
the local \mbhgal\ assumption and the empty square the offset expected from \cite[][ see \S~\ref{sec:bh_hypo}]{Nesvadba2011}.
We also report other QSO samples for comparison \citep[as indicated in
the legend; ][]{Willott2010, Urrutia2012}.  In addition, we compare our
specific accretion rates with predictions of two models for the growth
rate of \mbh=$10^9$\msun black holes similar to our mass estimates
\citep[dashed and solid thick lines, see text for details; ][]{Shankar2013}.}
\label{fig:z_vs_smdot}
\end{figure}

The specific black hole accretion rate (\smdot) corresponds to the inverse of
the mass doubling time for the black hole. Following \cite{Shankar2013},
\smdot\ can be defined as follows:

\begin{equation}
{\rm s}\dot{\rm M}_{\rm BH} = 2.5\times10^{-9} \frac{\lambda}{f}=\frac{\dot{\rm M}_{\rm BH}^{\rm acc}}{{\rm M}_{\rm BH}},
\label{eq:g}
\end{equation}
where $\lambda$ is the Eddington ratio (calculated in \S~\ref{sec:edd_rate}), 
and $f$ is the actual fraction of mass feeding the black
hole and increasing its mass, defined as $f$=$\epsilon$/(1-$\epsilon$), where
$\epsilon$ is the radiative efficiency factor (see Eq. \ref{eq:mdot}).
Fig. \ref{fig:z_vs_smdot} presents a mildly increasing value of \smdot\ 
with redshift, a flattening at $z$=4-5, and then a possible decline at $z$$>$6. 
This behaviour is very similar to what we observed in the sSFR (Fig.~\ref{fig:z_vs_smdot}).
The characteristic time for the growth of the black
hole is $t_S$$<$100\,Myr for $\lambda >$0.5 and $\epsilon$=0.1 
(see \S~\ref{sec:edd_rate} and Appendix D for related uncertainties).

How do these estimates compare to other classes of powerful AGN?
In order to compare our radio galaxies to similar objects, we also show
estimates of \smdot\ from various samples of quasars from the literature
and we use Eq. \ref{eq:g} in order to calculate \smdot\ from the
Eddington ratio for these samples. The high redshift sample from \cite{Willott2010}
provides the necessary constraints for high redshift optically selected quasars at 
$z$$\sim$6.2, while the sample from \cite{Urrutia2012} provides this information for the 
lower redshift ($z$$\sim$0.7) red quasars. Both of these samples
of quasars have similar specific accretion rates compared to high
redshift radio galaxies under a similar set of assumptions. So high
specific accretion rates appear to be a generic feature of bolometrically
luminous AGN whether or not they are radio loud.  

In addition, Fig. \ref{fig:z_vs_smdot} also compares our specific accretion rate estimates with
models for the growth of supermassive black holes \citep{Shankar2013}. The
models we are specifically comparing to are for the evolution of very
massive black holes, \mbh=$10^9$\msun, similar to what we think are
the best mass estimates for the black holes in these radio galaxies
\citep{Nesvadba2011}.  The models of \citet{Shankar2013} that
satisfy the most observational constraints are those that allow the
peak of an assumed Gaussian distribution of Eddington accretion rates
to evolve with redshift, G($z$), or with both redshift and black hole
mass, G($z$, M$_{\rm BH}$). These models for \mbh=$10^9$\msun\
fall far below our estimates, by 1-2 orders of magnitude. While
these models seek to reproduce the average growth rate of very massive
black holes, they fail to reproduce the intense growing phase of the
``quasar mode''.  However, black holes this massive are rare and do not
contribute significantly to the overall growth of black holes at high redshift.
However, it is also true that because of their extreme masses lying at the
exponential end of the mass function, they provide strong constraints
on any model.  This is especially true given their potential impact on
their host galaxies.  Having specific accretion rates off by orders of
magnitude means the impact of the AGN on its surroundings, whether it
be the surrounding interstellar, intra-cluster,
or intergalactic medium, will be greatly underestimated.

\subsection{Are the SMBHs outgrowing their hosts?}

\begin{figure}[t] \centering
\includegraphics[width=0.5\textwidth]{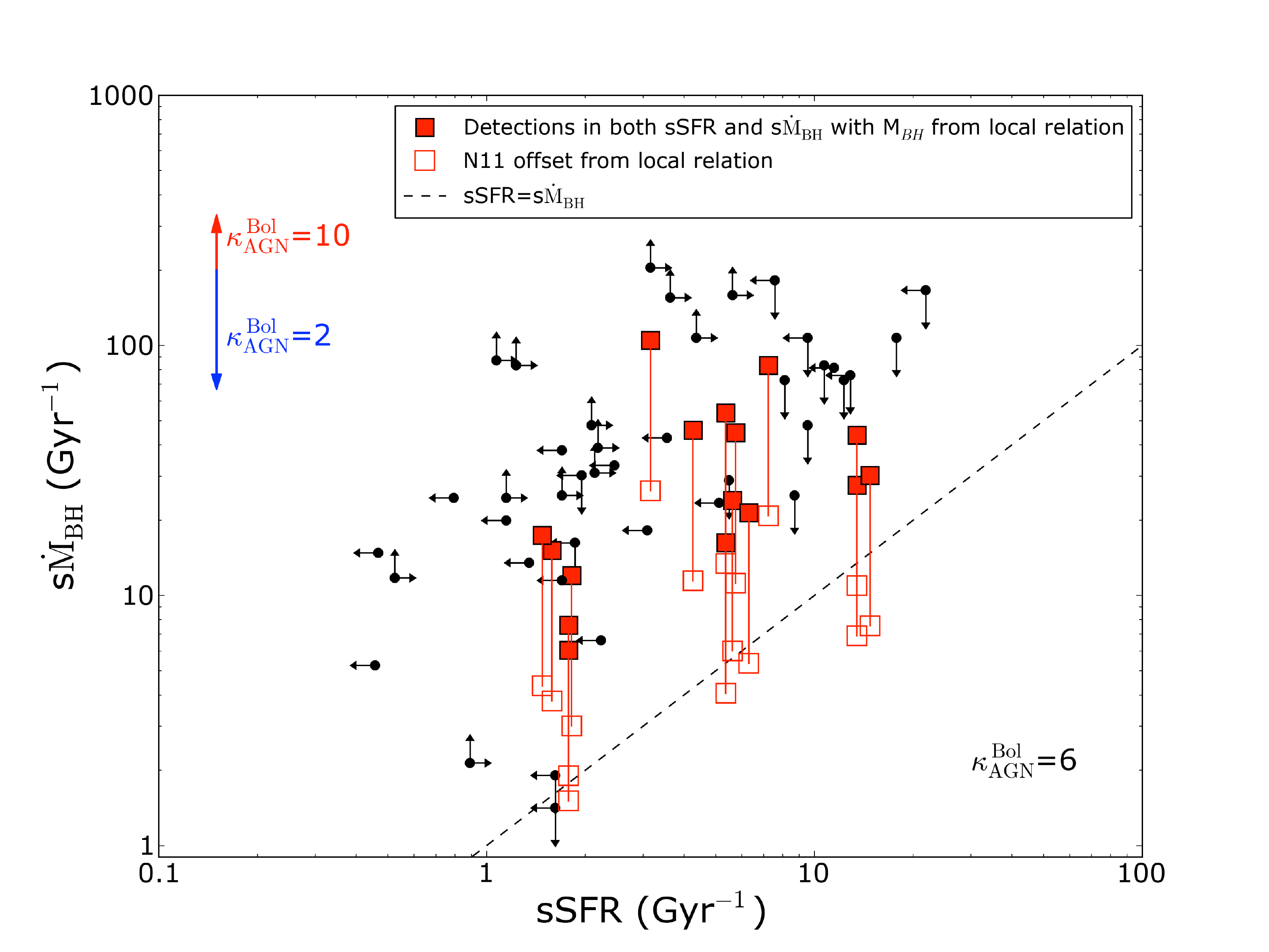}
\caption{The specific star-formation rate (sSFR, in Gyr$^{-1}$) versus the
specific black hole accretion rate, \smdot\ (in Gyr$^{-1}$).  The galaxies
with constrained estimates of both sSFR and \smdot\ are shown in red, while
the arrows indicated upper or lower limits for each or both quantities.
\smdot\ was estimated using the local \mbhbulge\ relation (filled red diamonds)
or using the N11 offset (empty red squares, see \S~\ref{sec:bh_hypo}).
The dashed line indicates where the \smdot=sSFR.}
\label{fig:smdot_vs_ssfr}
\end{figure}

The accretion rate of the BH compared to the SFR 
(Fig.~\ref{fig:liragn_vs_lirsb}), the relatively high BH accretion
rates relative to the Eddington limit (Fig.~\ref{fig:BH_hypo}),
and \smdot\ compared to sSFR (Fig.~\ref{fig:smdot_vs_ssfr}) all seem to
suggest that the super massive black holes in powerful radio galaxies
at high redshift may have out-grown and/or may be out-growing their host
galaxies.  Whether the black hole is too massive compared to its host,
or it is accreting at large Eddington ratio, the black hole of radio
galaxies is or will be (in a near future) off the \mbhbulge\ relation.
This offset is happening when the black holes are growing very rapidly,
with characteristic times of $<$100 Myr for doubling their masses.
At the same time the stellar mass of the hosts are also growing very rapidly, 
but apparently not rapidly enough (Fig.~\ref{fig:smdot_vs_ssfr}). Thus to
recover the local ratio of black hole and host mass as observed locally,
the host of radio galaxies need to ``catch up'' with their overly massive black
holes. We therefore have potential evidence for non-coeval growth of
the radio galaxies and their central black holes. In the process, black
holes appear to grow first, extremely quickly and efficiently, then the
host catch up to fall again the observed local ratio \mbh/$M_{gal}$.

Knowing the duration of the AGN phase would allow to estimate the time
lag, $R_{\rm lag}$, which is how long the host galaxy will need to grow at
its current rate to get back on the local ratio of stellar to black hole
mass. We can calculate this relative time lag assuming a simple model of
constant growth or at least a growth rate with a well defined average.
Let us define the final mass of the galaxy (or BH) after a episode of
growth as a linear equation:

\begin{equation}
M_{\rm gal}^{\rm final}=M_{\rm gal}^{\rm init} + {\dot{M}}_{\rm gal} \Delta T_{\rm gal},
\end{equation}
where $\Delta T_{\rm gal}$ is the duration of the star formation (or
fueling of the black hole, $\Delta T_{\rm BH}$), $\dot{M}_{\rm gal}$
is the star formation rate (or the mass accretion rate on to the black
hole, \mdotbh), and the initial, M$_{\rm gal}^{\rm init}$, and final
mass of the galaxy, M$_{\rm gal}^{\rm final}$ (or BH, M$_{\rm BH}^{\rm
init}$ and M$_{\rm BH}^{\rm final}$). We parameterize the growth of
galaxies and black holes in exactly the same linear way.  Assuming that the
galaxy and the black hole start and end on the ratio of their masses as
defined by the local stellar mass of the spheroid mass black hole mass
relationship, dividing the two equations allows us to estimate the ratio
of the duration of the star formation and the mass accretion onto the BH
that enables the galaxies to have the local M$_{\rm gal}$/M$_{\rm BH}$
ratio. Dividing these two equations and rearranging the terms gives:

\begin{equation}	
R_{\rm lag} = \frac{\Delta T_{\rm gal}}{\Delta T_{\rm BH}} \approx \frac{{\dot{M}}_{\rm BH}/{M}_{\rm BH}^{\rm init}}{ {\dot{M}}_{\rm gal}/{M}_{\rm gal}^{\rm init}} \approx 8.
\end{equation}
The relative time for the mass of the galaxy to catch up
with mass of the black hole such that it falls back onto the local
\mbhbulge\ relation, $R_{\rm lag}$ is simply the ratio between the
\smdot\ and sSFR. For our simple model, this
means if the black hole growth lasts for 20\,Myr, the typical lifetime of
a quasar phase \citep[e.g.][]{Steidel2002,Hopkins2005a}, the galaxy will
need at least 160\,Myr to catch-up with the black hole growth. This is
assuming that the actual measured growth rate is the average over the time
over which the growth occurred under our given framework, 
see \S~\ref{sec:sfr}---\ref{sec:smdotbh}. This simplistic model shows that galaxy
activity needs to be much longer than the AGN phase in order to catch up
after the relative rapid black hole growth. This would naturally explain
the observation that the black holes in radio galaxies lie preferentially
above the M$_{\rm BH}$-M$_{\rm gal}$ relation. It may be that already
the mass of the BH lies above the relation \citep{Nesvadba2011} which
would then require even more time for the host galaxy to catch up.

\subsection{The future evolution of powerful radio galaxies}

We have found that radio galaxies are growing their stellar populations
and supermassive black holes very rapidly. They are able to double their
respective masses in a few 100 Myr or less. Interestingly, it appears
that, relatively, the supermassive black hole is out-growing or has out-grown its
host galaxy.  Crudely speaking, the host galaxy will require
about an order-of-magnitude longer than the lifetime of the AGN for the
host galaxy to catch up.  ``Catching up'' in this context means how much
longer will it take for the galaxy at its current sSFR to have a mass
sufficient to land on the local \mbhbulge\ relation once the BH growth has 
slowed. We propose here to explore different scenarios making use 
of our previously calculated parameters:

\paragraph{ {\bf High redshift powerful radio galaxies will never land on the \mbhbulge\ relation.}} 
In the case of an over-massive black hole (compared to its host),
$R_{\rm lag}$ indicates that the black hole and its host are not growing at the same rate.
The gas supply on larger scales can satisfy the necessary condition to 
feed simultaneously both the black hole and its host galaxy over the same timescale
(though note the severe problems in terms of the physical processes required to bring the 
gas down to the central engine; \citealp{Alexander2012}). However, joint feeding implies 
that radio galaxies might never land on the local \mbhbulge\ relation. In the local Universe only a handful of deviant objects have been 
 observed \citep[e.g.][]{vandenBosch2012}. This scenario is plausible since powerful high redshift radio galaxies are rare objects 
(only few hundred have been observed out to $z$=5 so far).

\paragraph{ {\bf High redshift powerful radio galaxies will eventually land on the \mbhbulge\ relation.} }
 A variation of the previous scenario can be proposed. Indeed, the galaxies can experience their growth through 
mergers. Several merging scenarios can be instigated, both major/minor 
and gas poor/gas rich mergers. Major mergers are rare events but
they are expected to be mostly gas rich at high redshift as the gas fraction 
increases significantly \citep[e.g.][]{Tacconi2010}. Notably, in the case of a major merger, gas can
be efficiently brought to the innerpart of the galaxy ($<$1kpc) and probably
feeds the black hole and star formation simultaneously, allowing a new 
episode of black hole growth. The triggering
event of the radio galaxy episode is still an open discussion, but recent studies suggest
that major mergers can indeed play an important role \citep[e.g.][]{RamosAlmeida2013}. 
In the case of minor mergers, gas rich companions could form the stars
and be accreted within the cosmic time. This scenario is supported by some observational evidence thanks to
high resolution imaging with \hst\ \citep{Miley2006,Seymour2012}. This is also related
to the size evolution of galaxies as well as the
compactness of early-type galaxies at high redshift \citep[e.g.][]{Daddi2005,
vanDokkum2008, Delaye2013}, the change in the mass function with cosmic
time \citep{Ilbert2013}, the light profiles and elemental abundance
ratios in the outer regions of massive ellipticals \citep[][]{Huang2013,
Greene2013}, and the fact that at constant co-moving density, the
mass of massive early-type galaxies grew by about a factor of 4 over
approximately the last 10 Gyr \citep[e.g.][]{vanDokkum2010, Ilbert2013}. 
As HzRGs are sitting in dense environments, 
probably in the centre of proto-clusters \citep[e.g.][]{Wylezalek2013b}, they are likely to experience
an important series of minor dry mergers, consistent with the size evolution scenario.
Therefore, high redshift powerful radio galaxies will finally land on the \mbhbulge\ at $z$=0.

\paragraph{ {\bf High redshift powerful radio galaxies will land on the \mbhbulge\ relation but on a longer timescale.}}
In the case of black holes starting on the local relation and experiencing 
a fast, important growth, they will be significantly off the \mbhbulge\ relation in a relatively 
short timescale ($\sim10$\,Myr). The host, at the current sSFR, will roughly 
need 10 times more in order to ``catch up" with the black hole. Nevertheless,
several observations suggest the contrary. Indeed, to support such a high 
SFR, an important, continuous supply of gas is required. 
Such amounts of gas is presently available in the HzRG systems \citep[e.g.][]{Emonts2013}.
Nevertheless, this gas possesses properties which indicate that even if
a host growth is possible, it is unlikely to happen in this short timescale. Indeed,
HzRGs present copious outflows in ionized and neutral gas \citep{Nesvadba2006,Nesvadba2007}, 
gas with substantial angular momentum \citep{vanOjik1996,Humphrey2008} or gas 
in the close environment \citep{Ivison2012,Emonts2013}. Before participating to the host growth, this gas needs 
first to be driven into the galaxy. Our best guess is that this gas will actually 
participate to the host growth, but in a further episode of star formation. 
The $R_{\rm lag}$ is therefore a lower limit of the time lag between the black hole and its host. 

\paragraph{ {\bf High redshift powerful radio galaxies are examples of a symbiotic growth.}}
A variation of the previous scenario is by invoking a shorter timescale
for the important black hole growth ($<$1\,Myr), the host at the current 
star formation can ``keep up" with its black hole much more easily. Indeed,
typical starburst can last for tens of Myrs \citep[e.g.][]{Kennicutt1998}. 
Although convenient, this scenario imply that we are looking 
at the very peak of the AGN activity for all sources and not at the peak of their star formation. 
If indeed possible, this is unlikely to be the case, 
(i) given the large scatter seen, for instance, in Fig. \ref{fig:liragn_vs_lirsb}; and (ii) because the radio selection 
of our sources which is uncorrelated with the IR luminosity (\S~\ref{sec:ir_radio_cor}). Moreover, as 
mentioned earlier, the variability does induce scatter in our distributions but is not
playing a fundamental role on the average. \\

More observations at high resolution throughout the electromagnetic spectrum, 
and especially tighter constraints on the BH mass and the
mass of the spheroidal component of radio galaxies are needed to test
these different scenarios (see \S~\ref{sec:sfr}---\ref{sec:smdotbh} for our adopted assumptions and
Appendix D for a summary of the uncertainties)

\section{Conclusion}

We present new \herschel\ and sub-mm observations for the HeRG\'E
sample consisting on 70 powerful radio galaxies spanning
1$<$$z$$<$5.2. Complemented by other data sets, we now have continuous
coverage of the IR spectral energy distribution over the range from
16-870\mum.  All galaxies in our sample have integrated IR luminosities
L$_{\rm IR}$ $>$ 10$^{12}$\,\lsun, classifying them as ULIRGs, while half of all the sources at $z$$>$2 have
L$_{\rm IR}$$>$ 10$^{13}$ L$_{\sun}$ and are HyLIRGs.

We use the DecompIR code to decompose the IR SEDs of galaxies in
our sample in a robust and uniform way into an AGN and SB
components. To make these fits, we assume a single AGN template and a
variety of starburst templates.  Our results for the AGN contribution
are conservative in that we assumed a single template and it is
possible that this template could lead to an underestimate of its
contribution to the IR SED.  The estimated \liragn\ and \lirsb\ from
our decomposition imply both high black hole mass accretion rates
(1\msunyr$<$\mdotbh$<$100\msunyr) and vigorous on-going star formation
(100\msunyr$<$SFR$<$5000\msunyr).  Although no strong correlation is
detected between these rates, this result implies that both the black 
hole and its host galaxy are experiencing rapid growth, with the relative 
growth of the black hole exceeding that of the host galaxy.

Assuming empirical relations and basic physical assumptions, we estimate
\mbh\ from the stellar masses and infrared AGN luminosities. The black
holes appear overly massive compared to their hosts and are likely
accreting close to the Eddington limit ($\lambda$$\sim$1), similar to
estimates for radio quiet quasars. Alternatively, for lower Eddington
rates, the black holes are more massive than predicted by the local
M$_{BH}$-M$_{buldge}$ relationship.

We derive the specific growth properties, both the specific star
formation rate, sSFR, and the specific black hole mass accretion rate,
\smdot. Compared to galaxies that lie along the sSFR-stellar mass
relation at $z$$\ga$2.5  radio galaxies
appear to have higher sSFR. At $z$$\la$2.5, radio galaxies appear have the same
or perhaps lower sSFR generally.

We explore different scenarios for the future growth of radio galaxies. These 
scenarios are that high redshift powerful radio galaxies (i) will never land on the \mbhbulge\ relation; (ii) will land
on the local \mbhbulge\ relation, but at low redshift; (iii) will land on the \mbhbulge\
on a longer timescale than our estimated $R_{lag}$; or (iv) are indeed experiencing 
a symbiotic growth. However, observational evidence favours the scenario in which 
radio galaxies will land again on the \mbhbulge\ relation, but on a long timescale 
(most probably $>>$100 Myr).

\begin{acknowledgements} 
GD warmly thanks Clive Tadhunter and Rob Ivison for their comments allowing a significative
improvement of this paper. GD thanks also the referee for his comments helping
to clarify this paper. NS is the recipient of an Australian Research Council Future Fellowship. 
FEB acknowledges support from Basal-CATA PFB-06/2007 and 
CONICYT-Chile grants FONDECYT 1101024 and Anillo ACT1101. 
EI acknowledges funding from CONICYT/FONDECYT postdoctoral project N$^\circ$:3130504.
The work of D.S. was carried out at Jet Propulsion Laboratory, California Institute of Technology, under a contract with NASA.
Herschel is an ESA space observatory with science instruments provided
by European-led Principal Investigator consortia and with important
participation from NASA.  This work is based in part on observations made
with the {\it Spitzer Space Telescope}.  This work is based on observations
made with the APEX Telescope, based on the Chajnantor Plateau in Chile.
HIPE is a joint development by the Herschel Science Ground Segment
Consortium, consisting of ESA, the NASA Herschel Science Center, and
the HIFI, PACS, and SPIRE consortia.
{\it Facilities: Spitzer, Herschel, APEX}

\end{acknowledgements} 

\bibliographystyle{aa.bst}
\bibliography{Library}

\clearpage
\appendix

\section{Notes on sources}
\paragraph{{\it B2~0902+34} (WCD with 3 detections):}
This object is the only radio galaxy from our sample to be most likely dominated by synchrotron emission
\citep{Archibald2001}.  We therefore treat this galaxy as if it were
actually a member of the UL class for the purposes of fitting its
SED.

\paragraph{{\it 4C~23.56} (WD with 5 detections):}
\label{sec:4C2356}
This object is the prototypical case where the IR emission is dominated
by the emission from the AGN. There are other pieces of evidence
from other wavelengths to support this dominance.  For instance,
rest frame UV shows strong polarisation \citep{Cimatti1998}; the IRAC
colors are characteristic of sources dominated by AGN emission in rest
frame near-IR \citep[Fig. ][]{DeBreuck2010}; X-ray emission is also
prominent and suggestive of emission from an AGN.  This radio source
can be seen as having the most extreme AGN contribution to its SED in
our sample. We stress that the Mullaney AGN template reproduces well the
SED of  4C~23.56 without any modification. This indicates that the AGN 
DecompIR template can be a good representation of AGN emission in our sample.

\paragraph{{\it 4C~41.17} (WCD with 7 detections):}
Of course, with a radio galaxy dominated by its AGN in the infrared, it would
be interesting to have the opposite, a radio galaxy dominated with its IR
SED dominated by star formation.  {\it 4C~41.17} likely represents such
a case. This radio source has a SB dominated SED,
and can be reproduced well by the SB6 template. A more complete SED decomposition
confirms this results \citep{Rocca2013}. 

\section{AGN or SB dominated ?}

\begin{figure}[t] \centering
\includegraphics[width=0.5\textwidth]{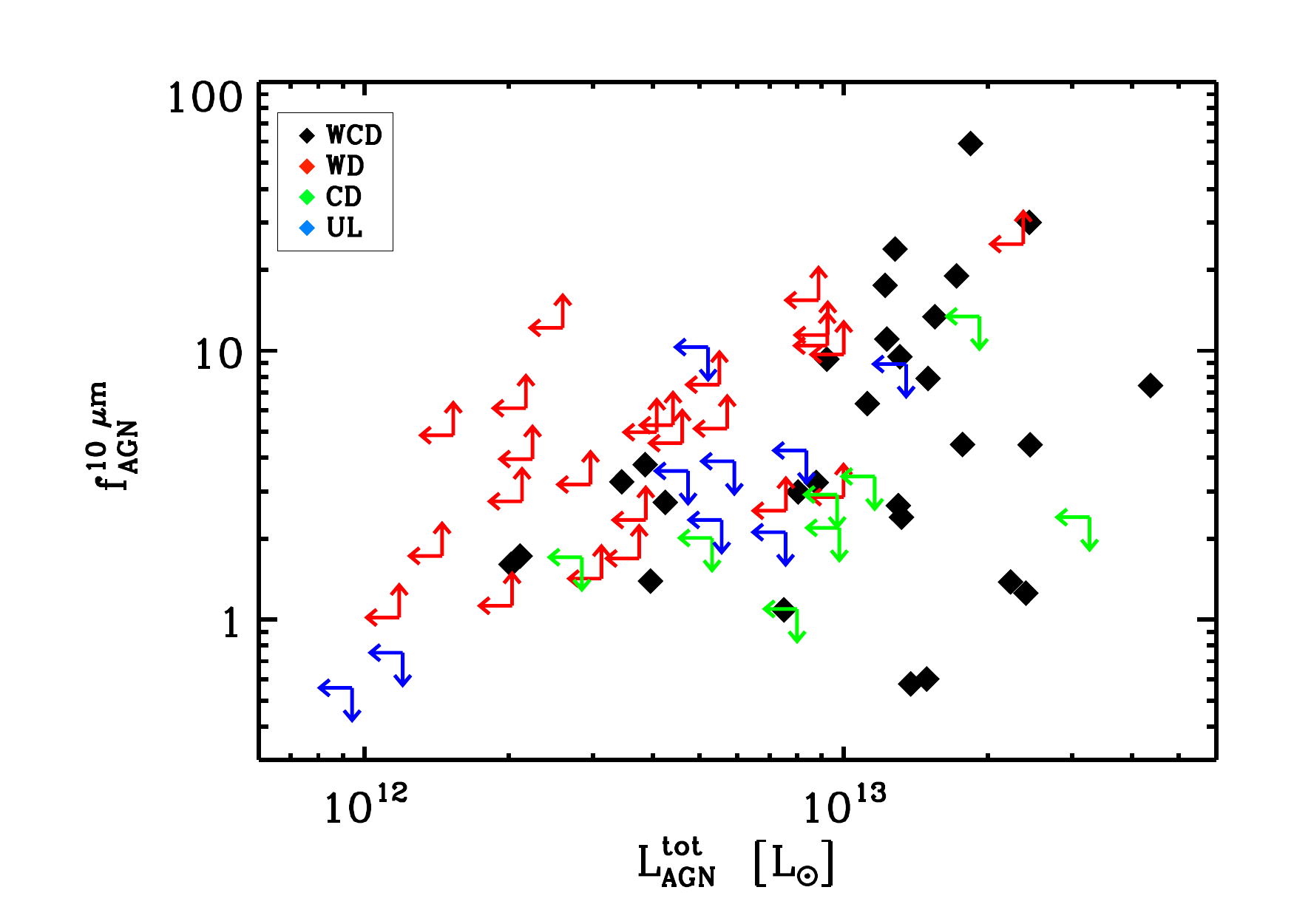}
\includegraphics[width=0.5\textwidth]{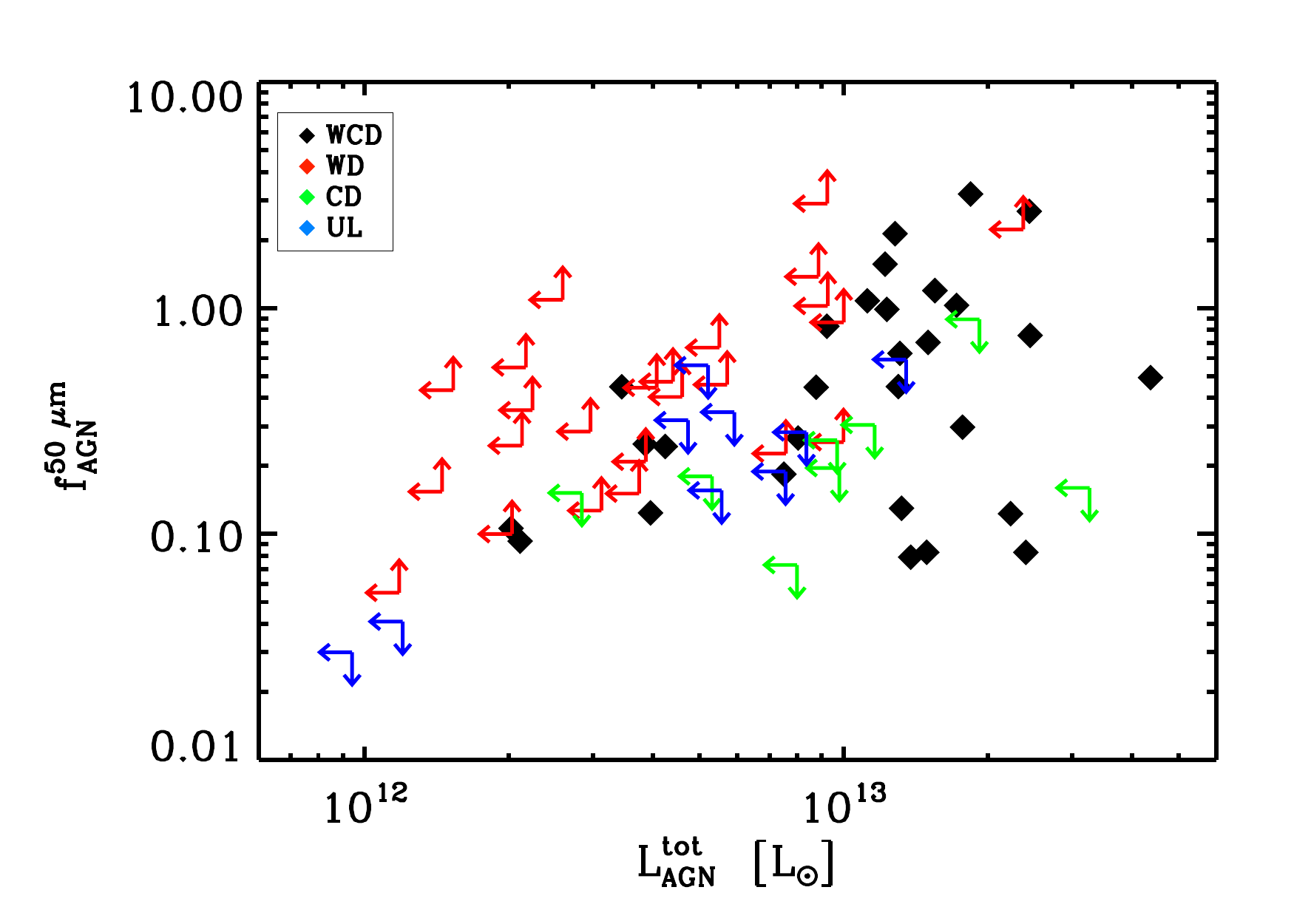}
\includegraphics[width=0.5\textwidth]{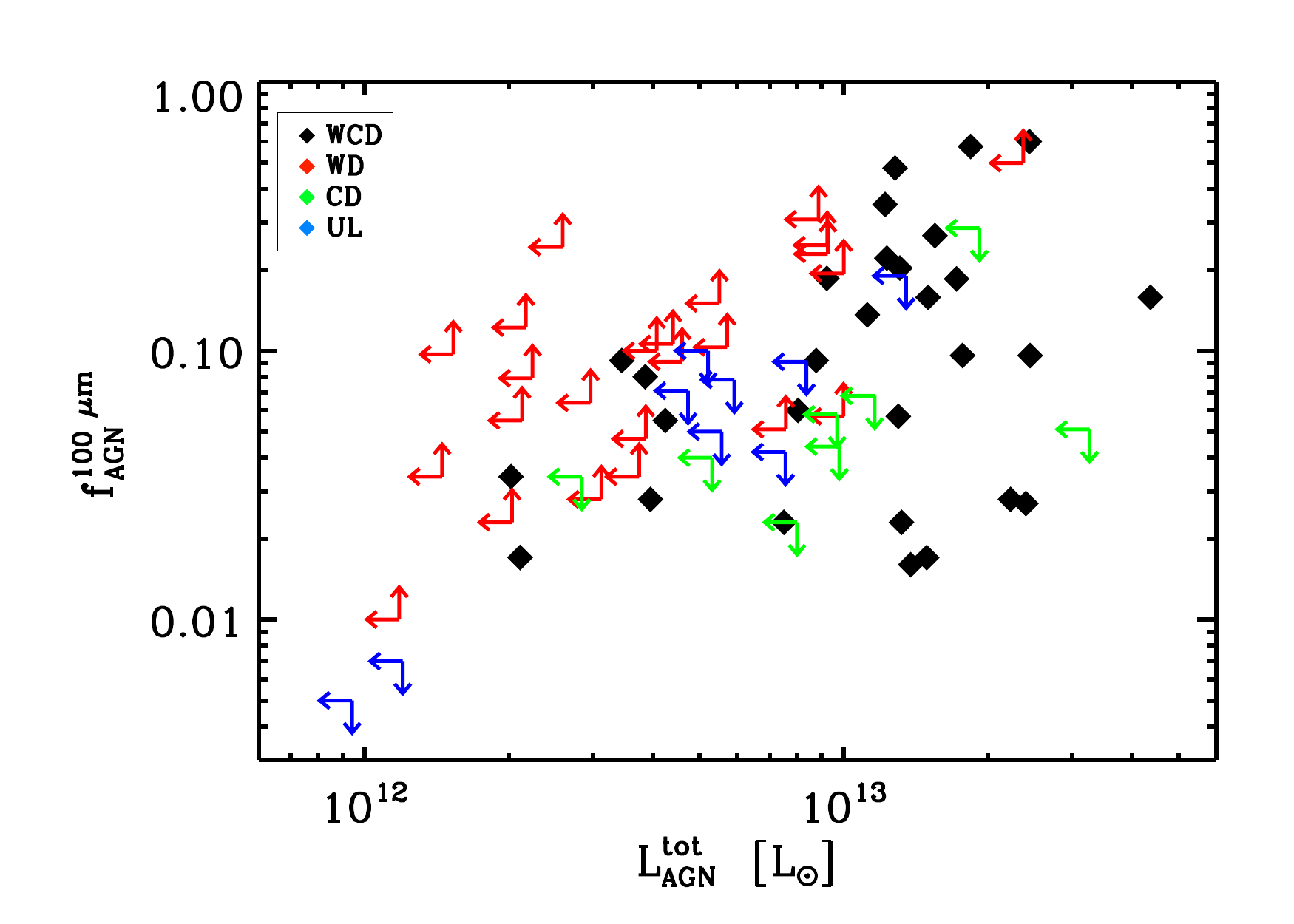}
\caption{From top to bottom, AGN fraction at 10, 50 and 100\,\mum\ 
against the total infrared luminosity. The colour and symbols corresponds to the class defined in \S~\ref{sec:class}} 
\label{fig:fagn_lirtot}
\end{figure}

We remind that $f_{\rm AGN}$ is defined as the ratio S$_{\rm AGN}$/S$_{\rm SB}$ 
where S is the flux of the AGN and the SB respectively, at 10, 50 and 100\mum.
Figure \ref{fig:fagn_lirtot} plots the $f^{10,50,100\mu \rm m}_{\rm AGN}$ fraction as a 
function of the total infrared luminosity, \lirtot\ (see \S~\ref{sec:tot_lir}). This 
fraction of AGN emission at 10,50 and 100\mum\ allow us to 
check whether the emission at the probed wavelength is dominated by AGN emission or not. 

The top plot shows that independently from the classification introduced 
in \S~\ref{sec:class}, the AGN contributes {\it at least} to 50\% of the flux 
at 10\mum. In contrast, at 100\mum\ (bottom plot), the AGN is generally 
at the $\la$10\% level. However, we can see that even at such long wavelengths, 
the contribution of the AGN can, from time to time, be exceptionally high (almost 50\%). 
This latter could refer to extreme objects such as {\it 4C~23.56} or to extended 
dust emission \citep{Dicken2010}. At 50\mum\ (middle plot), we can clearly see 
that AGN can be from dominant to completely negligible. We conclude 
that even trying to define classes to potentially differentiate between AGN and SB 
dominated objects from data is almost impossible for high redshift radio 
galaxies. Only a SED decomposition as presented in \S~\ref{sec:decompir} 
can finally settle this question. 

\section{Bolometric correction}
\label{sec:bol_corr}

In order to derive AGN intrinsic properties, the AGN bolometric
luminosity is needed.  Hard X-rays provide the best approximation to
the bolometric luminosity as most of the material along the
line-of-sight is optically transparent. We do not
possess X-rays measurement for our entire sample, we therefore make
use of the calculated infrared luminosities. Numerous attempts to
derive bolometric correction factor have been done 
\citep[e.g.][]{Elvis1994,Haas2003,Marconi2004,Krawczyk2013,Hao2014,Scott2014}. Although essential,
they are subject to significative variation from object to object. For our SED range
\cite{Elvis1994,Richards2006} provide some correction factor depending on the 
observed wavelength. This correction factor can be as small as 2 and
as high as 20 for X-rays, depending mainly on the morphology. We here
want an approximation of this factor for our sources.

First, we can use a purely geometrical approach. Assuming the torus around
AGN to be optically thick at all wavelengths, it absorbs light from across
the electromagnetic spectrum and re-radiates in IR. We can use the
statistical distribution of type 1 and type 2 AGN in the sky
\citep{Barthel1989}. The solid angle for an opening angle of 45\degree corresponds
to a factor $\sim$2. This is expected to be the minimum correction.

Second, we can assume that the nuclei in type 2 and type 1 AGN are
similar. Using a type 1 radio loud AGN template from \cite{Elvis1994}
and assuming that the total IR luminosity does not depend strongly on
orientation, the bolometric correction factor is $\sim$6. Doing the same exercise
with the \cite{Richards2006} template gives a correction factor of $\sim$5.

Third, some sources in our sample possess X-ray observations
\citep{Carilli2002}. Integrating the energy over X-rays, it appears
that X-rays does not present the most significative contribution to
the bolometric luminosity. 

As the radio emission is highly directional (i.e. subject to strong beaming effect), 
its inclusion in the bolometric factor is highly uncertain. Type 2 AGN SEDs 
show that the integrated radio energy is roughly at the X-ray level. Its 
contribution to the total energy should not be the most significant.

Estimation of the bolometric correction is rather difficult and uncertain. Nevertheless,
the geometric approach and the moderate contribution from X-rays and radio
indicates that most of the light comes from the UV-optical from the central AGN part
and the reprocessed light by the dust. Therefore, a factor of 6 seems appropriate
in the case of radio galaxies to convert \liragn\ to \lbolagn.

\section{Summary of the global uncertainties}

Since we are using various approximations throughout this paper which can have
an impact on this analysis, we summarise here each of these and discuss their
possible impact on our interpretation.

(i) The validity of the \mbhbulge\
relation at high redshift deserves some attention. The first
part of this relation is to consider the estimated stellar mass
as the mass of the bulge or spheroid of individual galaxies. \textit{HST}
observations have shown that radio galaxies have elliptical light profiles
\citep{vanBreugel1998,Pentericci1999, Pentericci2001,Zirm2003}. 
Nevertheless, these determinations represent the radially averaged or 
global best-fit light profile with moderate-to-low signal-to-noise and the
possible presence and contribution from substructure and heavily obscured
younger disk components cannot be excluded in the profile fitting \citep[see][]{Hatch2013}. After all,
our estimates of the star formation rates suggest that obscuration could
be important and since the gas supporting such intense star formation
would be highly dissipative and could easily be in a disk.  However,
the stellar masses estimated by \cite{DeBreuck2010} are measured in the
rest-frame $H$ band, minimizing the impact of extinction and also sampling
more appropriately the older population (modulo the contribution from
young super-giants). The measured mass can therefore be considered as the
total mass of the system and at least, in principle, sensitive to the older
generations of stars in the host galaxy \citep{Rocca2013}. Considering
the \mbhbulge\ relation itself, \cite{Jahnke2009} estimate that the
\mbh-M$_{\rm stel}$ relation shows little variation from $z$=1.4 to $z$=0.

(ii) The radiative efficiency of the accretion, $\epsilon$, is
not well-constrained and is certainly not a constant.  This factor
can vary from 0.06 to 0.42, related to the spin of the black hole
\citep{Krolik1999}. There are attempts to constrain the spin of radio
loud AGN in the literature. \cite{MartinezSansigre2011} show that
black hole spins tend to be lower at higher redshift even with the
presence of a bimodal distribution. As these constraints are quite poor
at high redshift, it is impossible to conclude on the possible value
of $\epsilon$, but perhaps a range of a few is reasonable (factor 3 at maximum).

(iii) The correction factor to estimate the bolometric luminosity,
$\kappa^{\rm Bol}_{\rm AGN}$, shows a
wide variety of possible values.  Pure geometric considerations imply
that $\kappa^{\rm Bol}_{\rm AGN}$$>$1.4 and is unlikely to be $>$10 for
the conversion of IR luminosity to bolometric luminosity (see Appendix A
for details).  We assumed $\kappa^{\rm Bol}_{\rm AGN}$=6 for the ensemble
of radio galaxies.  This correction is not expected to differ strongly
from this value as the energy is mostly radiated in IR in our object and
we have now a good coverage of this part of the SED.  However, a factor
as low as 2 is not unreasonable (see Appendix C).

(iv) The sSFR calculated are dependant on the canonical law used to 
transform \lirsb\ into SFR and the stellar mass. While the \cite{Kennicutt1998} relation seems to 
represent well most star forming galaxies, some discrepancies 
are expected as it is dependent of the star formation law. Indeed,
\cite{Calzetti2012} lists the impact of the approximation on the different
SFR indicators at various wavelength. This also depends on $\tau$ and the adopted IMF; the variation 
can be a factor of a few (up to $\sim$6). Also, the IMF can induce a factor 
of $\sim$2 in the stellar masses estimates \citep[e.g.][]{Marchesini2009}.
This effect will move the points horizontally in Figure \ref{fig:smdot_vs_ssfr}.

(v) The AGN SED can present a wide variation \citep[e.g.][]{Nenkova2008a,Fritz2006}. 
As mentioned, our generic AGN SED can miss a part of the extended flux
from the AGN heating of the NLR \citep[e.g.][]{Dicken2009,Pier1993}. However,
as we are dealing with integrated luminosities over the 8-1000\mum\ range,
the calculated \liragn\ is expected to differ strongly only with a drastic change of the AGN
SED which is unlikely (\S~\ref{sec:agn_template}). The most probable
case would be that we underestimated the AGN contribution, therefore,
all corresponding values will be increased by the same factor, increasing the
offset from the local \mbhbulge\ relation.


\begin{centering}
\section*{SEDs}



\pagestyle{empty}

\clearpage
\begin{figure*}[h]
  \caption{SEDs of the 70 radiogalaxies sorted by
  RA. IRS and MIPS data taken from \cite{DeBreuck2010}, PACS and
  SPIRE data in Table \ref{tab:herschel_flux} and sub-mm data in
  Table \ref{tab:submm_flux}. Filled diamonds, the firm
  detections ($>$3$\sigma$), open diamonds the weak detections
  (\weakdetect) and downward triangles the 3$\sigma$ upper
  limits. The red downward triangles mark to the most constraining
  upper limit. Continuous line represents for fitted components, depending
  on the class: AGN for WD, SB for CD and sum of AGN and SB for
  WCD (as marked in the figure legend). The 6 stamps correspond to
  the \mips1\ and the five \herschel\ observations when available, with
  north at the top, east at the left, centred on the radio coordinates
  of the radio galaxy. Each stamp covers 2x2 arcmin. We also overplot
  the IRS spectra when available for the source \citep{Seymour2008,Rawlings2013}.}
  \label{fig:sed} 
  
 \begin{centering}
  \begin{minipage}[c]{0.46\linewidth}
\begin{tabular}{c@{}c@{}c@{}c@{}c@{}c@{}}

      \multicolumn{6}{c}{\includegraphics[width=80mm,trim=30 0 20 0 ,clip=true]{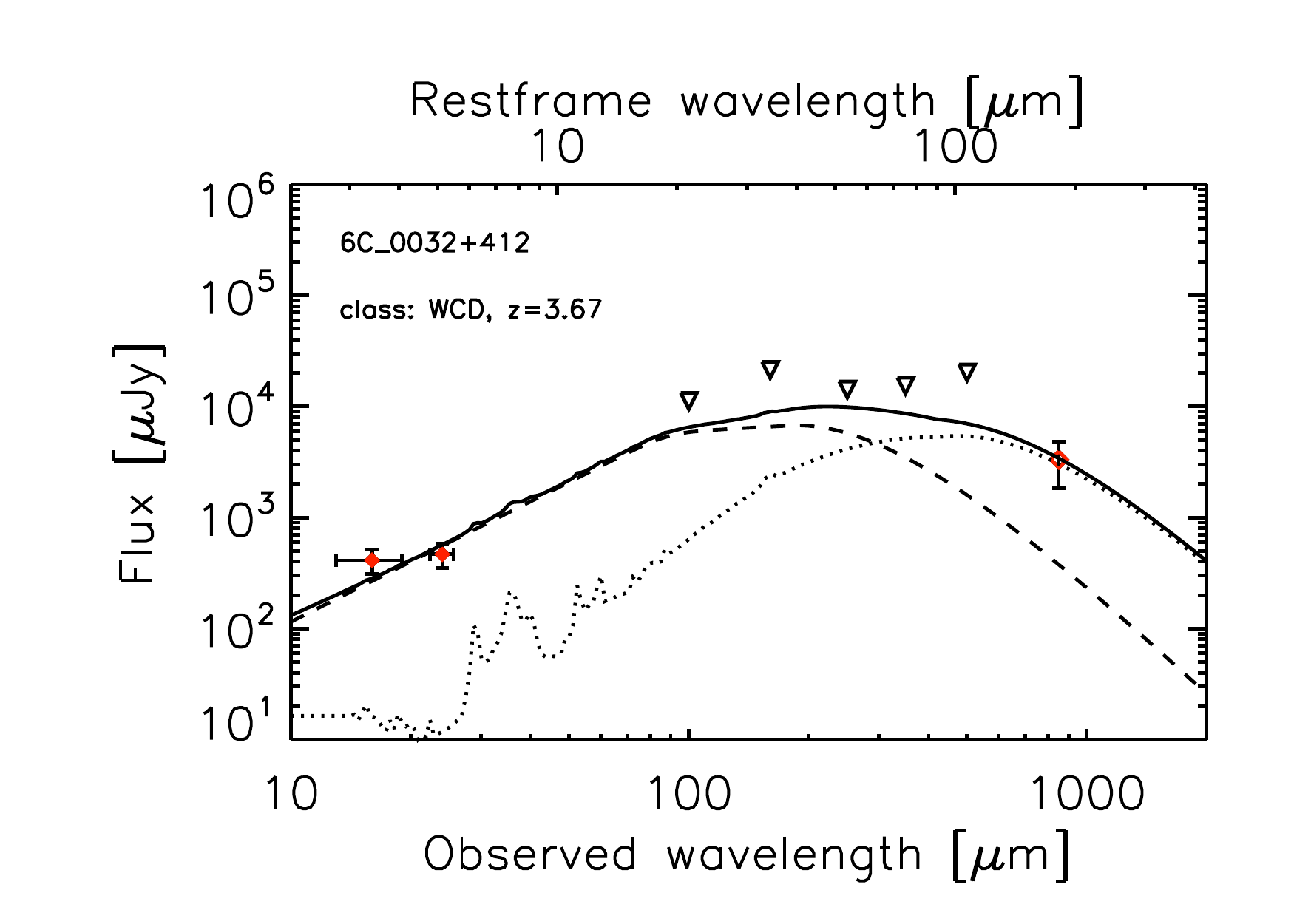}} \\
      24\mum & 100\mum & 160\mum & 250\mum & 350\mum & 500\mum \\
      \includegraphics[height=14mm,trim=70  -4 65 -10,clip=true]{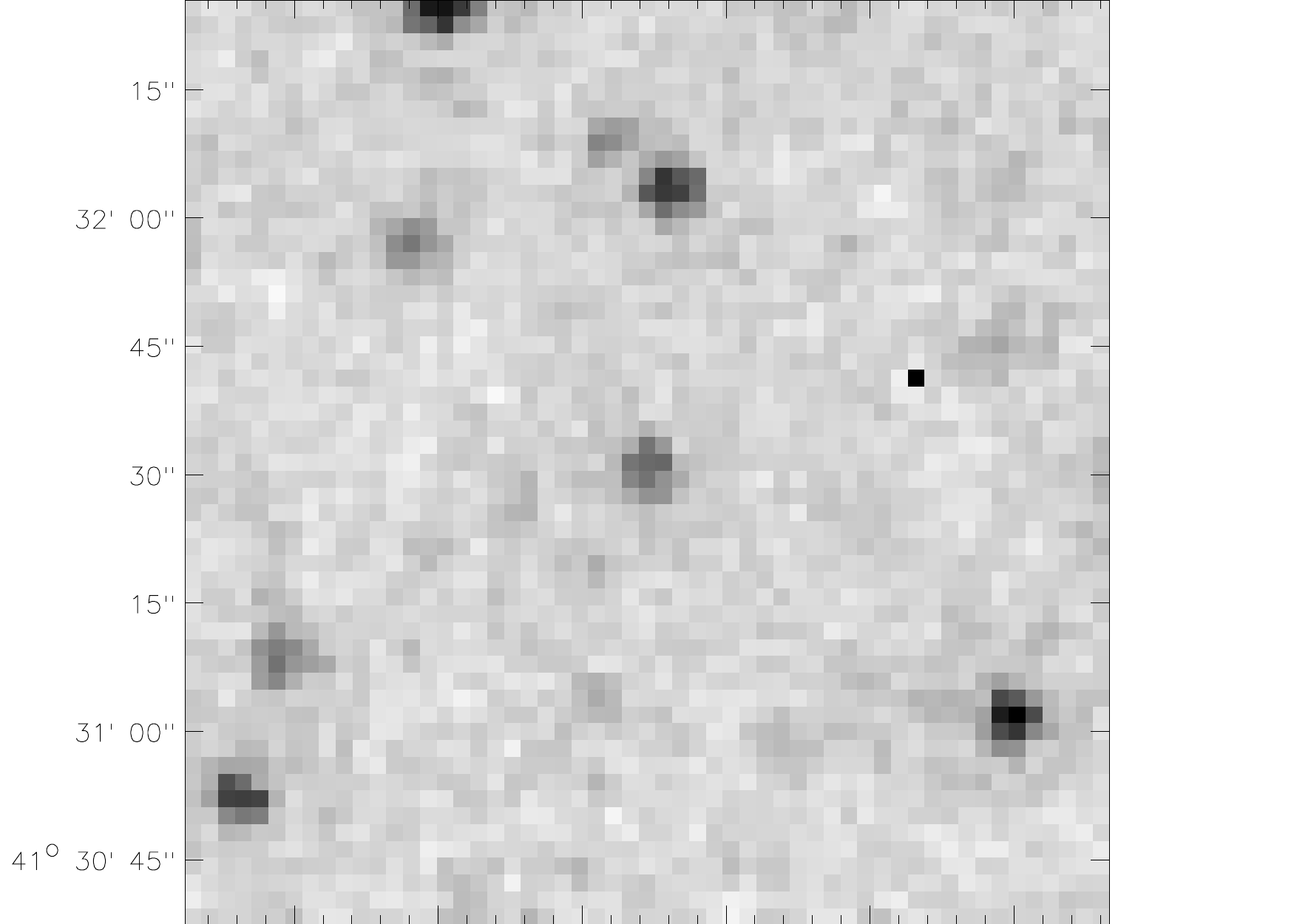} &
      \includegraphics[height=14mm,trim=70  -4 65 -10,clip=true]{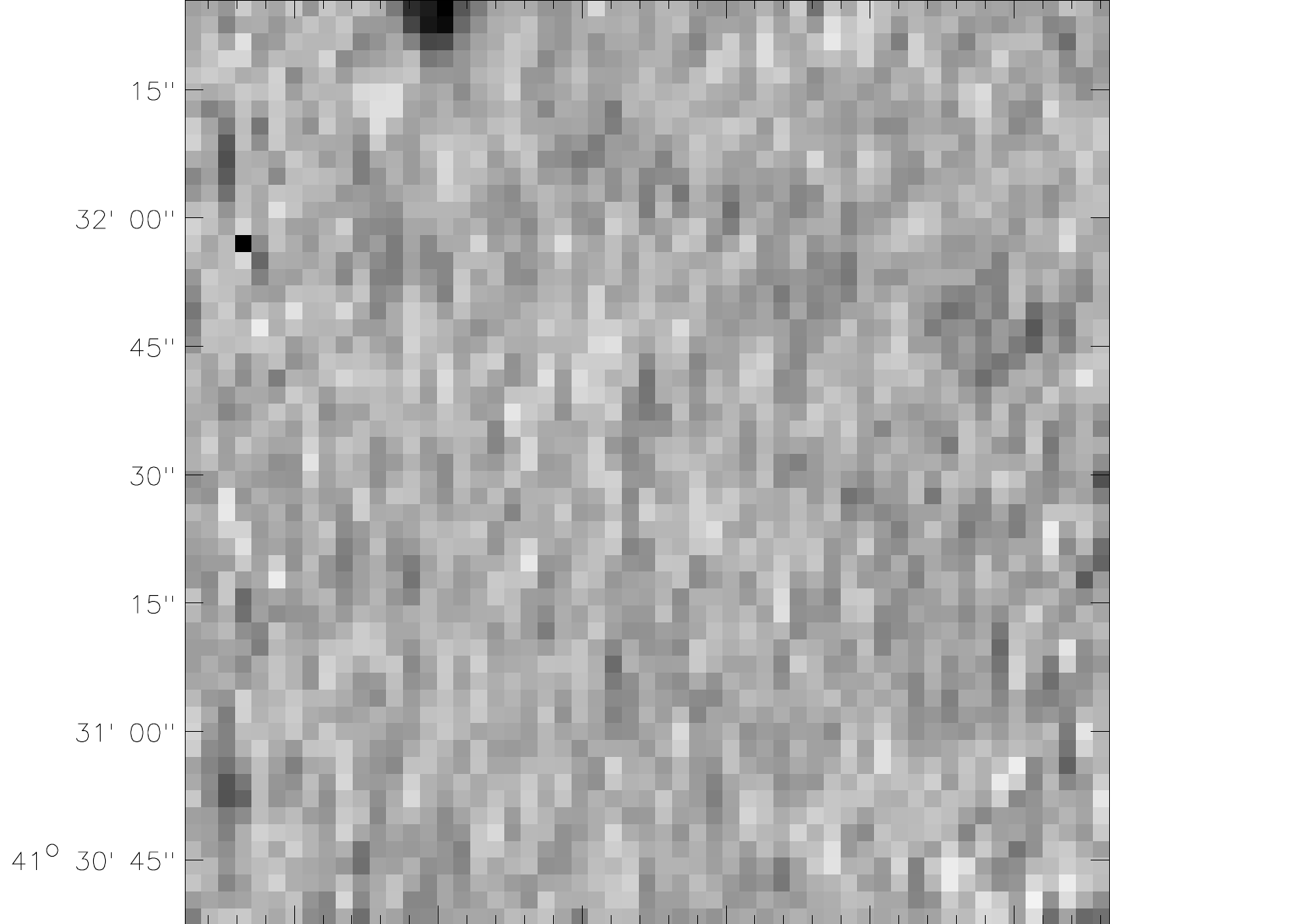} &
      \includegraphics[height=14mm,trim=70  -4 65 -10,clip=true]{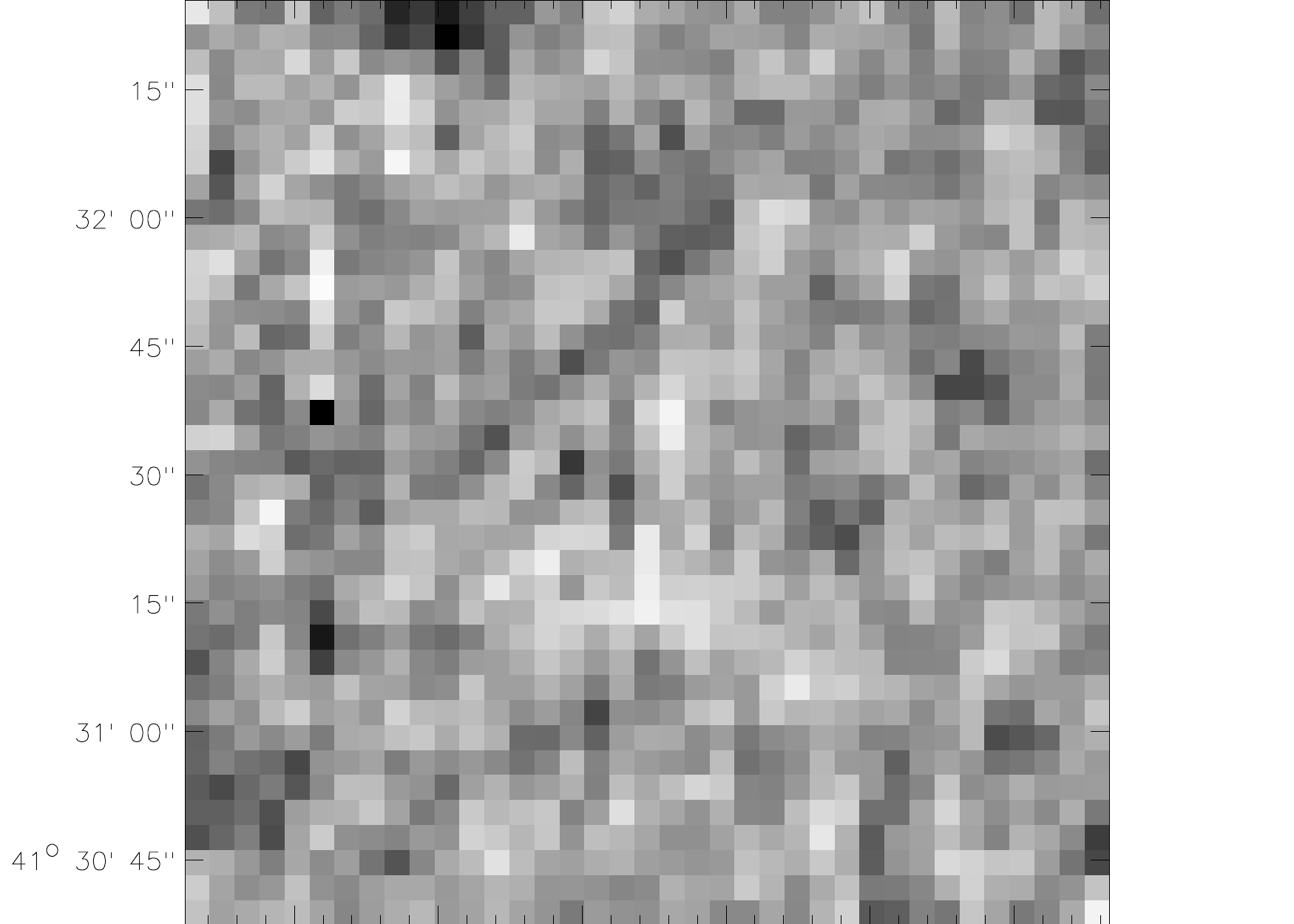} &
      \includegraphics[height=14mm,trim=70  -4 65 -10,clip=true]{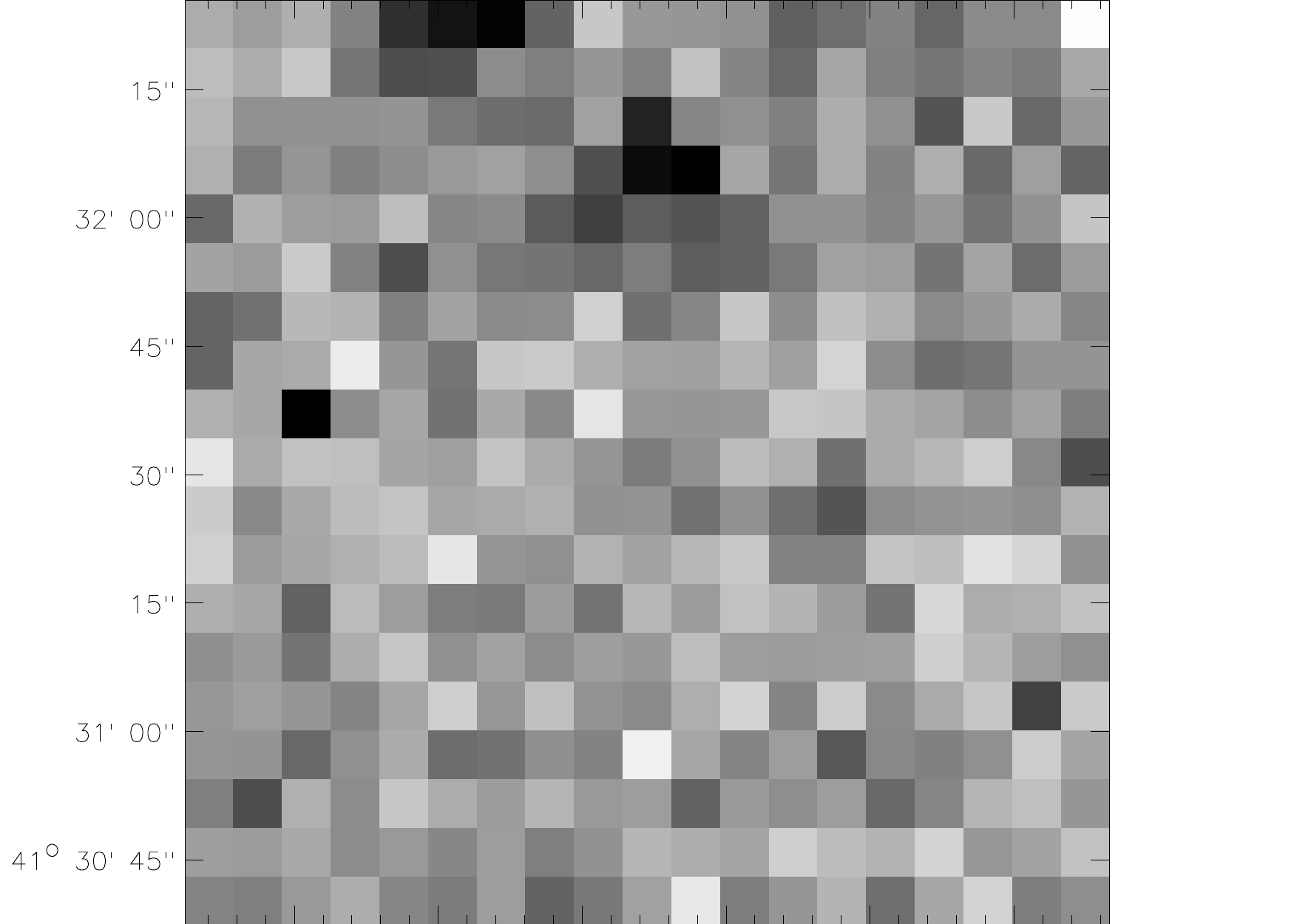}   &
      \includegraphics[height=14mm,trim=70  -4 65 -10,clip=true]{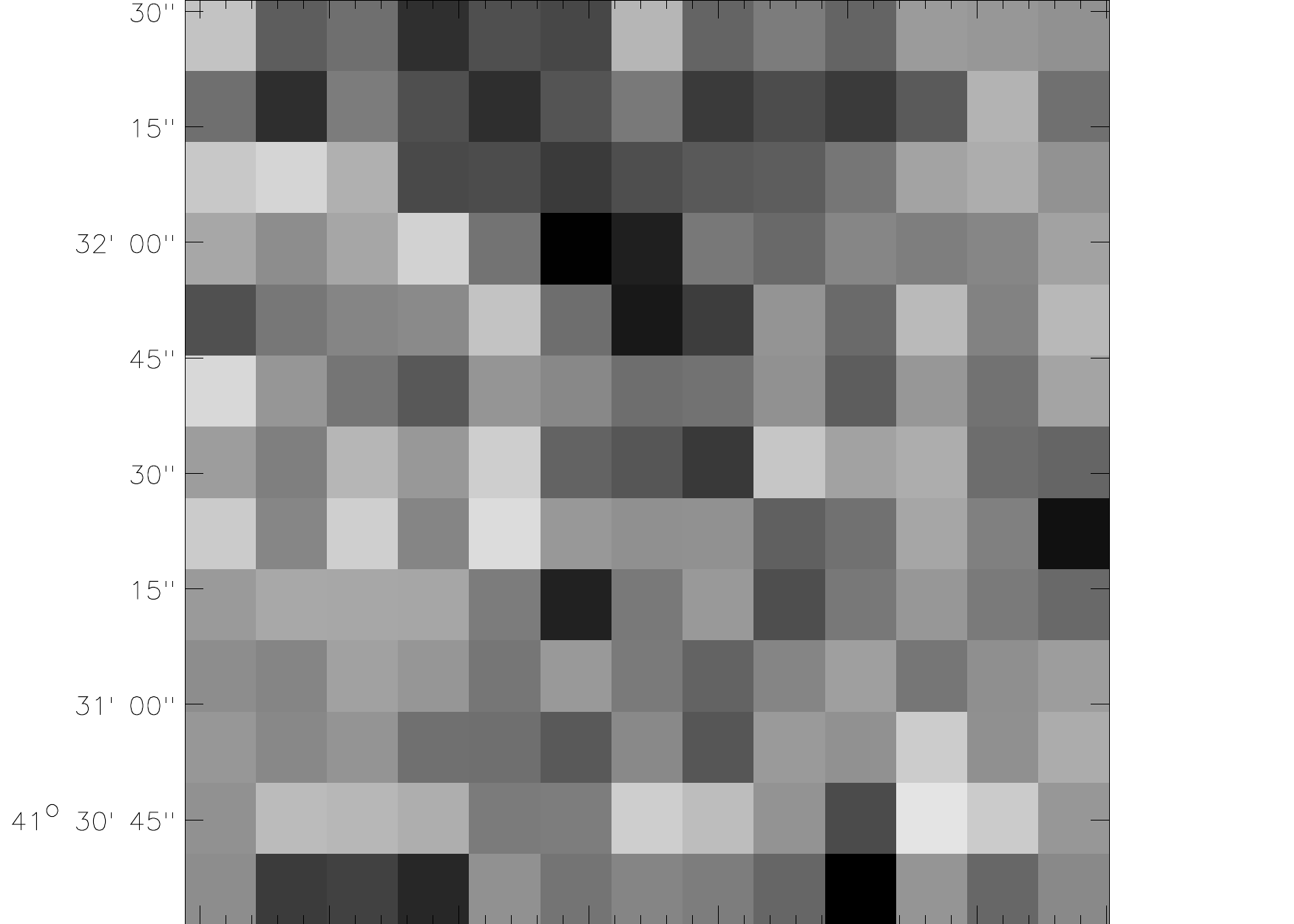}   &
      \includegraphics[height=14mm,trim=70  -4 65 -10,clip=true]{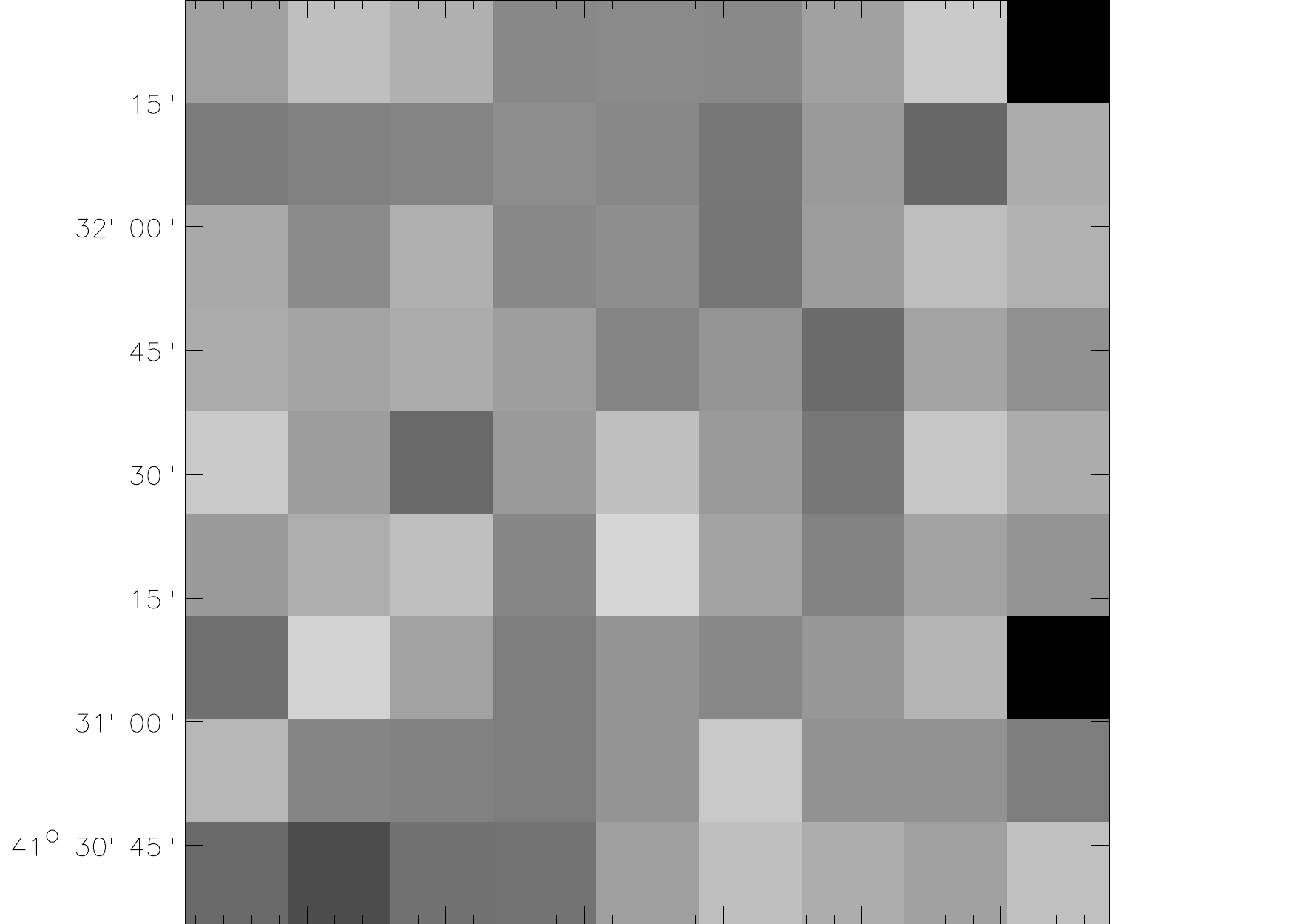} \\

      \multicolumn{6}{c}{\includegraphics[width=80mm,trim=30 0 20 0 ,clip=true]{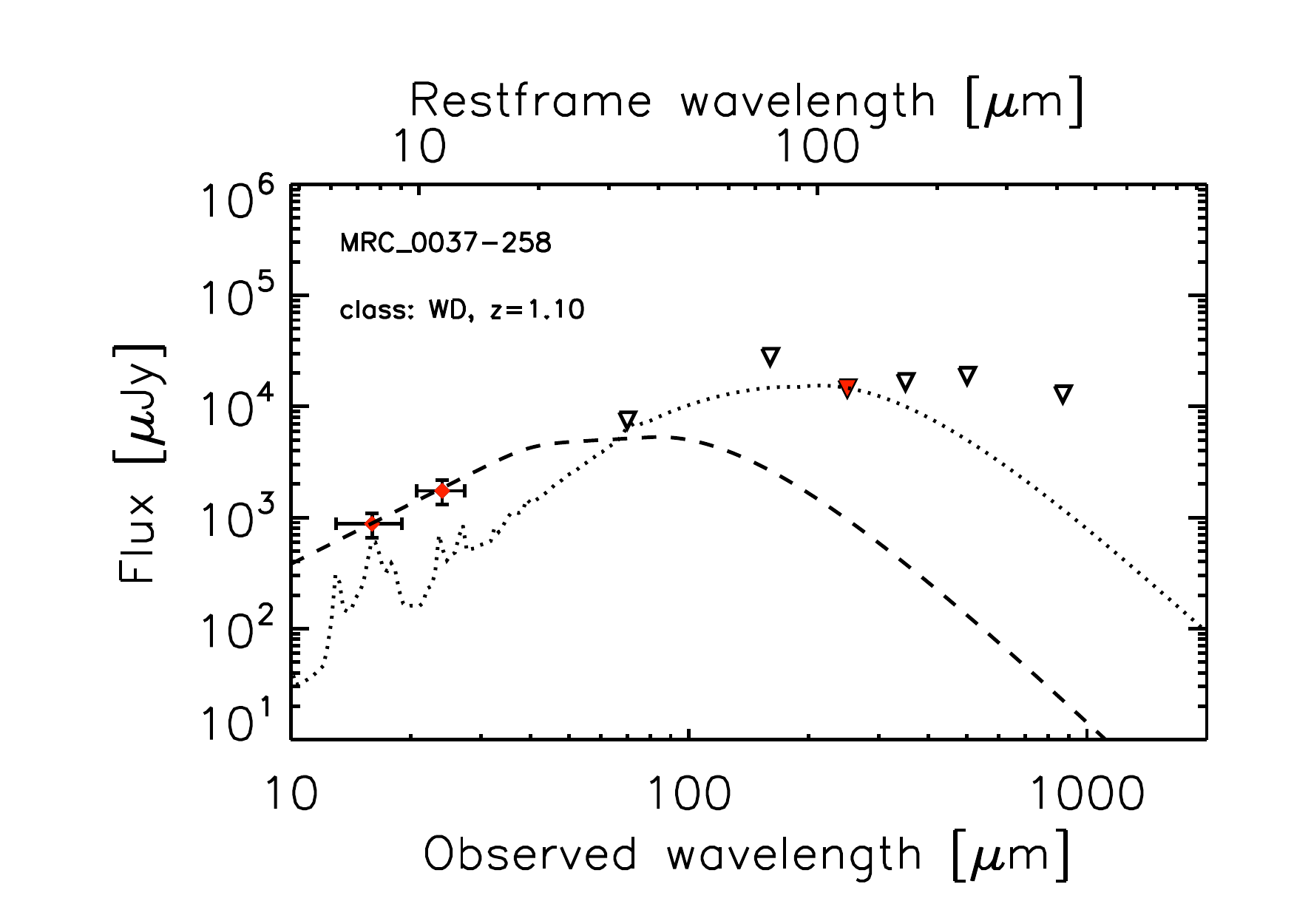}} \\
      24\mum & 70\mum & 160\mum & 250\mum & 350\mum & 500\mum \\
      \includegraphics[height=14mm,trim=70  -4 65 -10,clip=true]{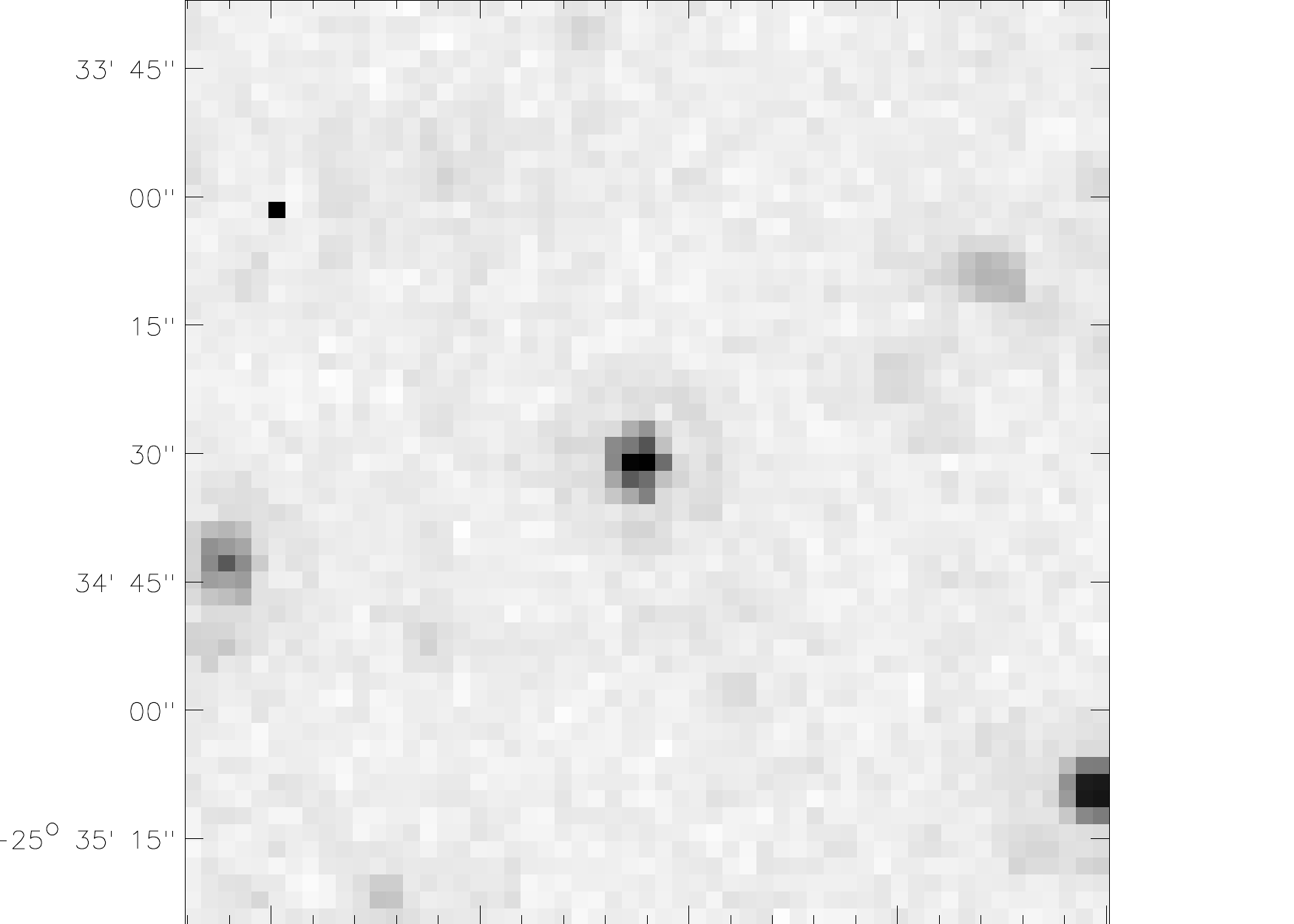} &
      \includegraphics[height=14mm,trim=70  -4 65 -10,clip=true]{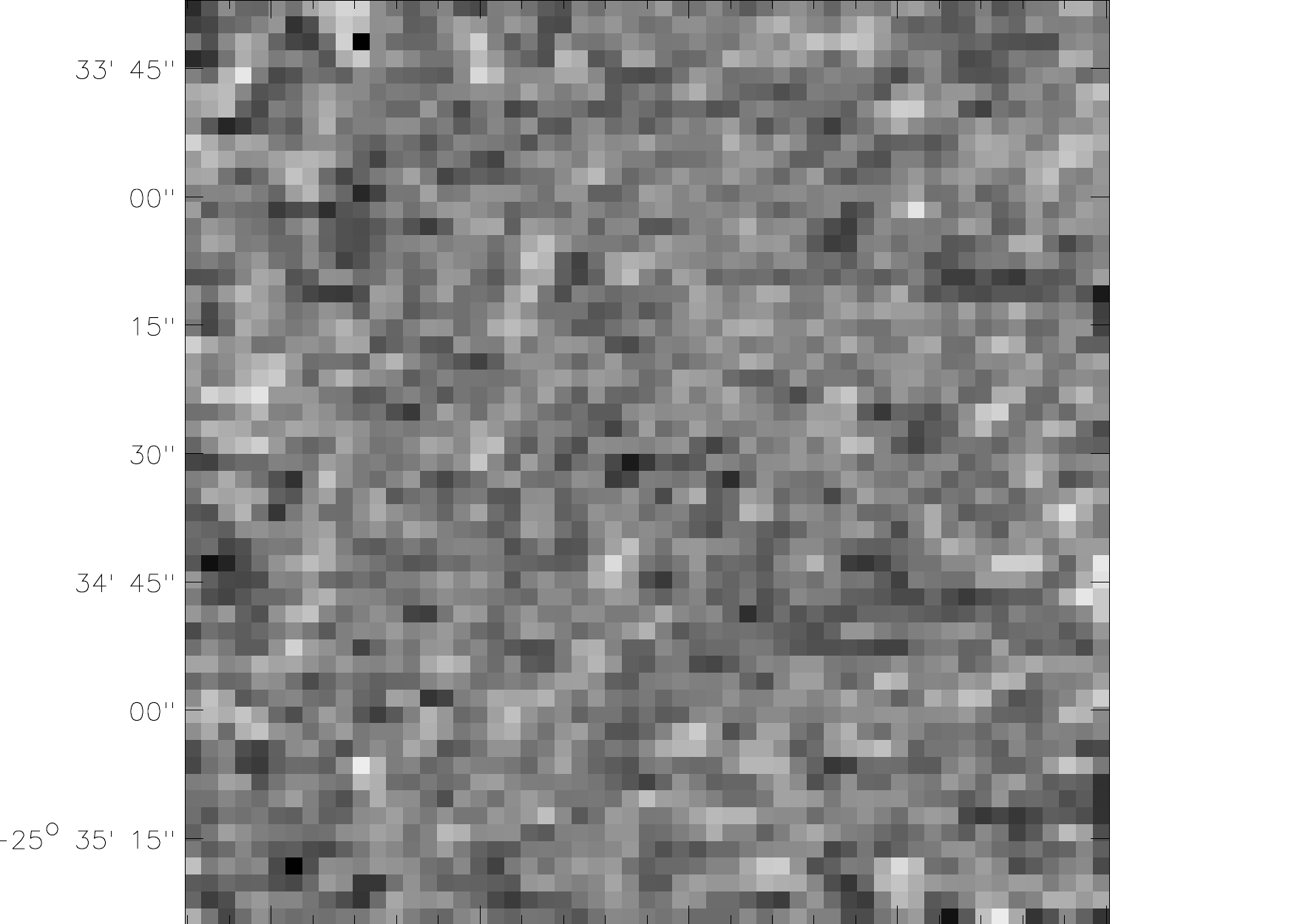} &
      \includegraphics[height=14mm,trim=70  -4 65 -10,clip=true]{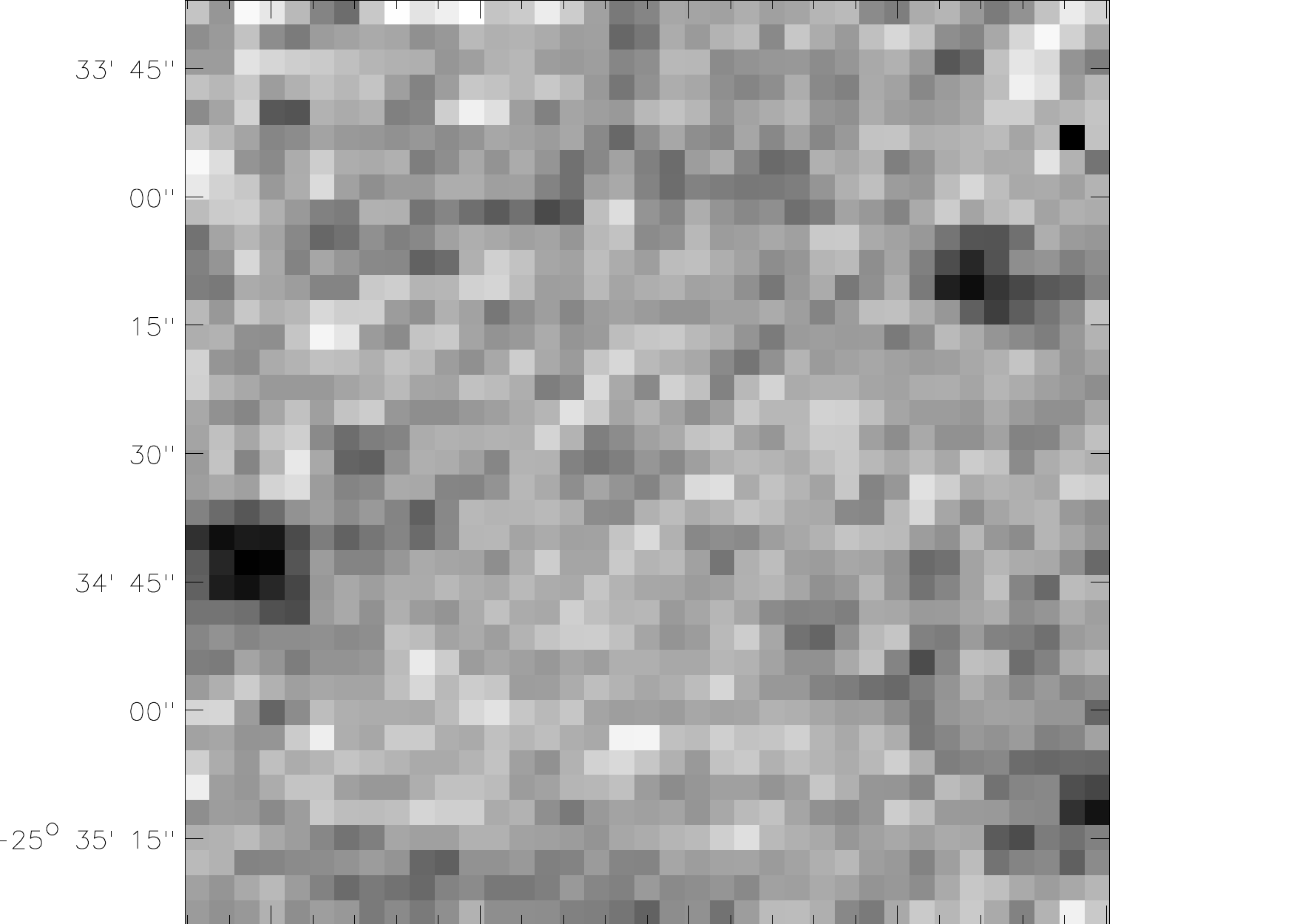} &
      \includegraphics[height=14mm,trim=70  -4 65 -10,clip=true]{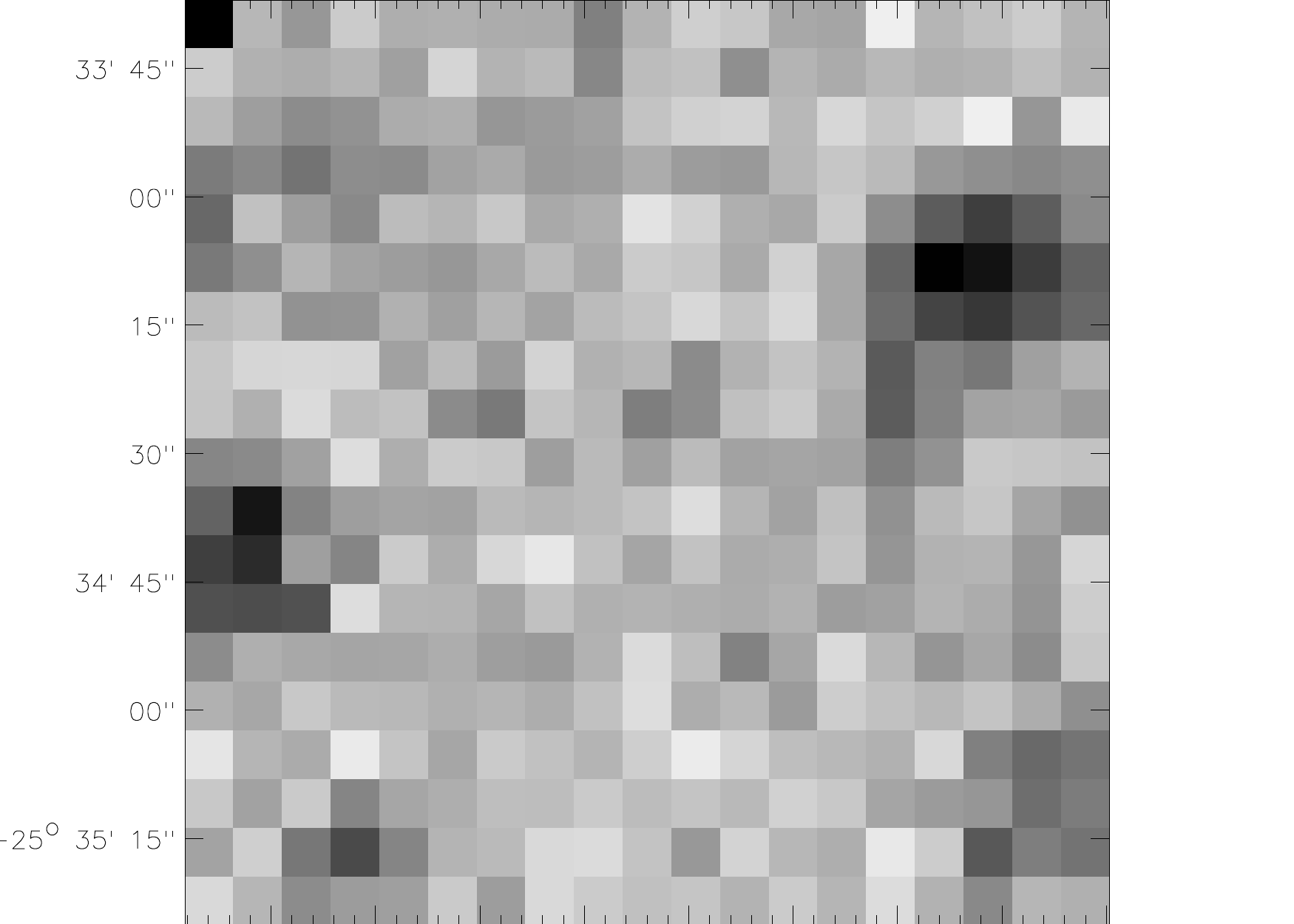}   &
      \includegraphics[height=14mm,trim=70  -4 65 -10,clip=true]{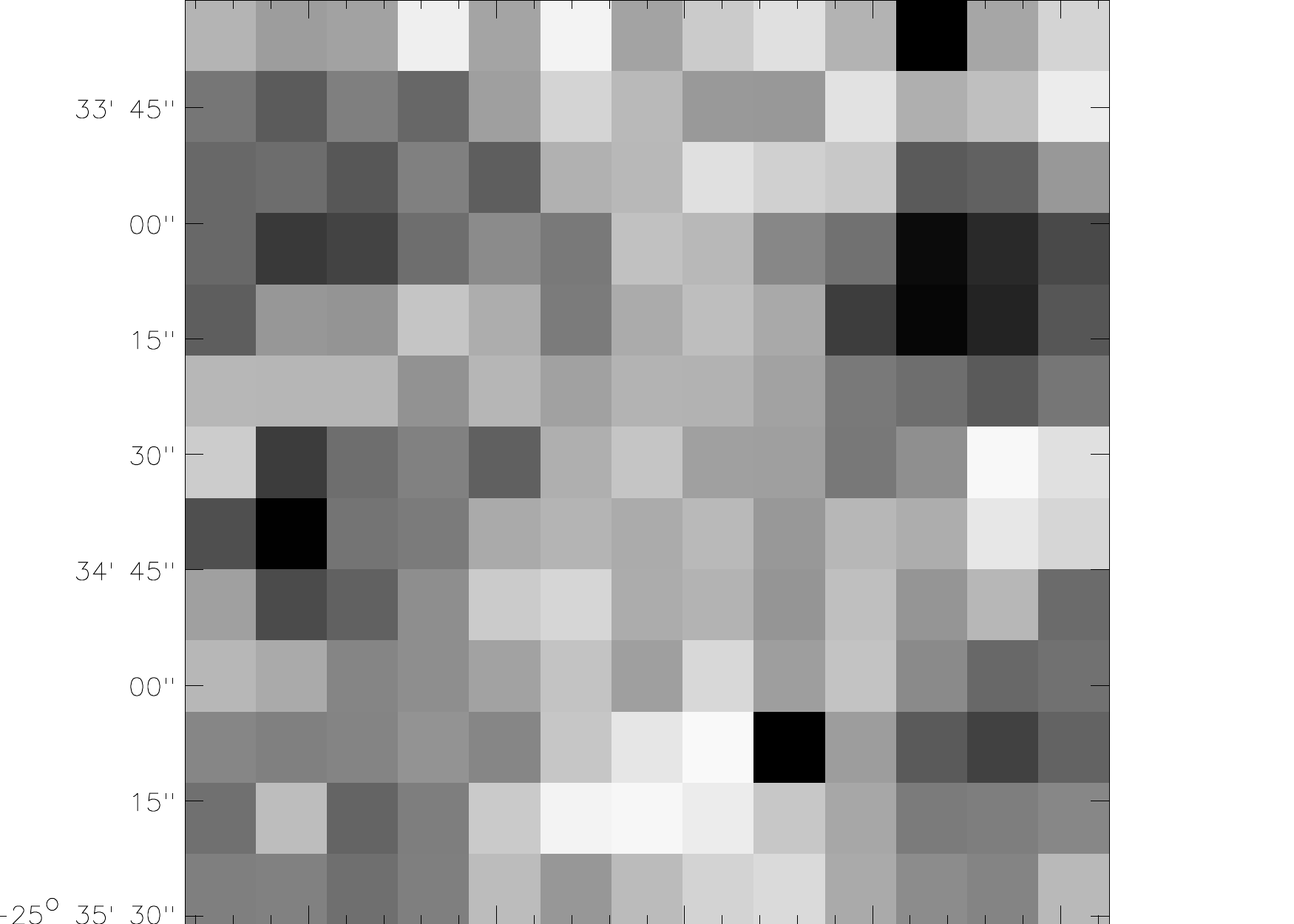}   &
      \includegraphics[height=14mm,trim=70  -4 65 -10,clip=true]{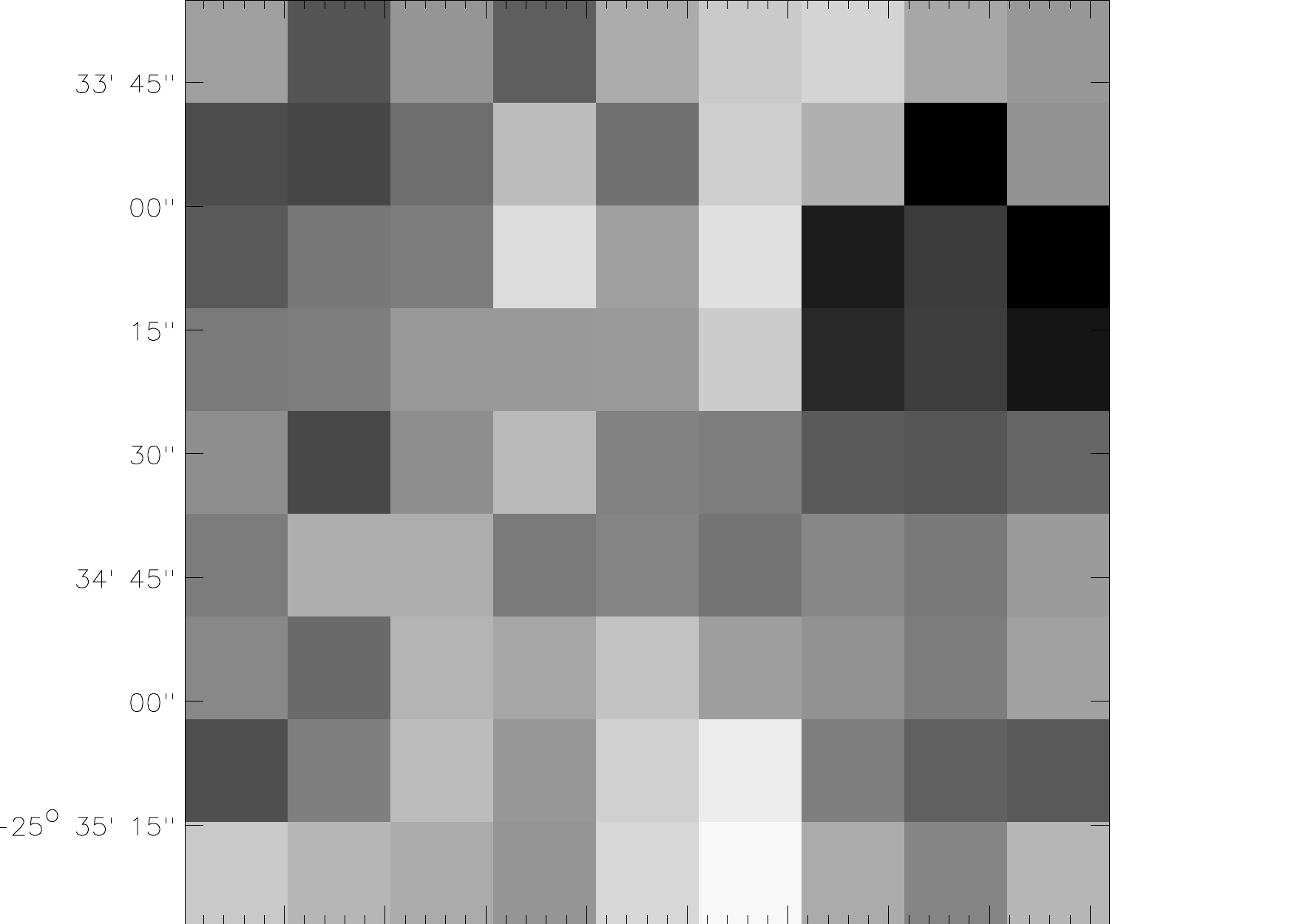} \\

\end{tabular}
\end{minipage}
\begin{minipage}[c]{0.46\linewidth} 
\begin{tabular}{c@{}c@{}c@{}c@{}c@{}c@{}}

      \multicolumn{6}{c}{\includegraphics[width=80mm,trim=30 0 20 0 ,clip=true]{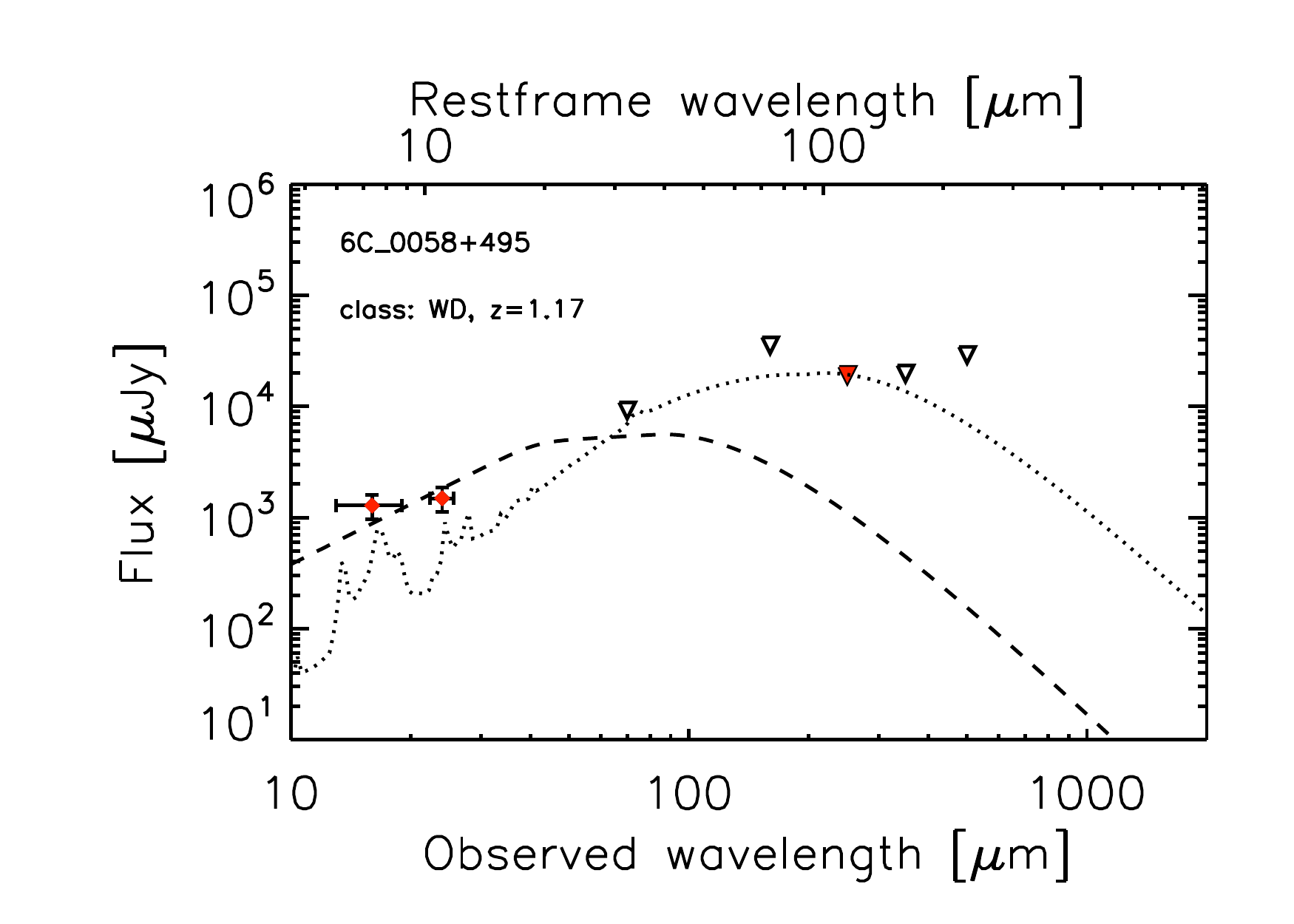}} \\
      24\mum & 70\mum & 160\mum & 250\mum & 350\mum & 500\mum \\
      \includegraphics[height=14mm,trim=70  -4 65 -10,clip=true]{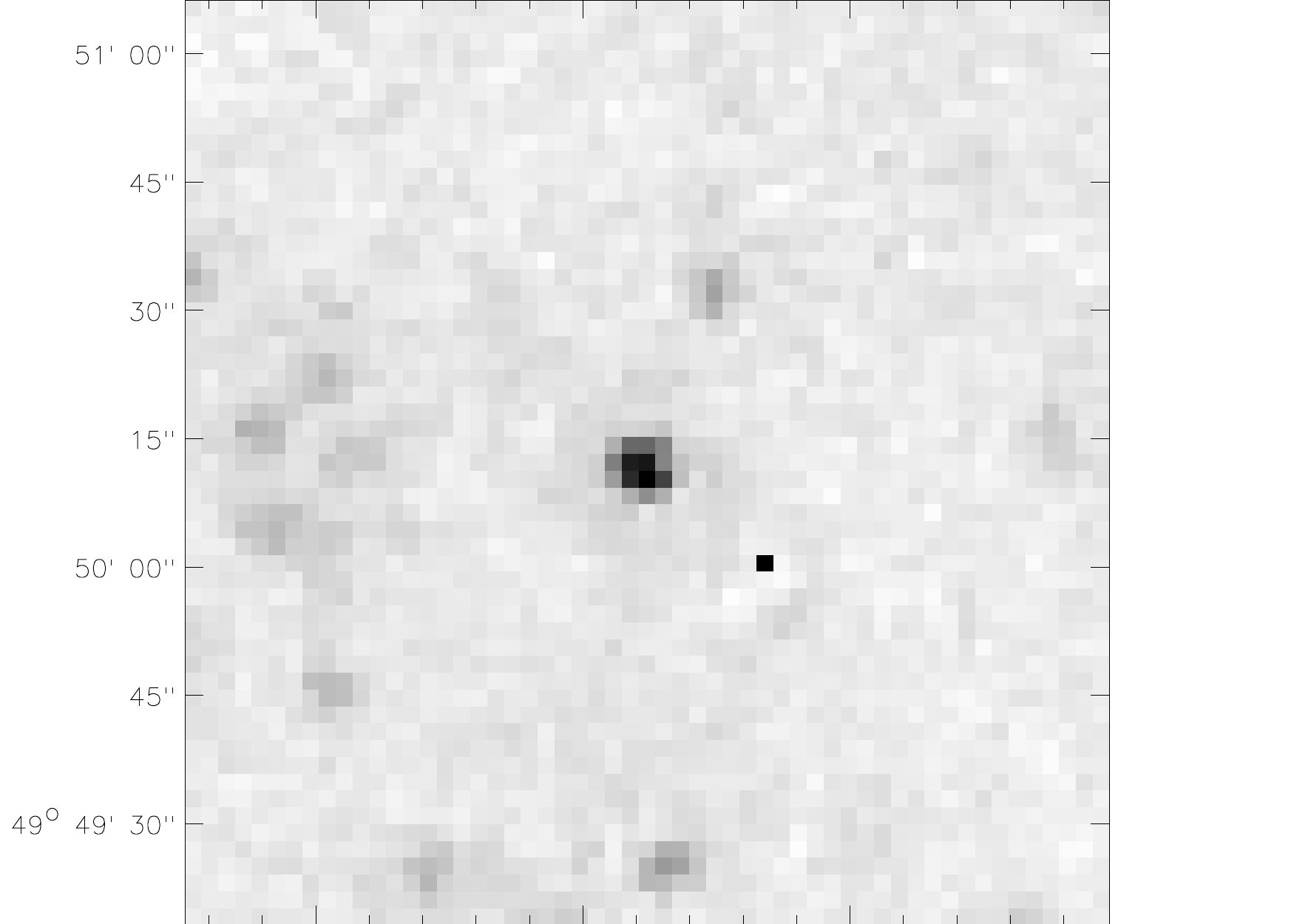} &
      \includegraphics[height=14mm,trim=70  -4 65 -10,clip=true]{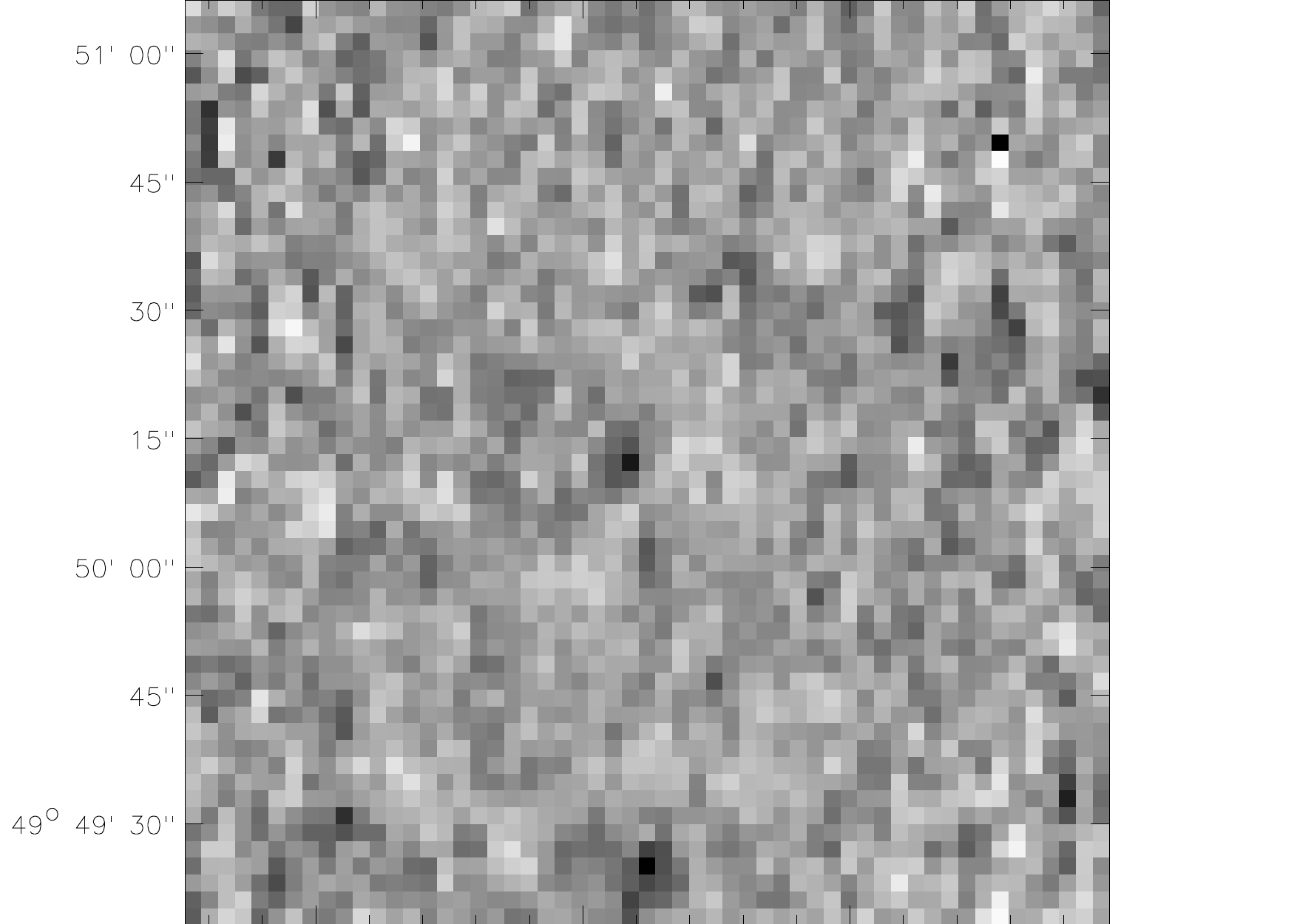} &
      \includegraphics[height=14mm,trim=70  -4 65 -10,clip=true]{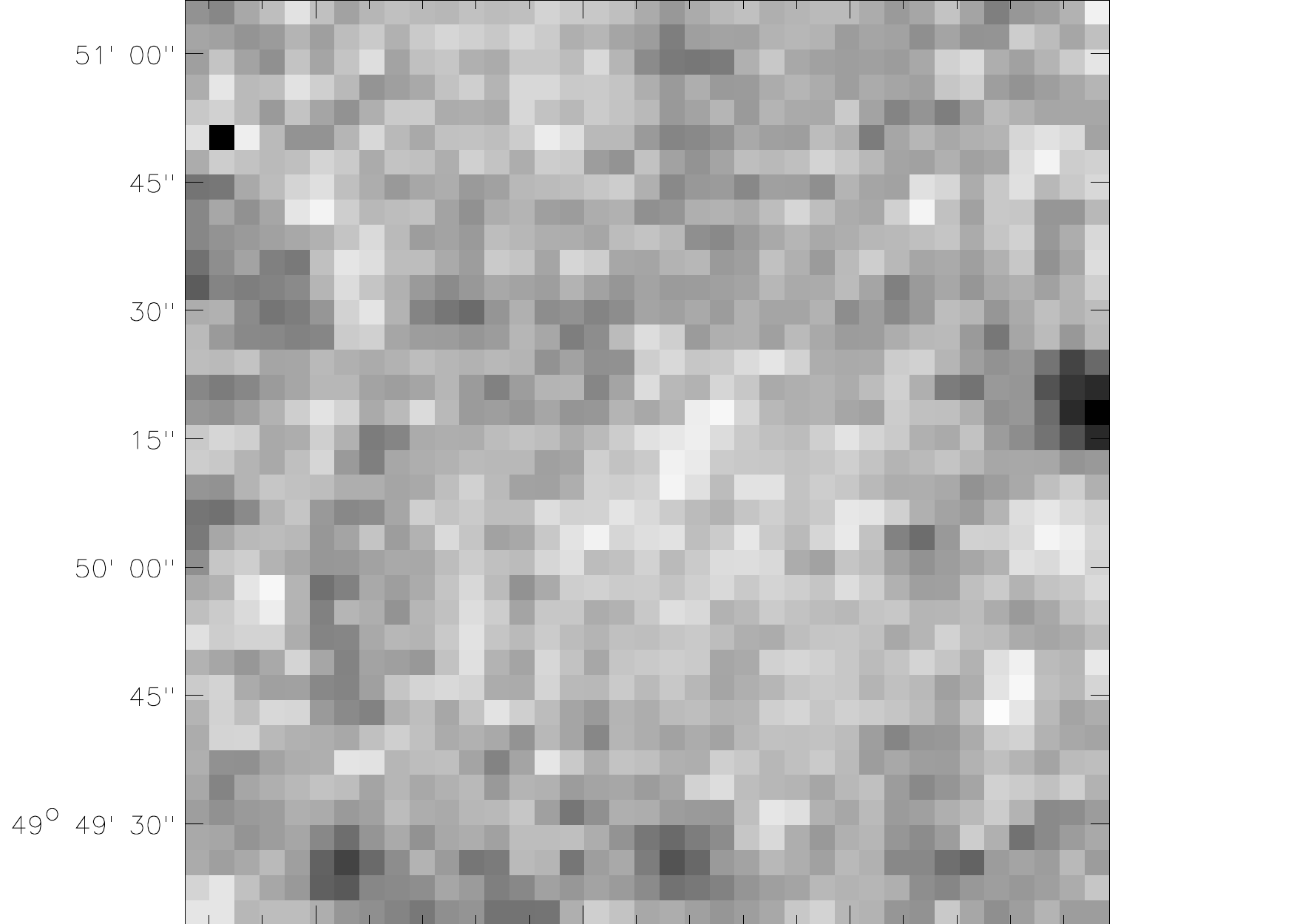} &
      \includegraphics[height=14mm,trim=70  -4 65 -10,clip=true]{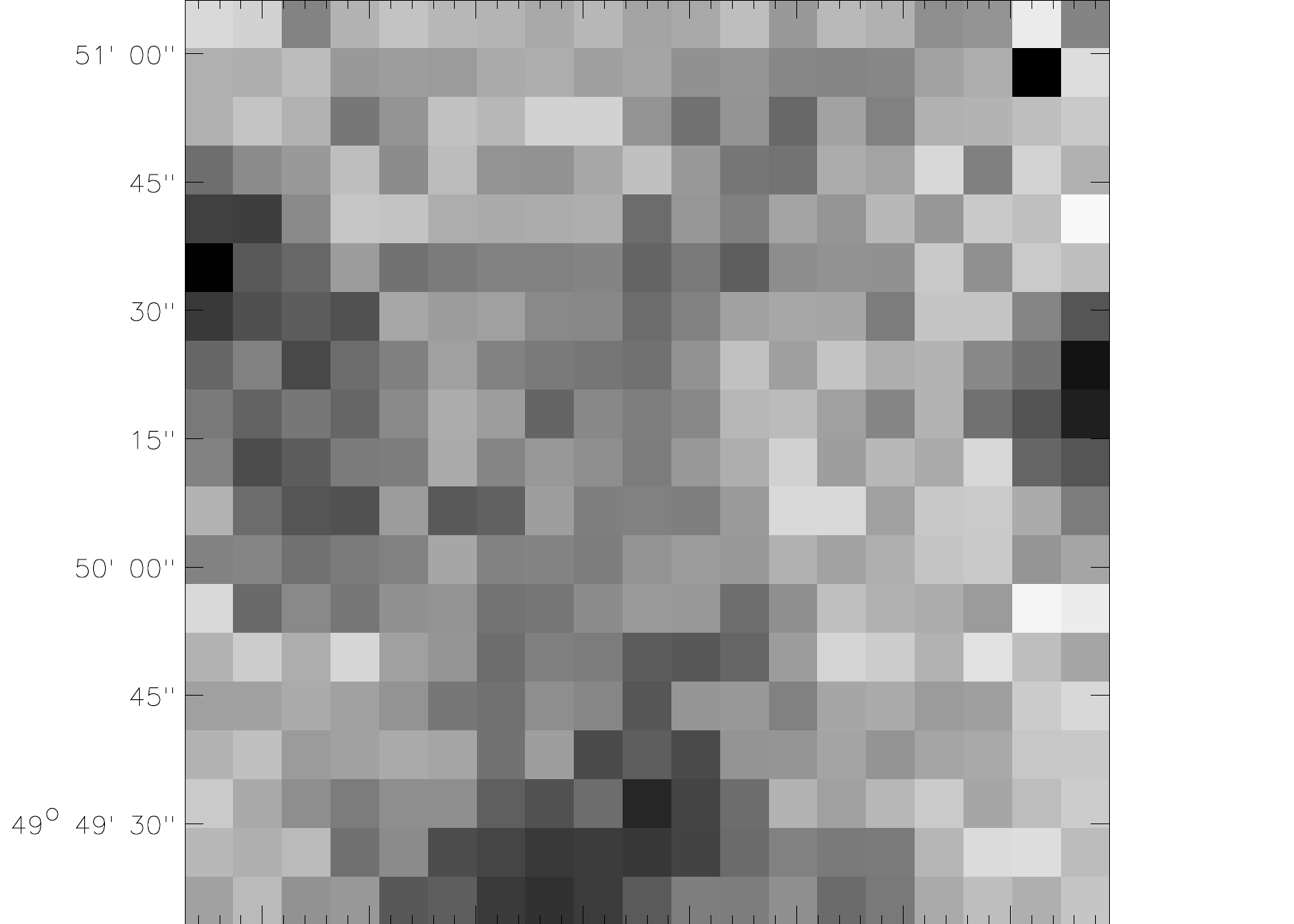}   &
      \includegraphics[height=14mm,trim=70  -4 65 -10,clip=true]{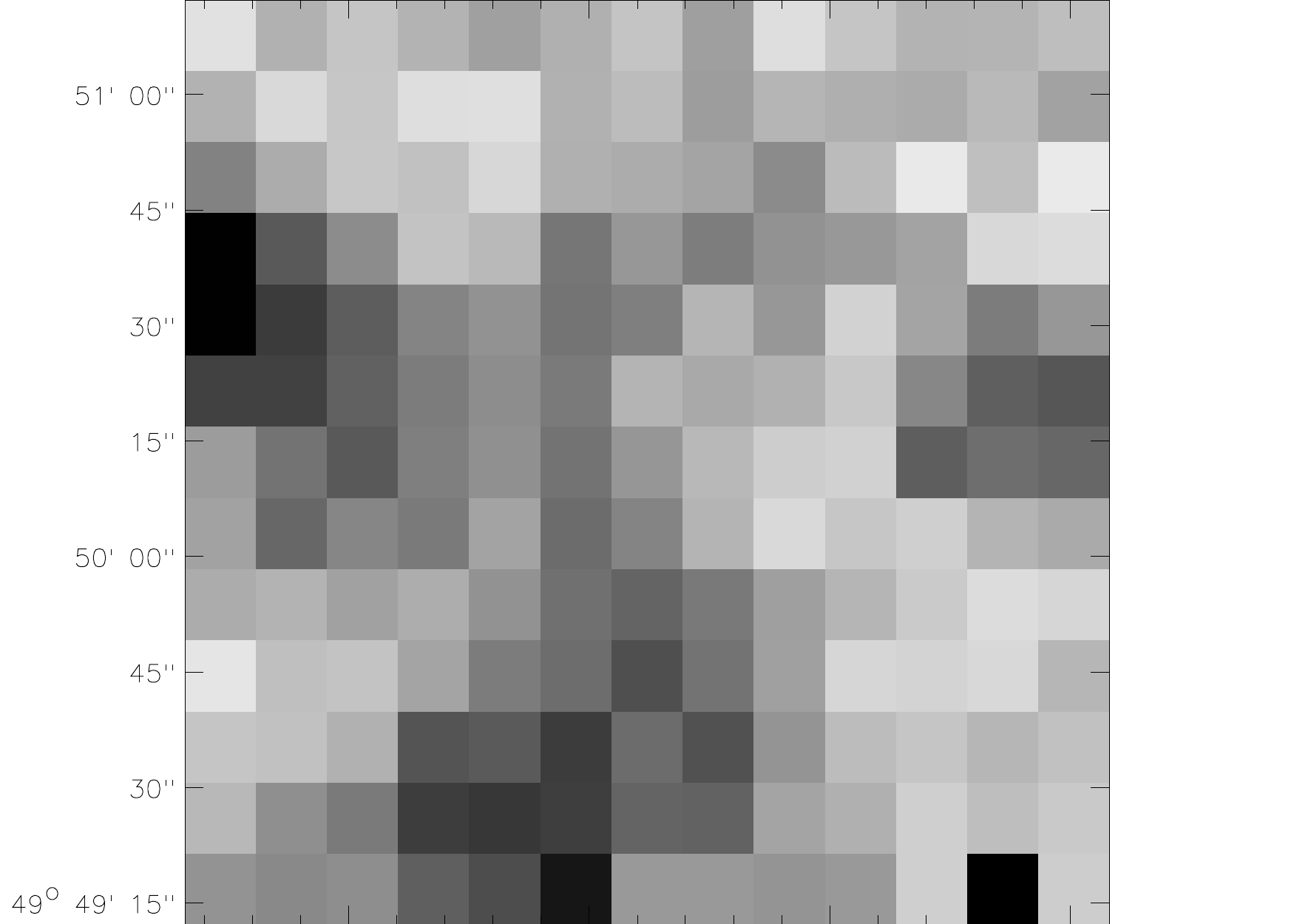}   &
      \includegraphics[height=14mm,trim=70  -4 65 -10,clip=true]{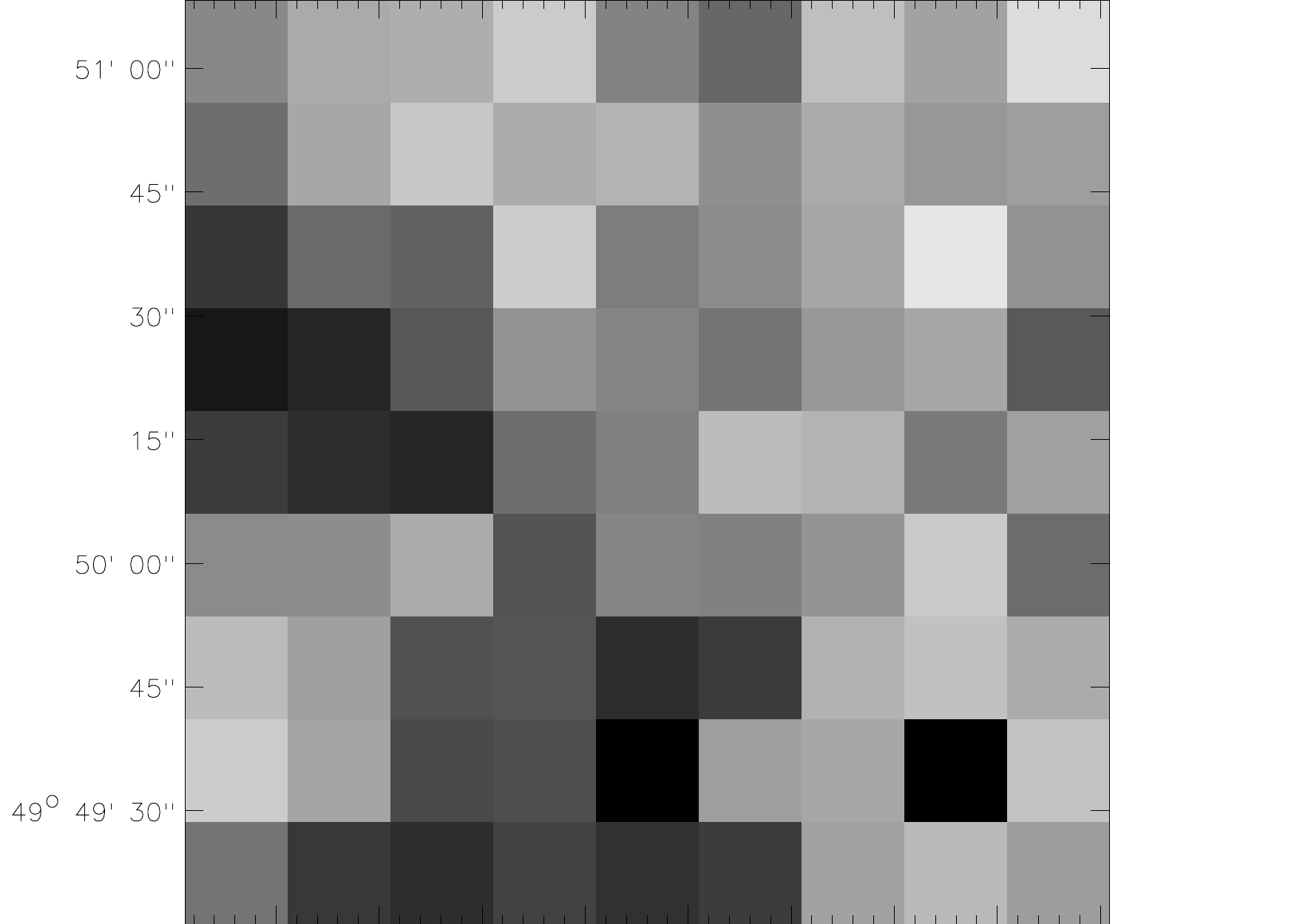} \\

      \multicolumn{6}{c}{\includegraphics[width=80mm,trim=30 0 20 0 ,clip=true]{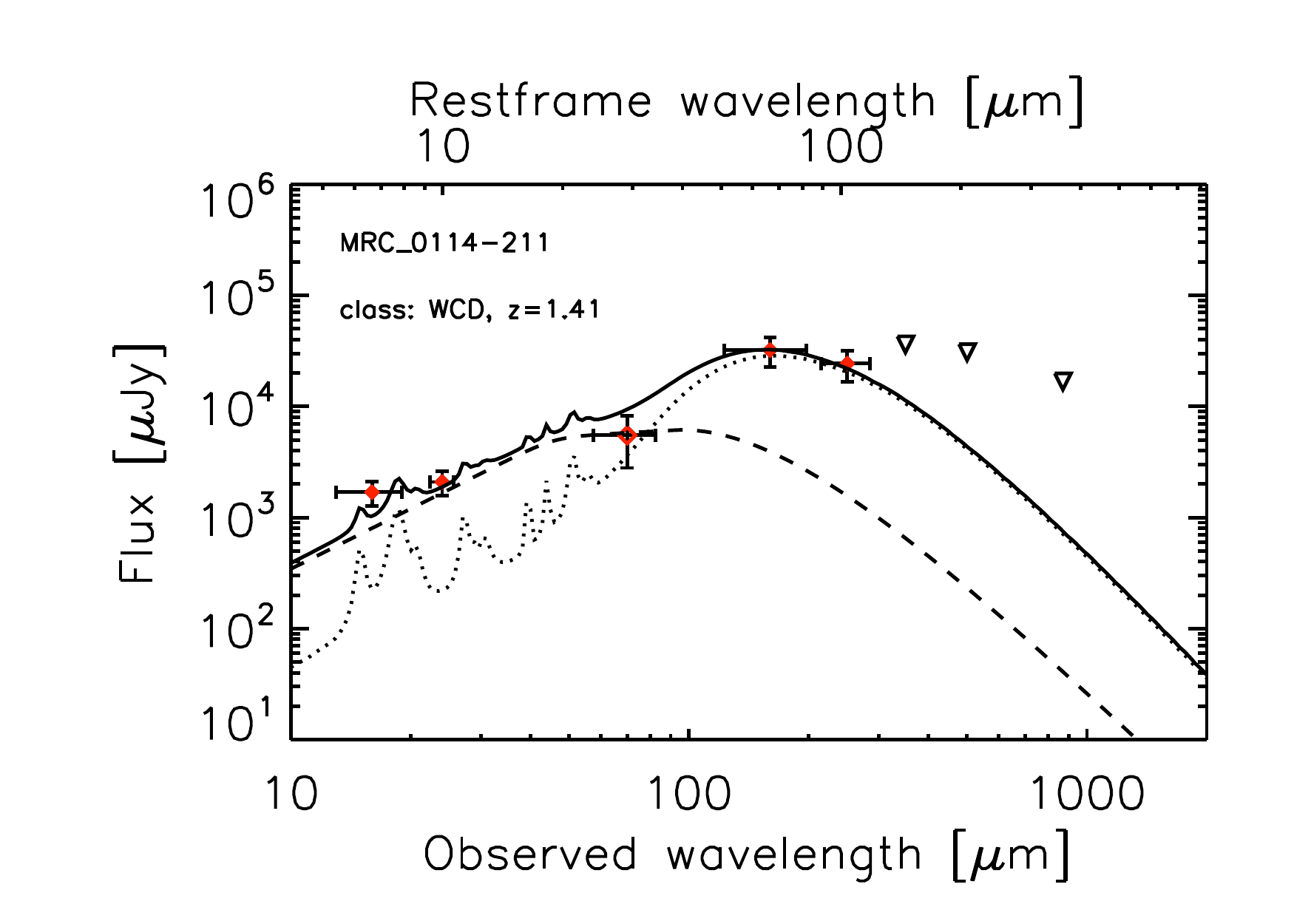}} \\
      24\mum & 70\mum & 160\mum & 250\mum & 350\mum & 500\mum \\
      \includegraphics[height=14mm,trim=70  -4 65 -10,clip=true]{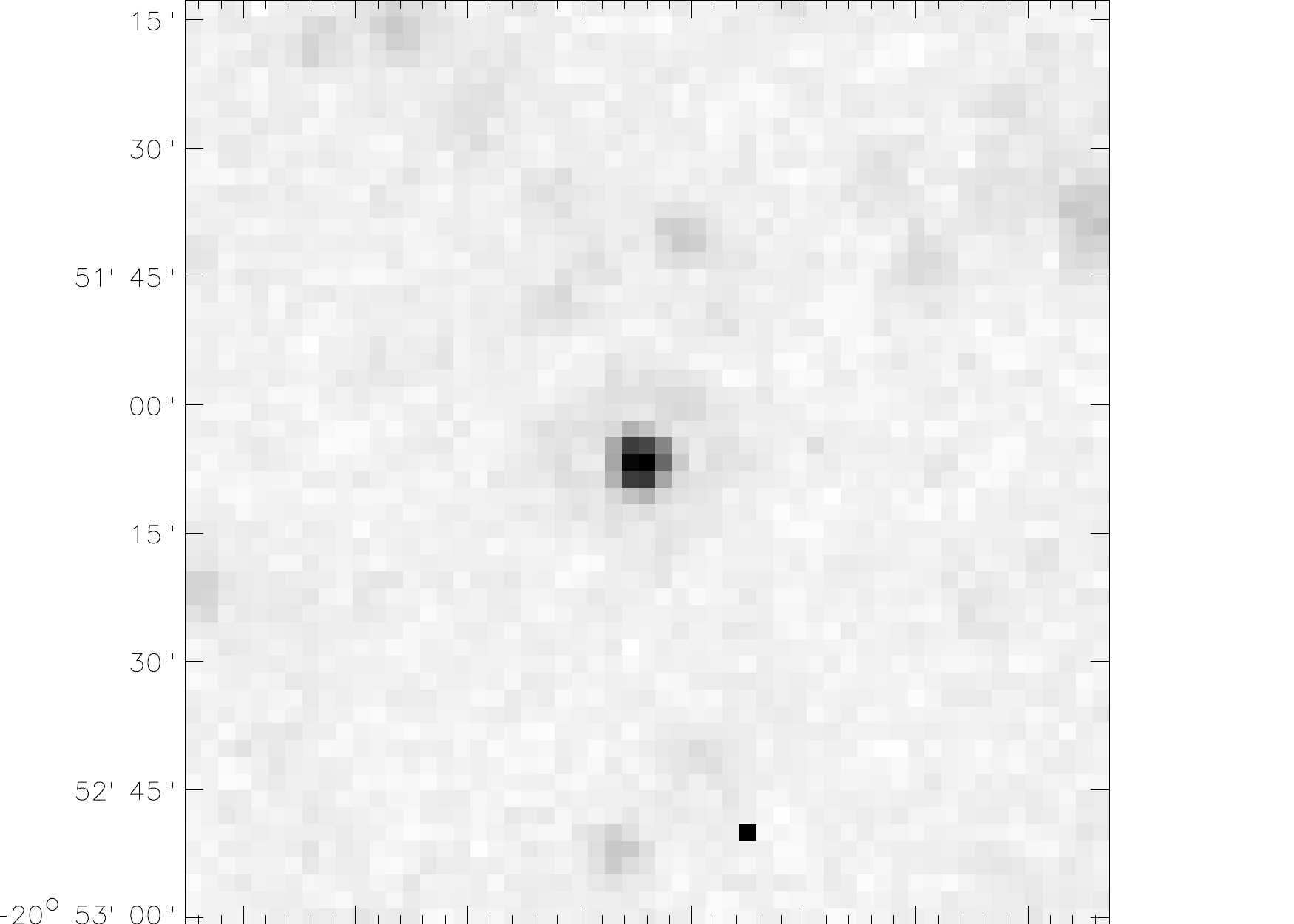} &
      \includegraphics[height=14mm,trim=70  -4 65 -10,clip=true]{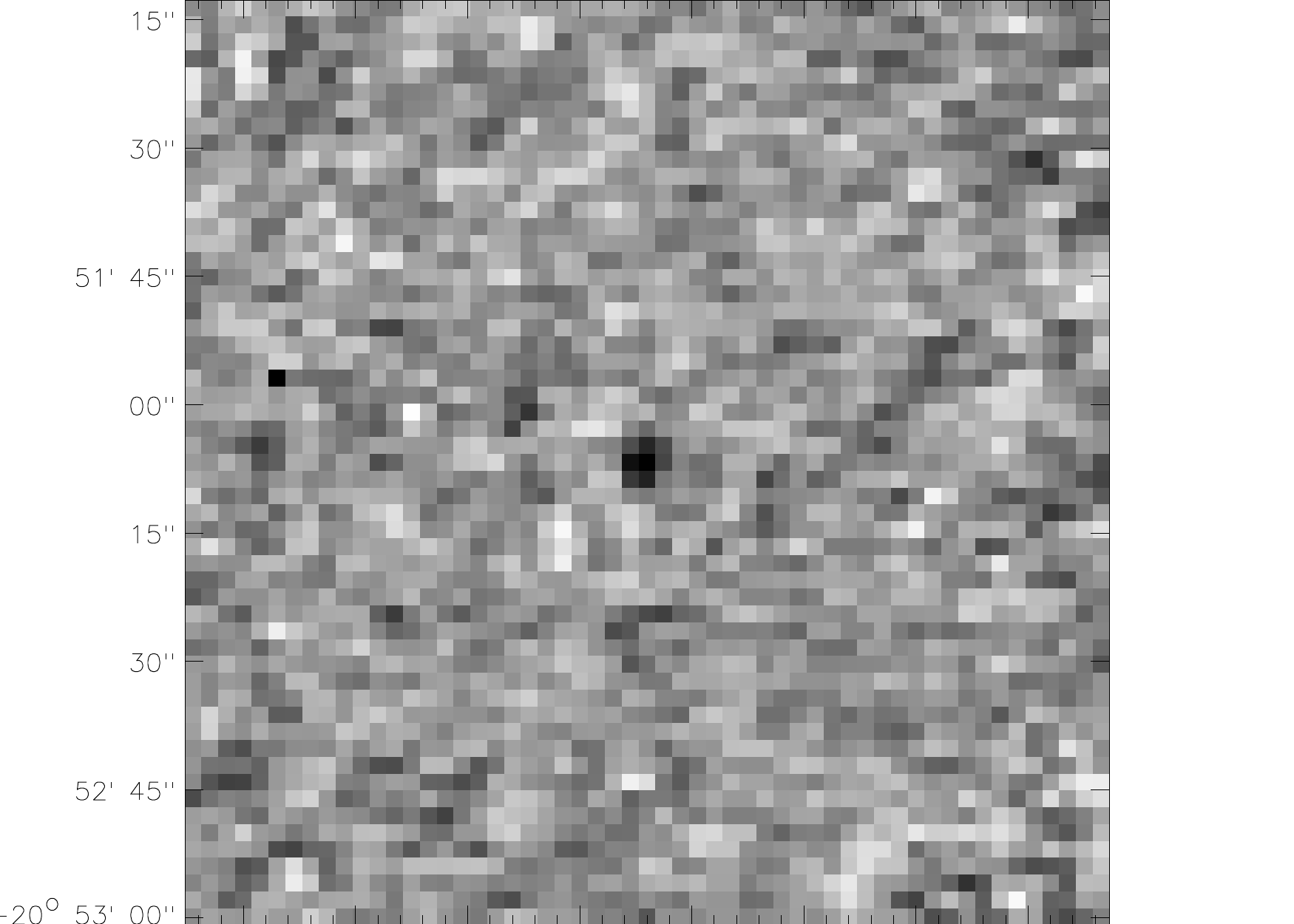} &
      \includegraphics[height=14mm,trim=70  -4 65 -10,clip=true]{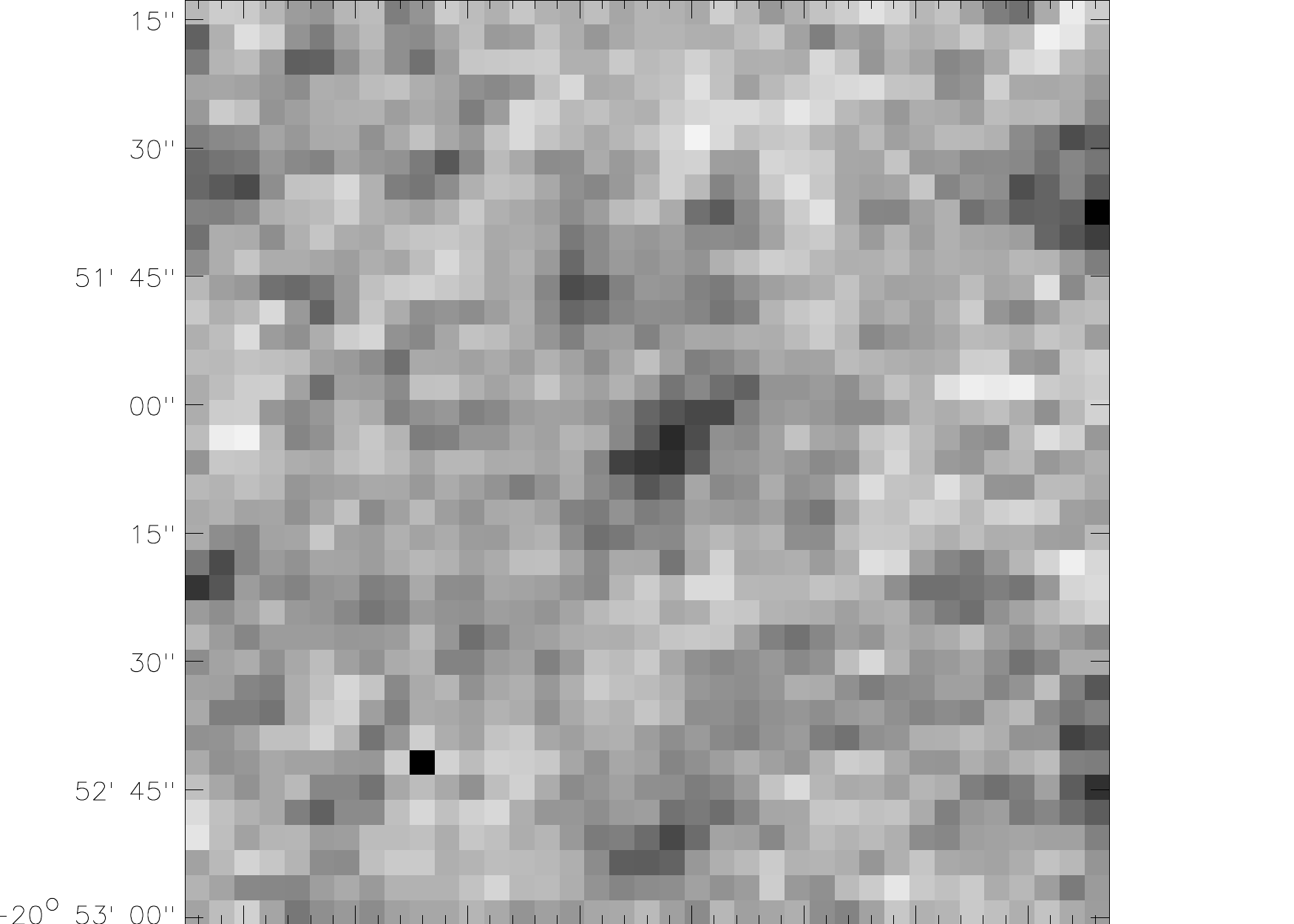} &
      \includegraphics[height=14mm,trim=70  -4 65 -10,clip=true]{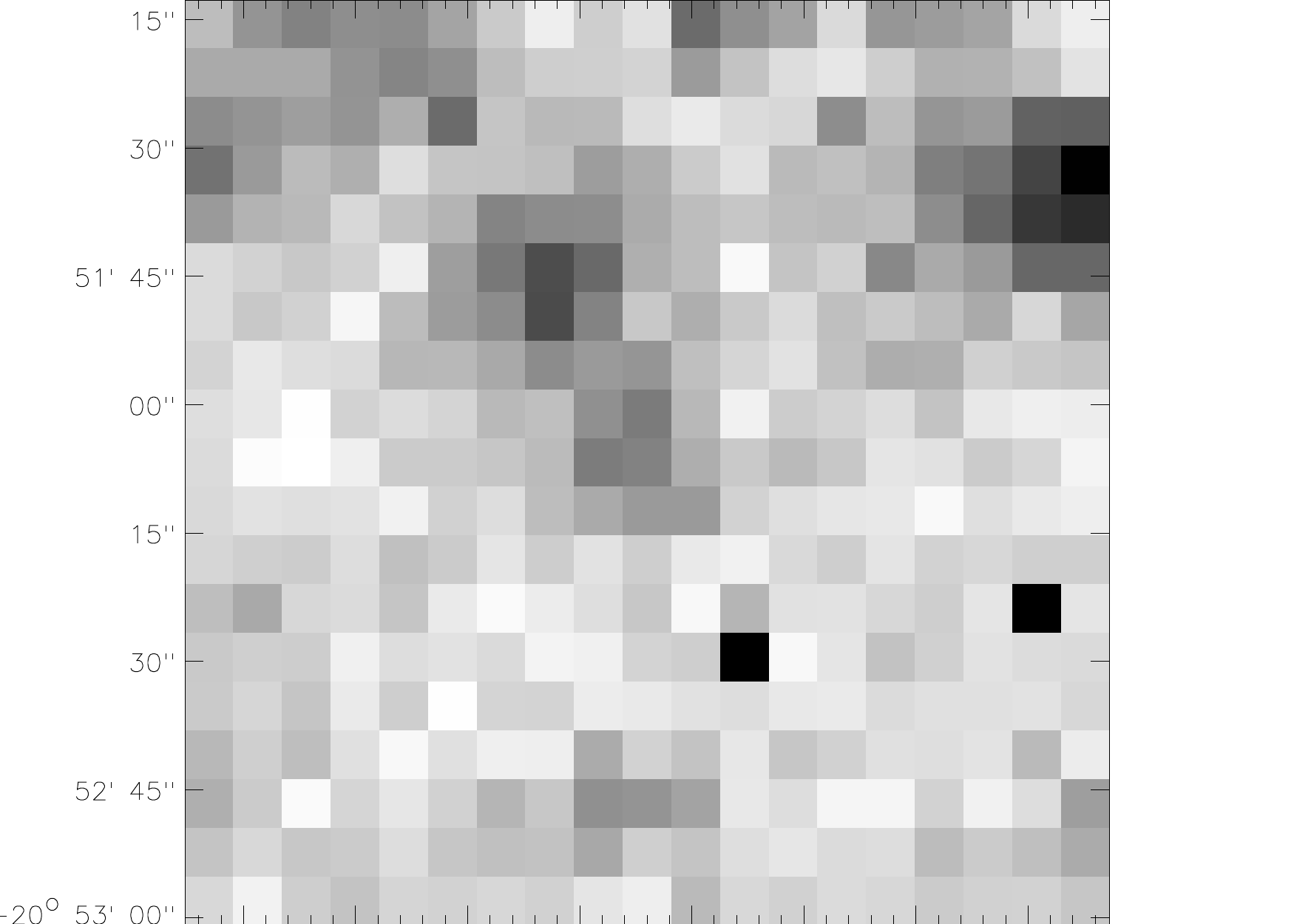}   &
      \includegraphics[height=14mm,trim=70  -4 65 -10,clip=true]{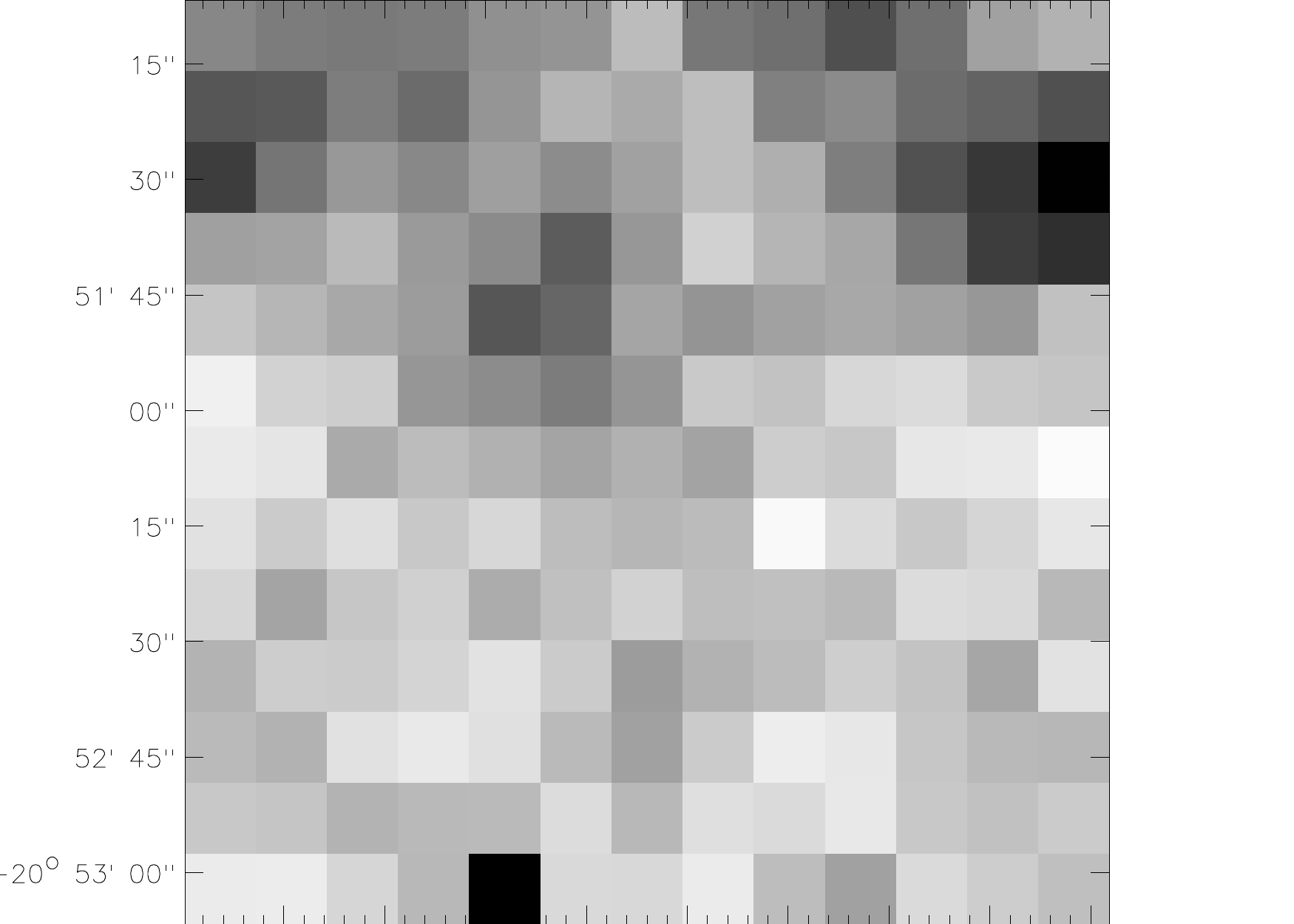}   &
      \includegraphics[height=14mm,trim=70  -4 65 -10,clip=true]{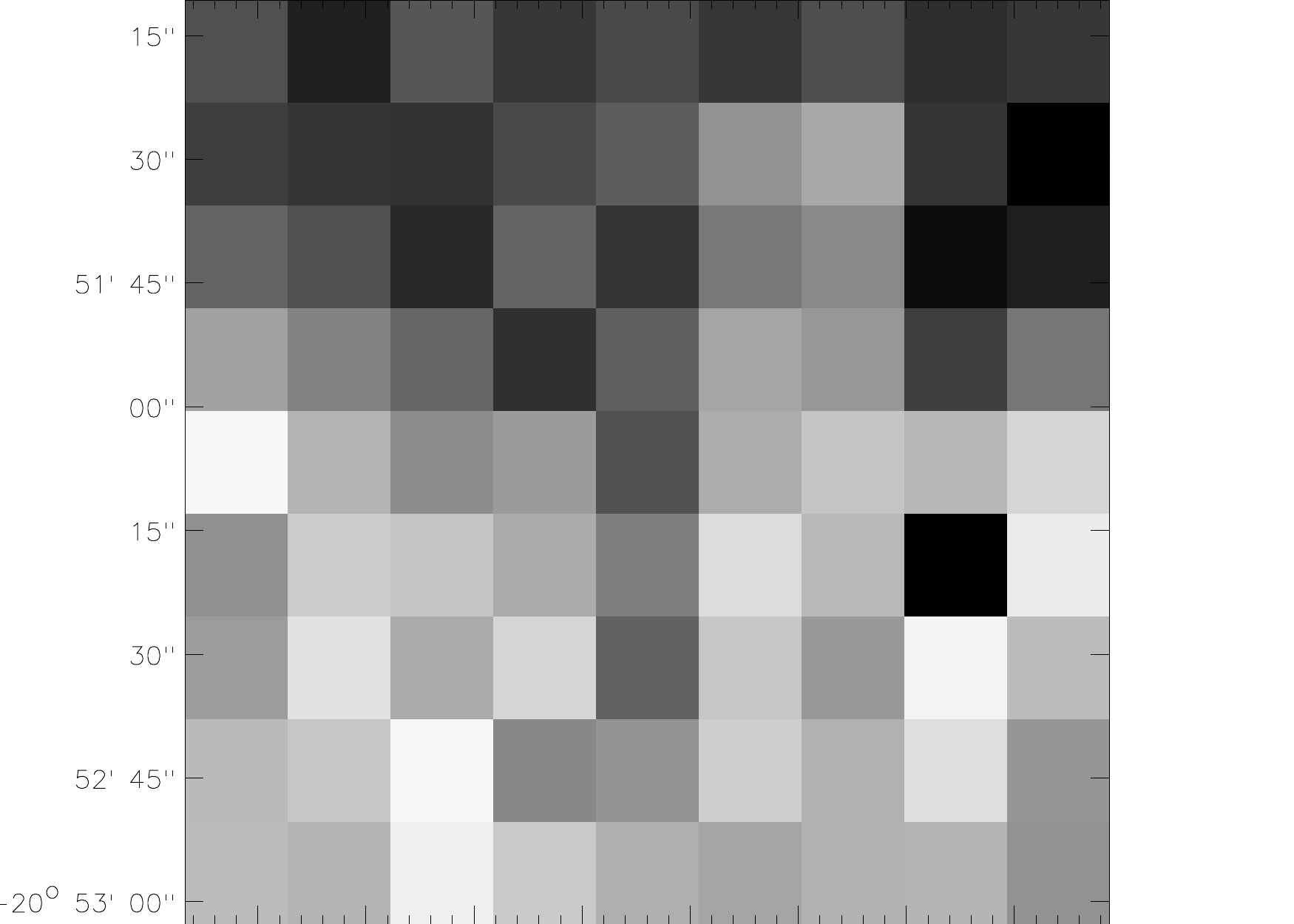} \\

    \end{tabular}
          \end{minipage}
  \end{centering}
\end{figure*}

\clearpage

\begin{figure*}[h]
  \begin{centering}
    \begin{minipage}[c]{0.46\linewidth}
    \begin{tabular}{c@{}c@{}c@{}c@{}c@{}c@{}}

      \multicolumn{6}{c}{\includegraphics[width=80mm,trim=30 0 20 0 ,clip=true]{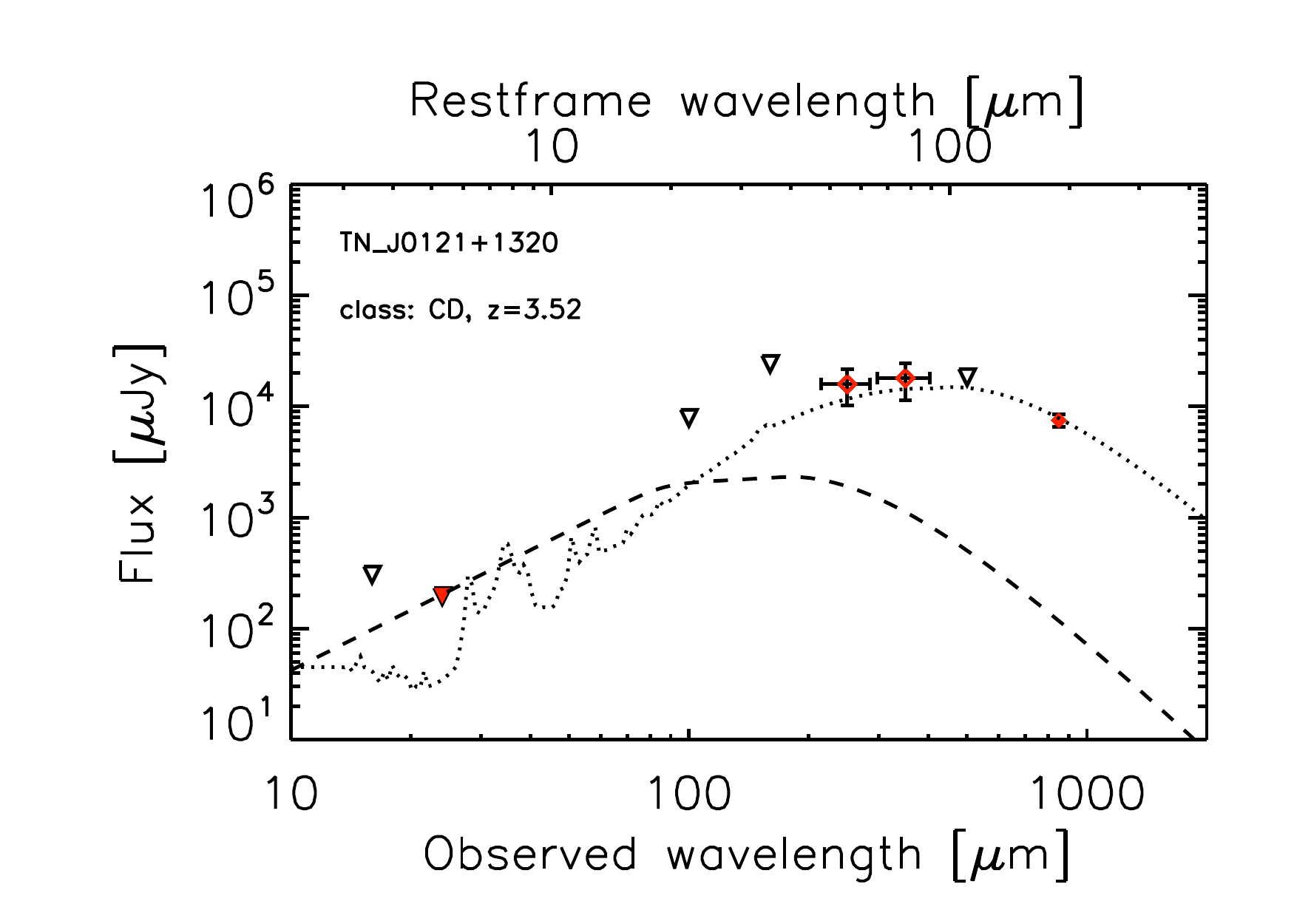}} \\
      24\mum & 100\mum & 160\mum & 250\mum & 350\mum & 500\mum \\
      \includegraphics[height=14mm,trim=70  -4 65 -10,clip=true]{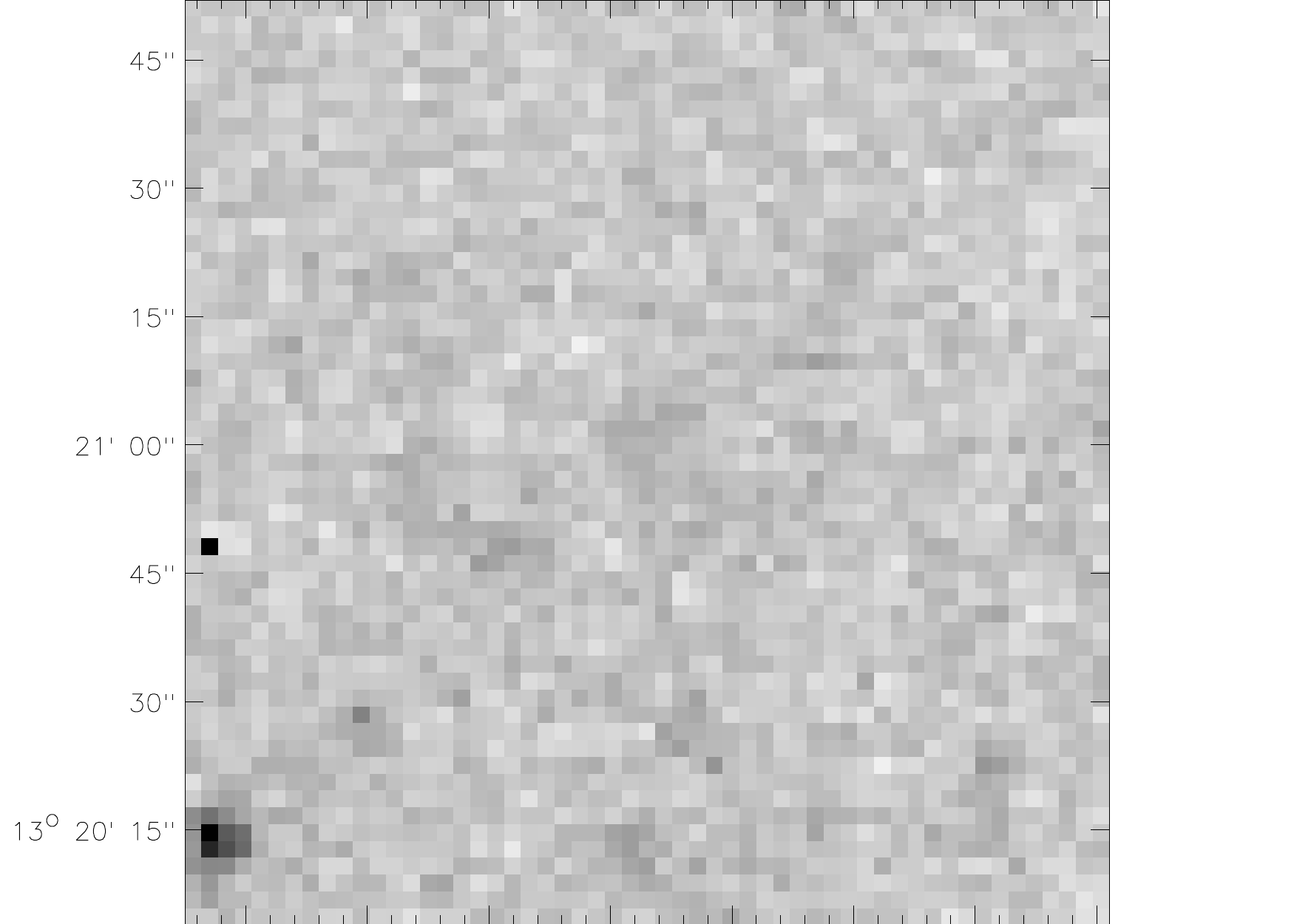} &
      \includegraphics[height=14mm,trim=70  -4 65 -10,clip=true]{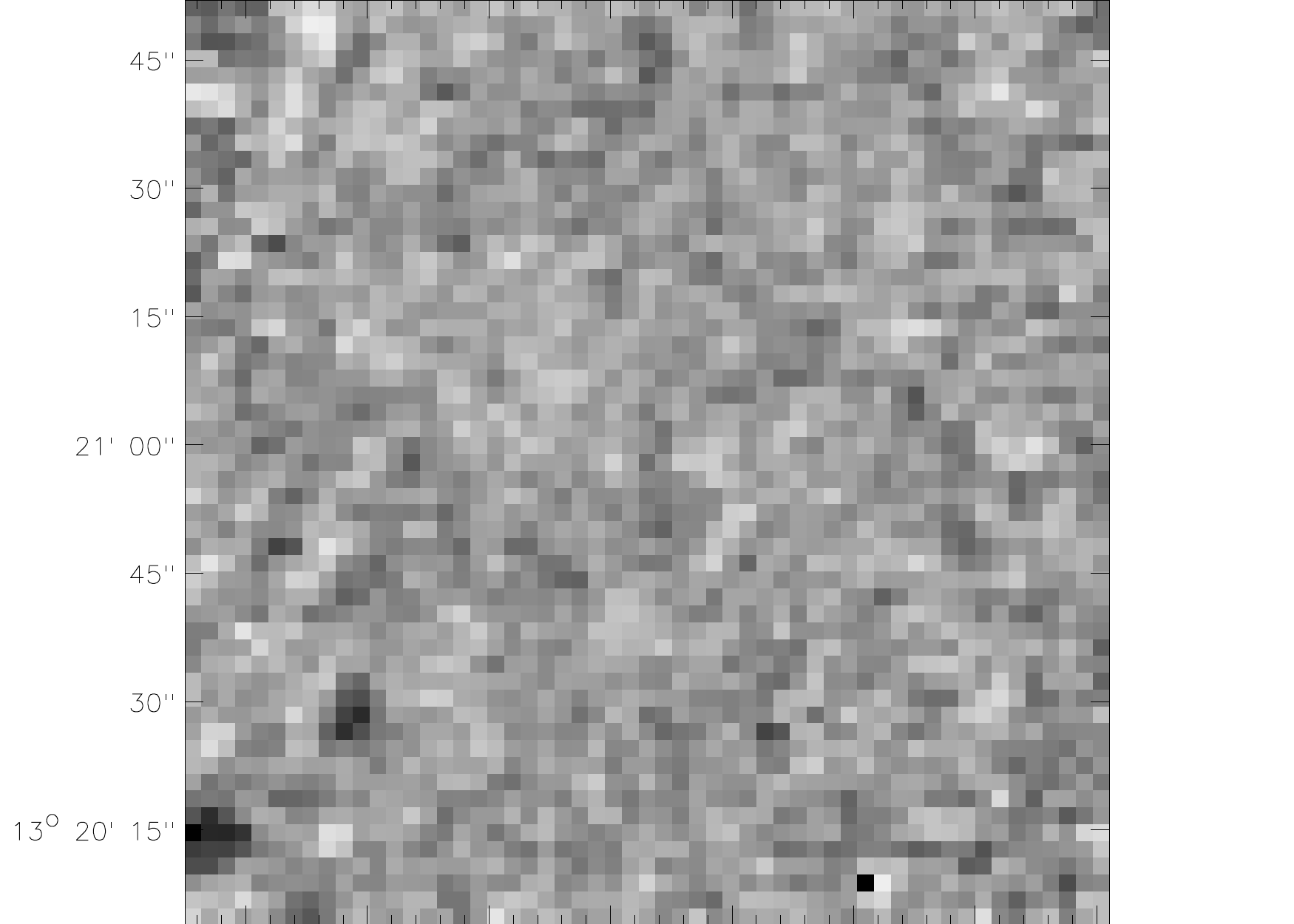} &
      \includegraphics[height=14mm,trim=70  -4 65 -10,clip=true]{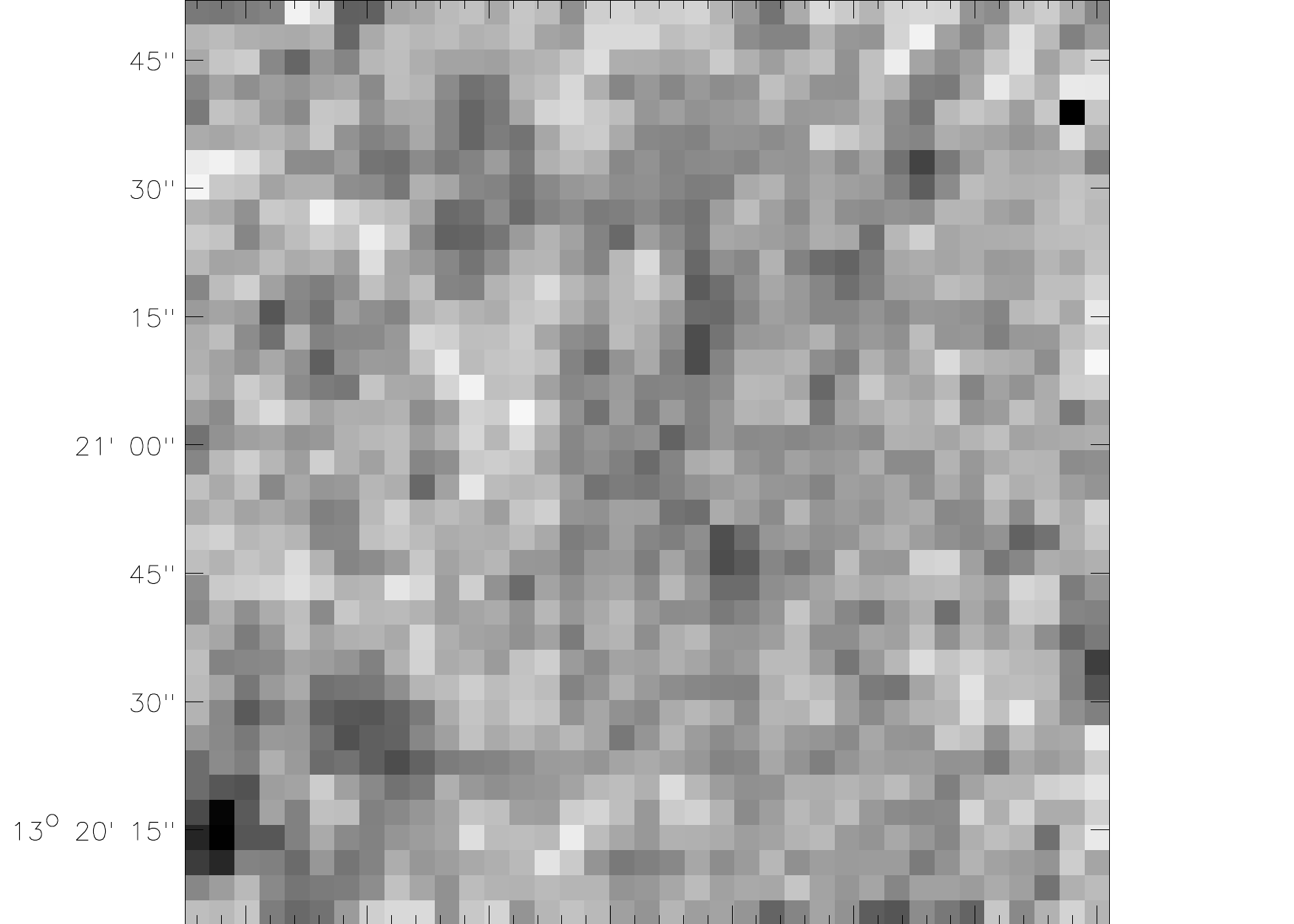} &
      \includegraphics[height=14mm,trim=70  -4 65 -10,clip=true]{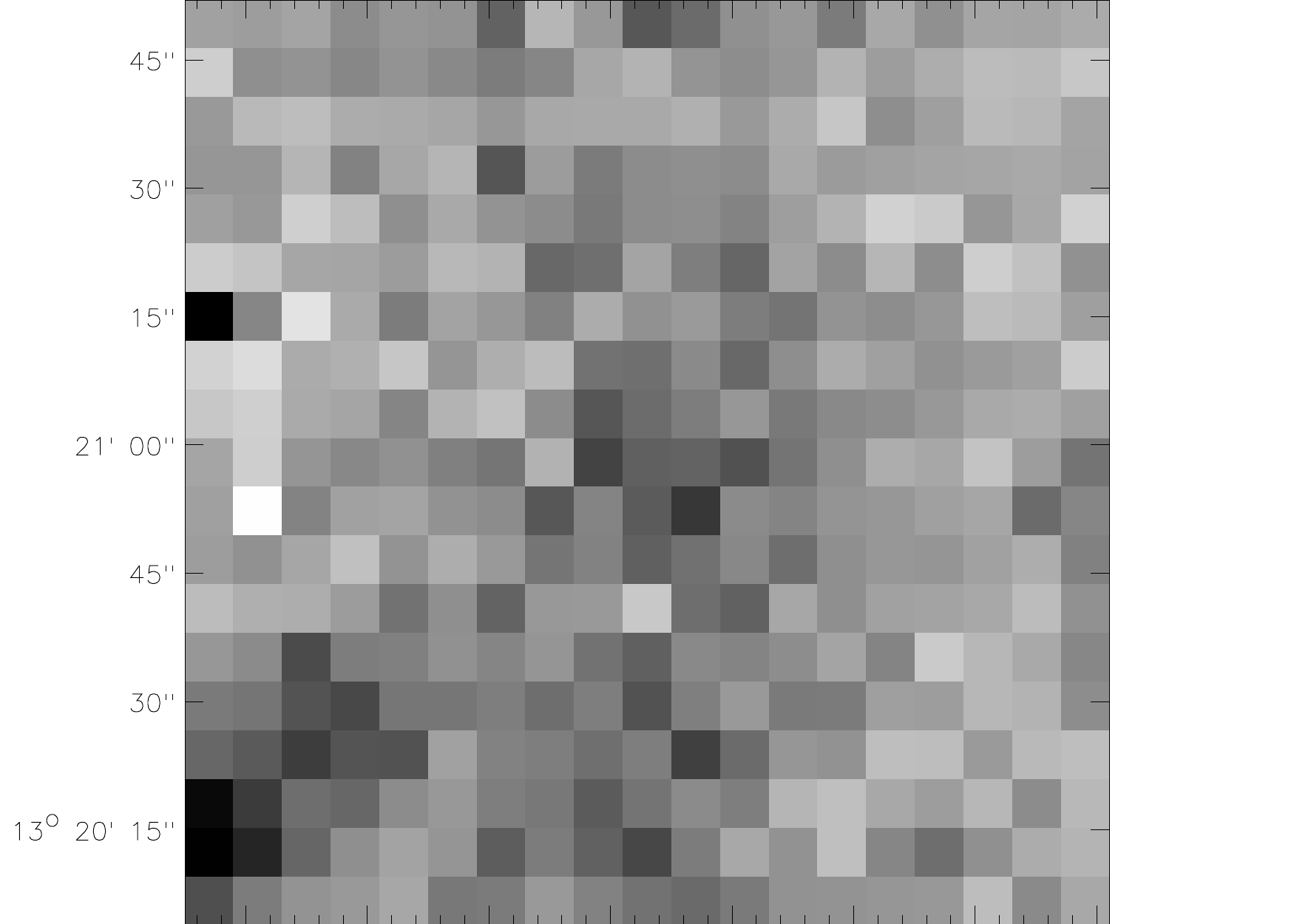}   &
      \includegraphics[height=14mm,trim=70  -4 65 -10,clip=true]{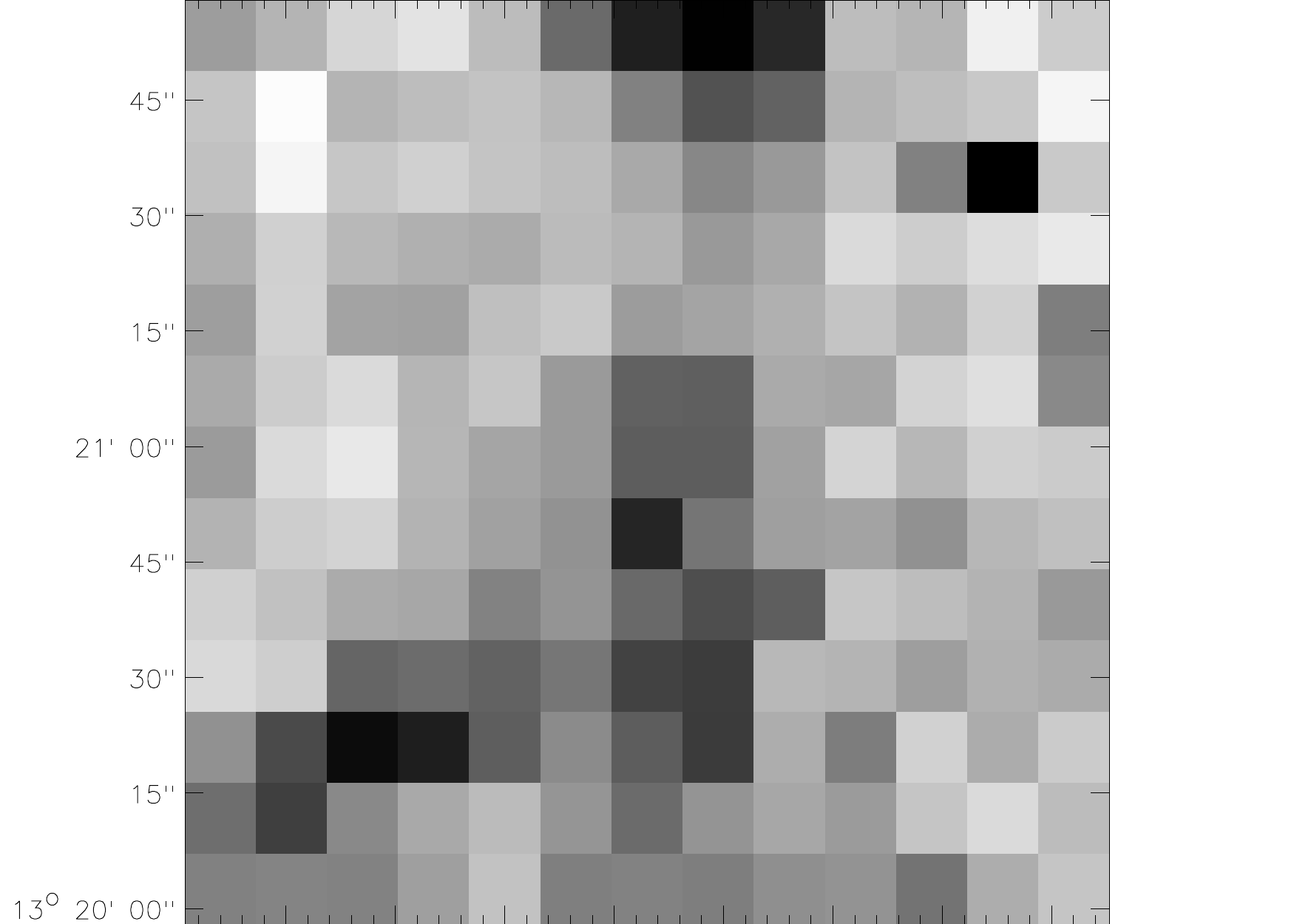}   &
      \includegraphics[height=14mm,trim=70  -4 65 -10,clip=true]{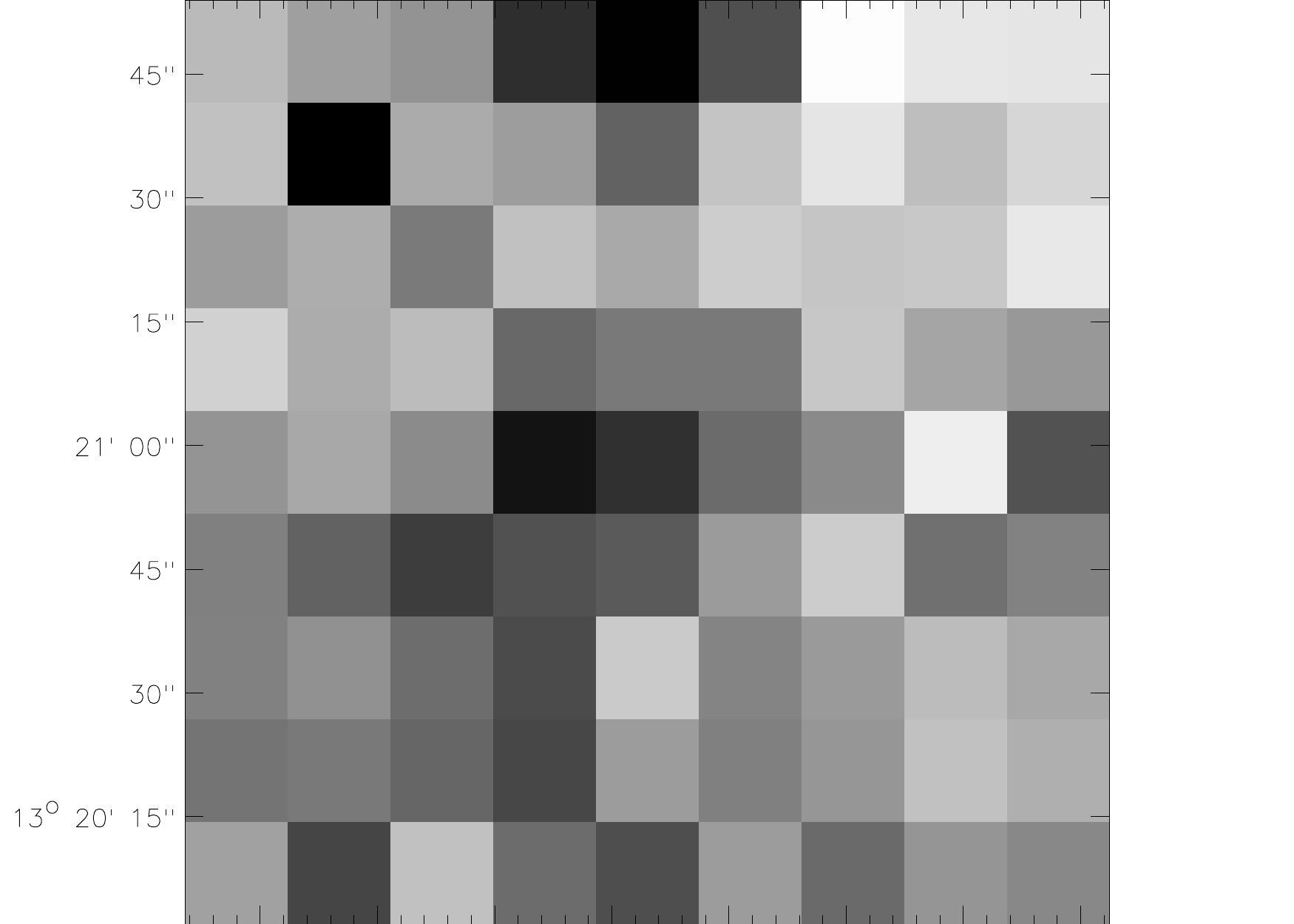} \\

      \multicolumn{6}{c}{\includegraphics[width=80mm,trim=30 0 20 0 ,clip=true]{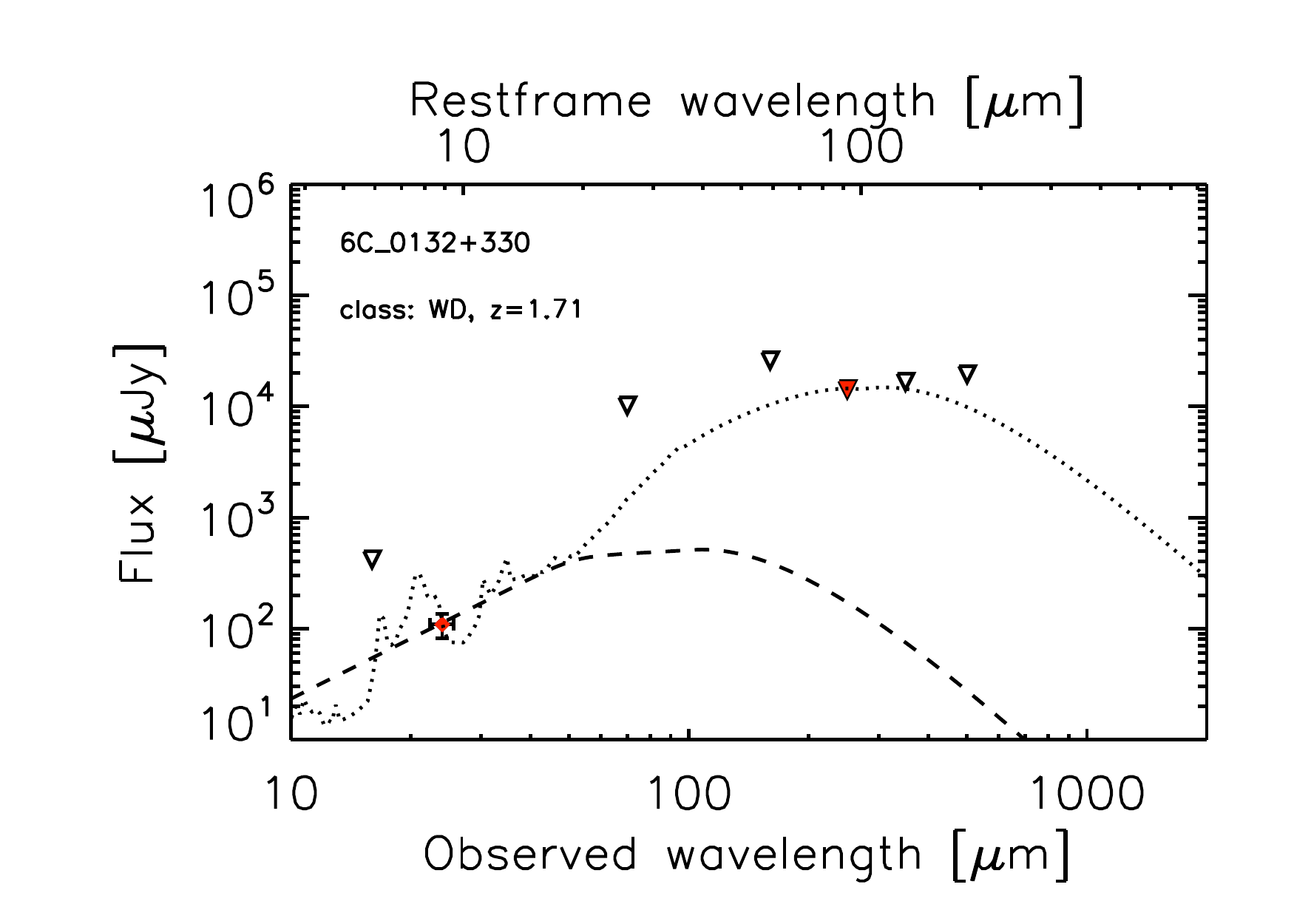}} \\
      24\mum & 70\mum & 160\mum & 250\mum & 350\mum & 500\mum \\
      \includegraphics[height=14mm,trim=70  -4 65 -10,clip=true]{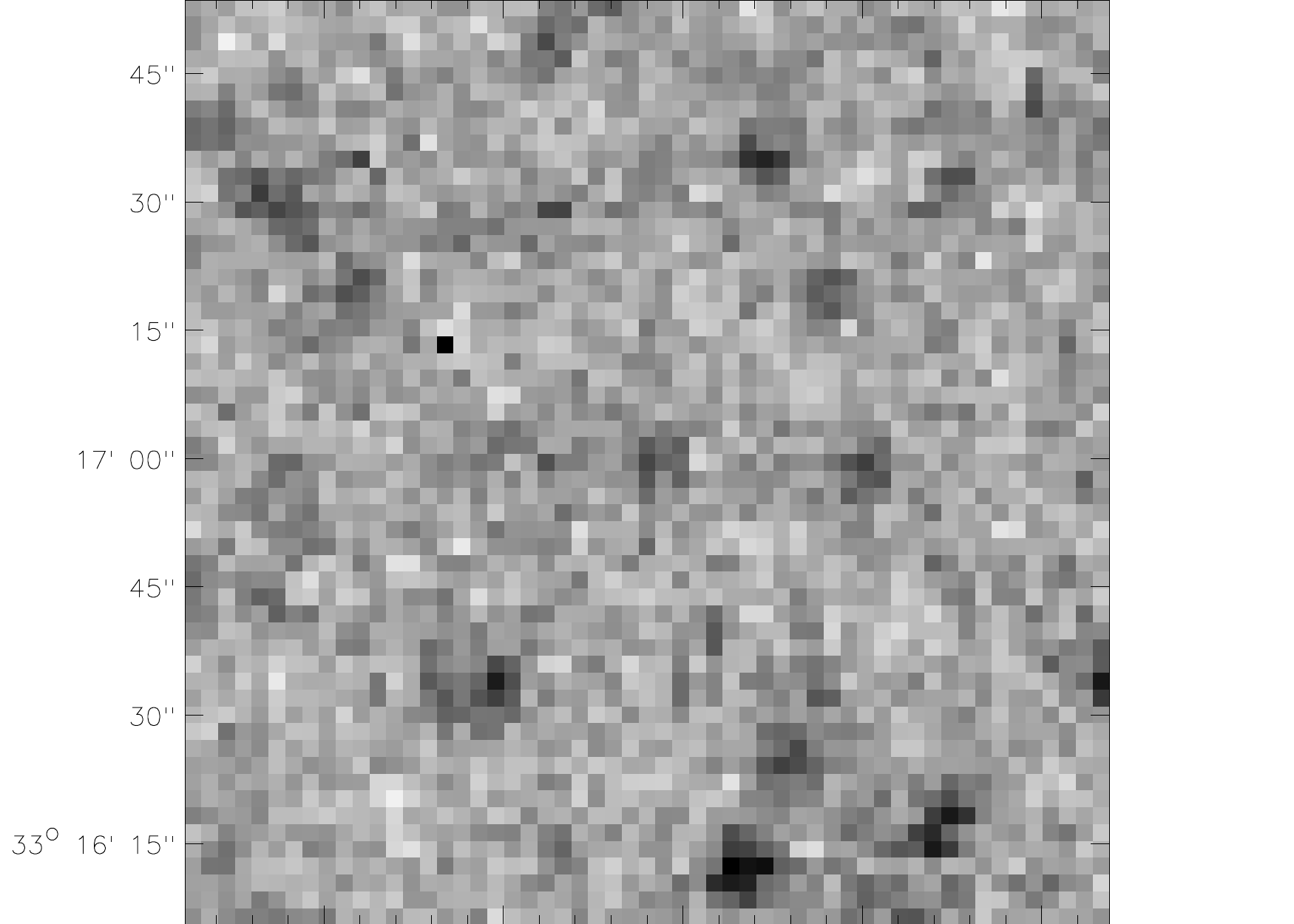} &
      \includegraphics[height=14mm,trim=70  -4 65 -10,clip=true]{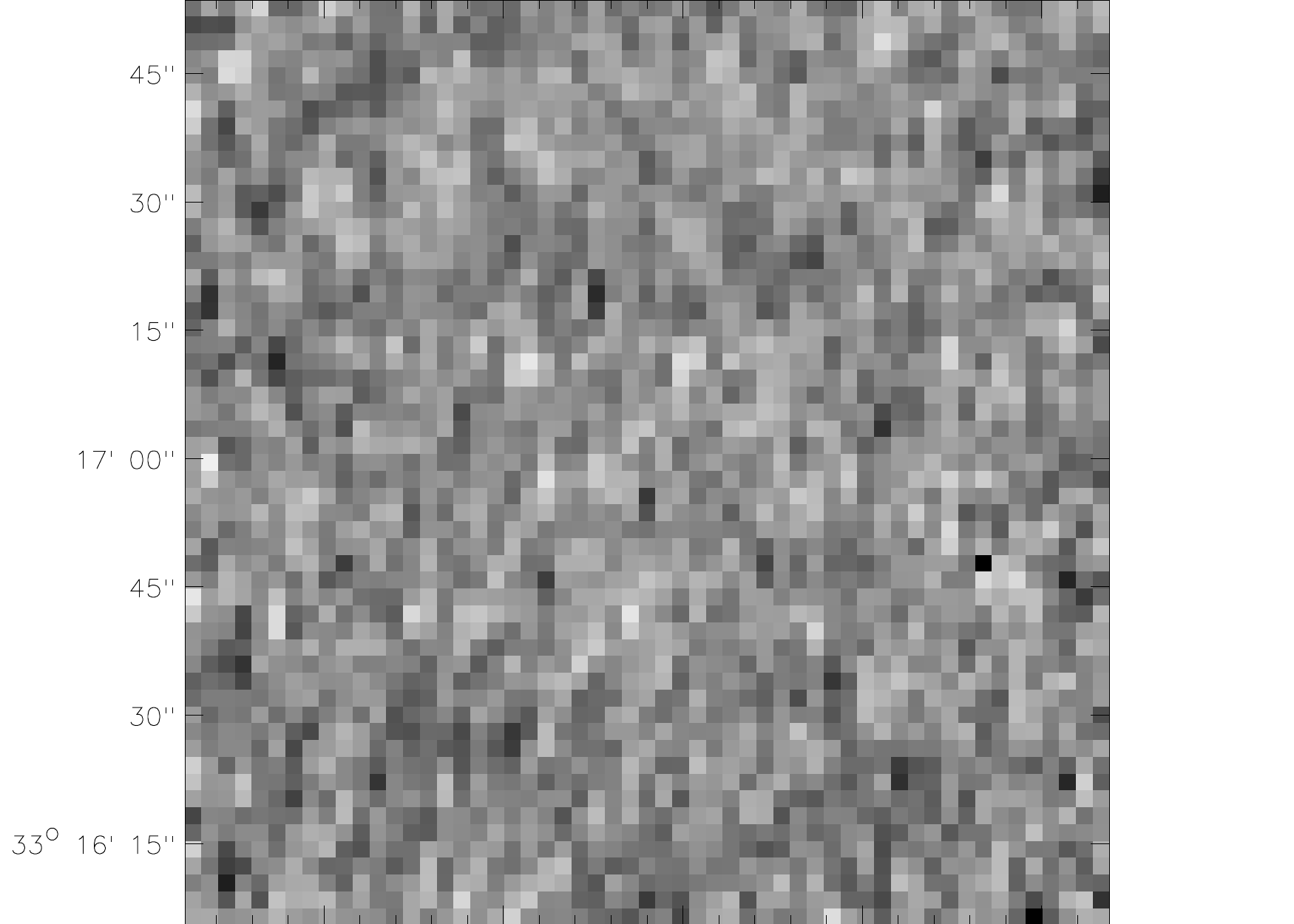} &
      \includegraphics[height=14mm,trim=70  -4 65 -10,clip=true]{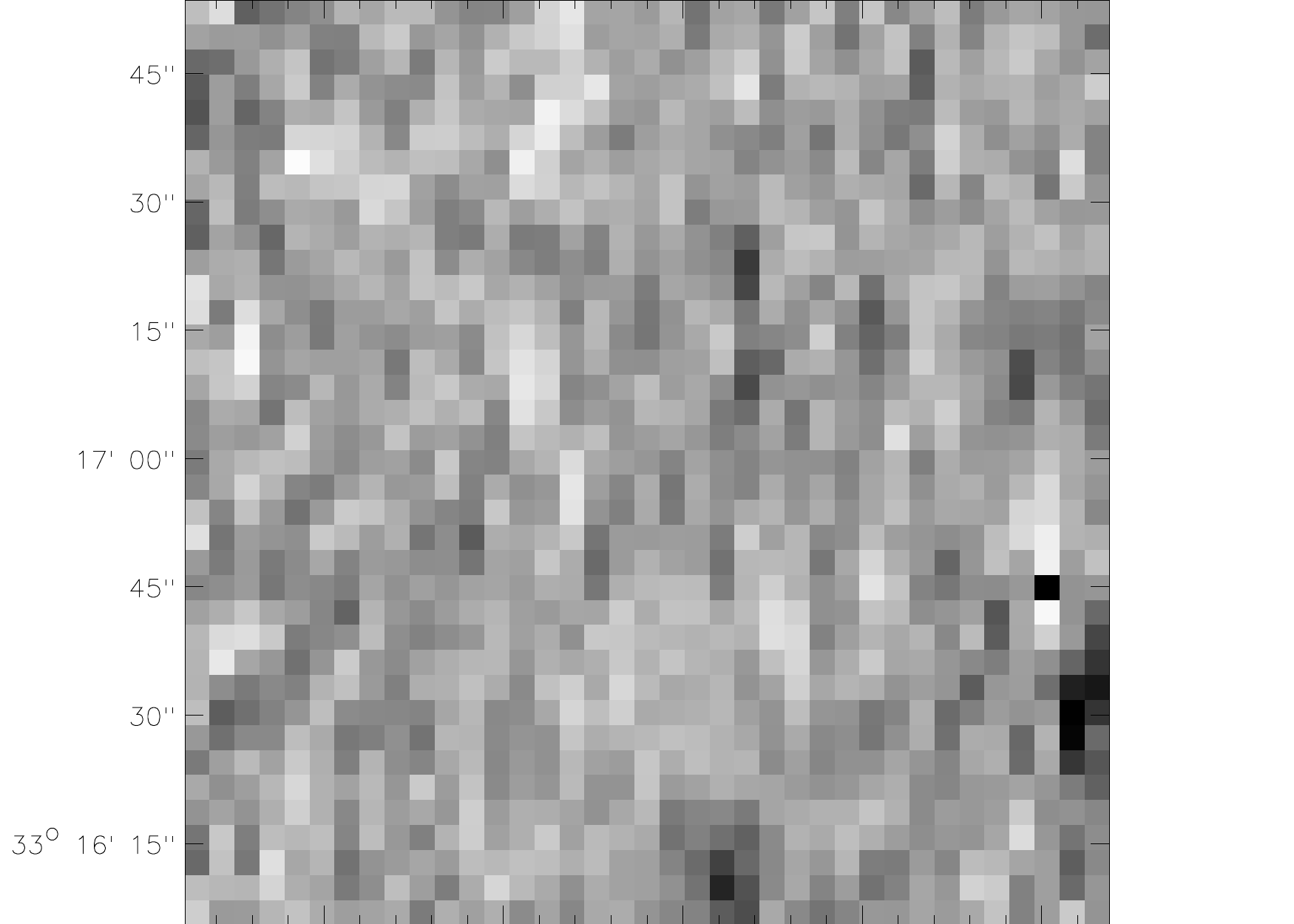} &
      \includegraphics[height=14mm,trim=70  -4 65 -10,clip=true]{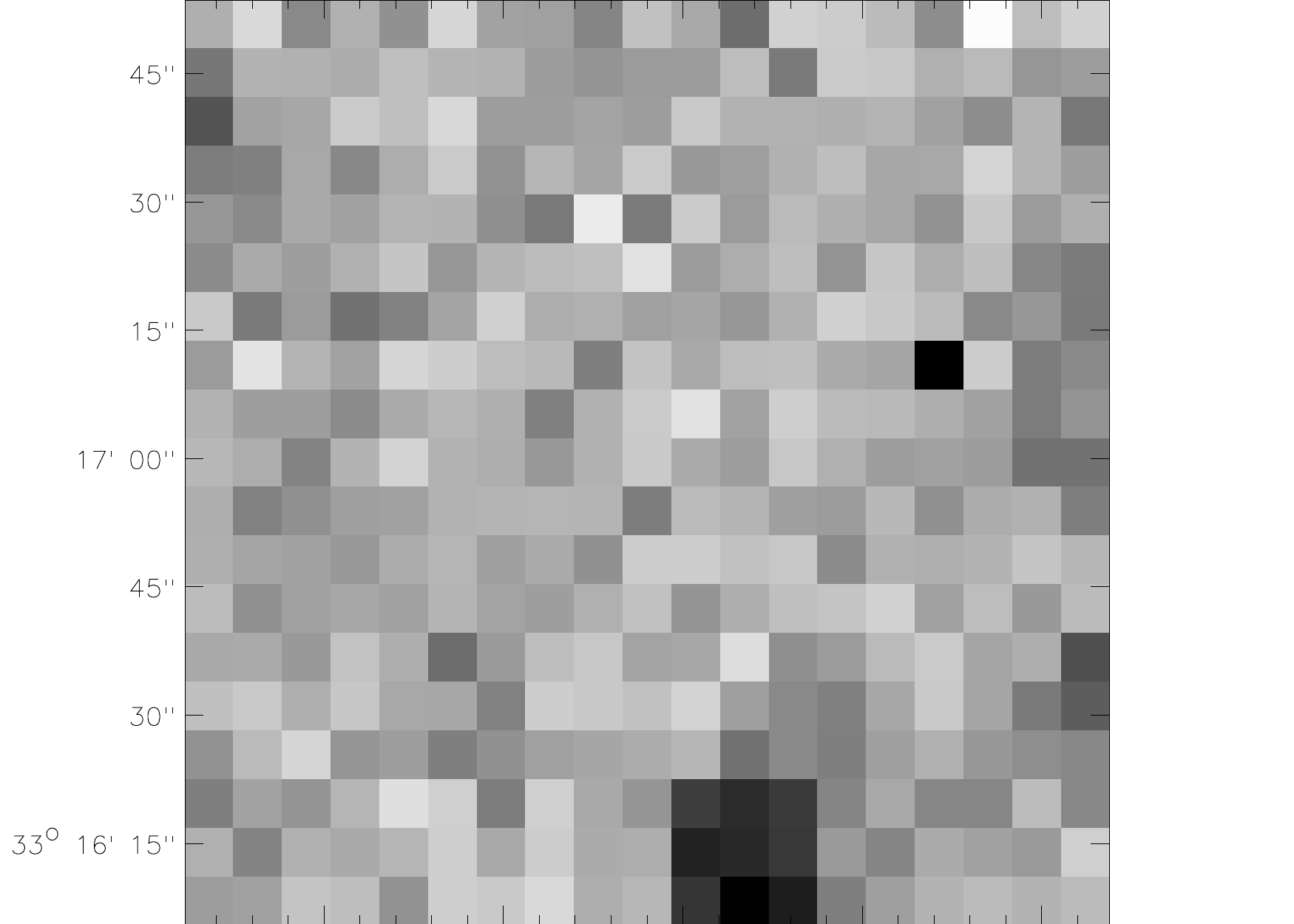}   &
      \includegraphics[height=14mm,trim=70  -4 65 -10,clip=true]{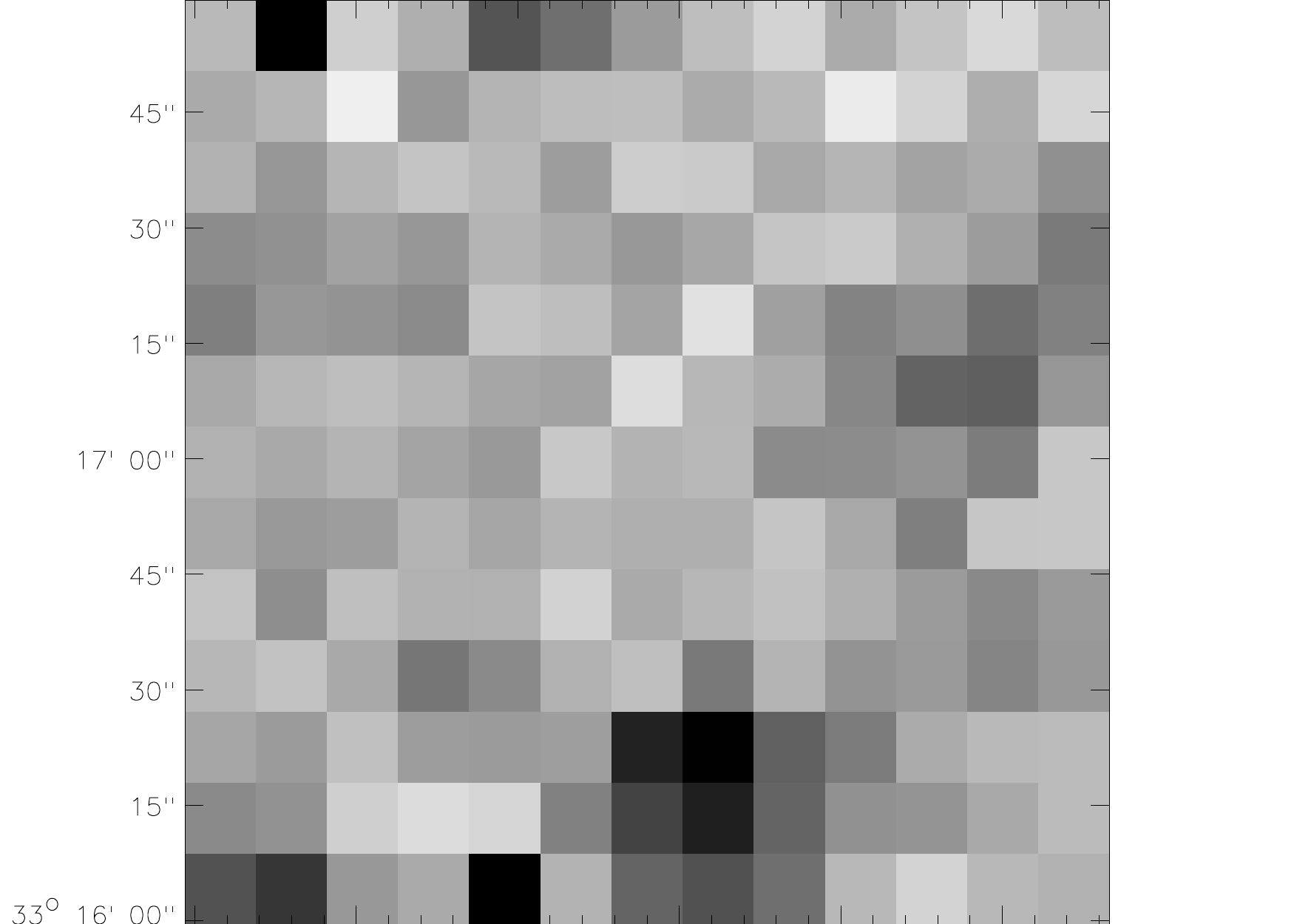}   &
      \includegraphics[height=14mm,trim=70  -4 65 -10,clip=true]{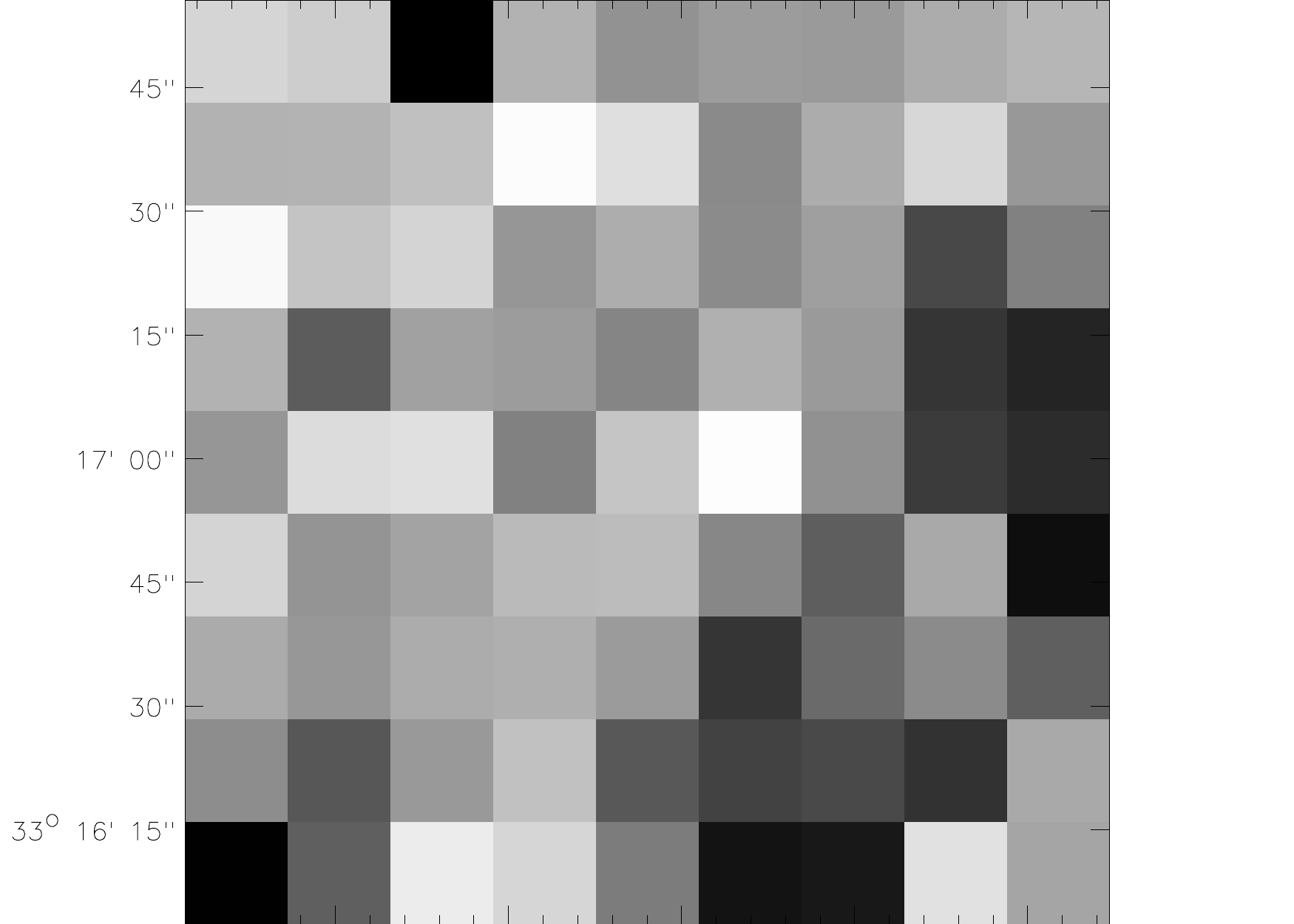} \\

      \multicolumn{6}{c}{\includegraphics[width=80mm,trim=30 0 20 0 ,clip=true]{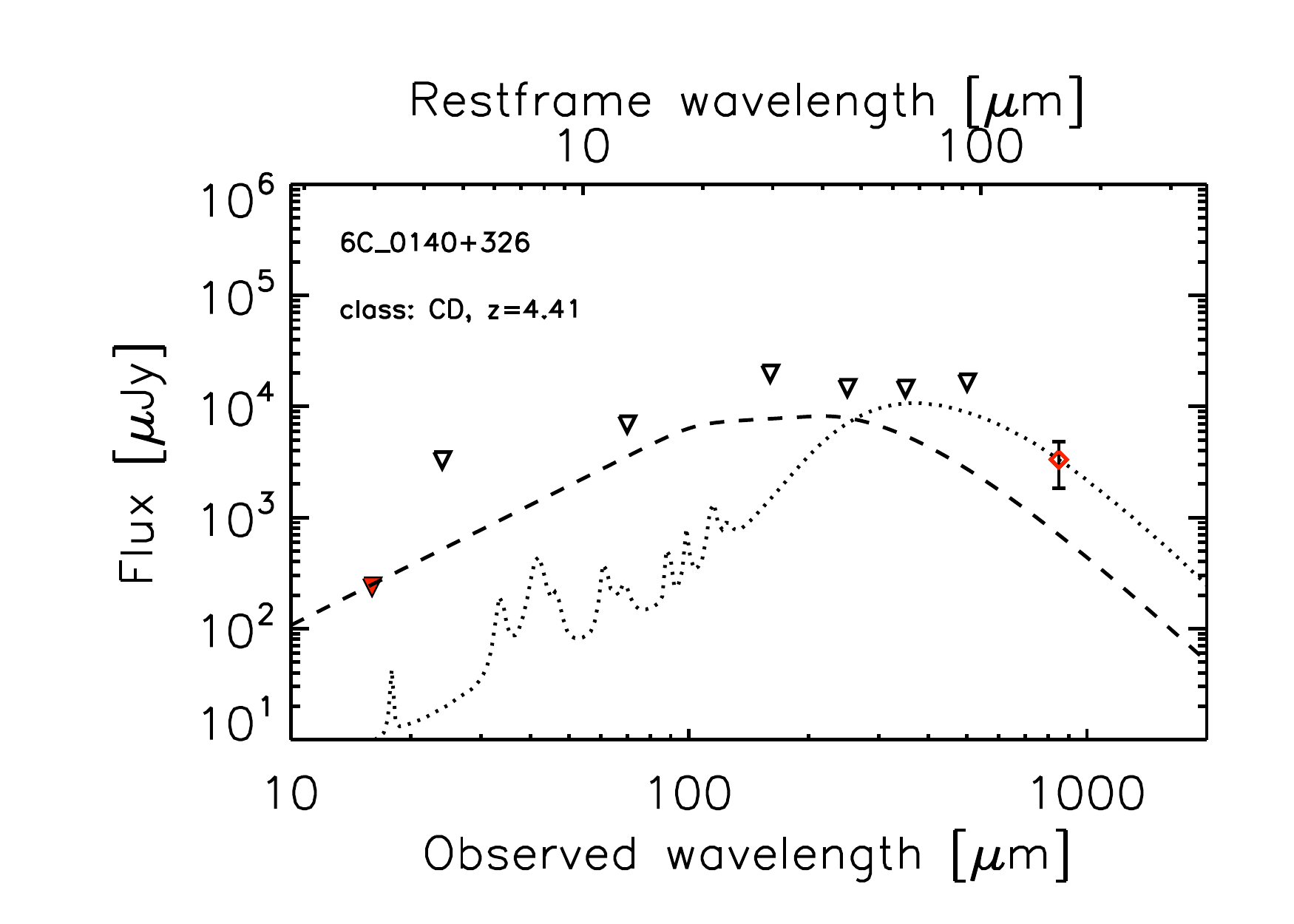}} \\
      24\mum & 70\mum & 160\mum & 250\mum & 350\mum & 500\mum \\
\end{tabular}
\end{minipage}
\begin{minipage}[c]{0.46\linewidth} 
    \begin{tabular}{c@{}c@{}c@{}c@{}c@{}c@{}}

      \multicolumn{6}{c}{\includegraphics[width=80mm,trim=30 0 20 0 ,clip=true]{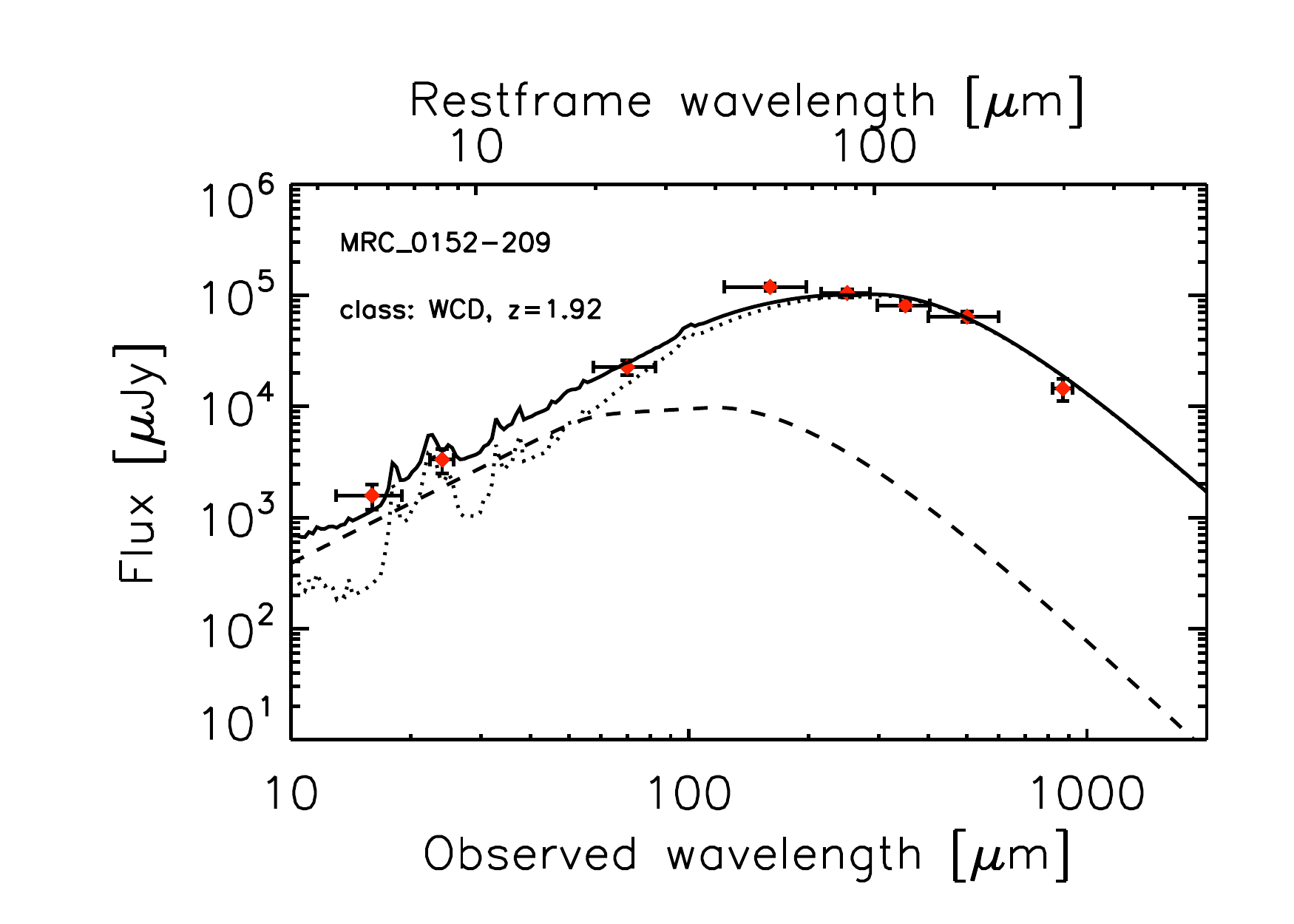}} \\
      24\mum & 70\mum & 160\mum & 250\mum & 350\mum & 500\mum \\
      \includegraphics[height=14mm,trim=70  -4 65 -10,clip=true]{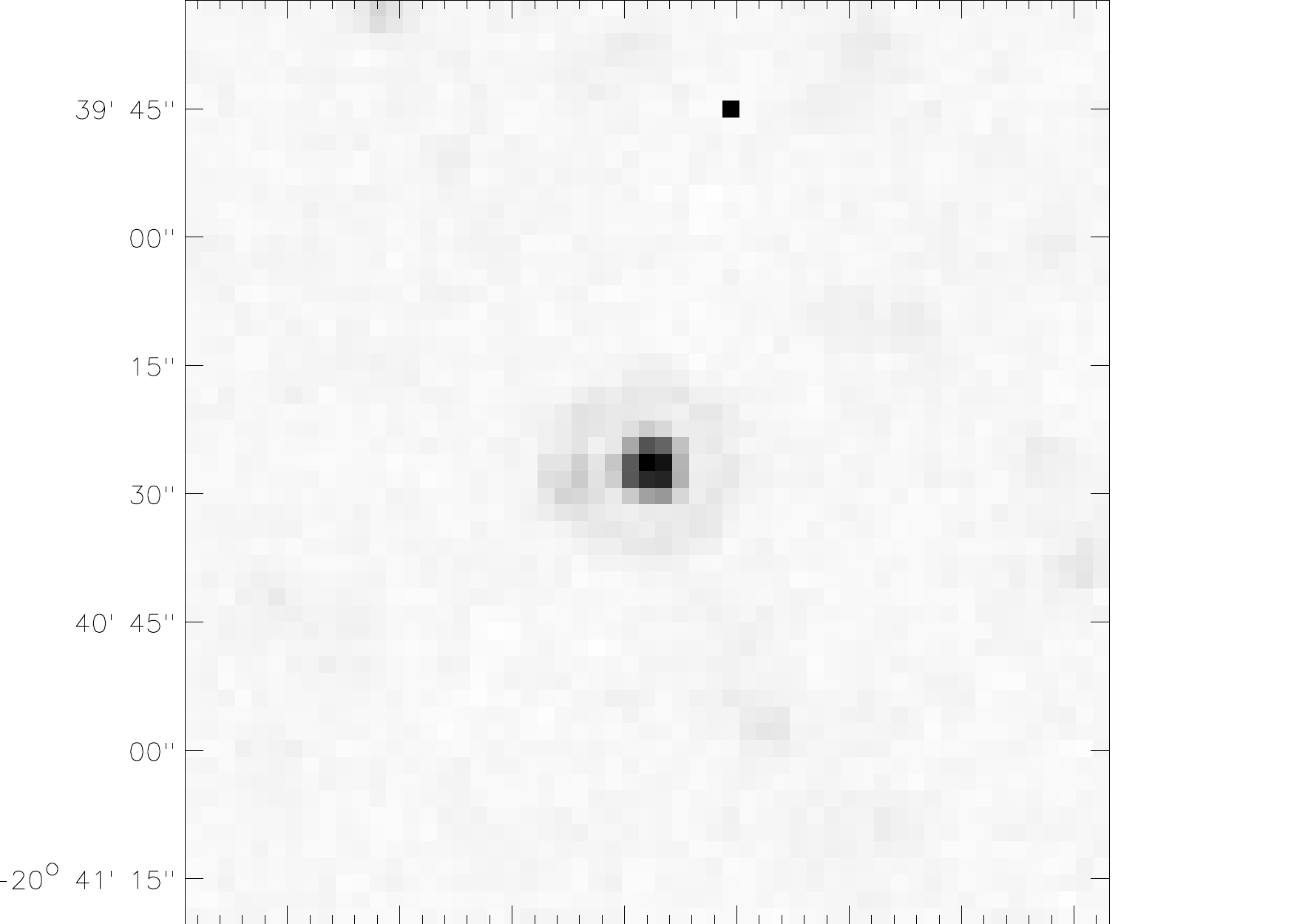} &
      \includegraphics[height=14mm,trim=70  -4 65 -10,clip=true]{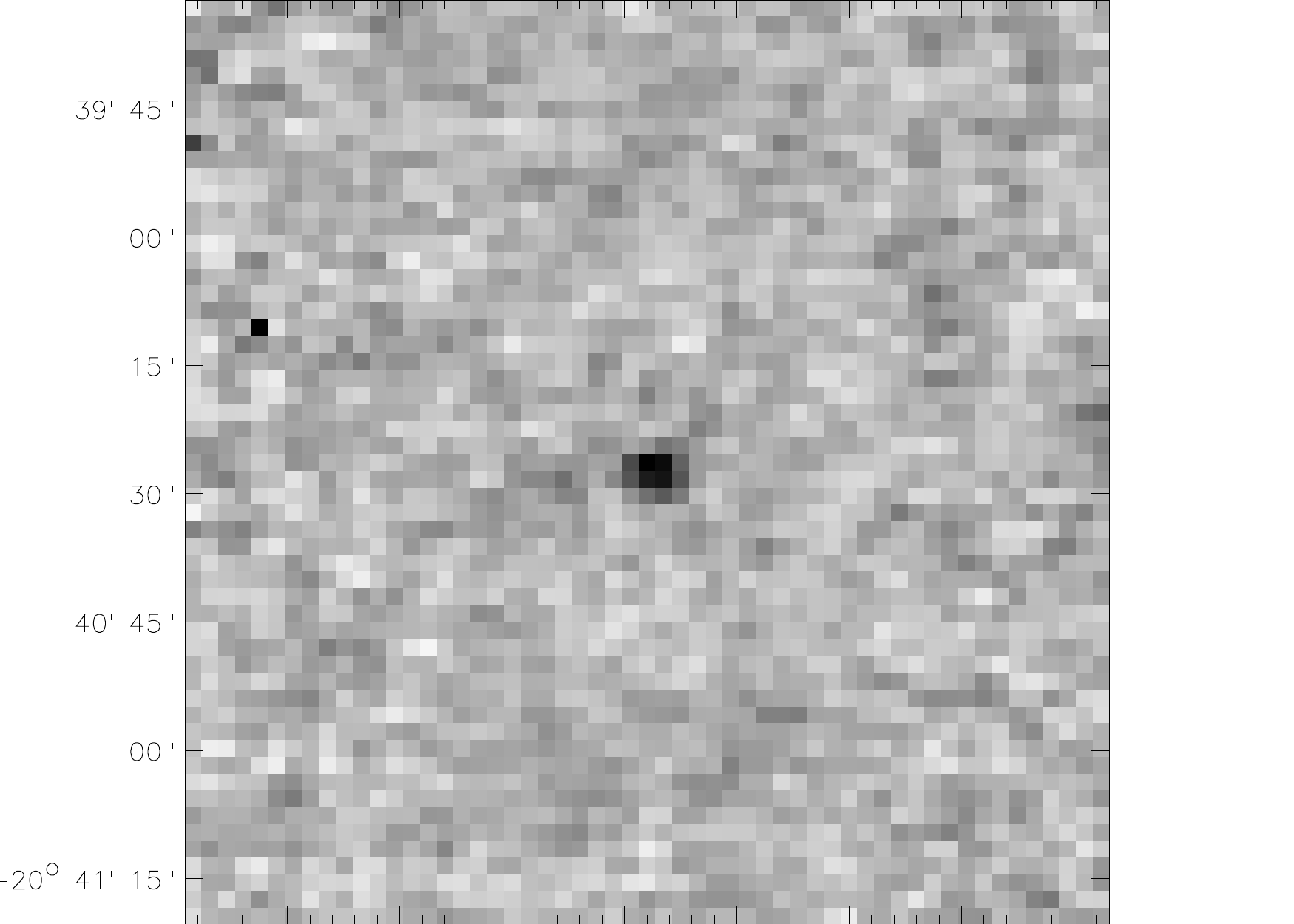} &
      \includegraphics[height=14mm,trim=70  -4 65 -10,clip=true]{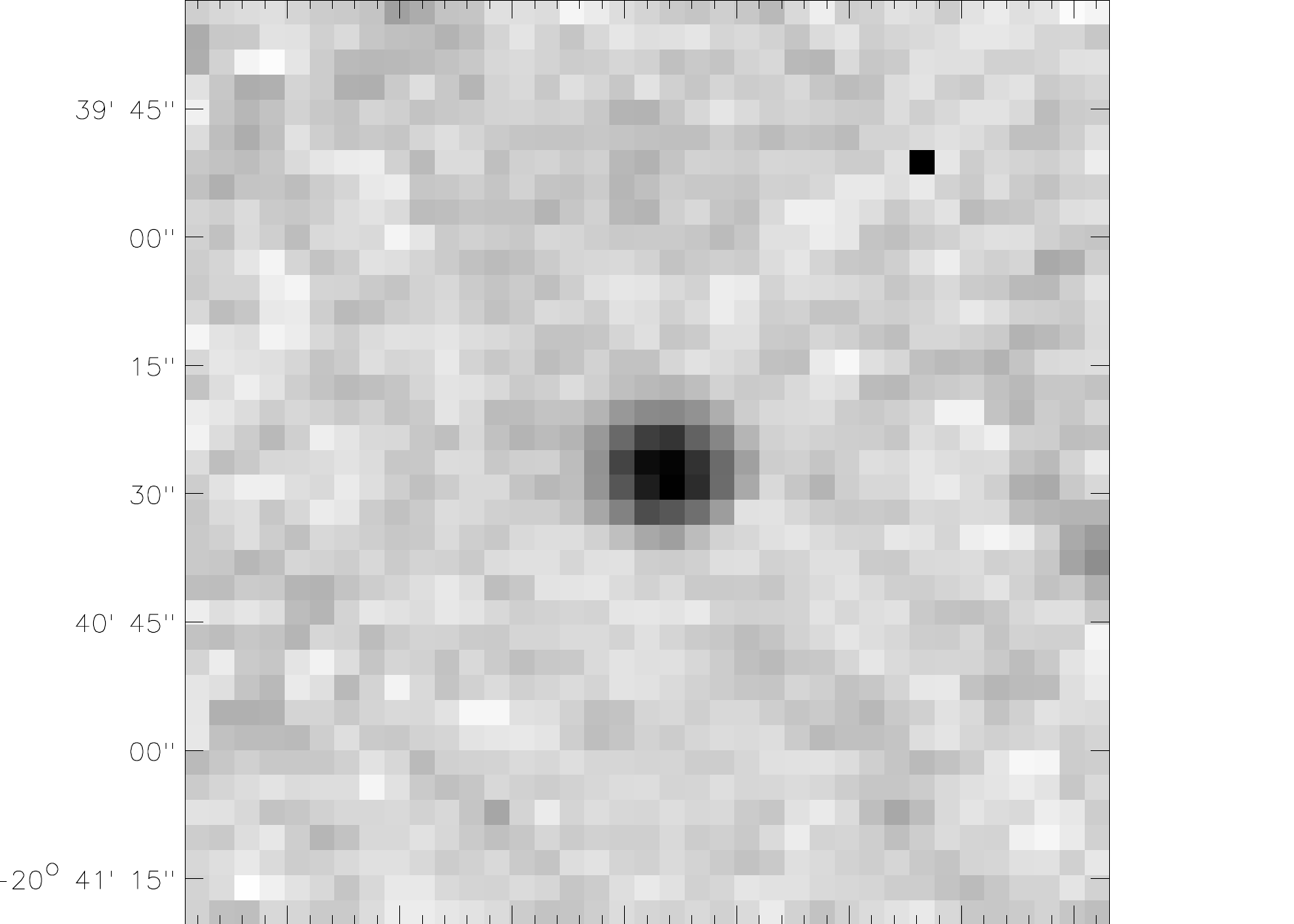} &
      \includegraphics[height=14mm,trim=70  -4 65 -10,clip=true]{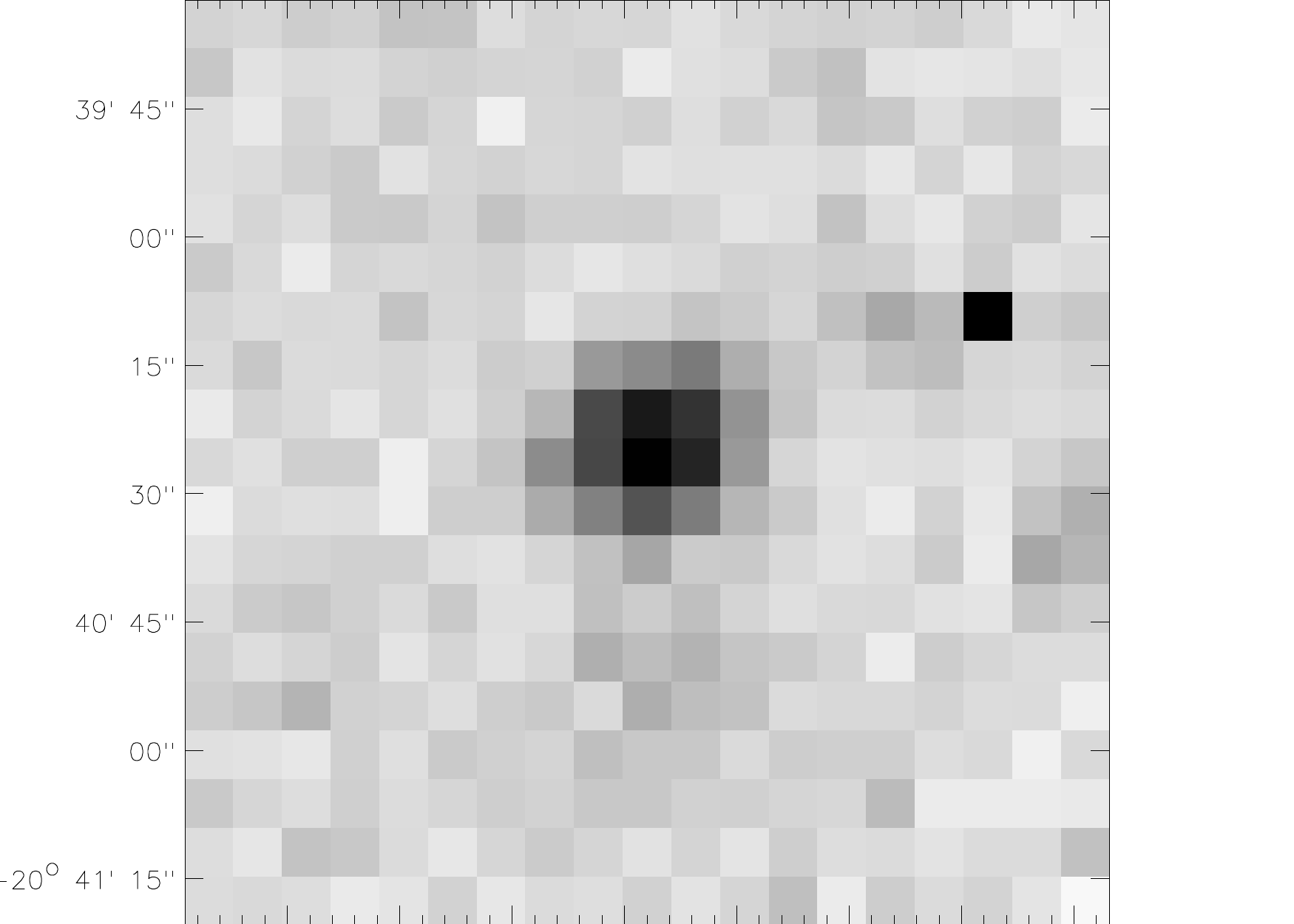}   &
      \includegraphics[height=14mm,trim=70  -4 65 -10,clip=true]{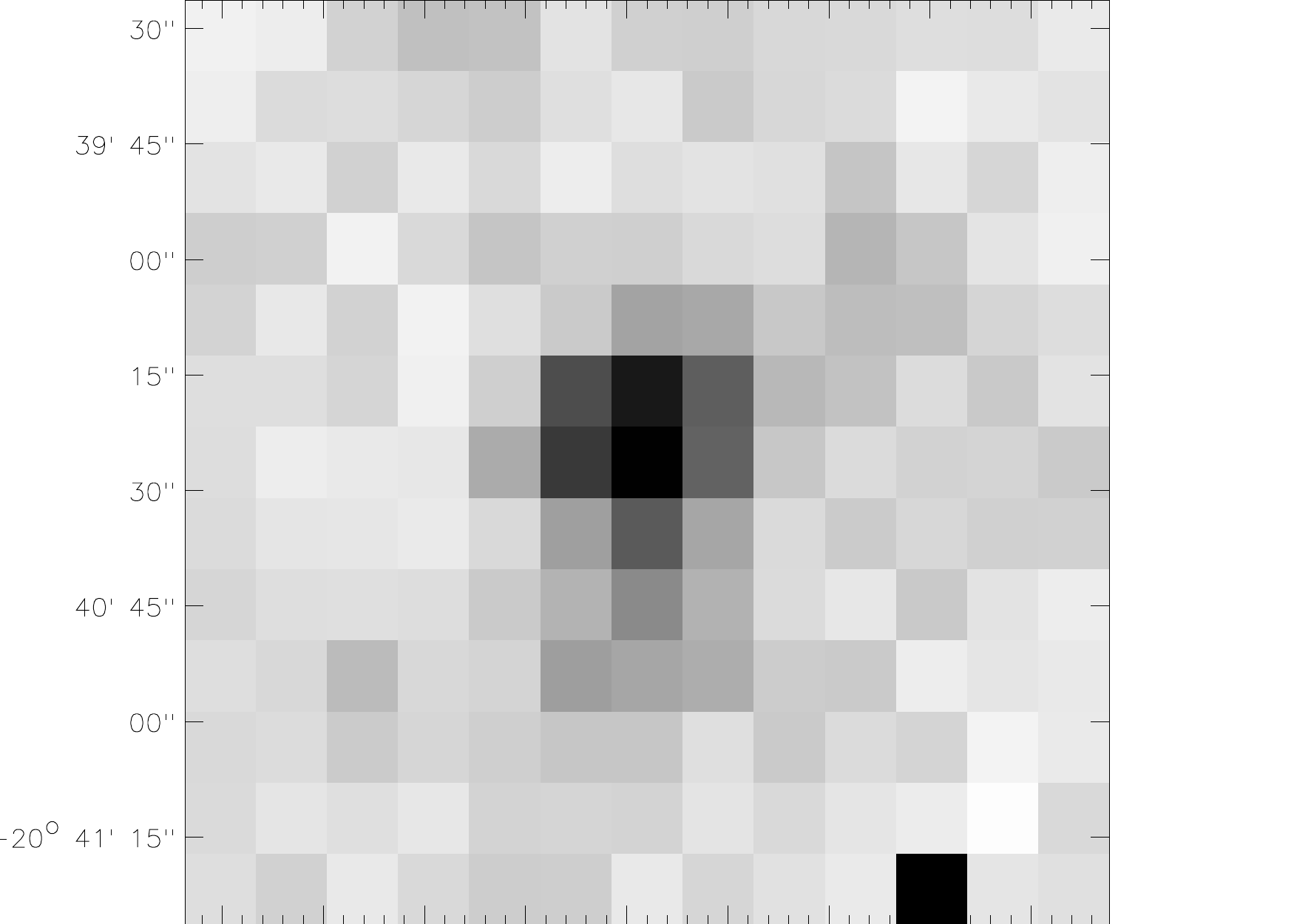}   &
      \includegraphics[height=14mm,trim=70  -4 65 -10,clip=true]{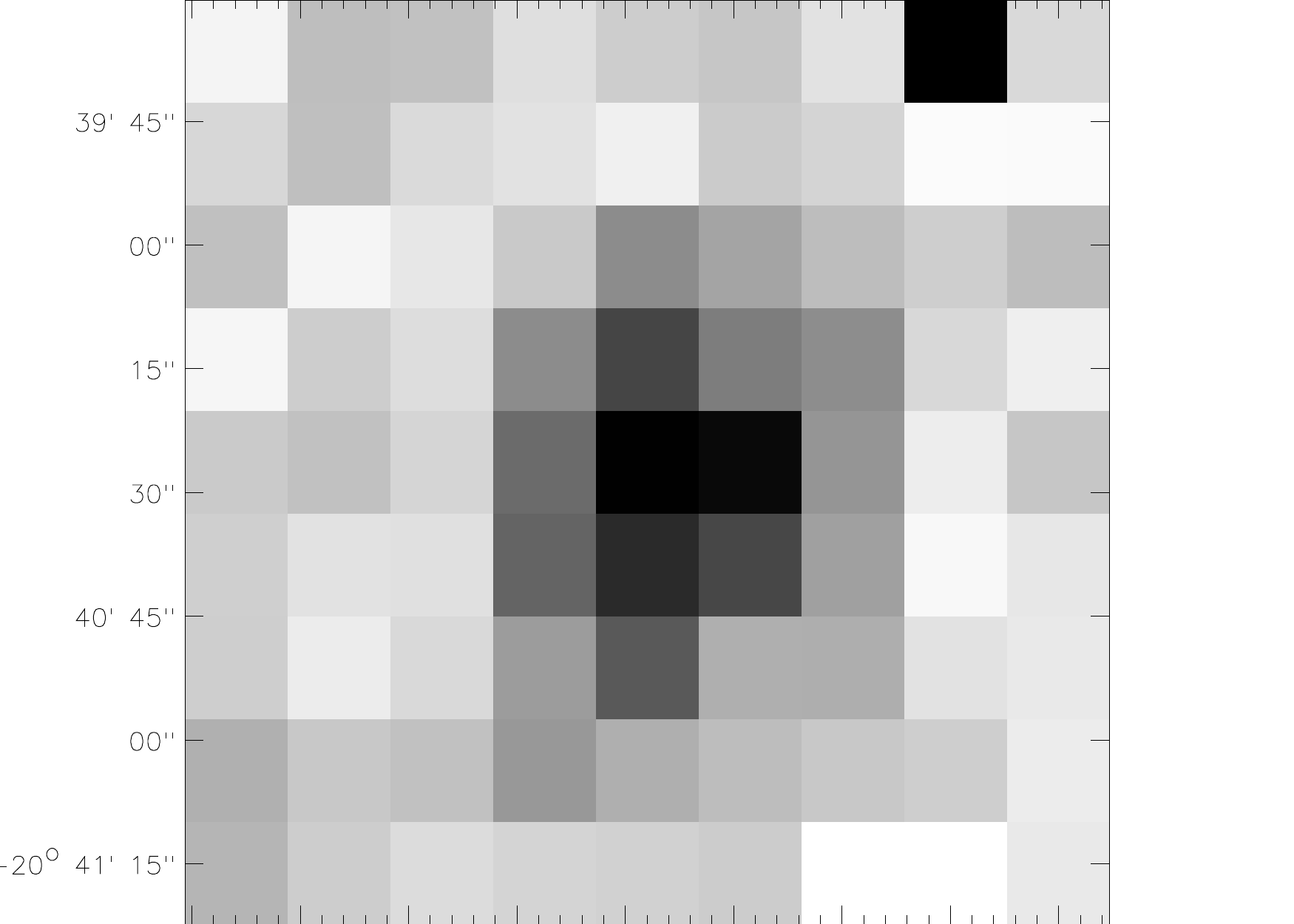} \\

      \multicolumn{6}{c}{\includegraphics[width=80mm,trim=30 0 20 0 ,clip=true]{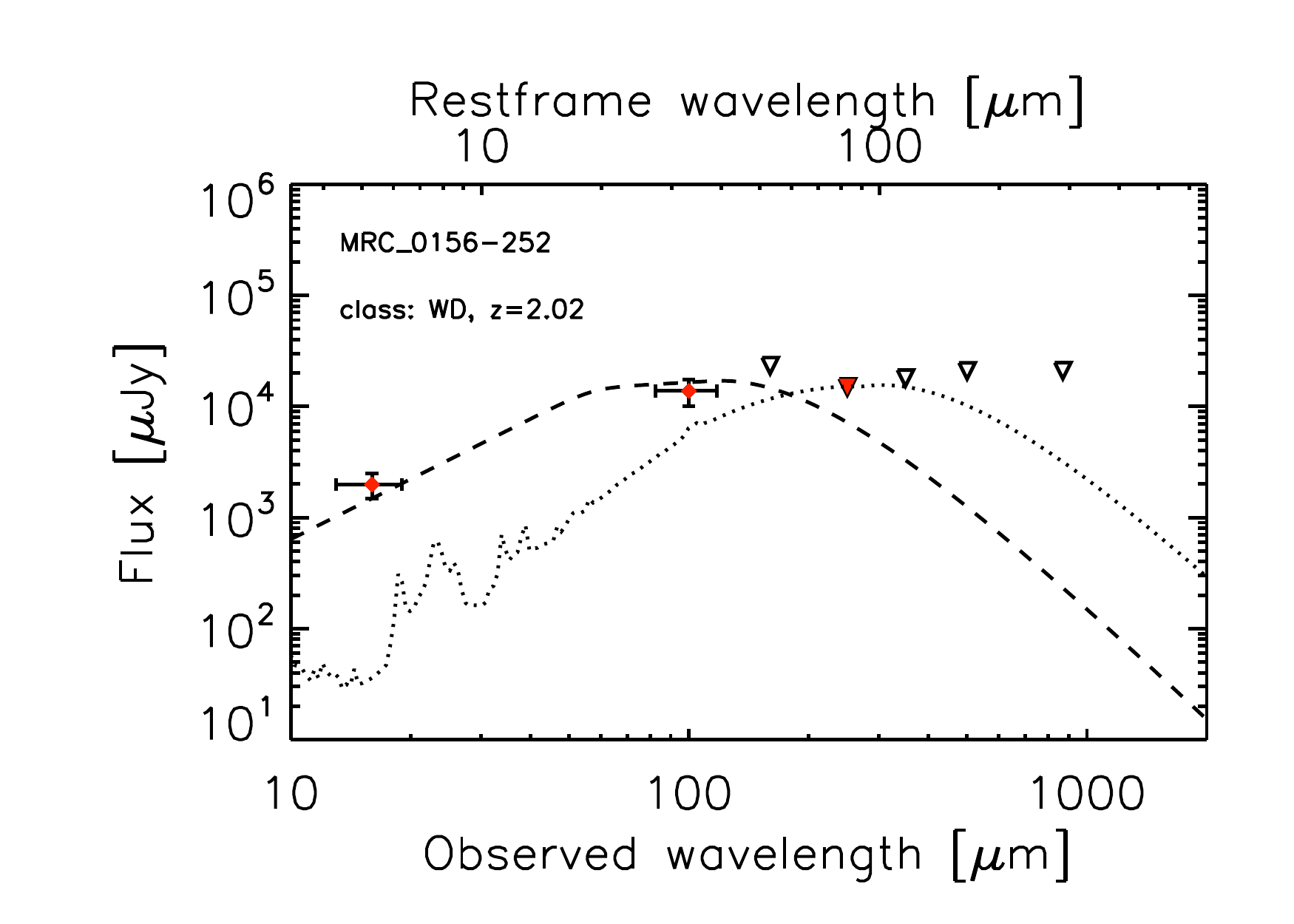}} \\
       & 100\mum & 160\mum & 250\mum & 350\mum & 500\mum \\
 &     \includegraphics[height=14mm,trim=70  -4 65 -10,clip=true]{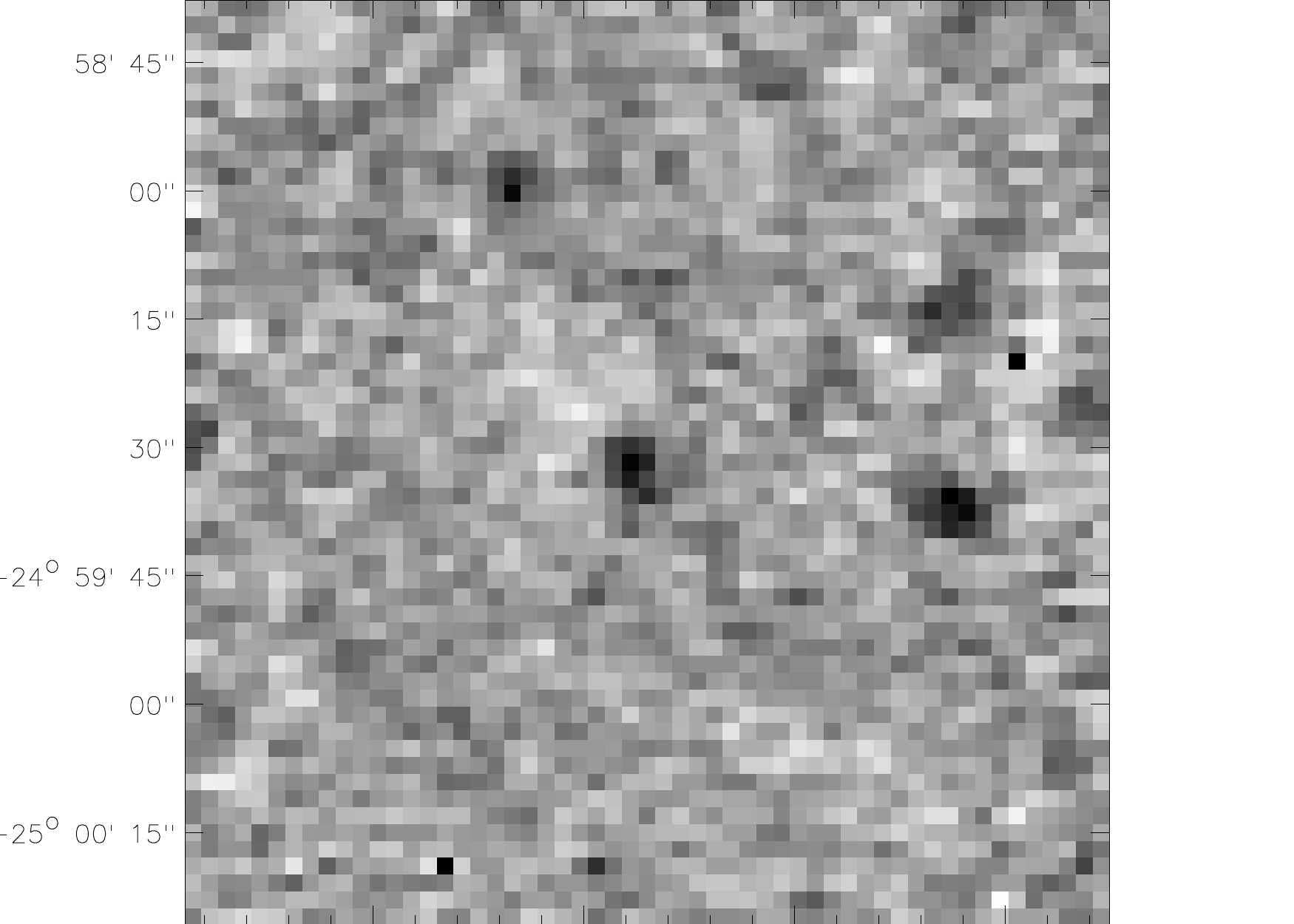} &
      \includegraphics[height=14mm,trim=70  -4 65 -10,clip=true]{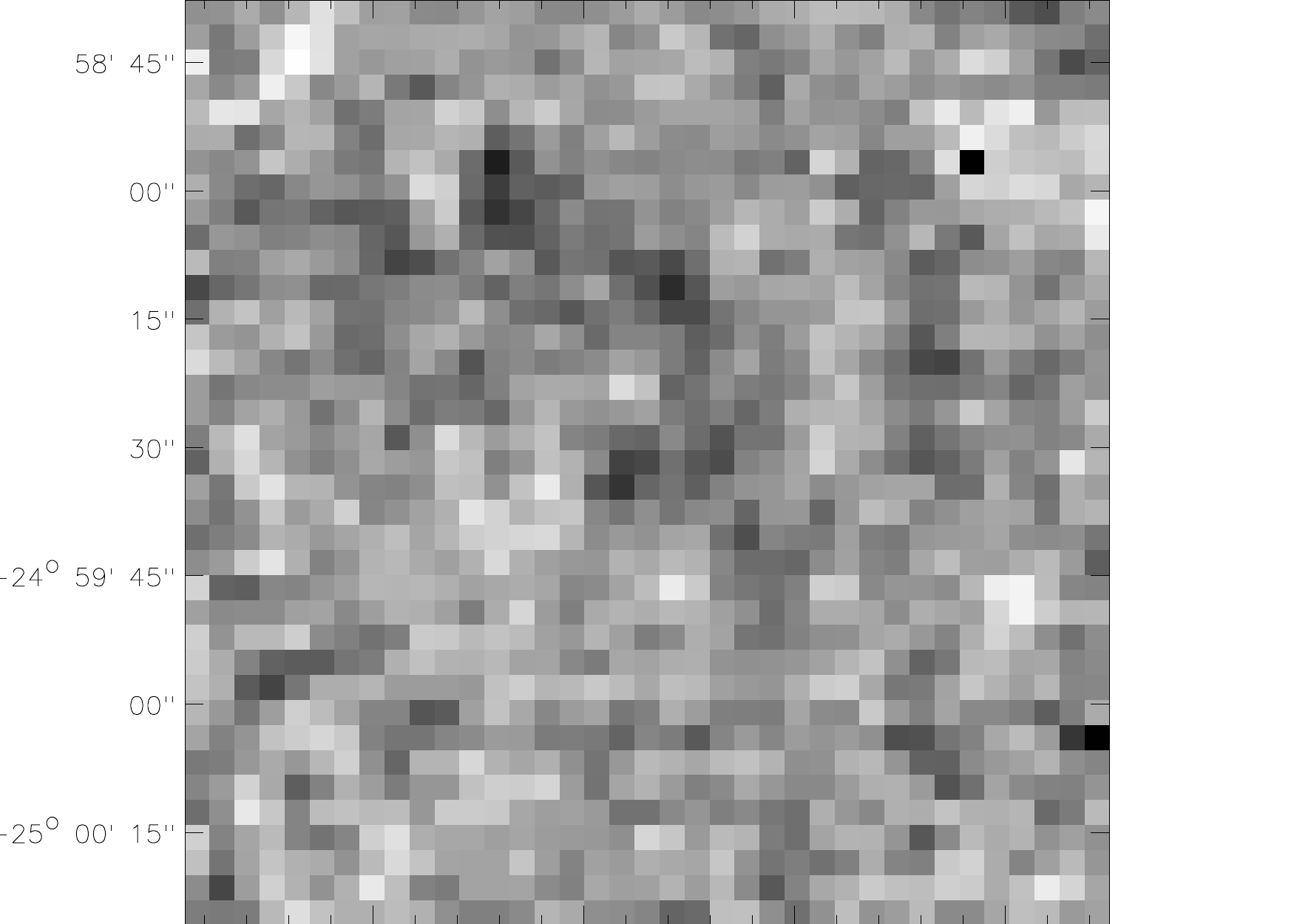} &
      \includegraphics[height=14mm,trim=70  -4 65 -10,clip=true]{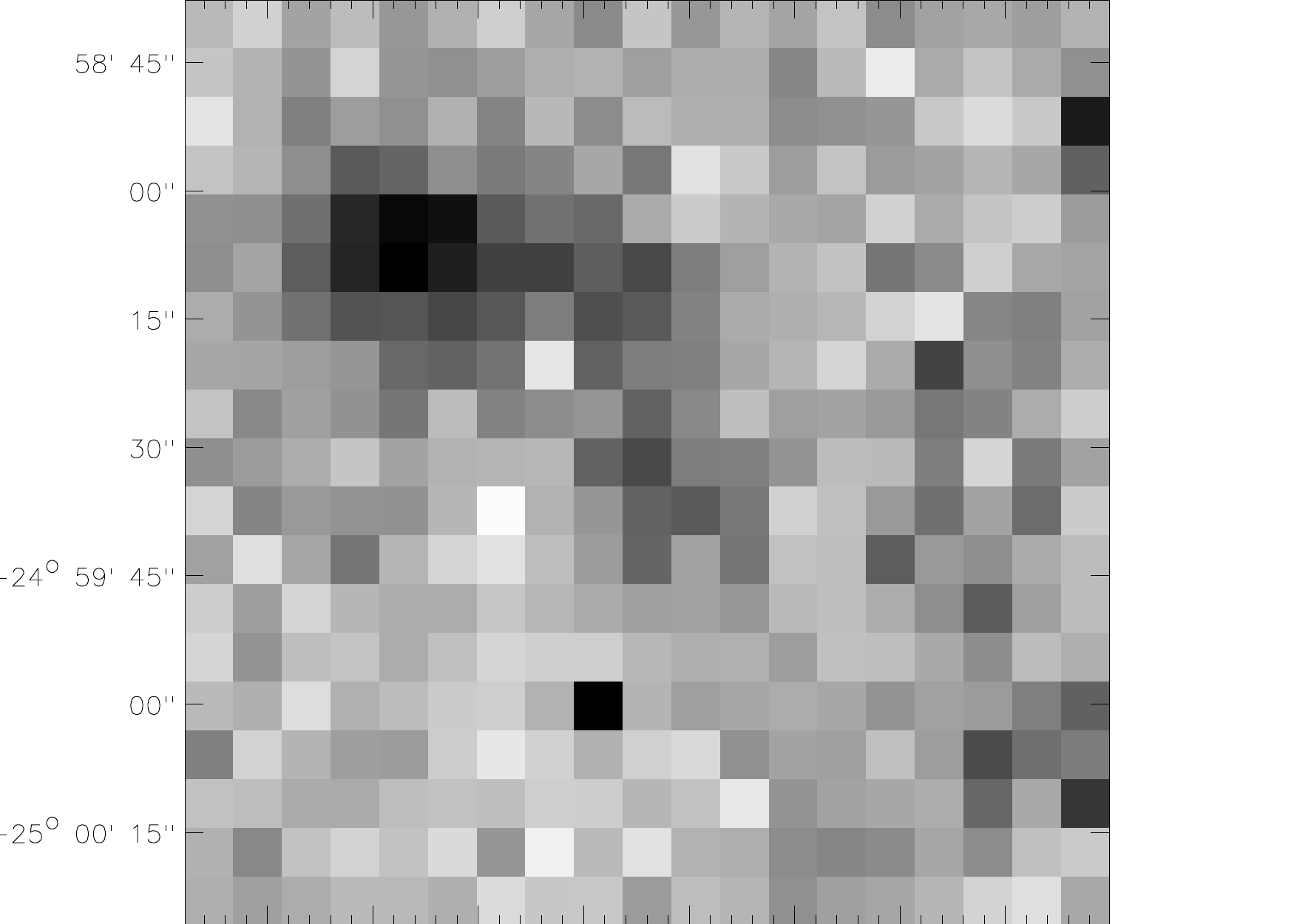}   &
      \includegraphics[height=14mm,trim=70  -4 65 -10,clip=true]{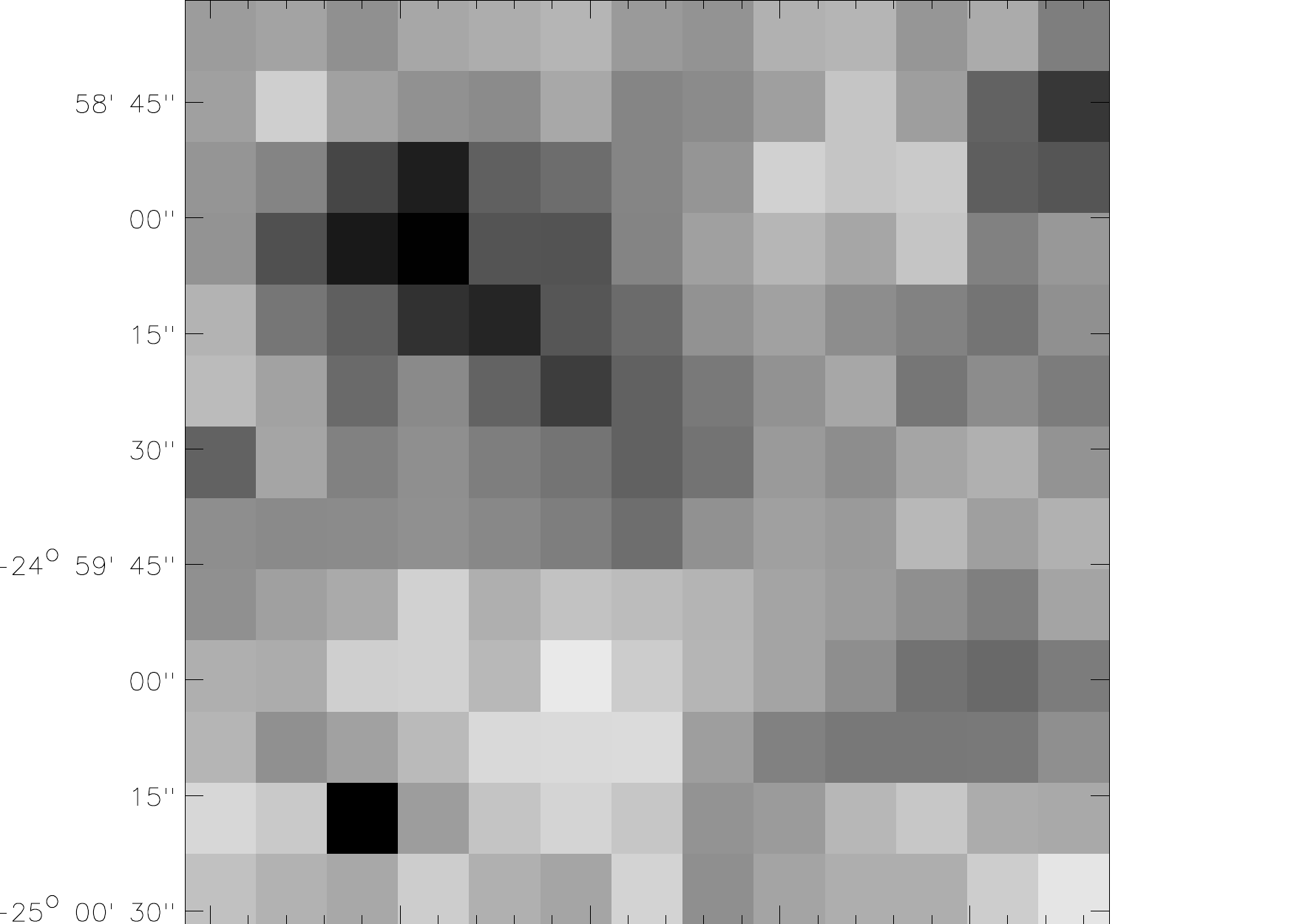}   &
      \includegraphics[height=14mm,trim=70  -4 65 -10,clip=true]{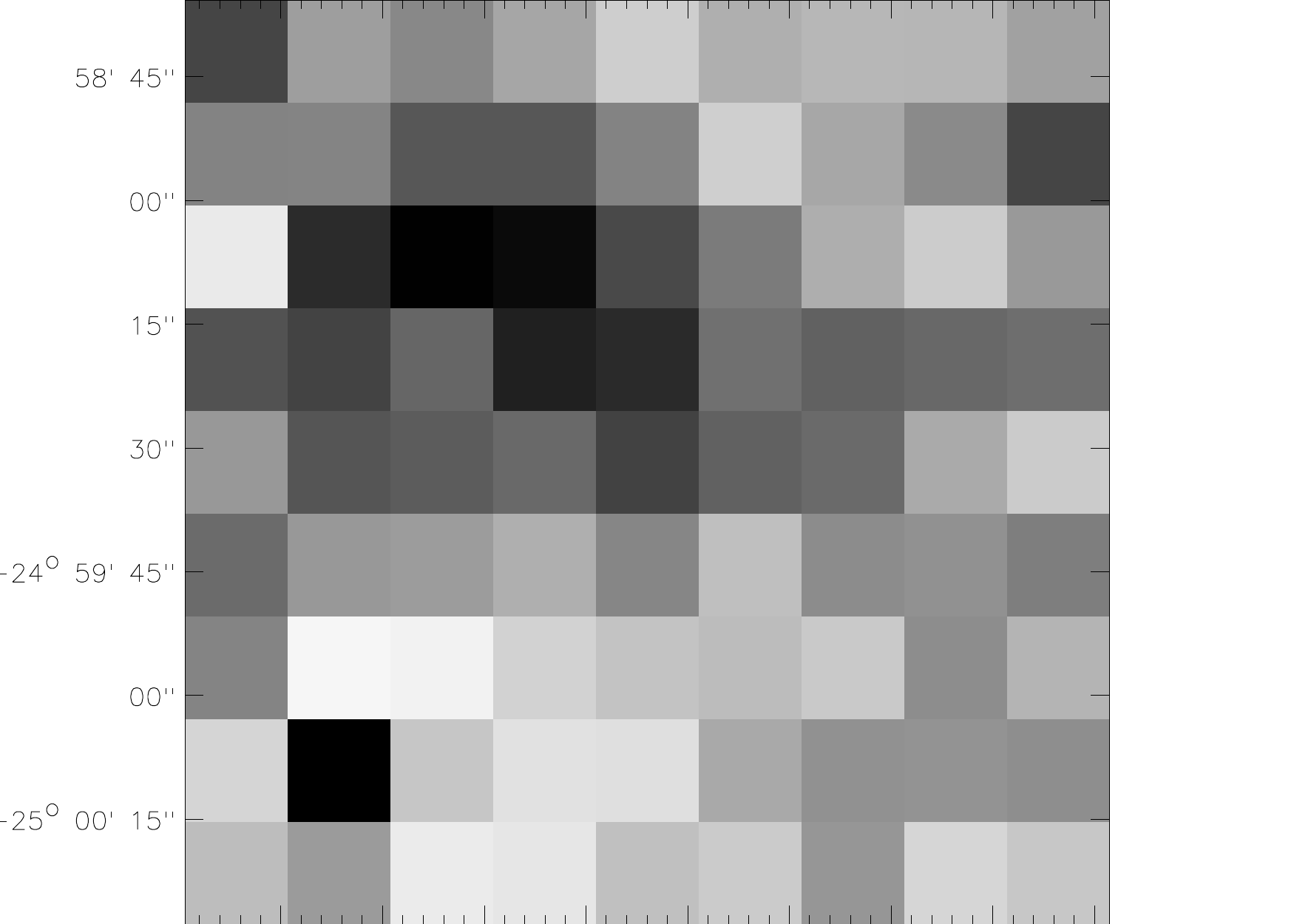} \\

      \multicolumn{6}{c}{\includegraphics[width=80mm,trim=30 0 20 0 ,clip=true]{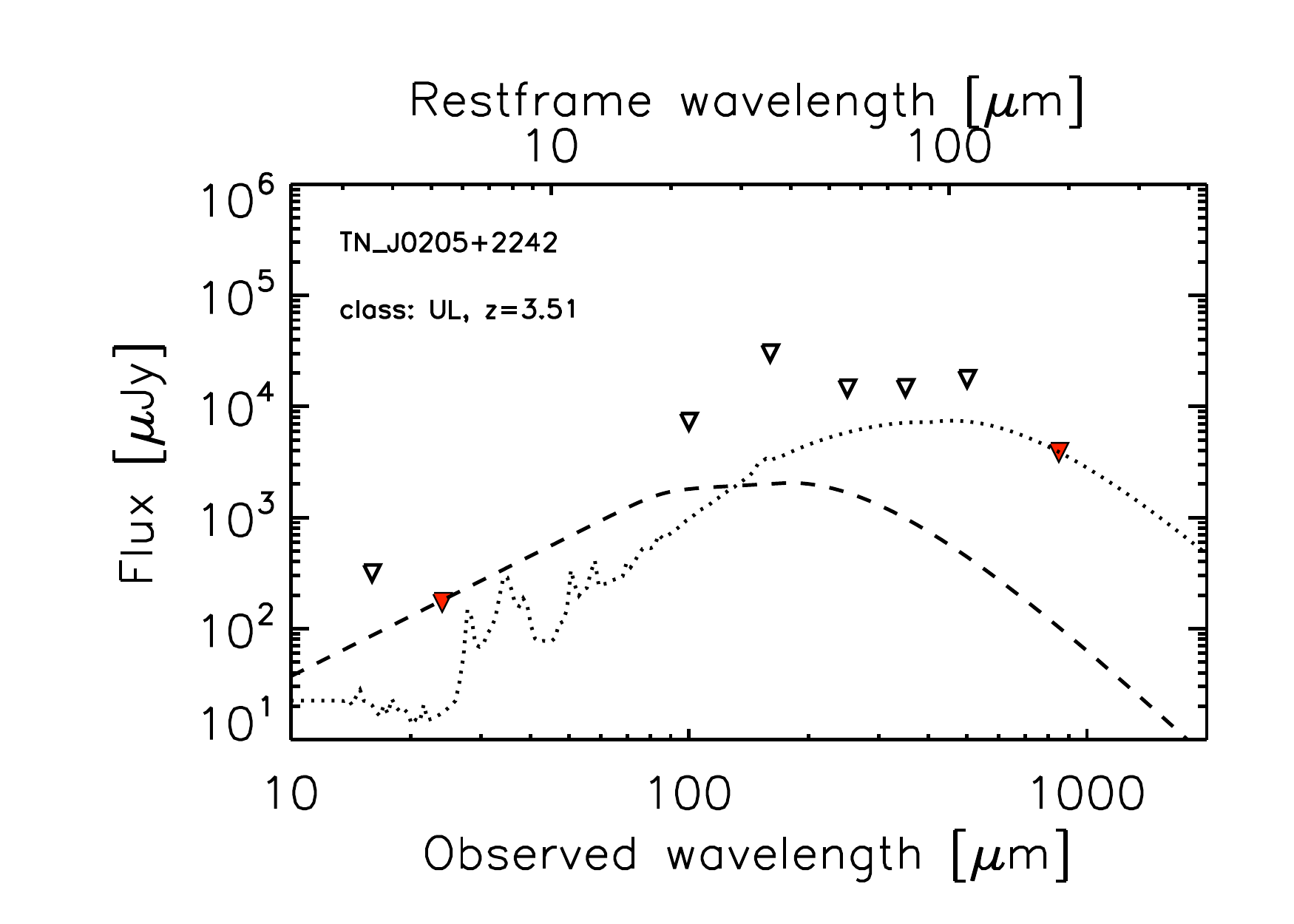}} \\
      24\mum & 100\mum & 160\mum & 250\mum & 350\mum & 500\mum \\
\end{tabular}
          \end{minipage}
  \end{centering}
\end{figure*}

\clearpage

\begin{figure*}[h]
  \begin{centering}
    \begin{minipage}[c]{0.46\linewidth}
    \begin{tabular}{c@{}c@{}c@{}c@{}c@{}c@{}}

      \multicolumn{6}{c}{\includegraphics[width=80mm,trim=30 0 20 0 ,clip=true]{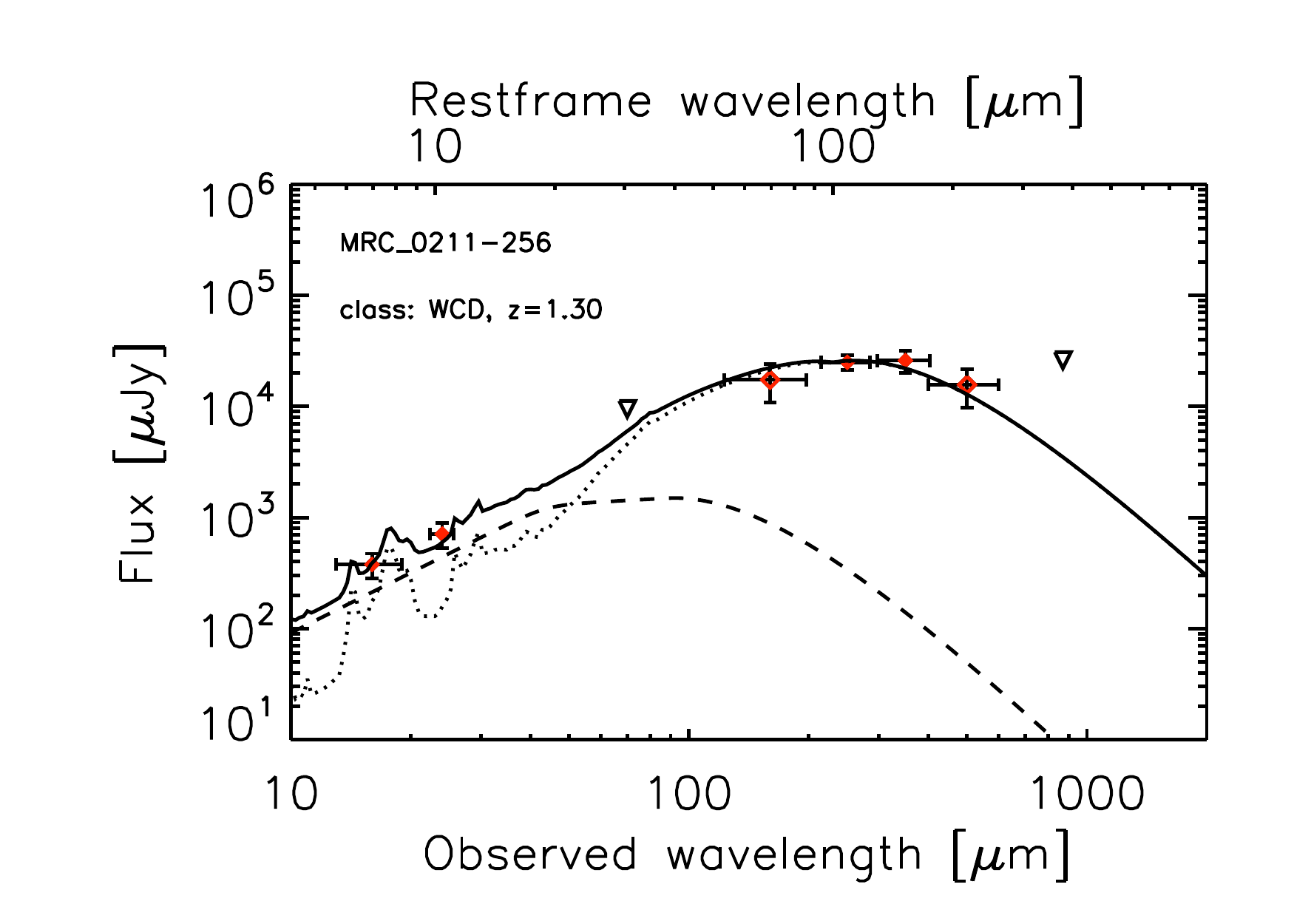}} \\
      24\mum & 70\mum & 160\mum & 250\mum & 350\mum & 500\mum \\
      \includegraphics[height=14mm,trim=70  -4 65 -10,clip=true]{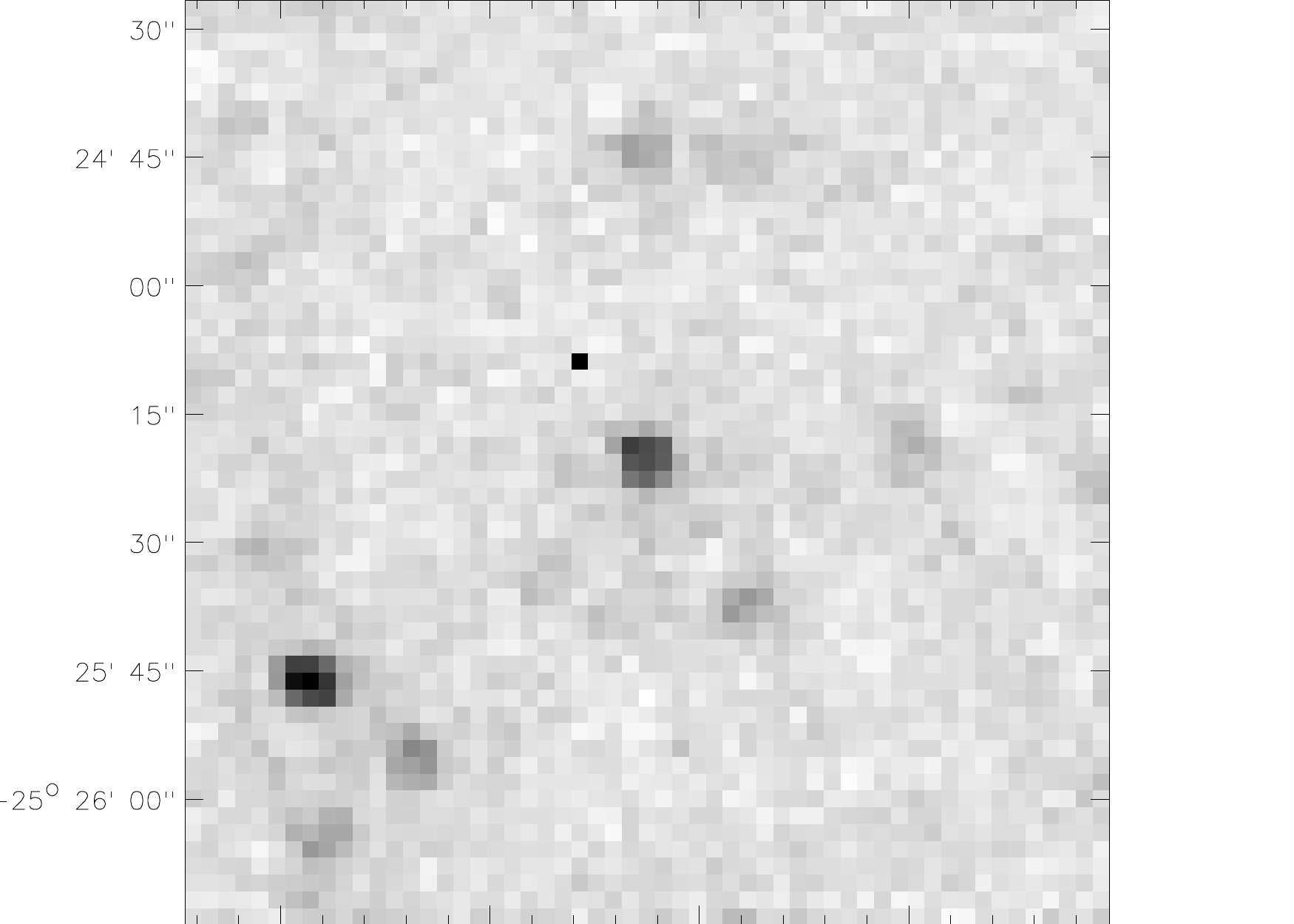} &
      \includegraphics[height=14mm,trim=70  -4 65 -10,clip=true]{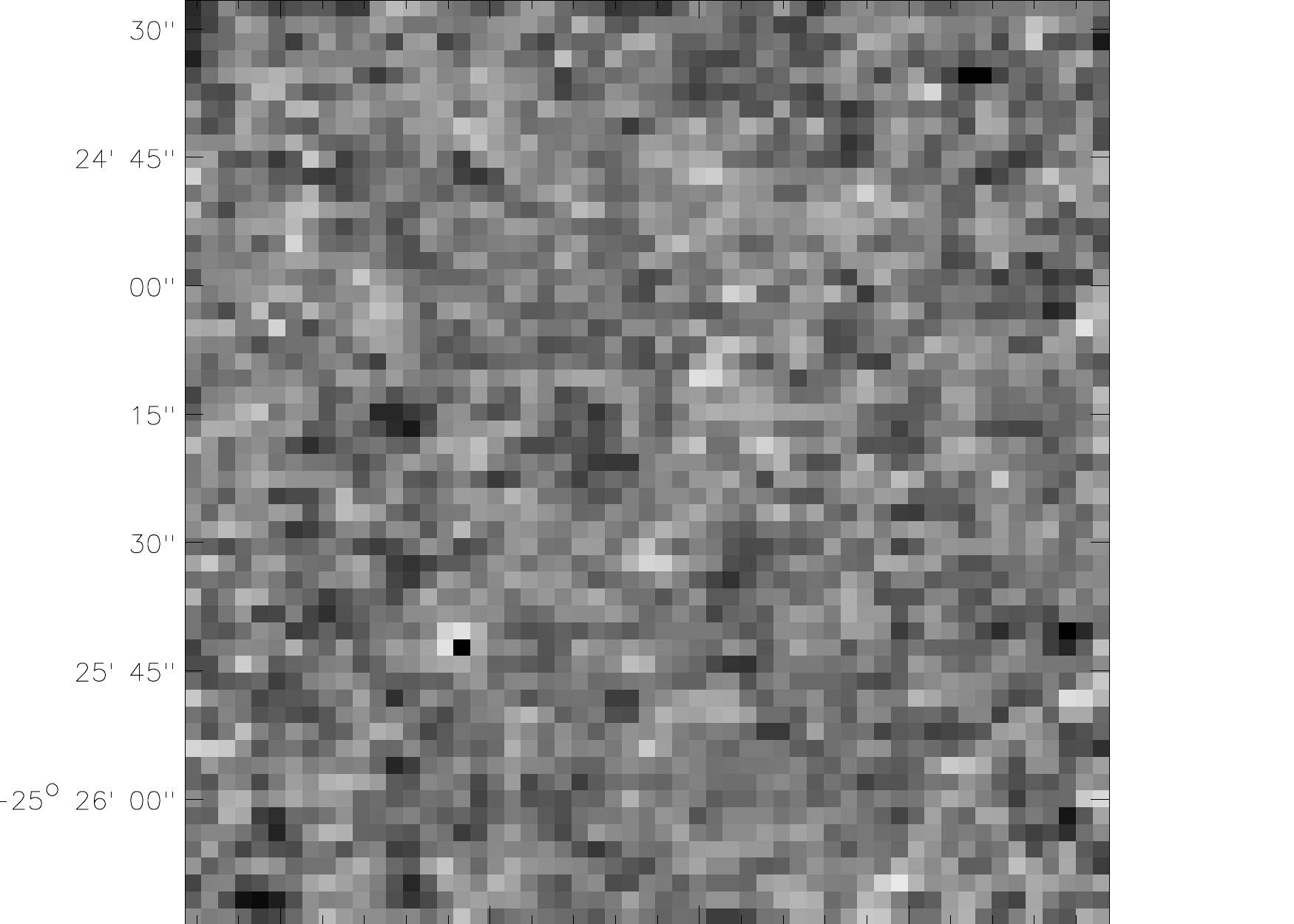} &
      \includegraphics[height=14mm,trim=70  -4 65 -10,clip=true]{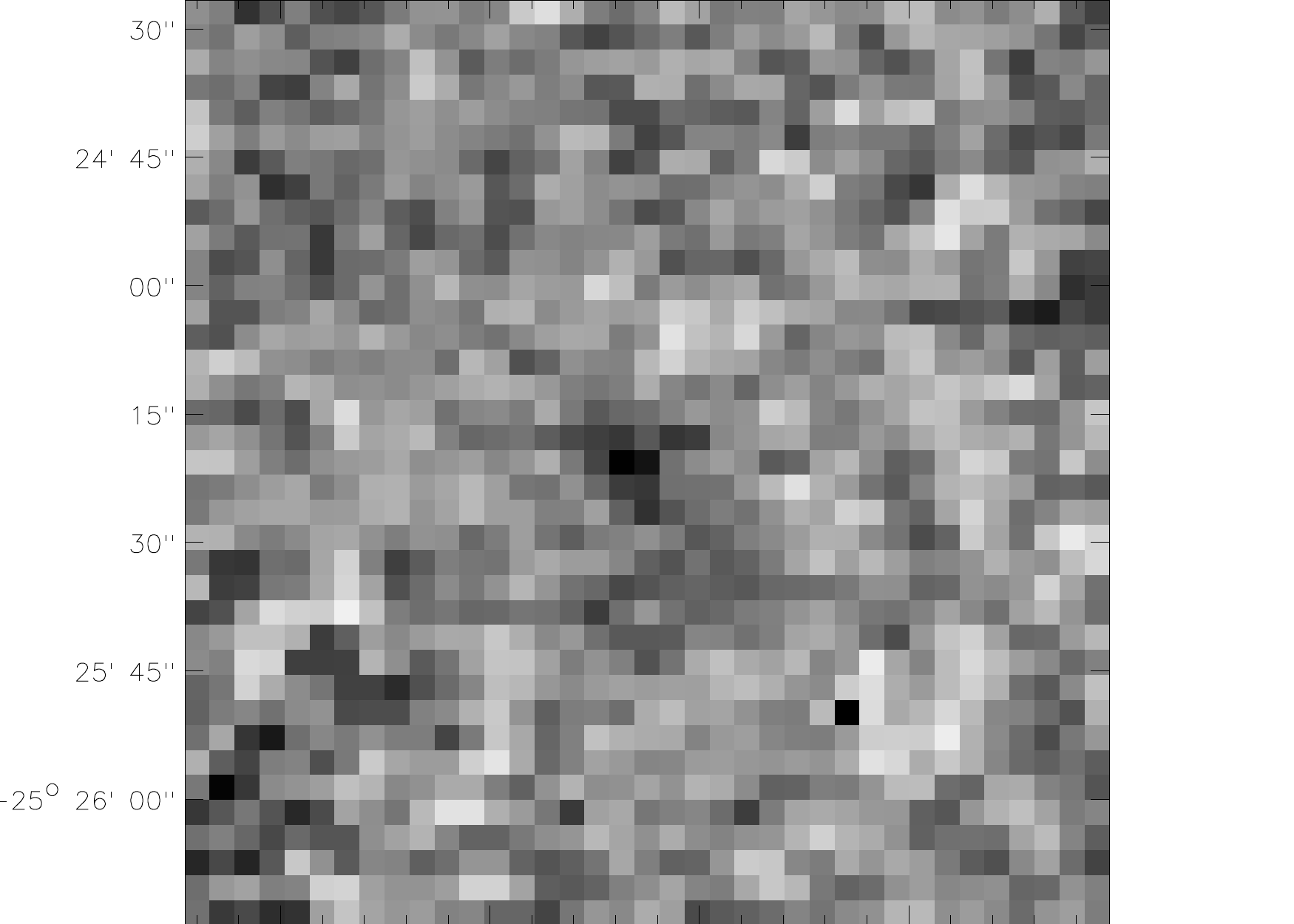} &
      \includegraphics[height=14mm,trim=70  -4 65 -10,clip=true]{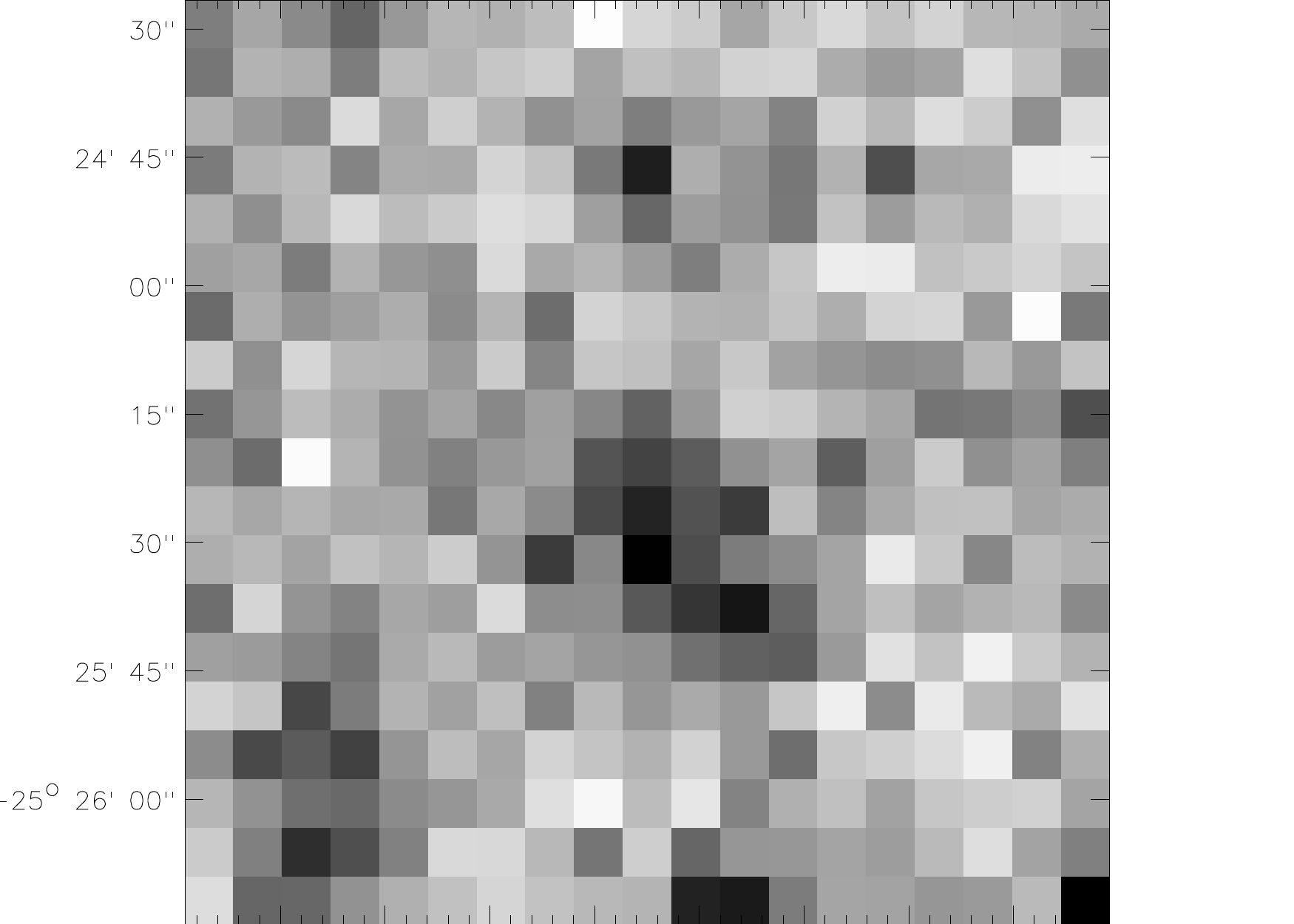}   &
      \includegraphics[height=14mm,trim=70  -4 65 -10,clip=true]{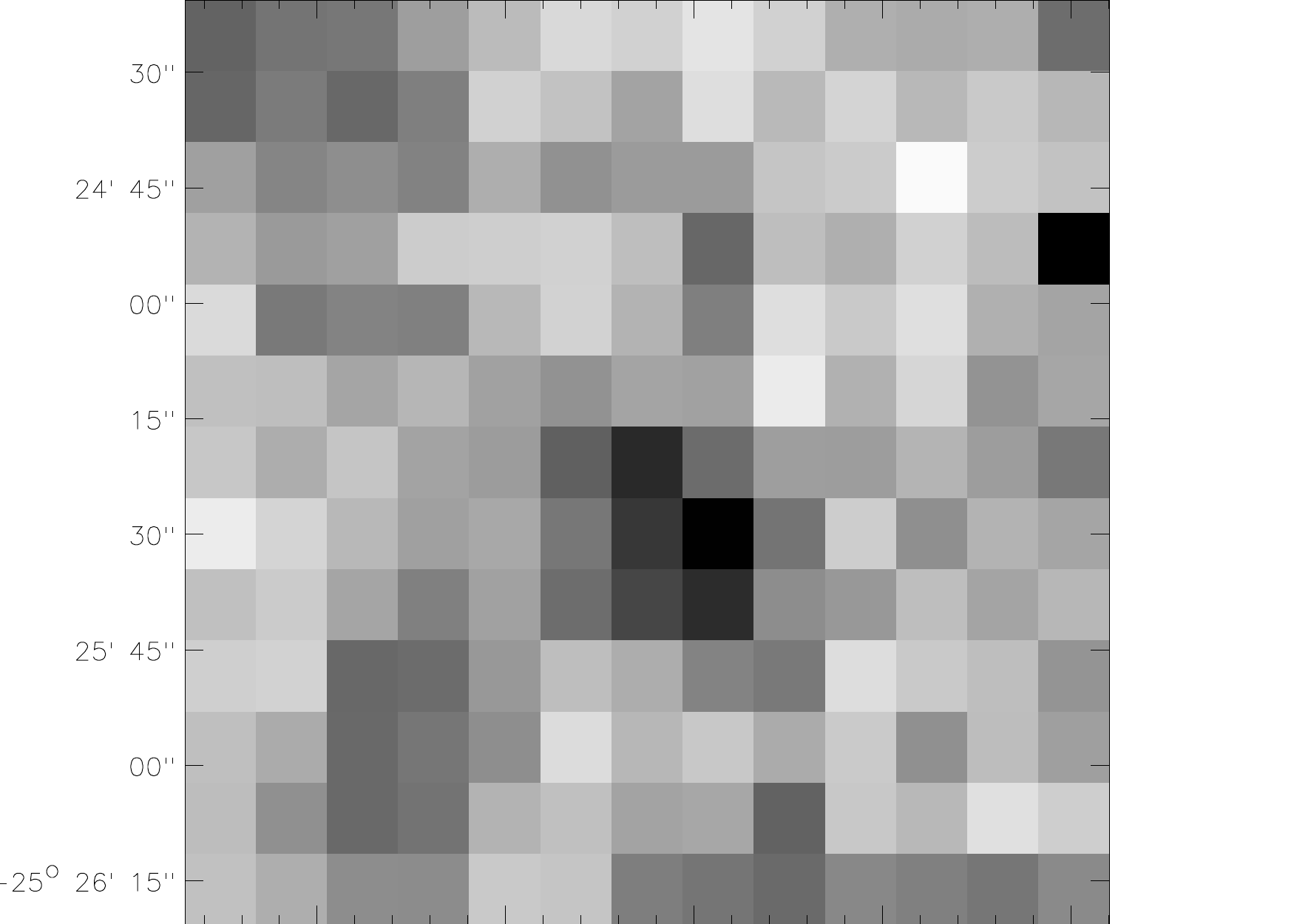}   &
      \includegraphics[height=14mm,trim=70  -4 65 -10,clip=true]{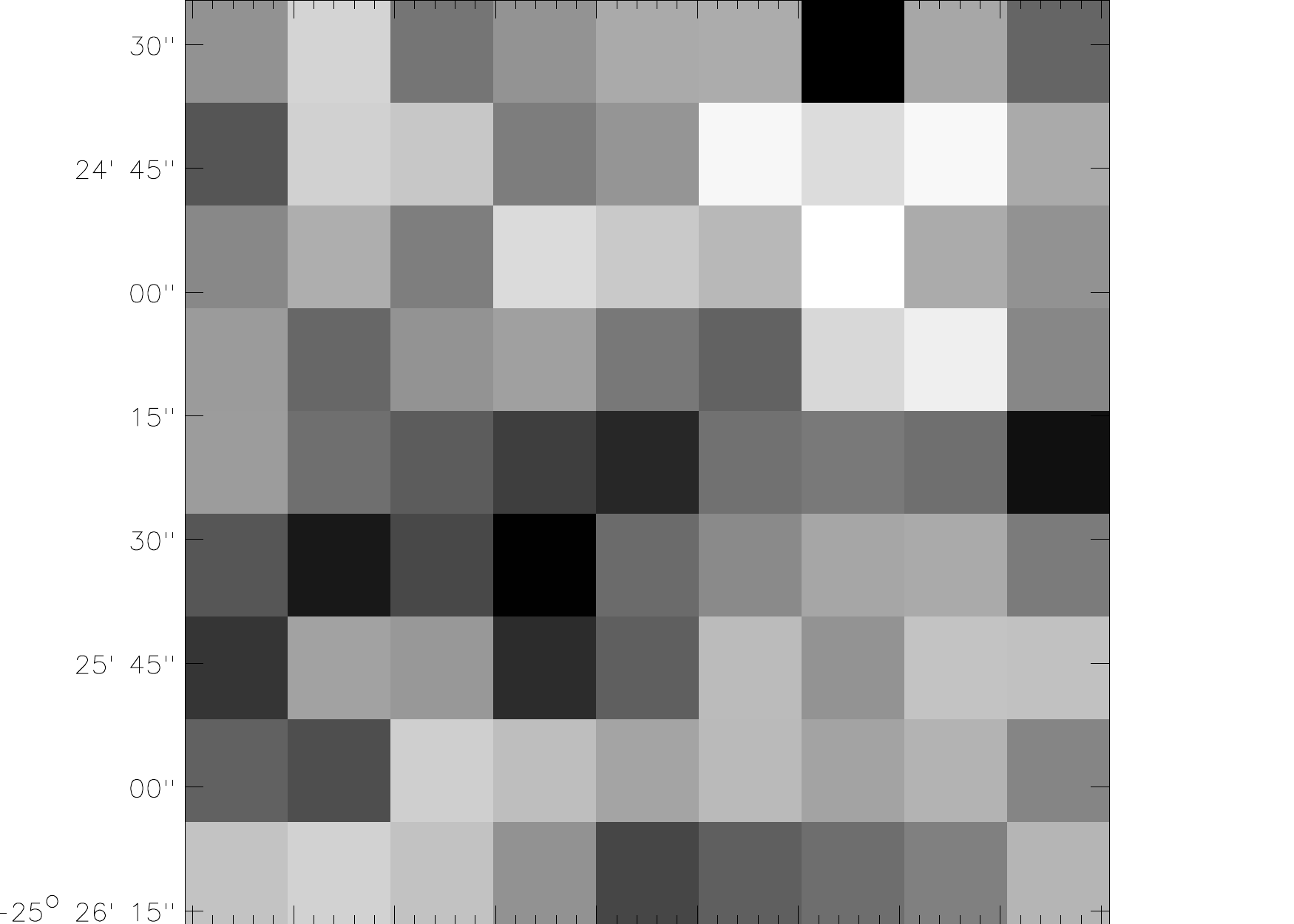} \\

      \multicolumn{6}{c}{\includegraphics[width=80mm,trim=30 0 20 0 ,clip=true]{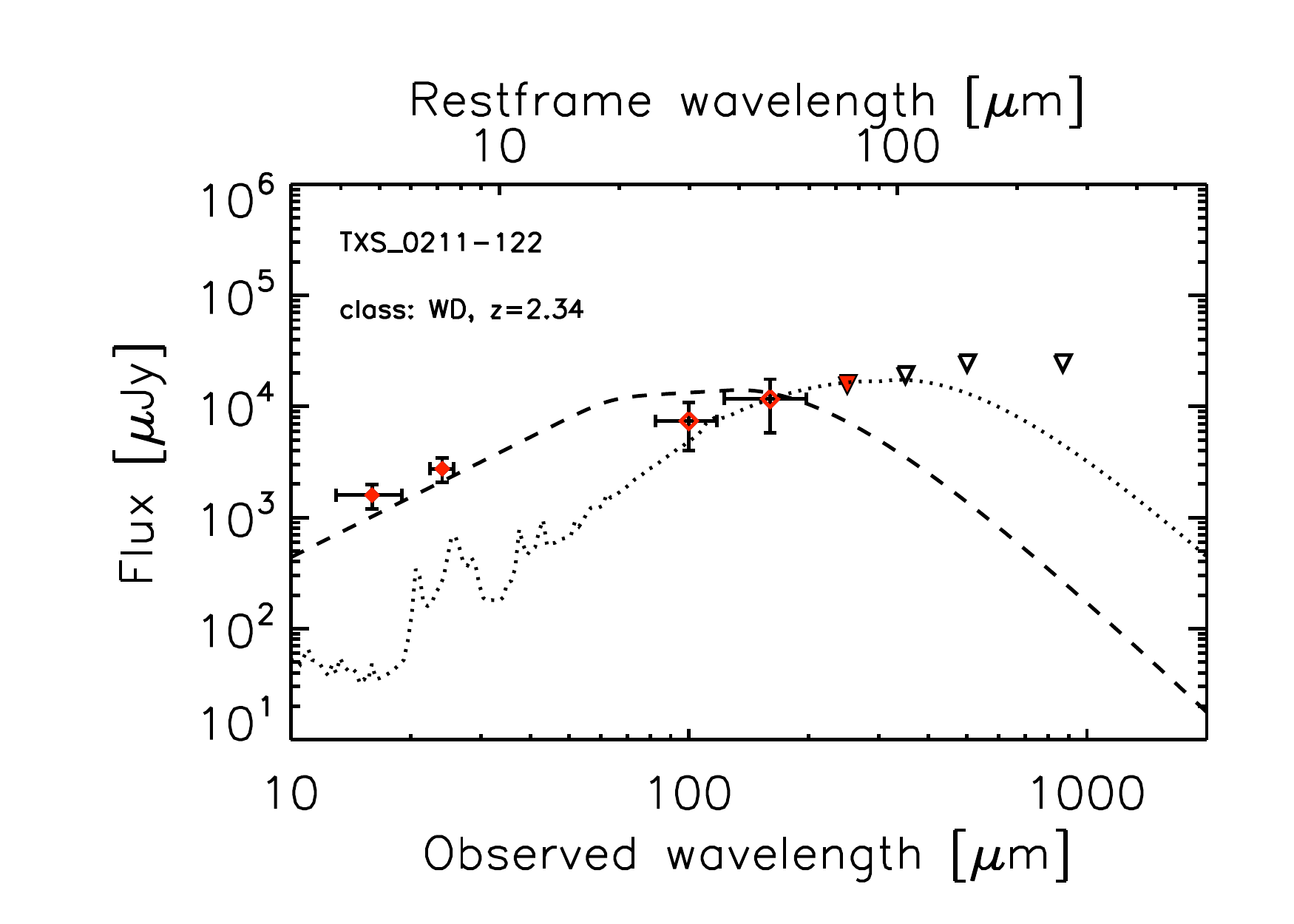}} \\
      24\mum & 100\mum & 160\mum & 250\mum & 350\mum & 500\mum \\
      \includegraphics[height=14mm,trim=70  -4 65 -10,clip=true]{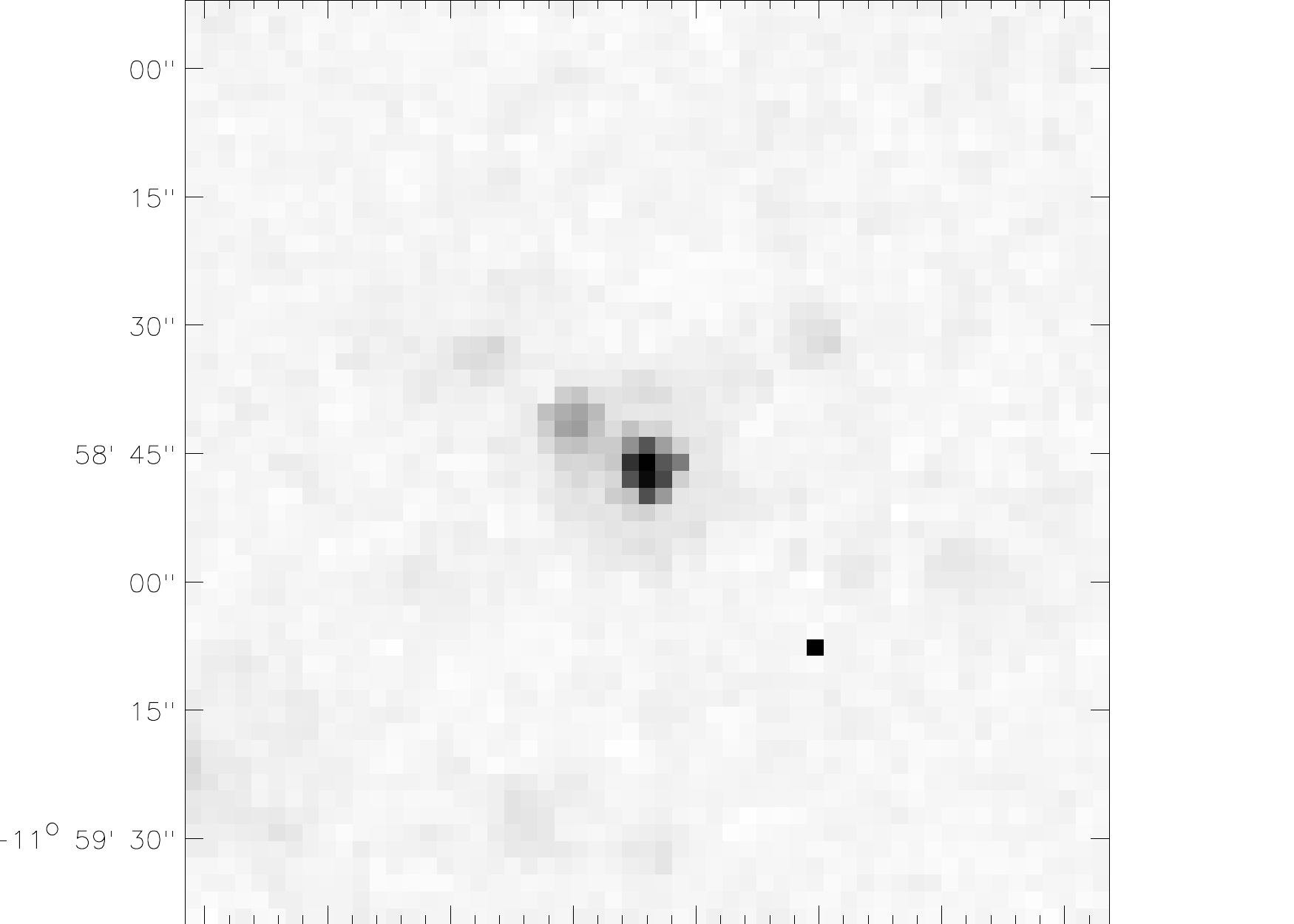} &
      \includegraphics[height=14mm,trim=70  -4 65 -10,clip=true]{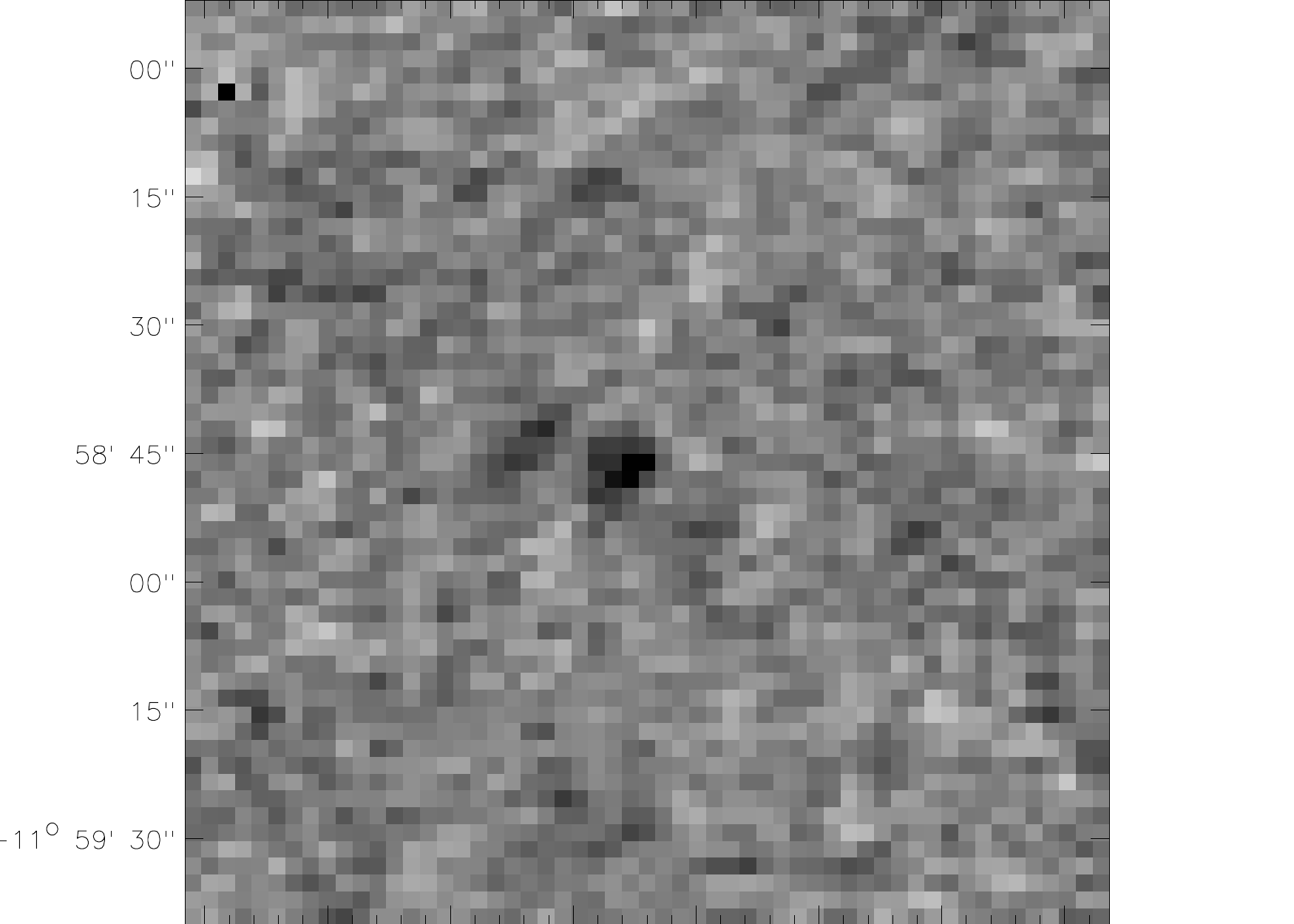} &
      \includegraphics[height=14mm,trim=70  -4 65 -10,clip=true]{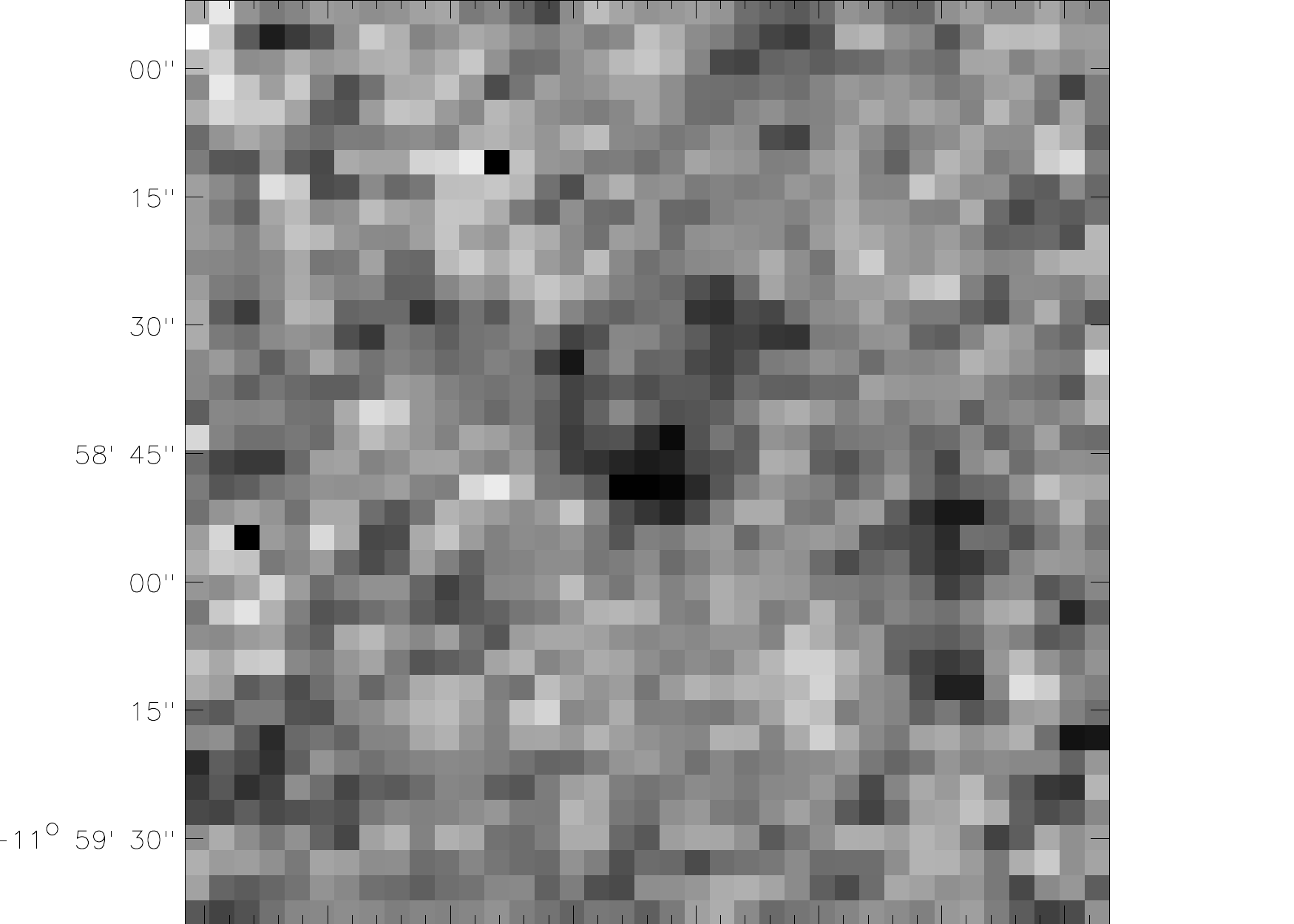} &
      \includegraphics[height=14mm,trim=70  -4 65 -10,clip=true]{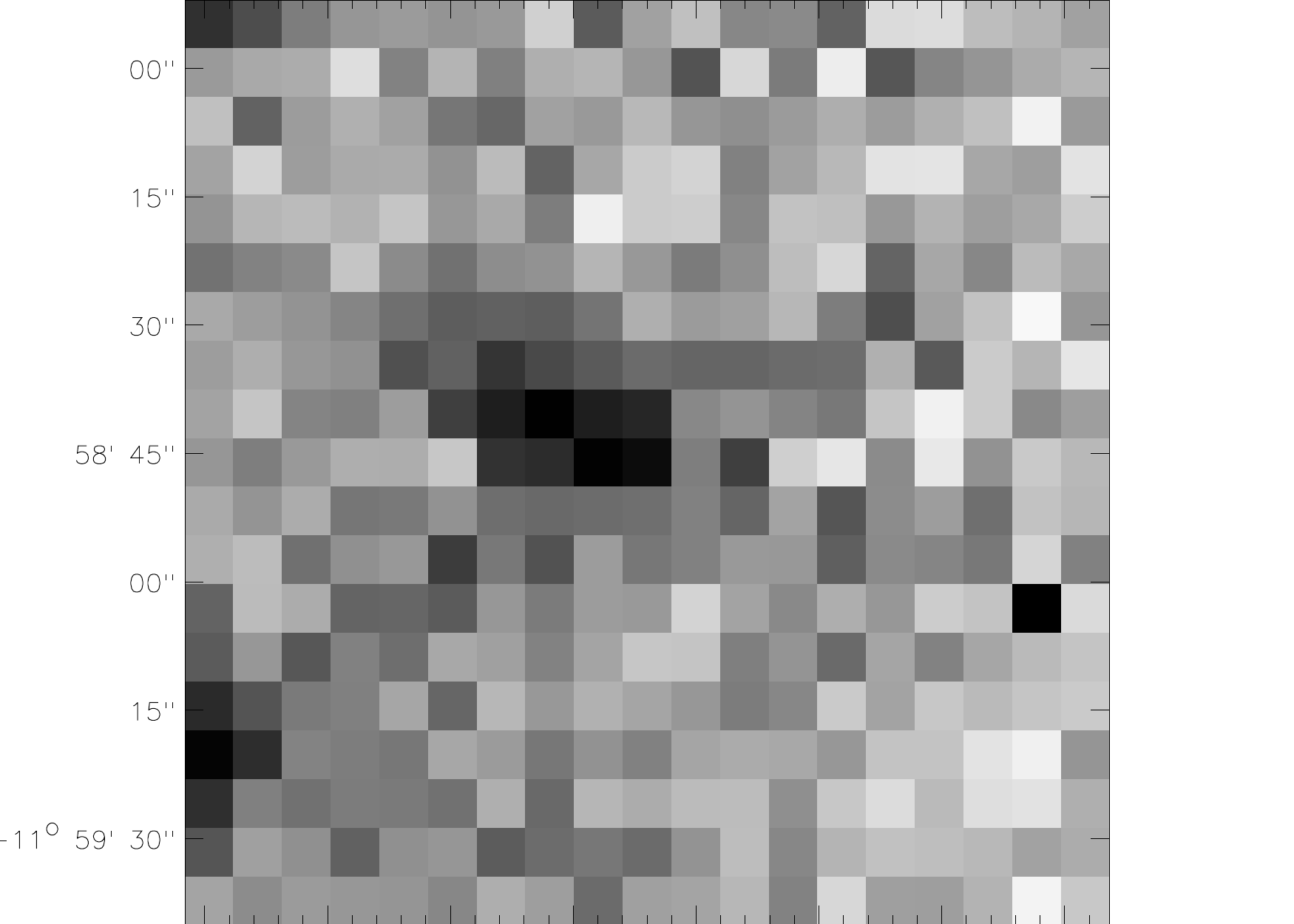}   &
      \includegraphics[height=14mm,trim=70  -4 65 -10,clip=true]{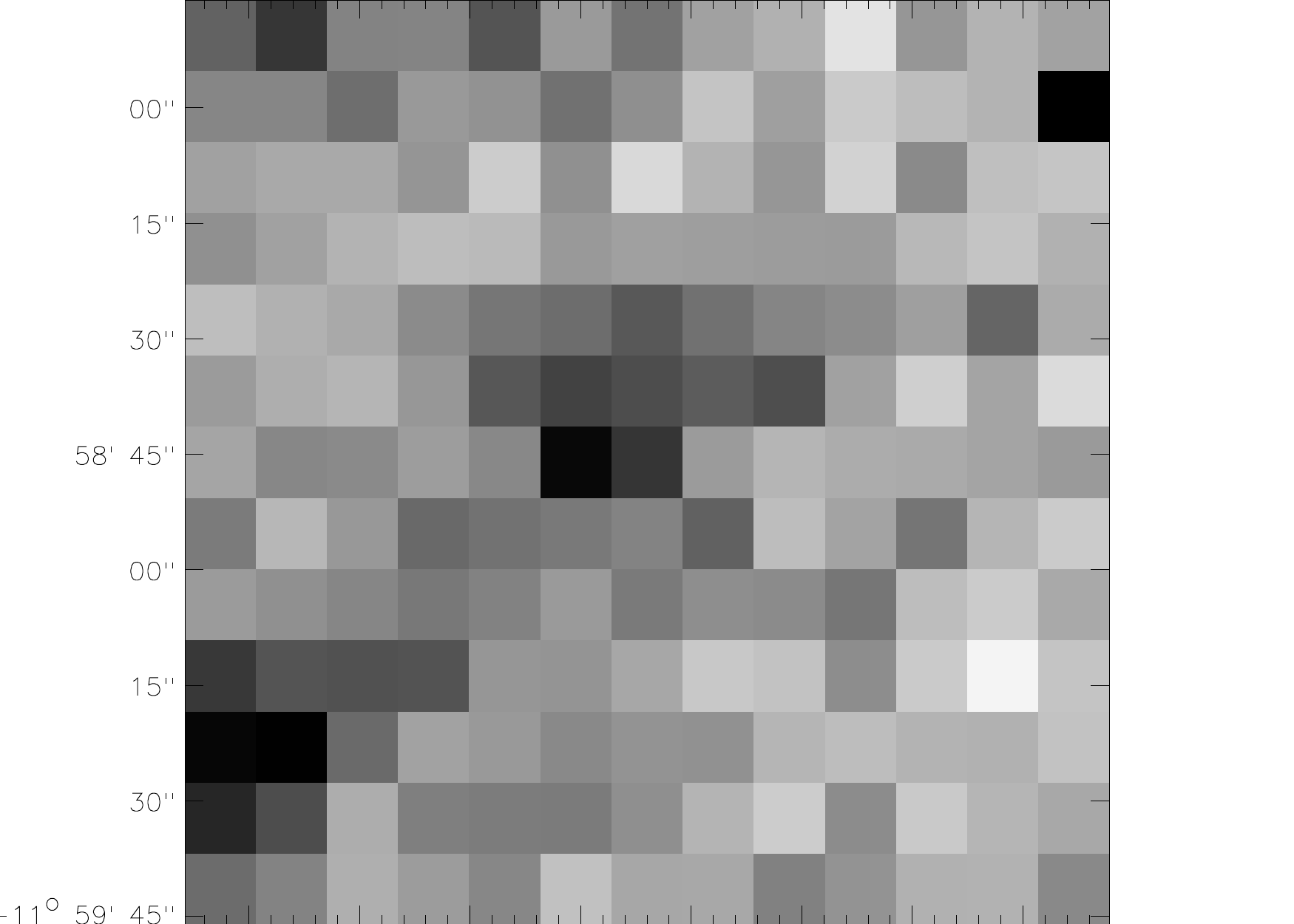}   &
      \includegraphics[height=14mm,trim=70  -4 65 -10,clip=true]{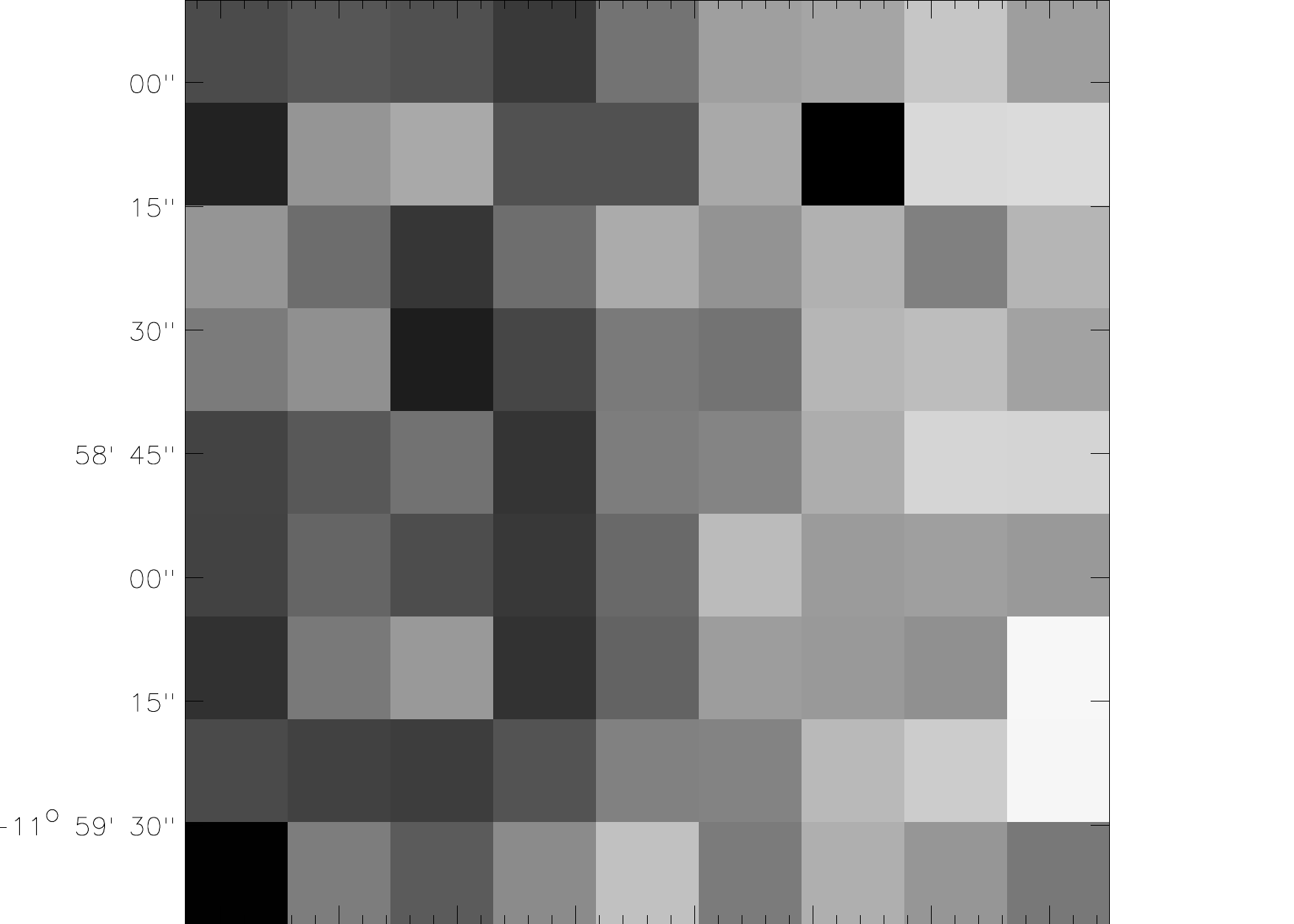} \\

      \multicolumn{6}{c}{\includegraphics[width=80mm,trim=30 0 20 0 ,clip=true]{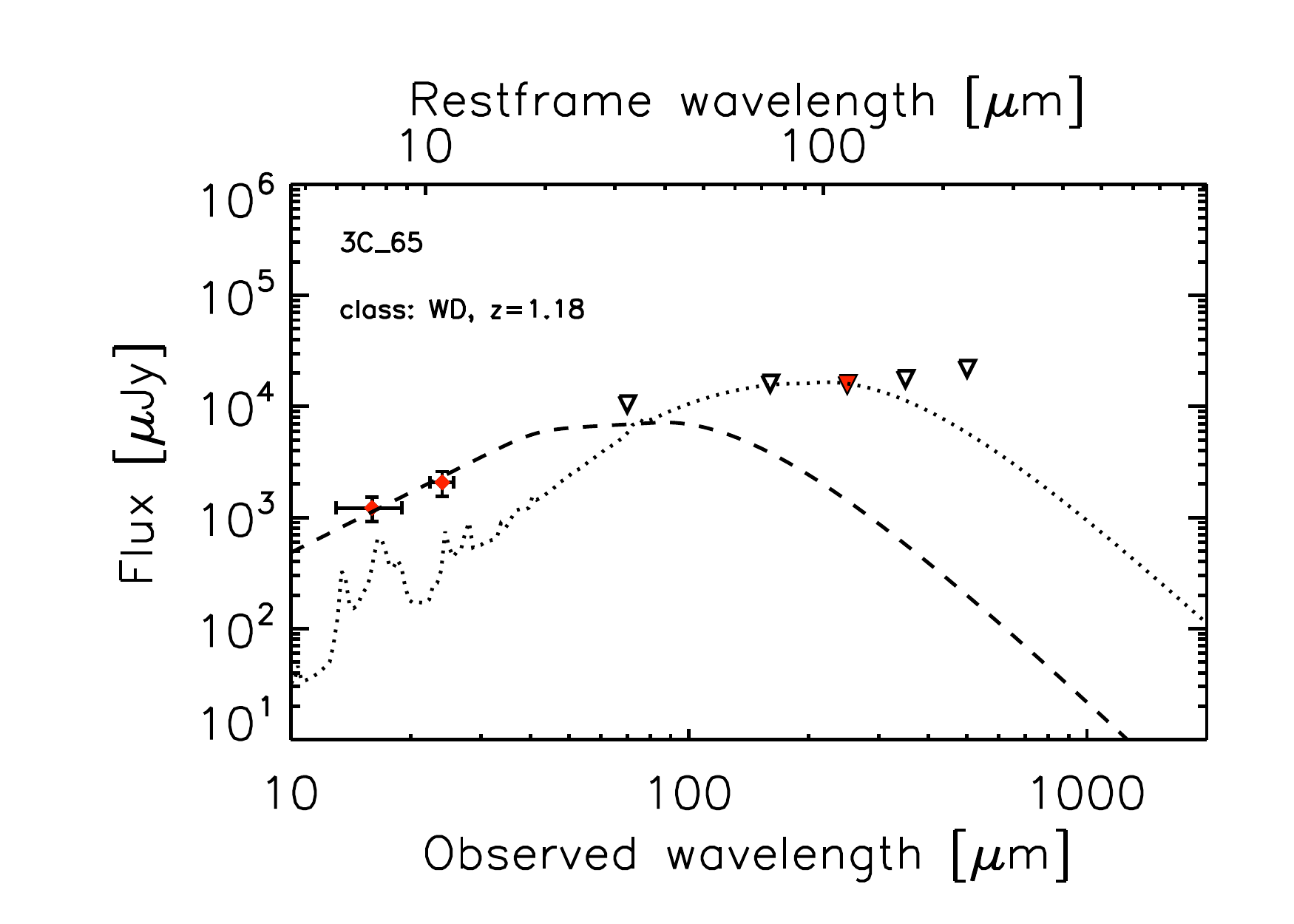}} \\
      24\mum & 70\mum & 160\mum & 250\mum & 350\mum & 500\mum \\
\end{tabular}
\end{minipage}
\begin{minipage}[c]{0.46\linewidth} 
    \begin{tabular}{c@{}c@{}c@{}c@{}c@{}c@{}}

      \multicolumn{6}{c}{\includegraphics[width=80mm,trim=30 0 20 0 ,clip=true]{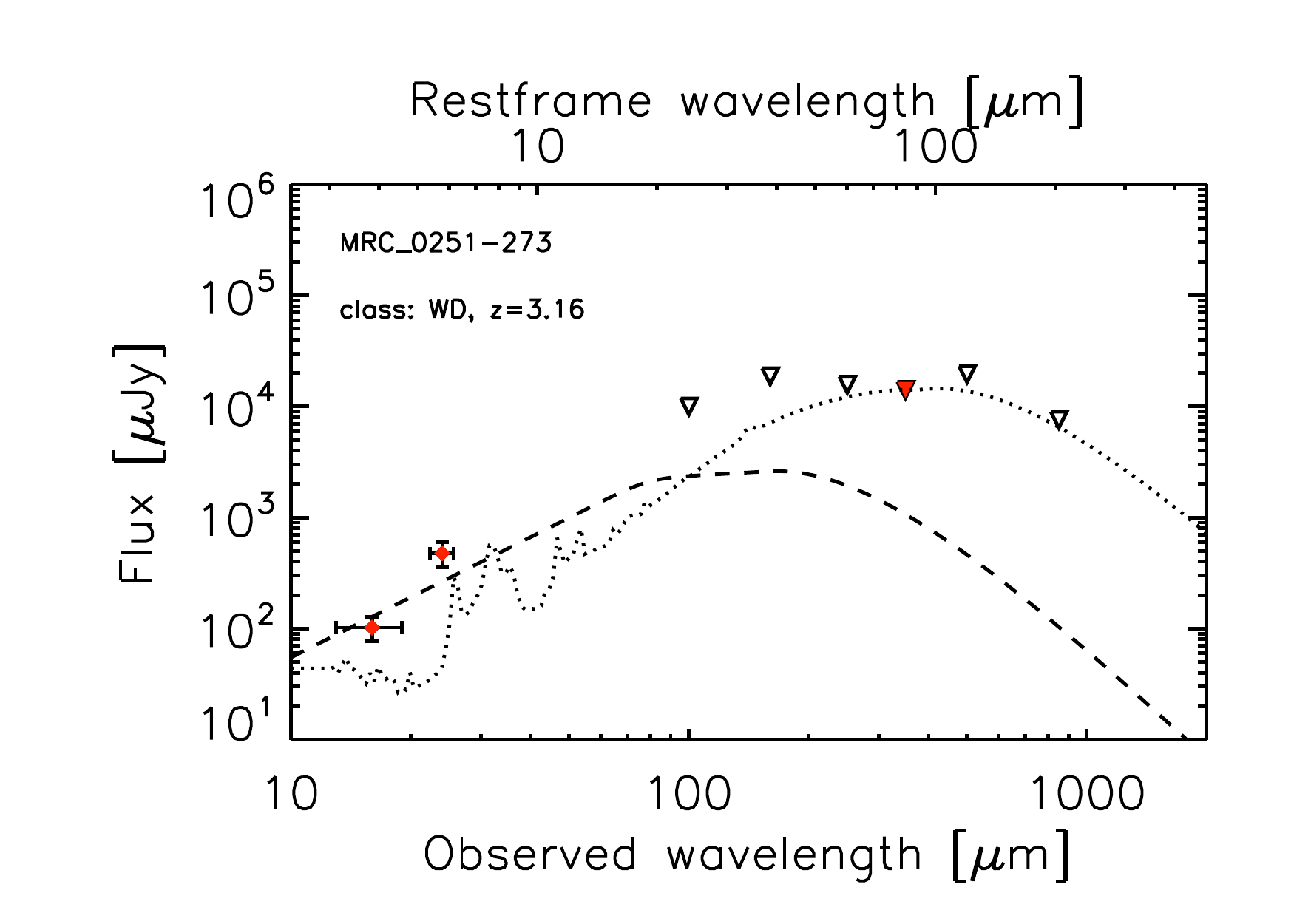}} \\
      24\mum & 100\mum & 160\mum & 250\mum & 350\mum & 500\mum \\
      \includegraphics[height=14mm,trim=70  -4 65 -10,clip=true]{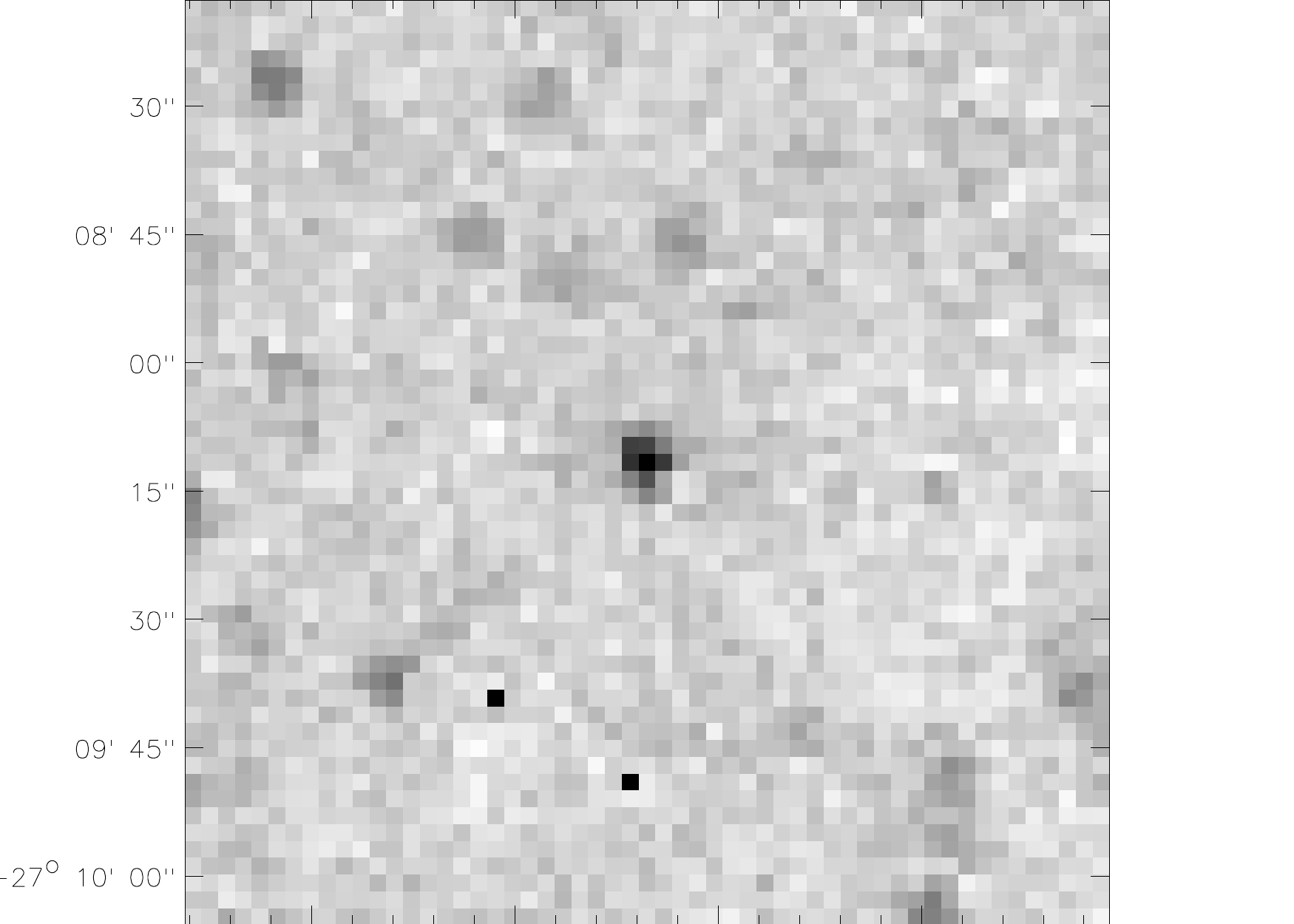} &
      \includegraphics[height=14mm,trim=70  -4 65 -10,clip=true]{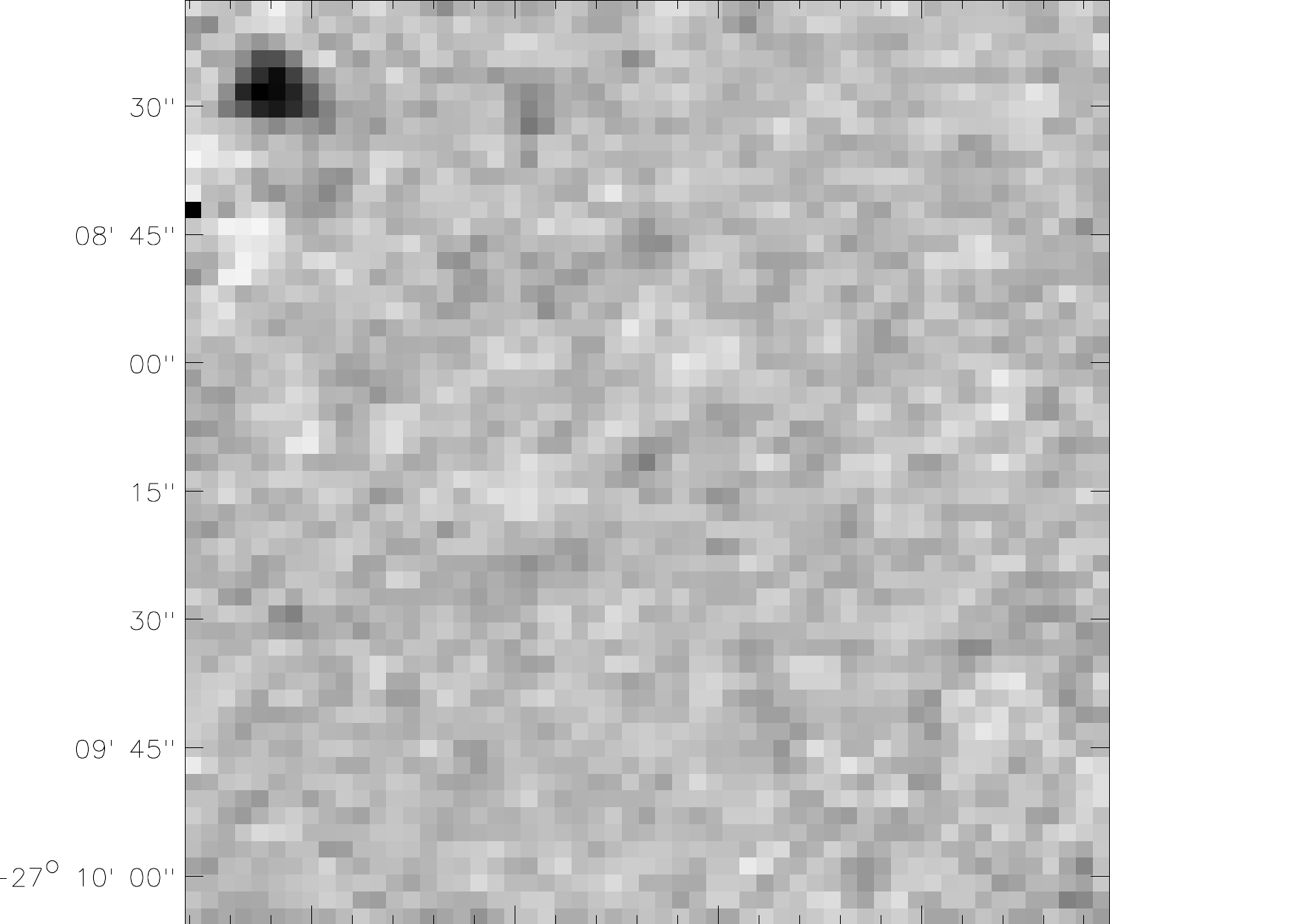} &
      \includegraphics[height=14mm,trim=70  -4 65 -10,clip=true]{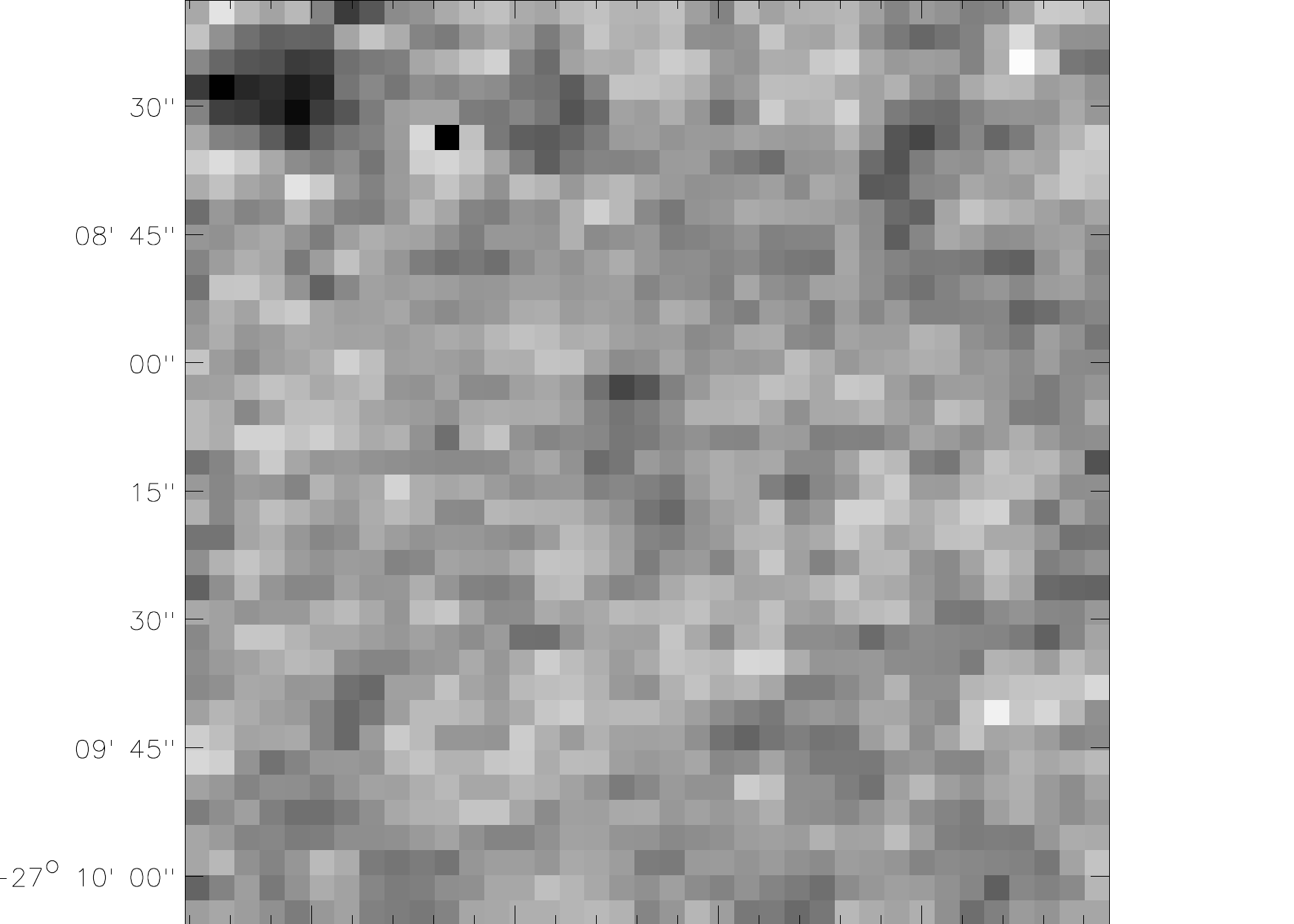} &
      \includegraphics[height=14mm,trim=70  -4 65 -10,clip=true]{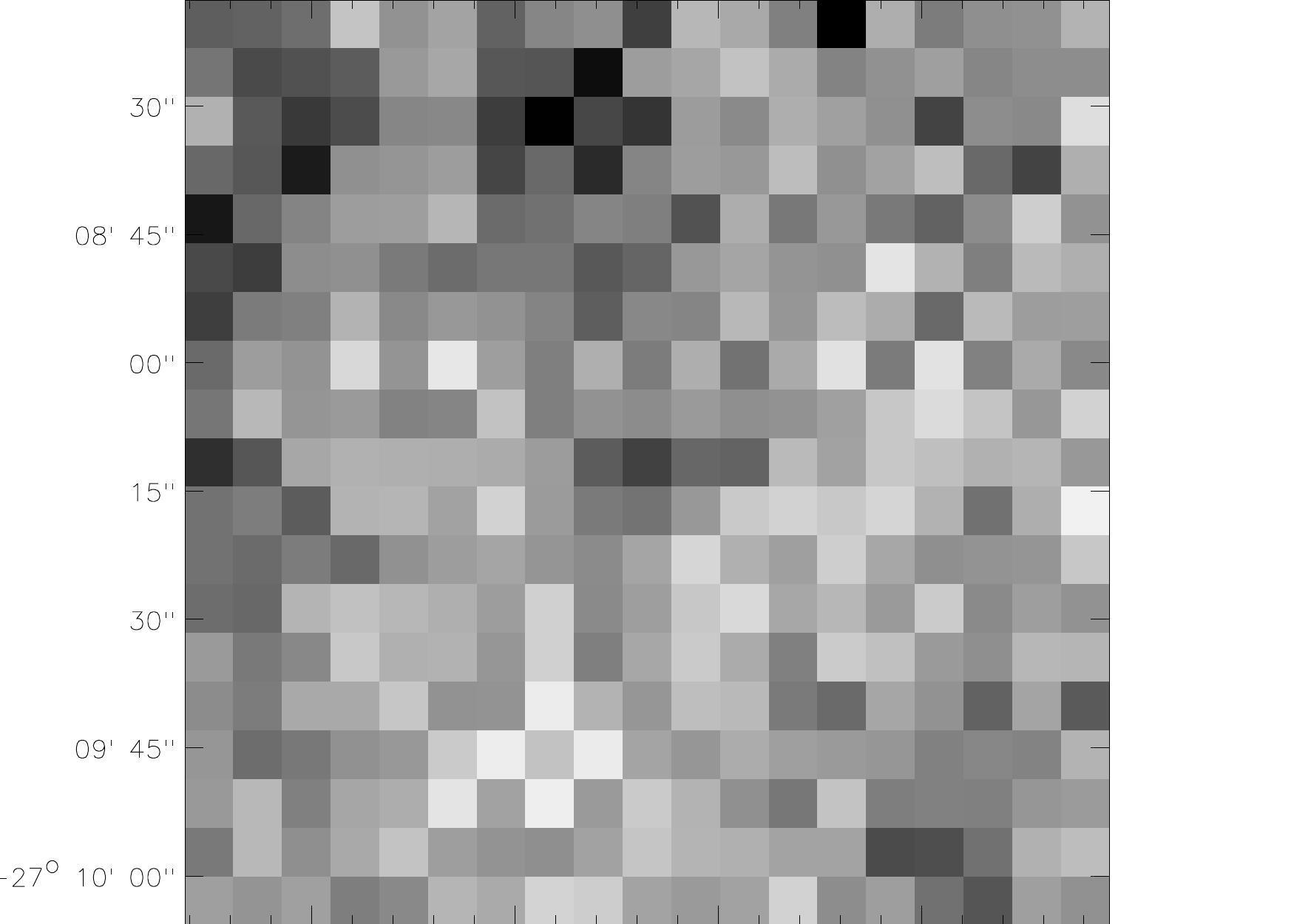}   &
      \includegraphics[height=14mm,trim=70  -4 65 -10,clip=true]{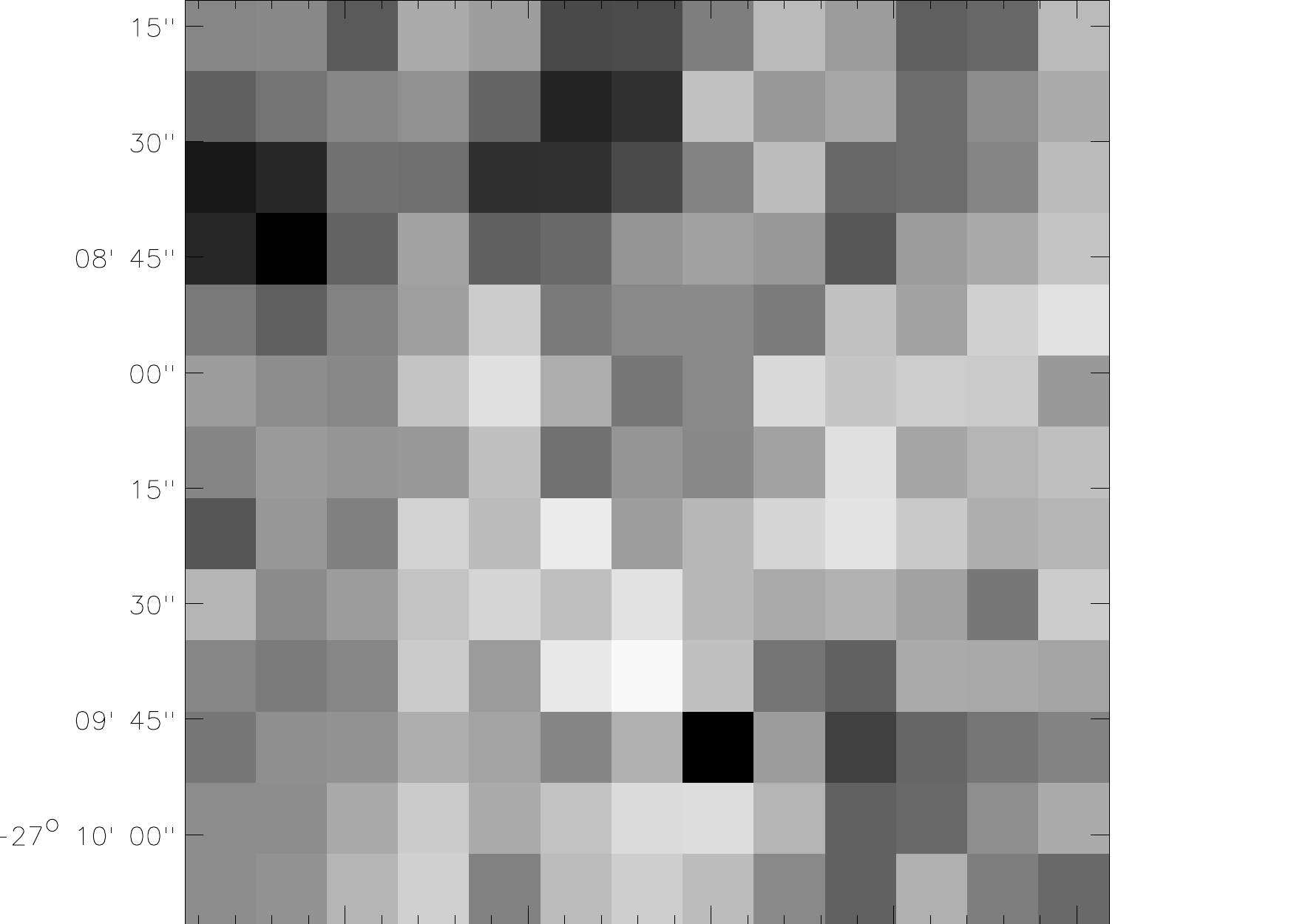}   &
      \includegraphics[height=14mm,trim=70  -4 65 -10,clip=true]{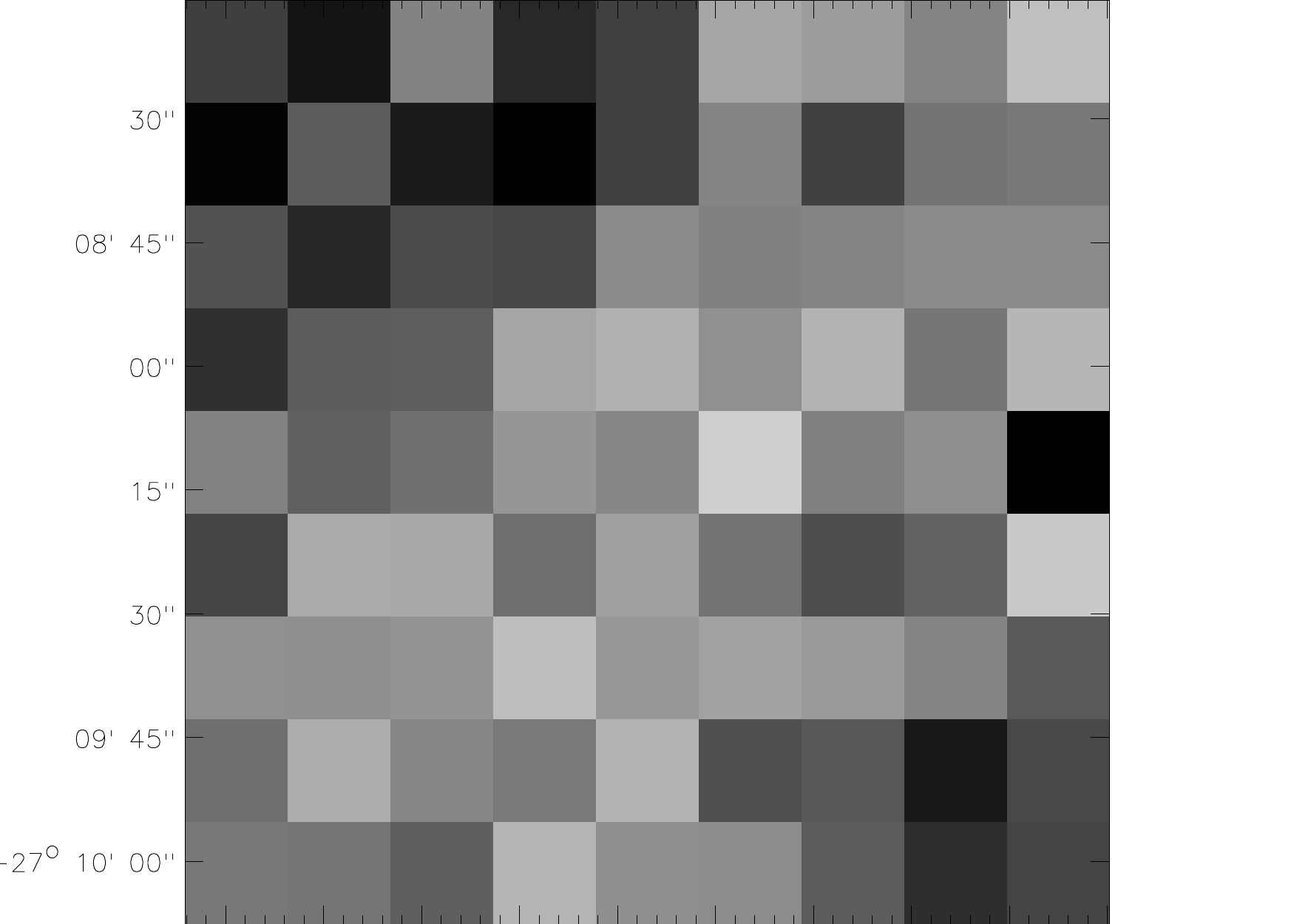} \\

      \multicolumn{6}{c}{\includegraphics[width=80mm,trim=30 0 20 0 ,clip=true]{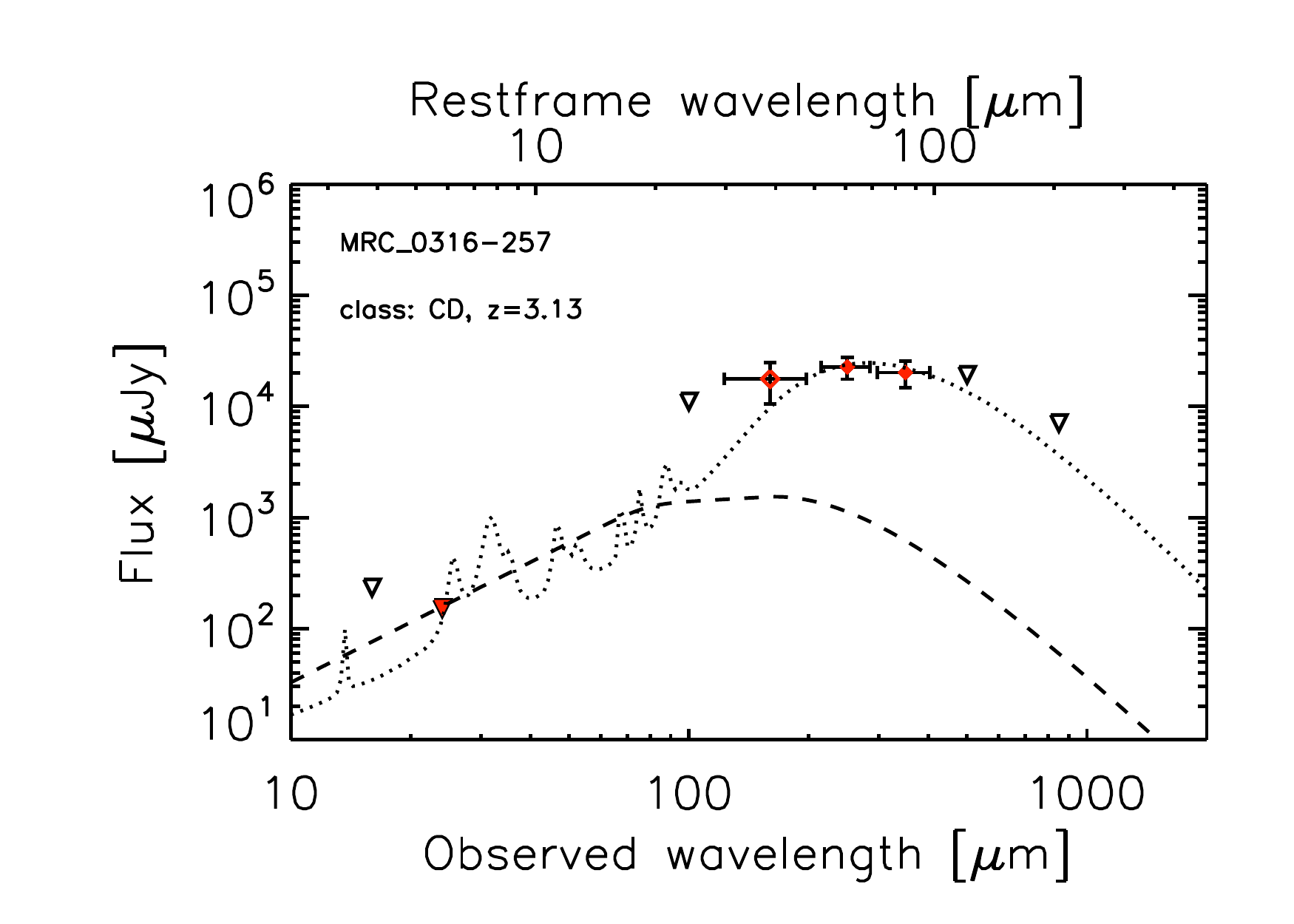}} \\
      24\mum & 100\mum & 160\mum & 250\mum & 350\mum & 500\mum \\
      \includegraphics[height=14mm,trim=70  -4 65 -10,clip=true]{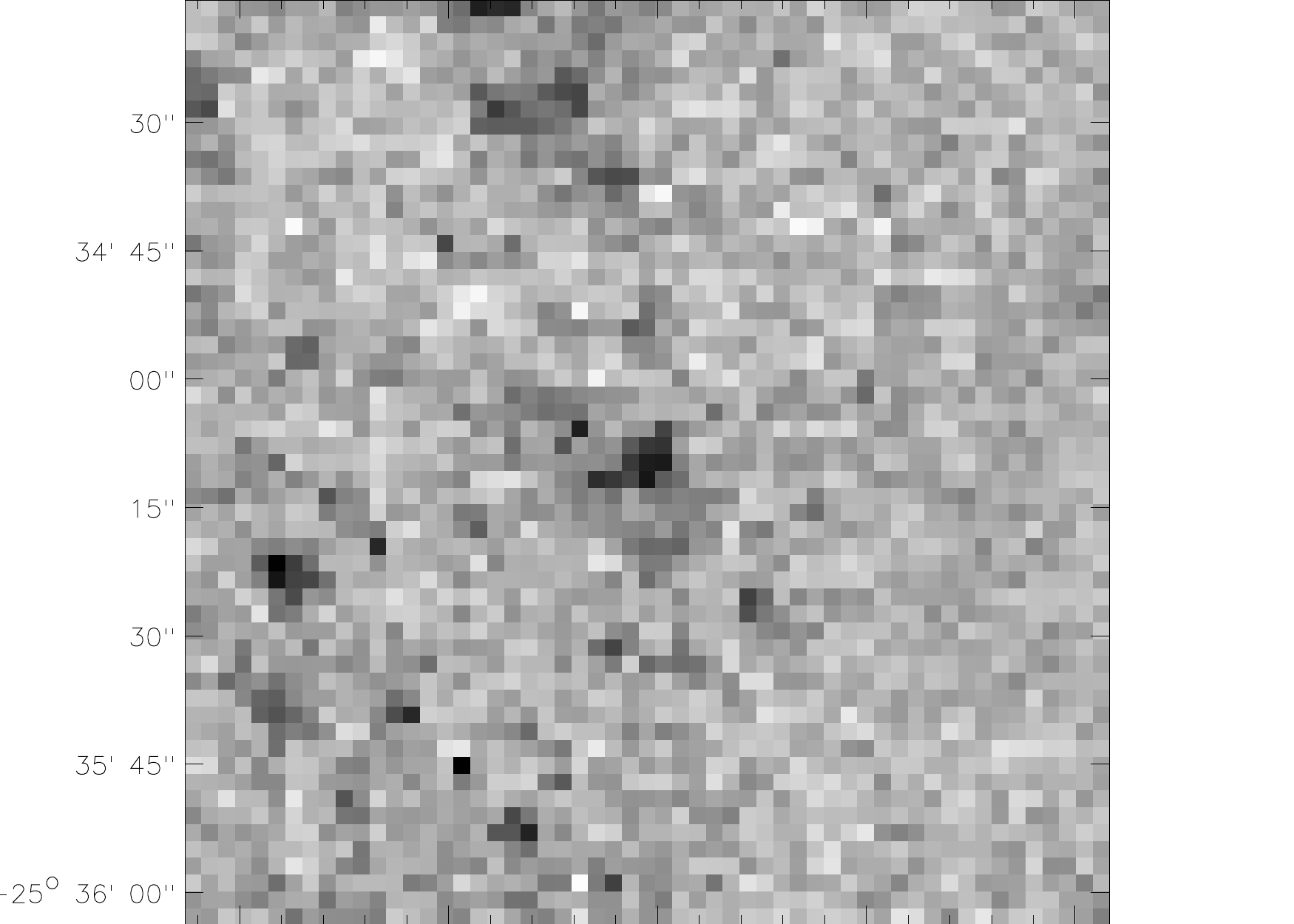} &
      \includegraphics[height=14mm,trim=70  -4 65 -10,clip=true]{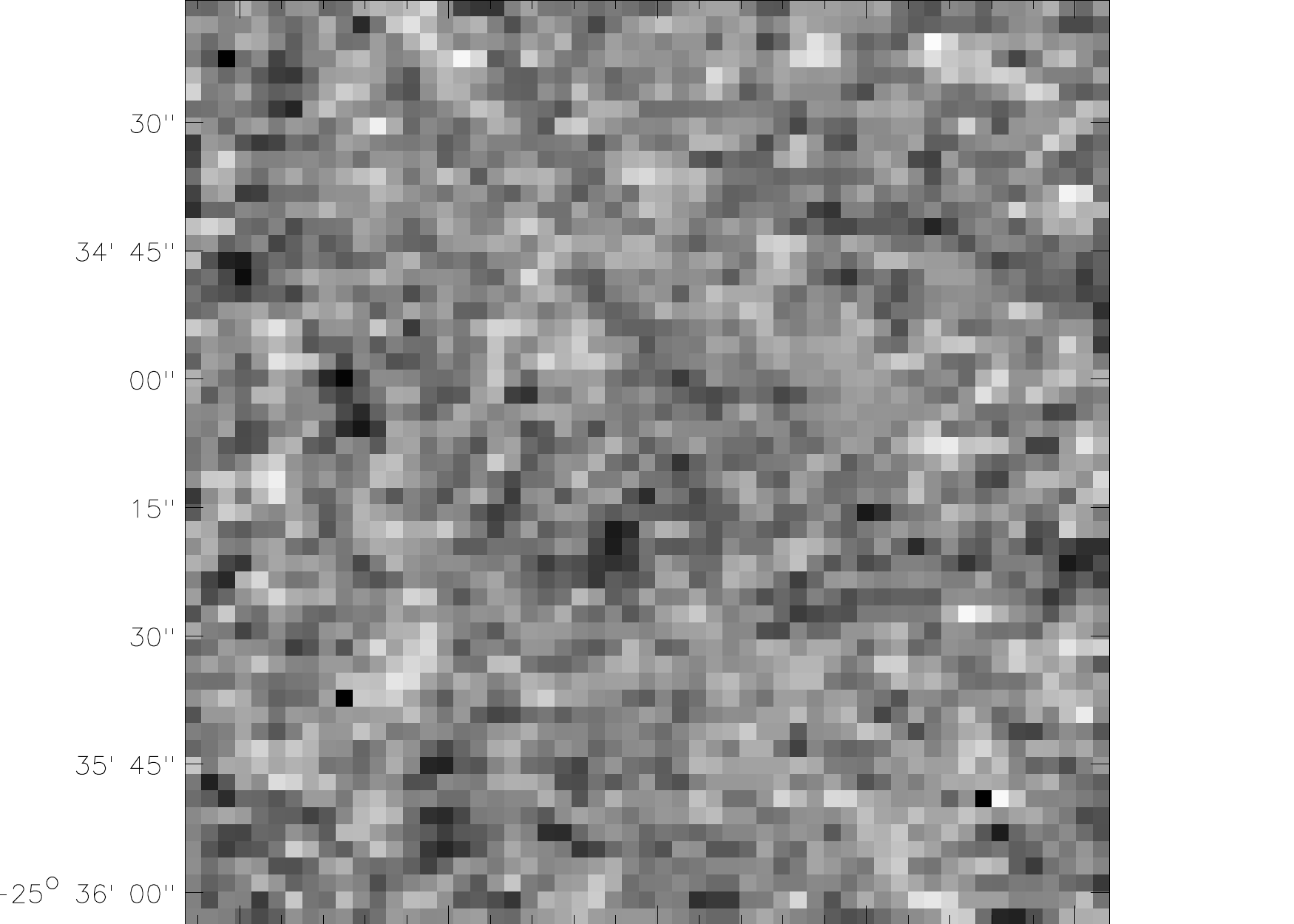} &
      \includegraphics[height=14mm,trim=70  -4 65 -10,clip=true]{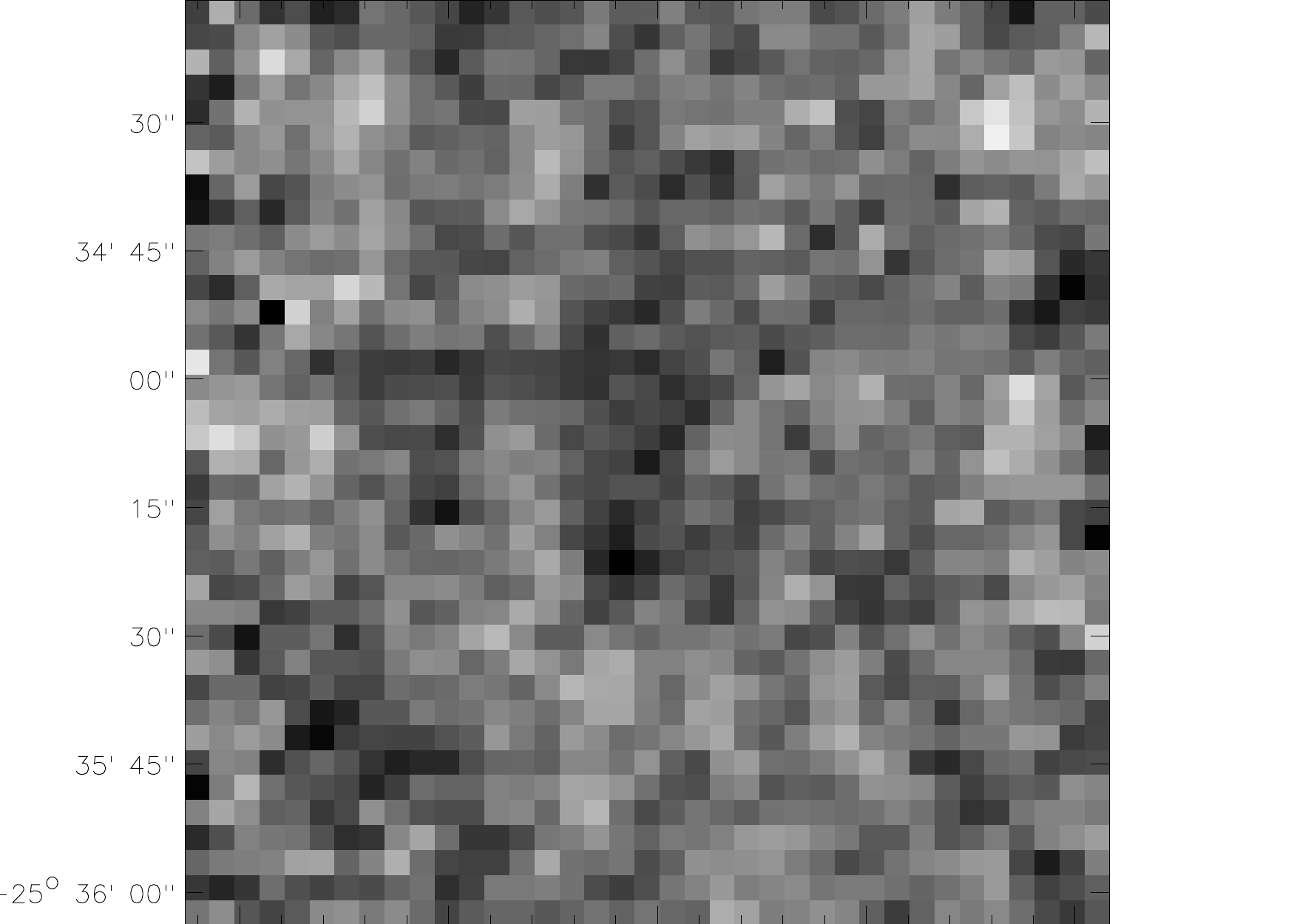} &
      \includegraphics[height=14mm,trim=70  -4 65 -10,clip=true]{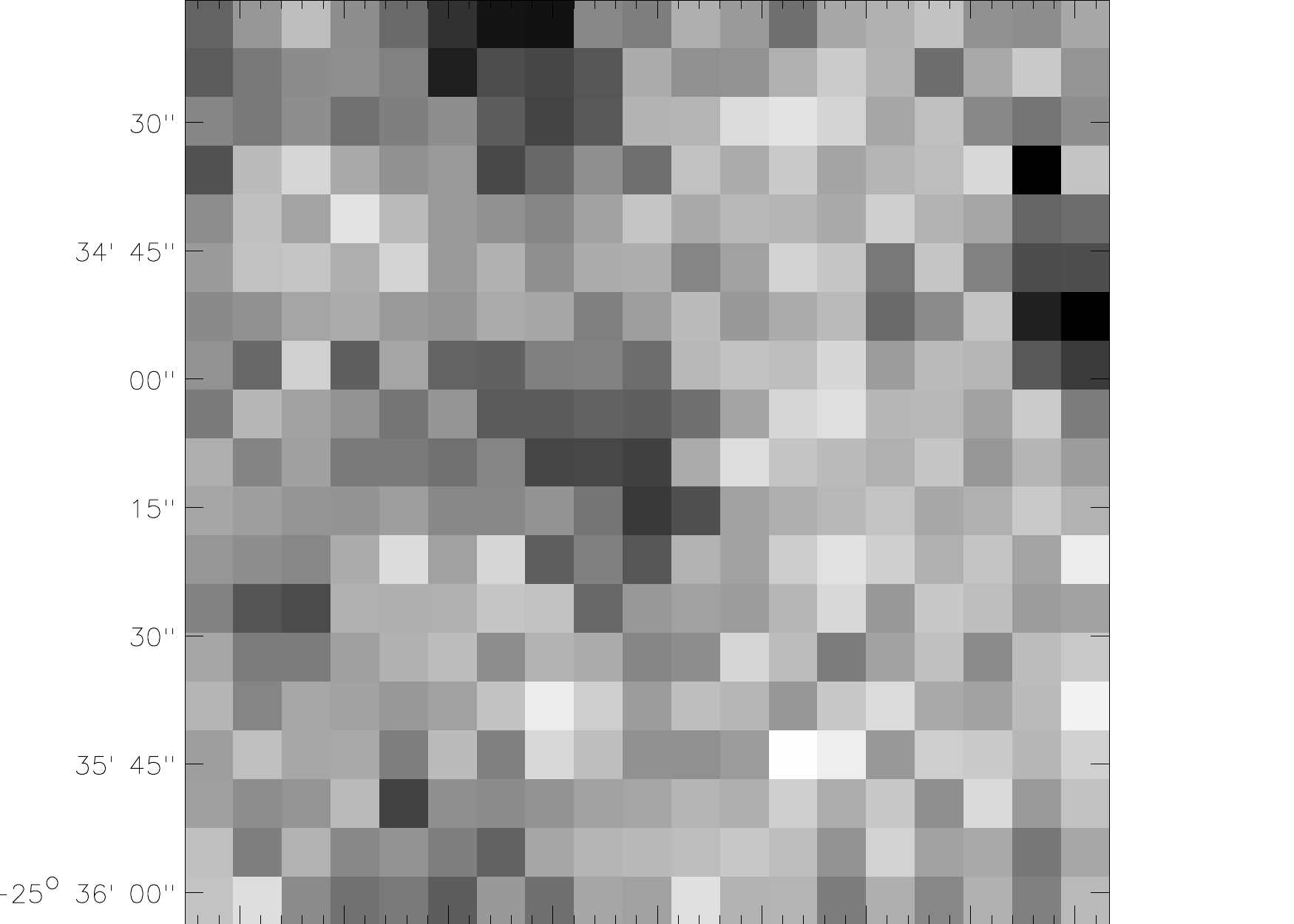}   &
      \includegraphics[height=14mm,trim=70  -4 65 -10,clip=true]{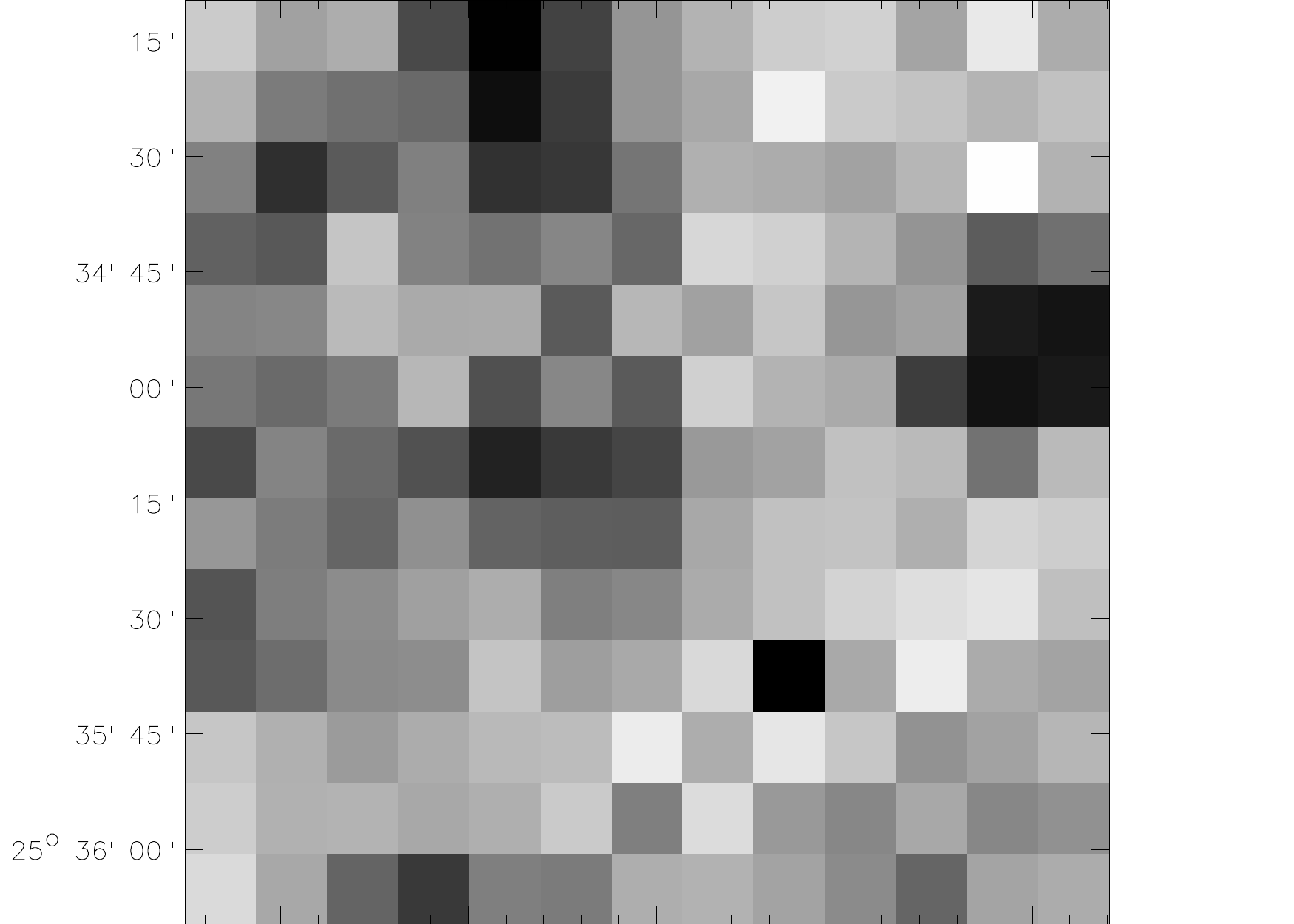}   &
      \includegraphics[height=14mm,trim=70  -4 65 -10,clip=true]{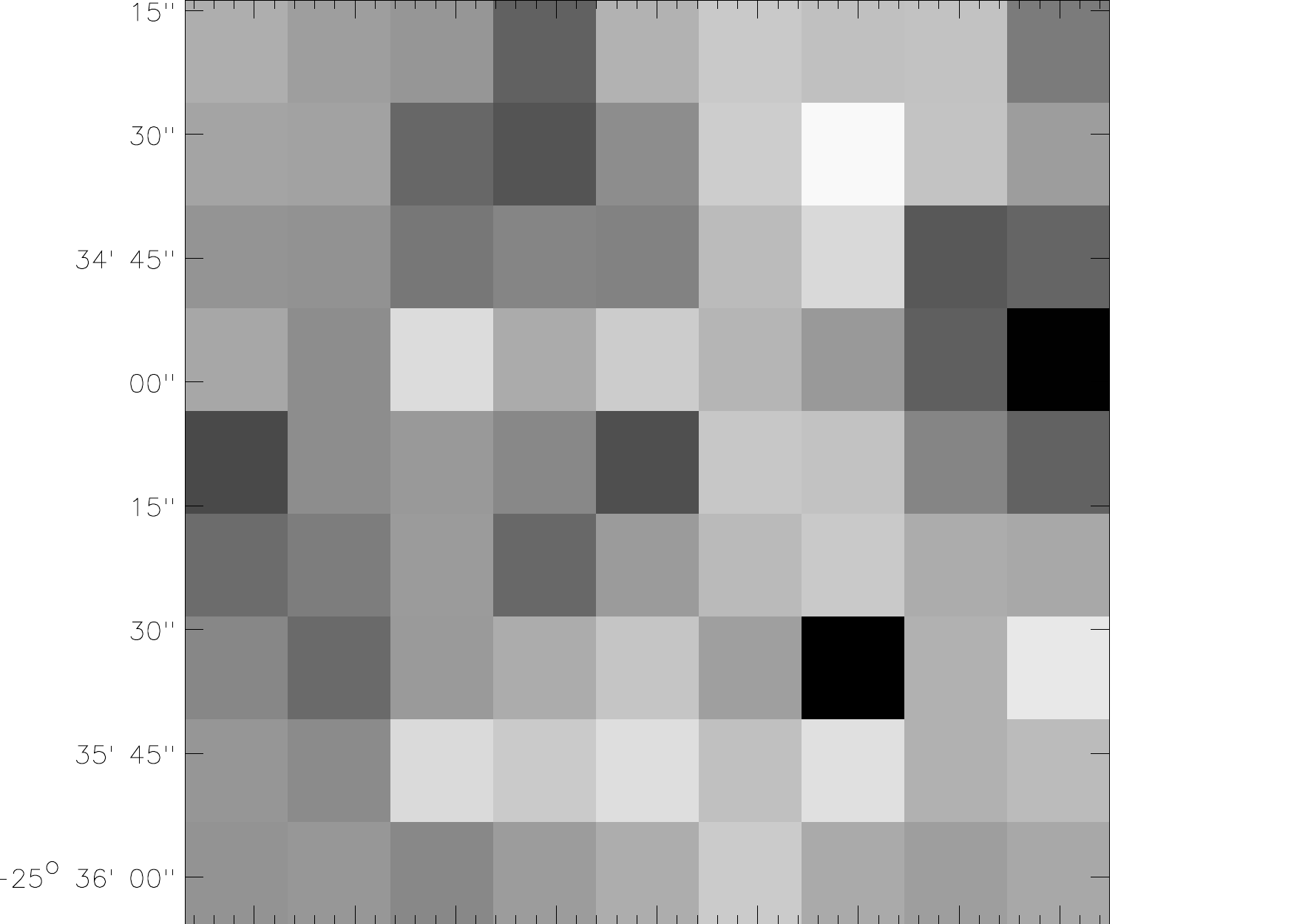} \\

      \multicolumn{6}{c}{\includegraphics[width=80mm,trim=30 0 20 0 ,clip=true]{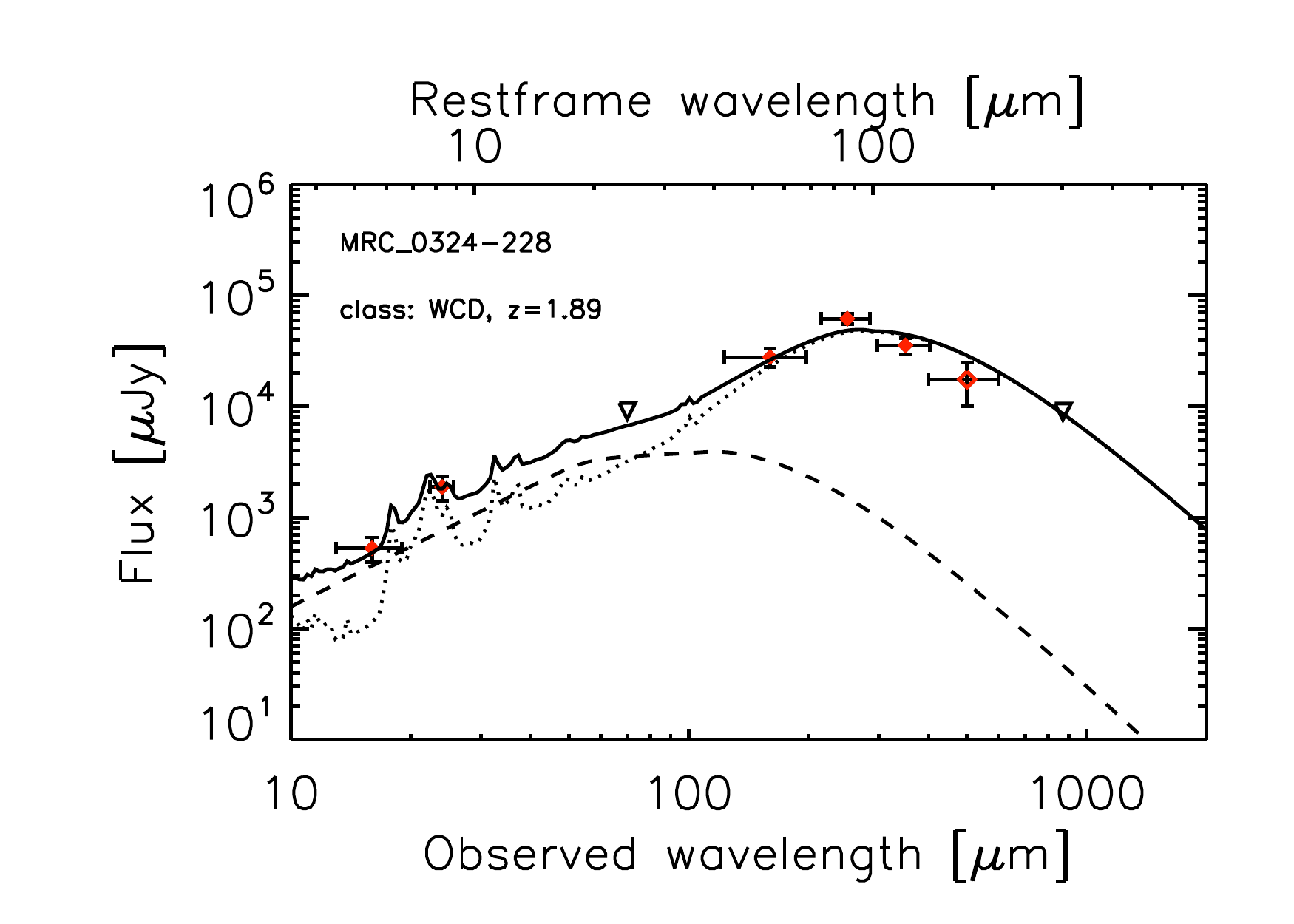}} \\
      24\mum & 70\mum & 160\mum & 250\mum & 350\mum & 500\mum \\
\end{tabular}
          \end{minipage}
  \end{centering}
\end{figure*}

\clearpage

\begin{figure*}[h]
  \begin{centering}
    \begin{minipage}[c]{0.46\linewidth}
    \begin{tabular}{c@{}c@{}c@{}c@{}c@{}c@{}}

      \multicolumn{6}{c}{\includegraphics[width=80mm,trim=30 0 20 0 ,clip=true]{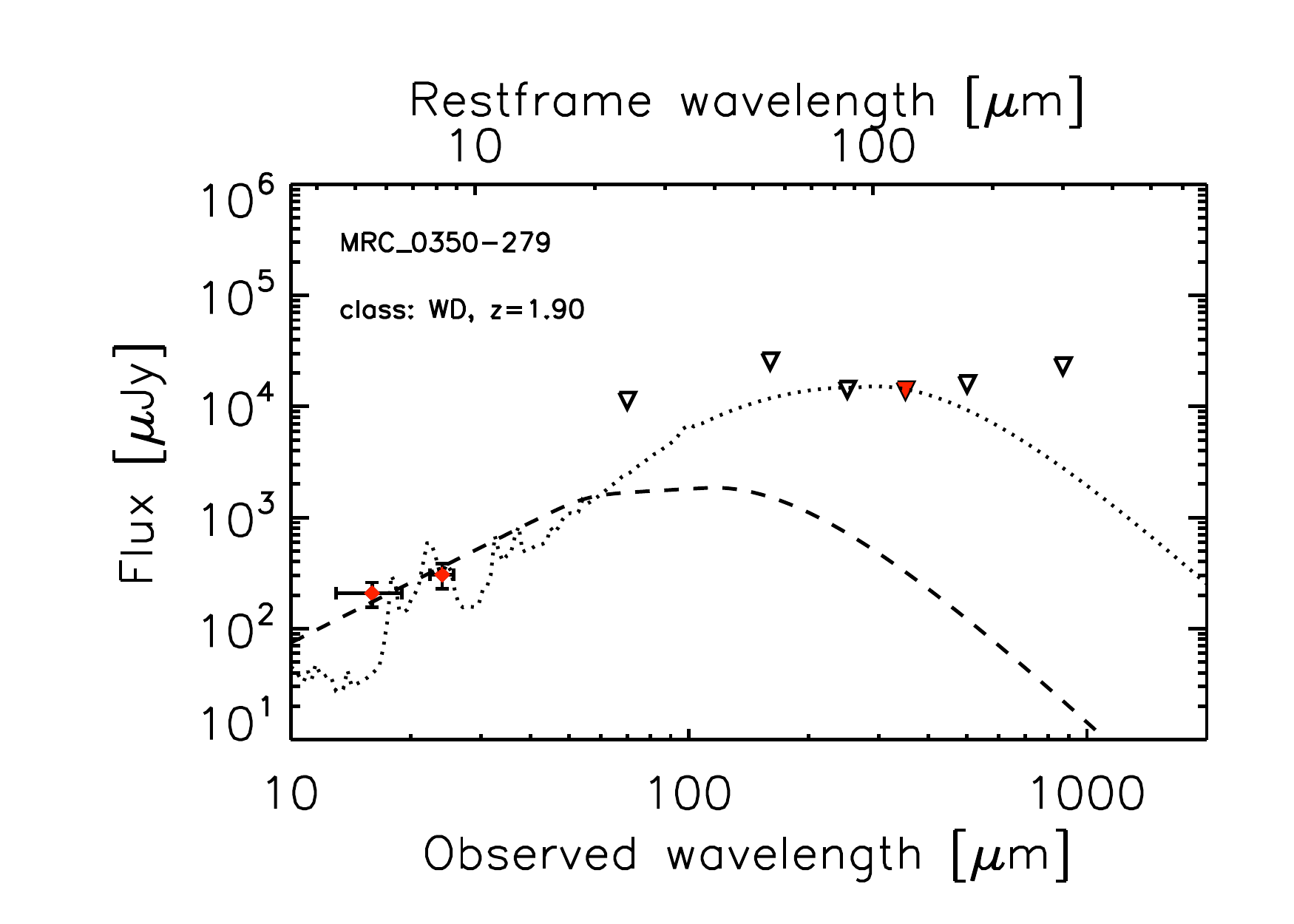}} \\
      24\mum & 70\mum & 160\mum & 250\mum & 350\mum & 500\mum \\
      \includegraphics[height=14mm,trim=70  -4 65 -10,clip=true]{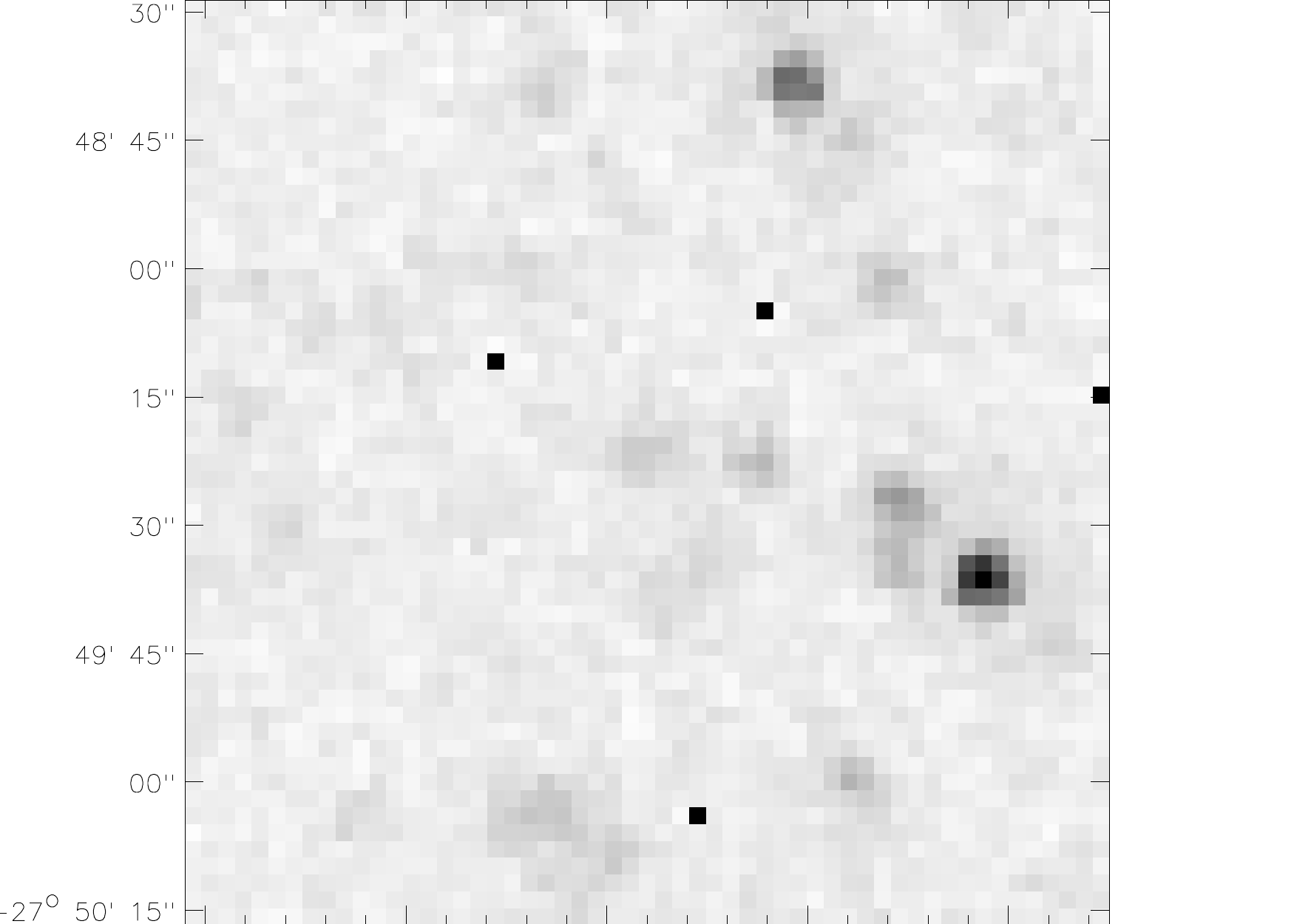} &
      \includegraphics[height=14mm,trim=70  -4 65 -10,clip=true]{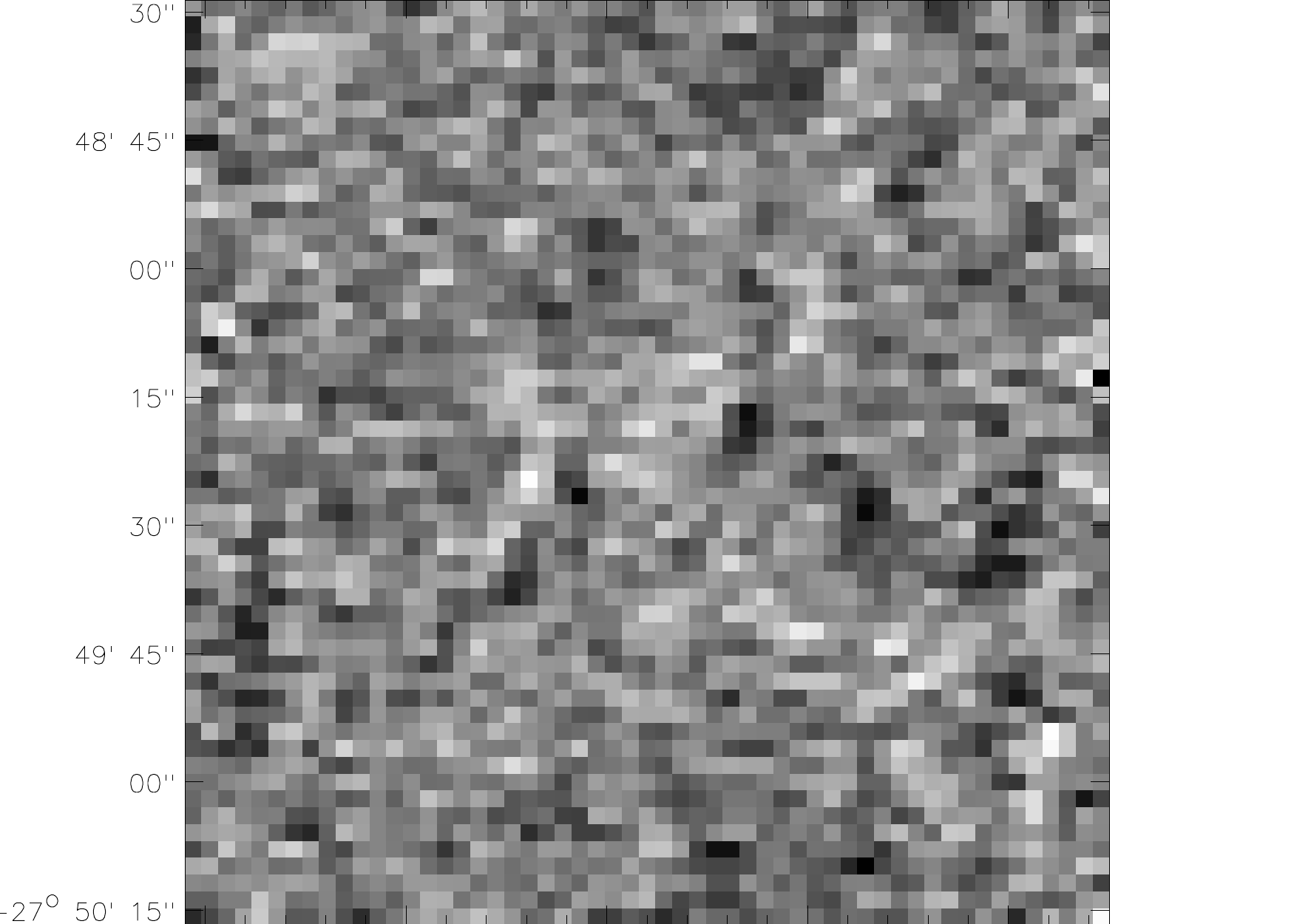} &
      \includegraphics[height=14mm,trim=70  -4 65 -10,clip=true]{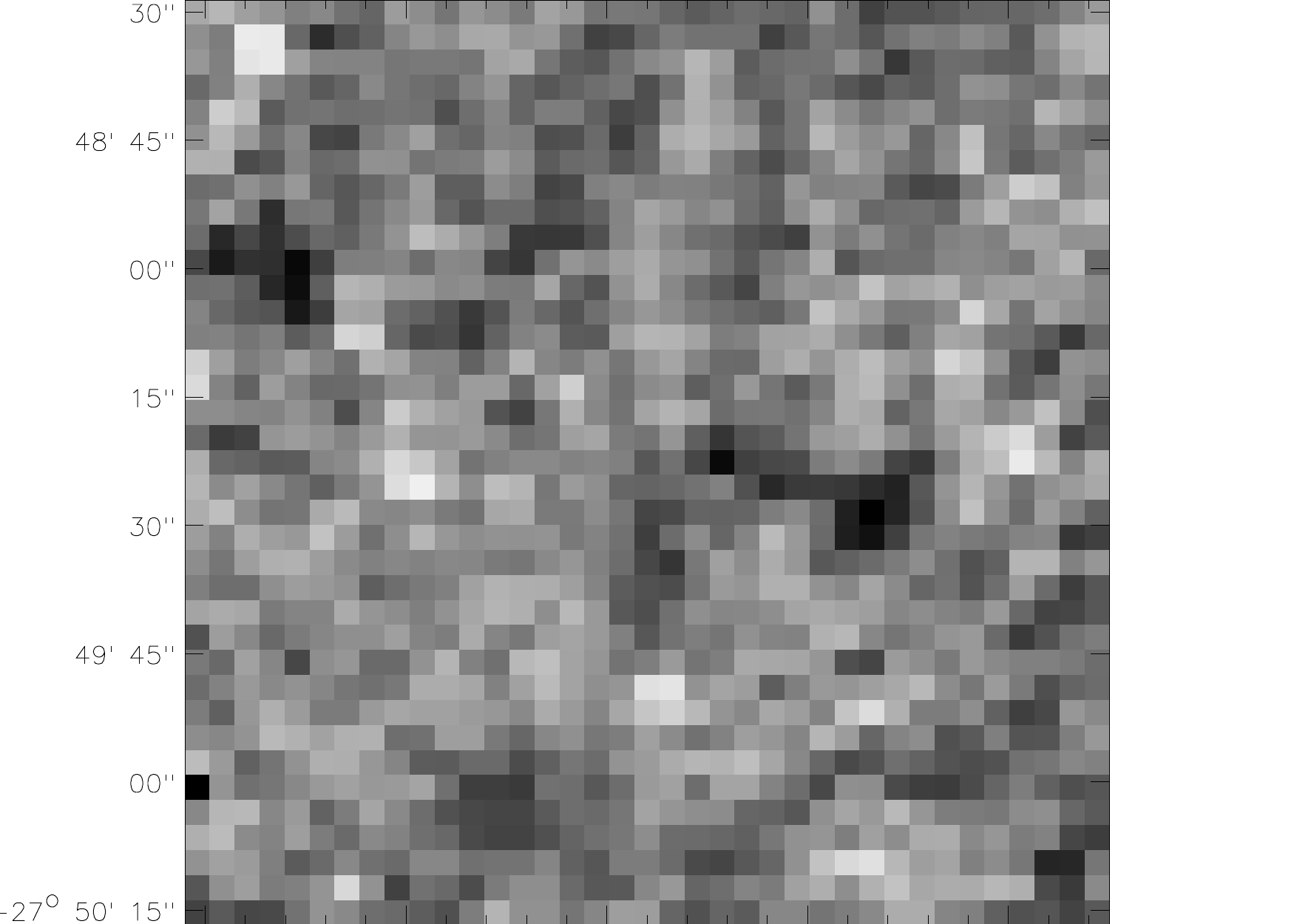} &
      \includegraphics[height=14mm,trim=70  -4 65 -10,clip=true]{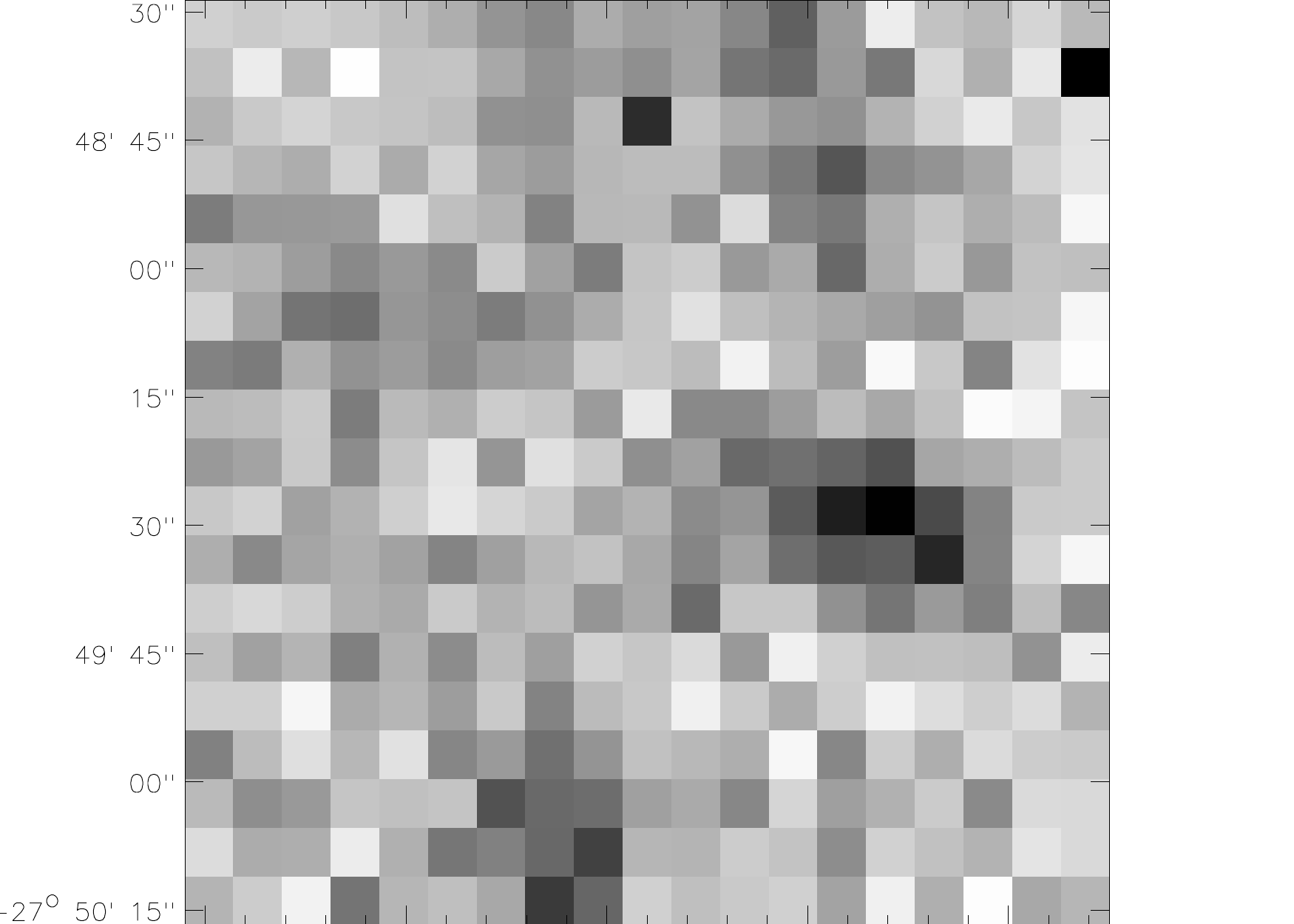}   &
      \includegraphics[height=14mm,trim=70  -4 65 -10,clip=true]{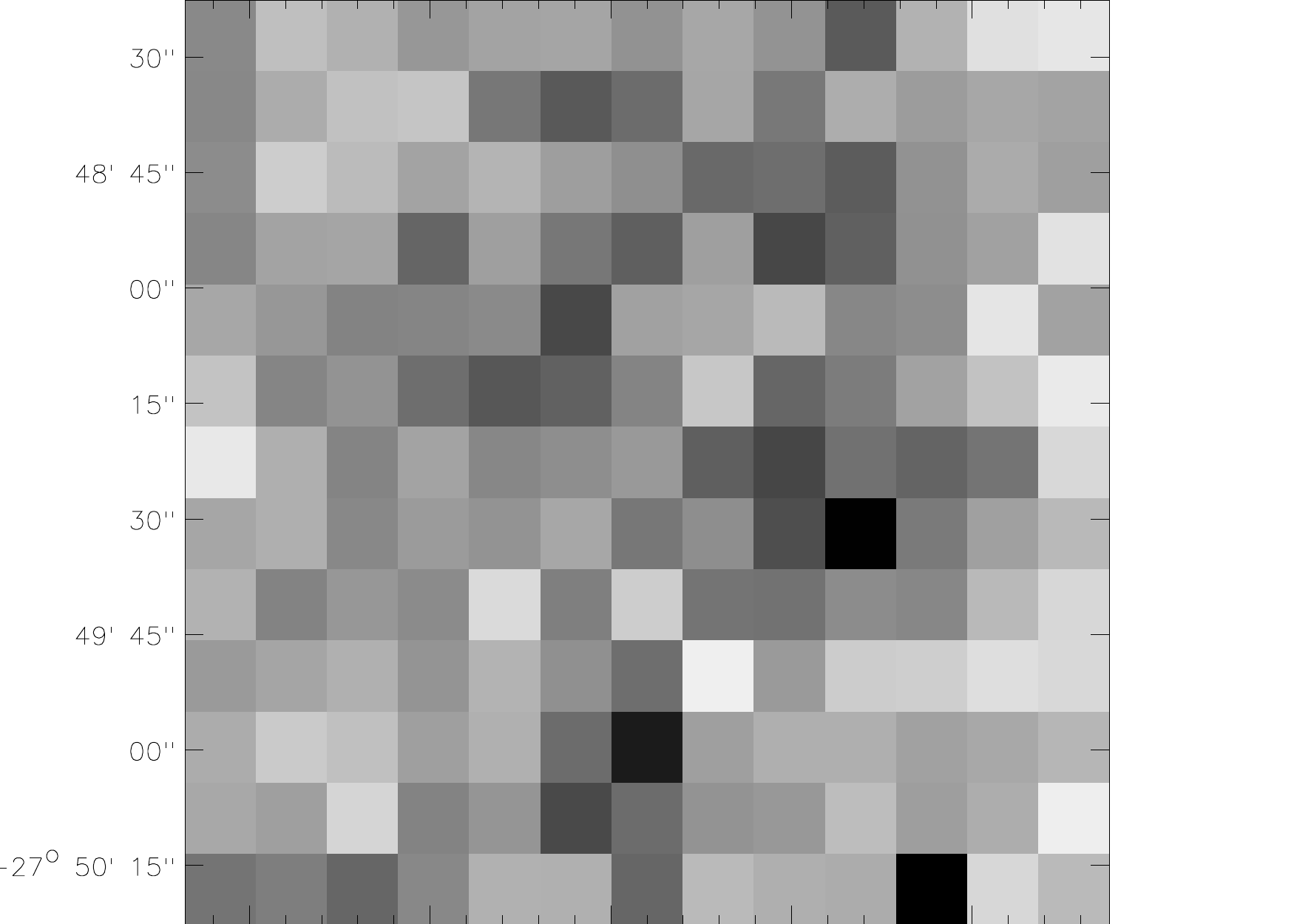}   &
      \includegraphics[height=14mm,trim=70  -4 65 -10,clip=true]{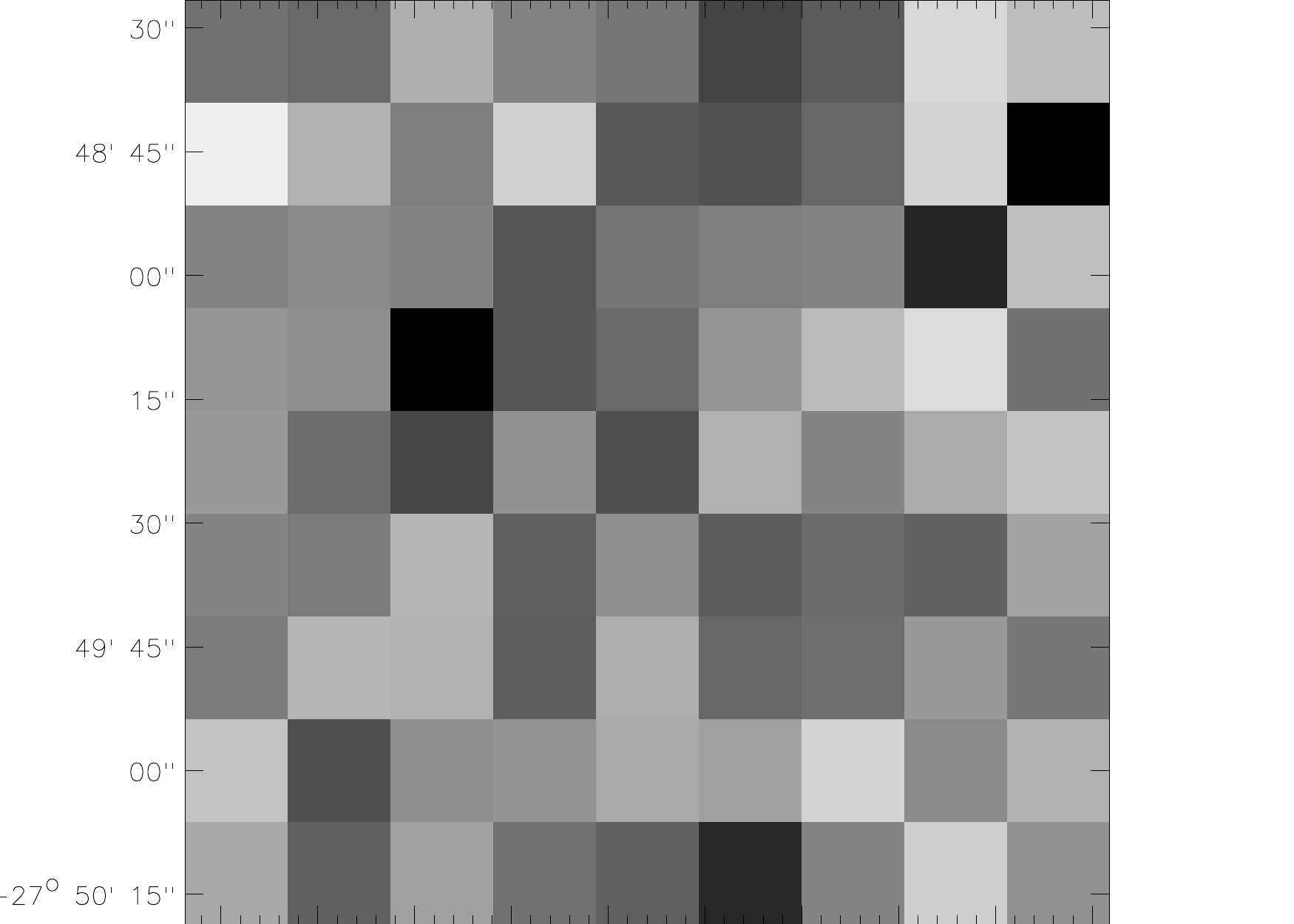} \\

      \multicolumn{6}{c}{\includegraphics[width=80mm,trim=30 0 20 0 ,clip=true]{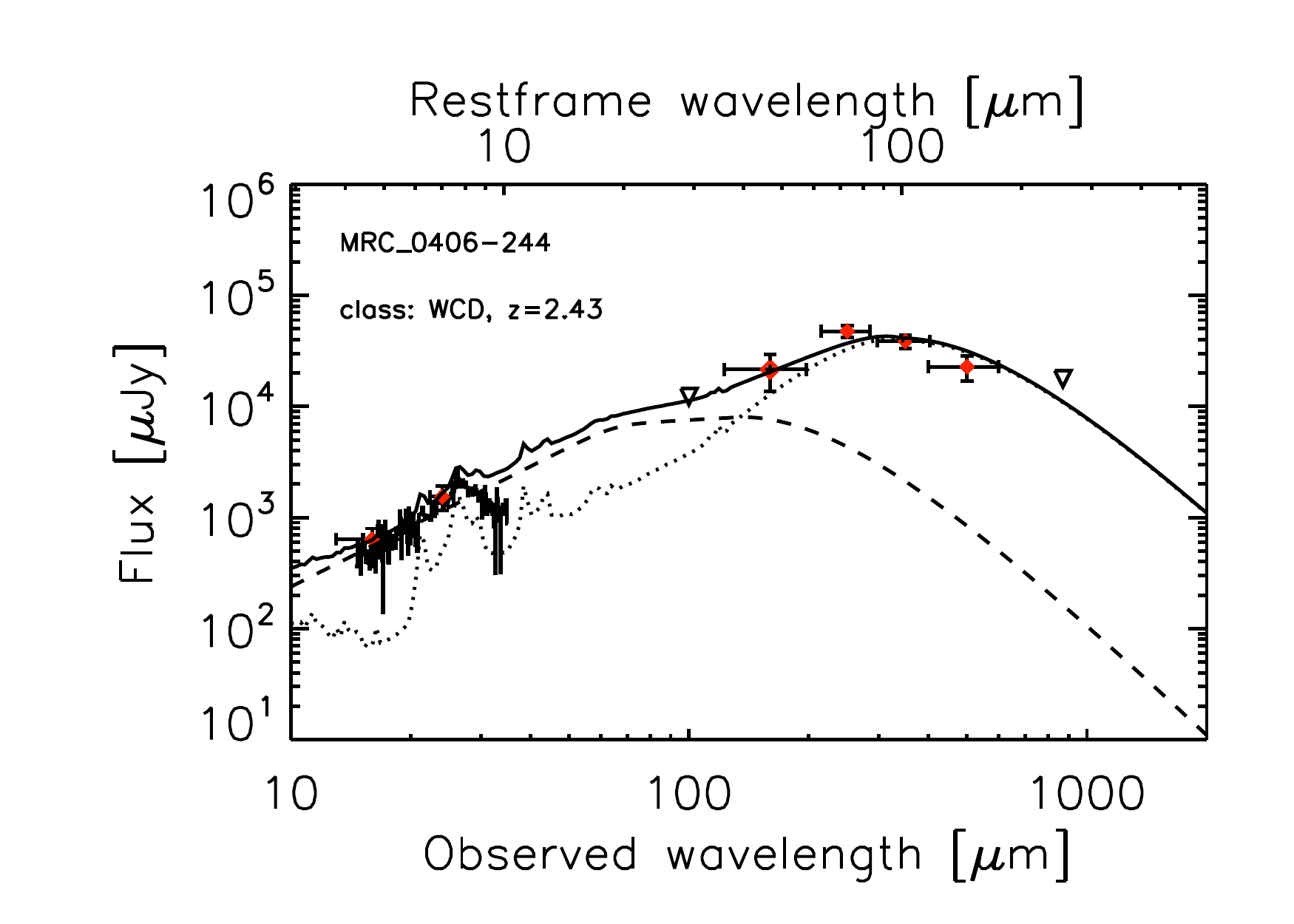}} \\
      24\mum & 100\mum & 160\mum & 250\mum & 350\mum & 500\mum \\
      \includegraphics[height=14mm,trim=70  -4 65 -10,clip=true]{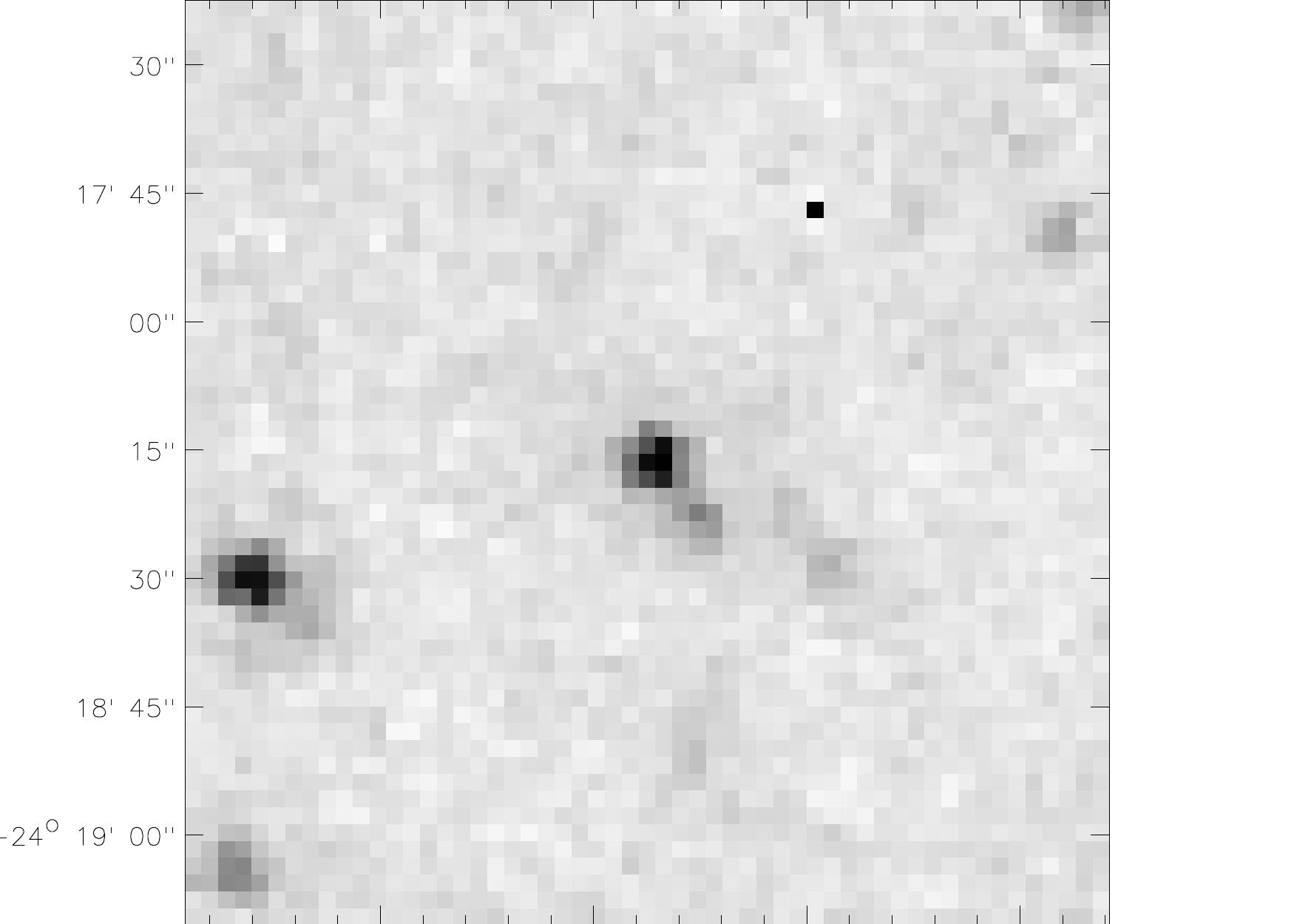} &
      \includegraphics[height=14mm,trim=70  -4 65 -10,clip=true]{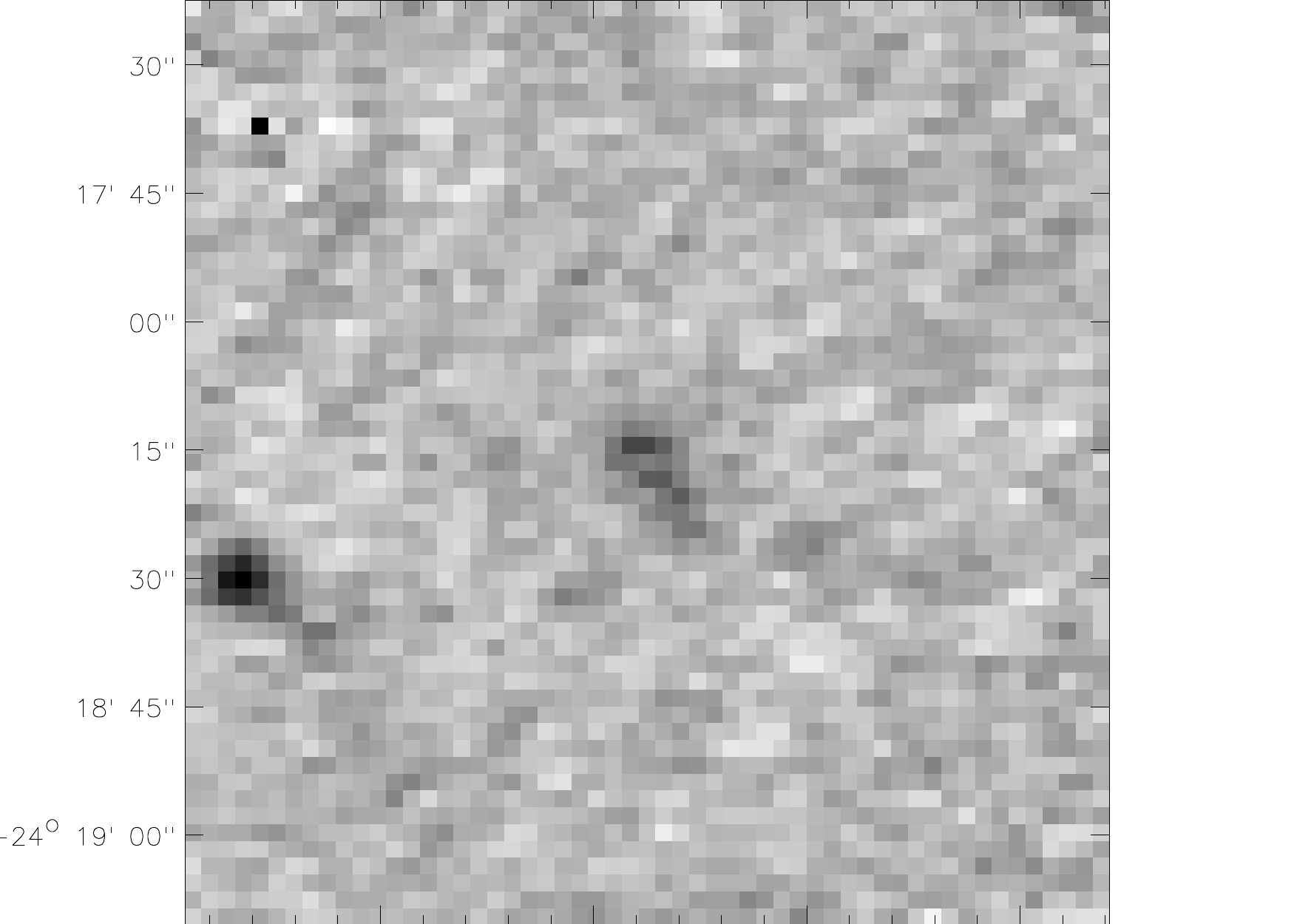} &
      \includegraphics[height=14mm,trim=70  -4 65 -10,clip=true]{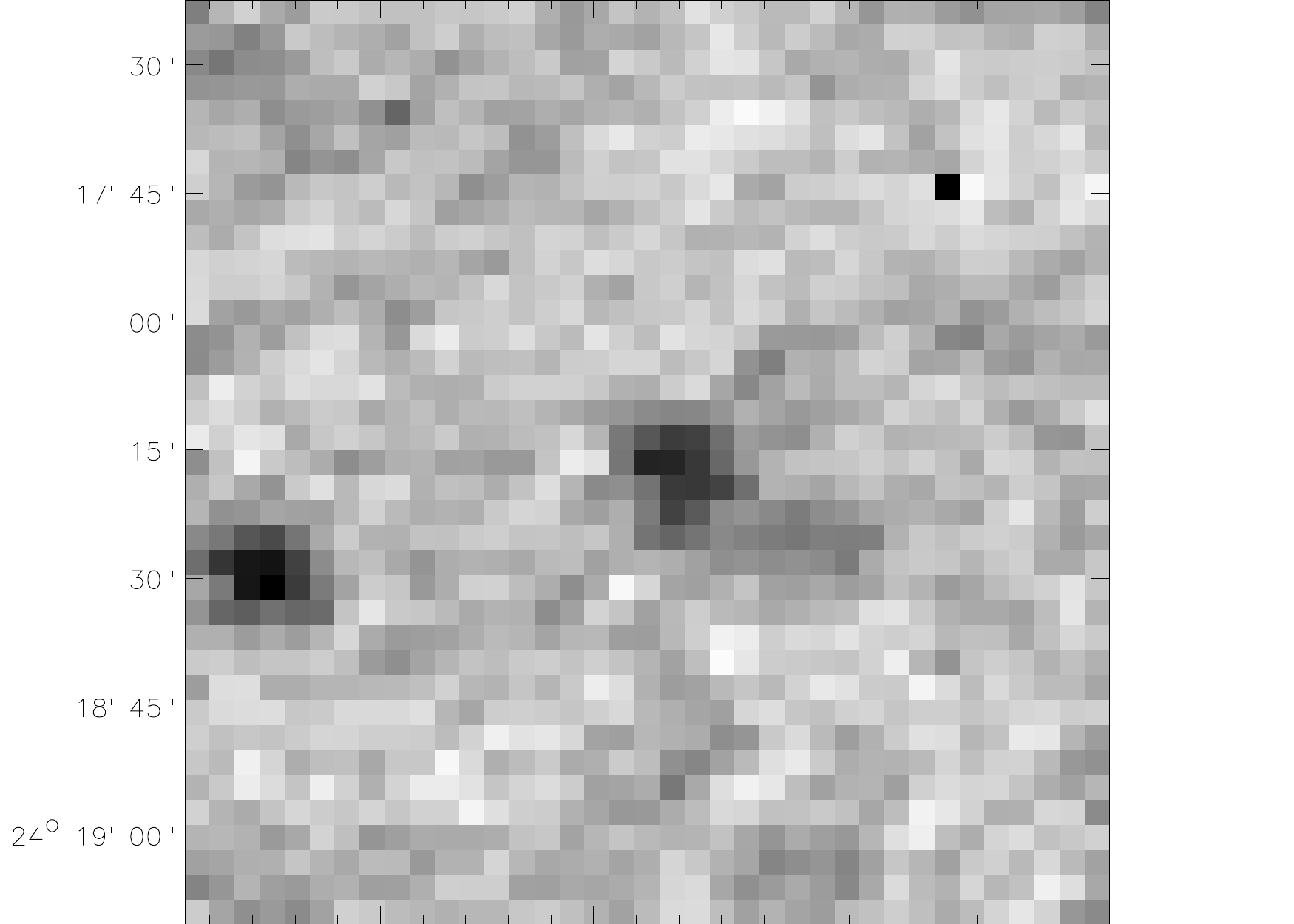} &
      \includegraphics[height=14mm,trim=70  -4 65 -10,clip=true]{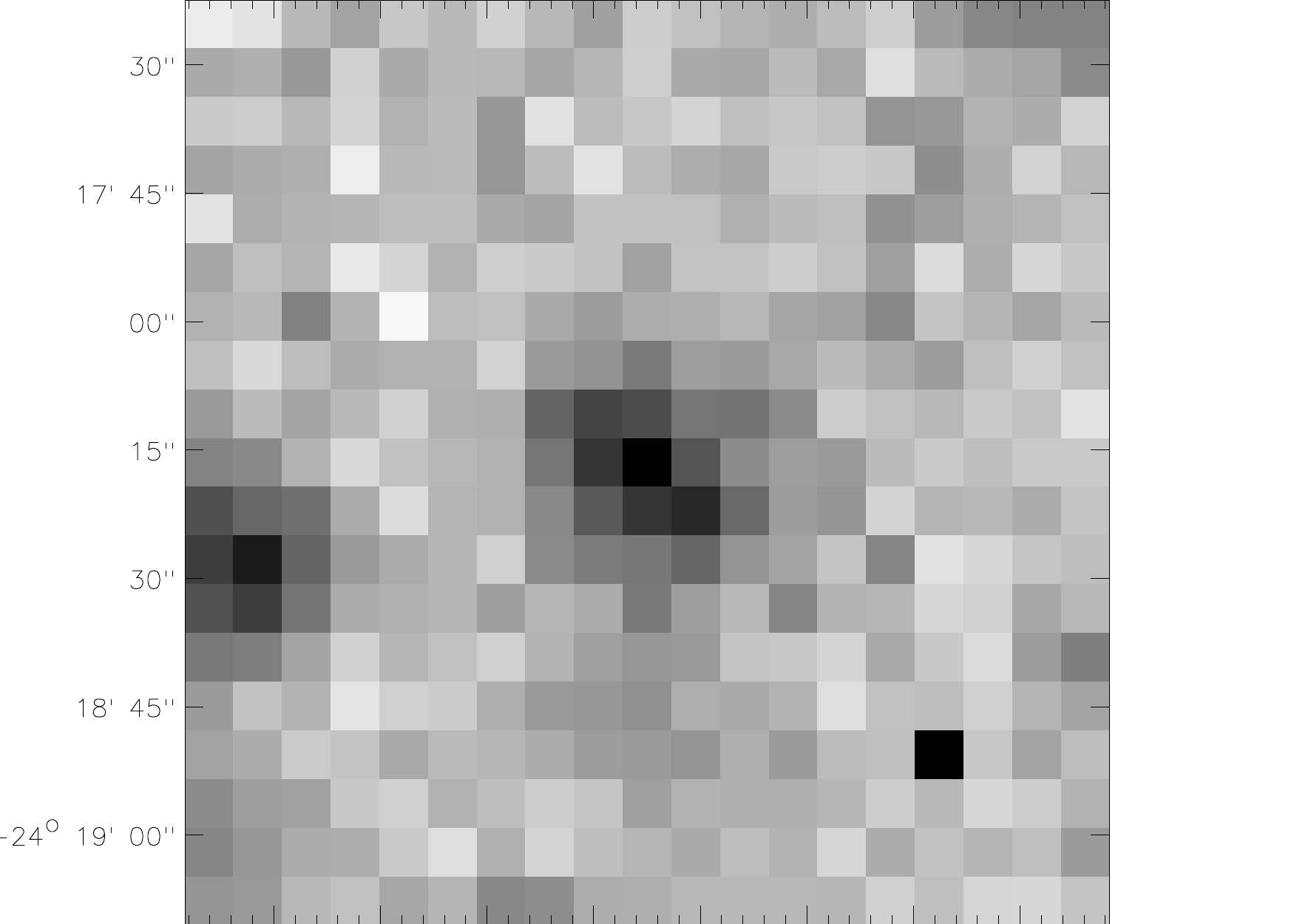}   &
      \includegraphics[height=14mm,trim=70  -4 65 -10,clip=true]{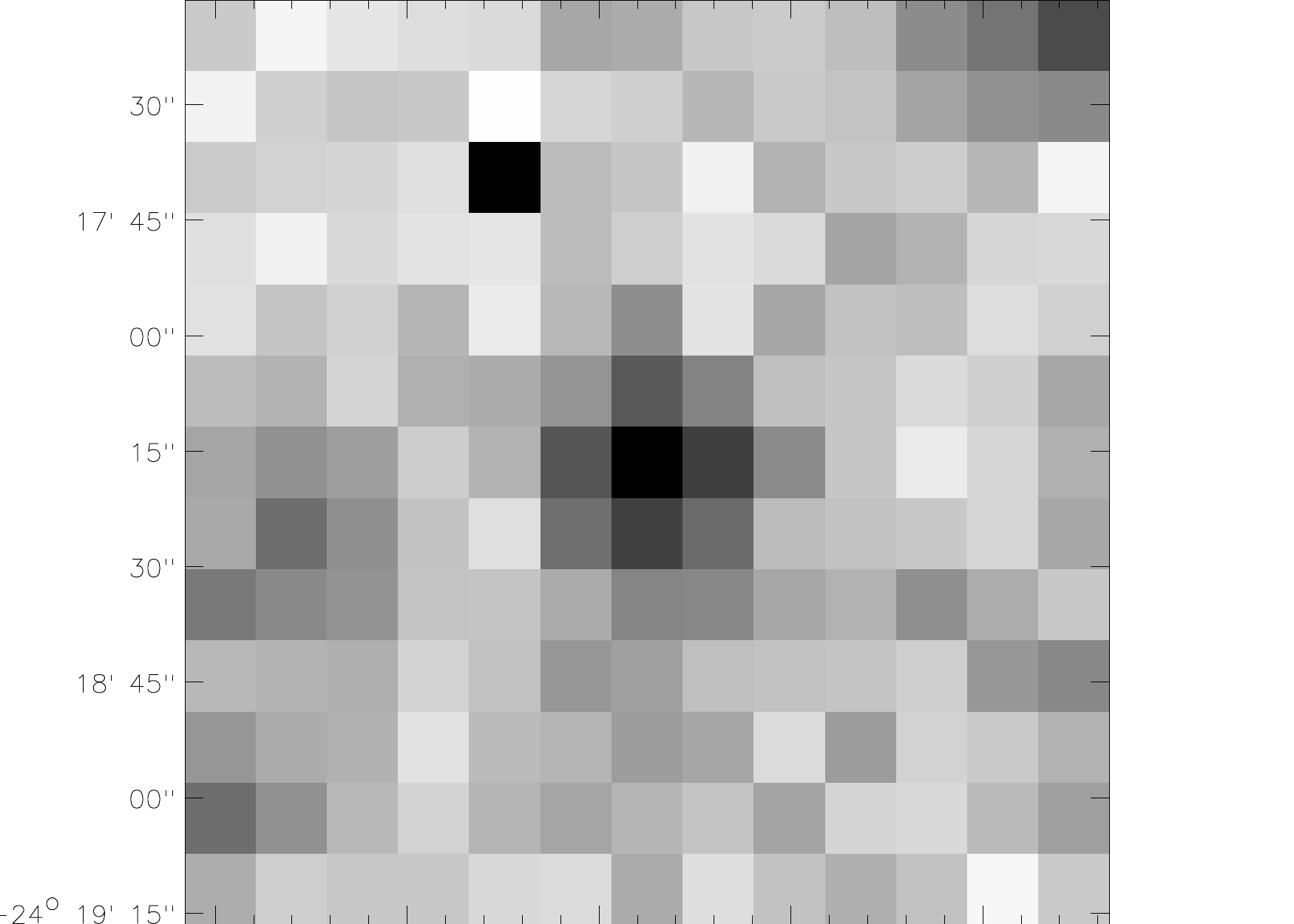}   &
      \includegraphics[height=14mm,trim=70  -4 65 -10,clip=true]{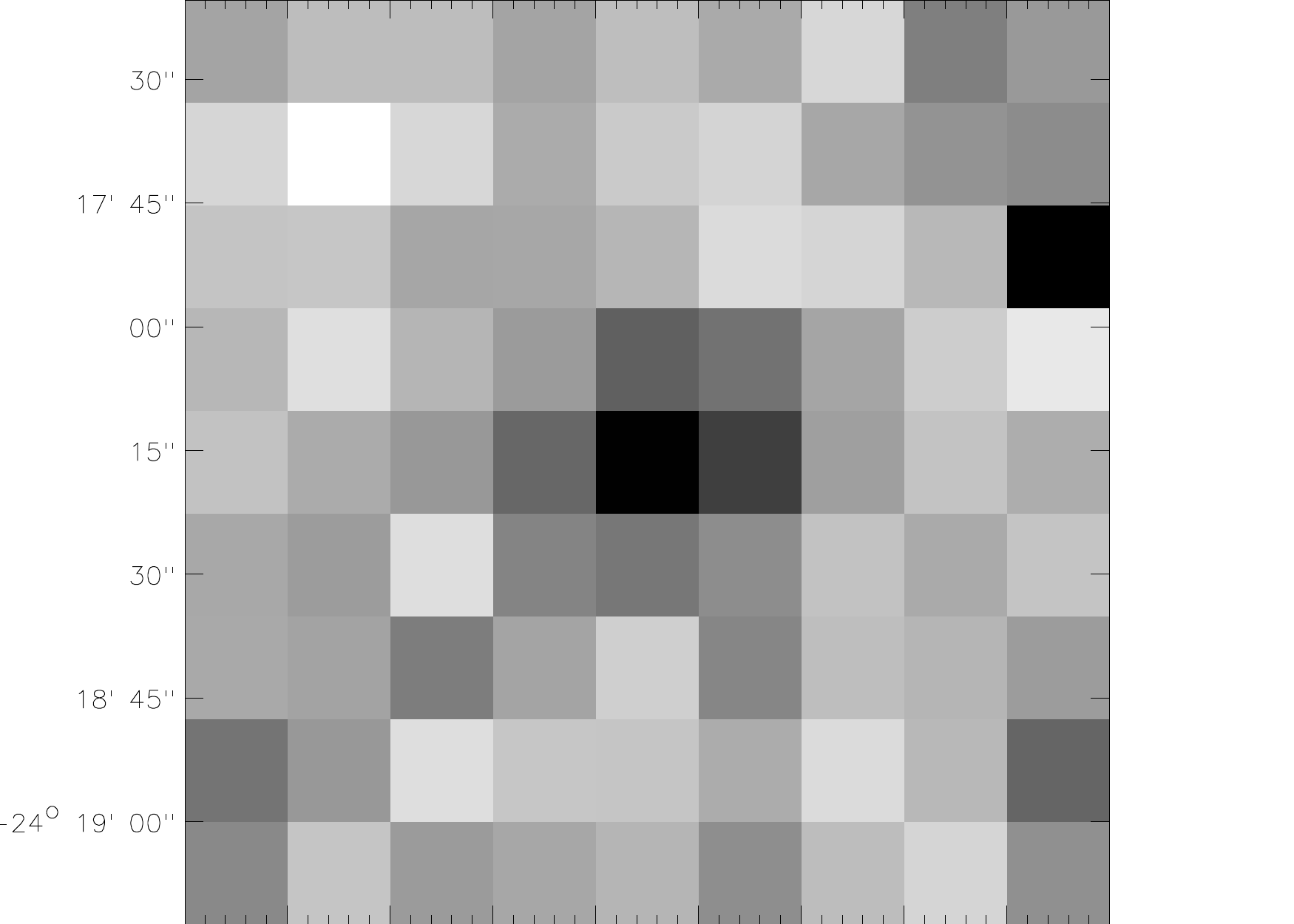} \\

      \multicolumn{6}{c}{\includegraphics[width=80mm,trim=30 0 20 0 ,clip=true]{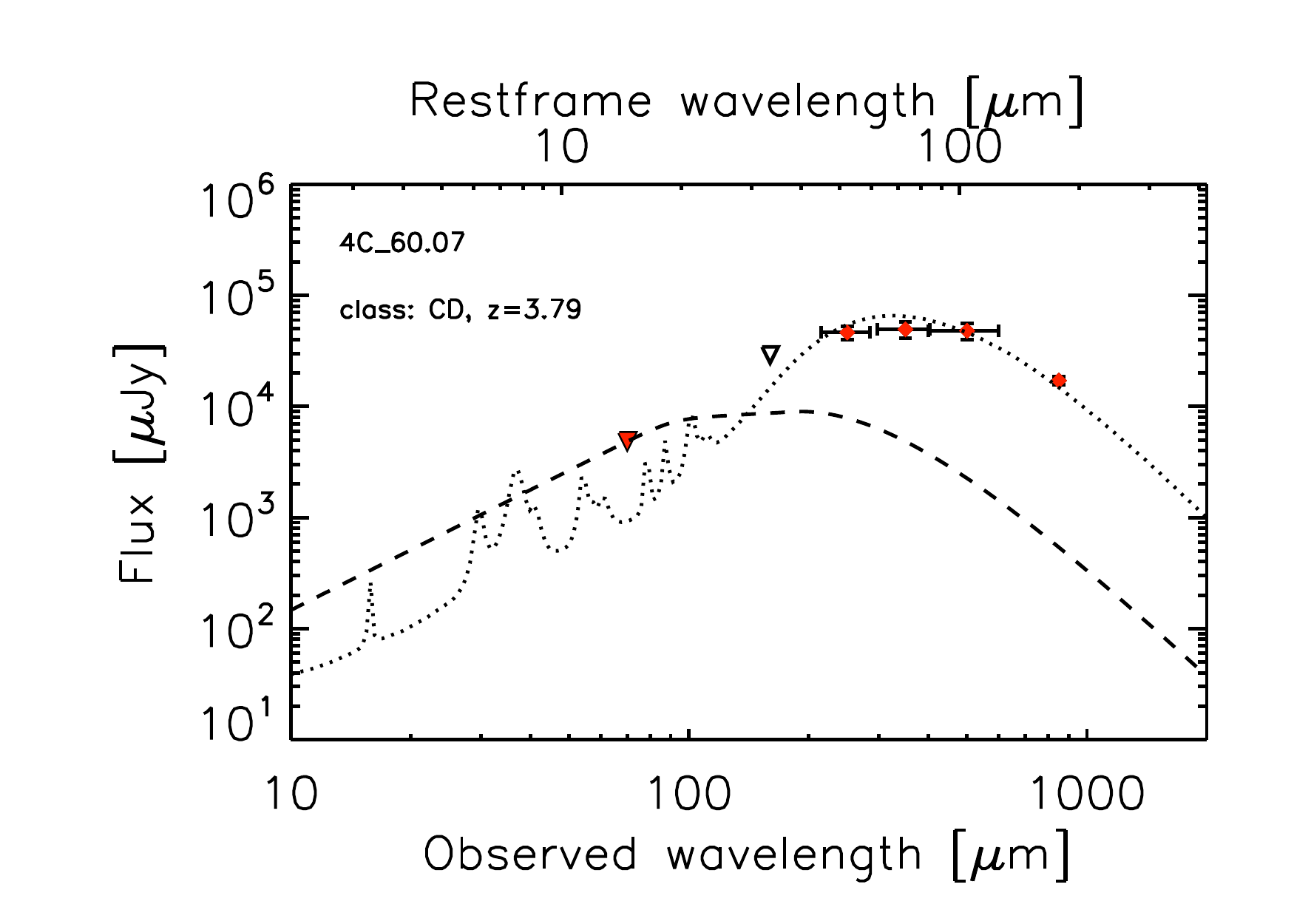}} \\
      24\mum & 70\mum & 160\mum & 250\mum & 350\mum & 500\mum \\
\end{tabular}
\end{minipage}
\begin{minipage}[c]{0.46\linewidth} 
    \begin{tabular}{c@{}c@{}c@{}c@{}c@{}c@{}}

      \multicolumn{6}{c}{\includegraphics[width=80mm,trim=30 0 20 0 ,clip=true]{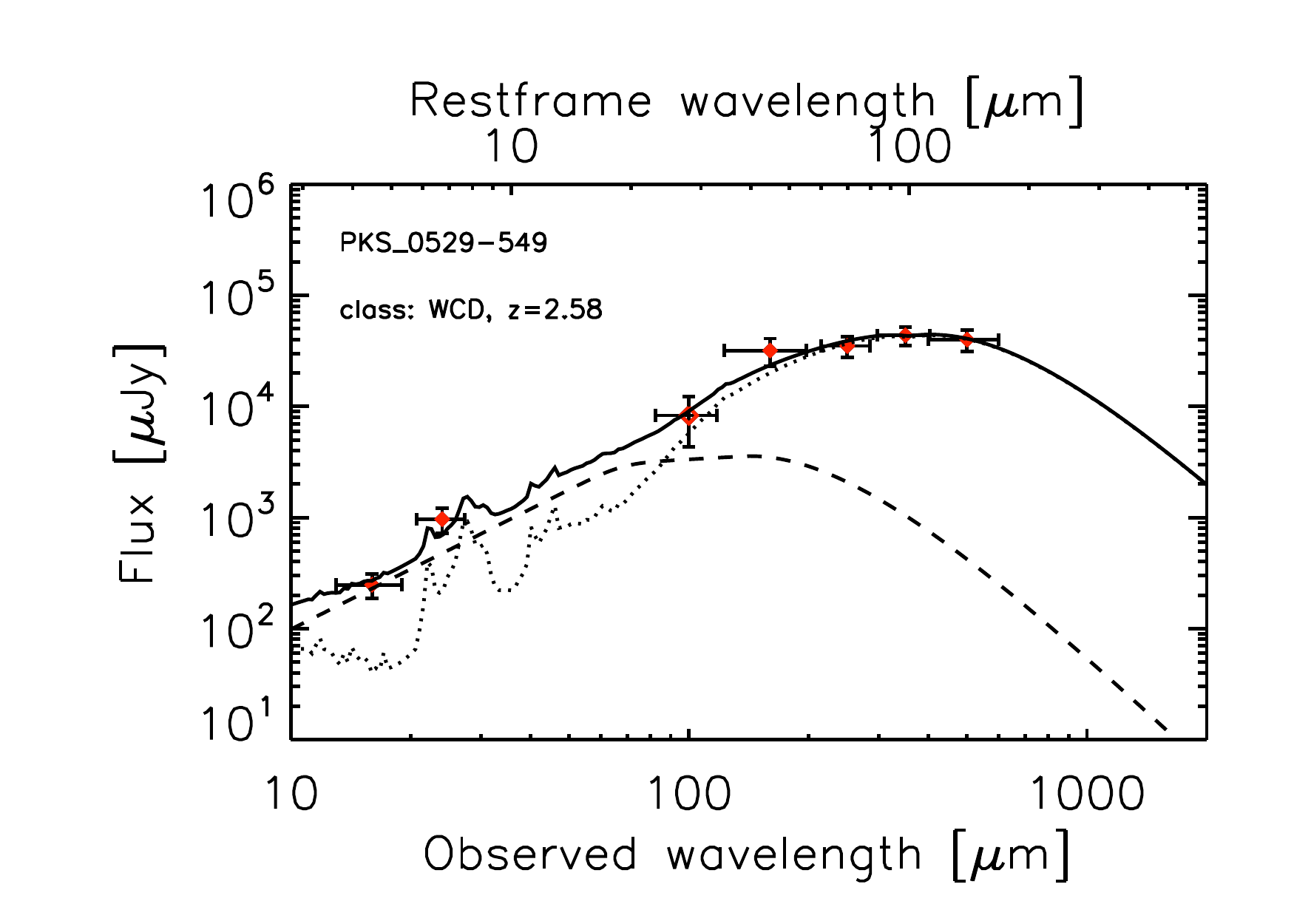}} \\
      24\mum & 100\mum & 160\mum & 250\mum & 350\mum & 500\mum \\
      \includegraphics[height=14mm,trim=70  -4 65 -10,clip=true]{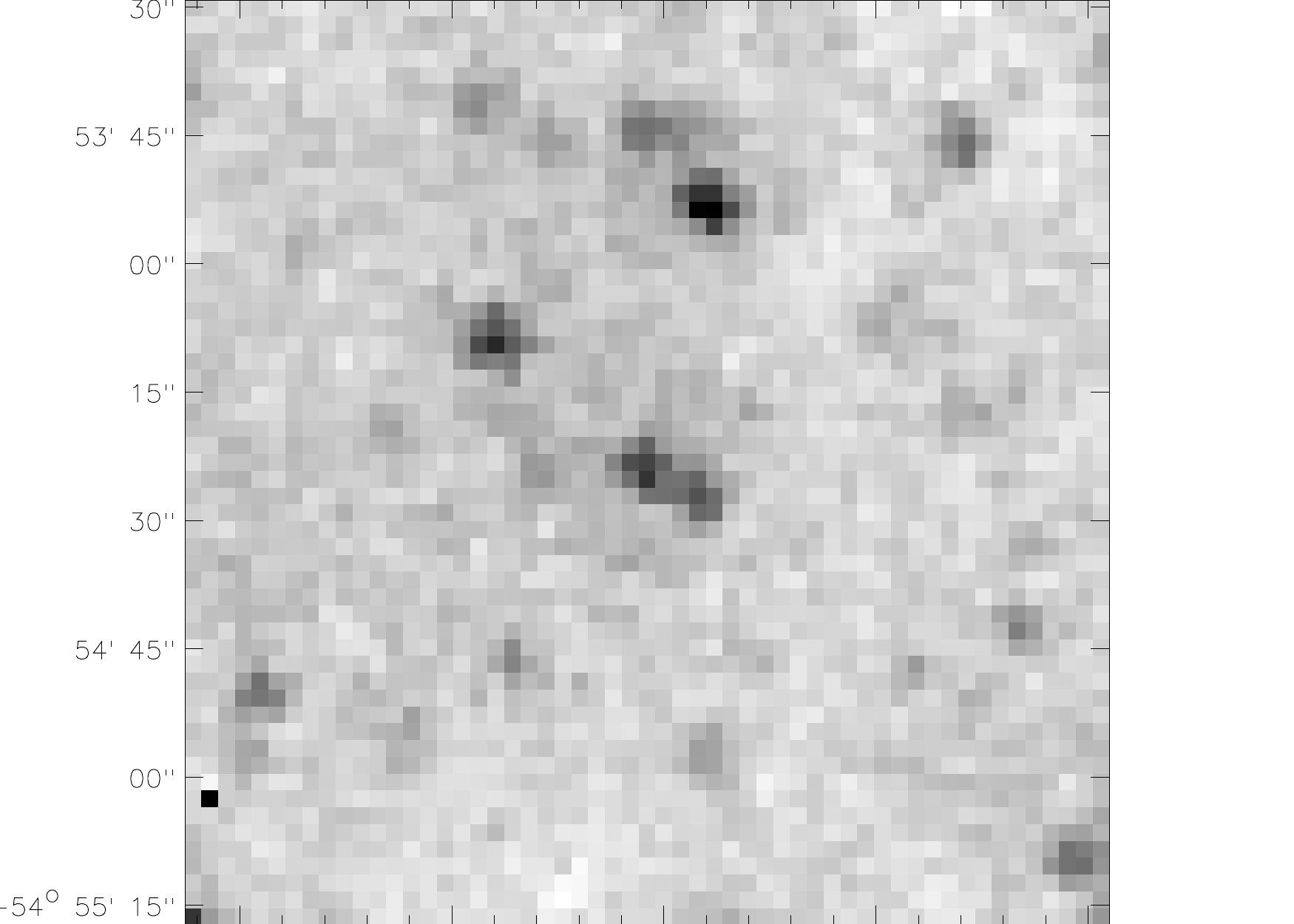} &
      \includegraphics[height=14mm,trim=70  -4 65 -10,clip=true]{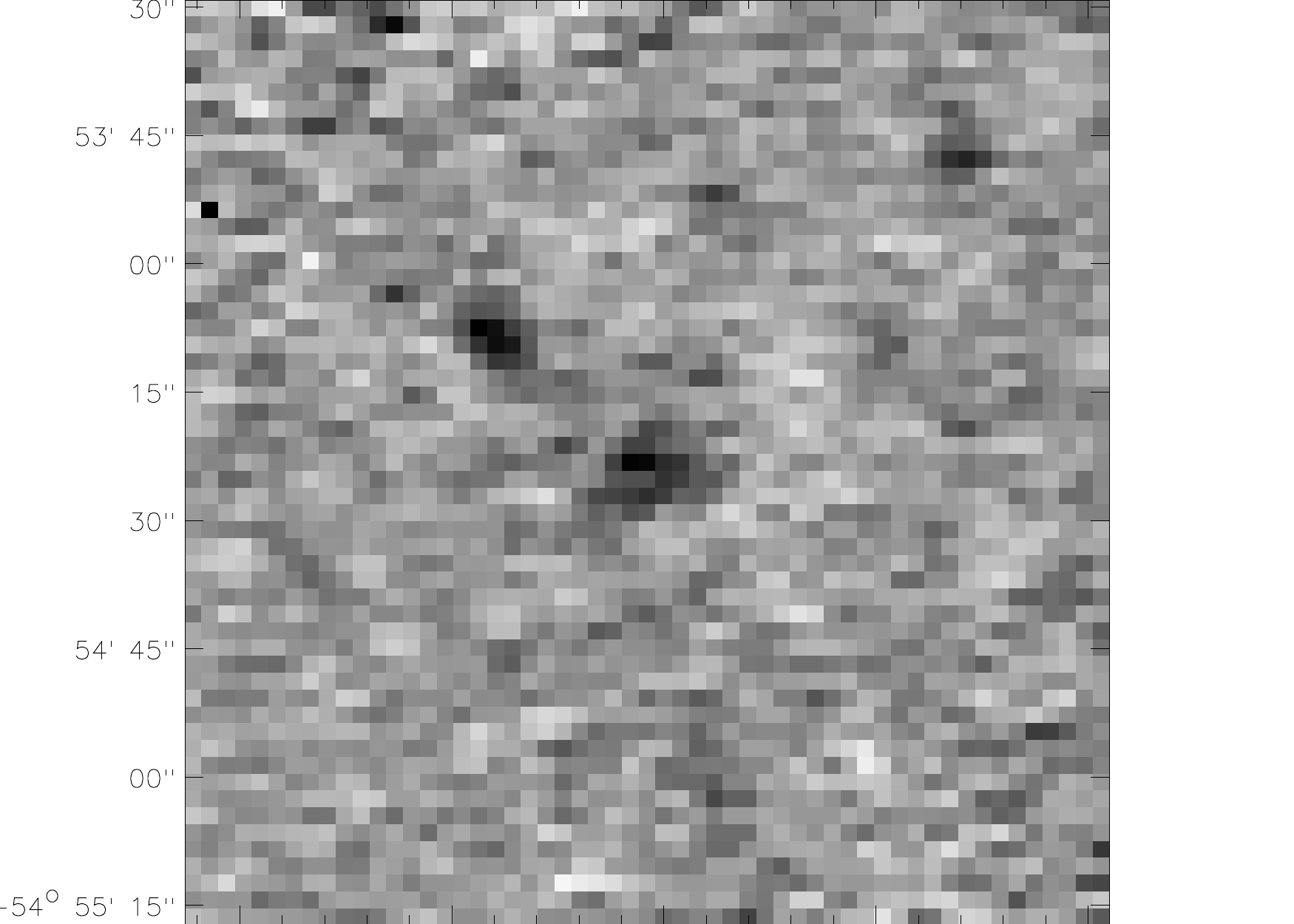} &
      \includegraphics[height=14mm,trim=70  -4 65 -10,clip=true]{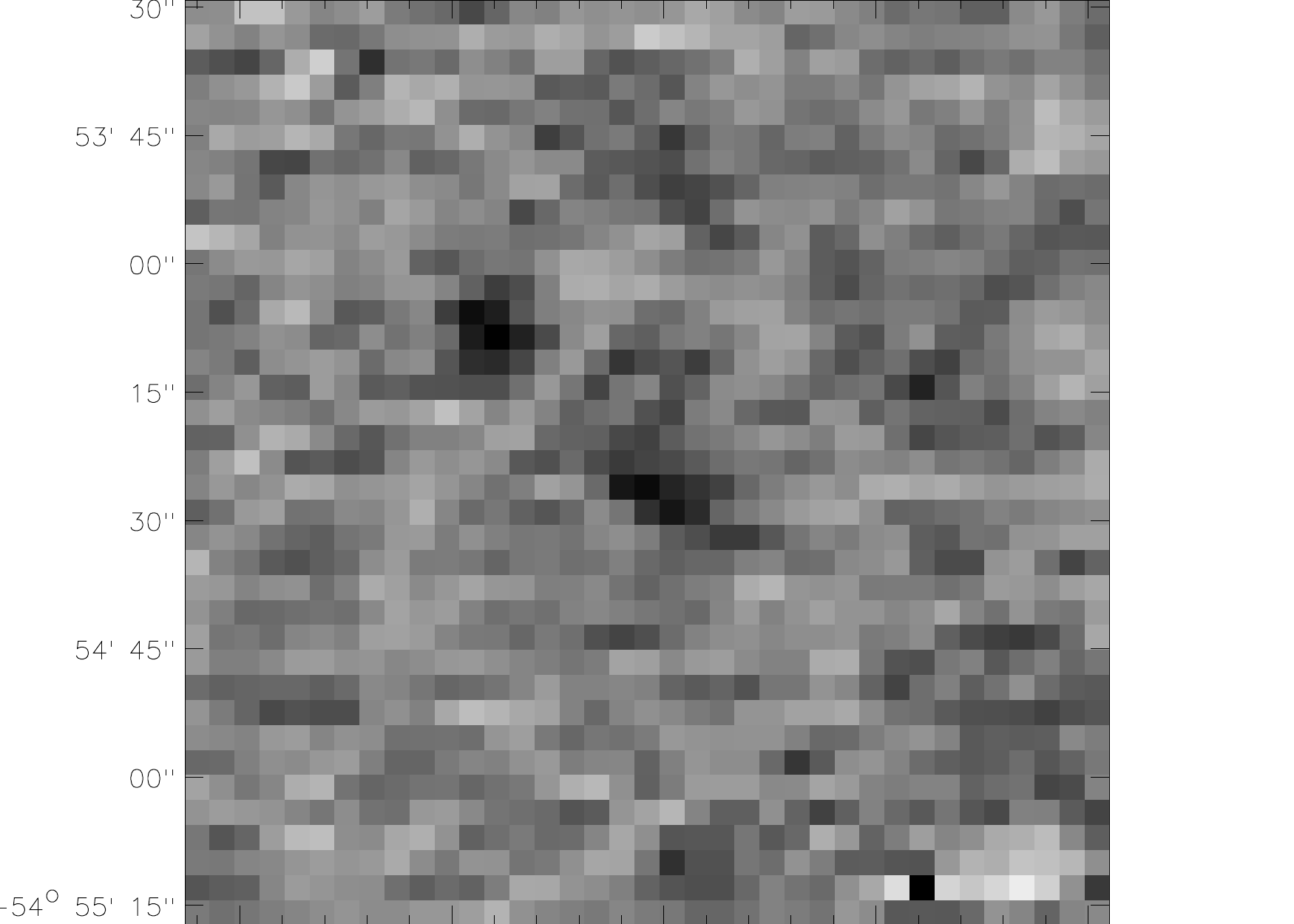} &
      \includegraphics[height=14mm,trim=70  -4 65 -10,clip=true]{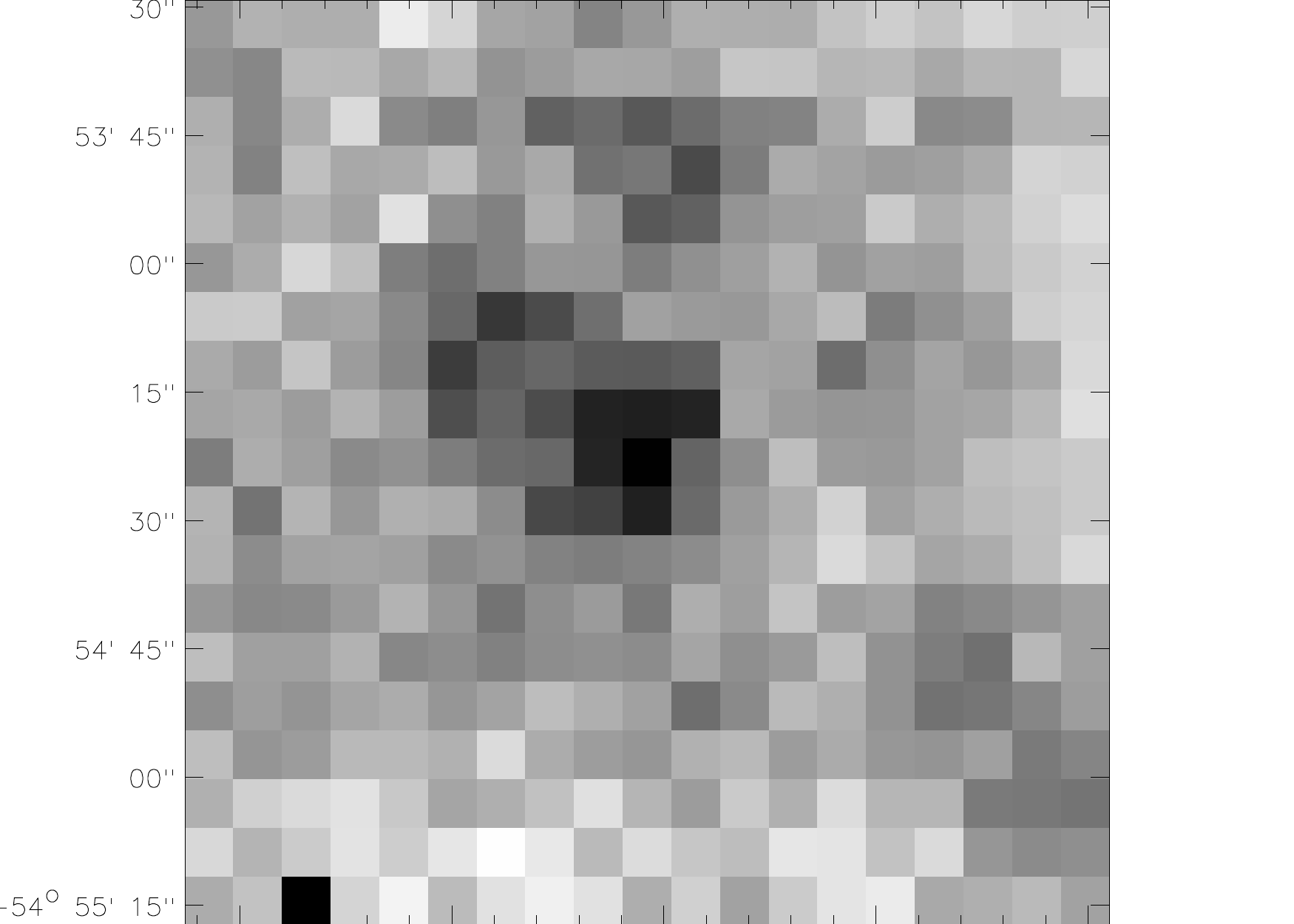}   &
      \includegraphics[height=14mm,trim=70  -4 65 -10,clip=true]{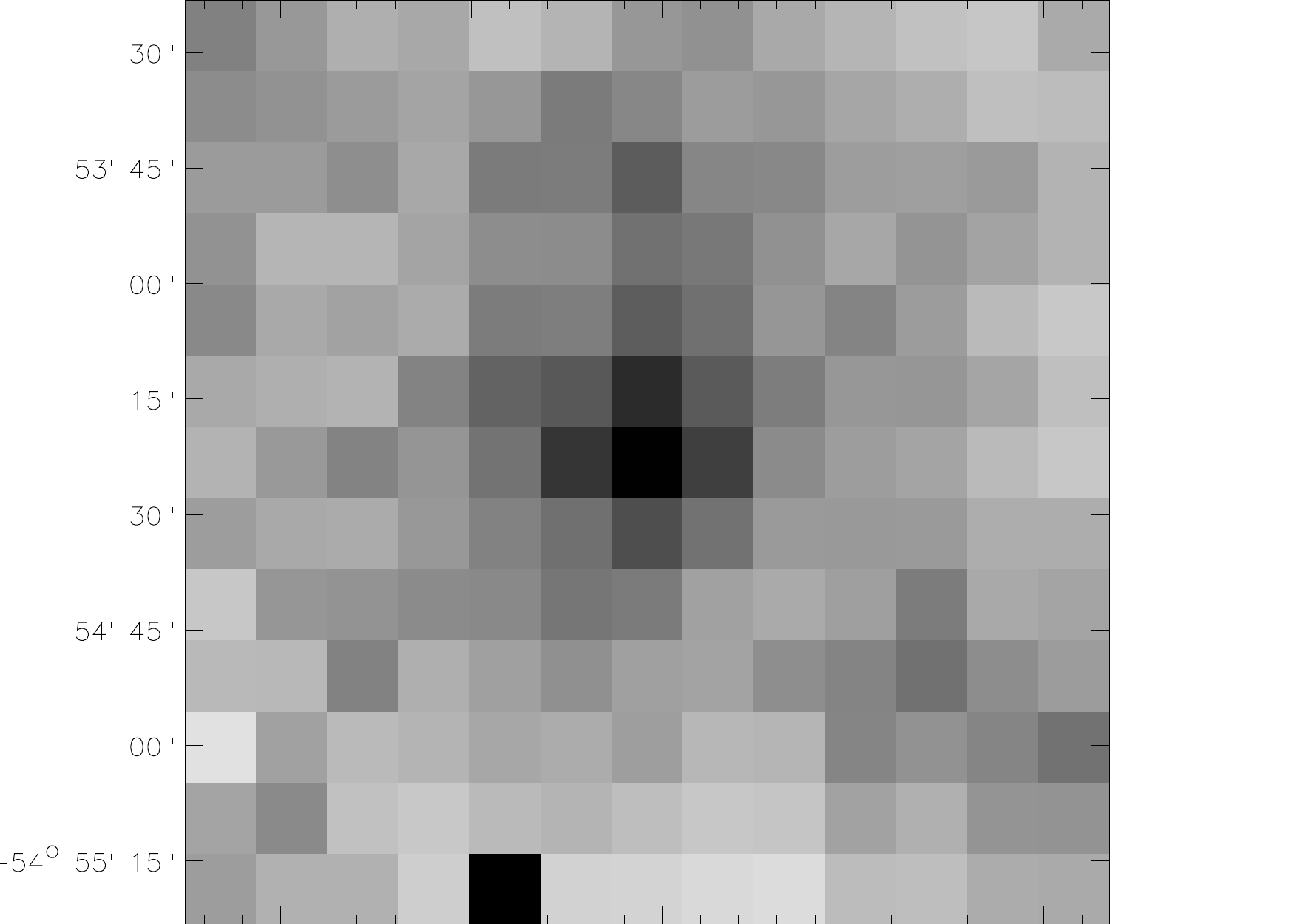}   &
      \includegraphics[height=14mm,trim=70  -4 65 -10,clip=true]{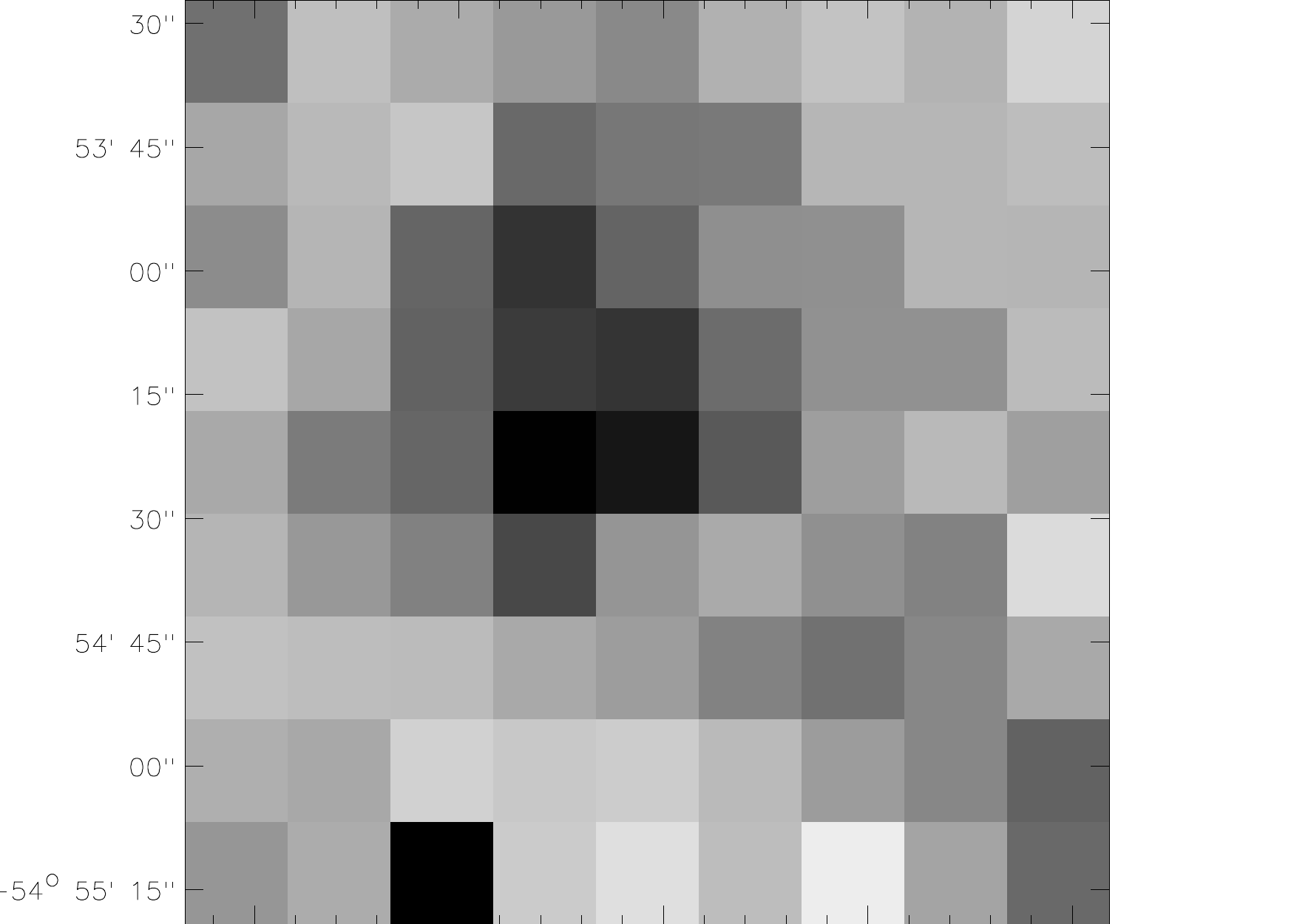} \\

      \multicolumn{6}{c}{\includegraphics[width=80mm,trim=30 0 20 0 ,clip=true]{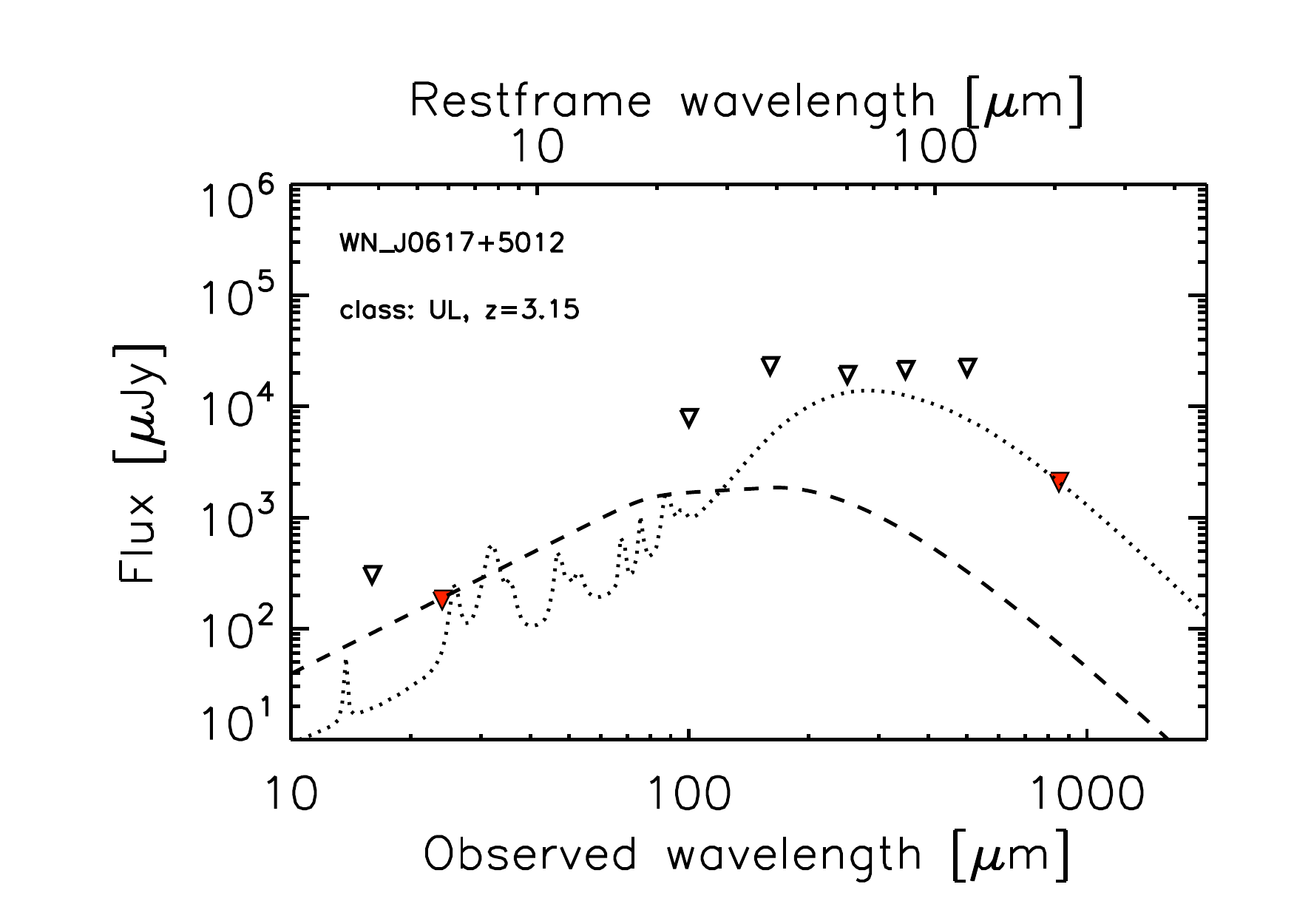}} \\
      24\mum & 100\mum & 160\mum & 250\mum & 350\mum & 500\mum \\
      \includegraphics[height=14mm,trim=70  -4 65 -10,clip=true]{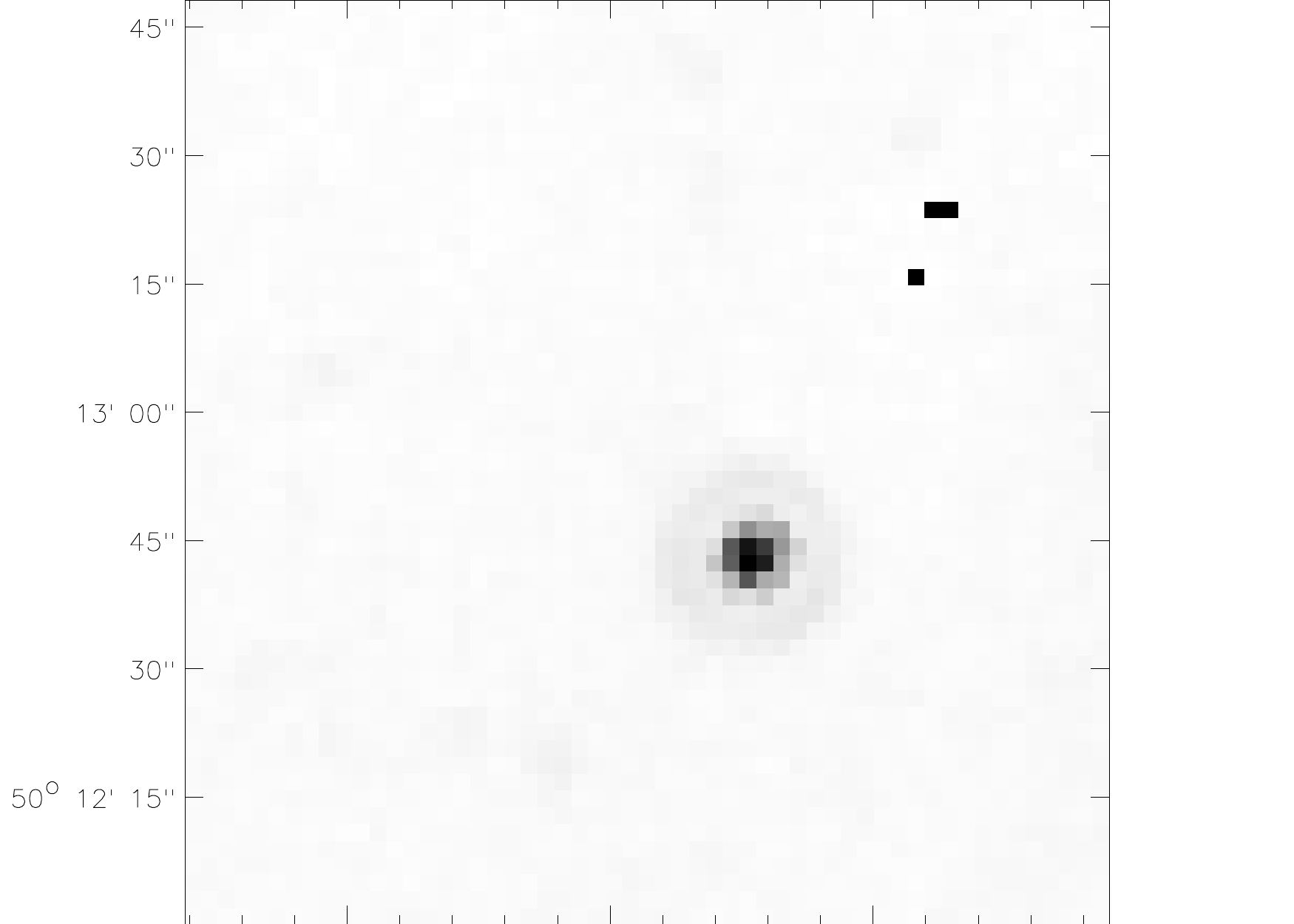} &
      \includegraphics[height=14mm,trim=70  -4 65 -10,clip=true]{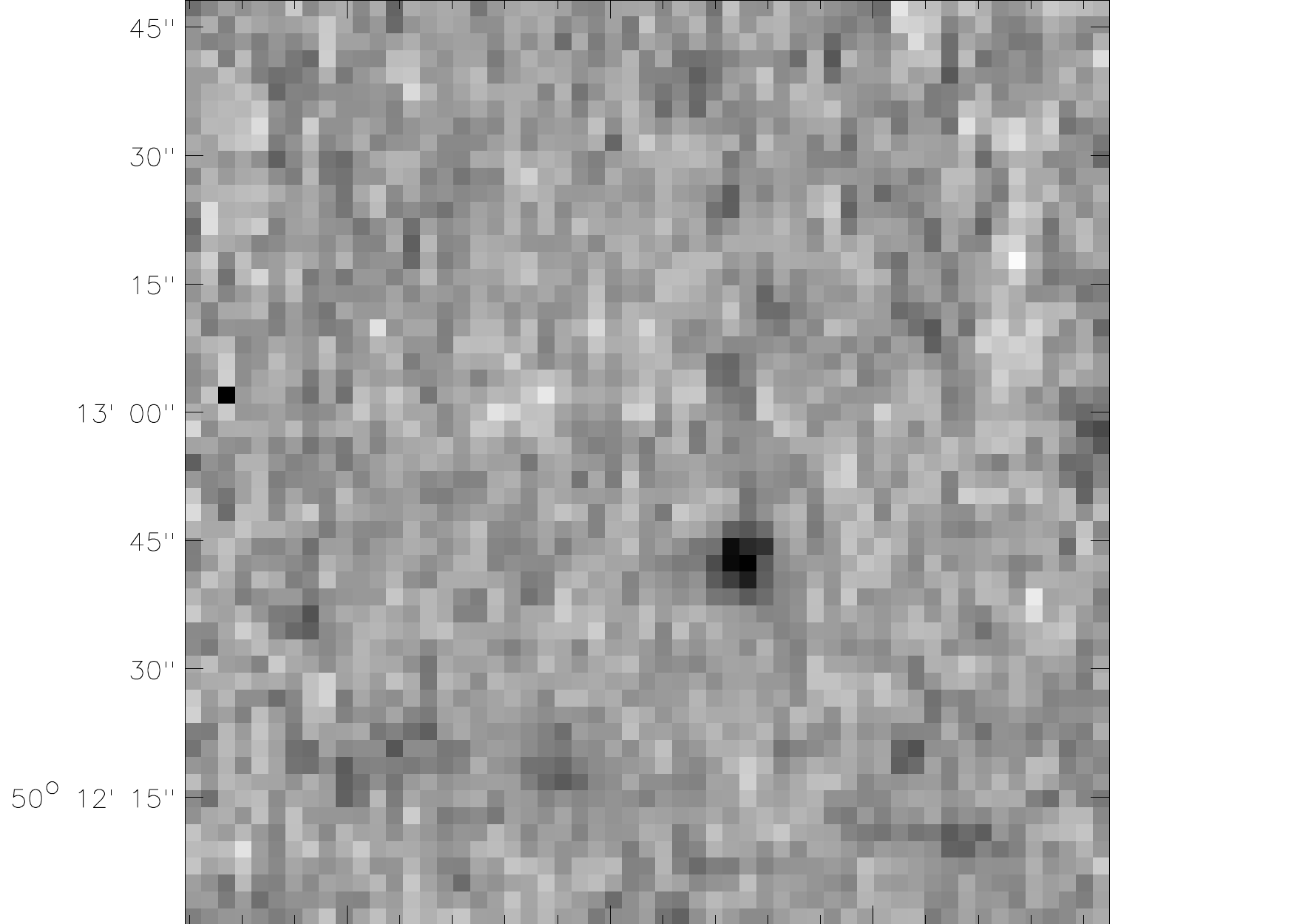} &
      \includegraphics[height=14mm,trim=70  -4 65 -10,clip=true]{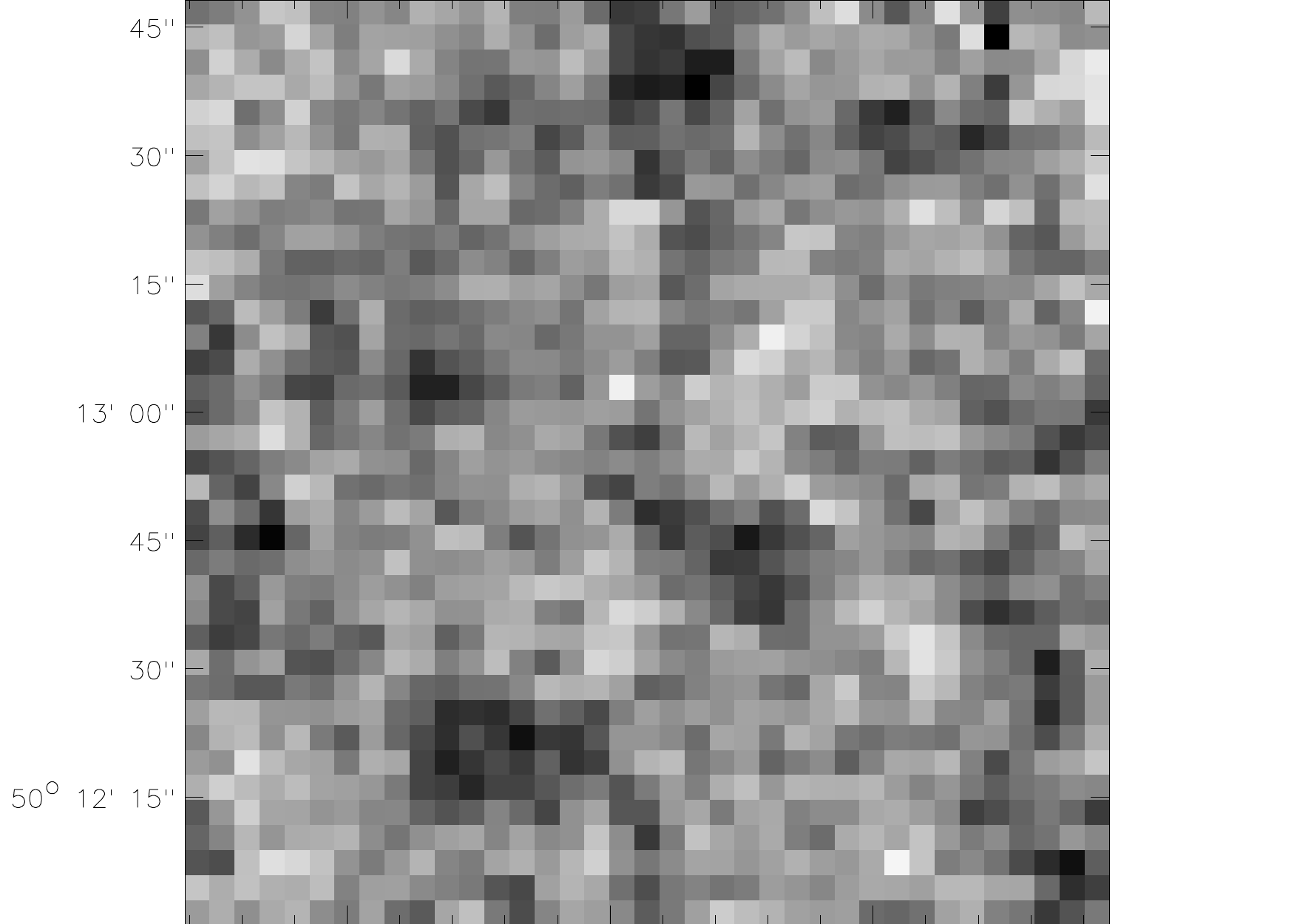} &
      \includegraphics[height=14mm,trim=70  -4 65 -10,clip=true]{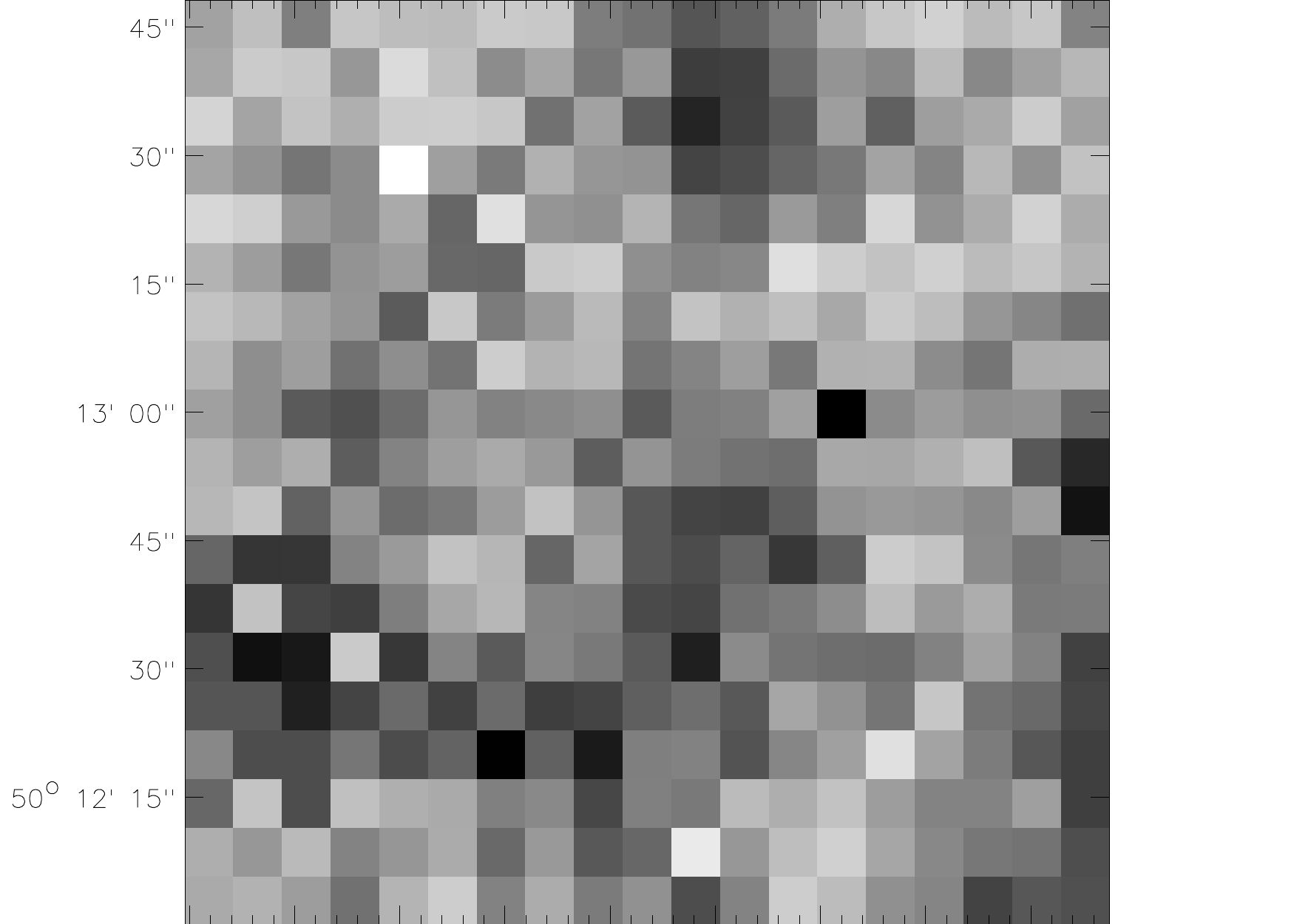}   &
      \includegraphics[height=14mm,trim=70  -4 65 -10,clip=true]{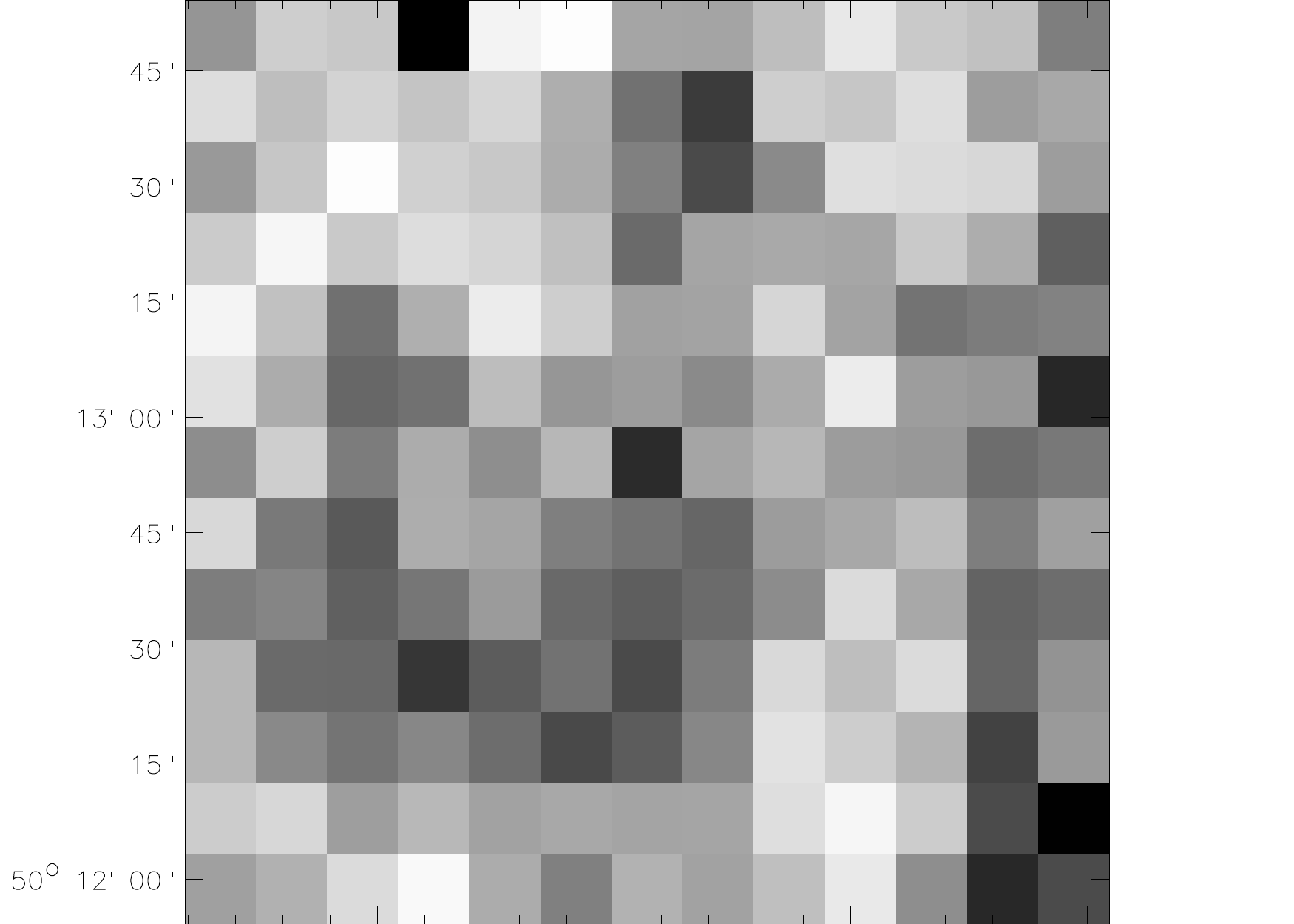}   &
      \includegraphics[height=14mm,trim=70  -4 65 -10,clip=true]{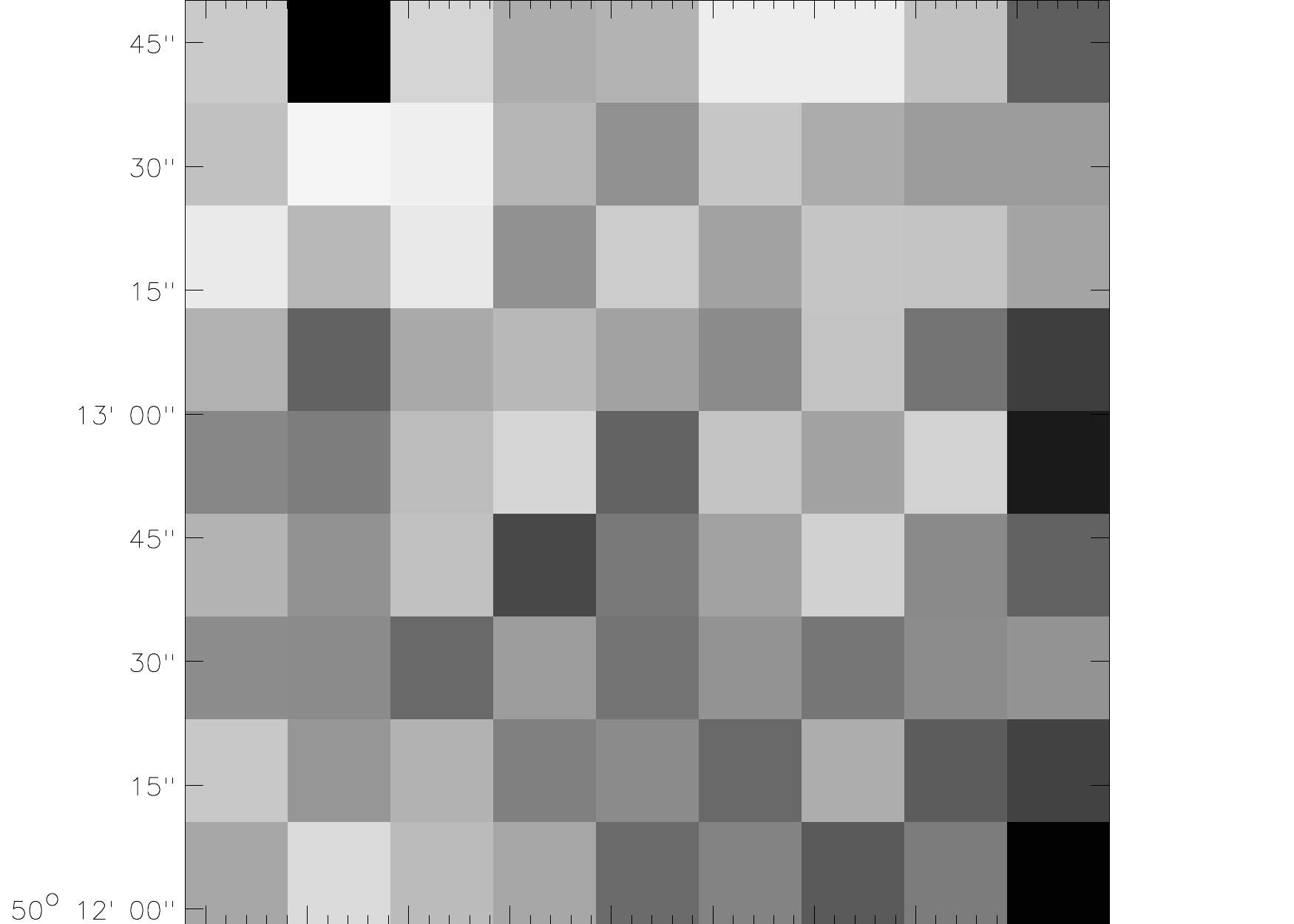} \\

      \multicolumn{6}{c}{\includegraphics[width=80mm,trim=30 0 20 0 ,clip=true]{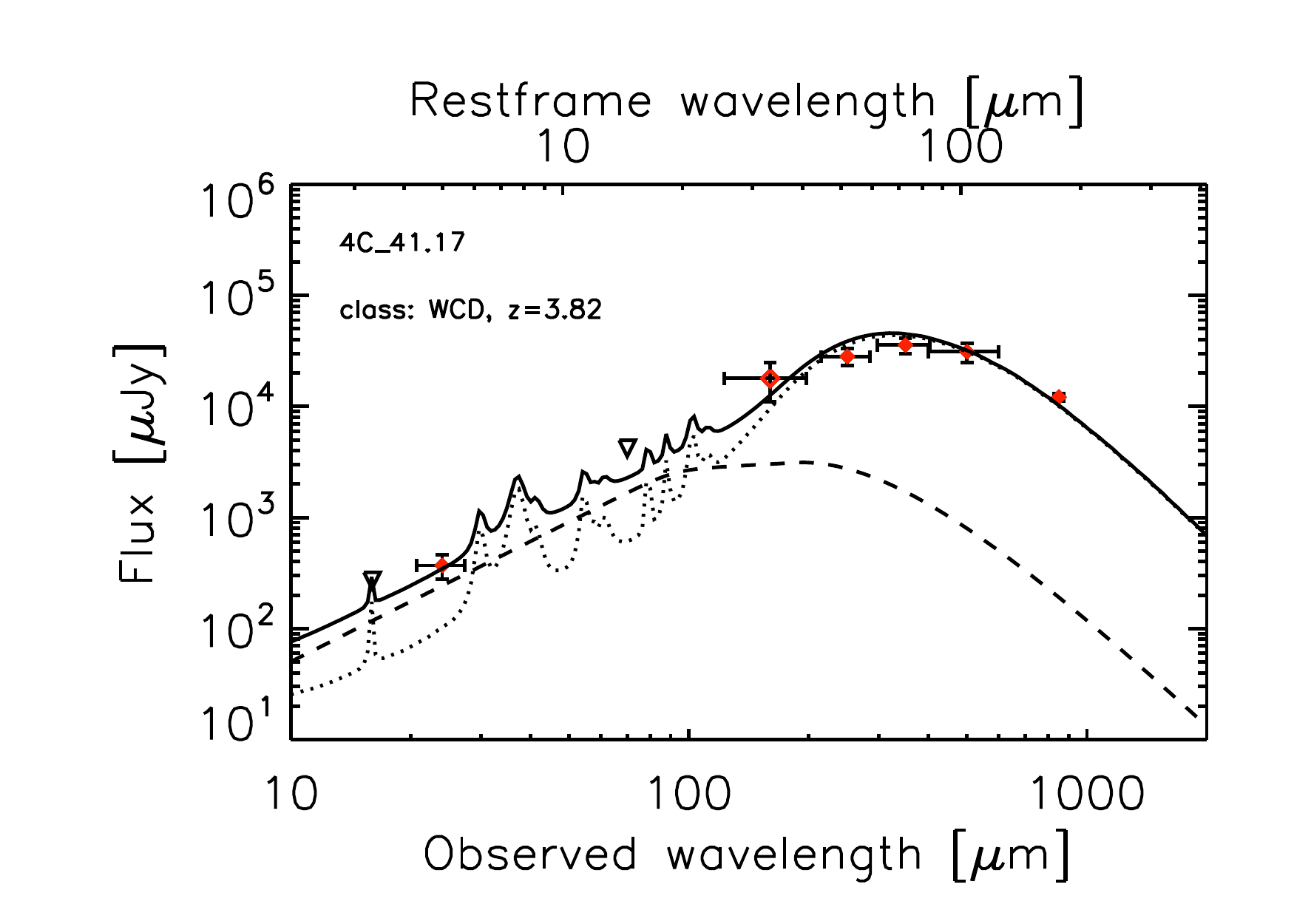}} \\
      24\mum & 70\mum & 160\mum & 250\mum & 350\mum & 500\mum \\
\end{tabular}
          \end{minipage}
  \end{centering}
\end{figure*}

\clearpage

\begin{figure*}[h]
  \begin{centering}
    \begin{minipage}[c]{0.46\linewidth}
    \begin{tabular}{c@{}c@{}c@{}c@{}c@{}c@{}}

      \multicolumn{6}{c}{\includegraphics[width=80mm,trim=30 0 20 0 ,clip=true]{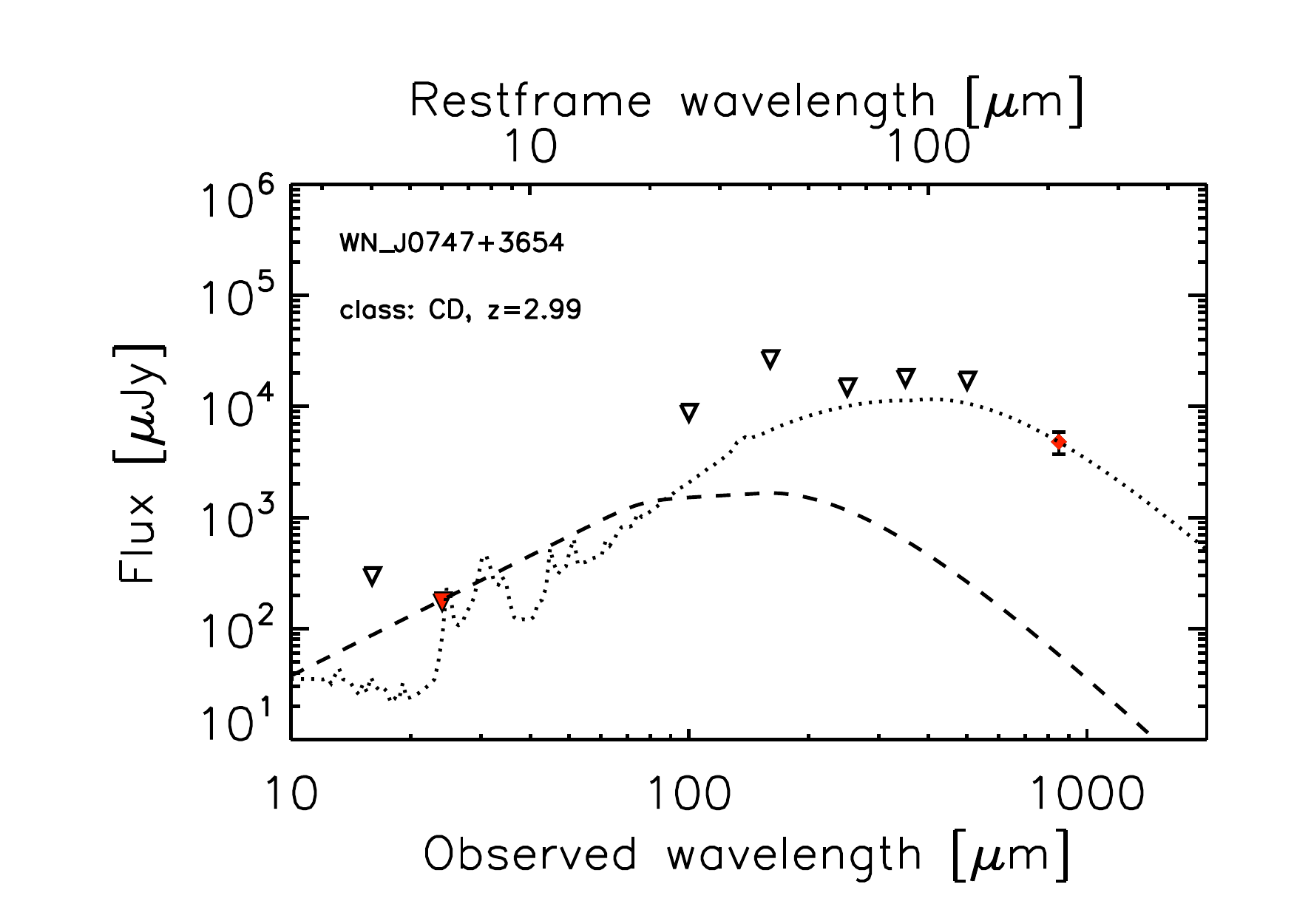}} \\
      24\mum & 100\mum & 160\mum & 250\mum & 350\mum & 500\mum \\
      \includegraphics[height=14mm,trim=70  -4 65 -10,clip=true]{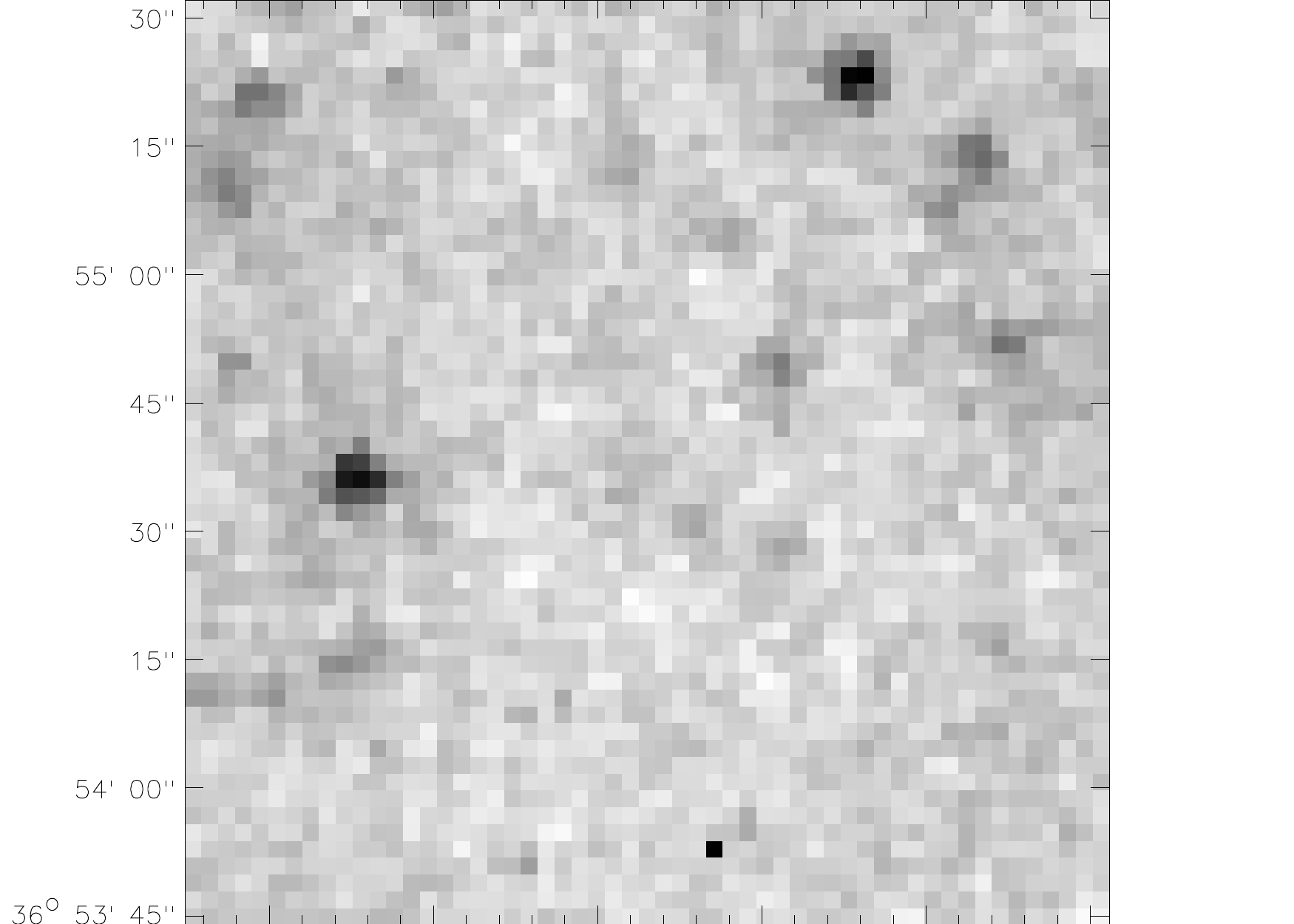} &
      \includegraphics[height=14mm,trim=70  -4 65 -10,clip=true]{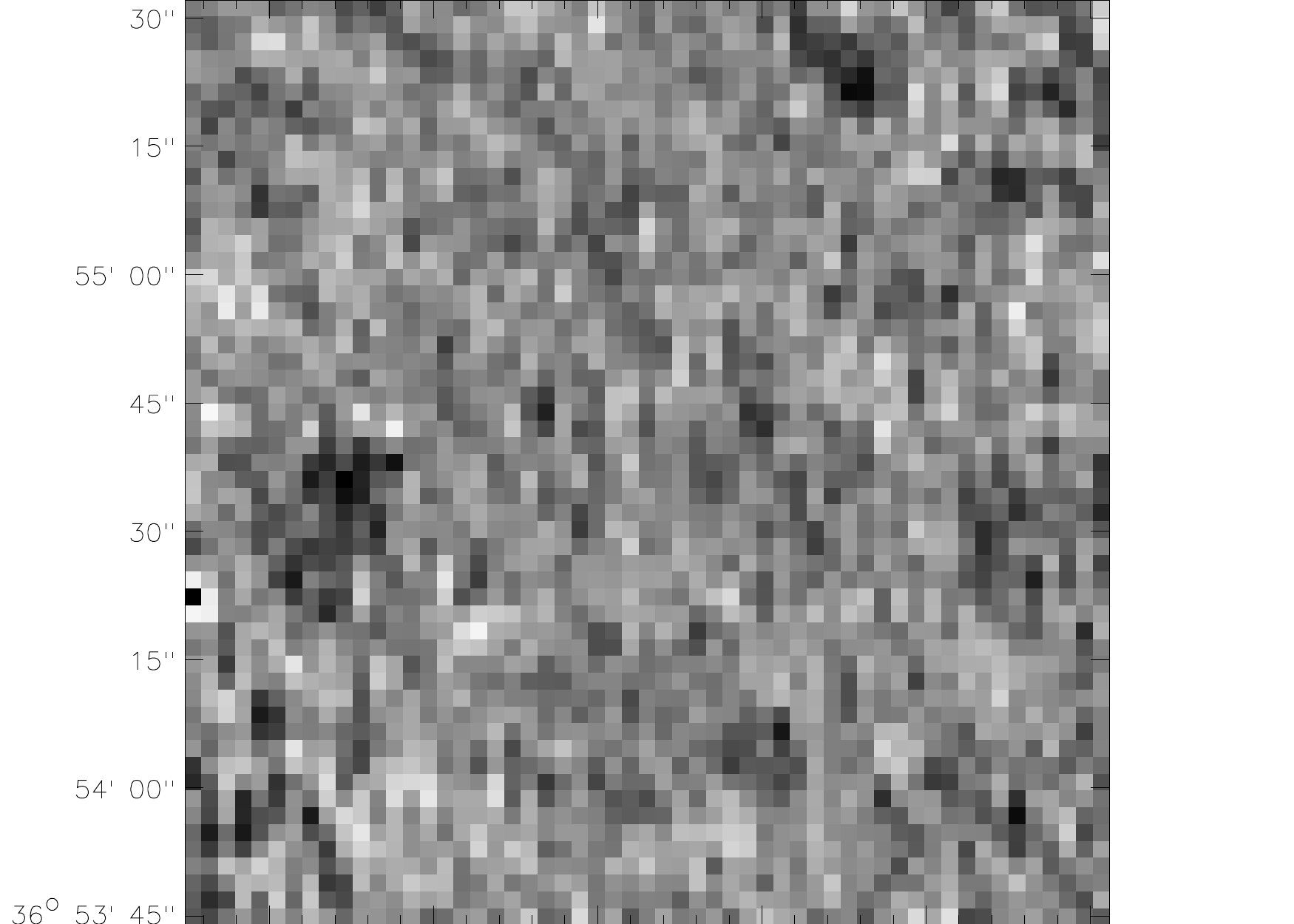} &
      \includegraphics[height=14mm,trim=70  -4 65 -10,clip=true]{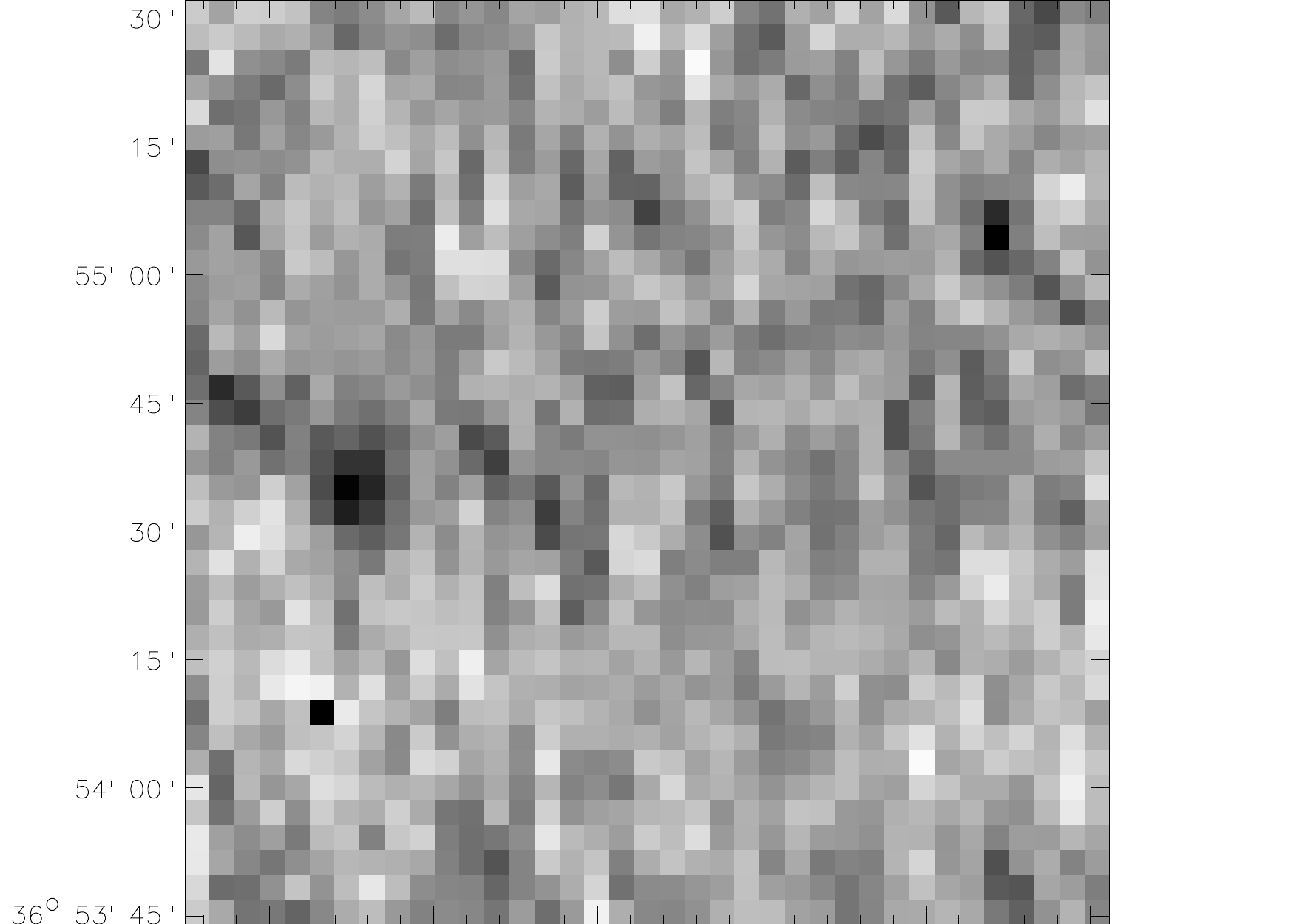} &
      \includegraphics[height=14mm,trim=70  -4 65 -10,clip=true]{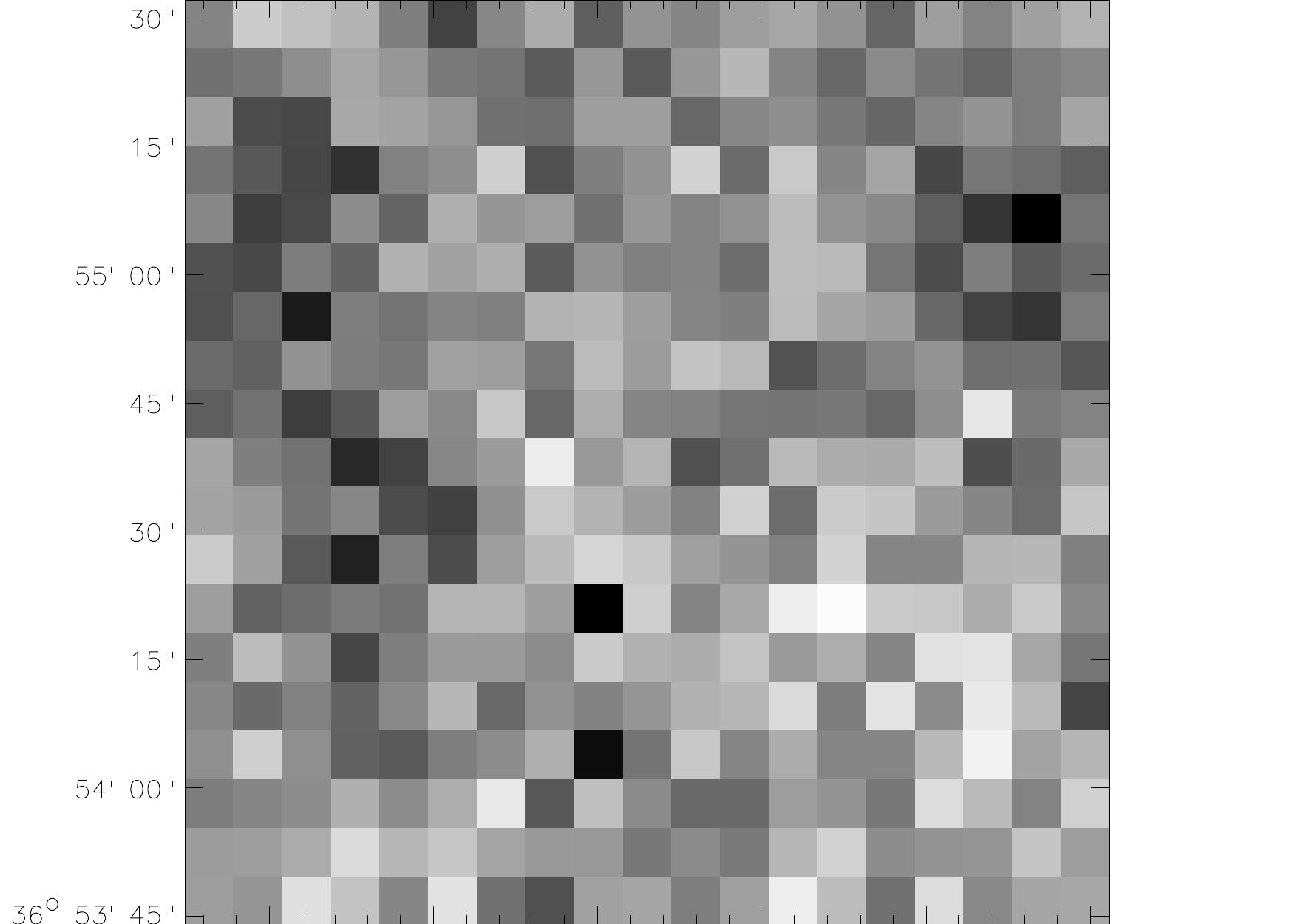}   &
      \includegraphics[height=14mm,trim=70  -4 65 -10,clip=true]{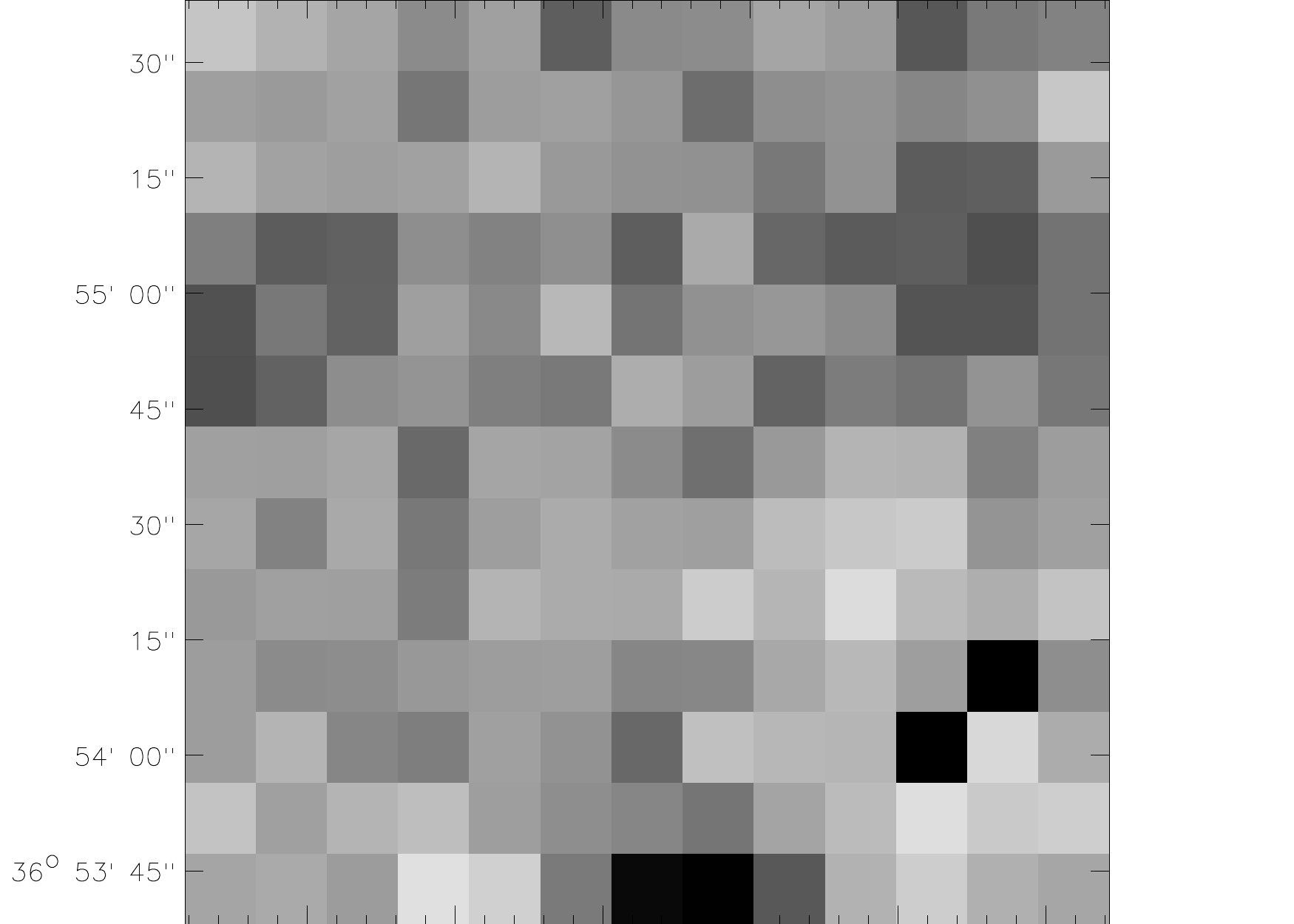}   &
      \includegraphics[height=14mm,trim=70  -4 65 -10,clip=true]{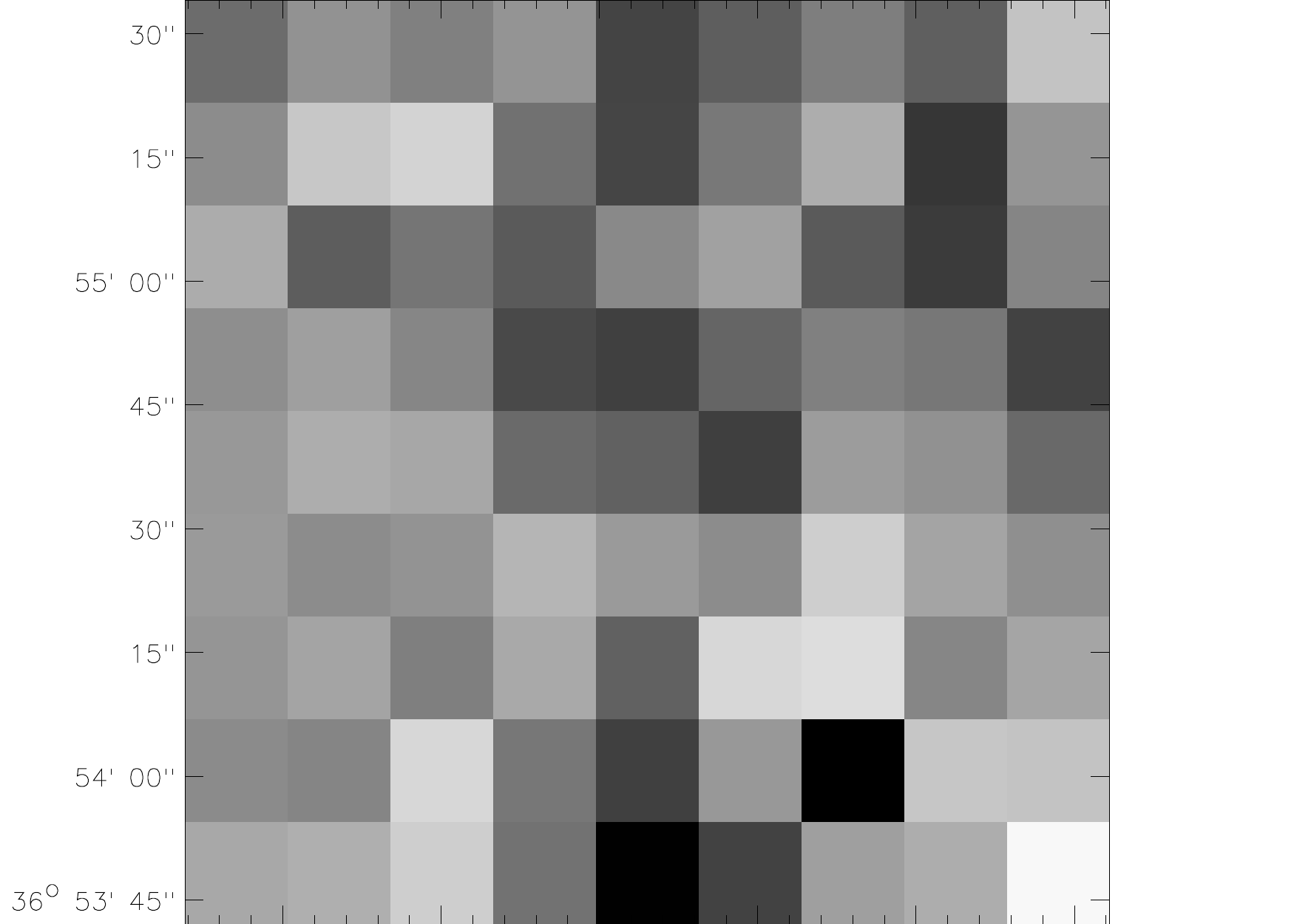} \\

      \multicolumn{6}{c}{\includegraphics[width=80mm,trim=30 0 20 0 ,clip=true]{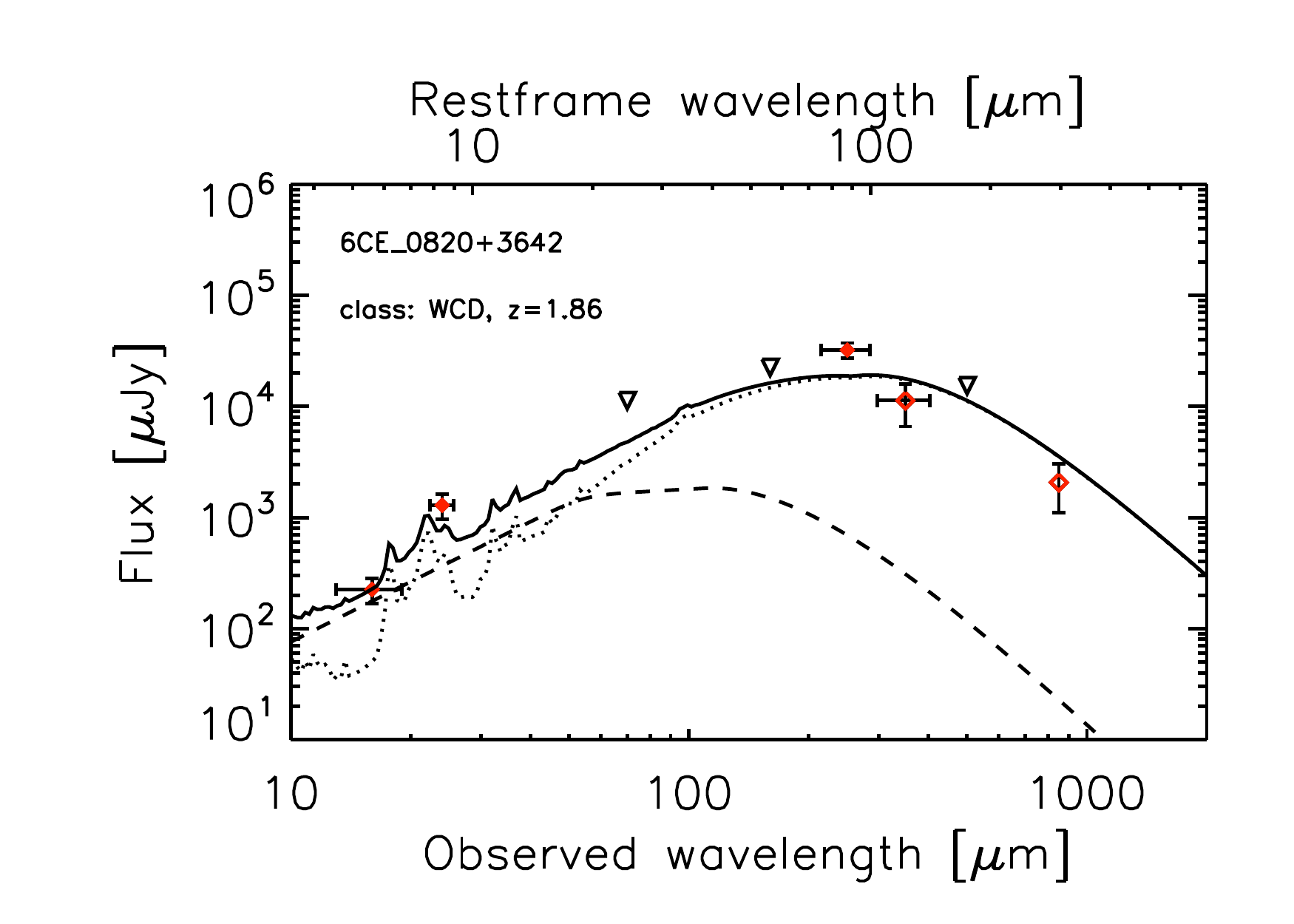}} \\
      24\mum & 70\mum & 160\mum & 250\mum & 350\mum & 500\mum \\
      \includegraphics[height=14mm,trim=70  -4 65 -10,clip=true]{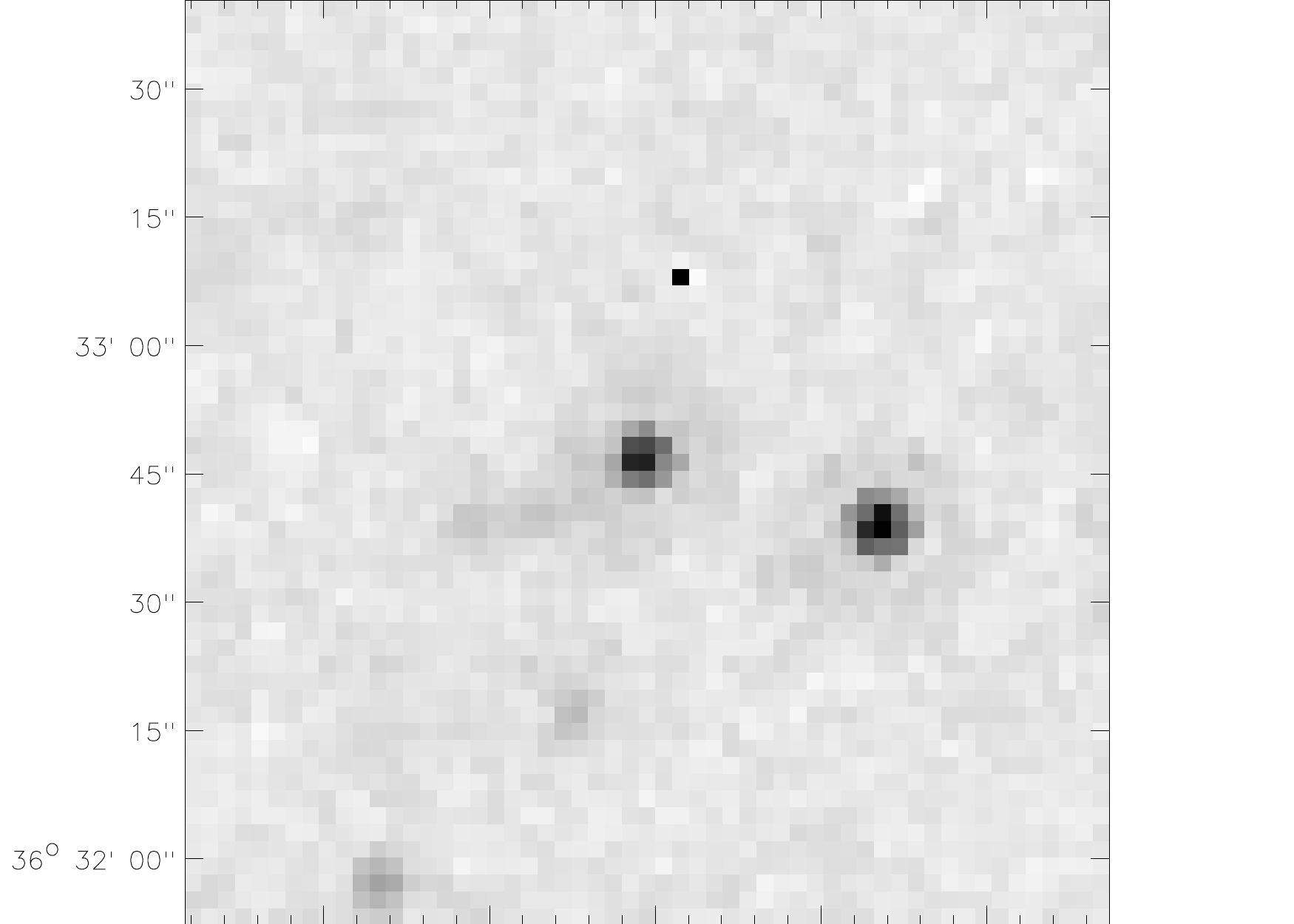} &
      \includegraphics[height=14mm,trim=70  -4 65 -10,clip=true]{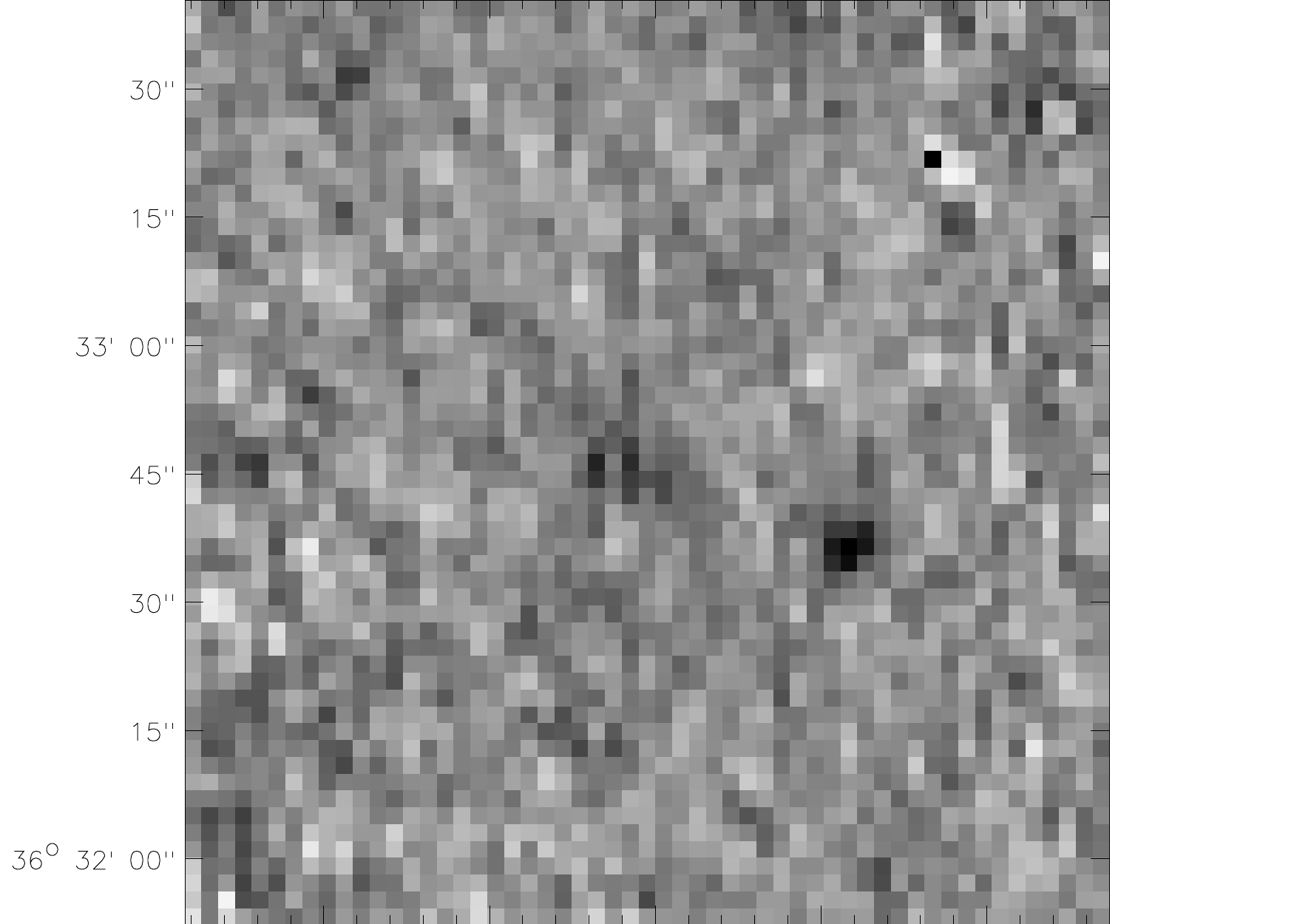} &
      \includegraphics[height=14mm,trim=70  -4 65 -10,clip=true]{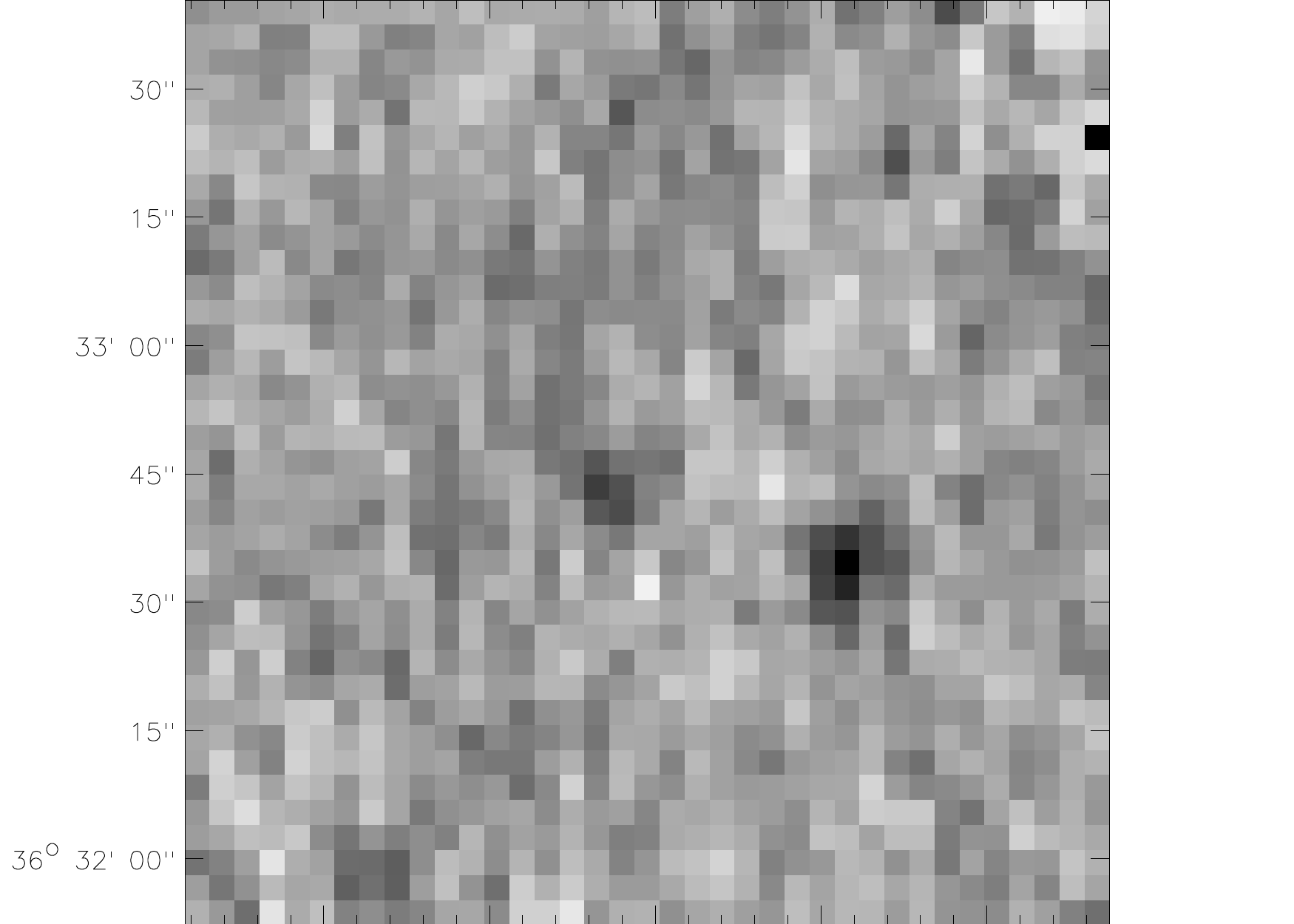} &
      \includegraphics[height=14mm,trim=70  -4 65 -10,clip=true]{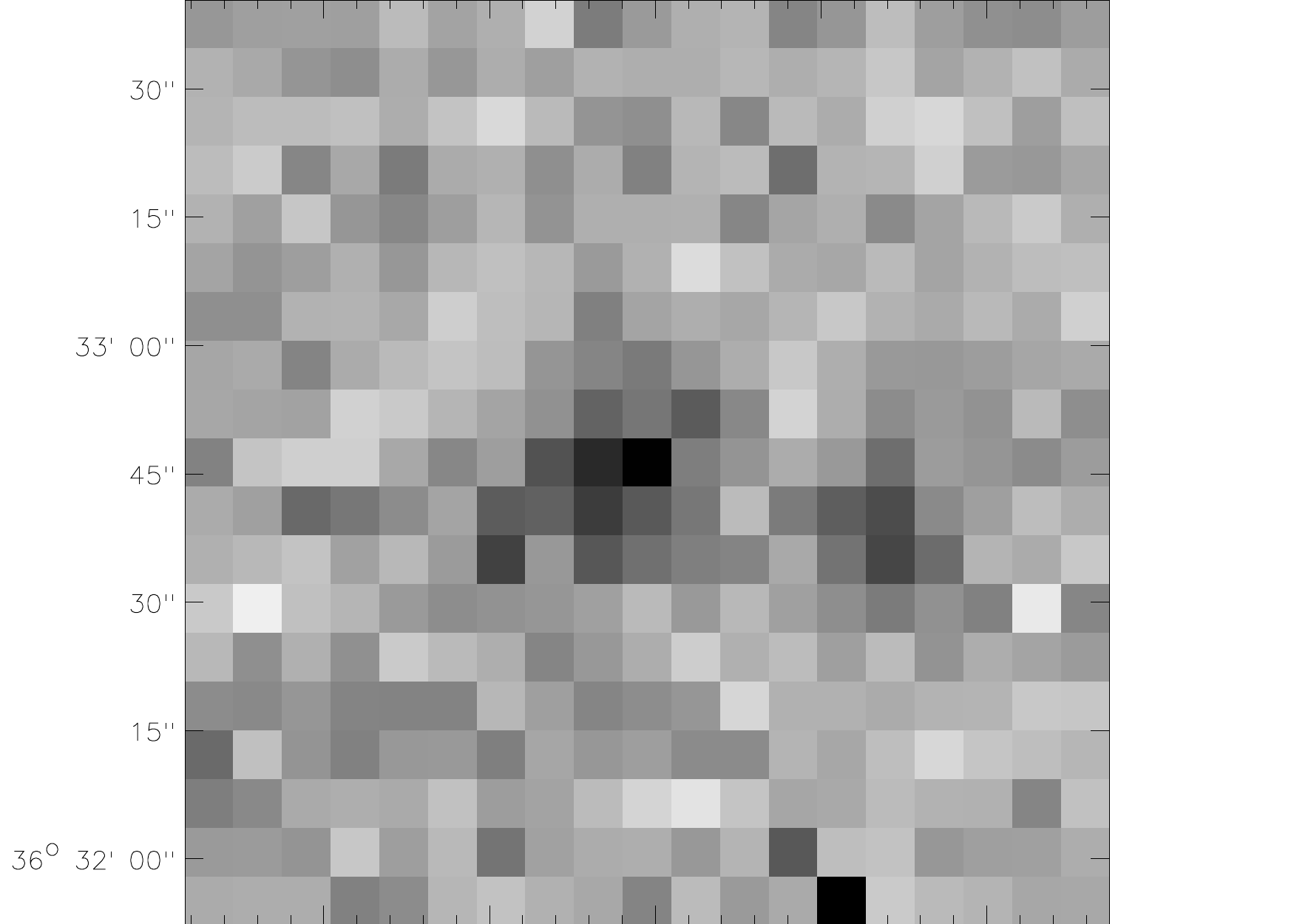}   &
      \includegraphics[height=14mm,trim=70  -4 65 -10,clip=true]{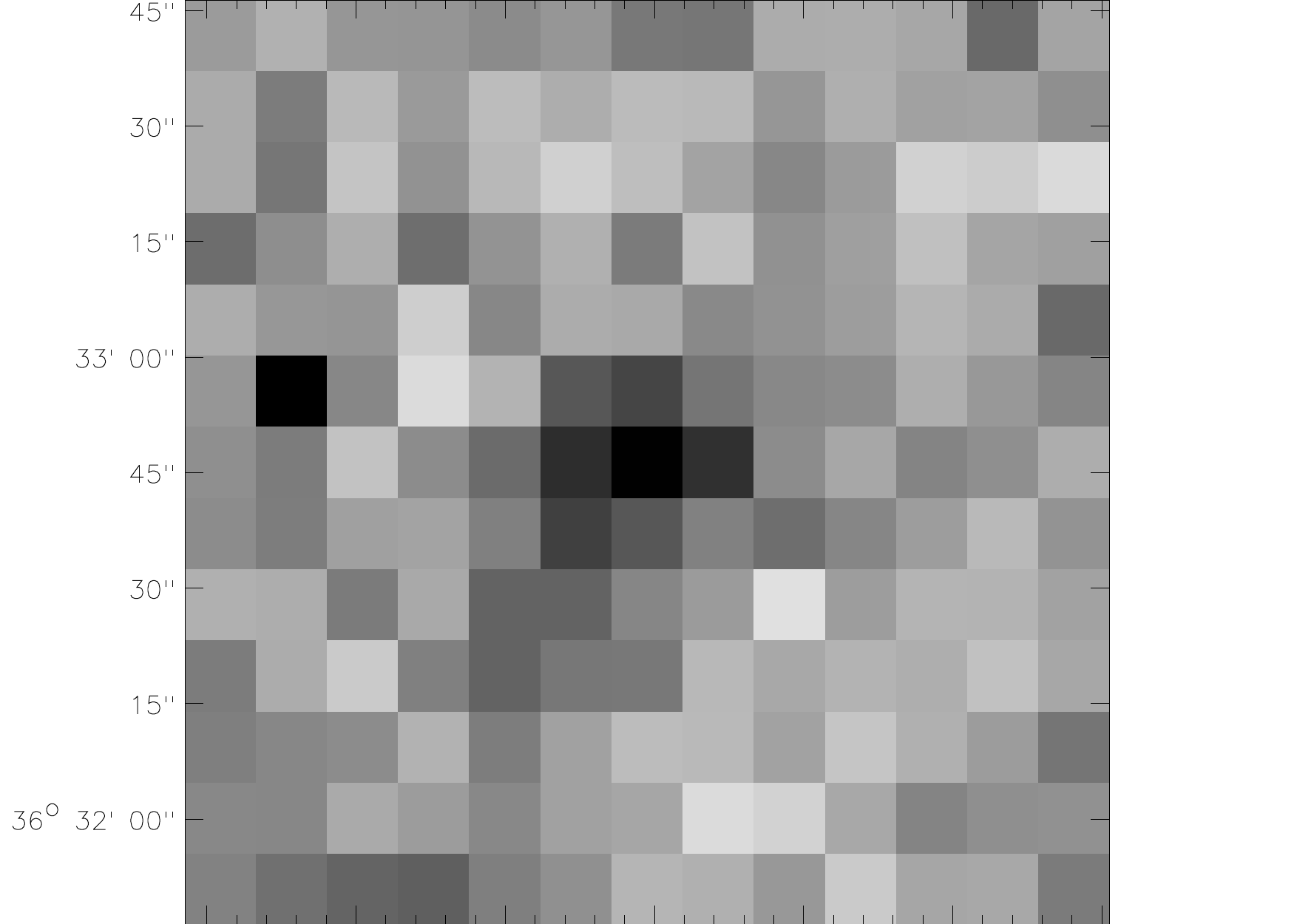}   &
      \includegraphics[height=14mm,trim=70  -4 65 -10,clip=true]{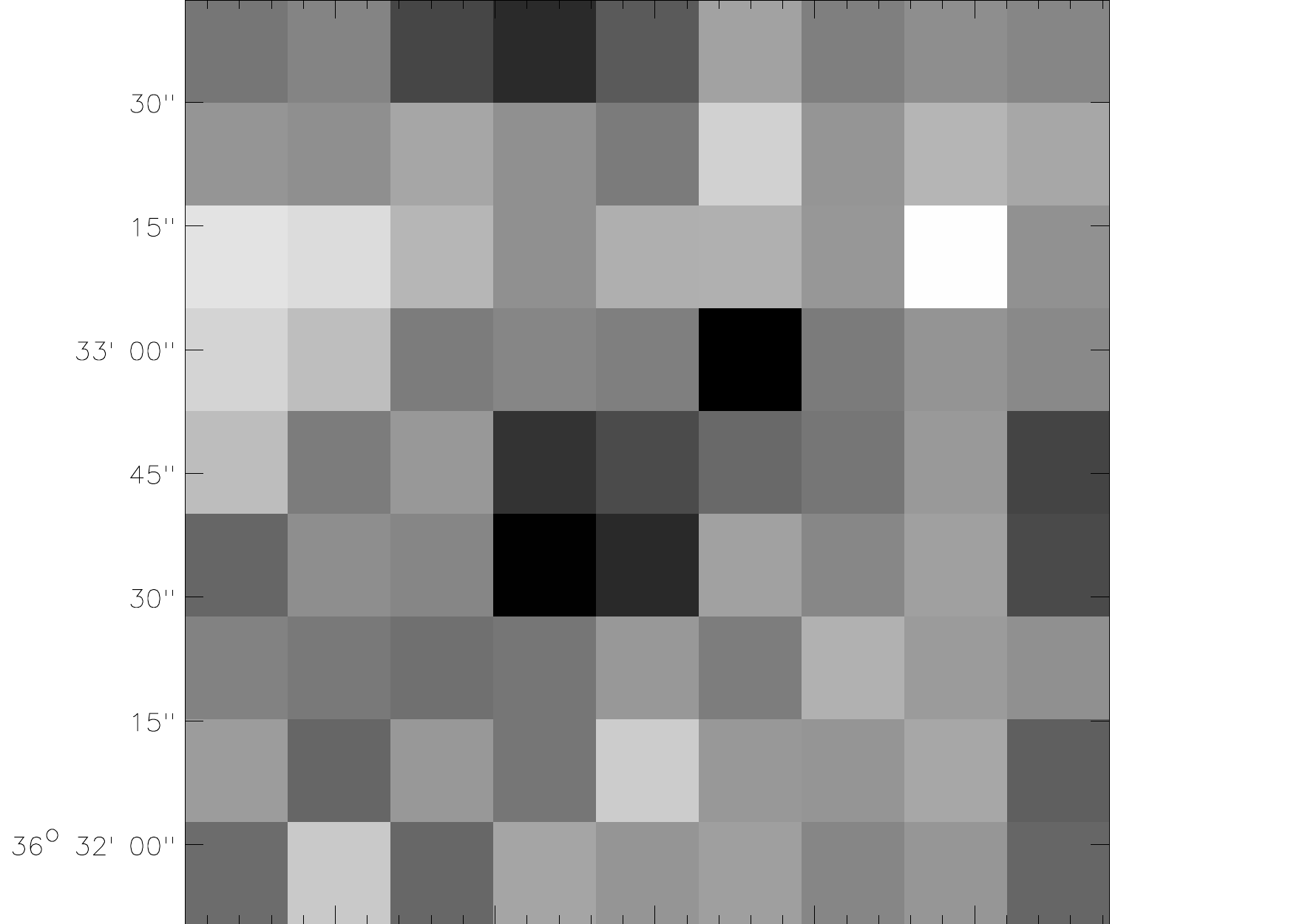} \\

      \multicolumn{6}{c}{\includegraphics[width=80mm,trim=30 0 20 0 ,clip=true]{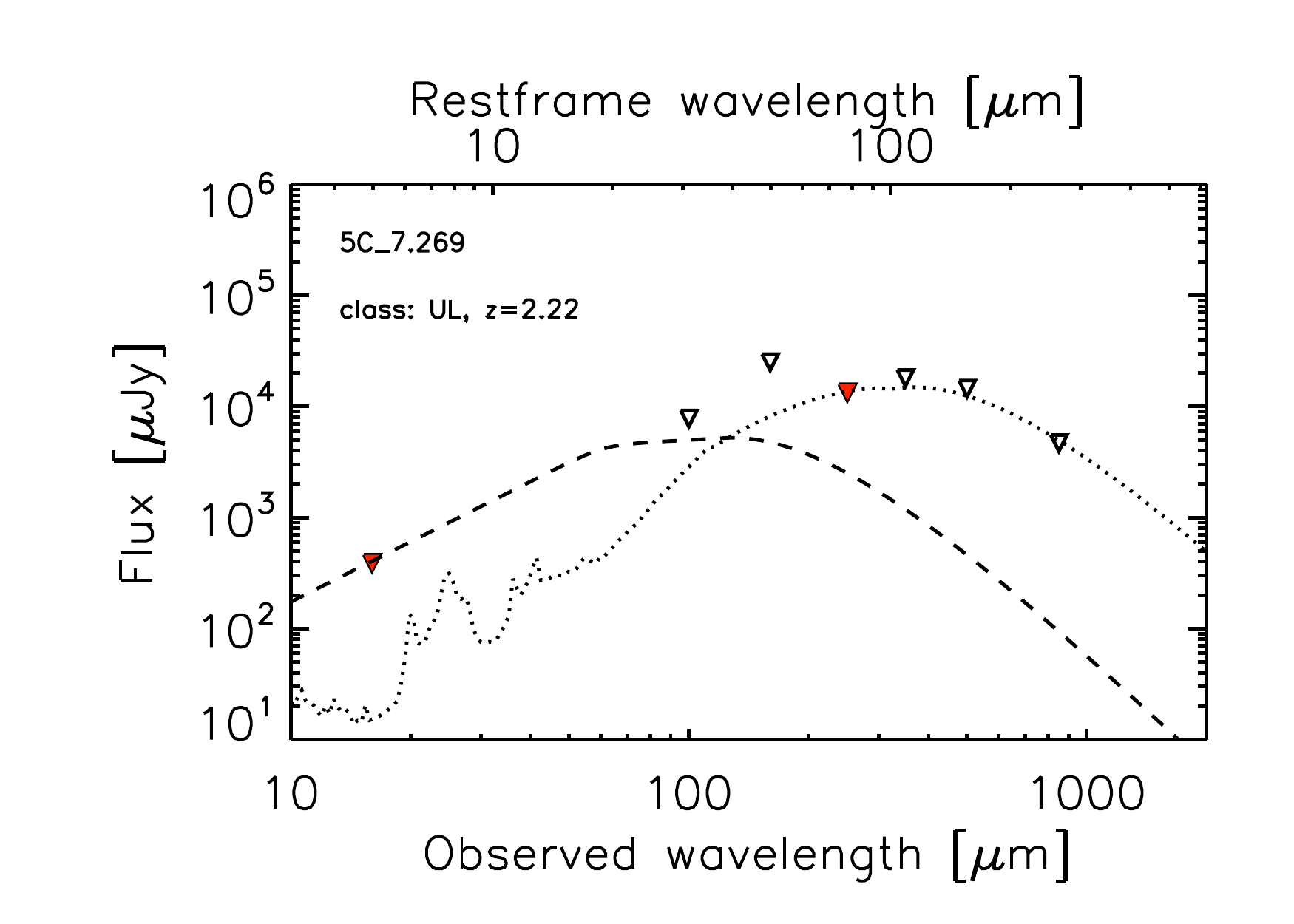}} \\
      24\mum & 100\mum & 160\mum & 250\mum & 350\mum & 500\mum \\
\end{tabular}
\end{minipage}
\begin{minipage}[c]{0.46\linewidth} 
    \begin{tabular}{c@{}c@{}c@{}c@{}c@{}c@{}}

      \multicolumn{6}{c}{\includegraphics[width=80mm,trim=30 0 20 0 ,clip=true]{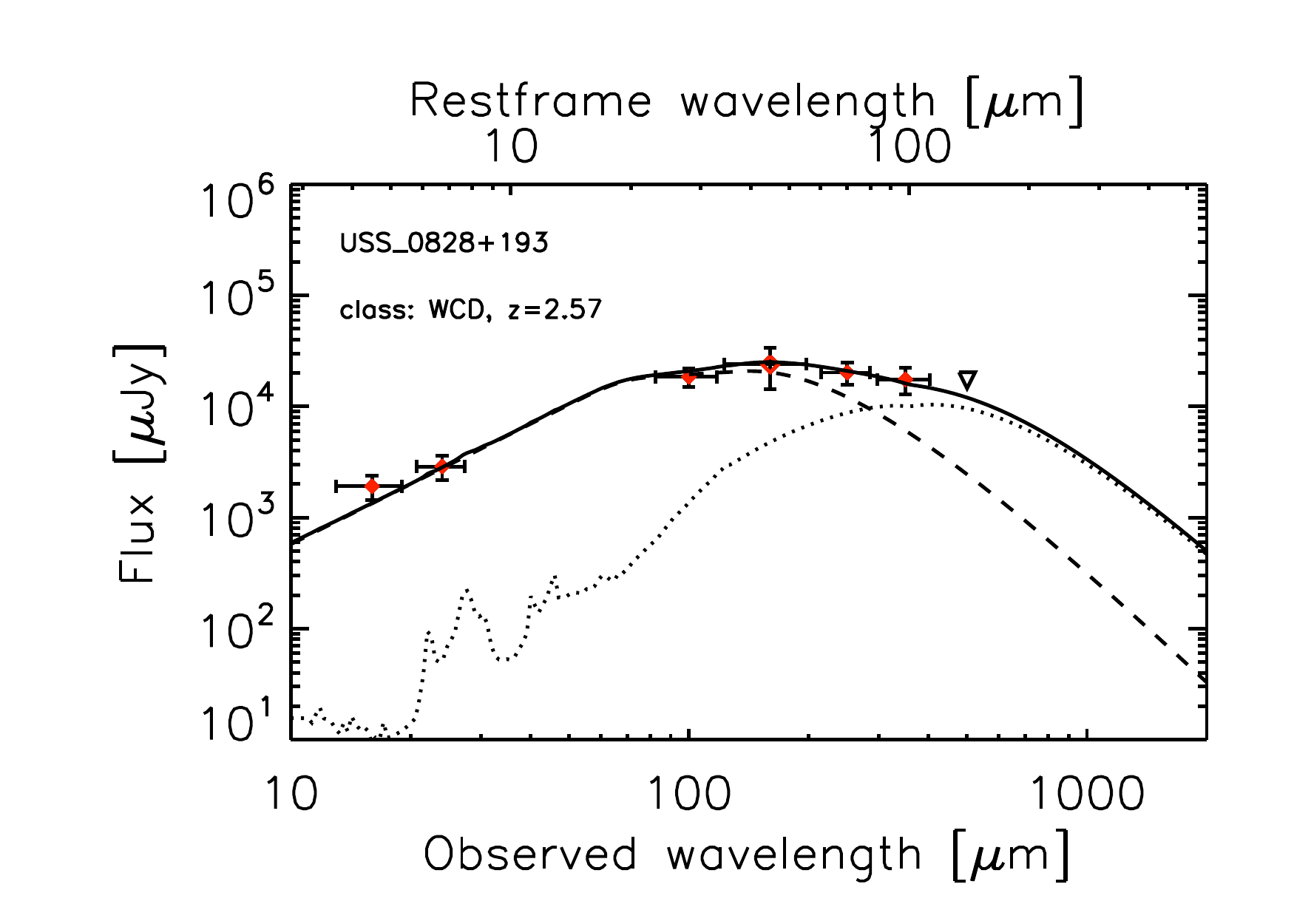}} \\
      24\mum & 100\mum & 160\mum & 250\mum & 350\mum & 500\mum \\
      \includegraphics[height=14mm,trim=70  -4 65 -10,clip=true]{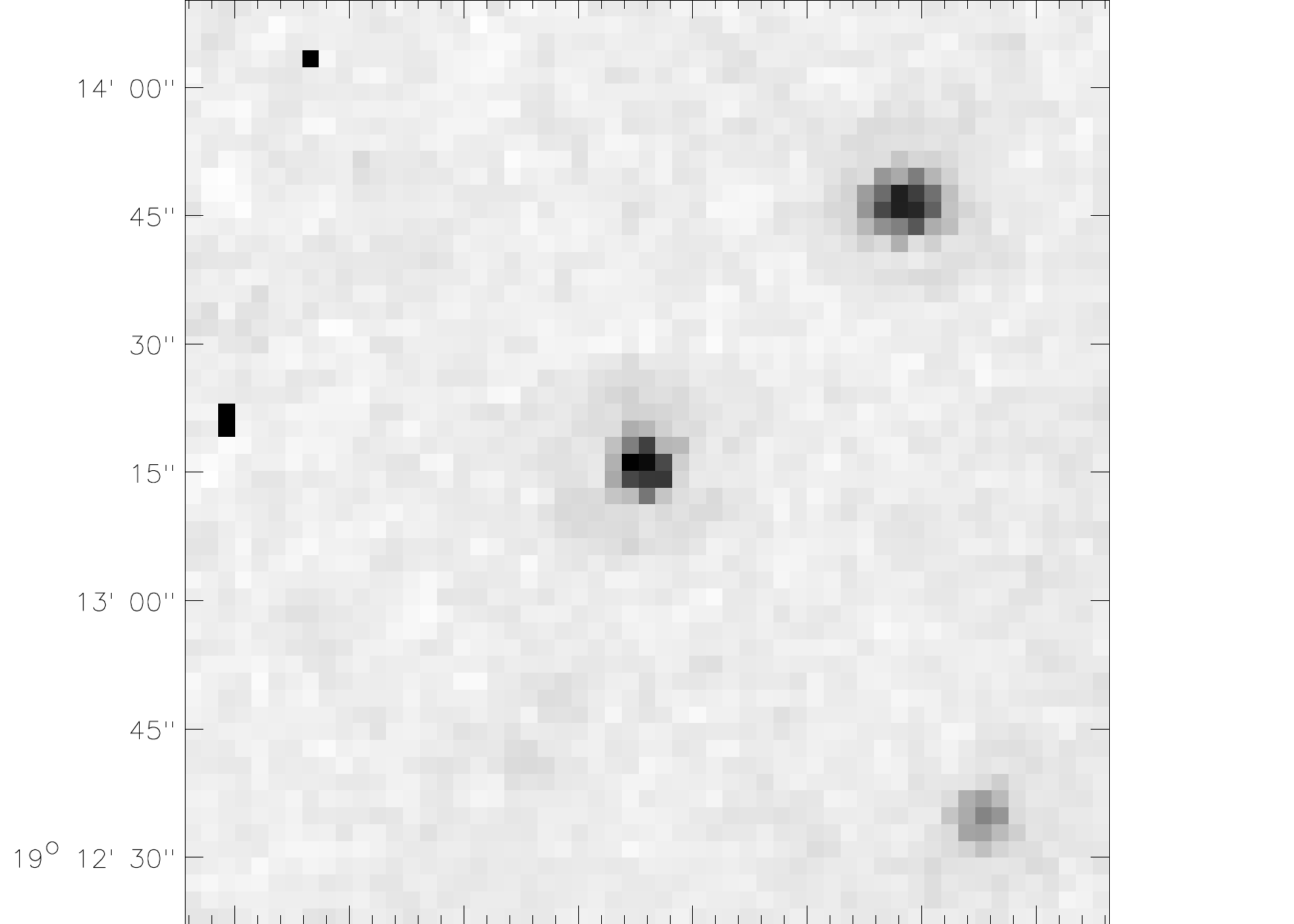} &
      \includegraphics[height=14mm,trim=70  -4 65 -10,clip=true]{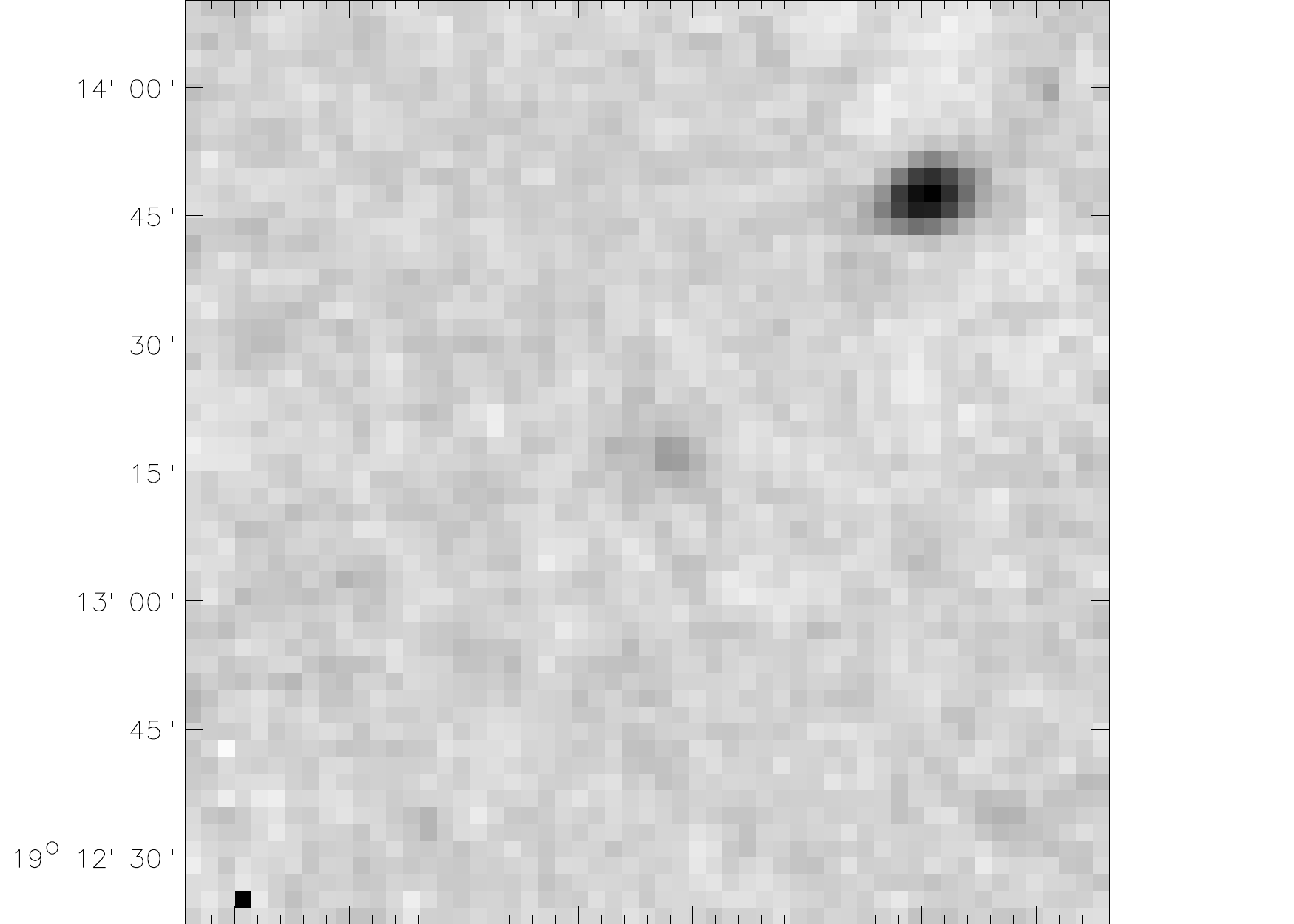} &
      \includegraphics[height=14mm,trim=70  -4 65 -10,clip=true]{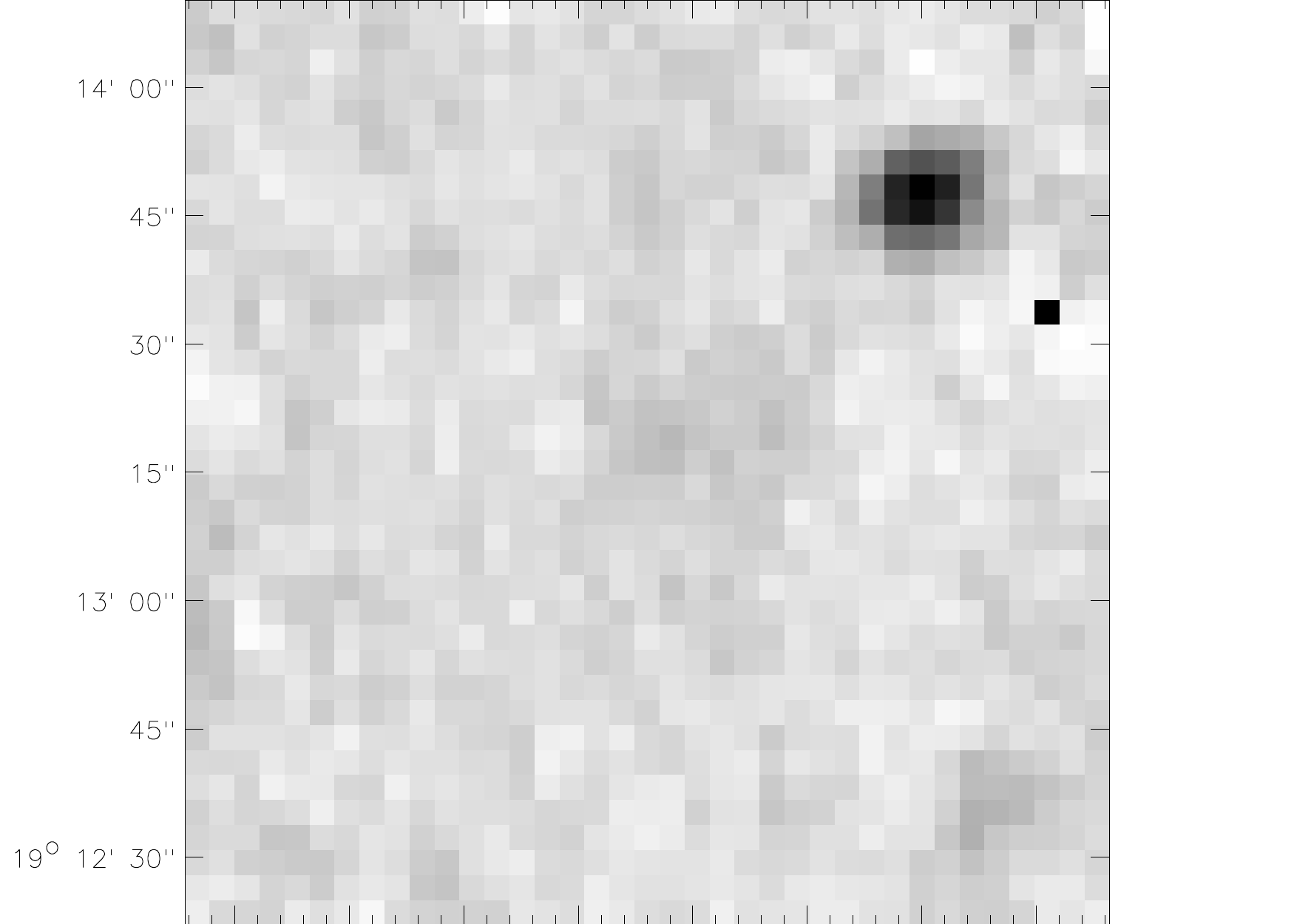} &
      \includegraphics[height=14mm,trim=70  -4 65 -10,clip=true]{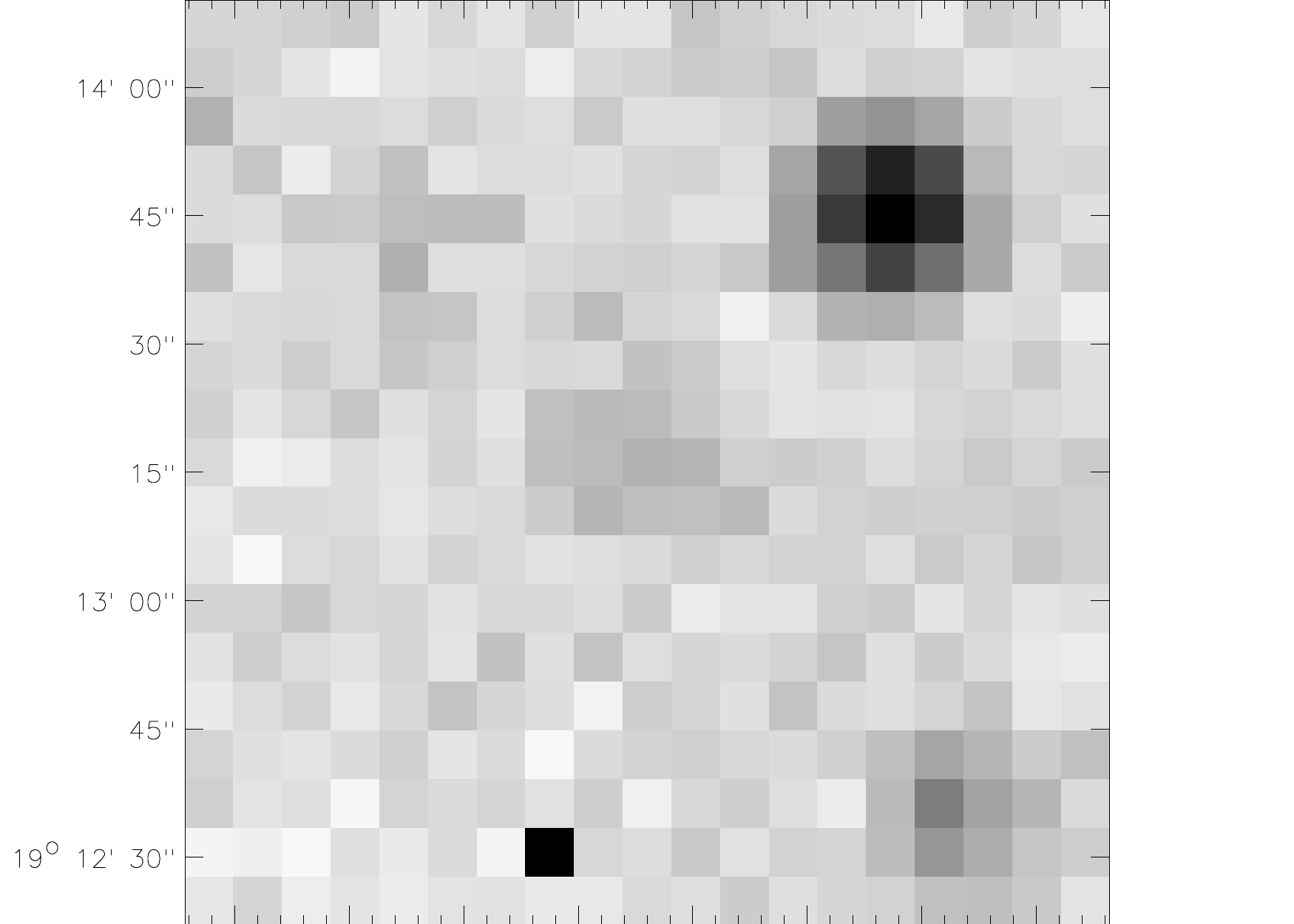}   &
      \includegraphics[height=14mm,trim=70  -4 65 -10,clip=true]{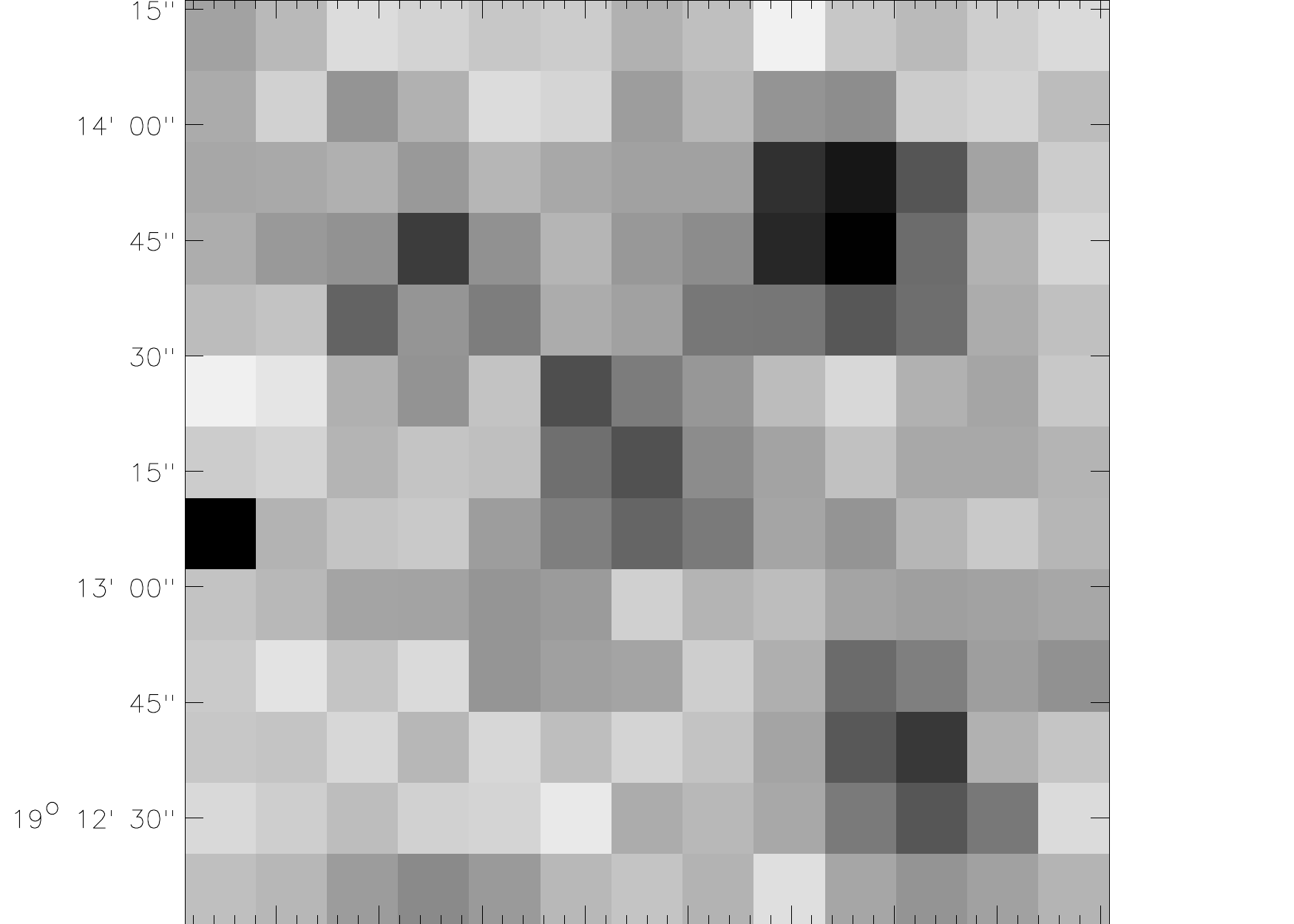}   &
      \includegraphics[height=14mm,trim=70  -4 65 -10,clip=true]{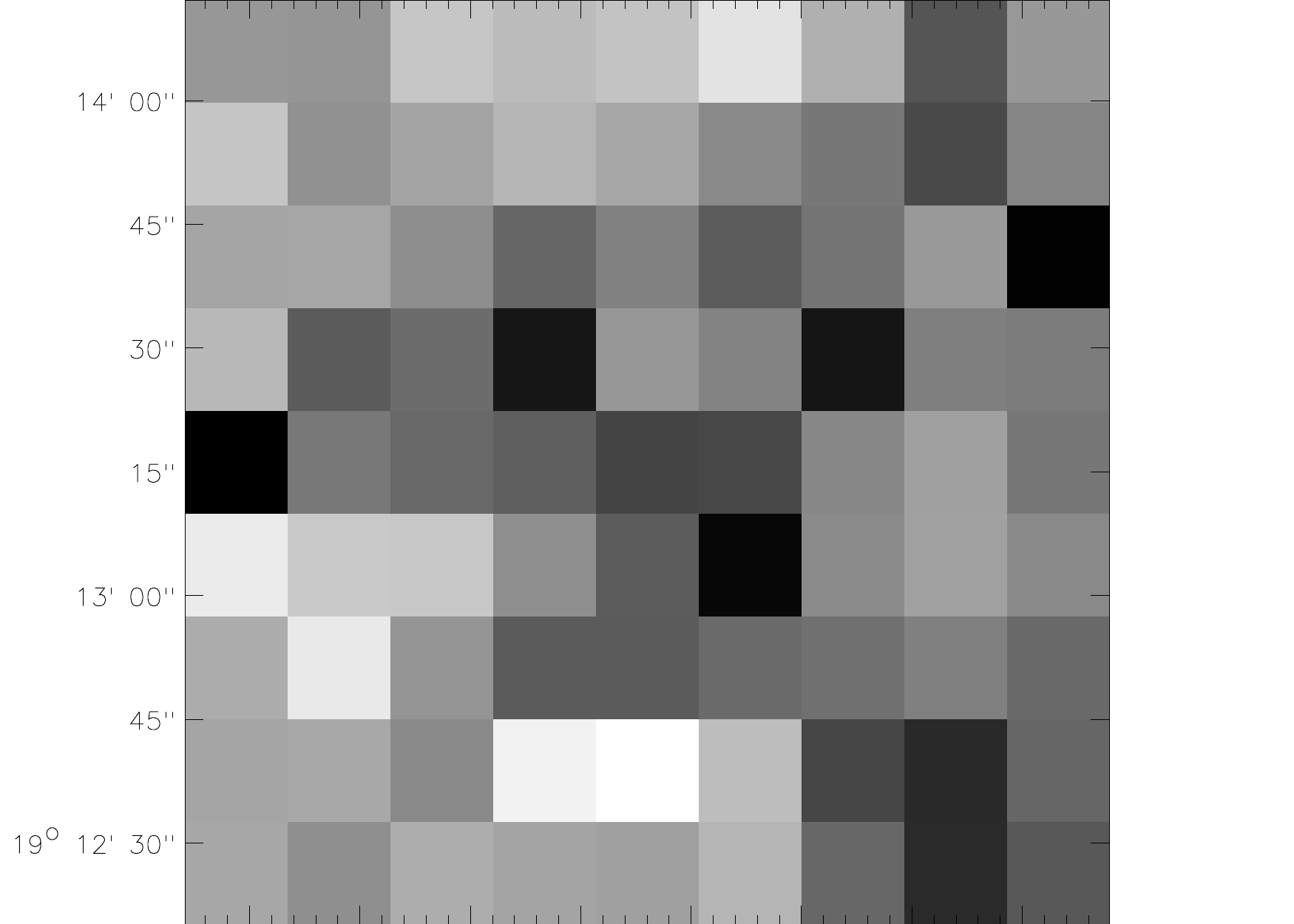} \\

      \multicolumn{6}{c}{\includegraphics[width=80mm,trim=30 0 20 0 ,clip=true]{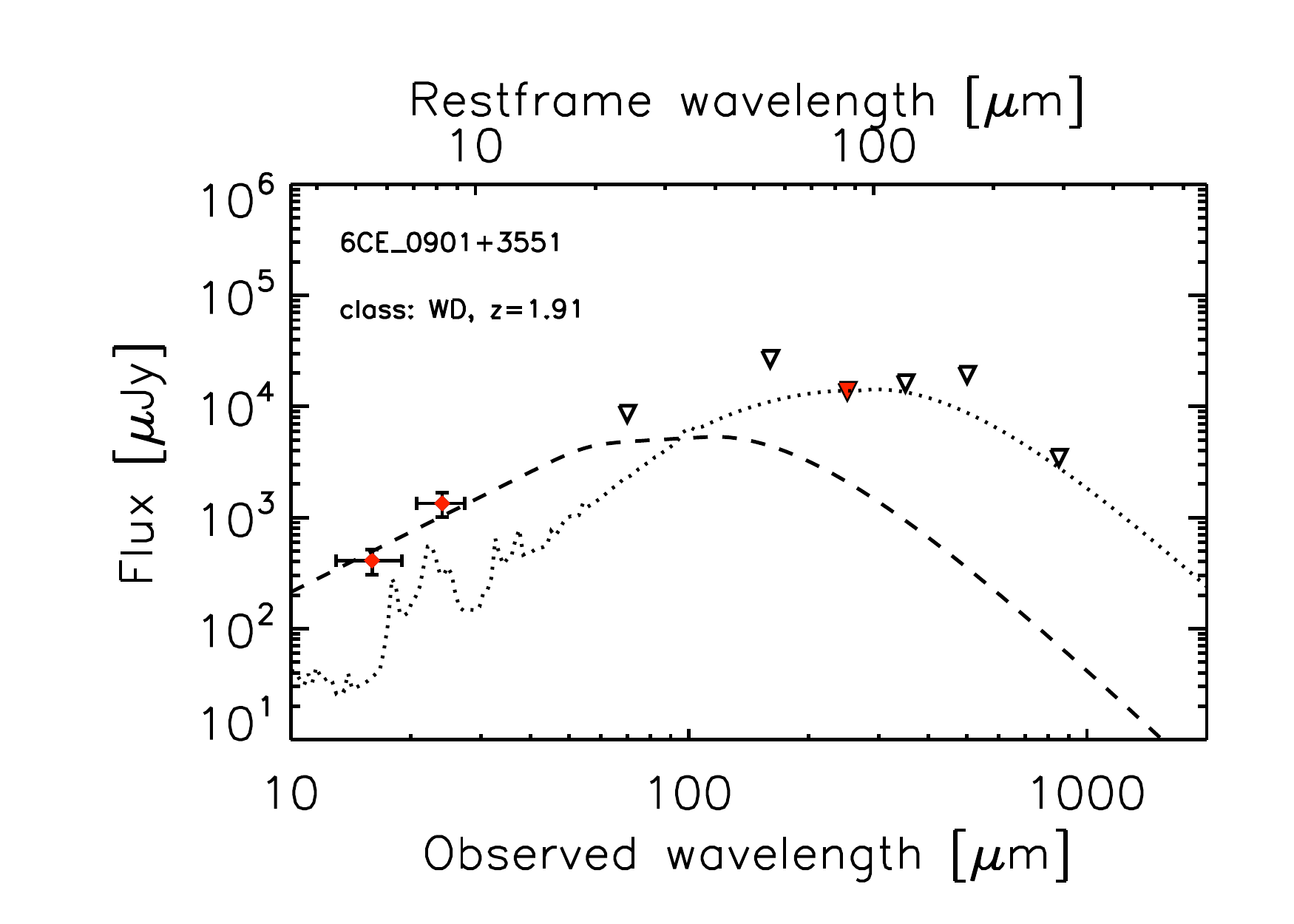}} \\
      24\mum & 70\mum & 160\mum & 250\mum & 350\mum & 500\mum \\
      \includegraphics[height=14mm,trim=70  -4 65 -10,clip=true]{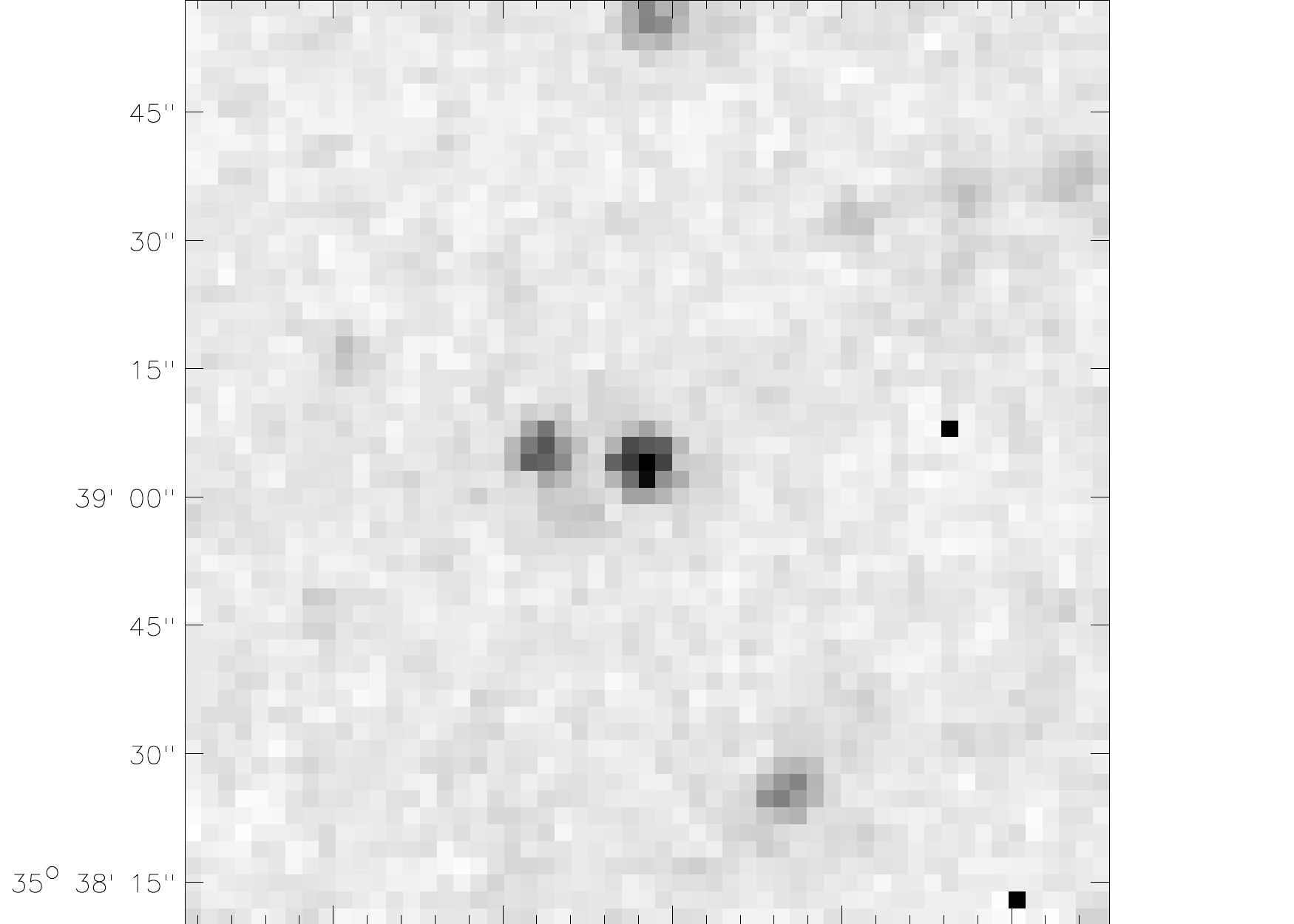} &
      \includegraphics[height=14mm,trim=70  -4 65 -10,clip=true]{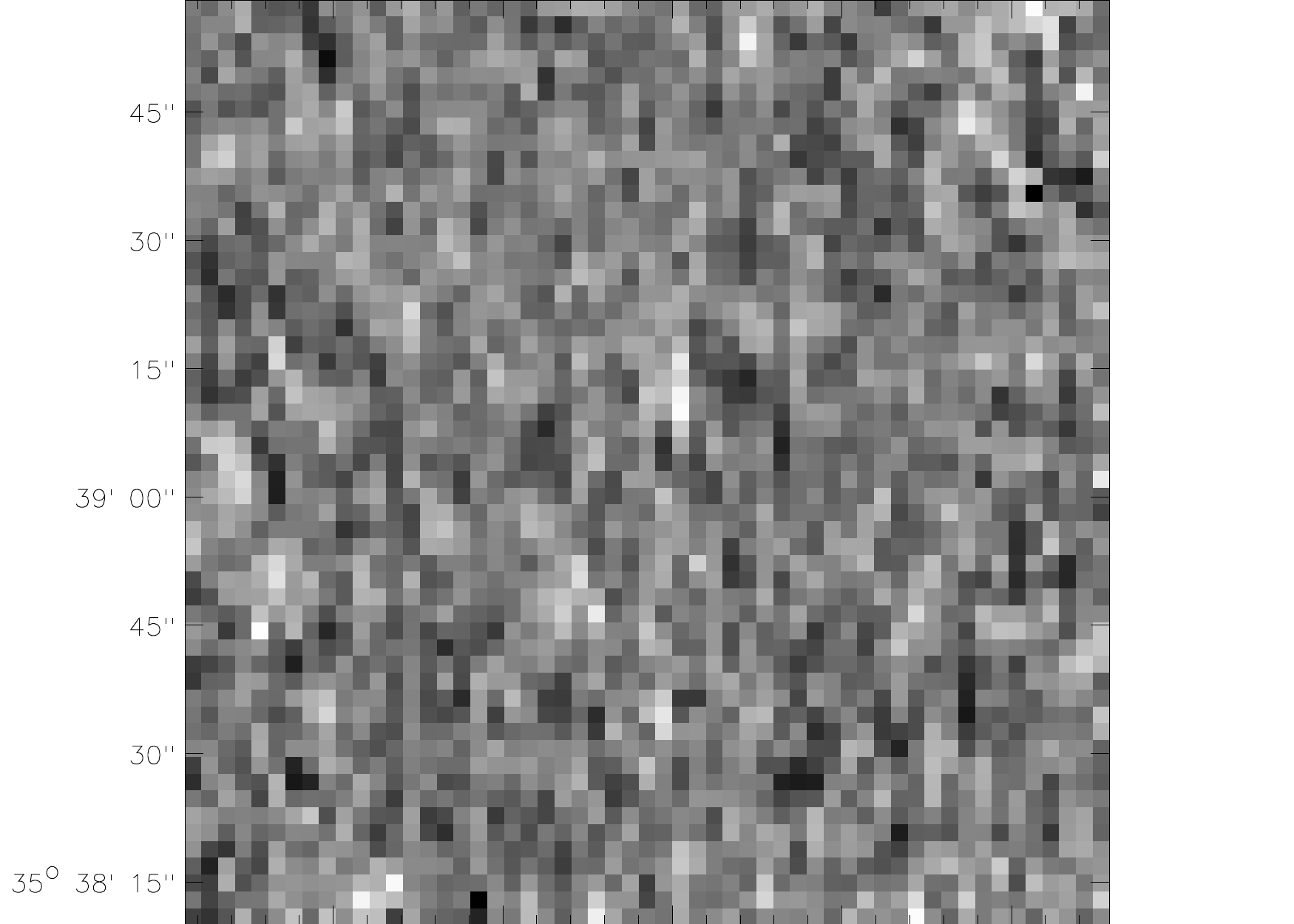} &
      \includegraphics[height=14mm,trim=70  -4 65 -10,clip=true]{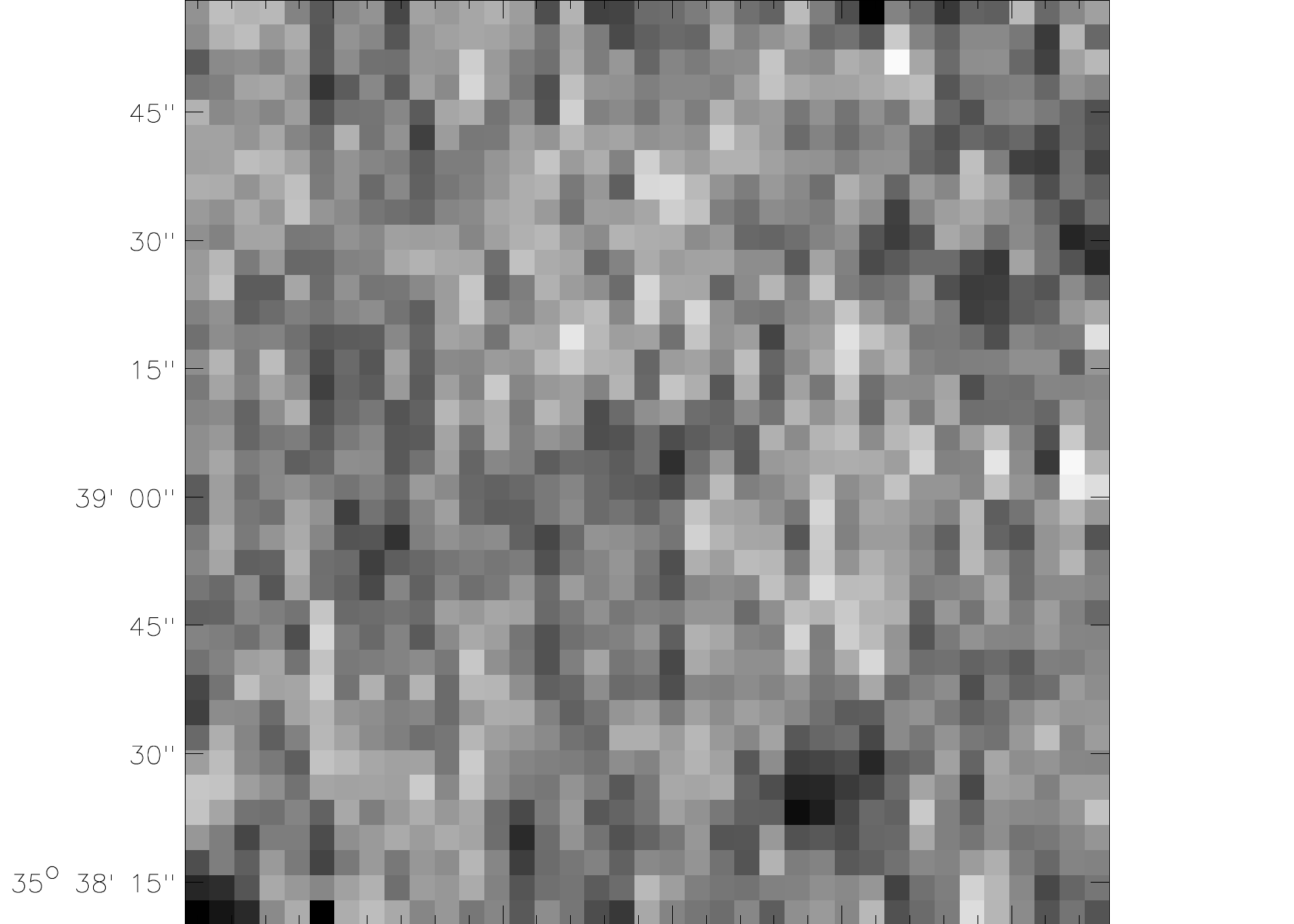} &
      \includegraphics[height=14mm,trim=70  -4 65 -10,clip=true]{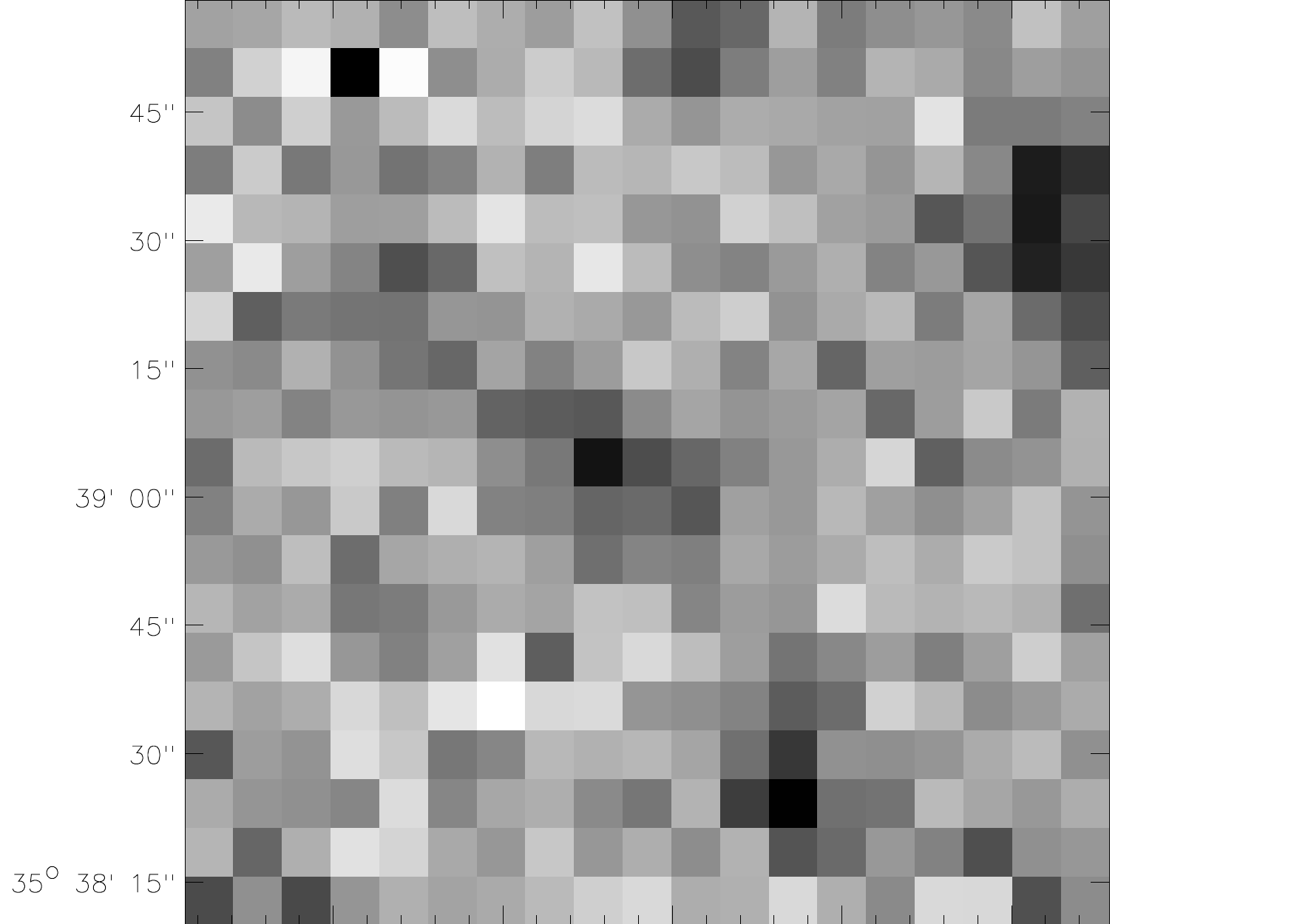}   &
      \includegraphics[height=14mm,trim=70  -4 65 -10,clip=true]{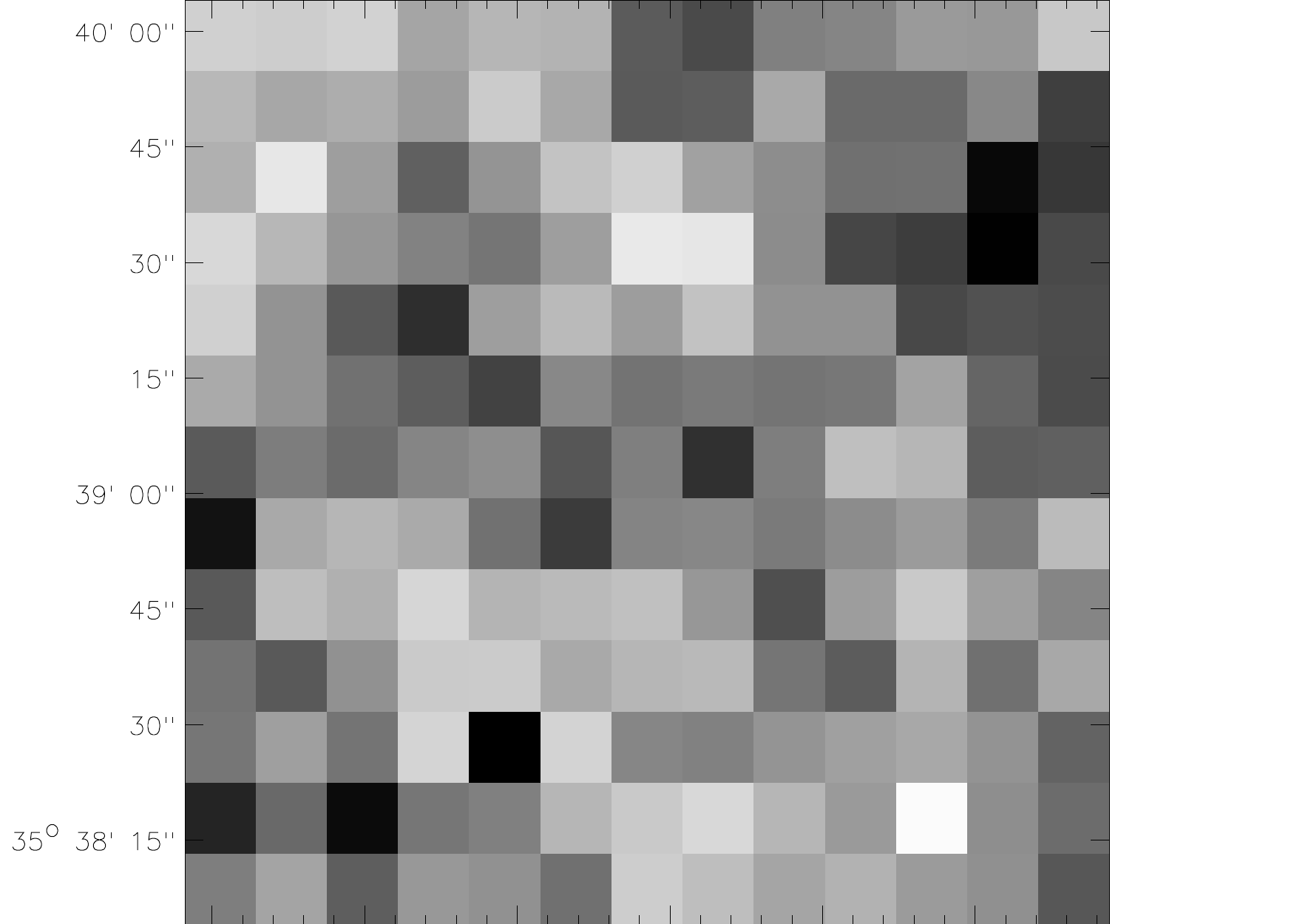}   &
      \includegraphics[height=14mm,trim=70  -4 65 -10,clip=true]{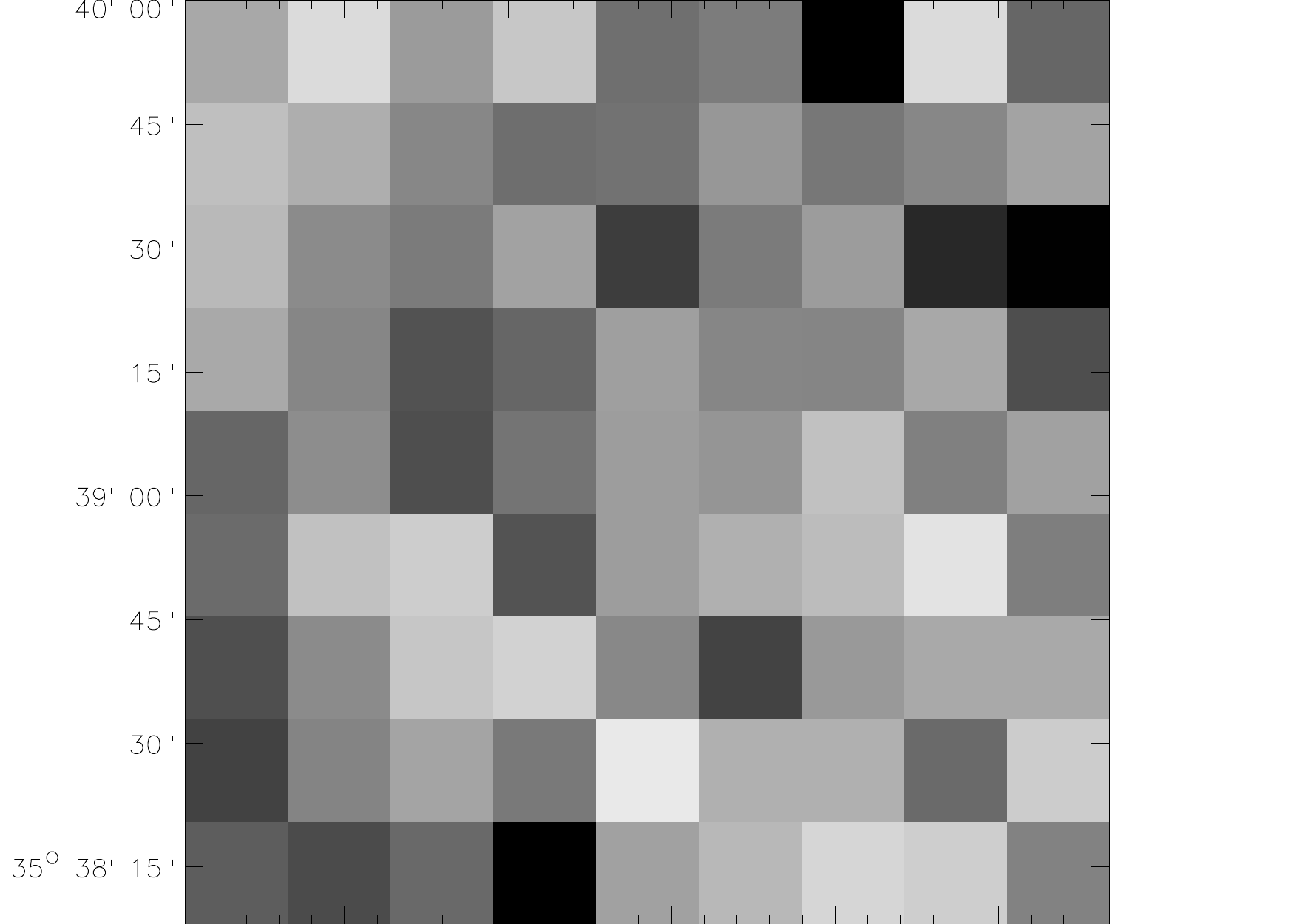} \\

      \multicolumn{6}{c}{\includegraphics[width=80mm,trim=30 0 20 0 ,clip=true]{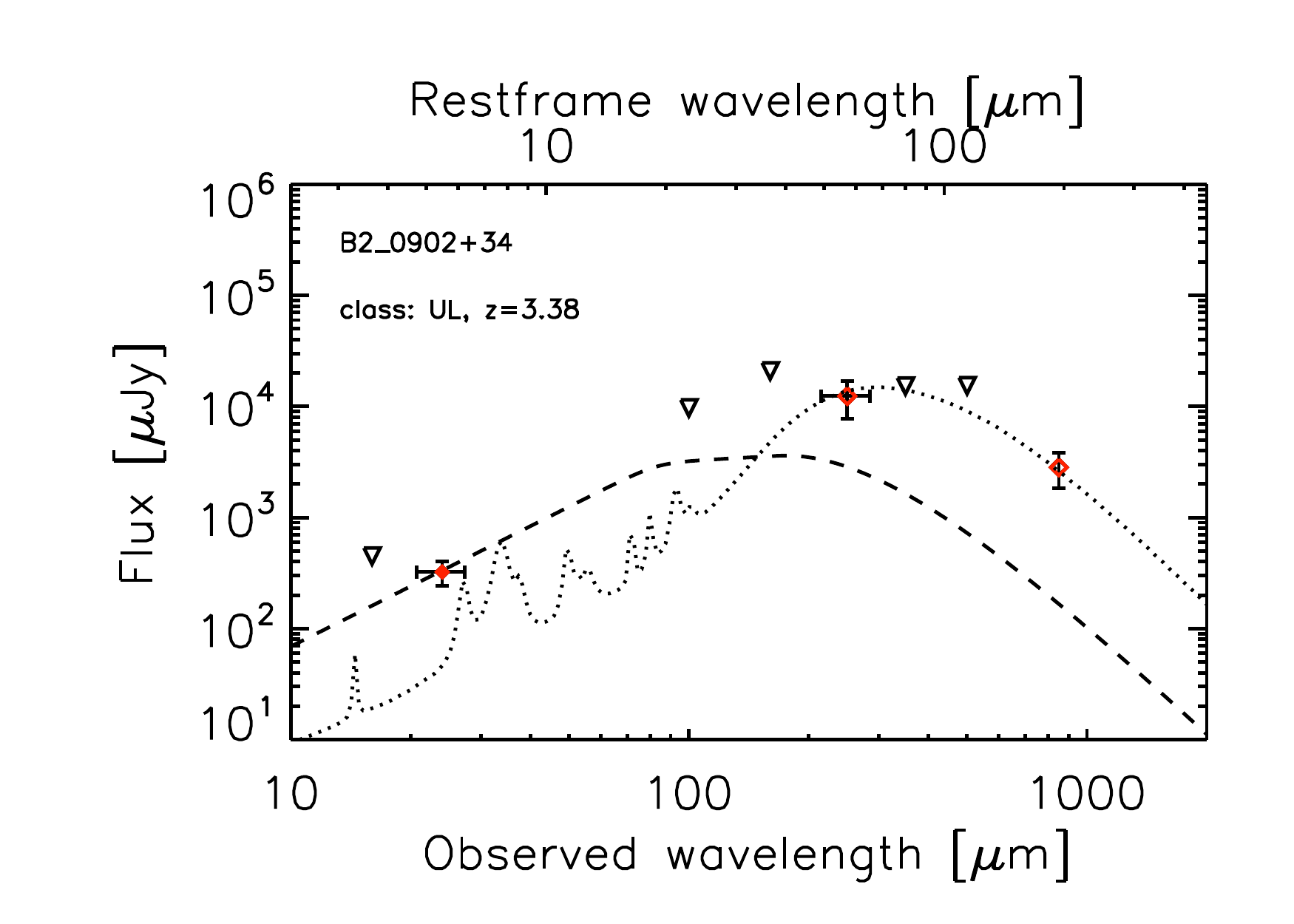}} \\
      24\mum & 100\mum & 160\mum & 250\mum & 350\mum & 500\mum \\
\end{tabular}
          \end{minipage}
  \end{centering}
\end{figure*}

\clearpage

\begin{figure*}[h]
  \begin{centering}
    \begin{minipage}[c]{0.46\linewidth}
    \begin{tabular}{c@{}c@{}c@{}c@{}c@{}c@{}}

      \multicolumn{6}{c}{\includegraphics[width=80mm,trim=30 0 20 0 ,clip=true]{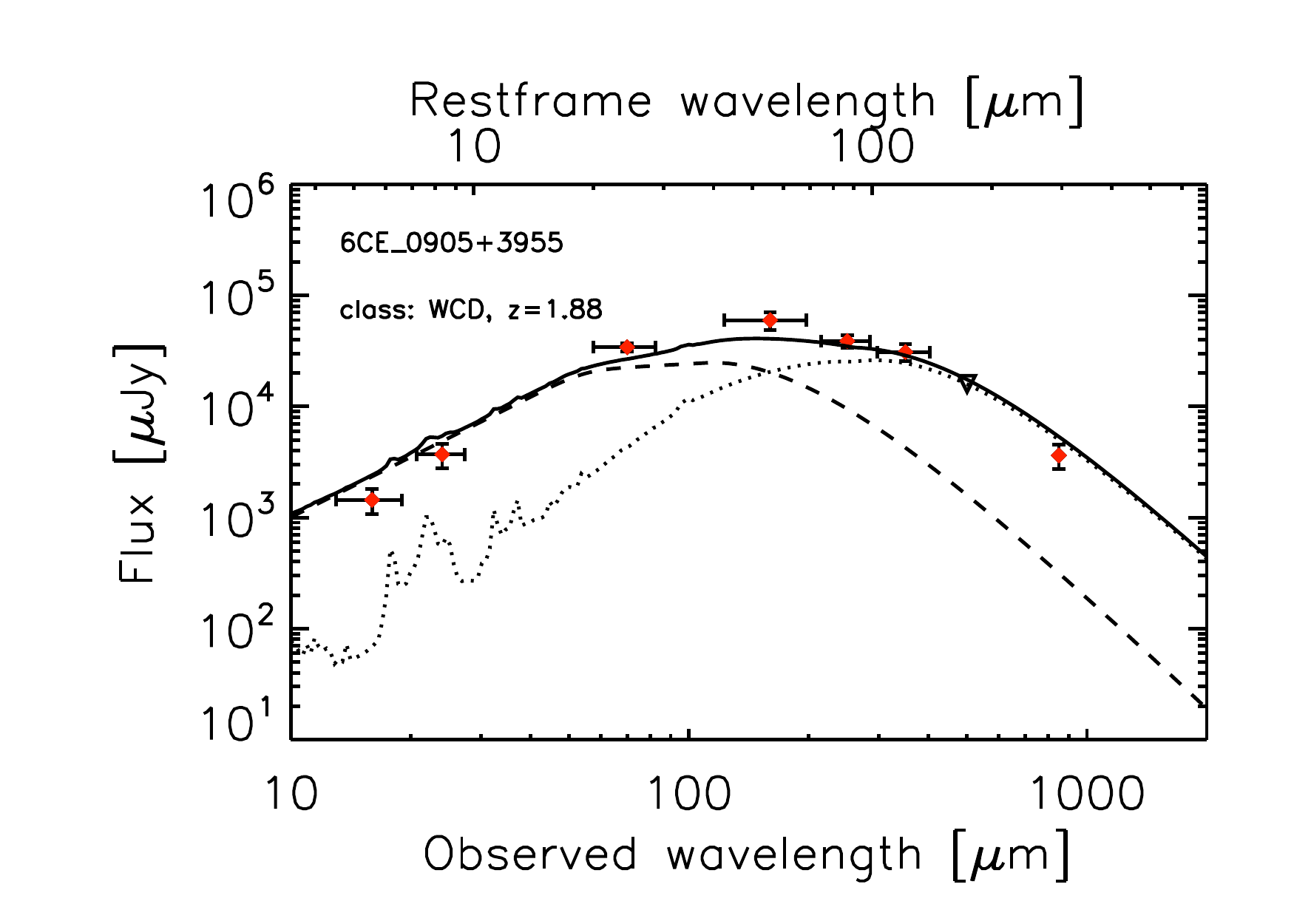}} \\
      24\mum & 100\mum & 160\mum & 250\mum & 350\mum & 500\mum \\
      \includegraphics[height=14mm,trim=70  -4 65 -10,clip=true]{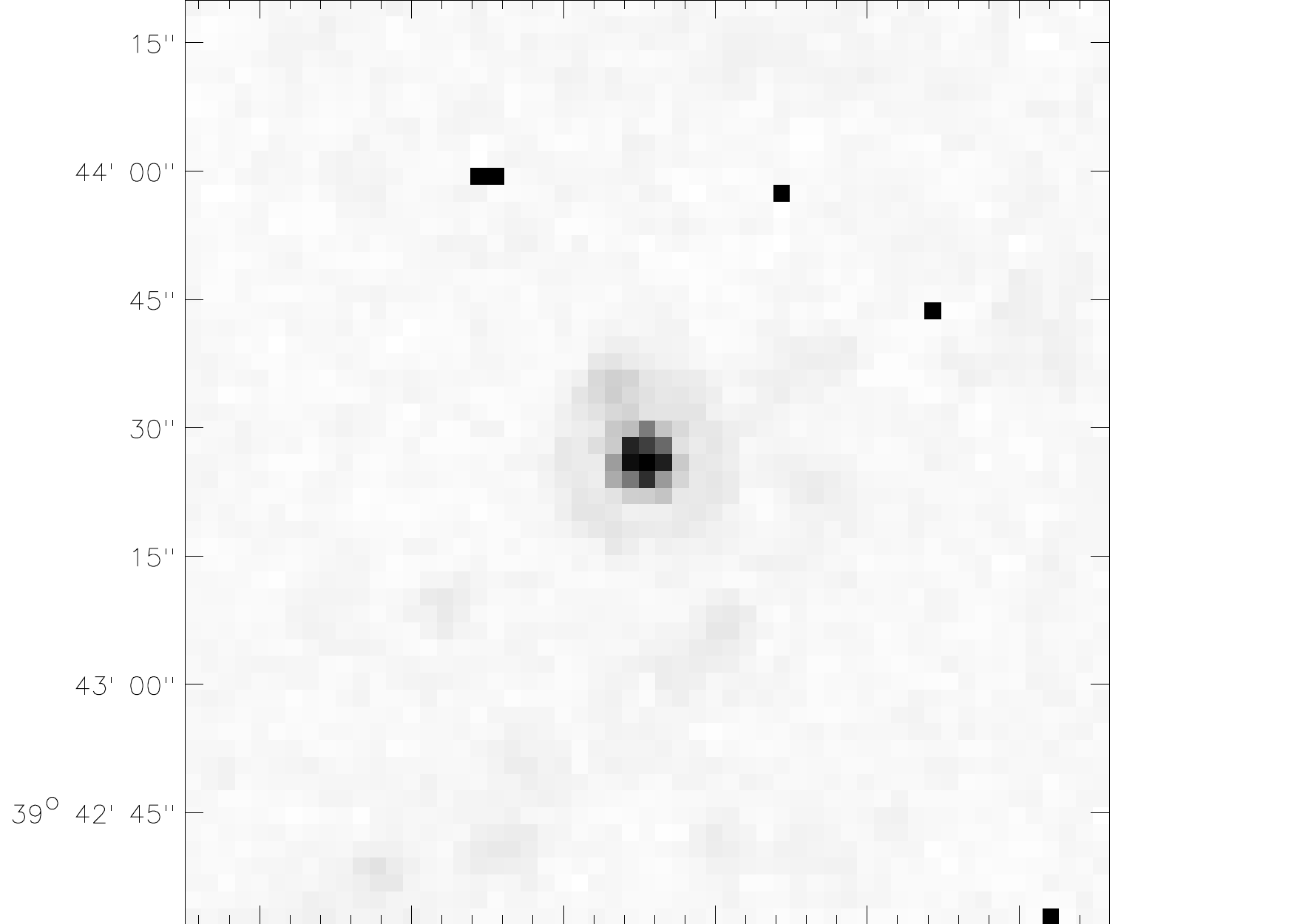} &
      \includegraphics[height=14mm,trim=70  -4 65 -10,clip=true]{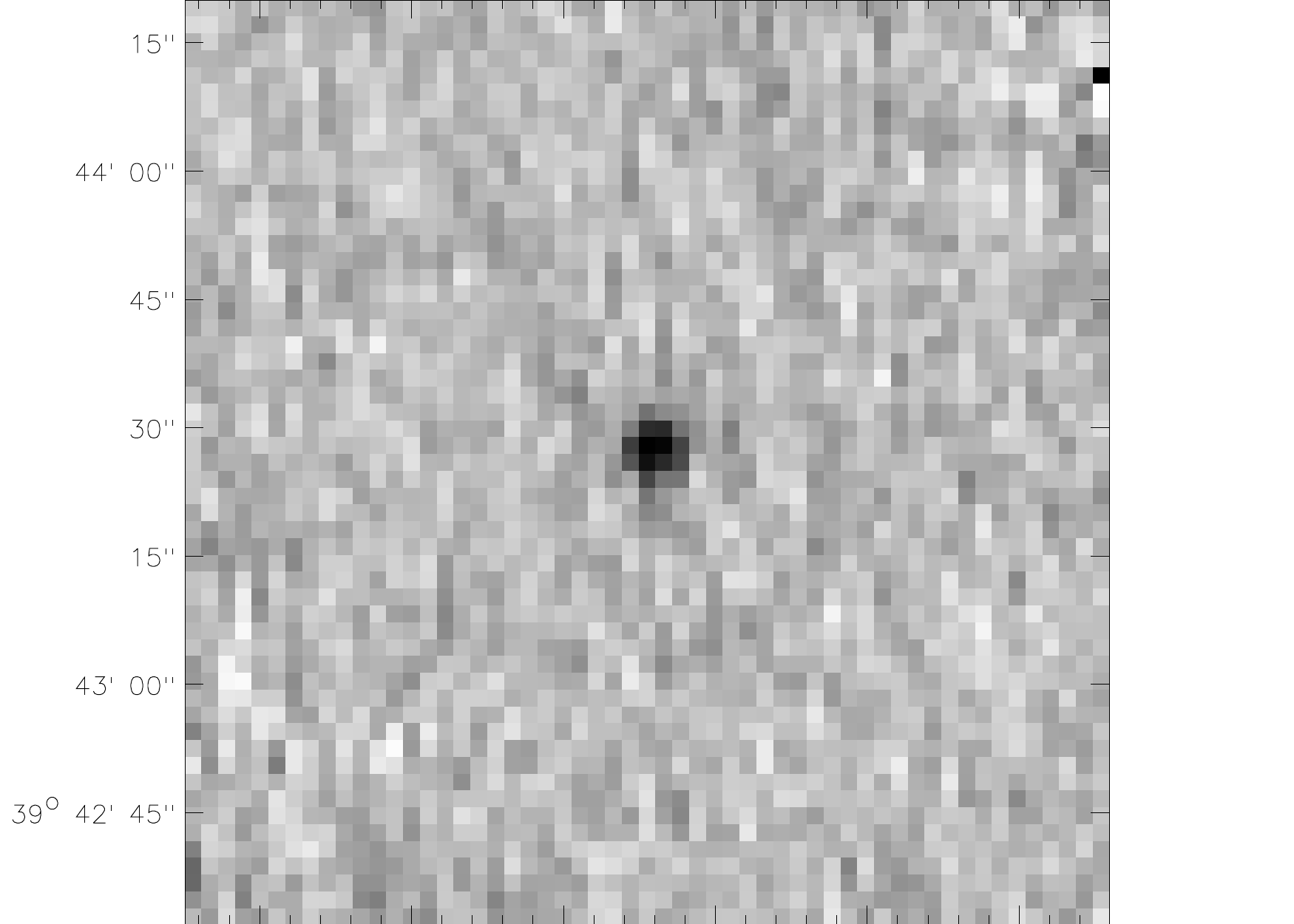} &
      \includegraphics[height=14mm,trim=70  -4 65 -10,clip=true]{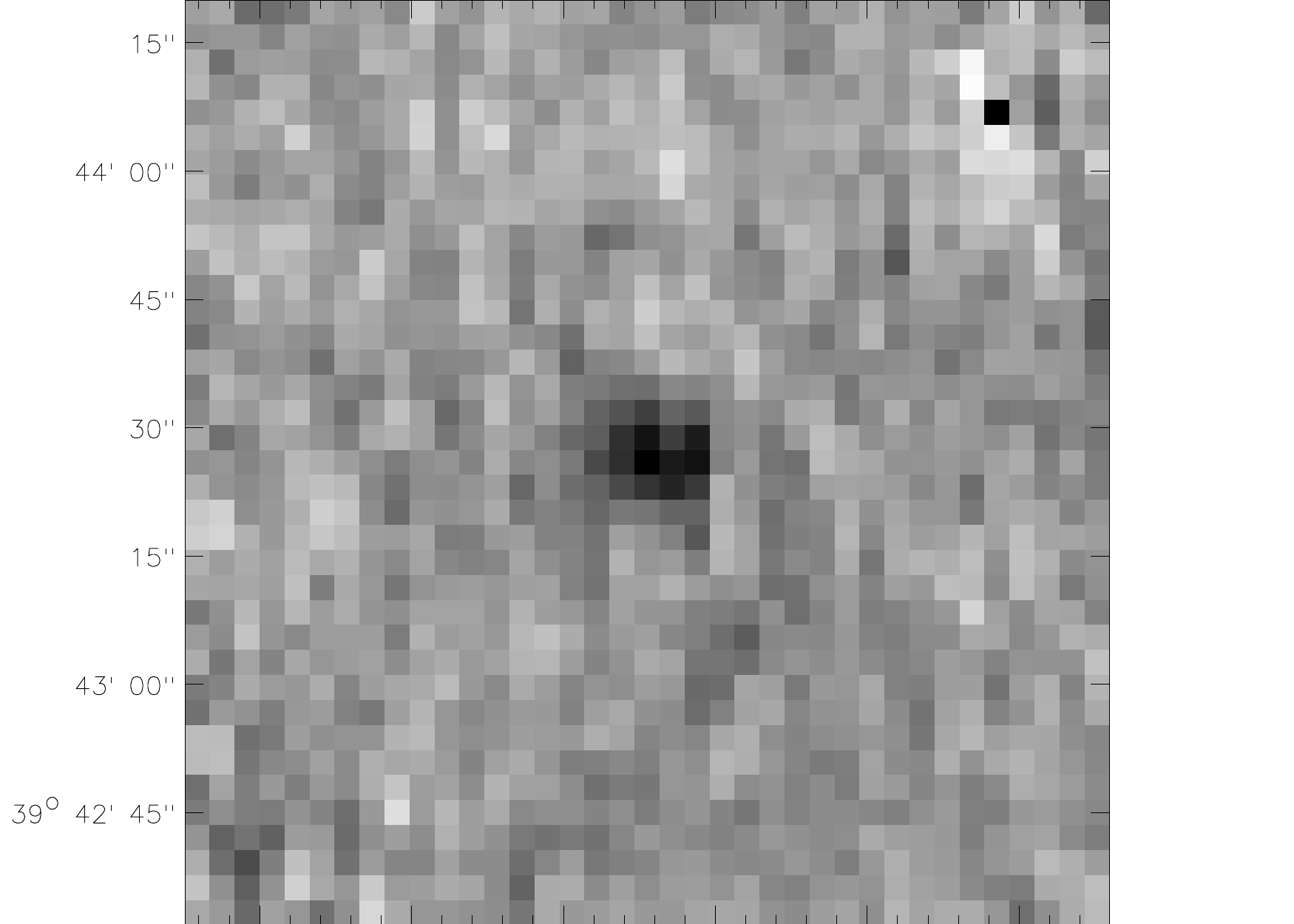} &
      \includegraphics[height=14mm,trim=70  -4 65 -10,clip=true]{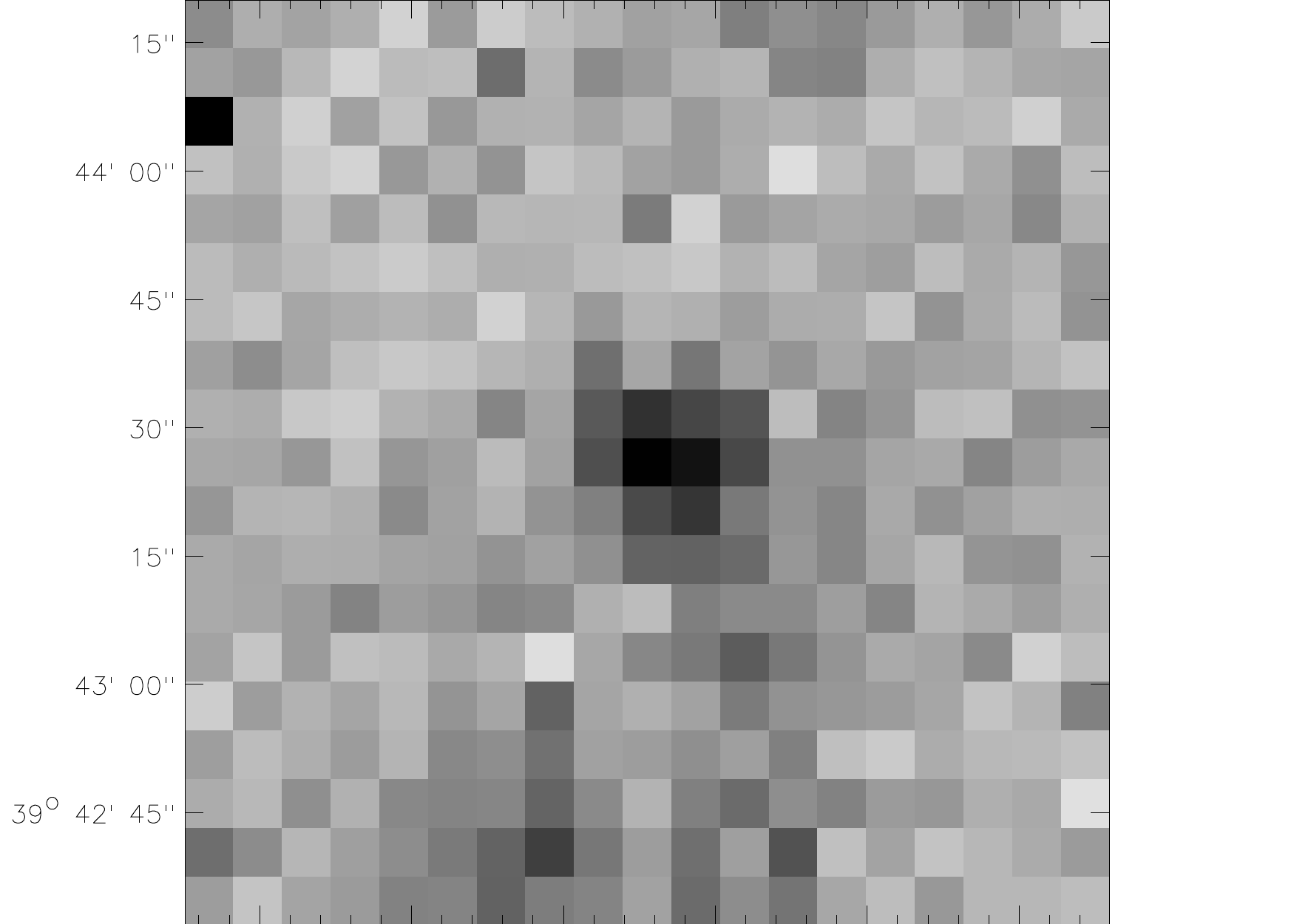}   &
      \includegraphics[height=14mm,trim=70  -4 65 -10,clip=true]{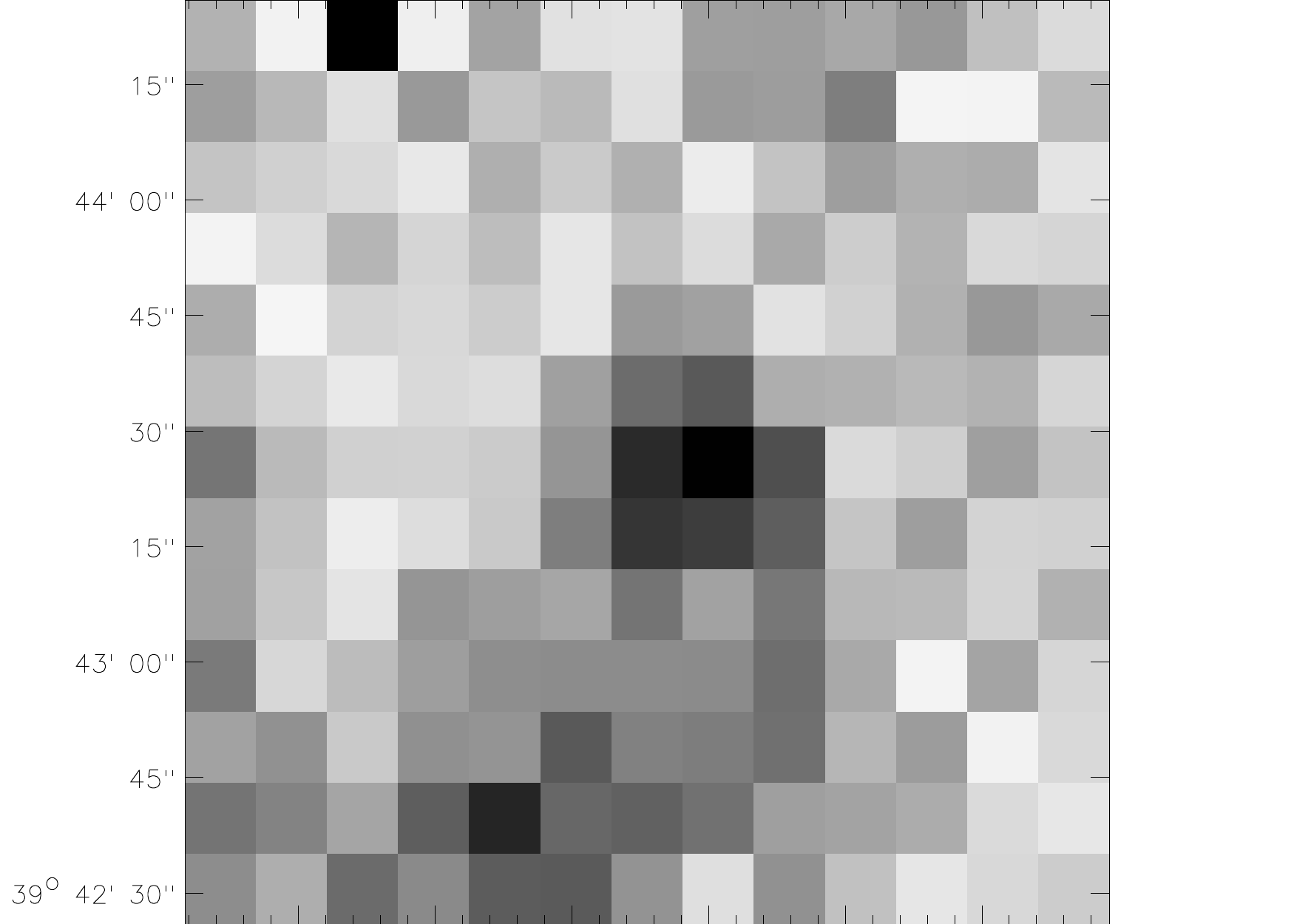}   &
      \includegraphics[height=14mm,trim=70  -4 65 -10,clip=true]{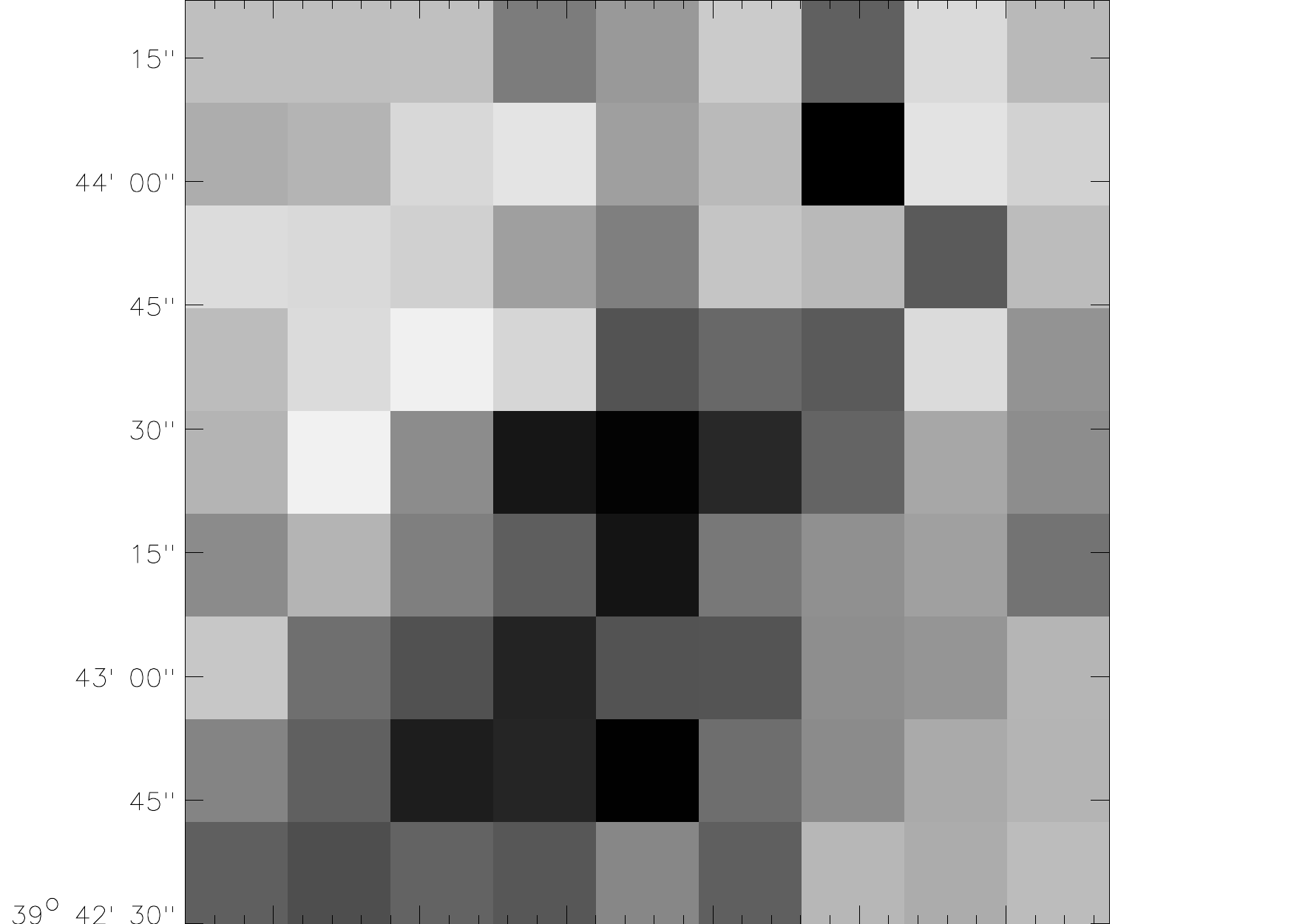} \\

      \multicolumn{6}{c}{\includegraphics[width=80mm,trim=30 0 20 0 ,clip=true]{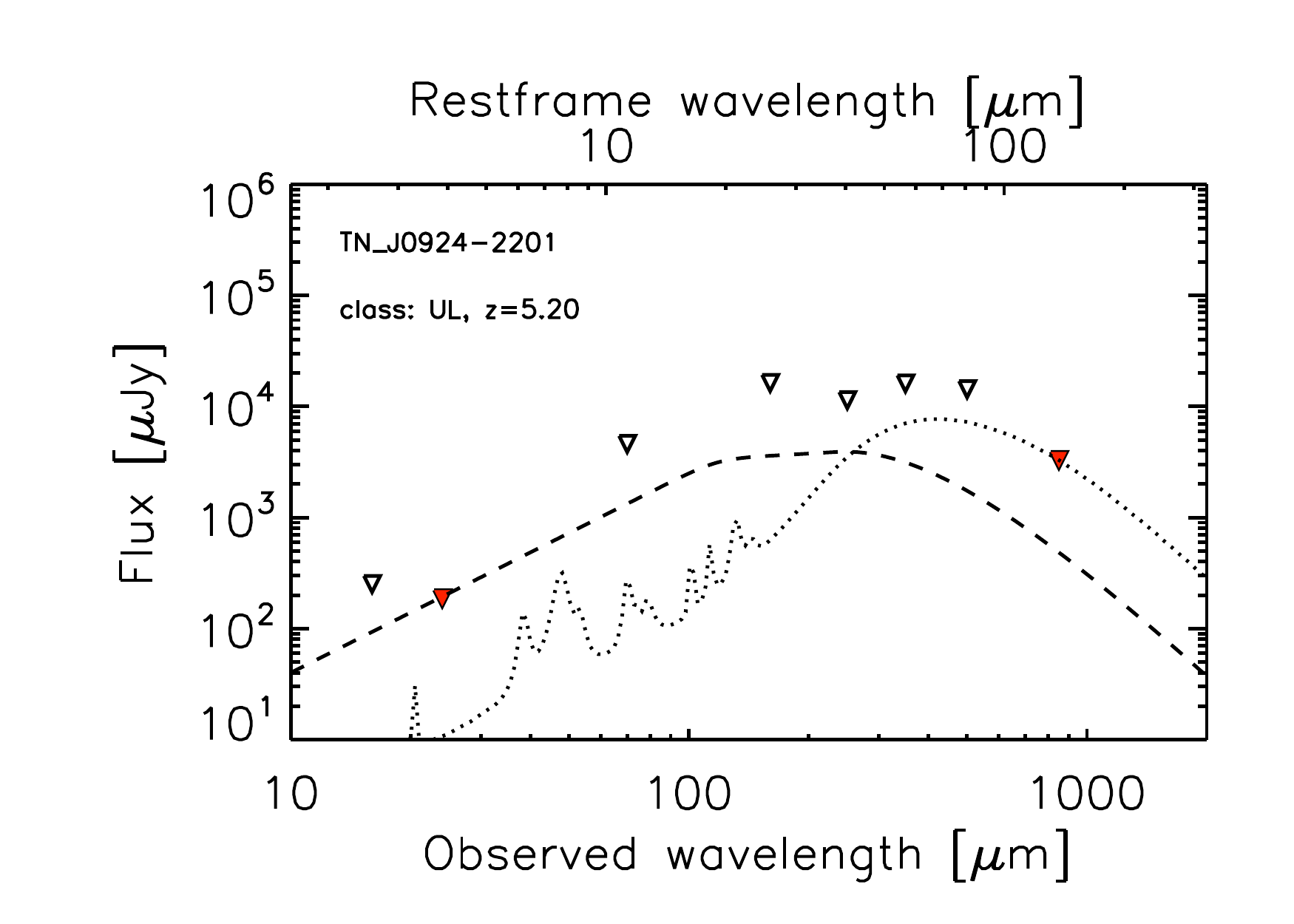}} \\
      24\mum & 70\mum & 160\mum & 250\mum & 350\mum & 500\mum \\
      \includegraphics[height=14mm,trim=70  -4 65 -10,clip=true]{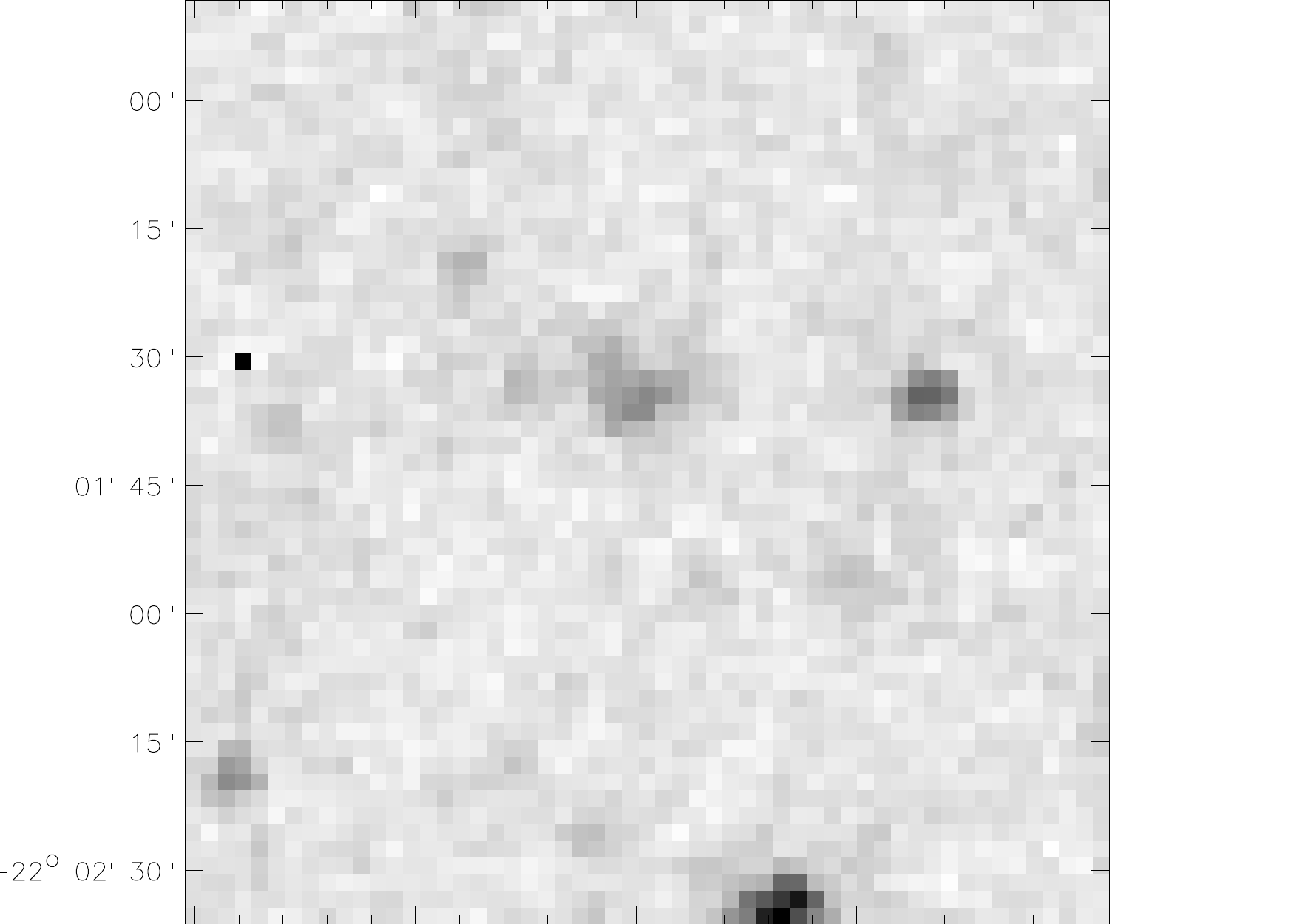} &
      \includegraphics[height=14mm,trim=70  -4 65 -10,clip=true]{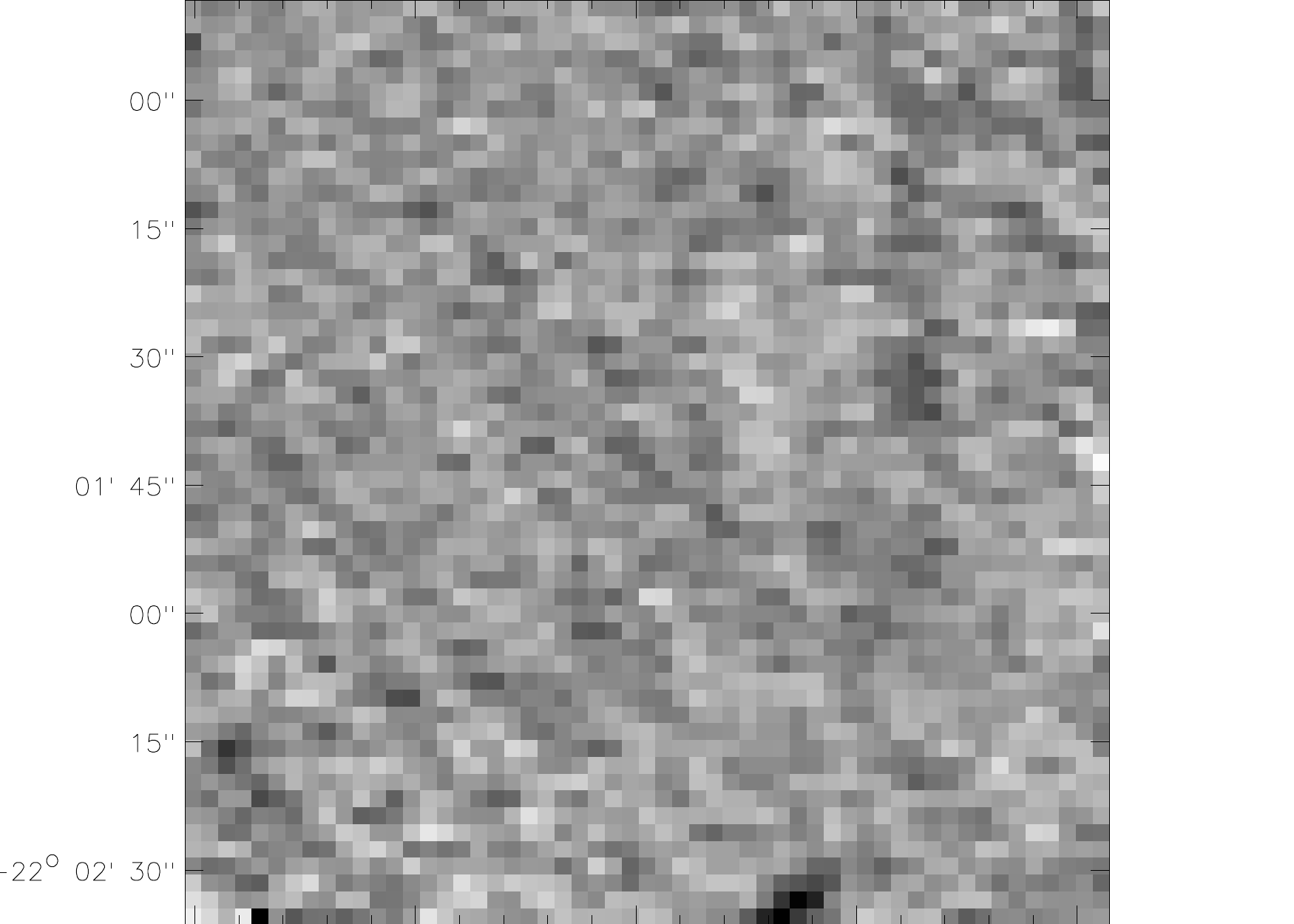} &
      \includegraphics[height=14mm,trim=70  -4 65 -10,clip=true]{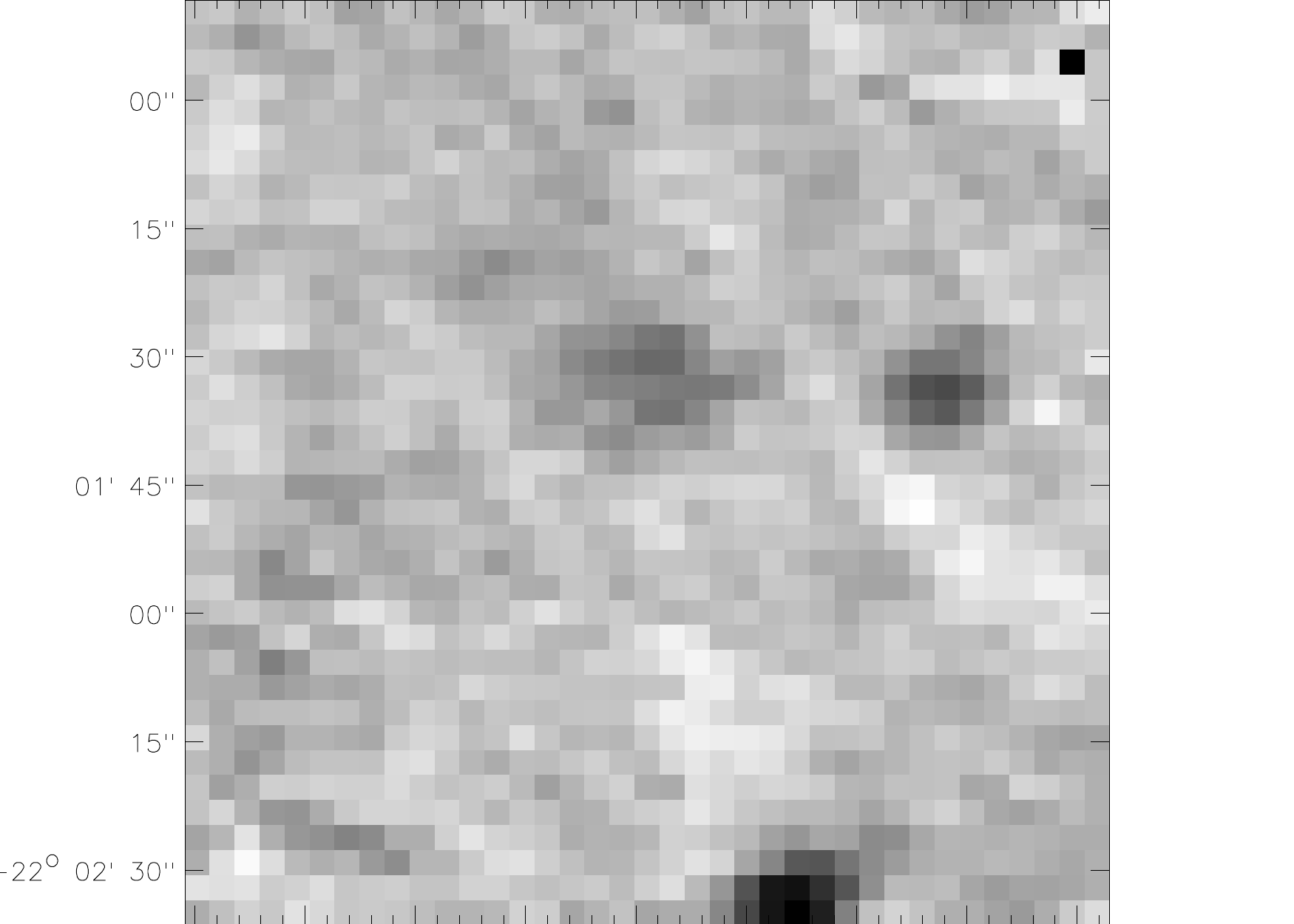} &
      \includegraphics[height=14mm,trim=70  -4 65 -10,clip=true]{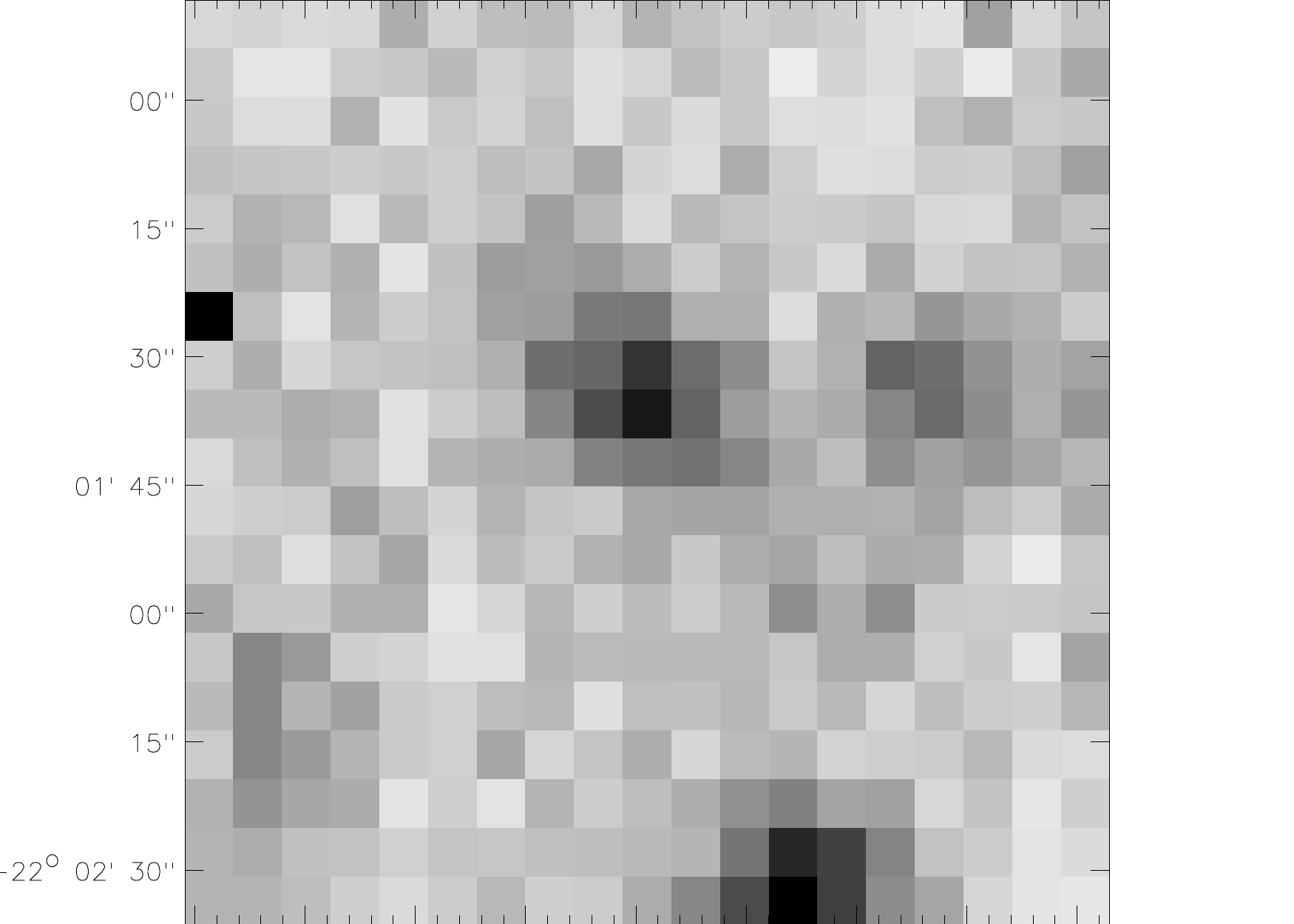}   &
      \includegraphics[height=14mm,trim=70  -4 65 -10,clip=true]{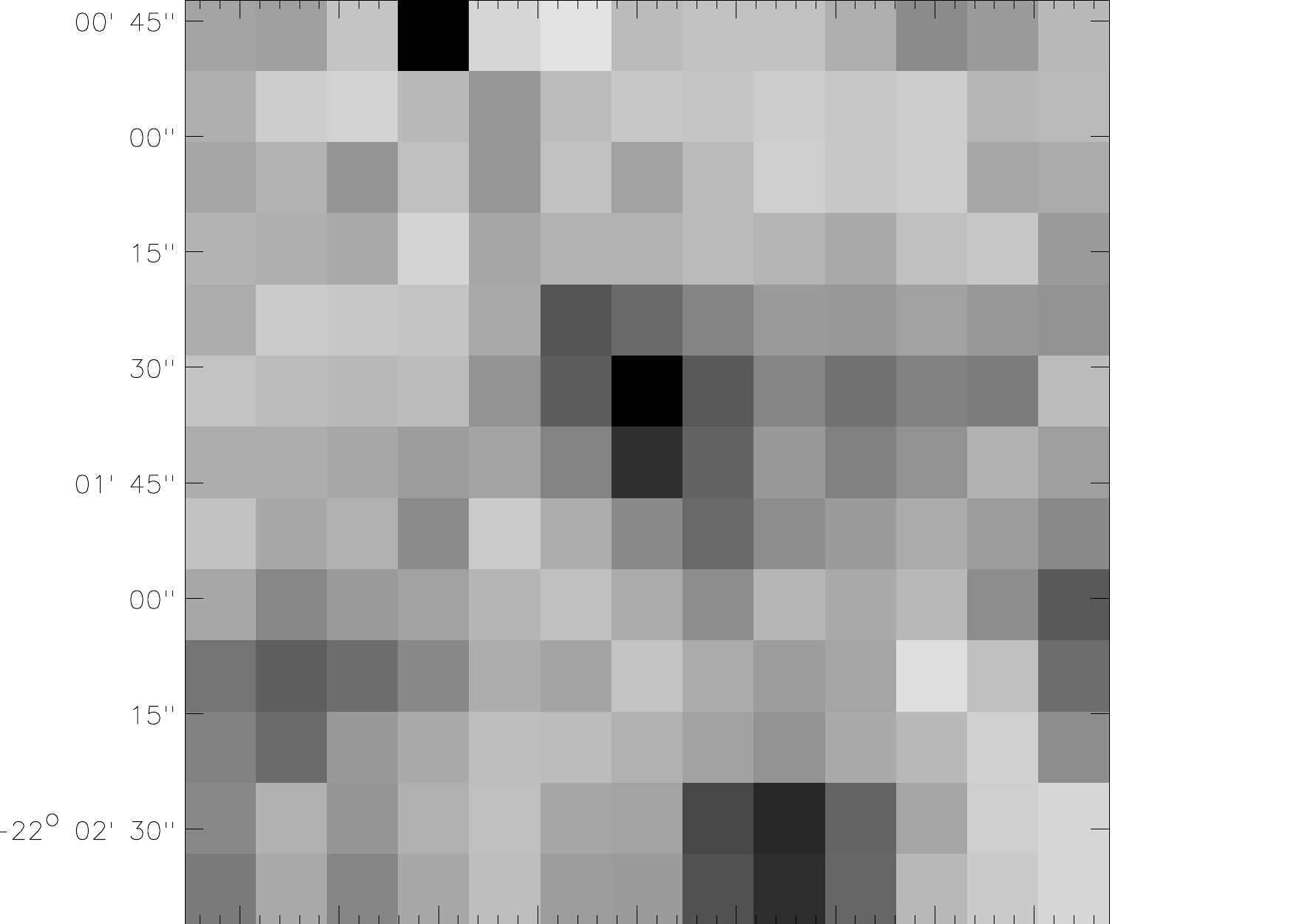}   &
      \includegraphics[height=14mm,trim=70  -4 65 -10,clip=true]{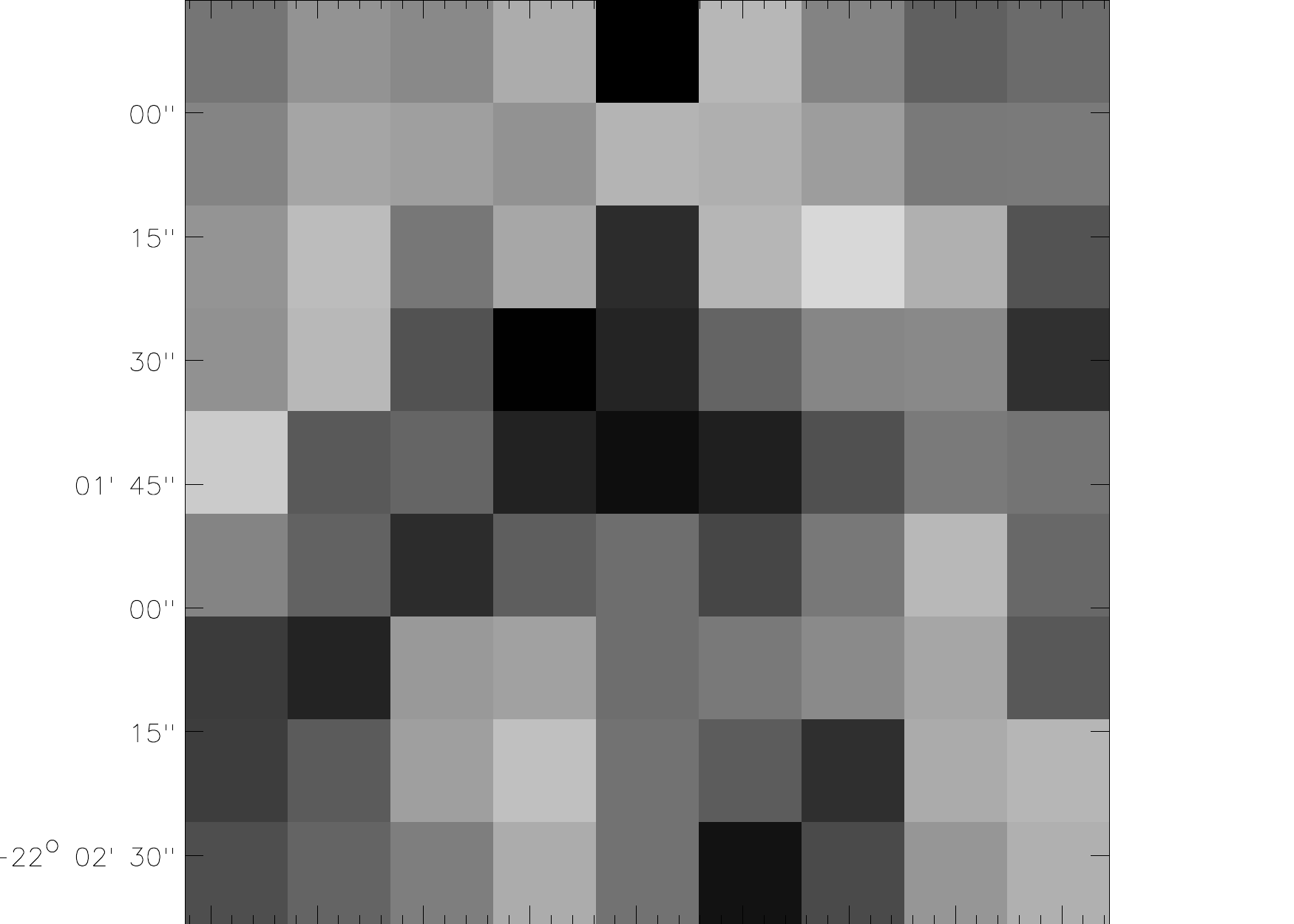} \\

      \multicolumn{6}{c}{\includegraphics[width=80mm,trim=30 0 20 0 ,clip=true]{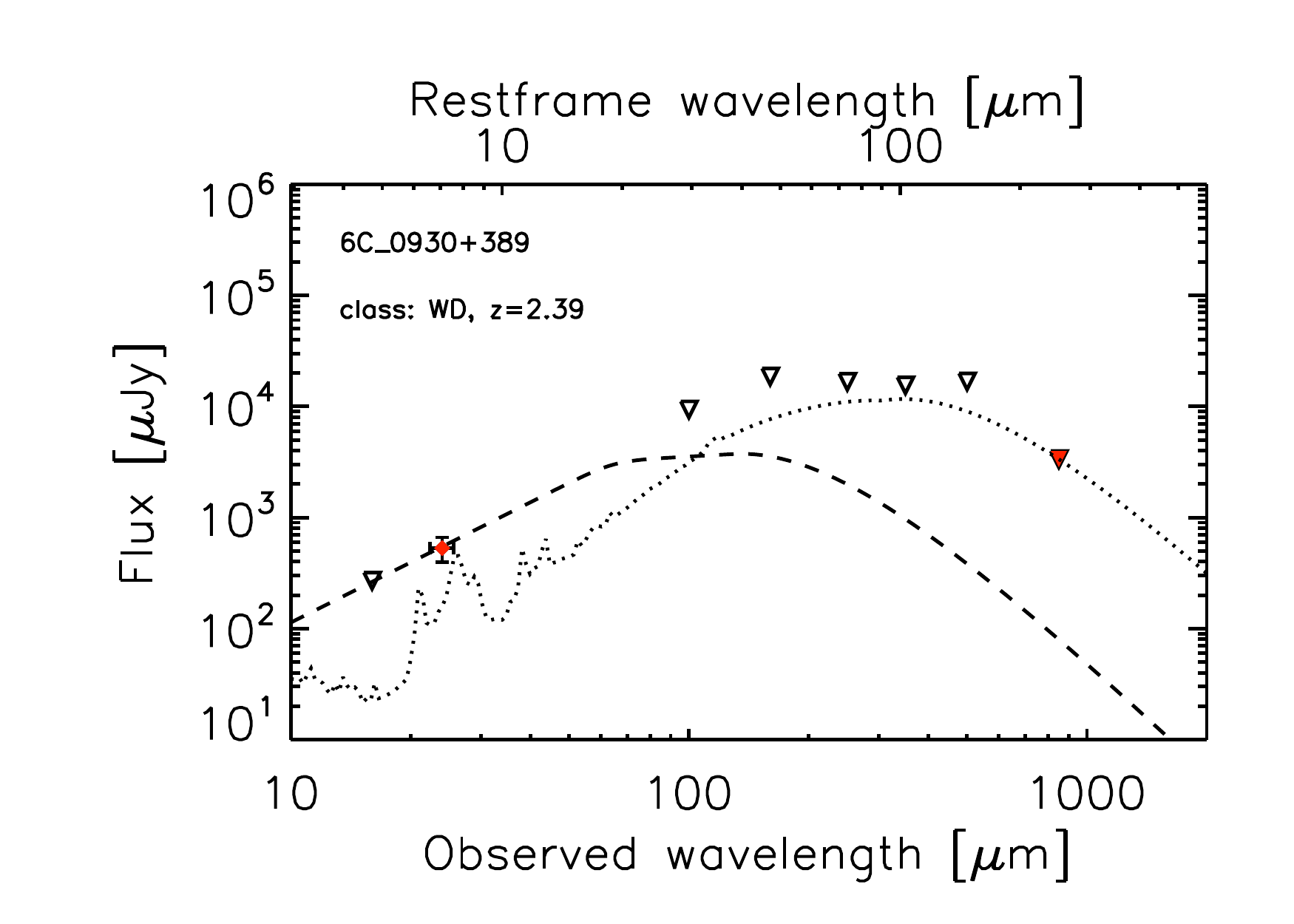}} \\
      24\mum & 100\mum & 160\mum & 250\mum & 350\mum & 500\mum \\
\end{tabular}
\end{minipage}
\begin{minipage}[c]{0.46\linewidth} 
\begin{tabular}{c@{}c@{}c@{}c@{}c@{}c@{}}

      \multicolumn{6}{c}{\includegraphics[width=80mm,trim=30 0 20 0 ,clip=true]{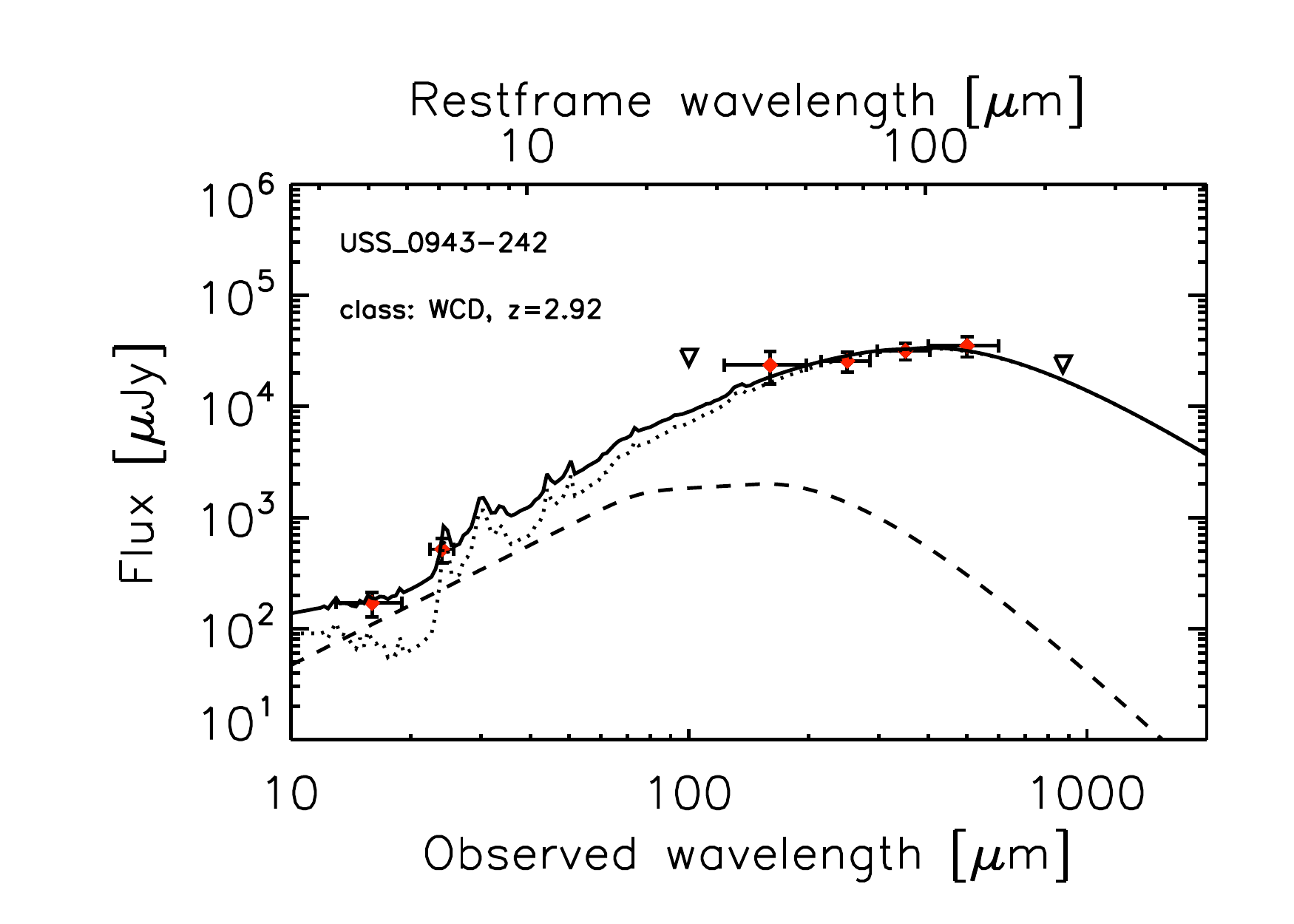}} \\
      24\mum & 100\mum & 160\mum & 250\mum & 350\mum & 500\mum \\
      \includegraphics[height=14mm,trim=70  -4 65 -10,clip=true]{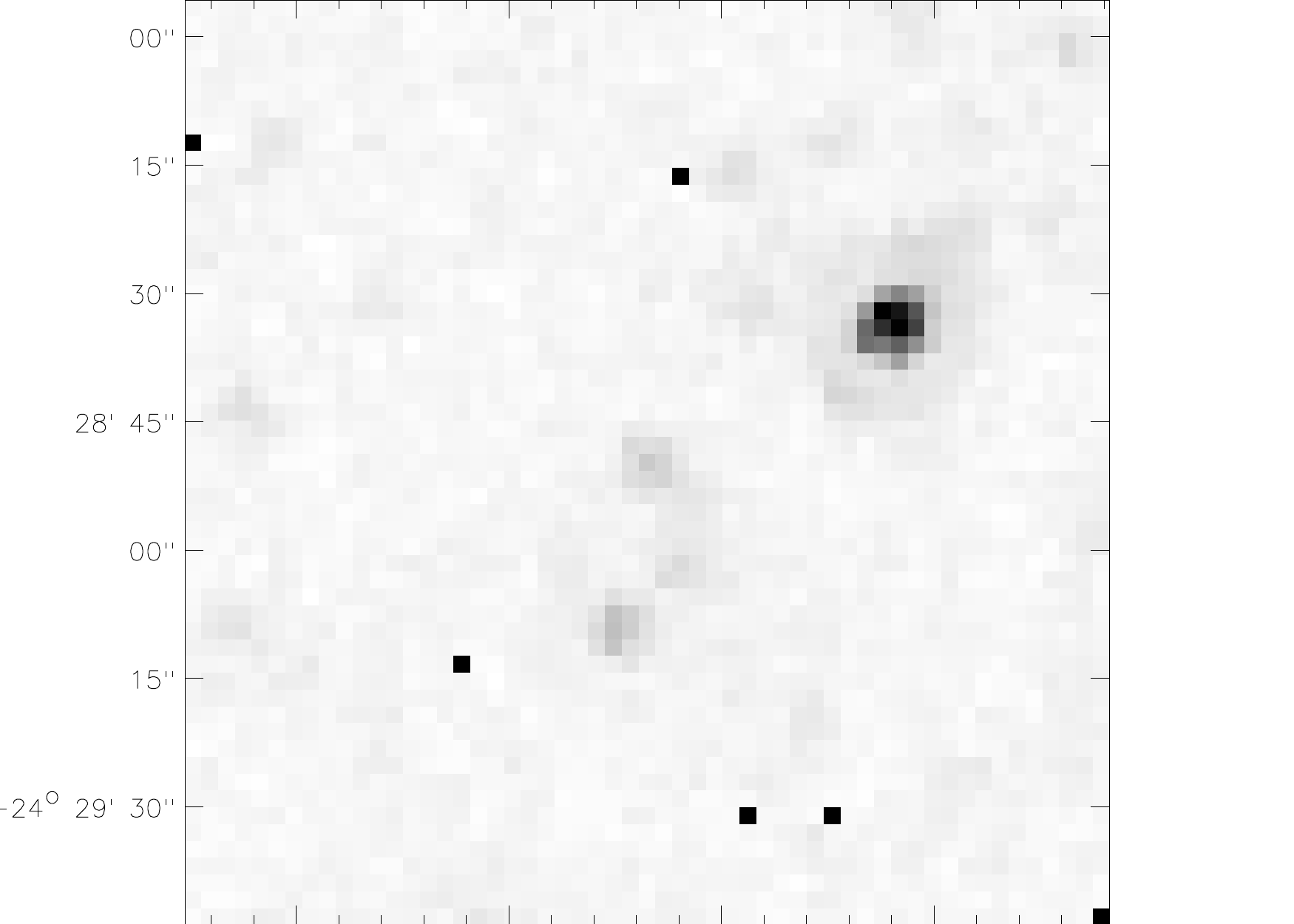} &
      \includegraphics[height=14mm,trim=70  -4 65 -10,clip=true]{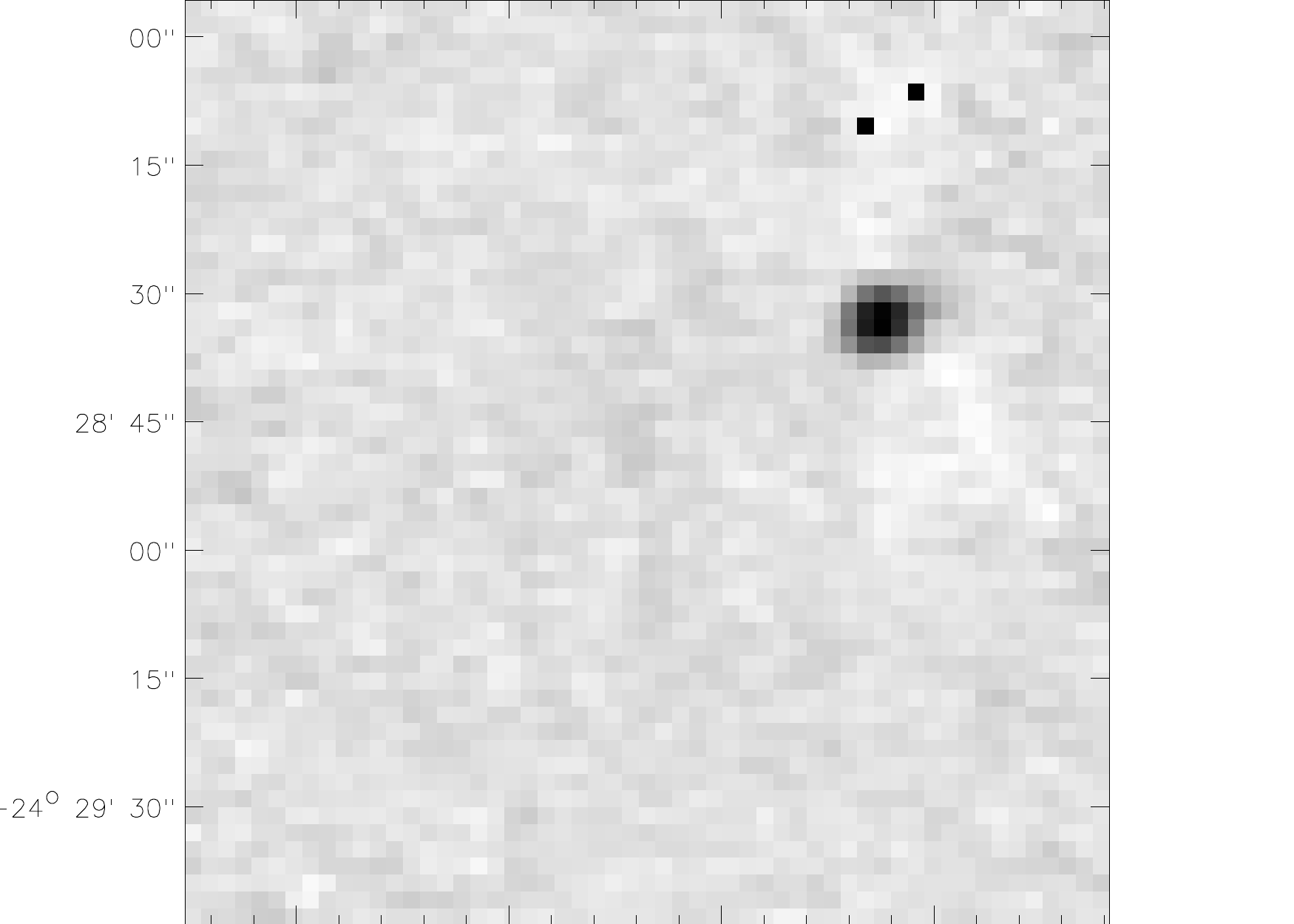} &
      \includegraphics[height=14mm,trim=70  -4 65 -10,clip=true]{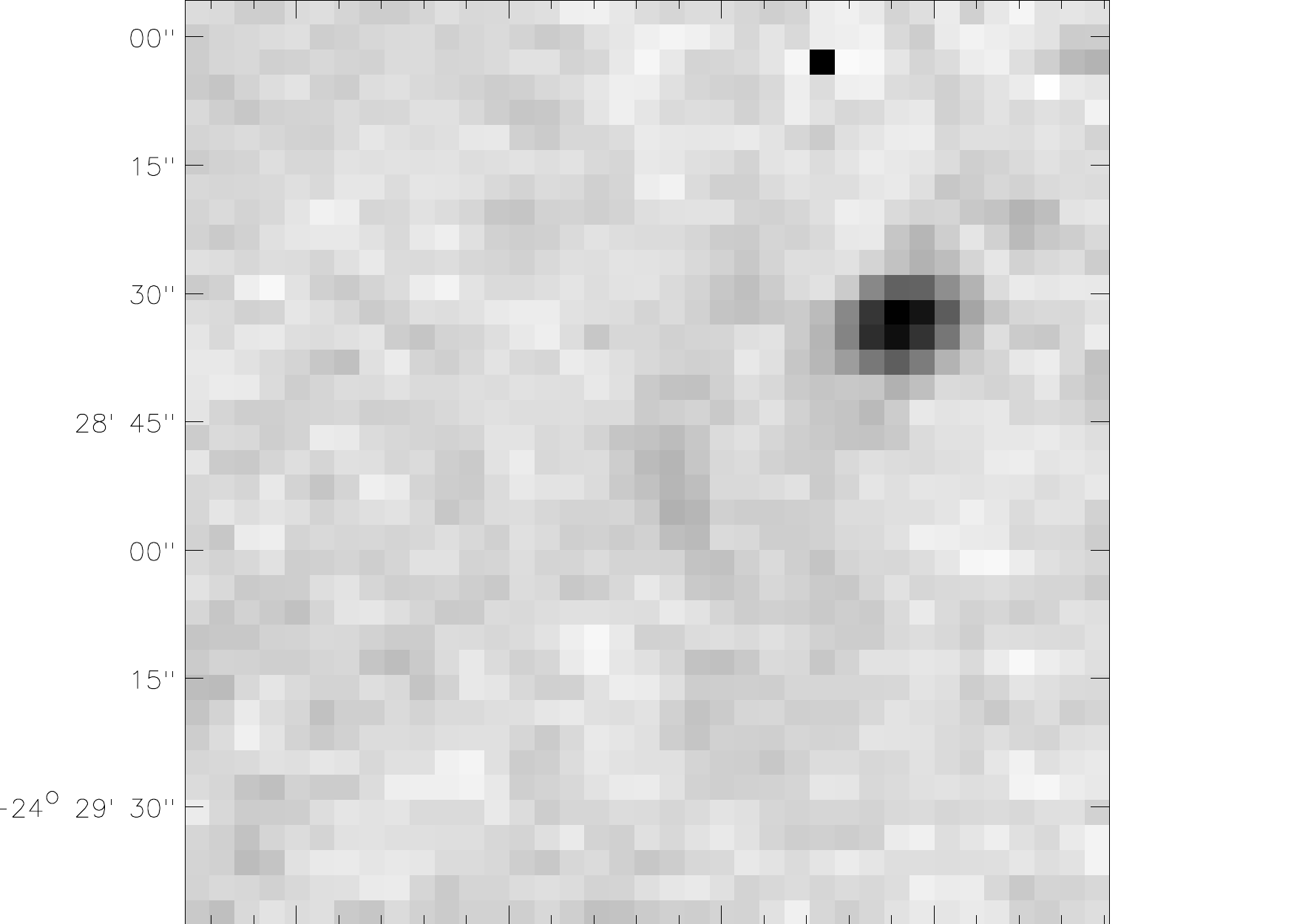} &
      \includegraphics[height=14mm,trim=70  -4 65 -10,clip=true]{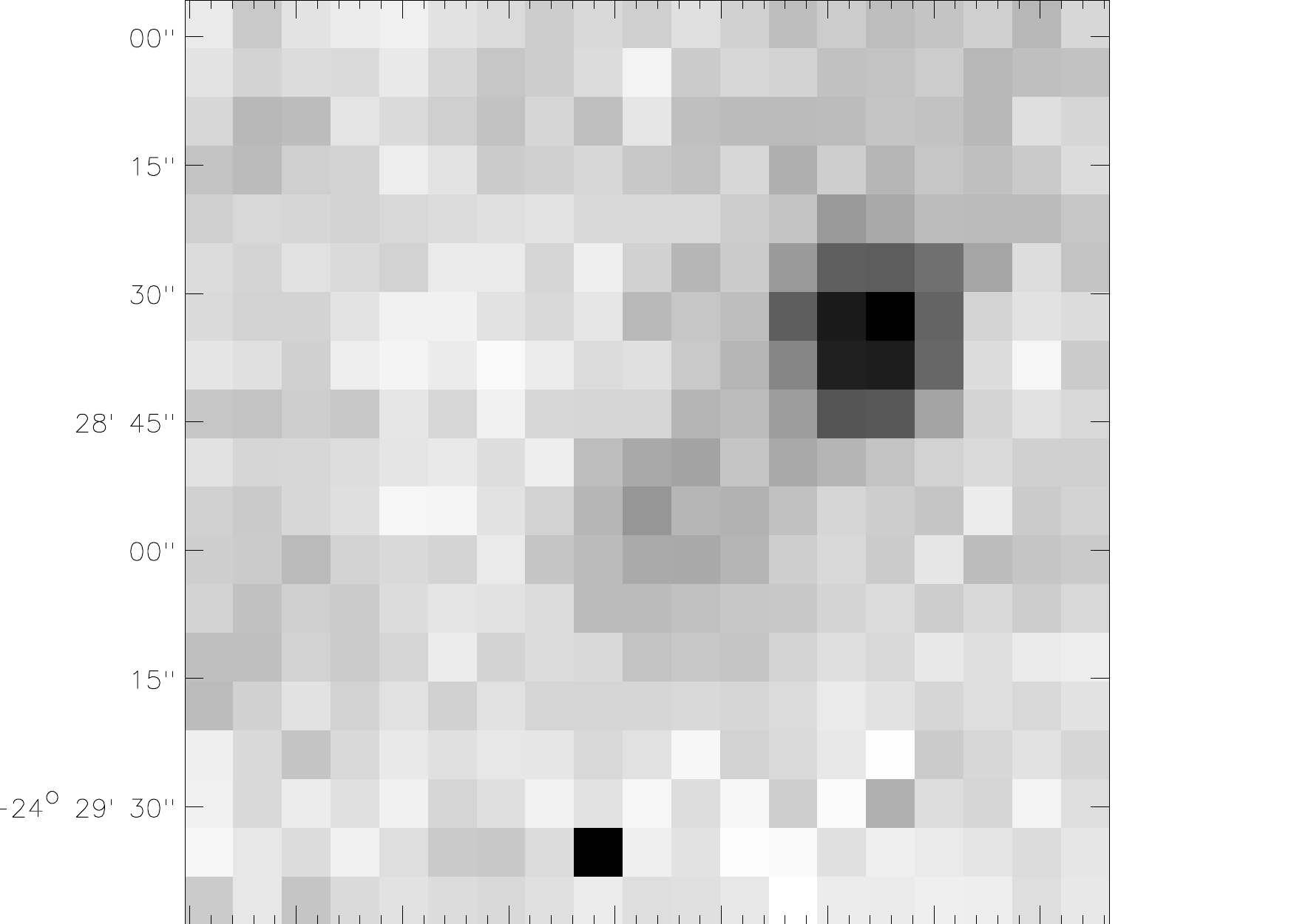}   &
      \includegraphics[height=14mm,trim=70  -4 65 -10,clip=true]{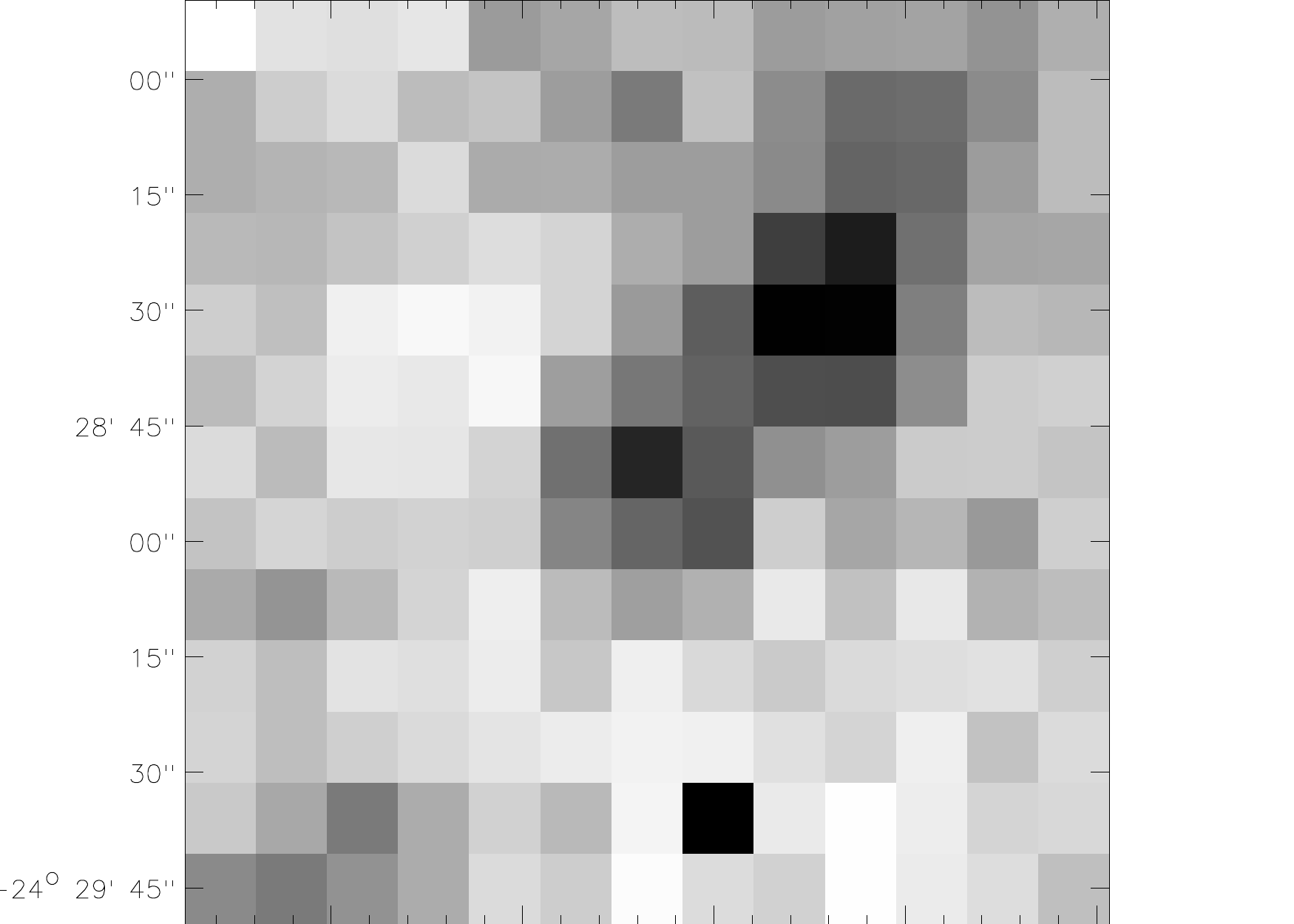}   &
      \includegraphics[height=14mm,trim=70  -4 65 -10,clip=true]{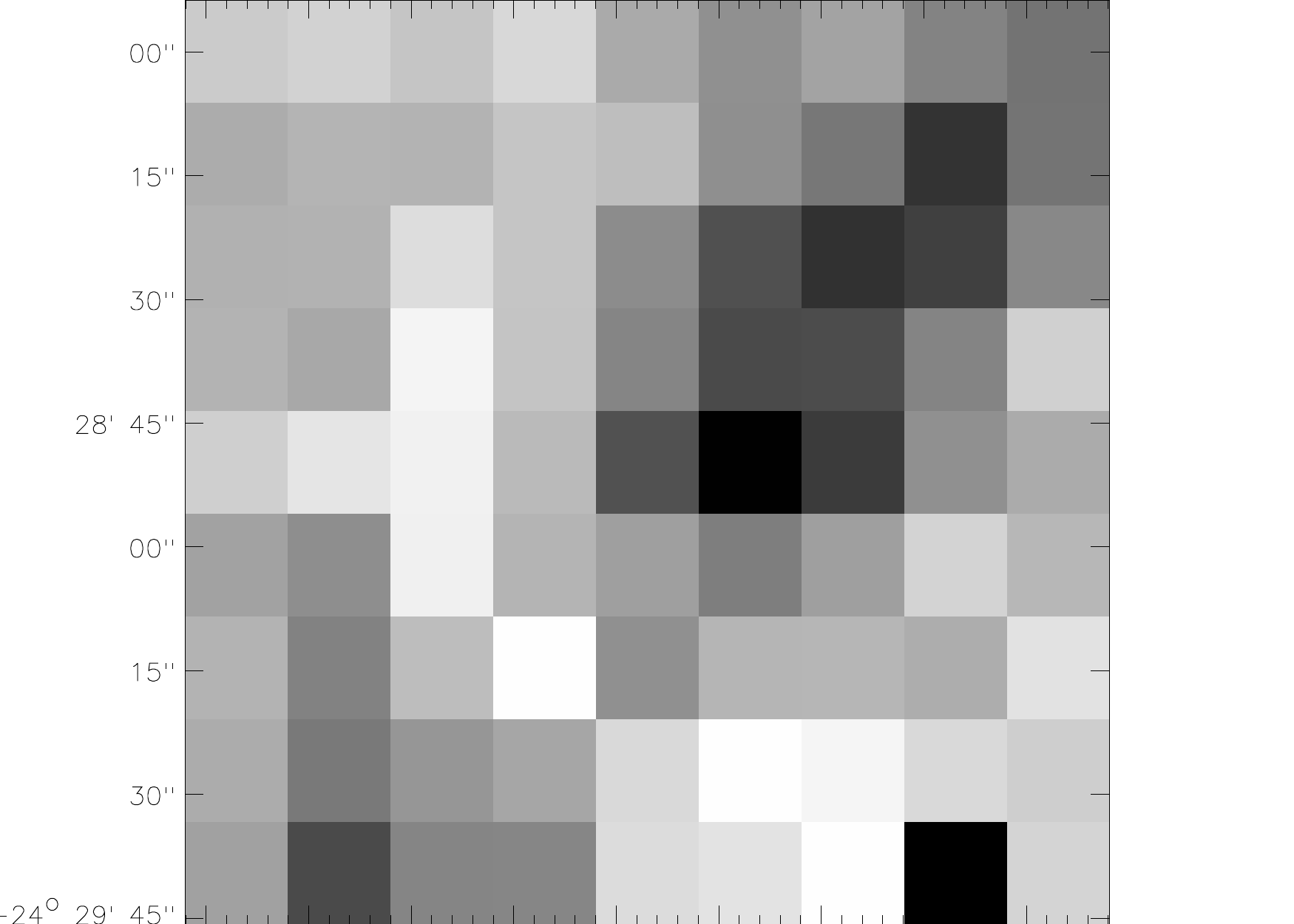} \\

      \multicolumn{6}{c}{\includegraphics[width=80mm,trim=30 0 20 0 ,clip=true]{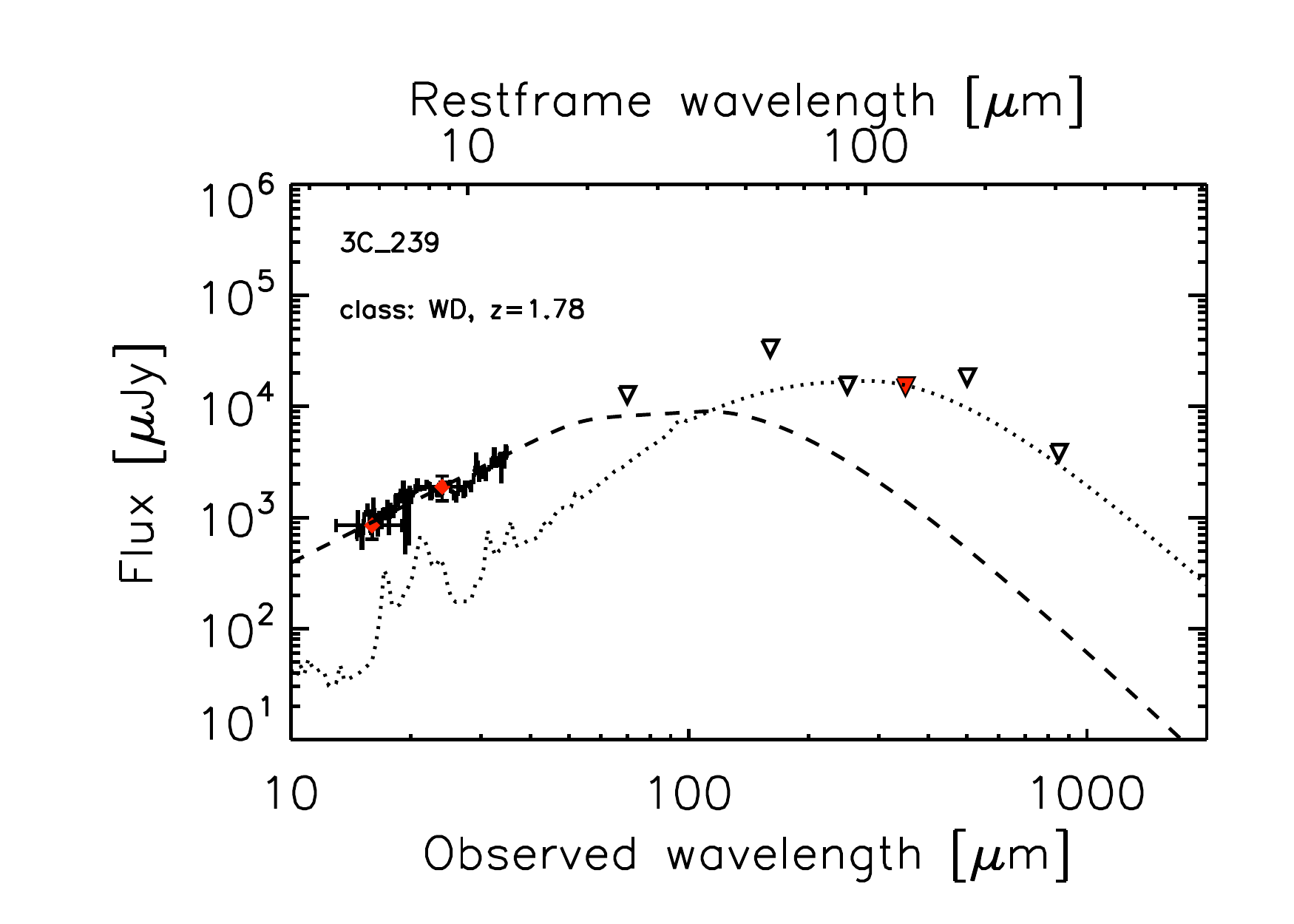}} \\
      24\mum & 70\mum & 160\mum & 250\mum & 350\mum & 500\mum \\
      \includegraphics[height=14mm,trim=70  -4 65 -10,clip=true]{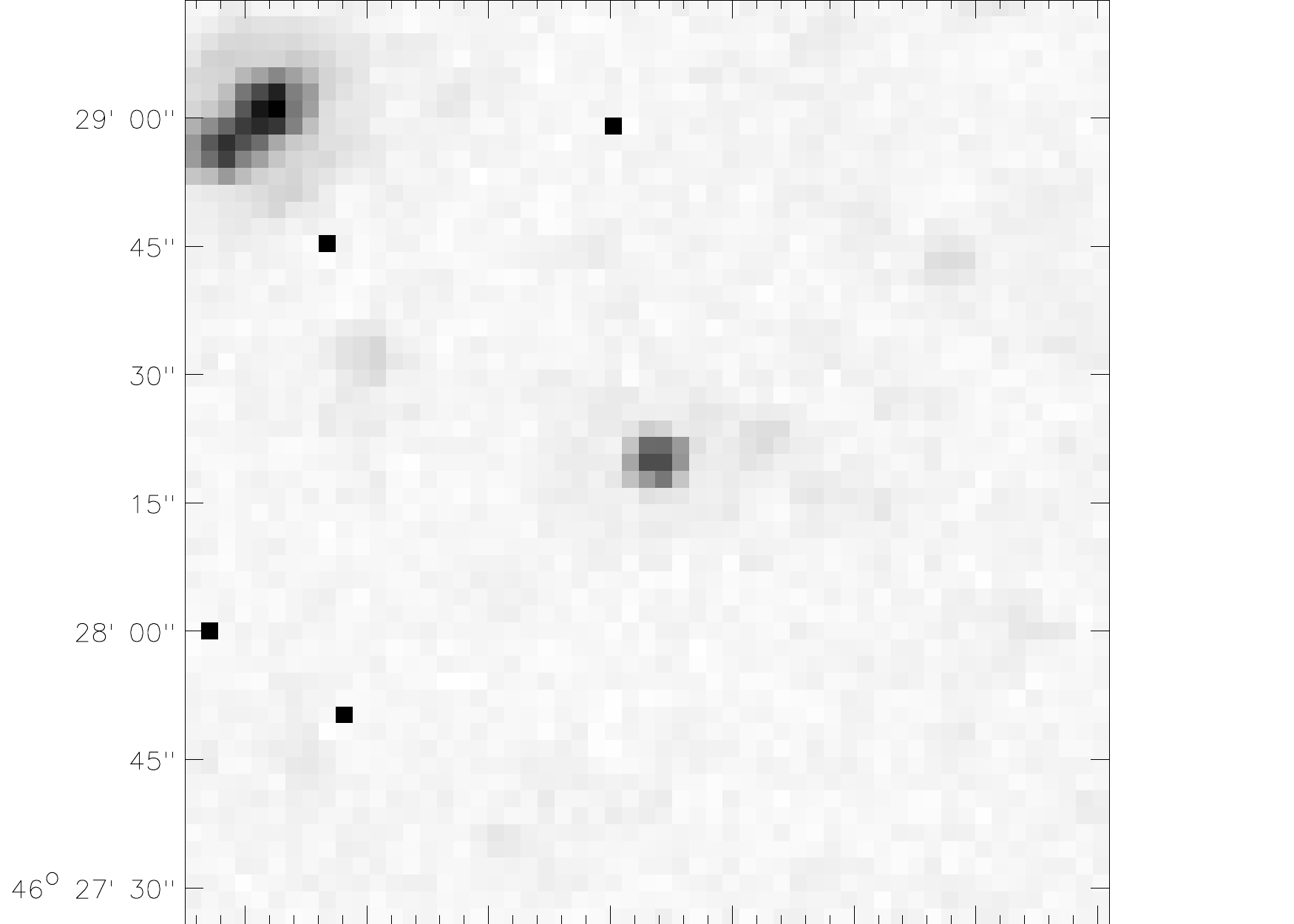} &
      \includegraphics[height=14mm,trim=70  -4 65 -10,clip=true]{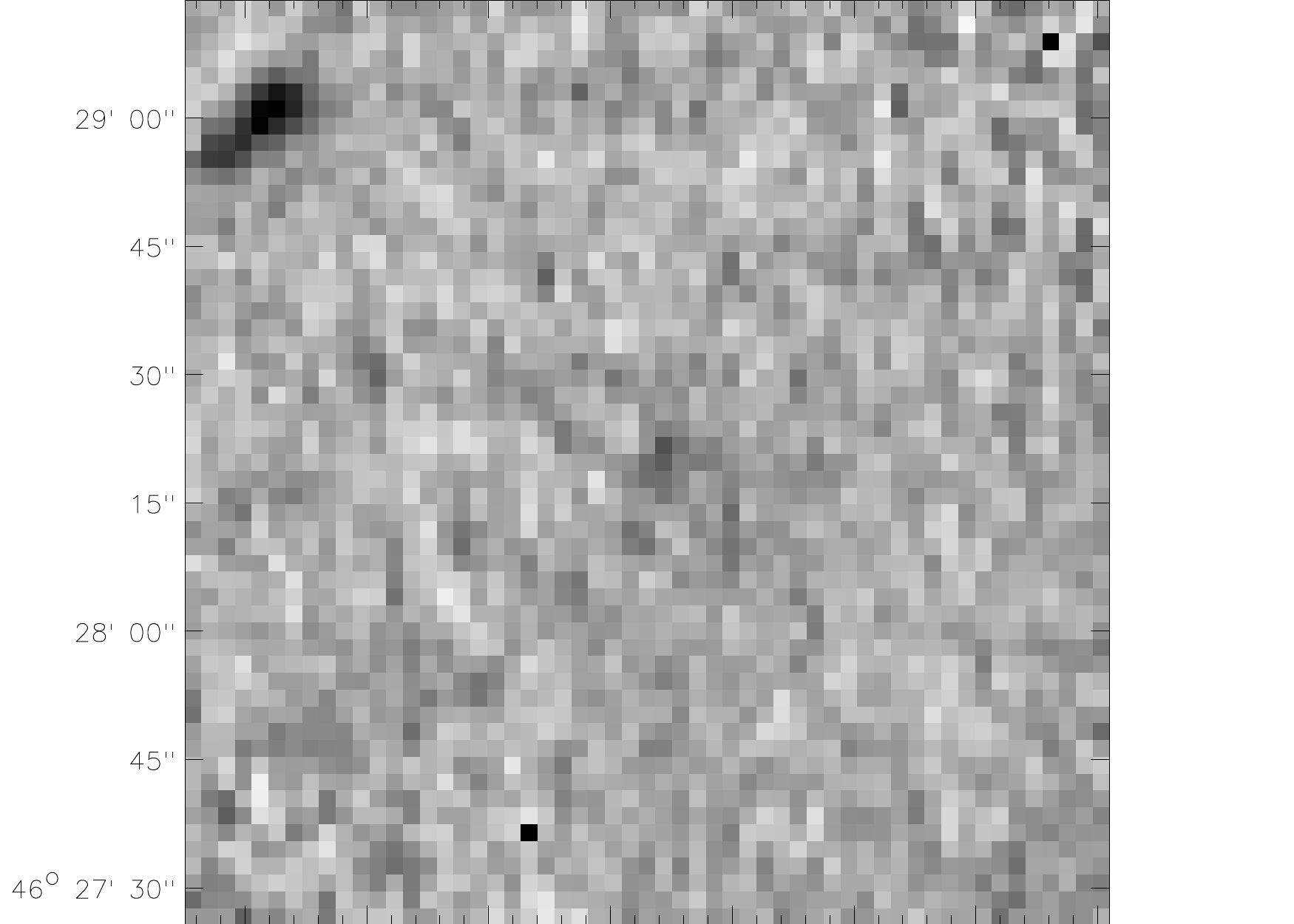} &
      \includegraphics[height=14mm,trim=70  -4 65 -10,clip=true]{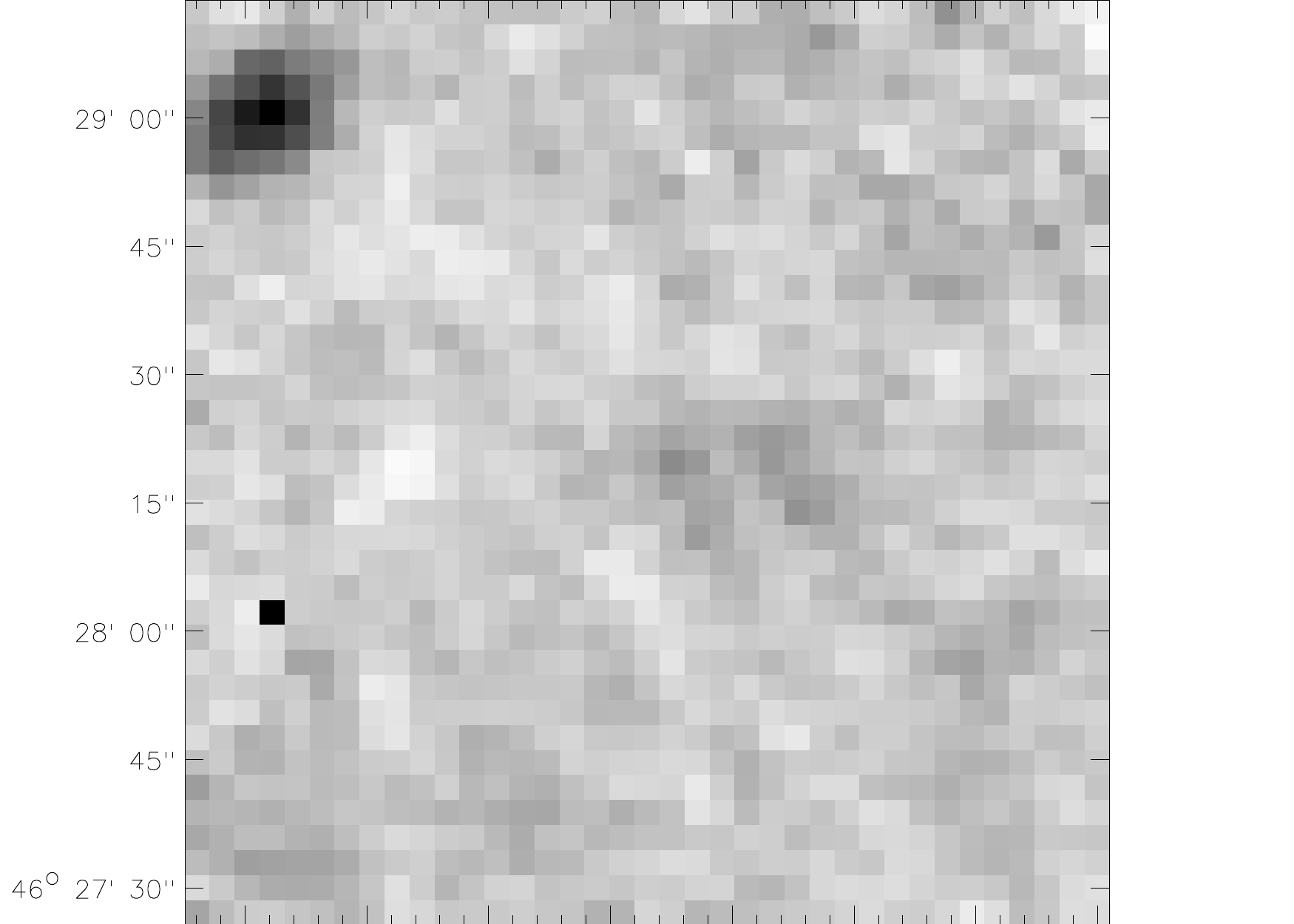} &
      \includegraphics[height=14mm,trim=70  -4 65 -10,clip=true]{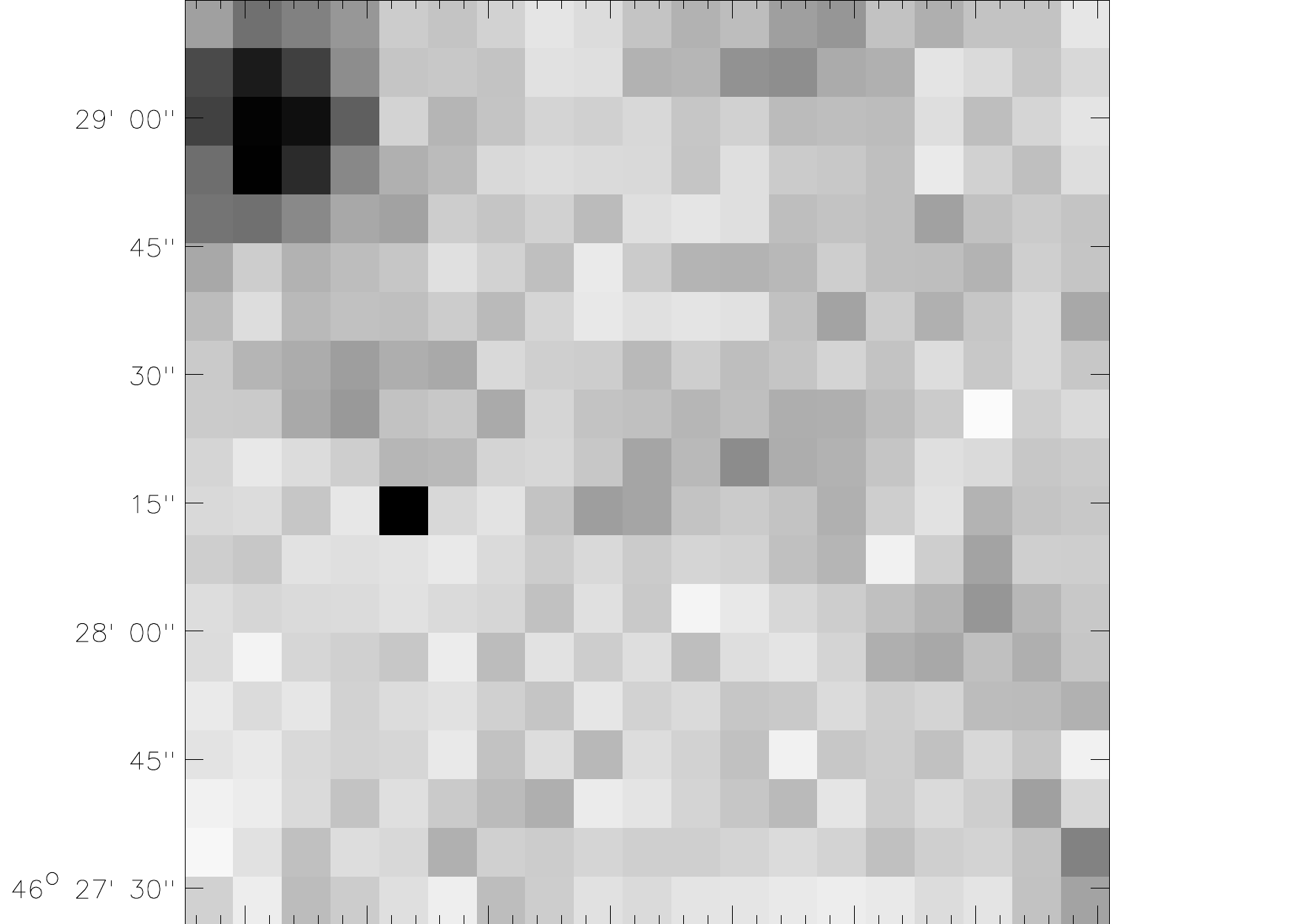}   &
      \includegraphics[height=14mm,trim=70  -4 65 -10,clip=true]{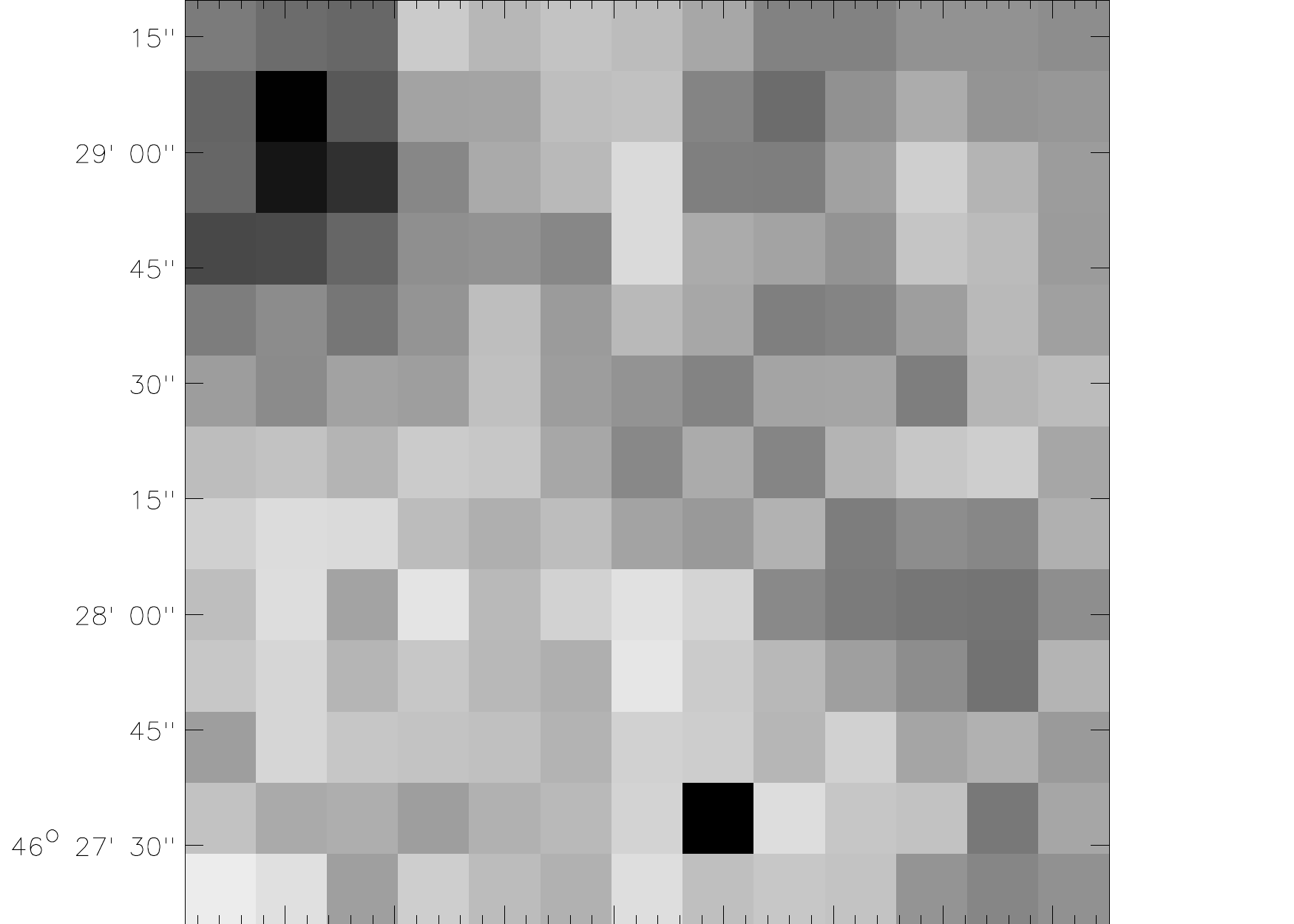}   &
      \includegraphics[height=14mm,trim=70  -4 65 -10,clip=true]{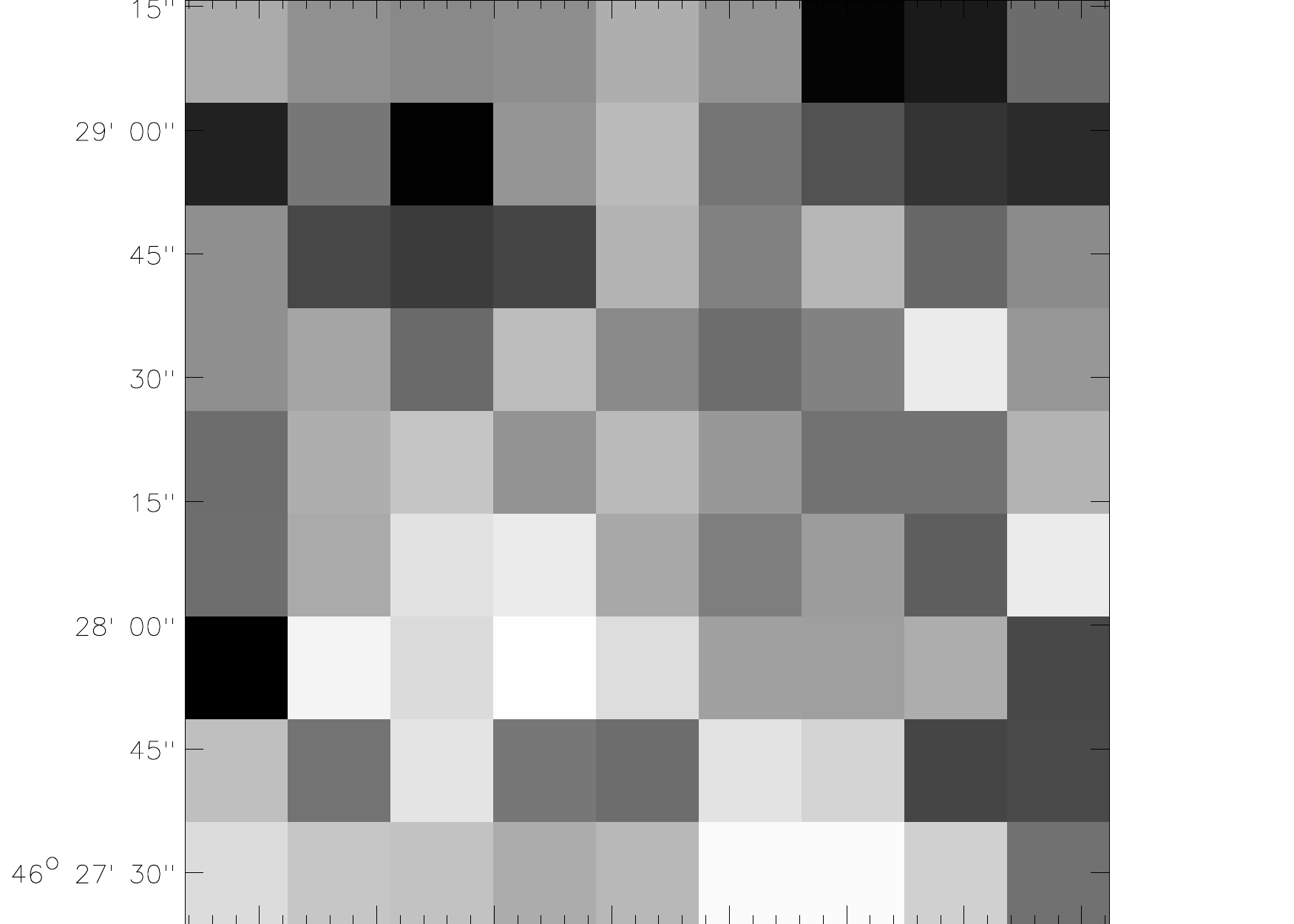} \\

      \multicolumn{6}{c}{\includegraphics[width=80mm,trim=30 0 20 0 ,clip=true]{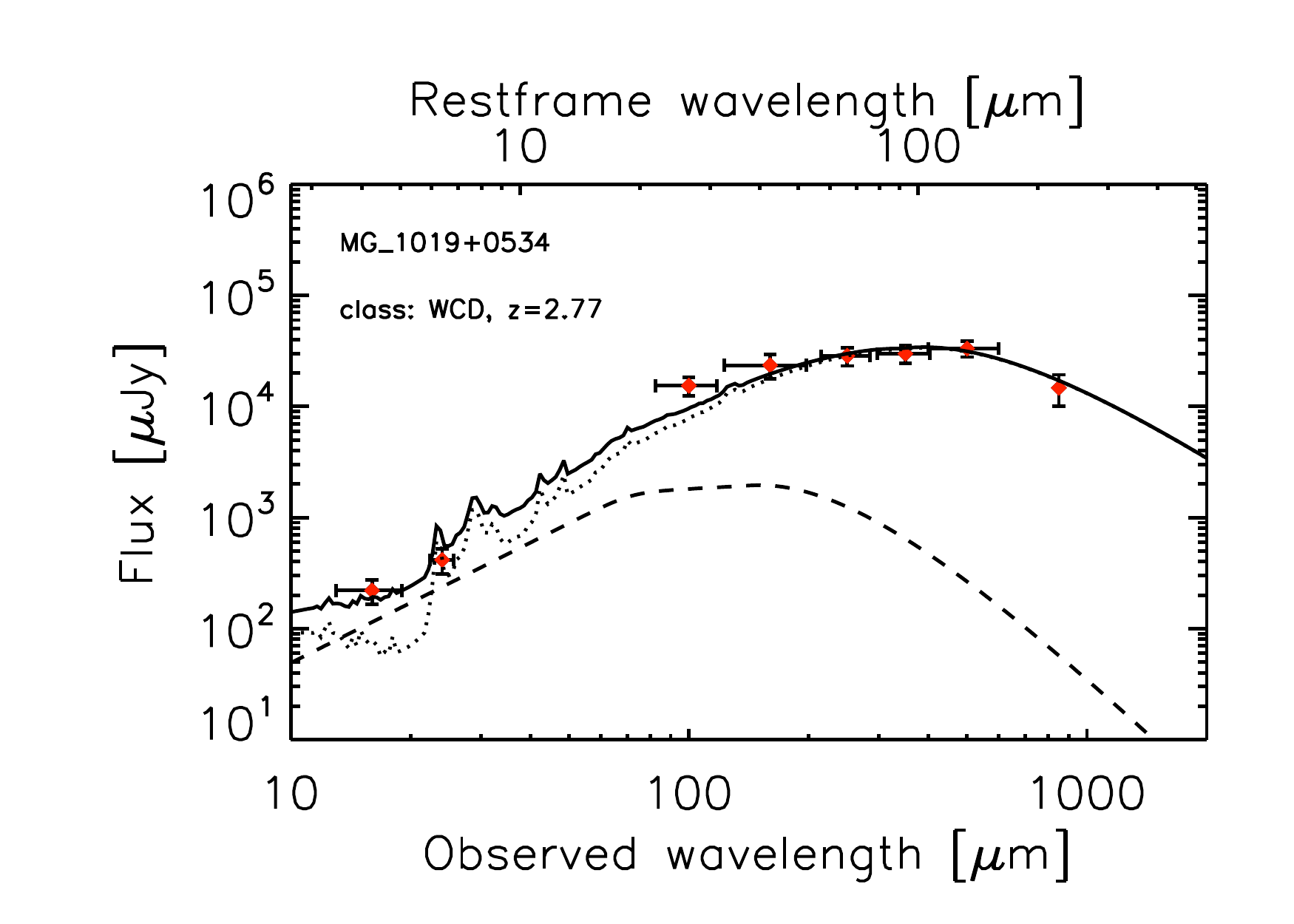}} \\
      24\mum & 100\mum & 160\mum & 250\mum & 350\mum & 500\mum \\
\end{tabular}
          \end{minipage}
  \end{centering}
\end{figure*}

\clearpage

\begin{figure*}[h]
  \begin{centering}
    \begin{minipage}[c]{0.46\linewidth}
    \begin{tabular}{c@{}c@{}c@{}c@{}c@{}c@{}}

      \multicolumn{6}{c}{\includegraphics[width=80mm,trim=30 0 20 0 ,clip=true]{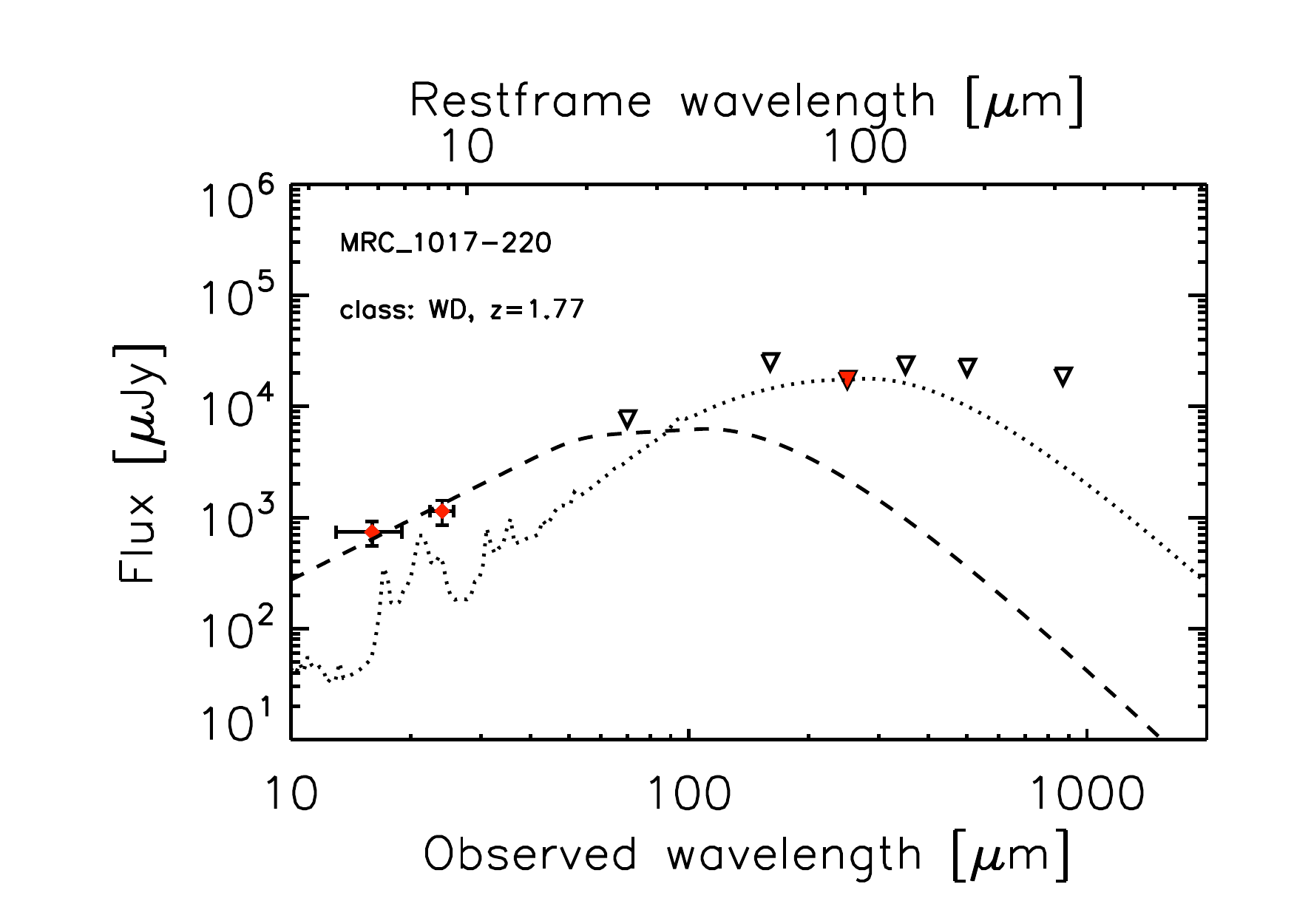}} \\
      24\mum & 70\mum & 160\mum & 250\mum & 350\mum & 500\mum \\
      \includegraphics[height=14mm,trim=70  -4 65 -10,clip=true]{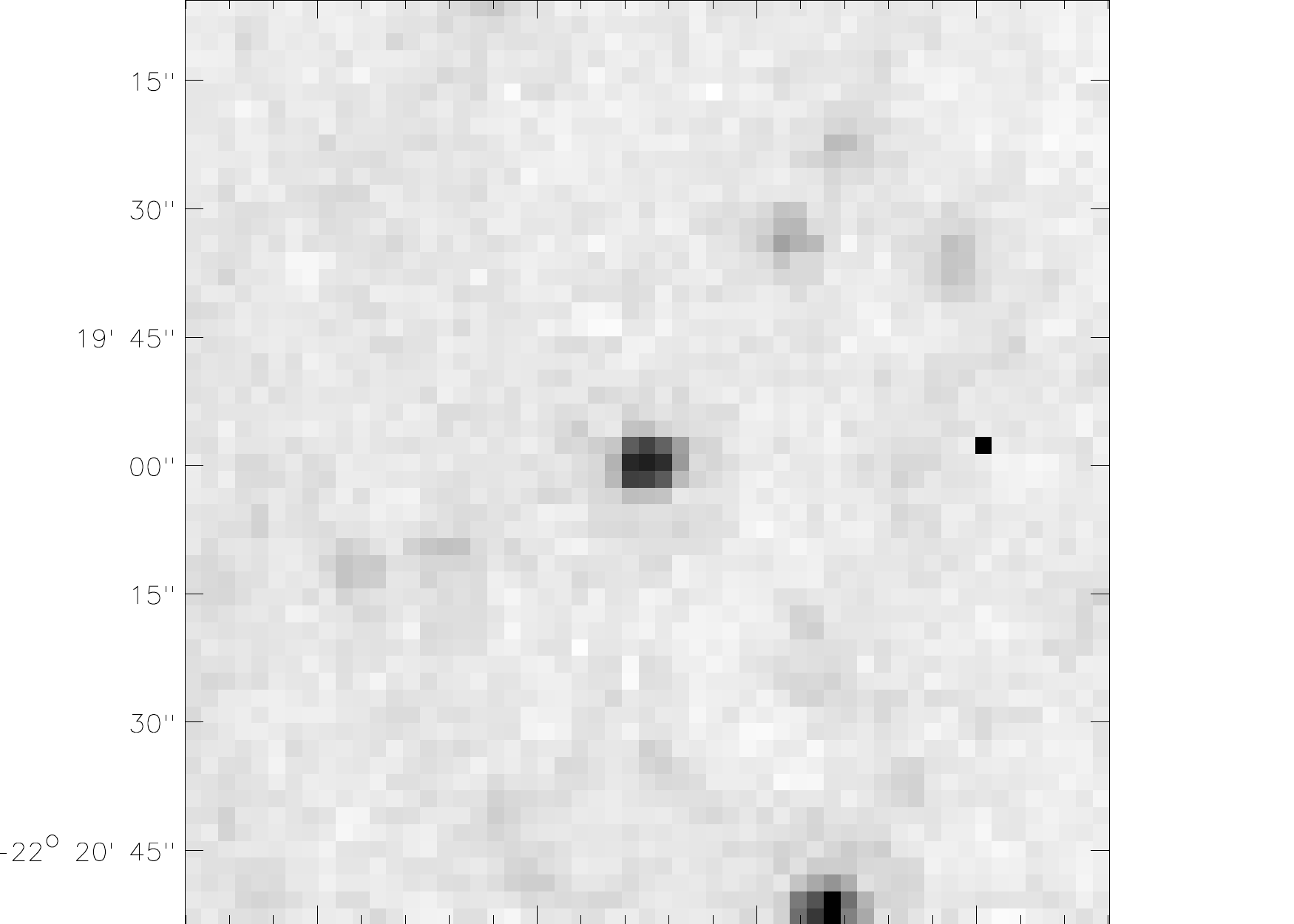} &
      \includegraphics[height=14mm,trim=70  -4 65 -10,clip=true]{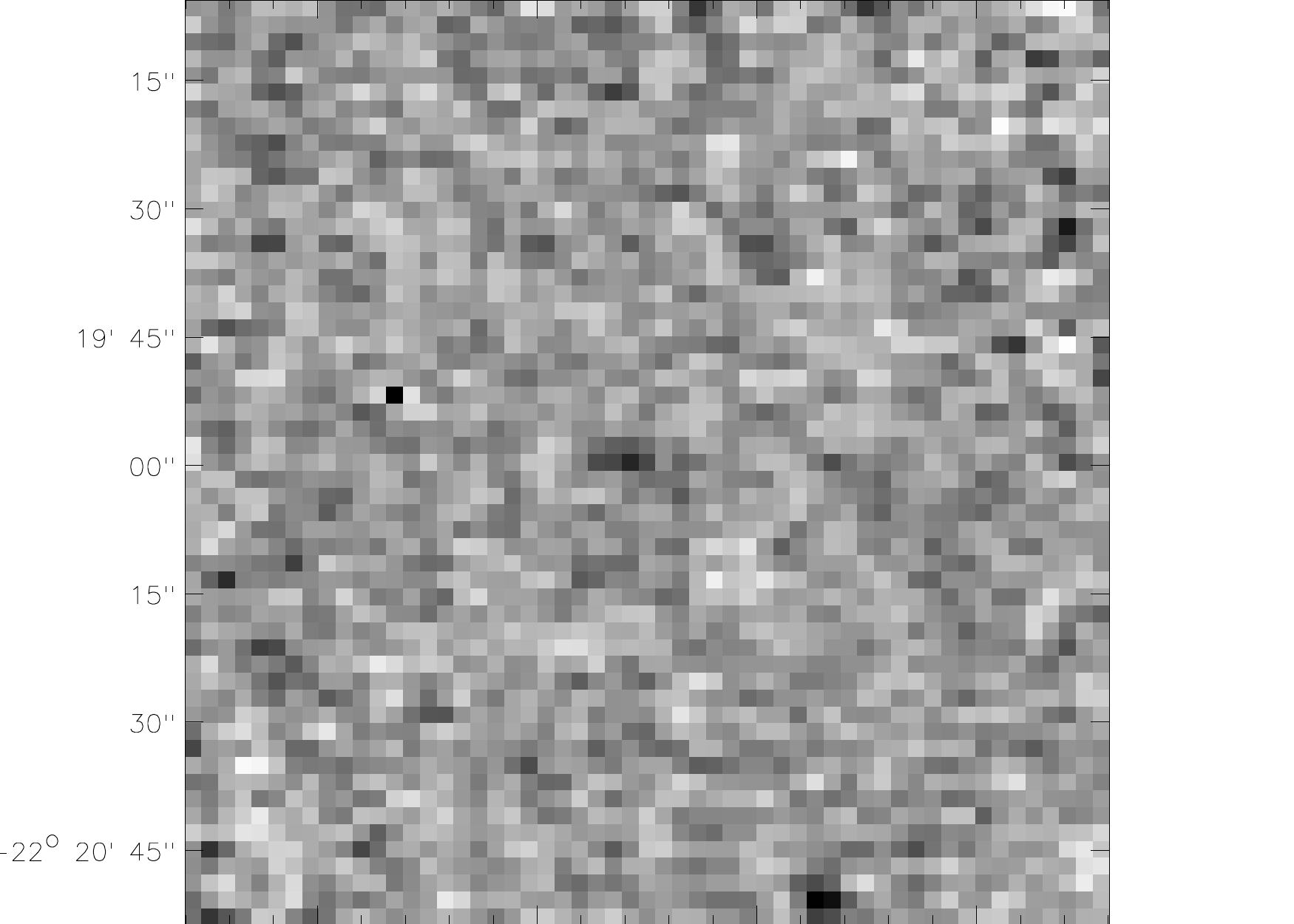} &
      \includegraphics[height=14mm,trim=70  -4 65 -10,clip=true]{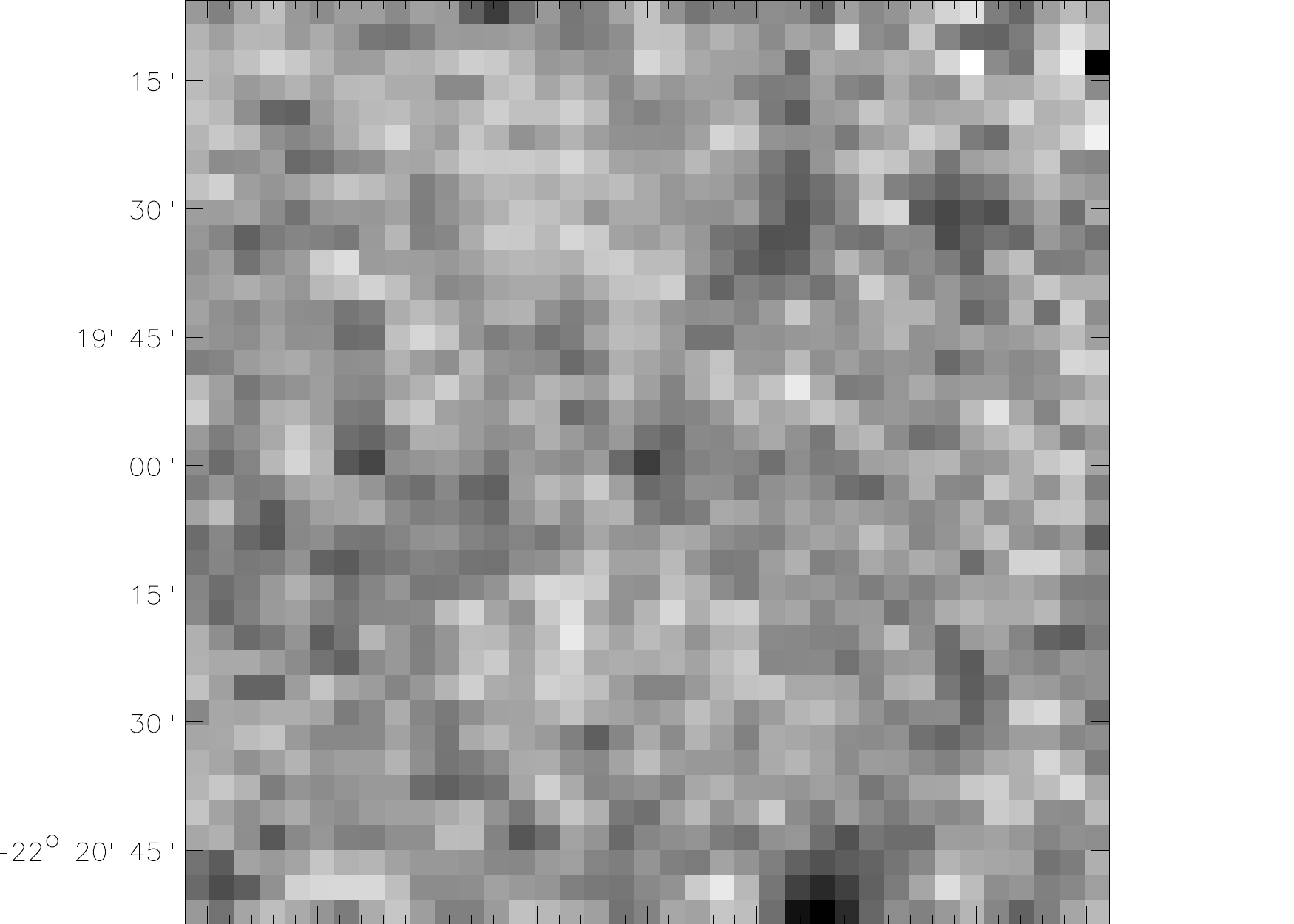} &
      \includegraphics[height=14mm,trim=70  -4 65 -10,clip=true]{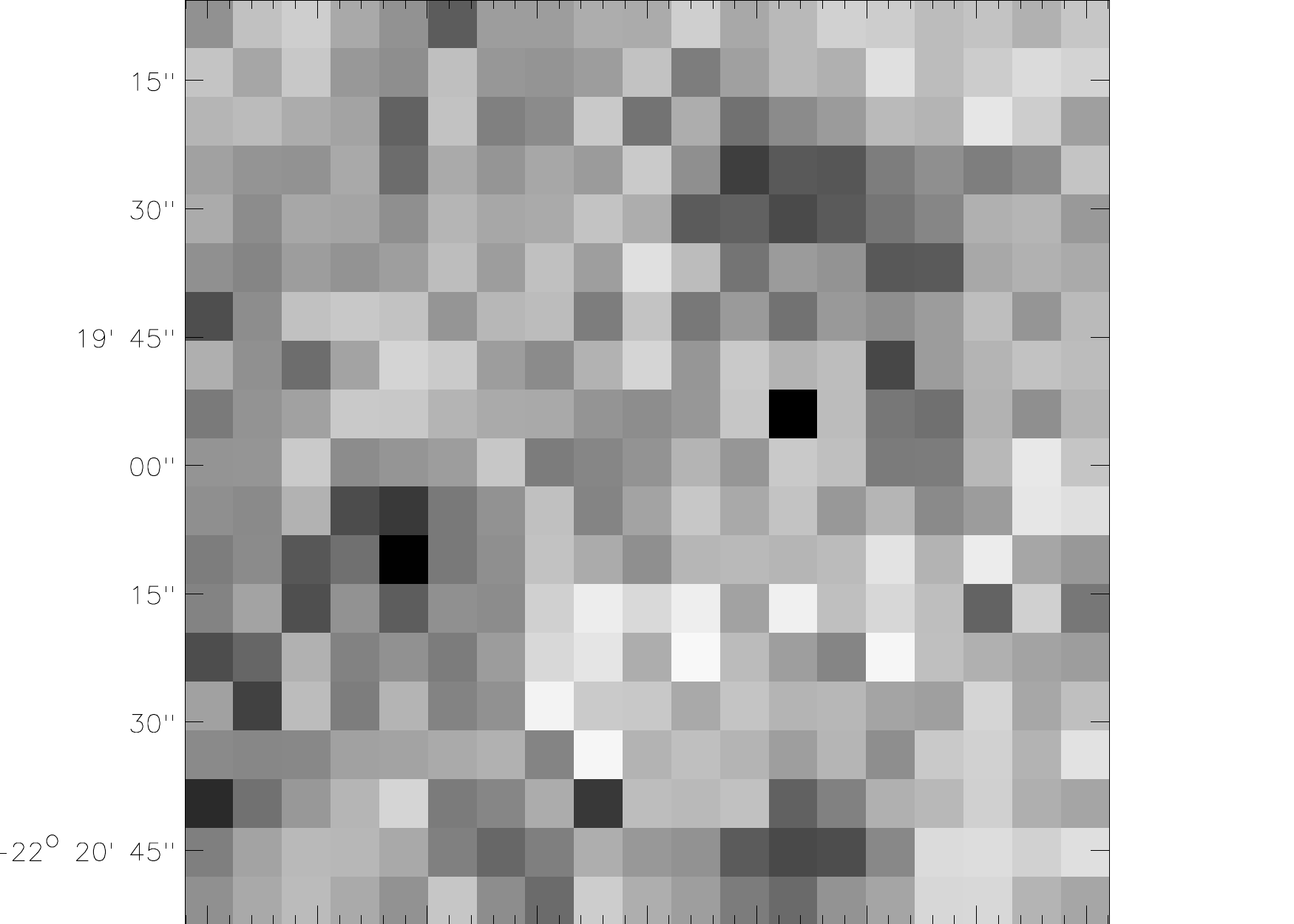}   &
      \includegraphics[height=14mm,trim=70  -4 65 -10,clip=true]{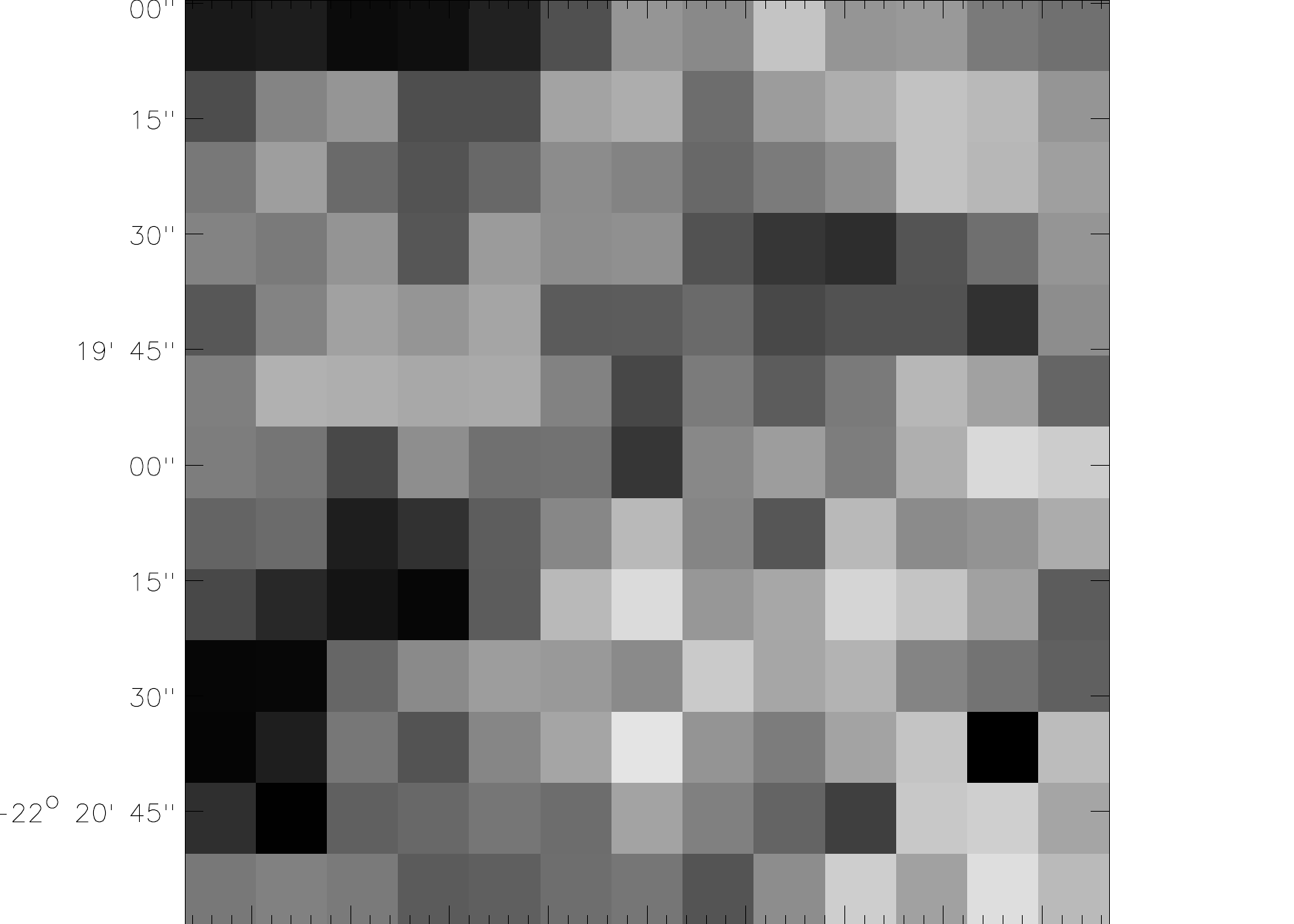}   &
      \includegraphics[height=14mm,trim=70  -4 65 -10,clip=true]{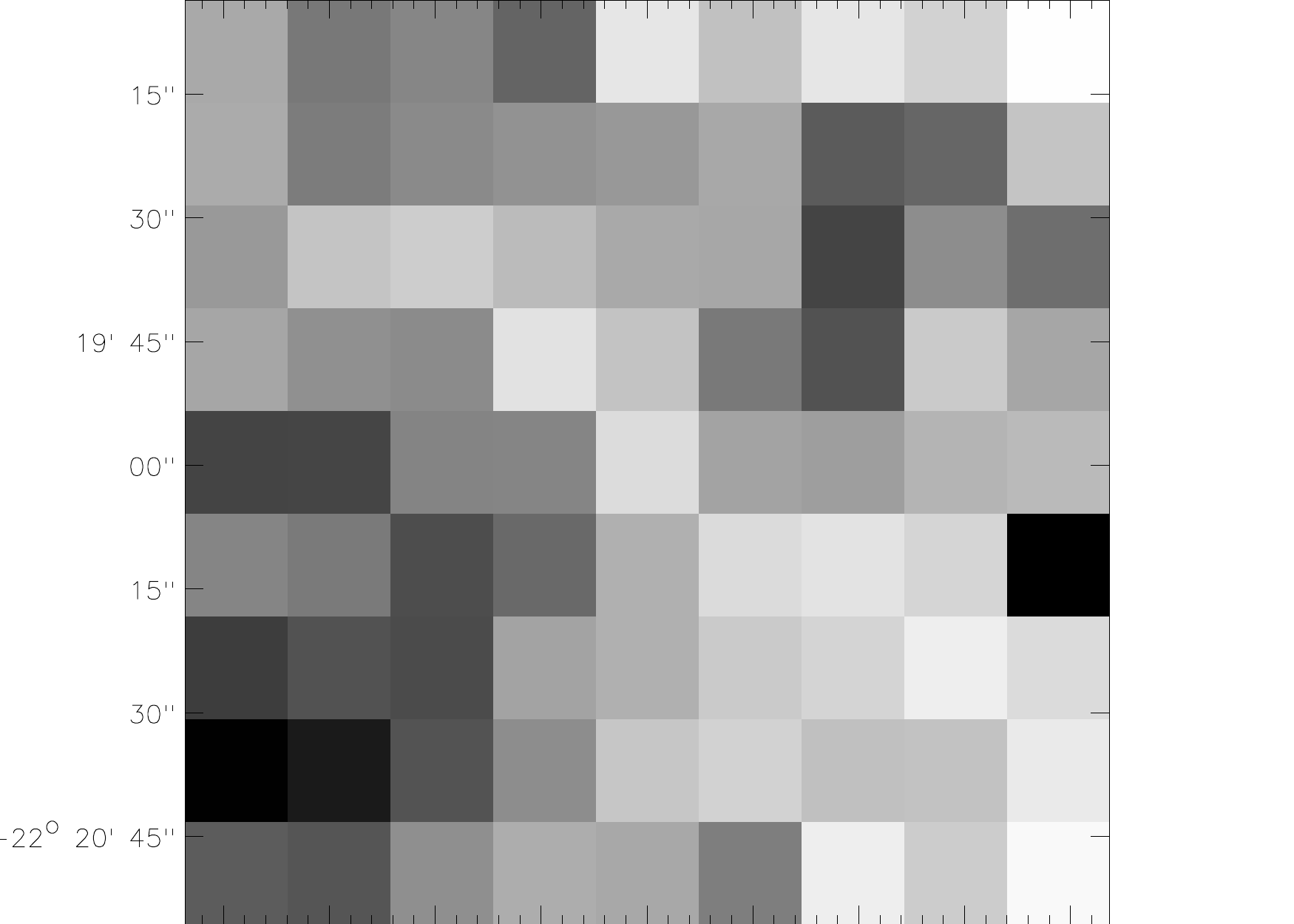} \\

      \multicolumn{6}{c}{\includegraphics[width=80mm,trim=30 0 20 0 ,clip=true]{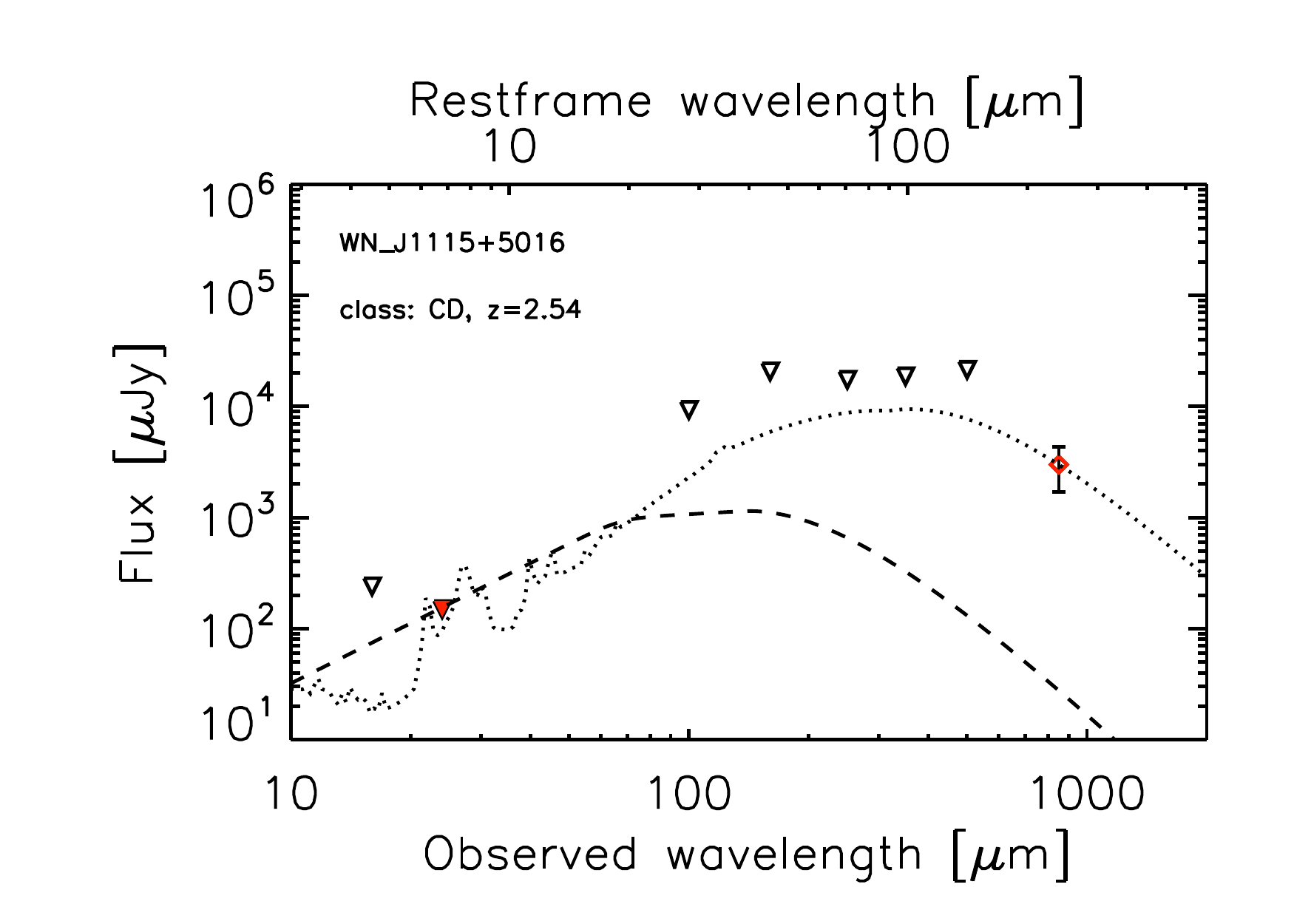}} \\
      24\mum & 100\mum & 160\mum & 250\mum & 350\mum & 500\mum \\
      \includegraphics[height=14mm,trim=70  -4 65 -10,clip=true]{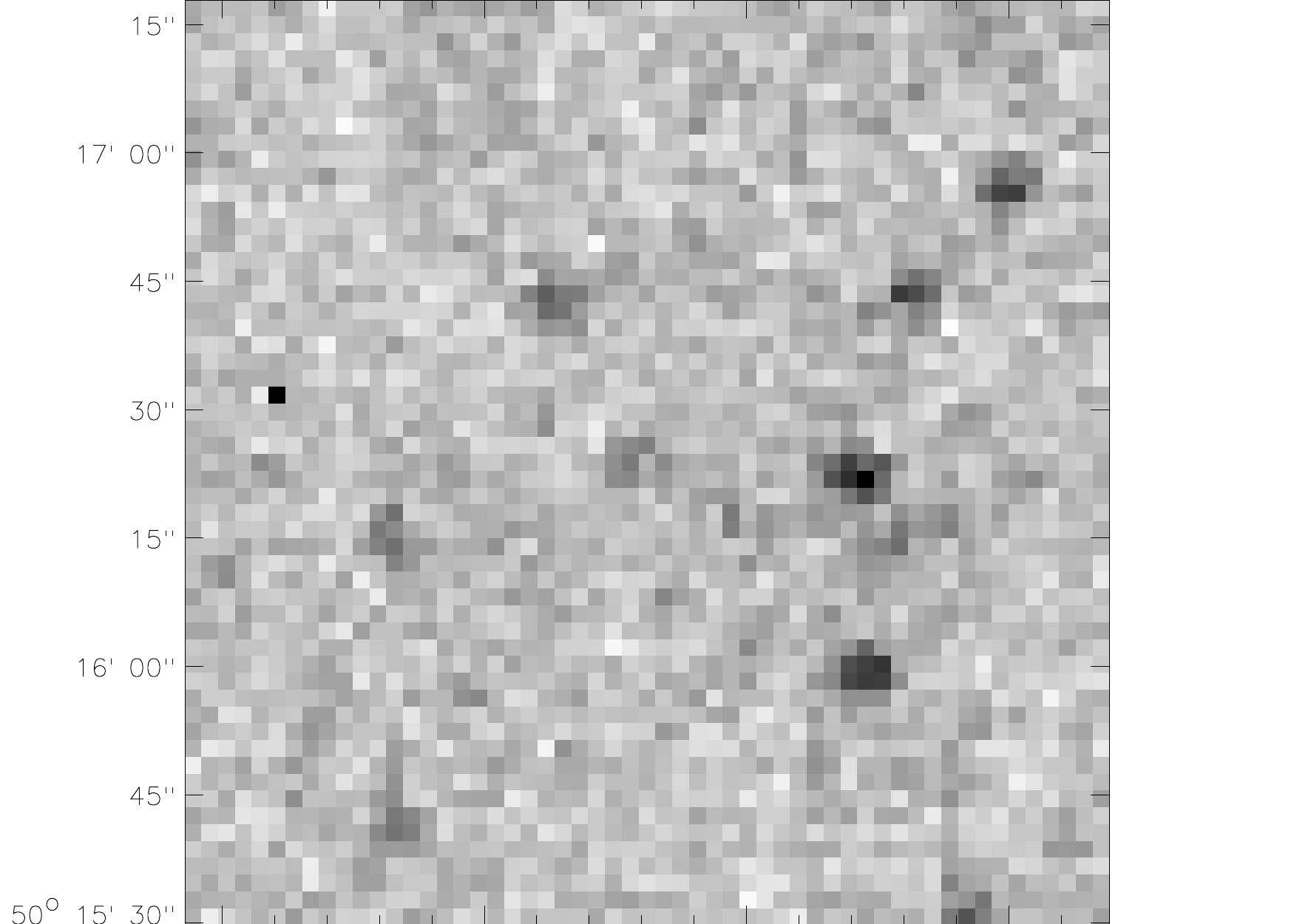} &
      \includegraphics[height=14mm,trim=70  -4 65 -10,clip=true]{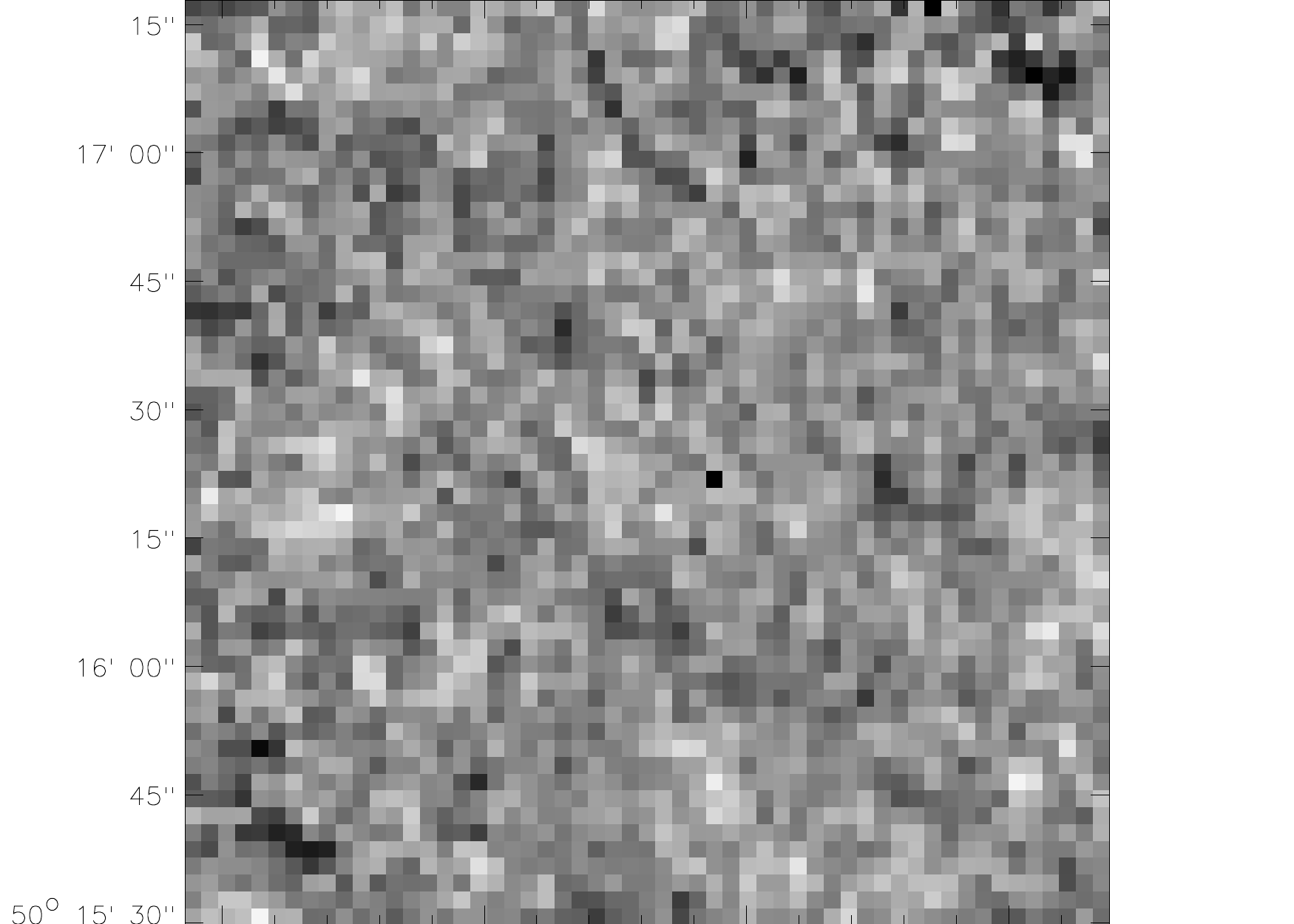} &
      \includegraphics[height=14mm,trim=70  -4 65 -10,clip=true]{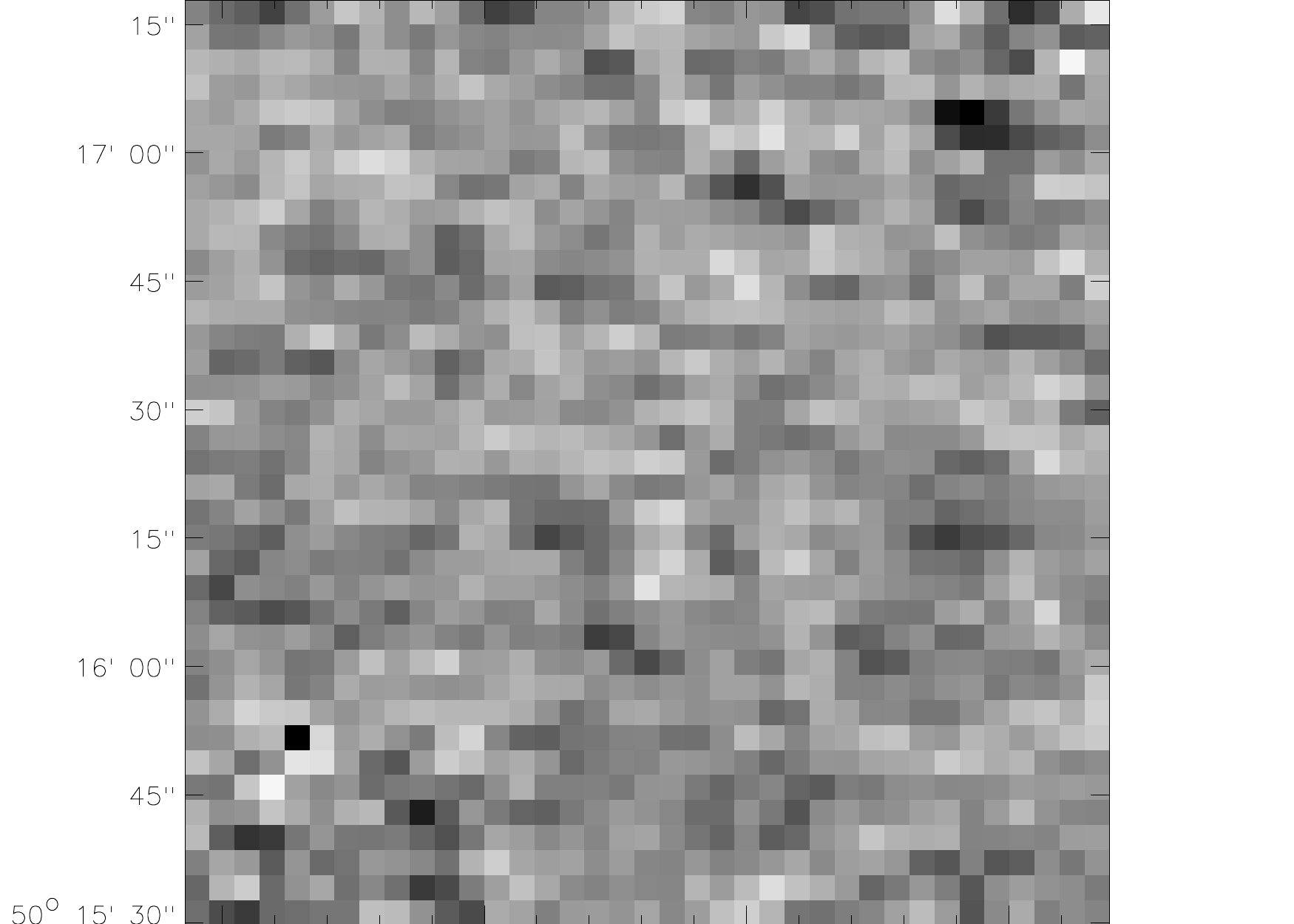} &
      \includegraphics[height=14mm,trim=70  -4 65 -10,clip=true]{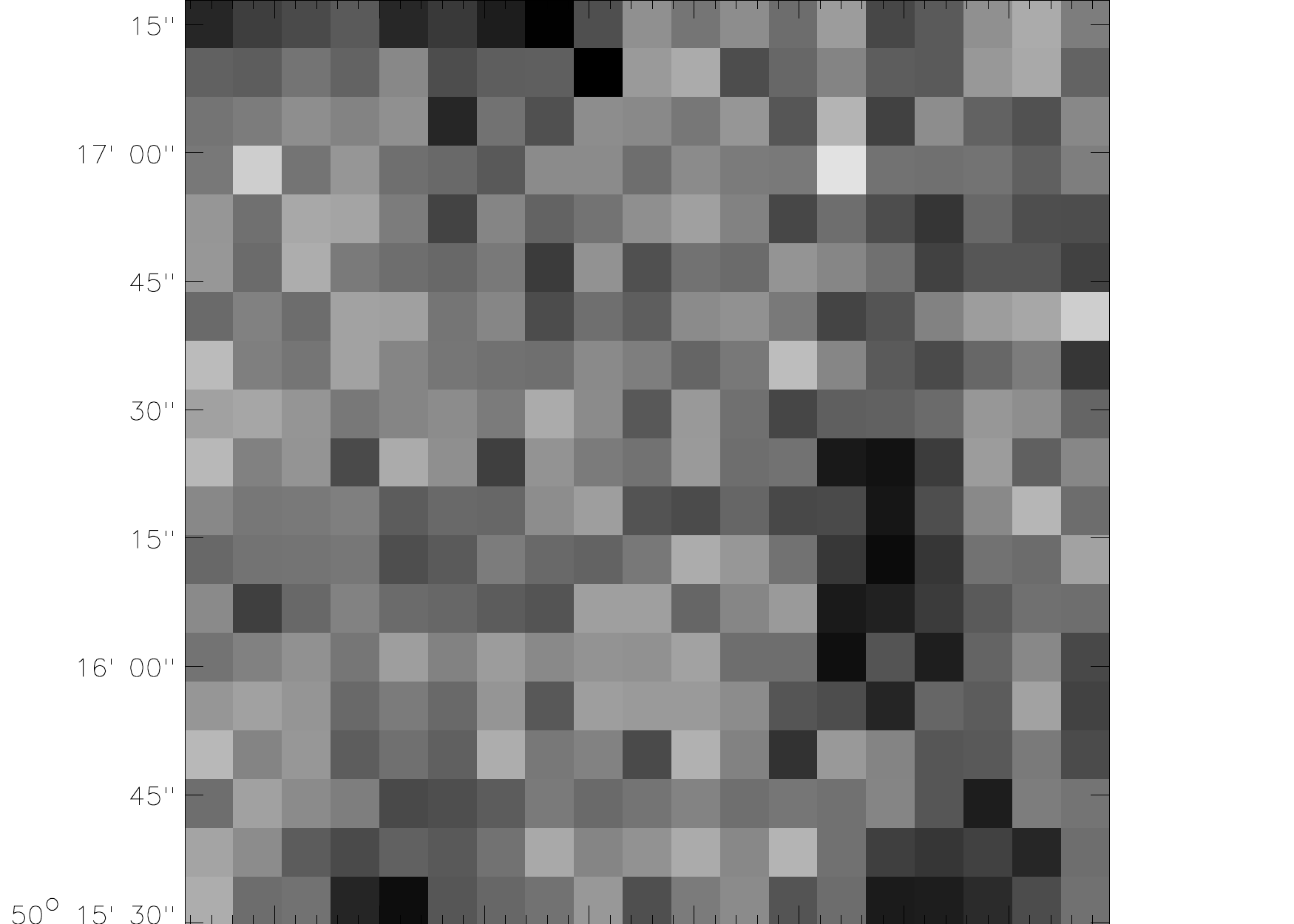}   &
      \includegraphics[height=14mm,trim=70  -4 65 -10,clip=true]{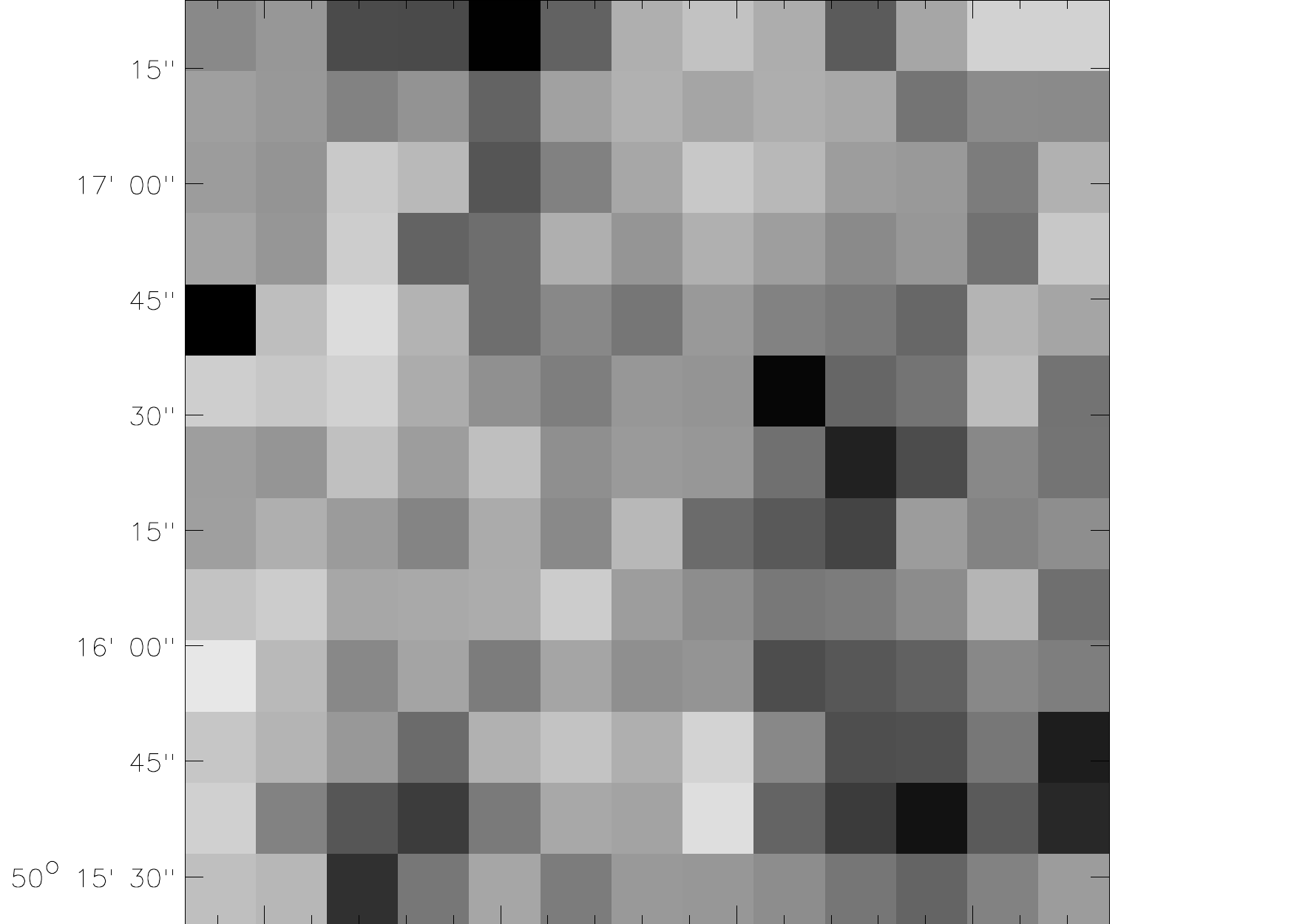}   &
      \includegraphics[height=14mm,trim=70  -4 65 -10,clip=true]{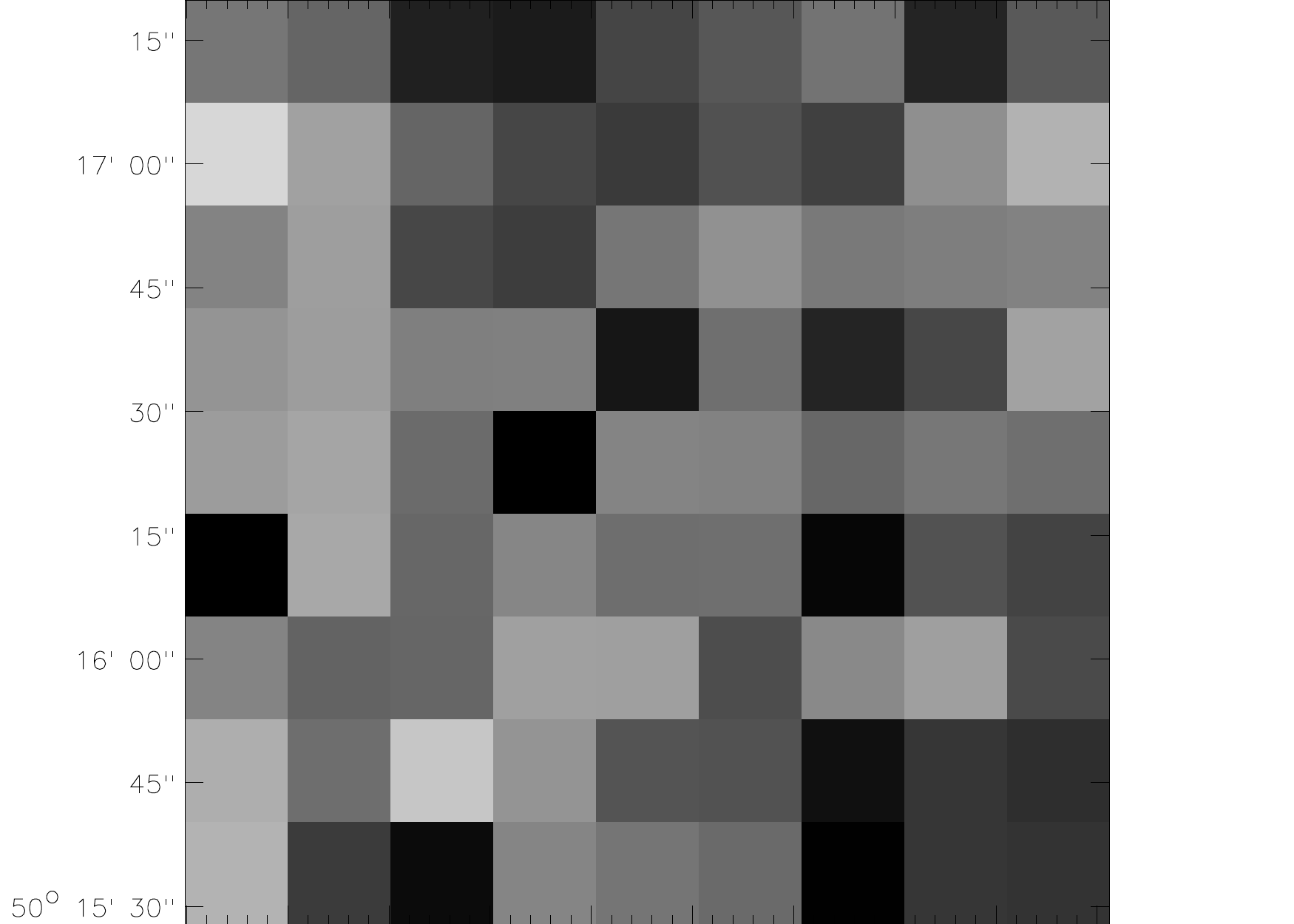} \\

      \multicolumn{6}{c}{\includegraphics[width=80mm,trim=30 0 20 0 ,clip=true]{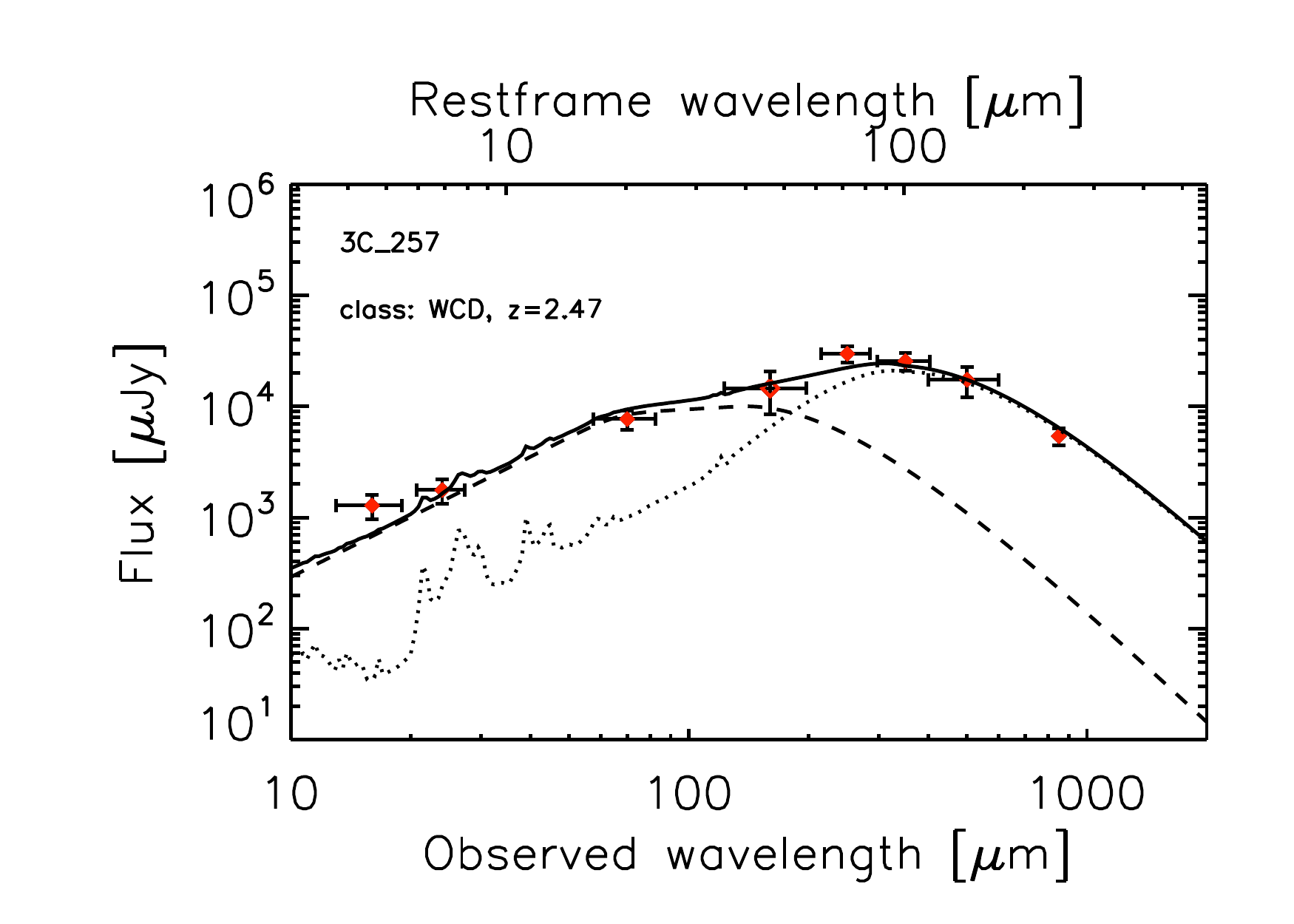}} \\
      24\mum & 70\mum & 160\mum & 250\mum & 350\mum & 500\mum \\
\end{tabular}
\end{minipage}
\begin{minipage}[c]{0.46\linewidth} 
\begin{tabular}{c@{}c@{}c@{}c@{}c@{}c@{}}

      \multicolumn{6}{c}{\includegraphics[width=80mm,trim=30 0 20 0 ,clip=true]{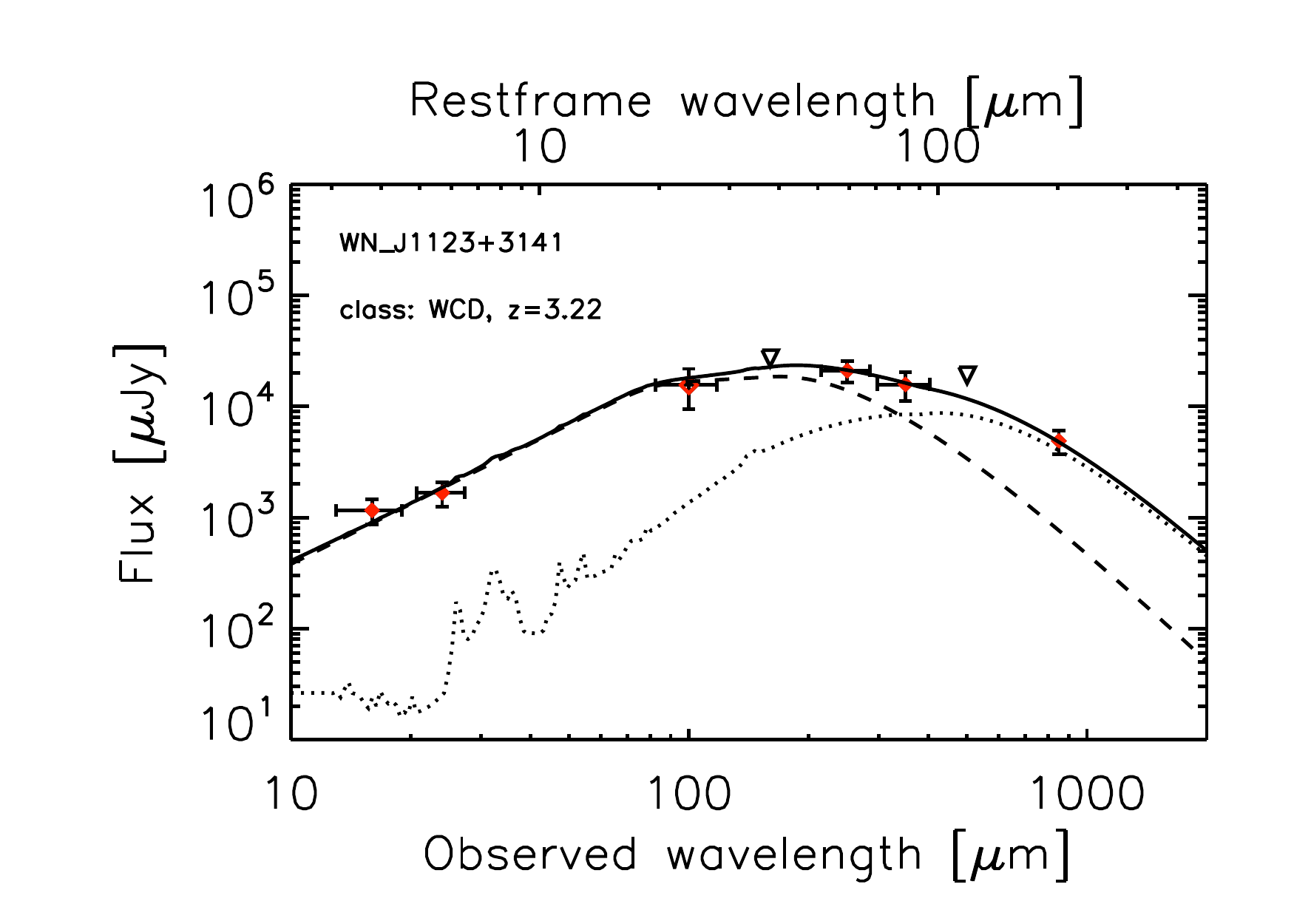}} \\
      24\mum & 100\mum & 160\mum & 250\mum & 350\mum & 500\mum \\
      \includegraphics[height=14mm,trim=70  -4 65 -10,clip=true]{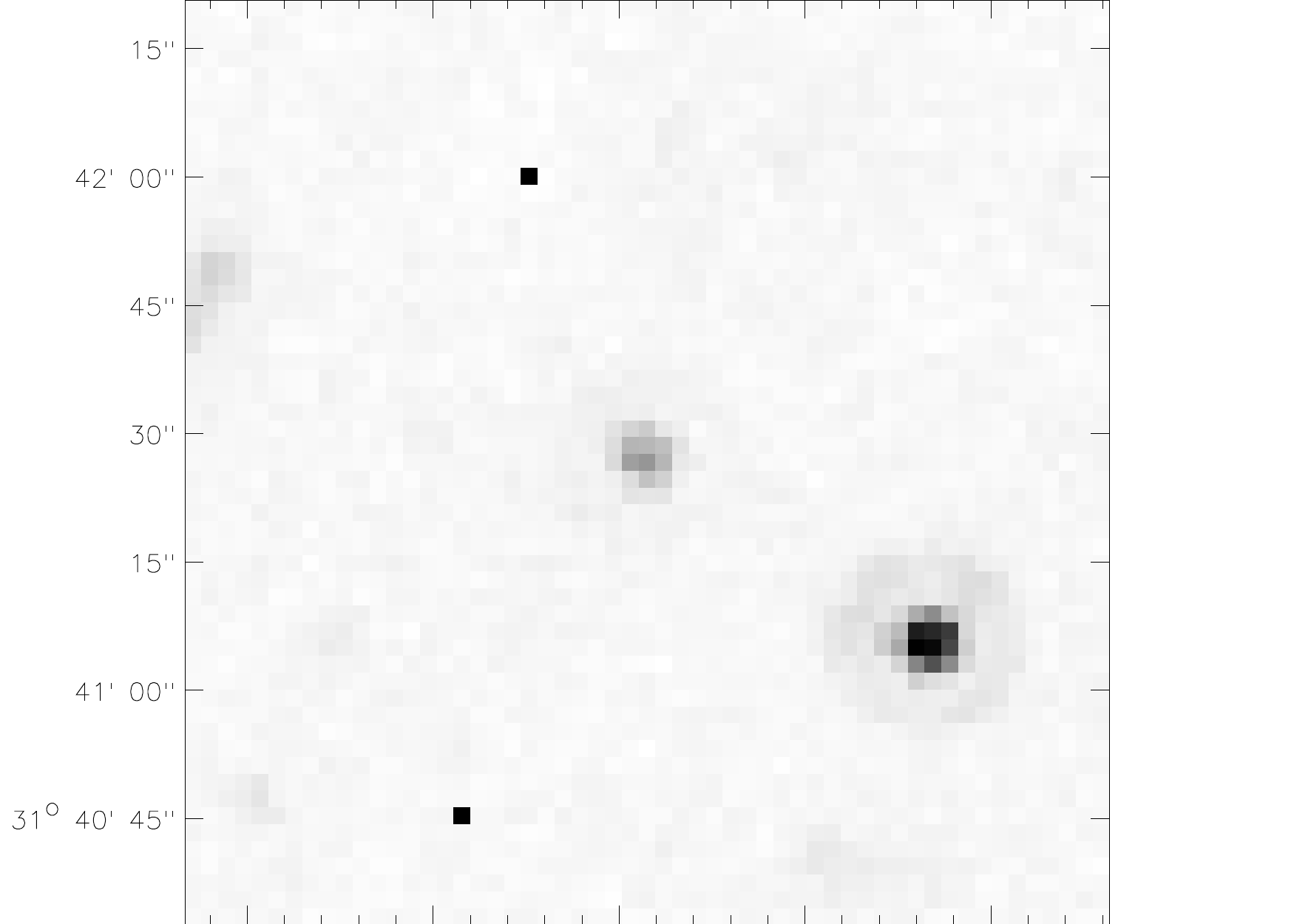} &
      \includegraphics[height=14mm,trim=70  -4 65 -10,clip=true]{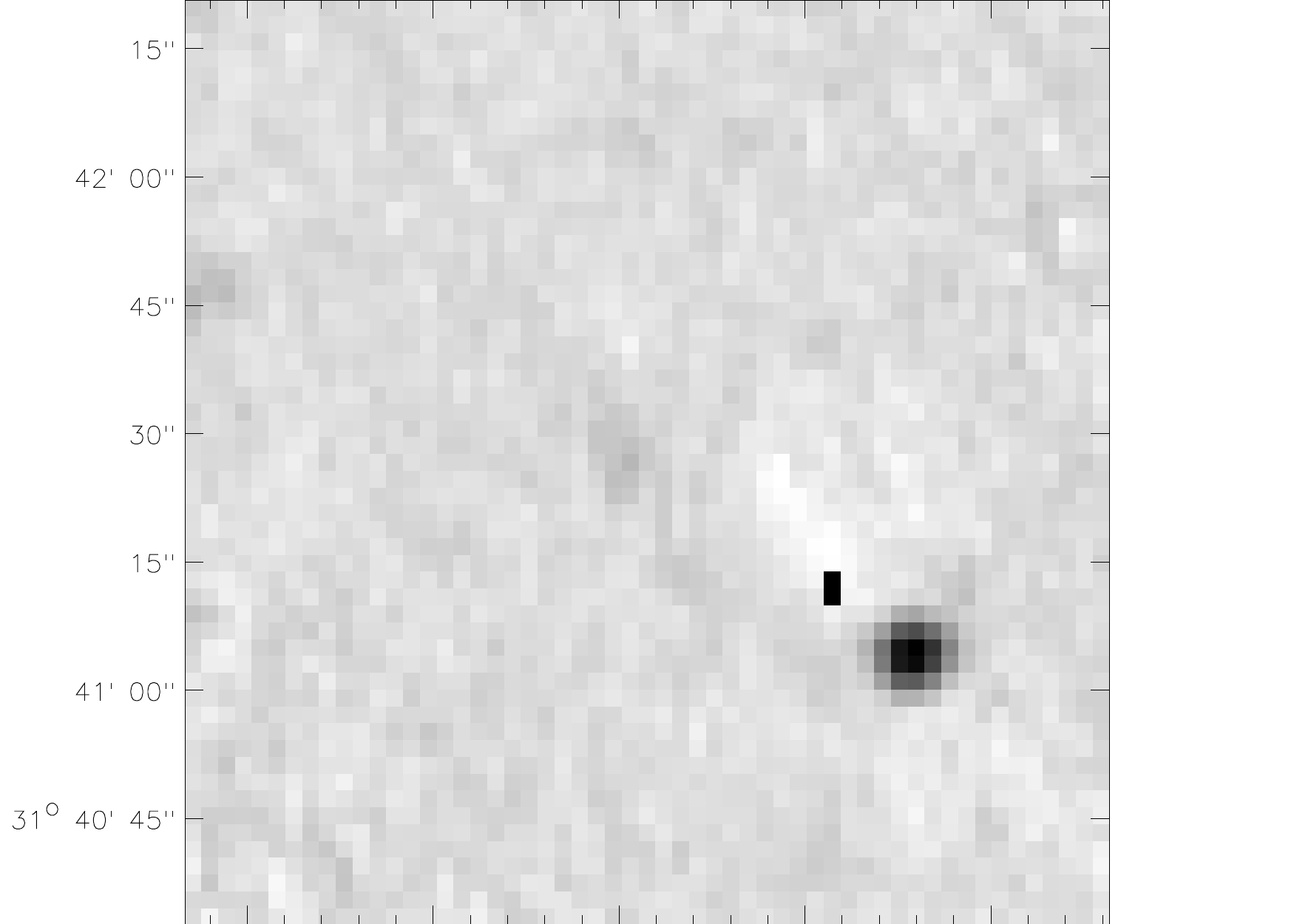} &
      \includegraphics[height=14mm,trim=70  -4 65 -10,clip=true]{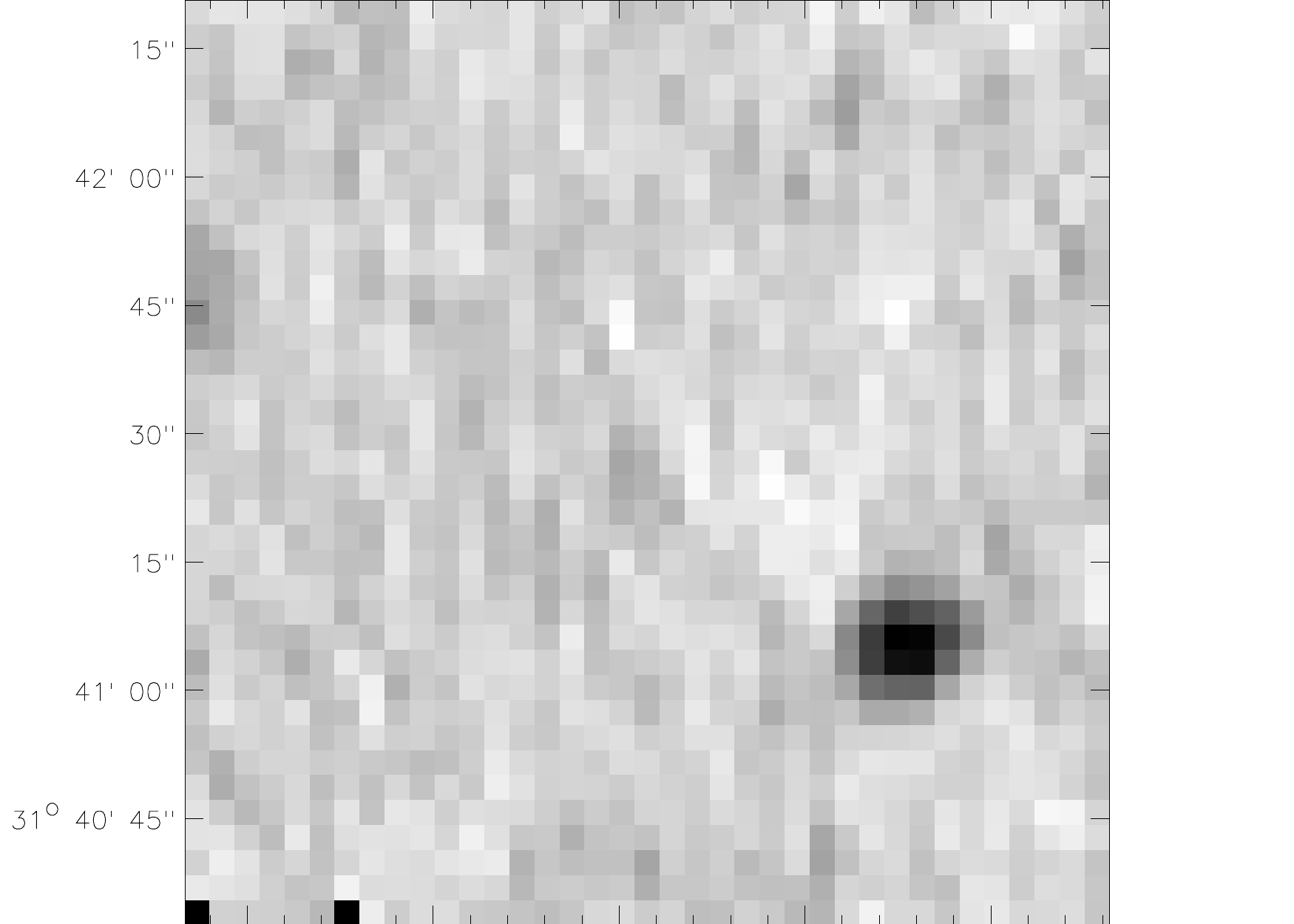} &
      \includegraphics[height=14mm,trim=70  -4 65 -10,clip=true]{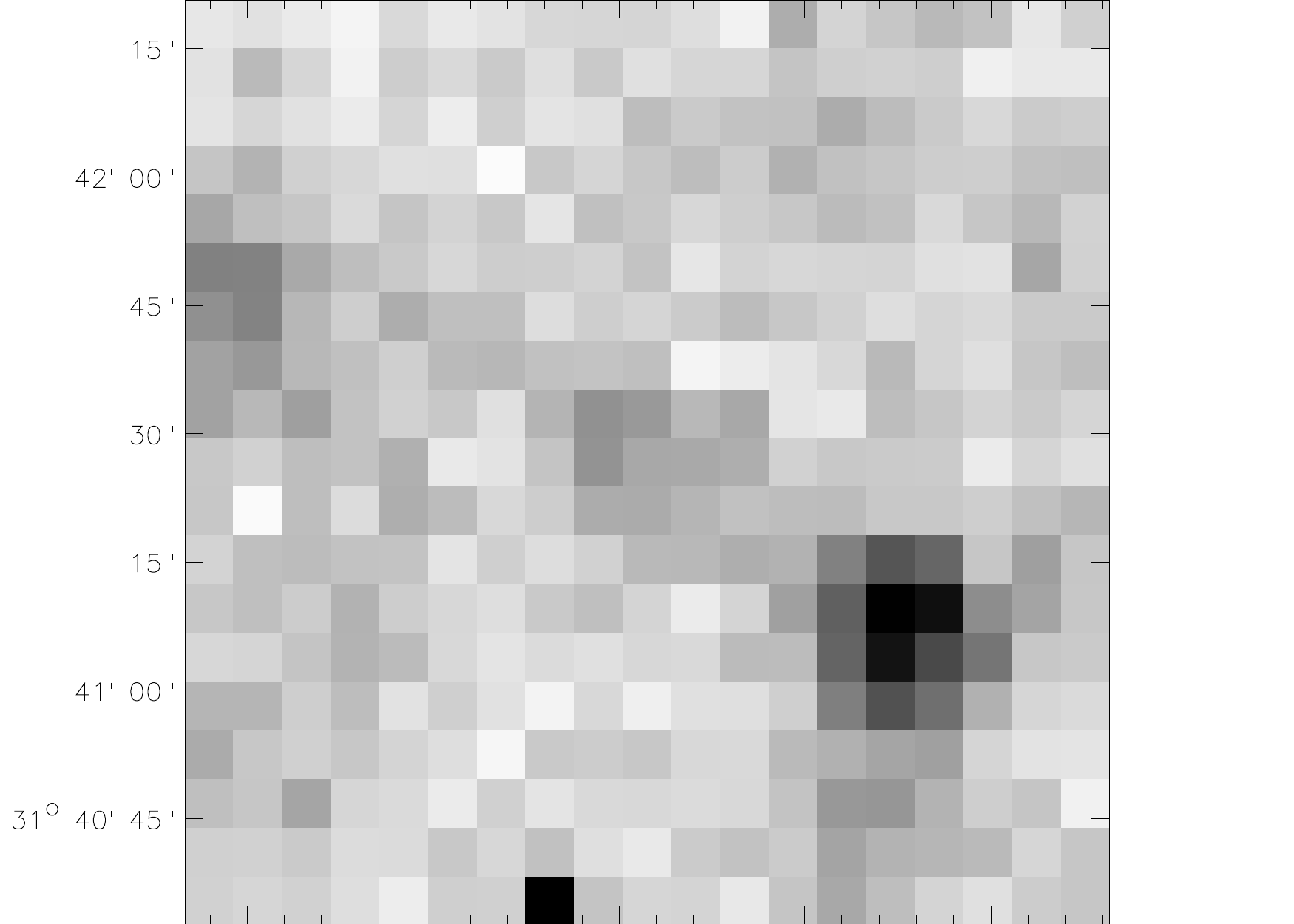}   &
      \includegraphics[height=14mm,trim=70  -4 65 -10,clip=true]{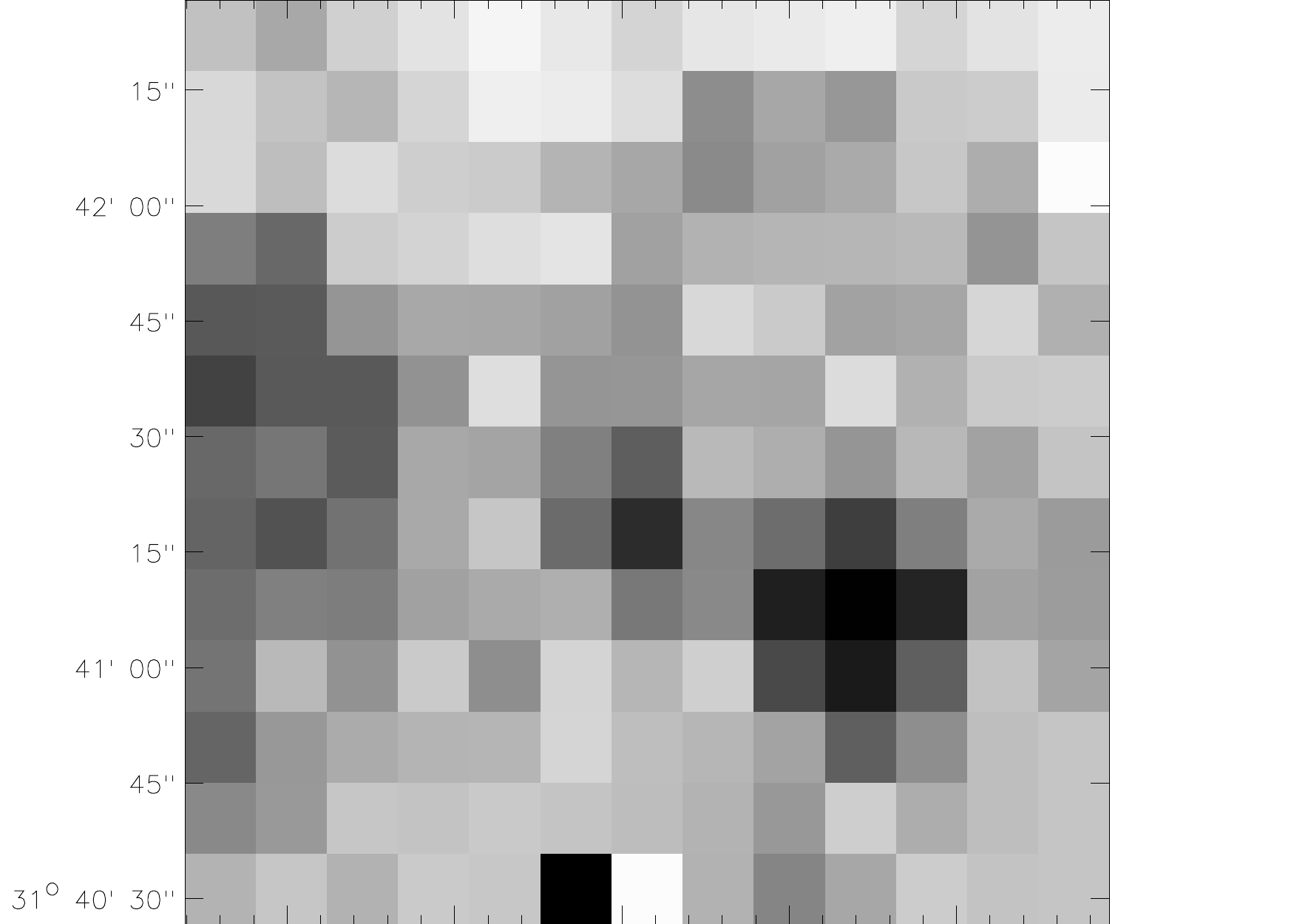}   &
      \includegraphics[height=14mm,trim=70  -4 65 -10,clip=true]{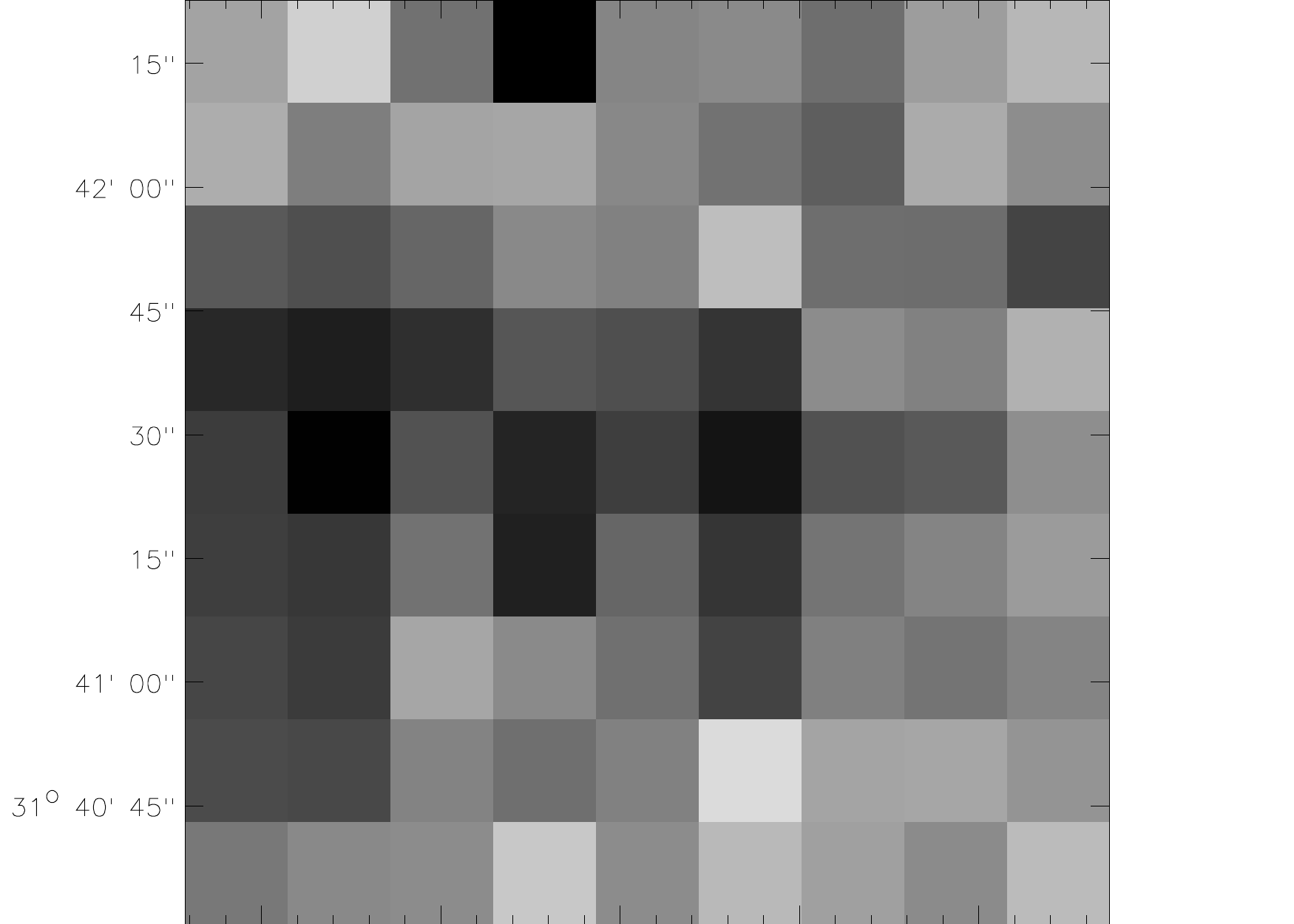} \\

      \multicolumn{6}{c}{\includegraphics[width=80mm,trim=30 0 20 0 ,clip=true]{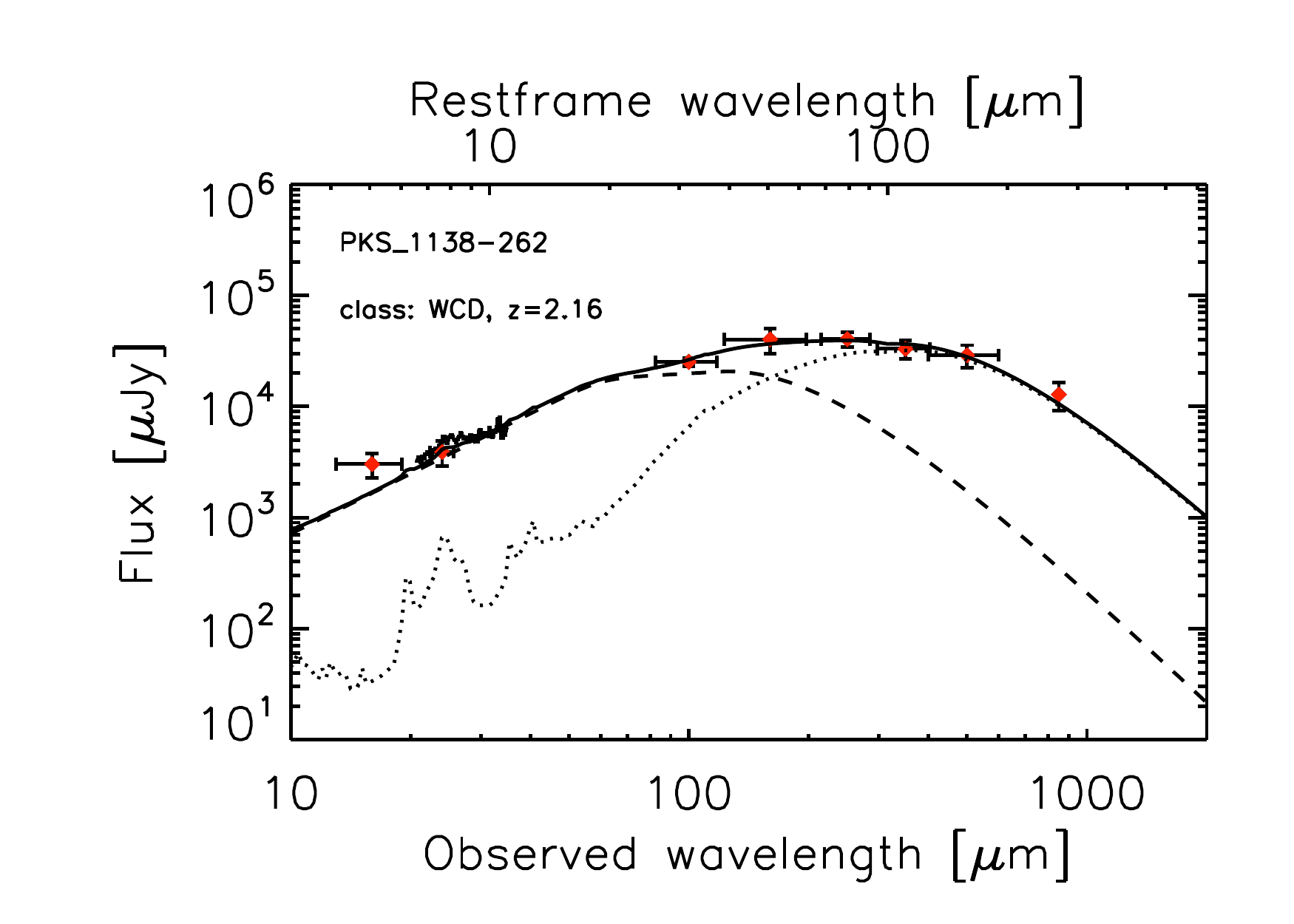}} \\
      24\mum & 100\mum & 160\mum & 250\mum & 350\mum & 500\mum \\
      \includegraphics[height=14mm,trim=70  -4 65 -10,clip=true]{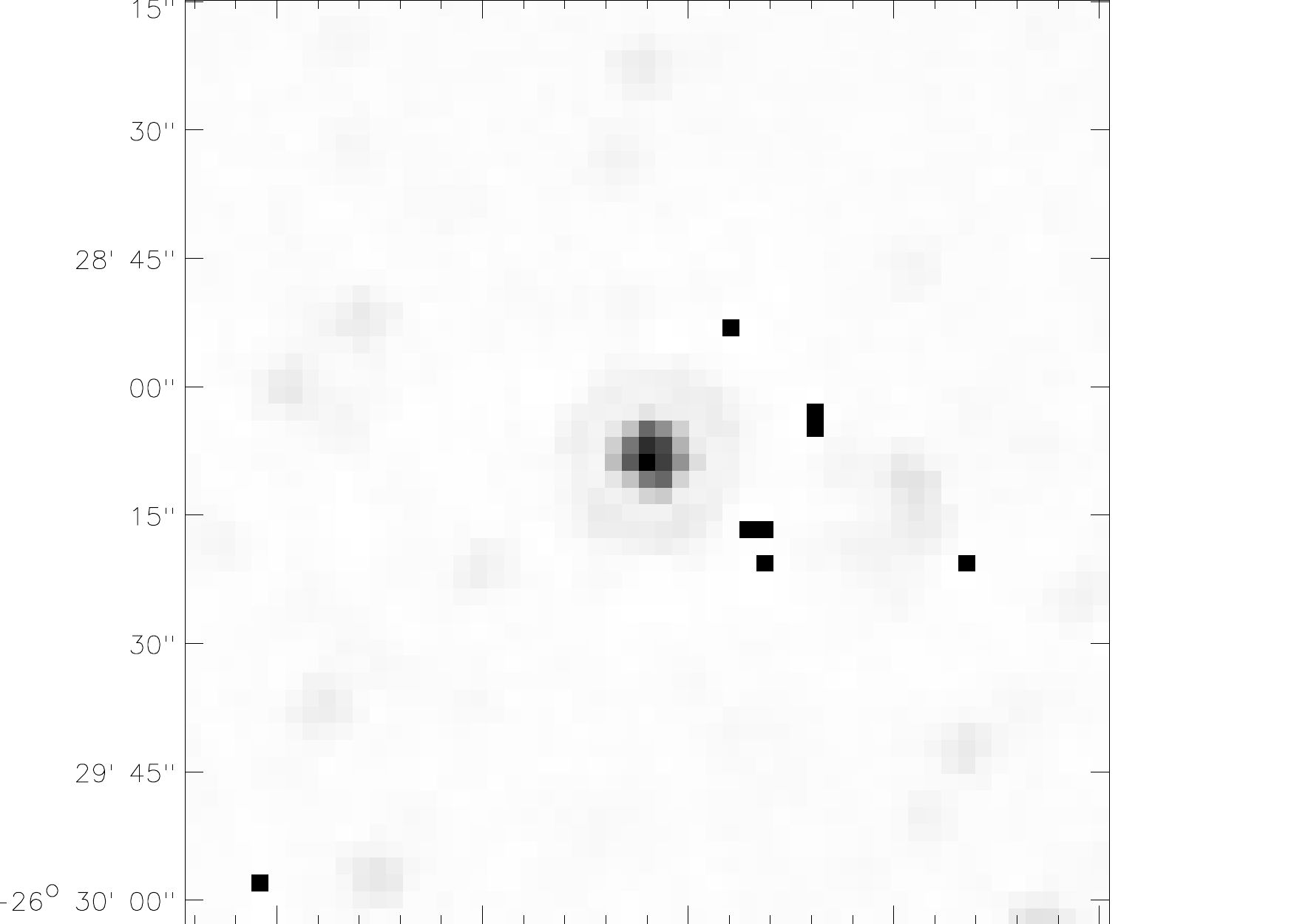} &
      \includegraphics[height=14mm,trim=70  -4 65 -10,clip=true]{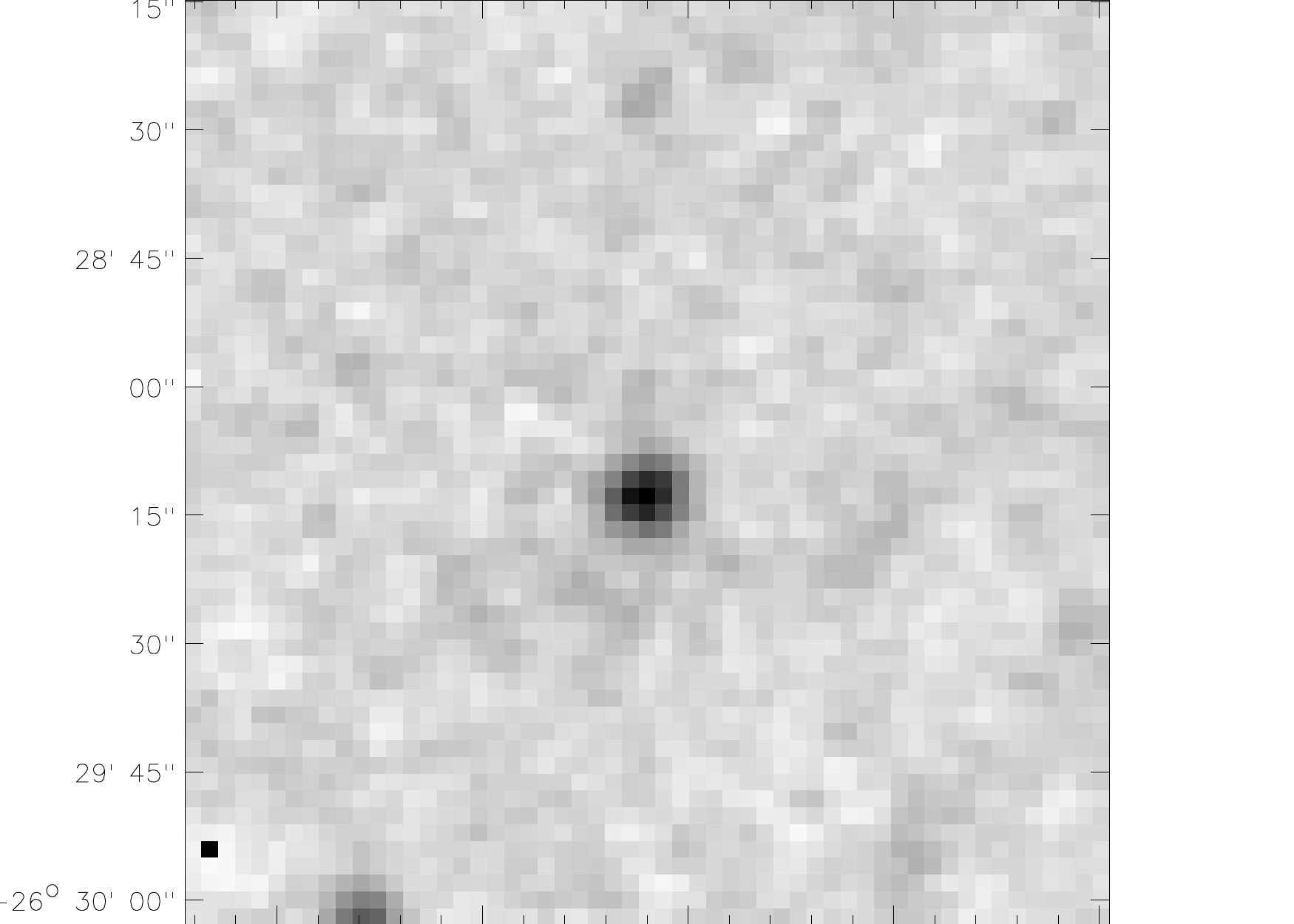} &
      \includegraphics[height=14mm,trim=70  -4 65 -10,clip=true]{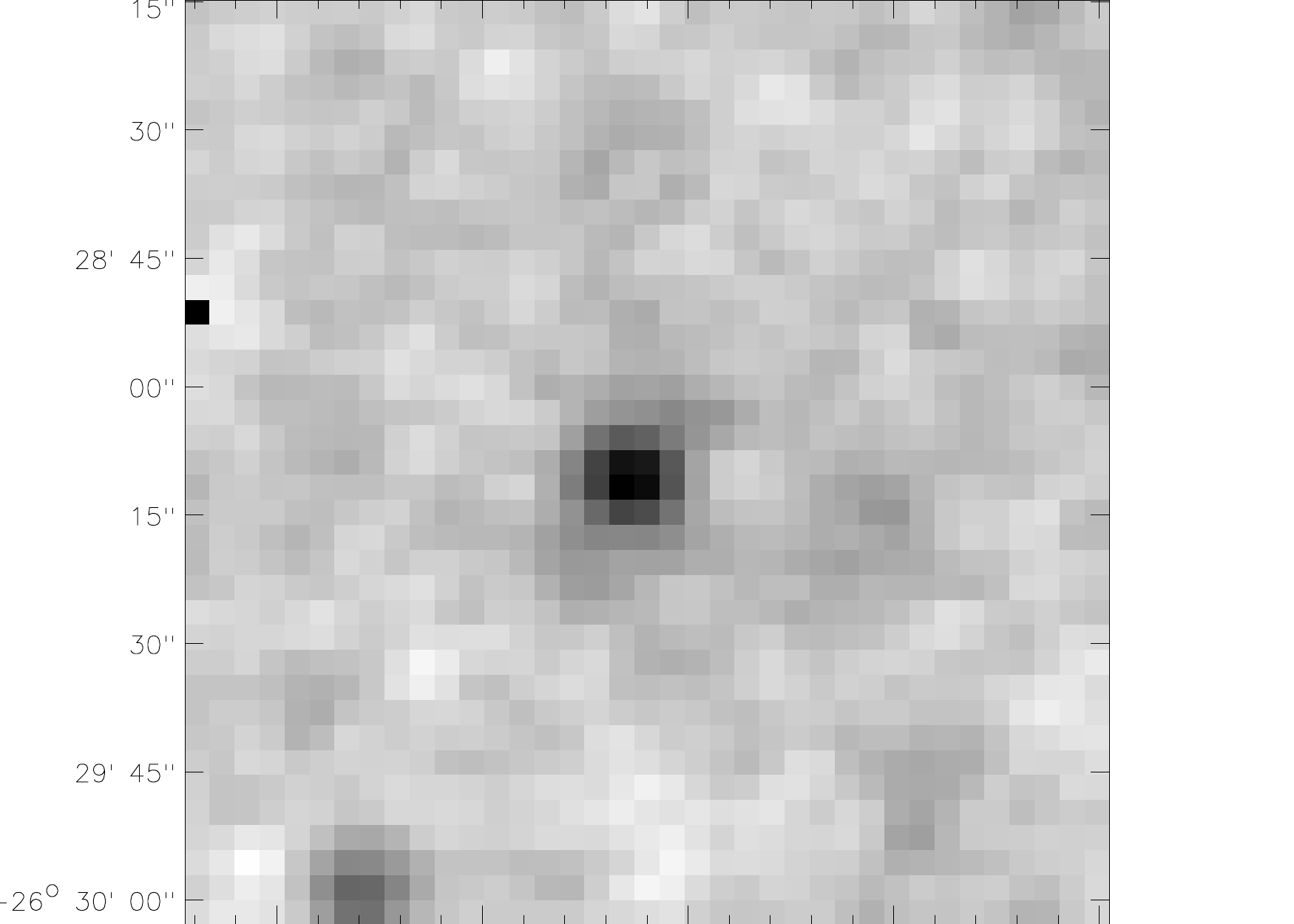} &
      \includegraphics[height=14mm,trim=70  -4 65 -10,clip=true]{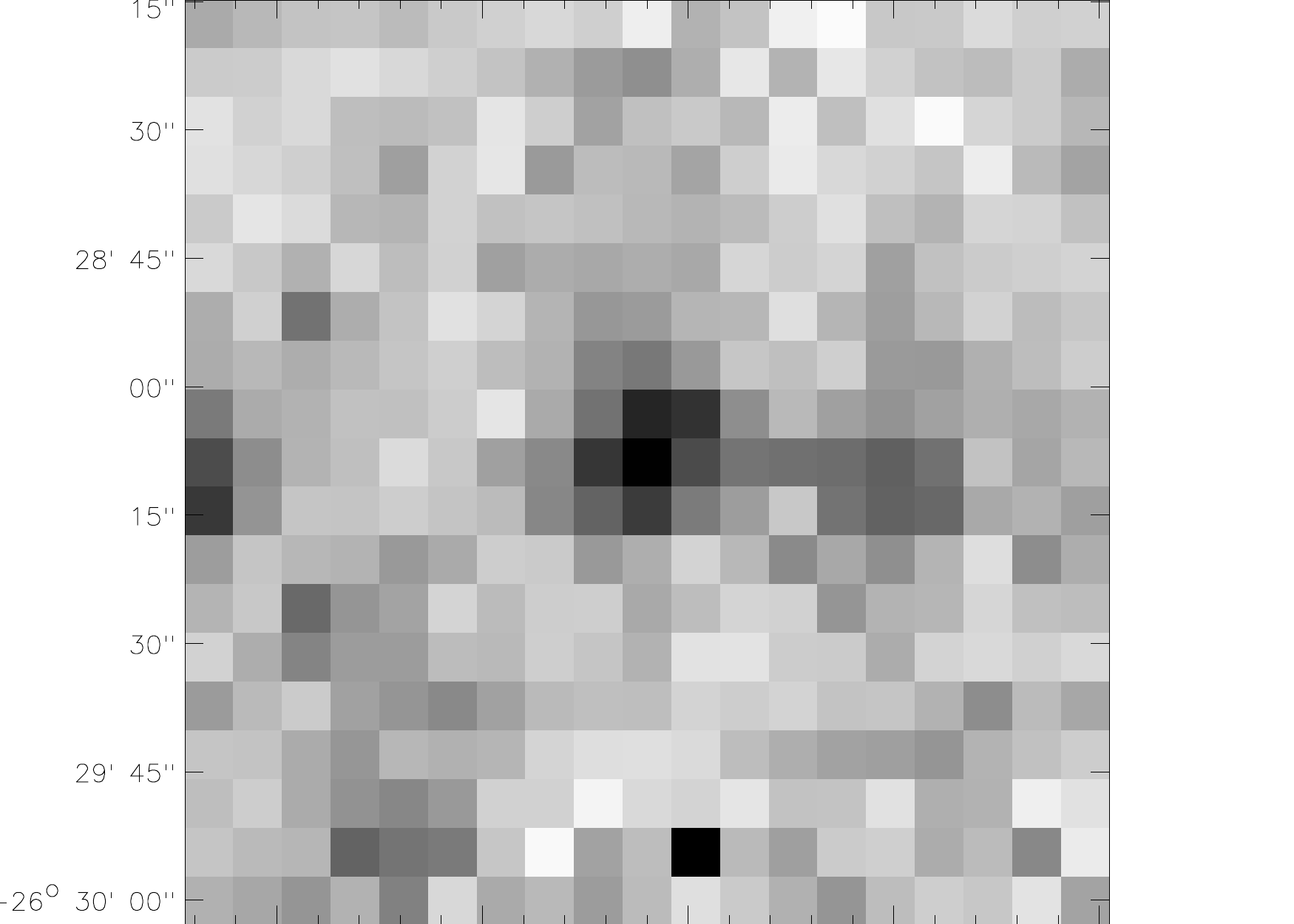}   &
      \includegraphics[height=14mm,trim=70  -4 65 -10,clip=true]{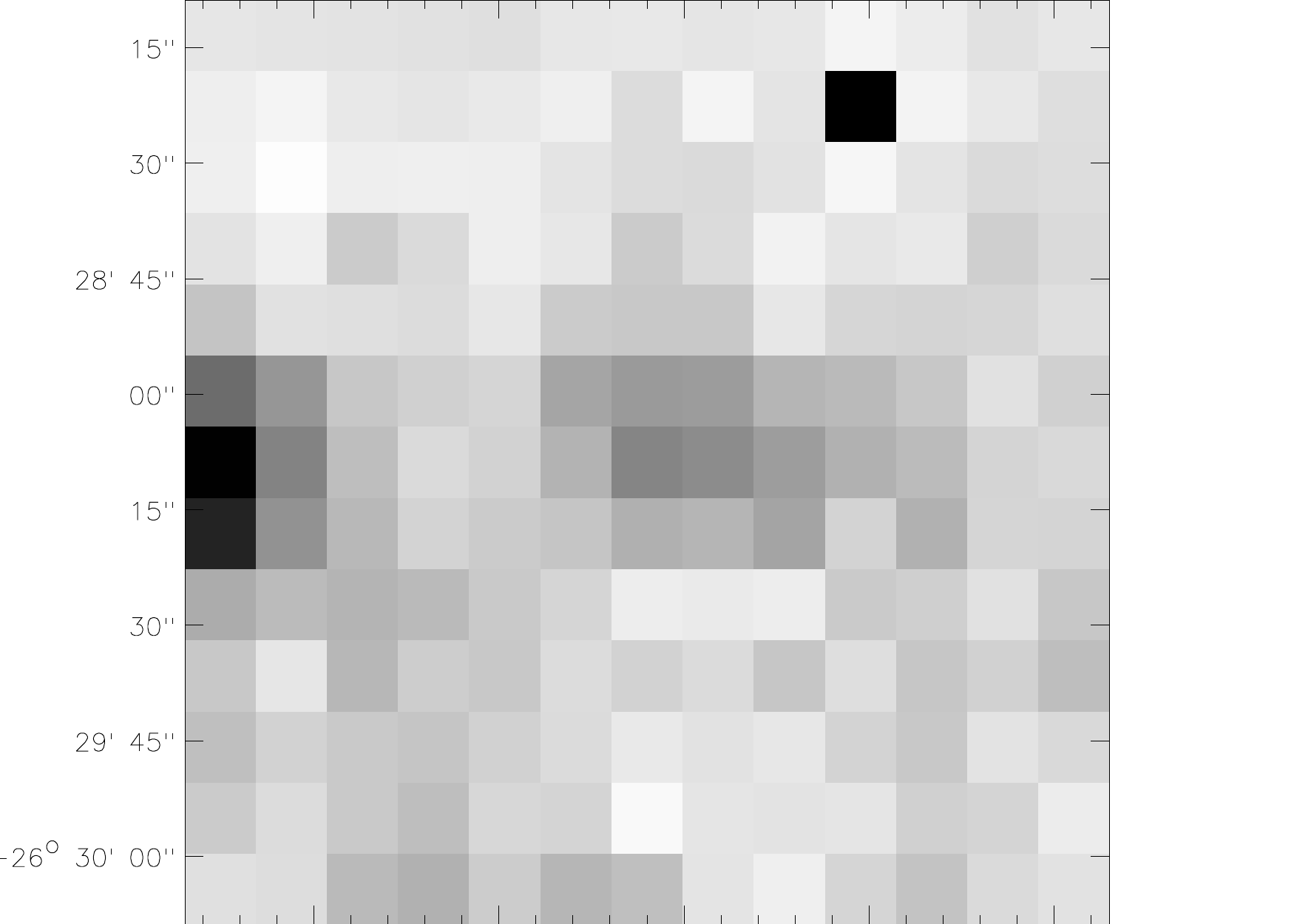}   &
      \includegraphics[height=14mm,trim=70  -4 65 -10,clip=true]{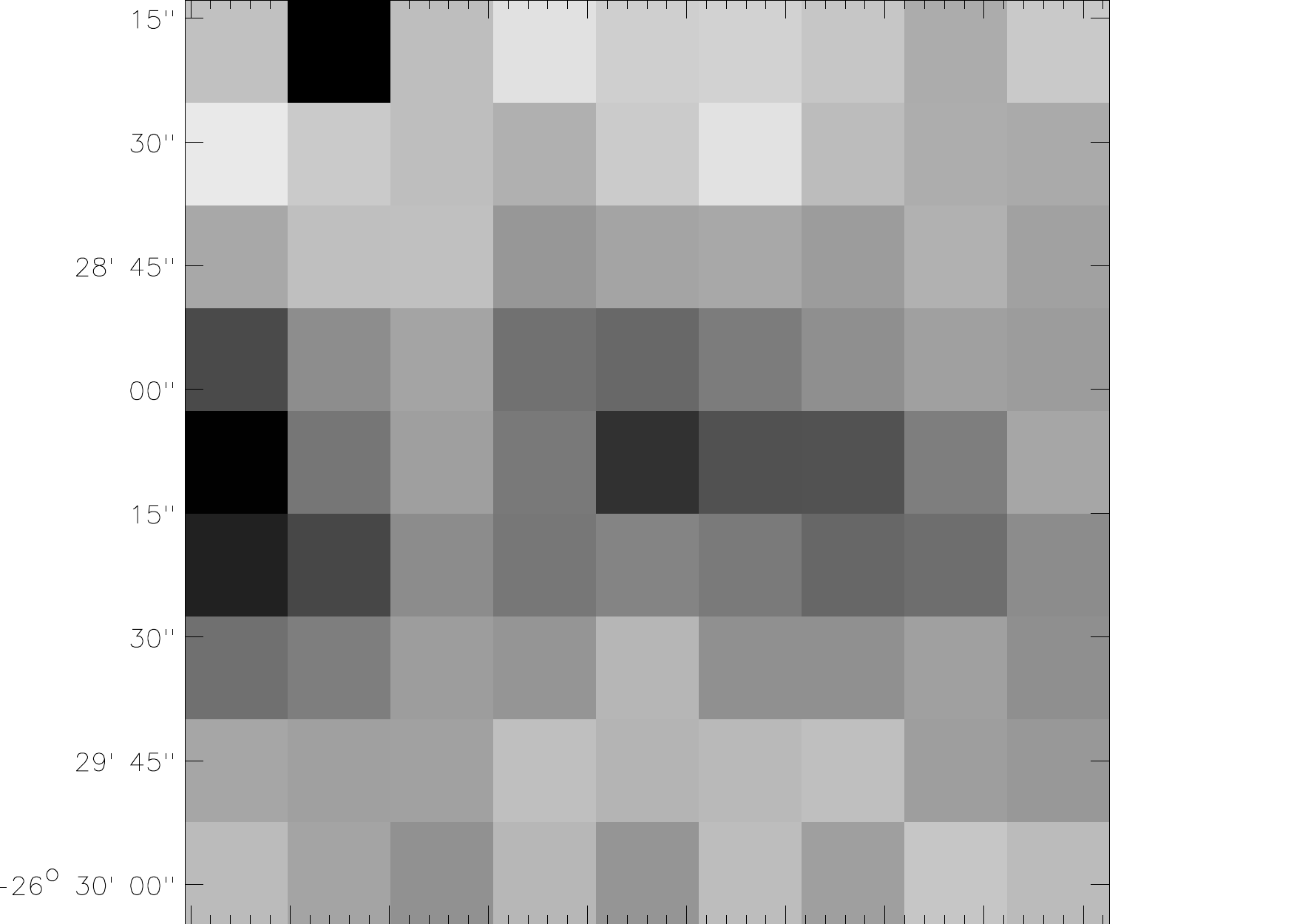} \\

      \multicolumn{6}{c}{\includegraphics[width=80mm,trim=30 0 20 0 ,clip=true]{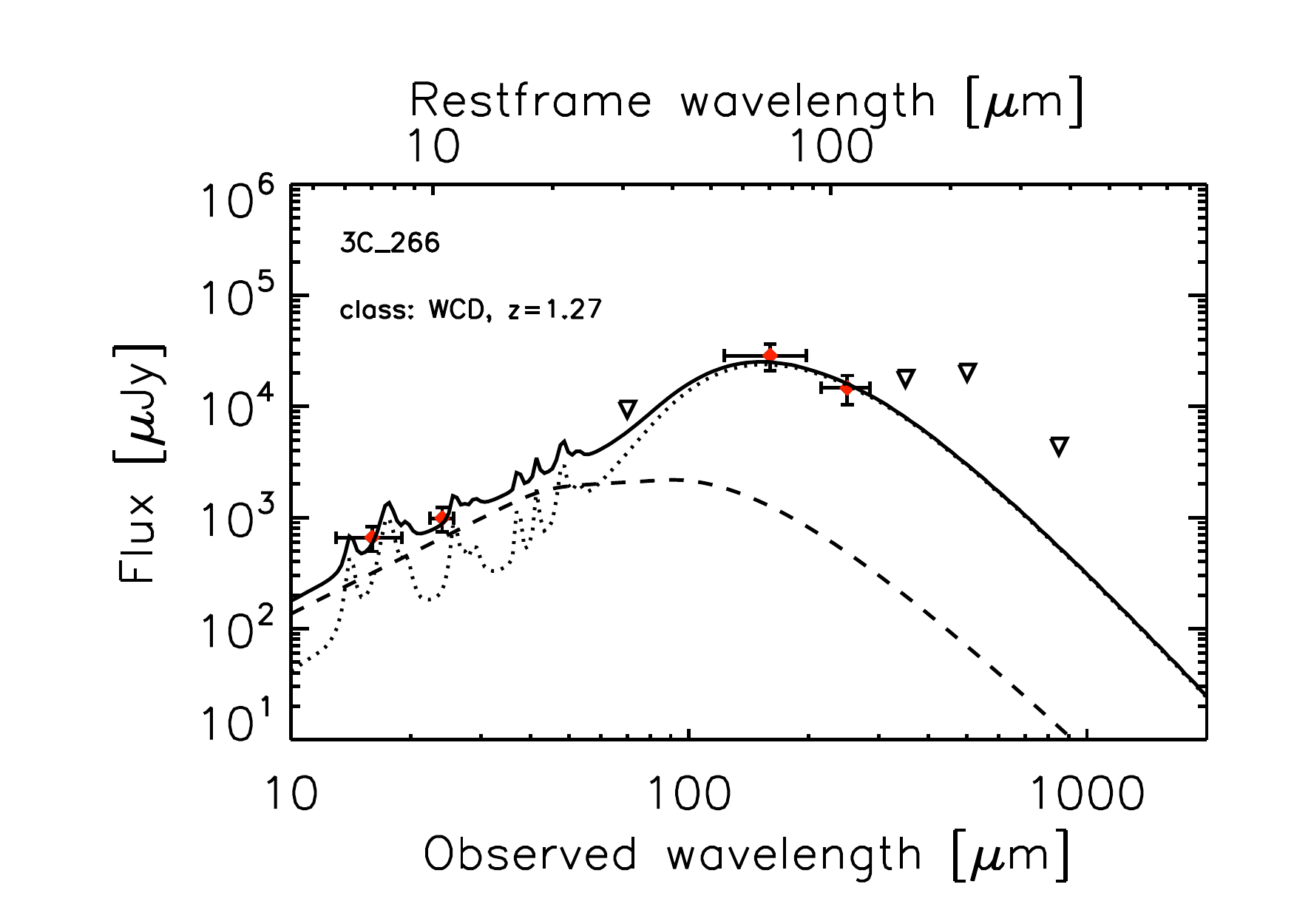}} \\
      24\mum & 70\mum & 160\mum & 250\mum & 350\mum & 500\mum \\
\end{tabular}
          \end{minipage}
  \end{centering}
\end{figure*}

\clearpage

\begin{figure*}[h]
  \begin{centering}
    \begin{minipage}[c]{0.46\linewidth}
    \begin{tabular}{c@{}c@{}c@{}c@{}c@{}c@{}}

      \multicolumn{6}{c}{\includegraphics[width=80mm,trim=30 0 20 0 ,clip=true]{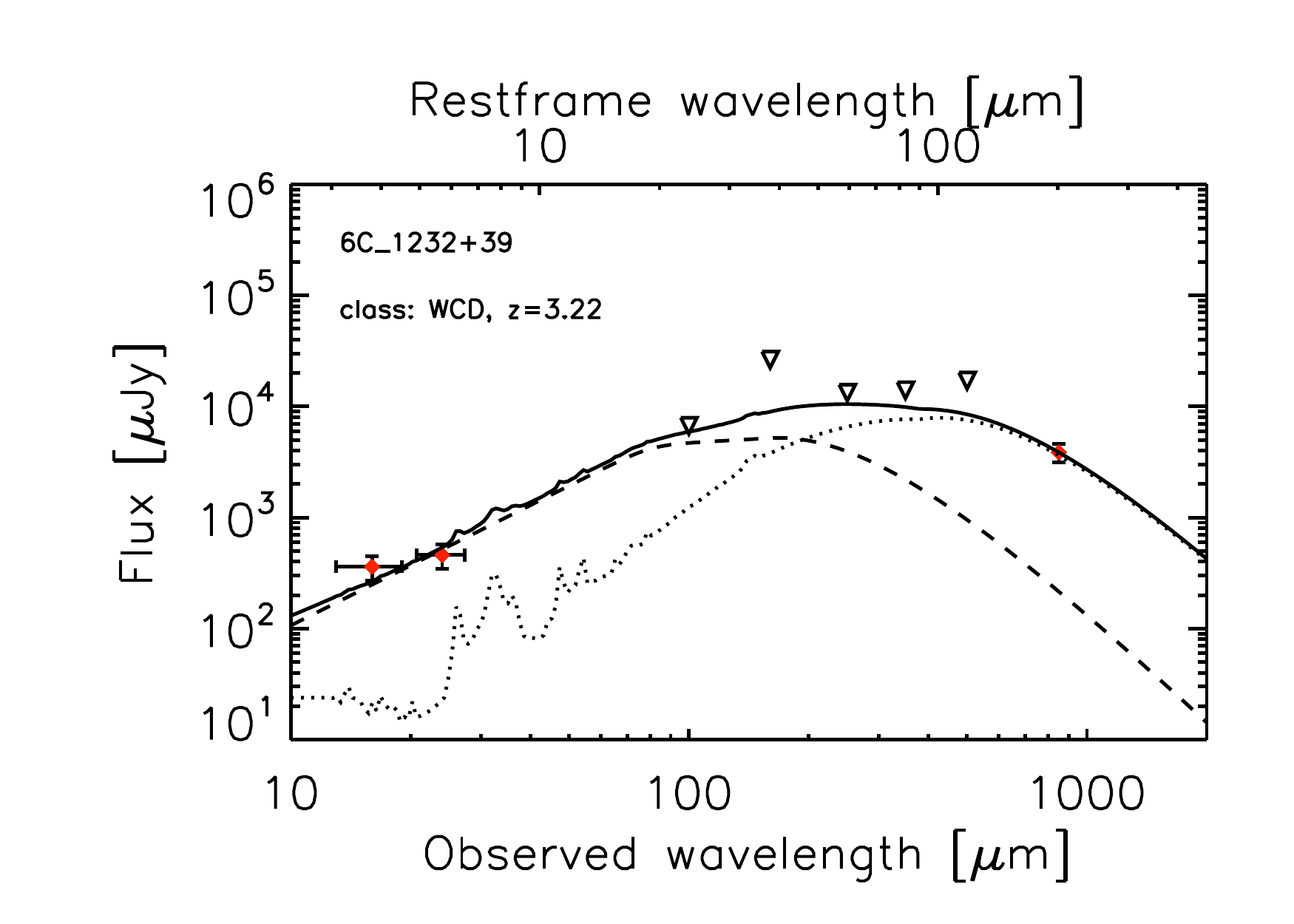}} \\
      24\mum & 100\mum & 160\mum & 250\mum & 350\mum & 500\mum \\
      \includegraphics[height=14mm,trim=70  -4 65 -10,clip=true]{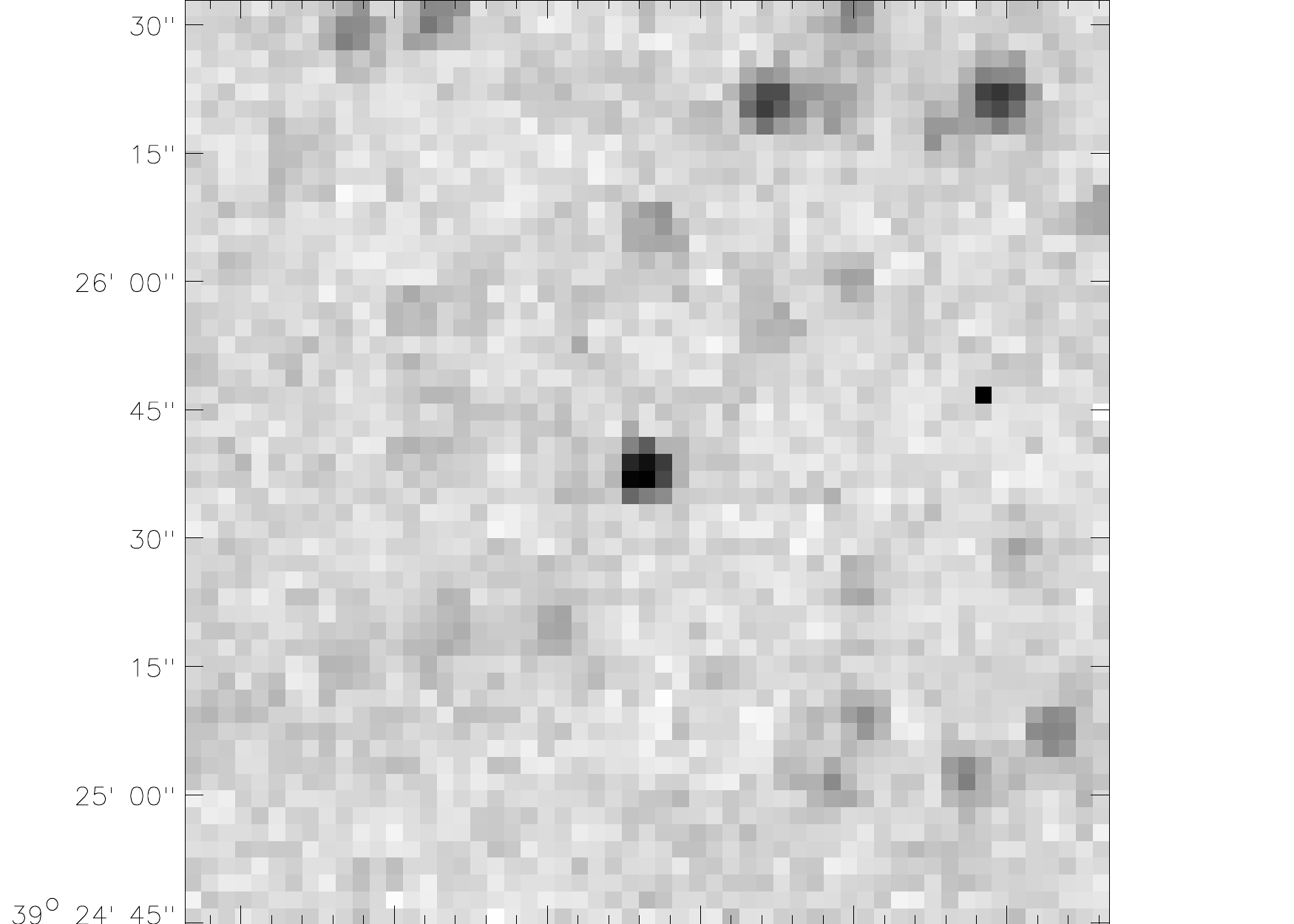} &
      \includegraphics[height=14mm,trim=70  -4 65 -10,clip=true]{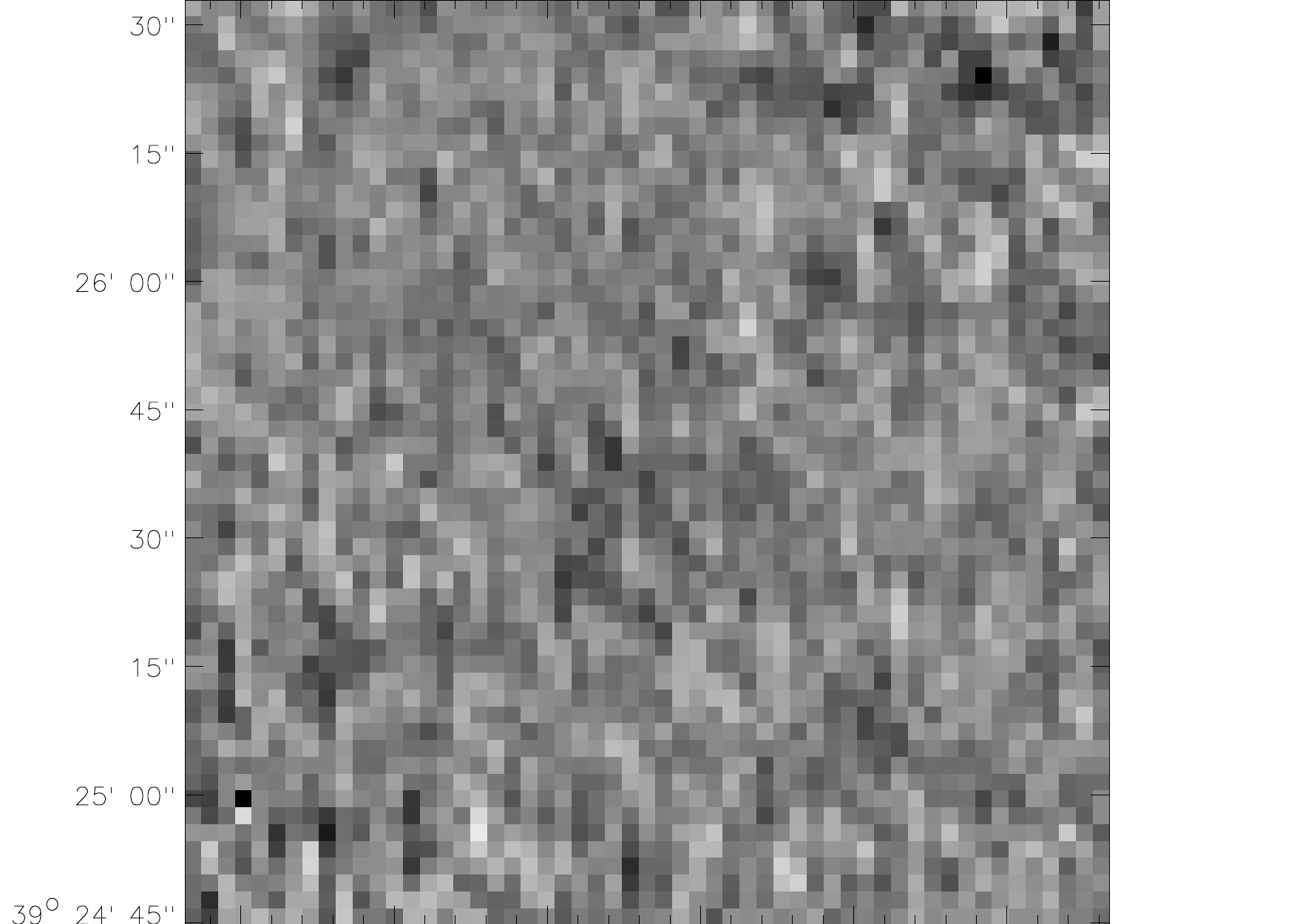} &
      \includegraphics[height=14mm,trim=70  -4 65 -10,clip=true]{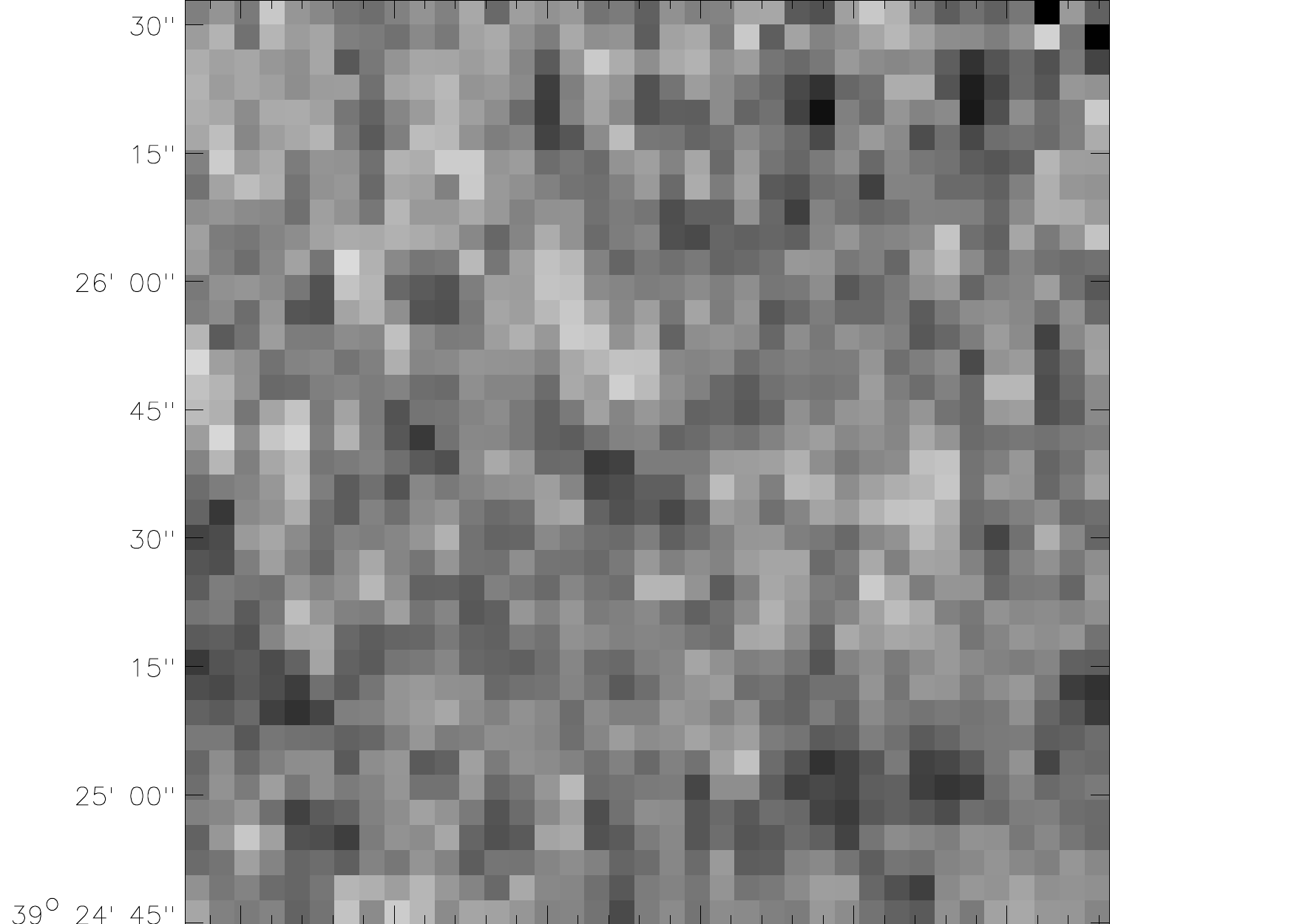} &
      \includegraphics[height=14mm,trim=70  -4 65 -10,clip=true]{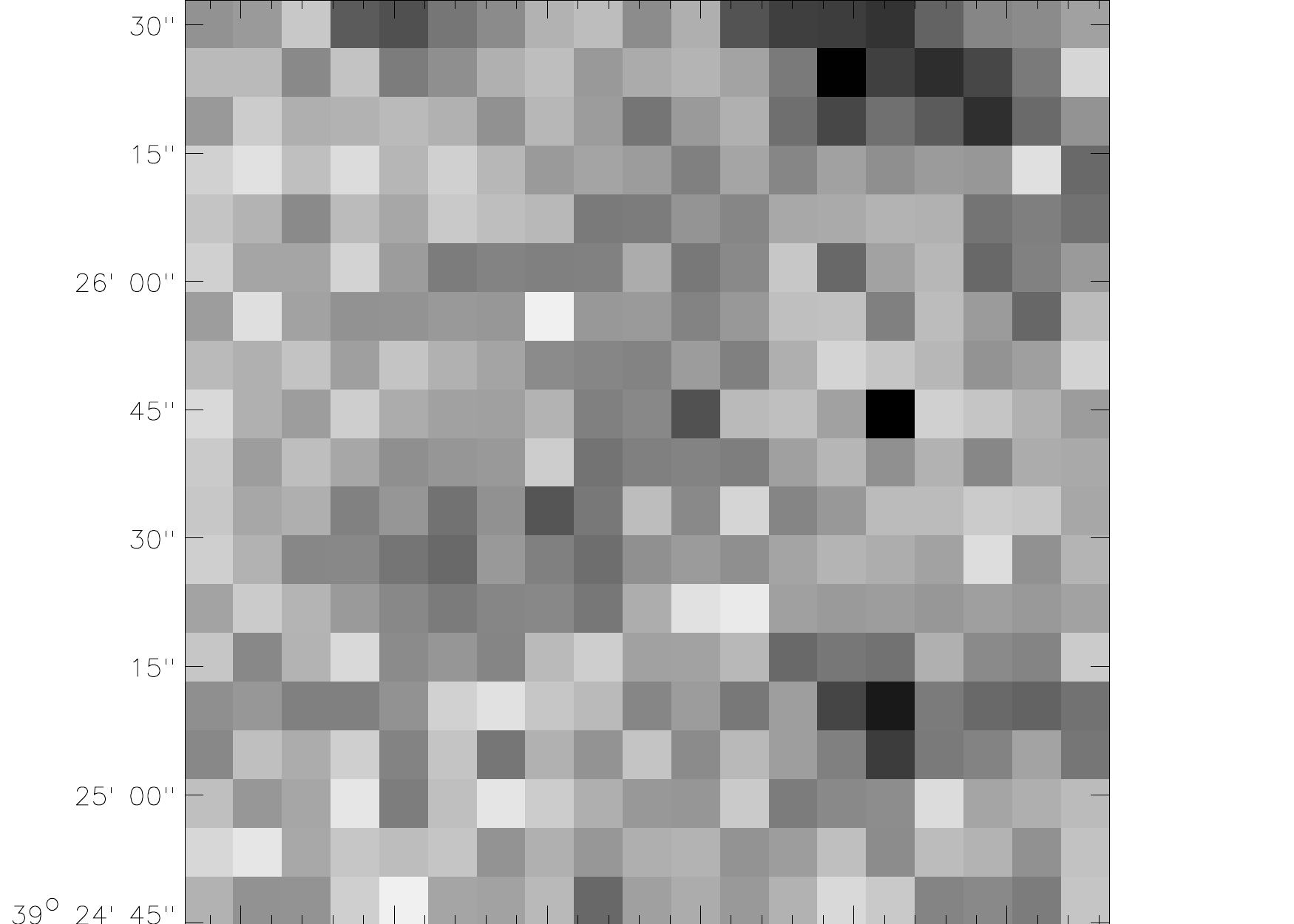}   &
      \includegraphics[height=14mm,trim=70  -4 65 -10,clip=true]{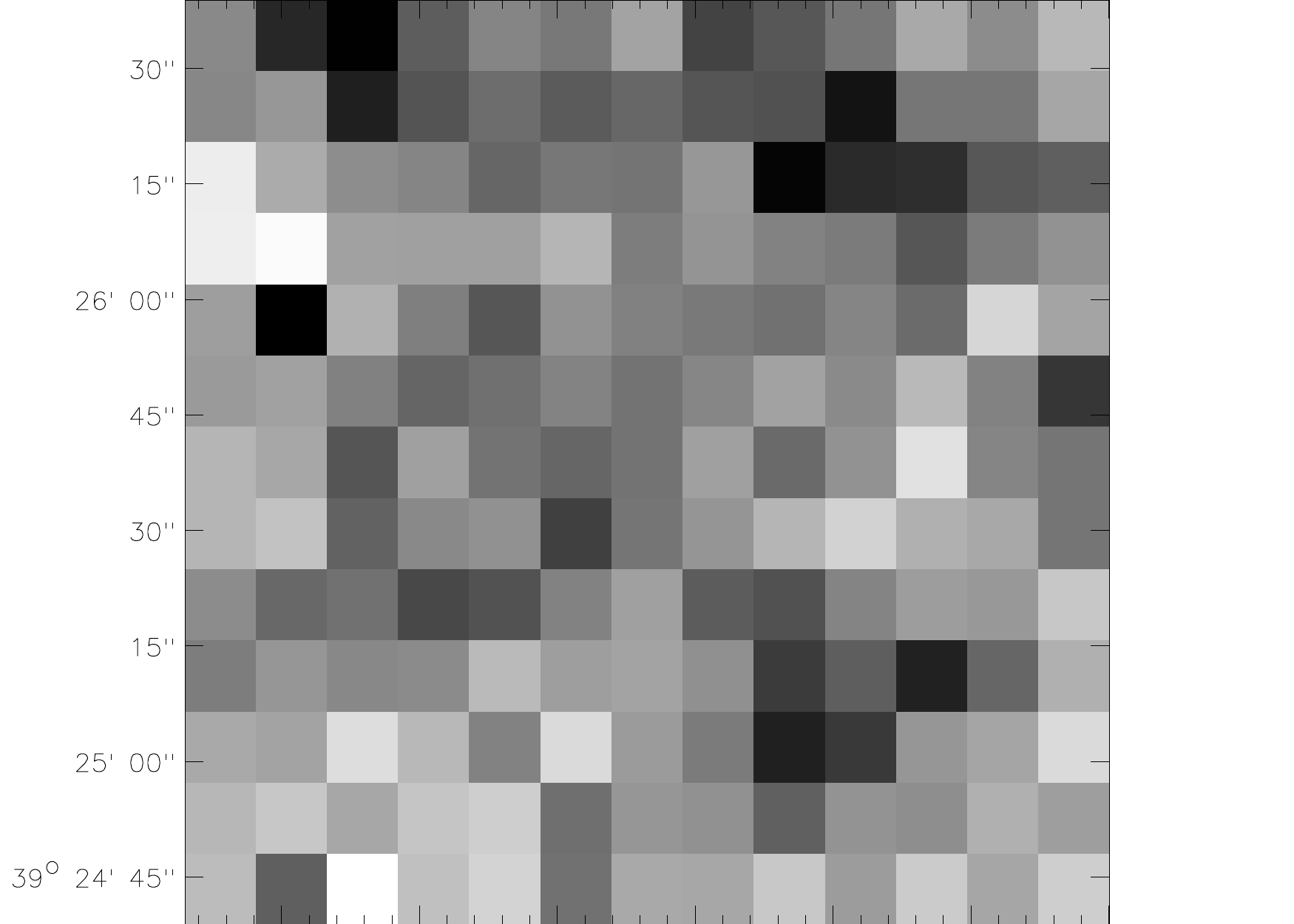}   &
      \includegraphics[height=14mm,trim=70  -4 65 -10,clip=true]{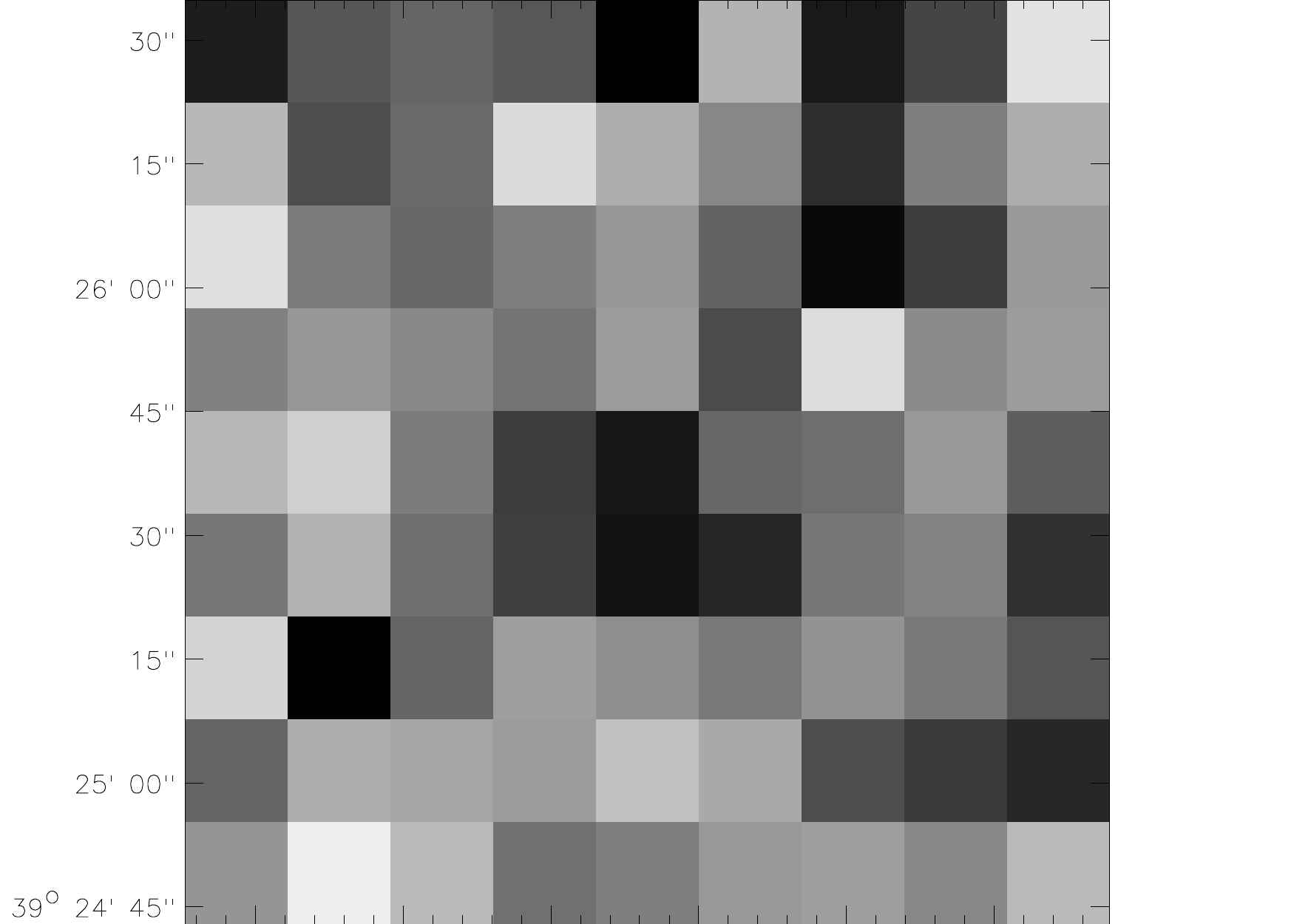} \\

      \multicolumn{6}{c}{\includegraphics[width=80mm,trim=30 0 20 0 ,clip=true]{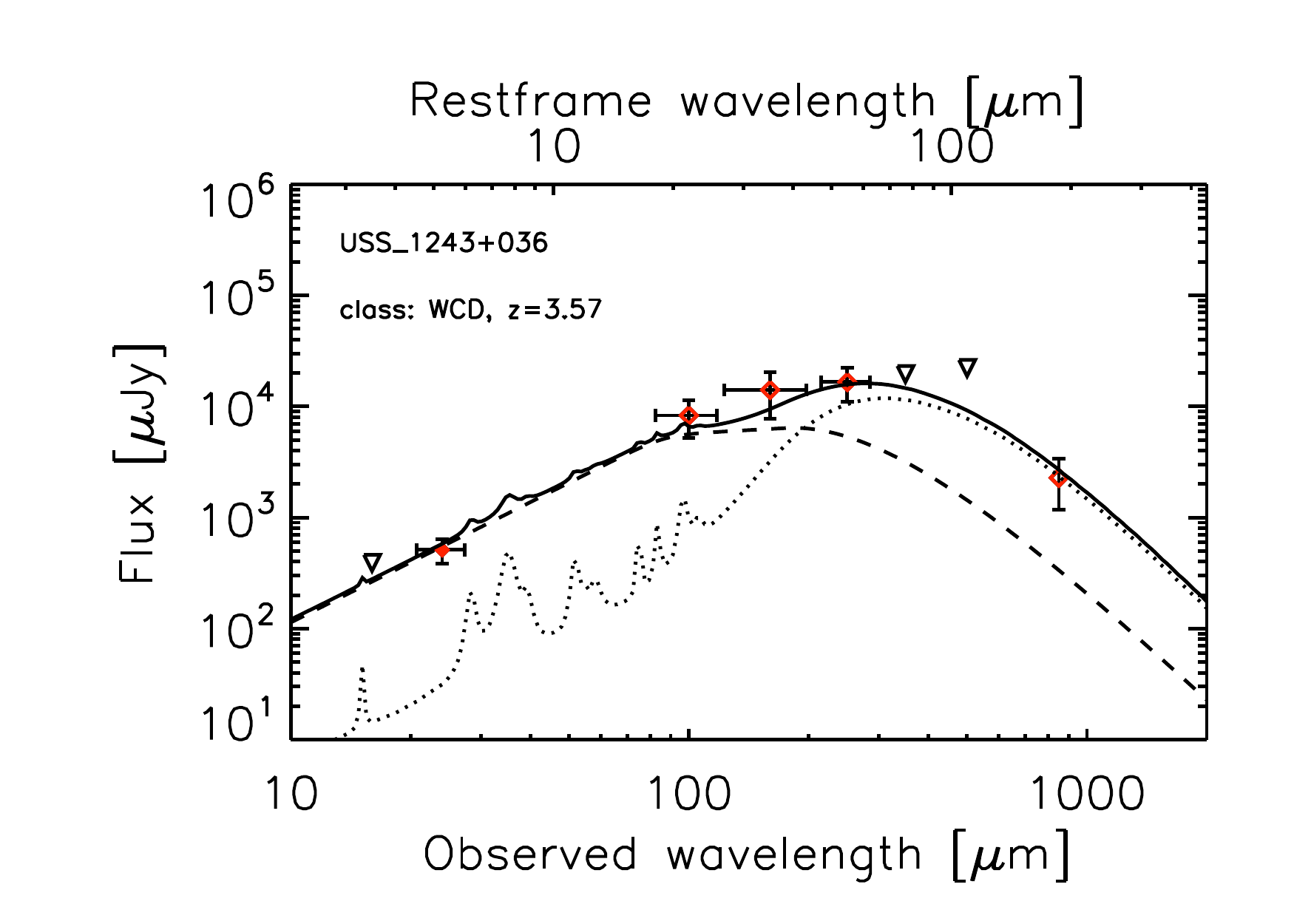}} \\
      24\mum & 100\mum & 160\mum & 250\mum & 350\mum & 500\mum \\
      \includegraphics[height=14mm,trim=70  -4 65 -10,clip=true]{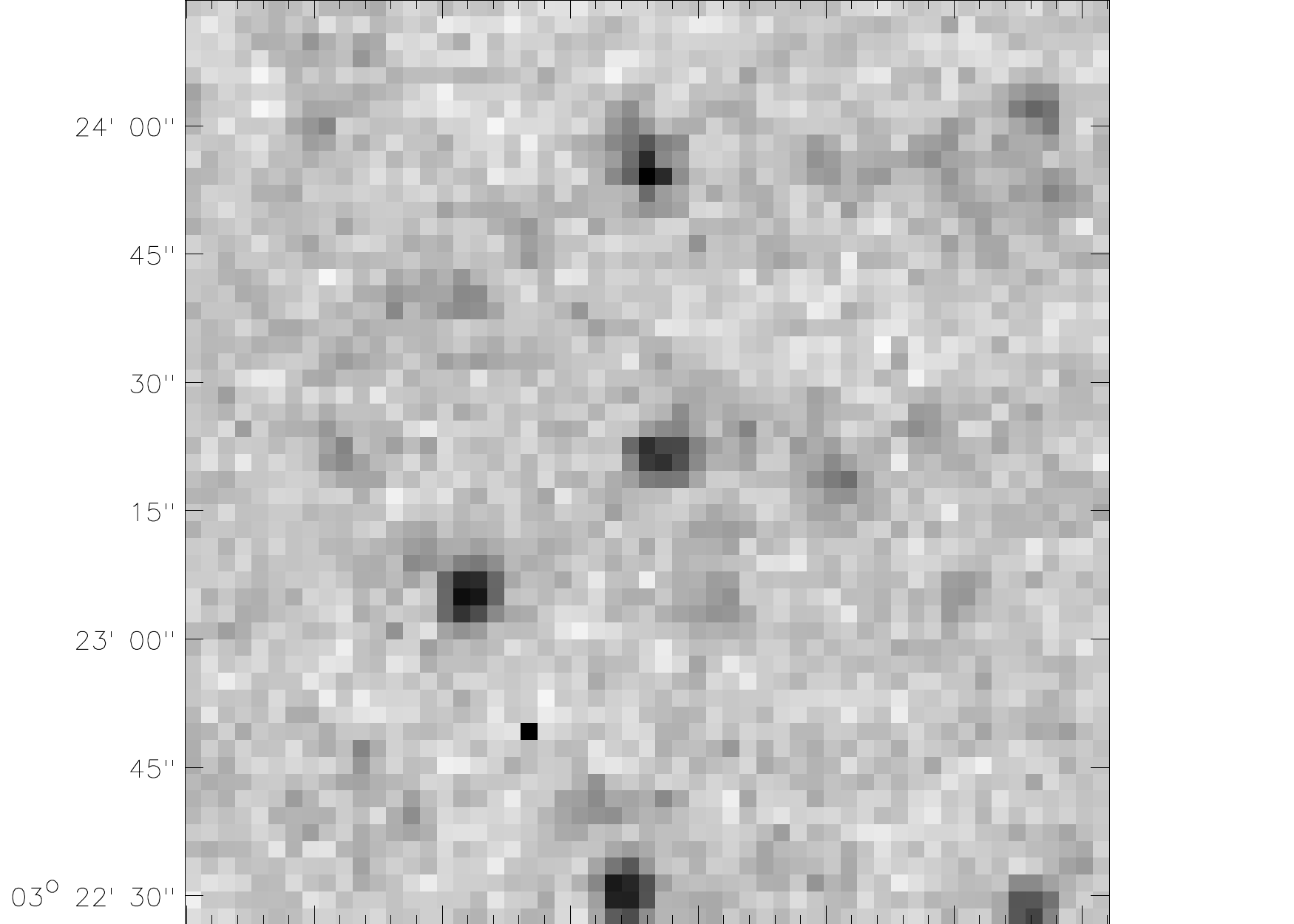} &
      \includegraphics[height=14mm,trim=70  -4 65 -10,clip=true]{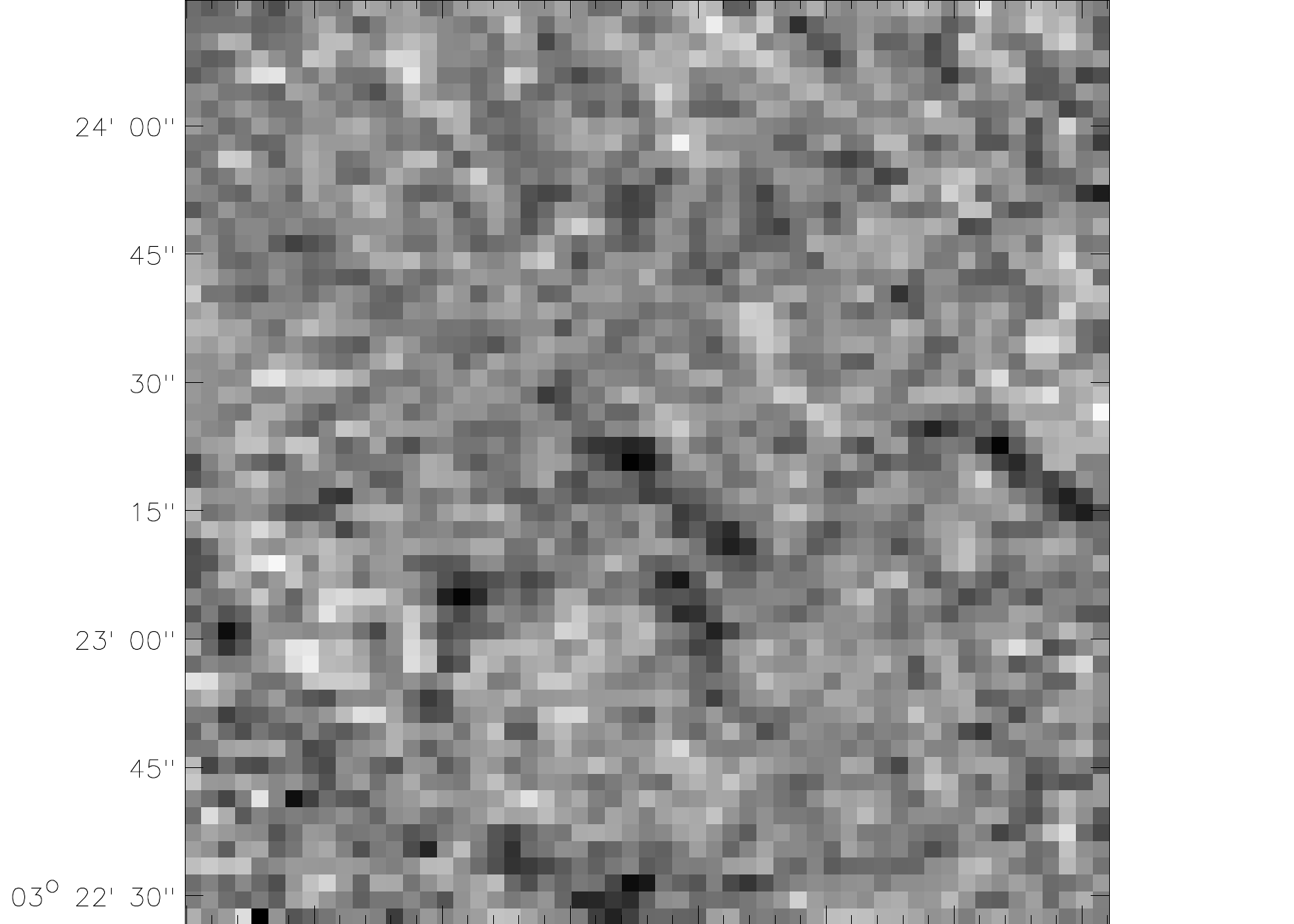} &
      \includegraphics[height=14mm,trim=70  -4 65 -10,clip=true]{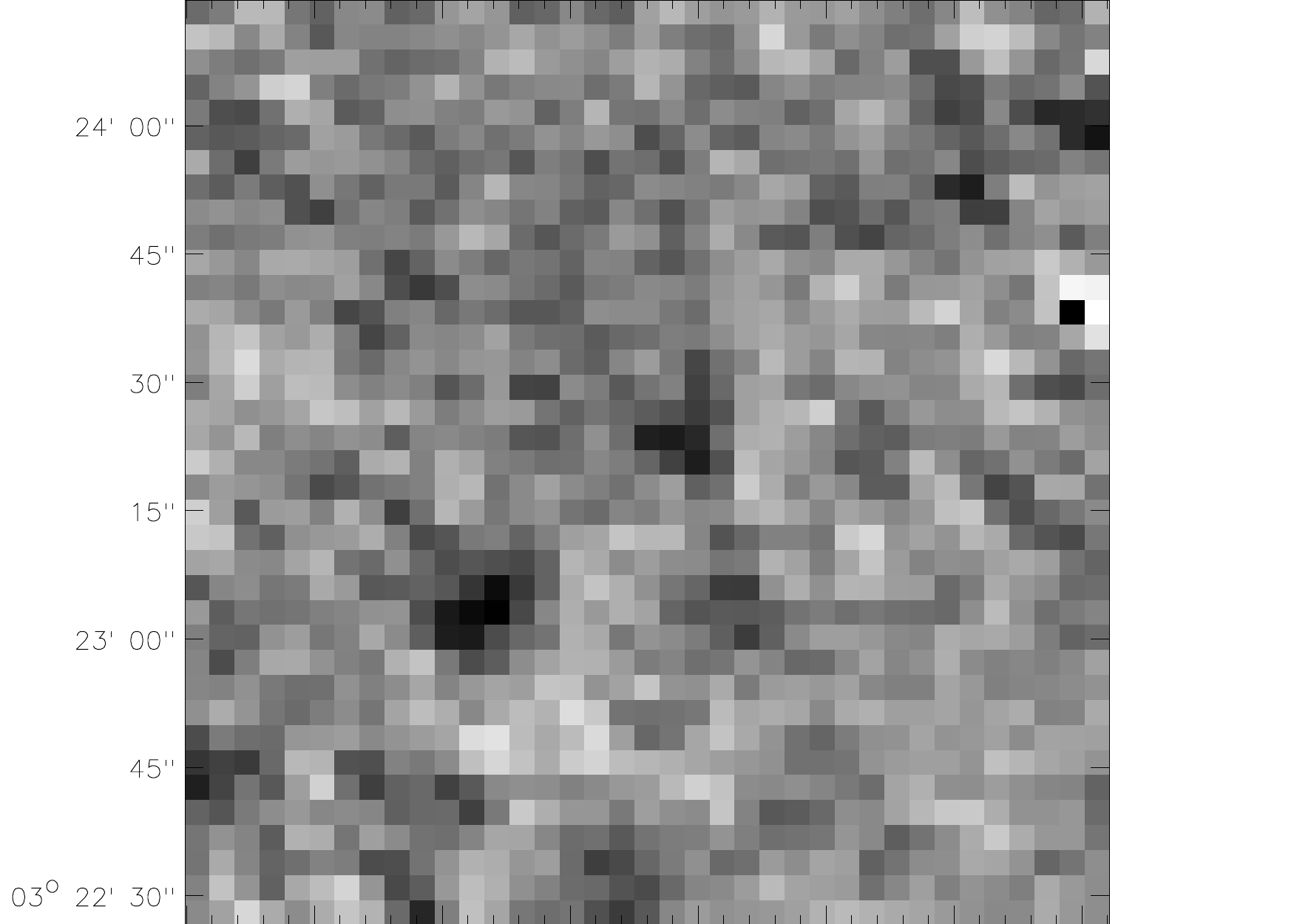} &
      \includegraphics[height=14mm,trim=70  -4 65 -10,clip=true]{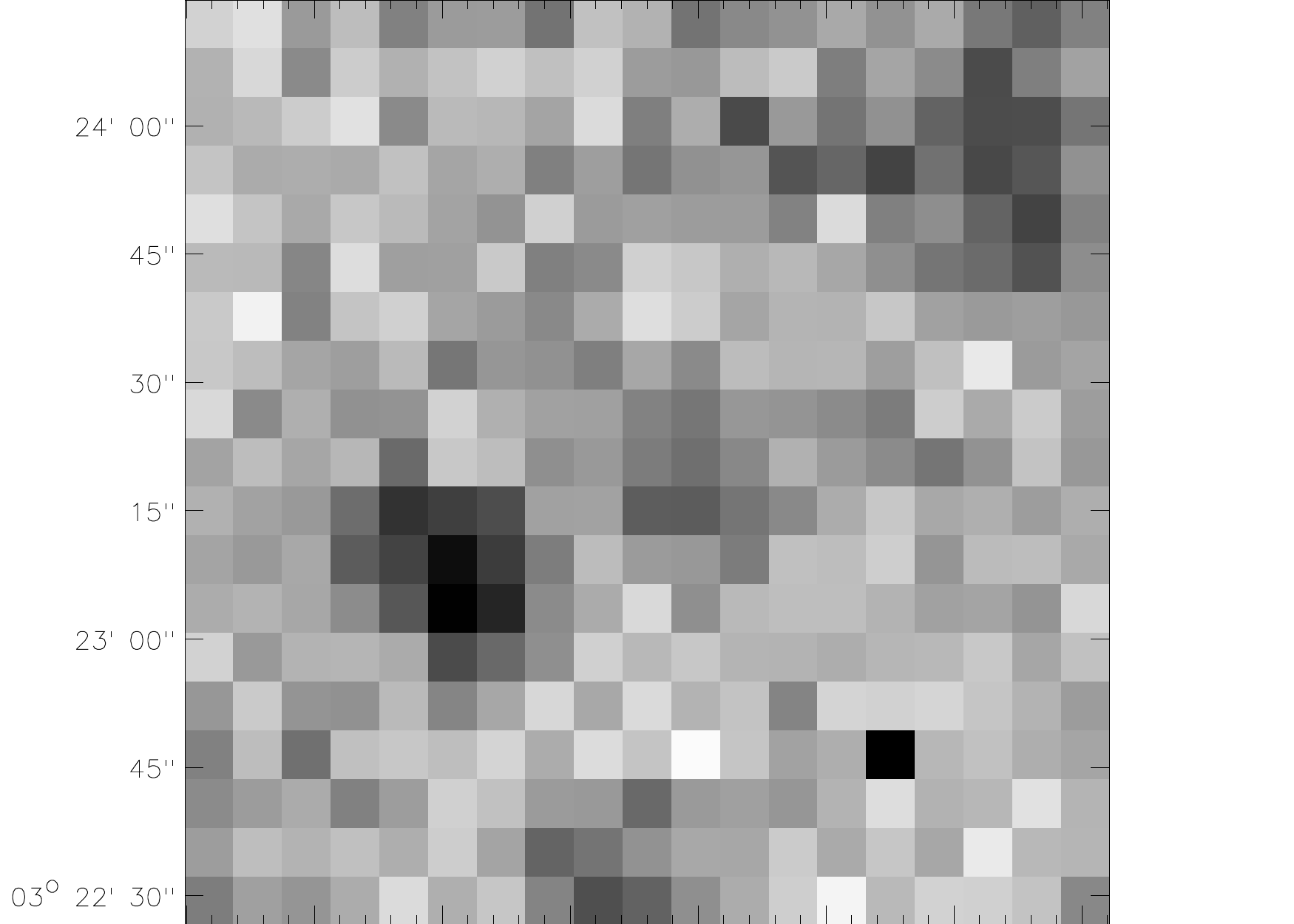}   &
      \includegraphics[height=14mm,trim=70  -4 65 -10,clip=true]{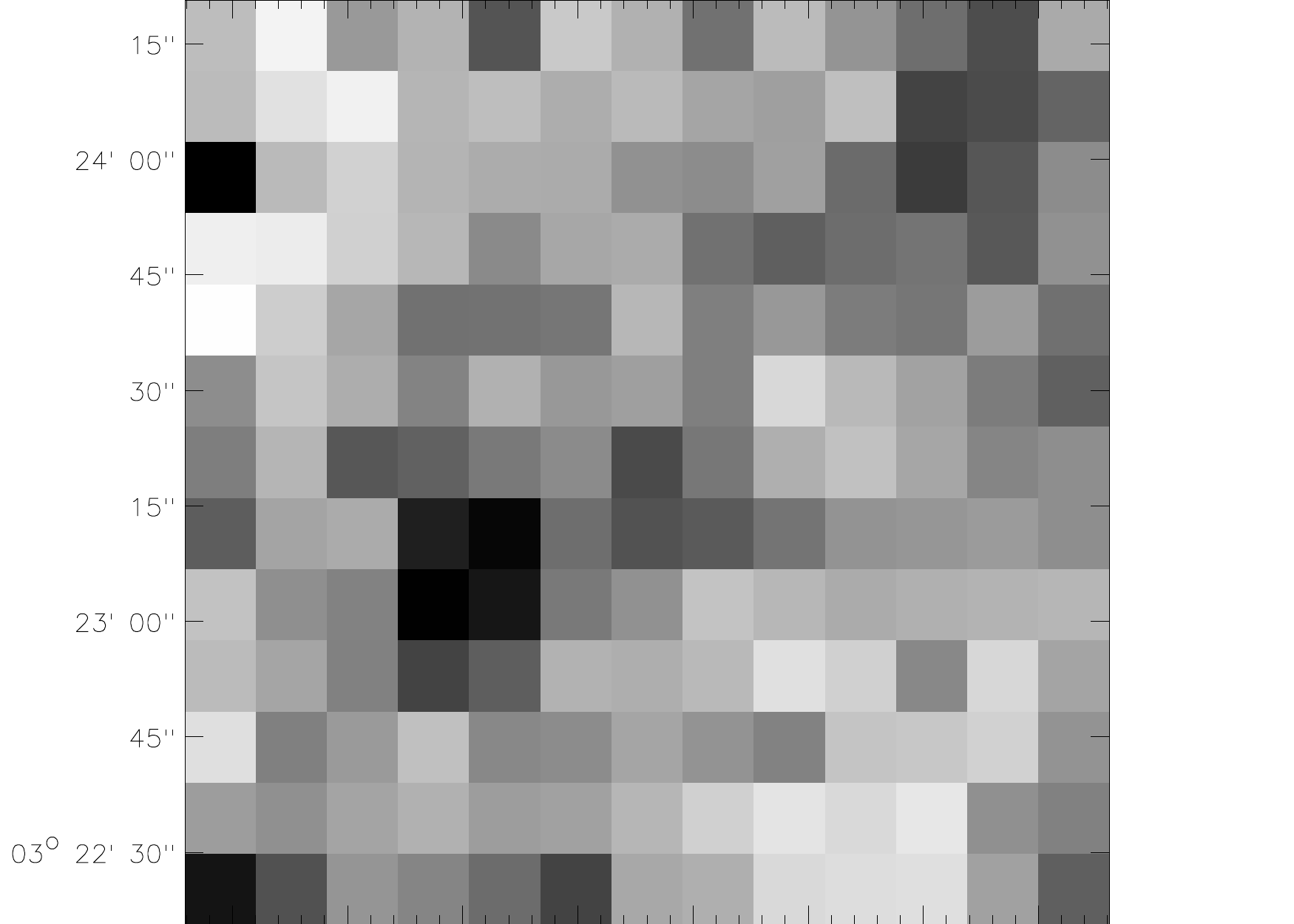}   &
      \includegraphics[height=14mm,trim=70  -4 65 -10,clip=true]{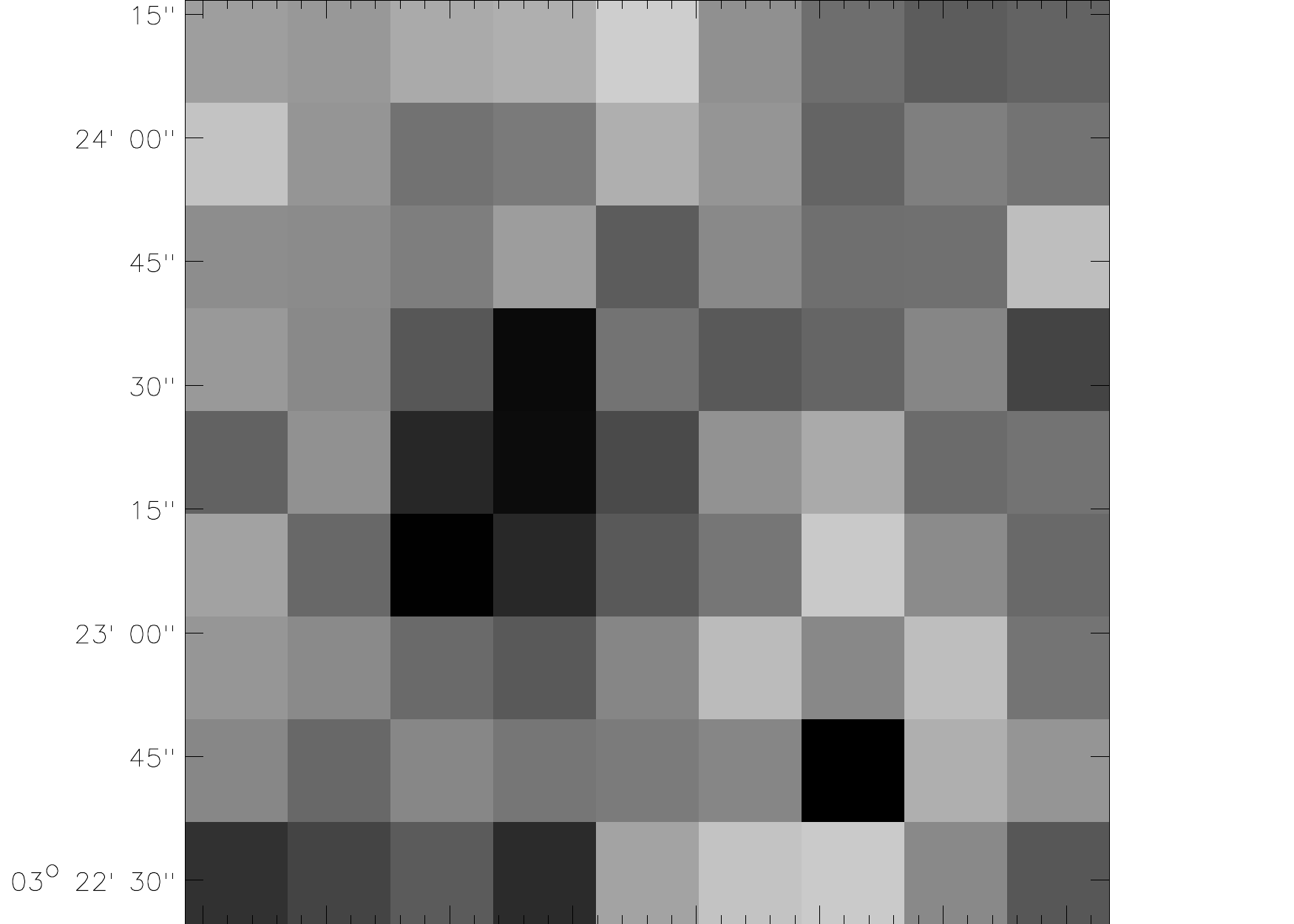} \\

      \multicolumn{6}{c}{\includegraphics[width=80mm,trim=30 0 20 0 ,clip=true]{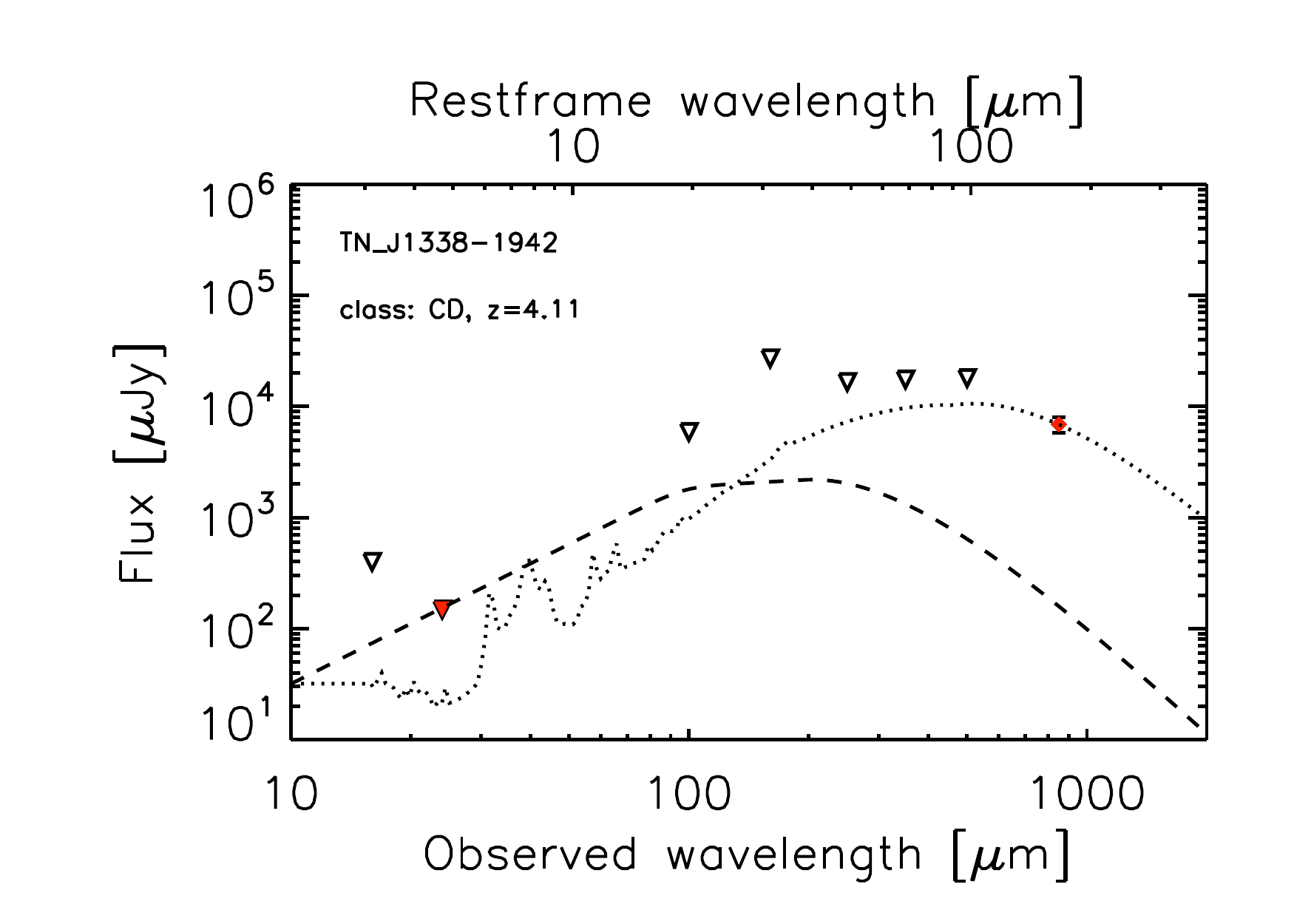}} \\
      24\mum & 100\mum & 160\mum & 250\mum & 350\mum & 500\mum \\
\end{tabular}
\end{minipage}
\begin{minipage}[c]{0.46\linewidth} 
\begin{tabular}{c@{}c@{}c@{}c@{}c@{}c@{}}

      \multicolumn{6}{c}{\includegraphics[width=80mm,trim=30 0 20 0 ,clip=true]{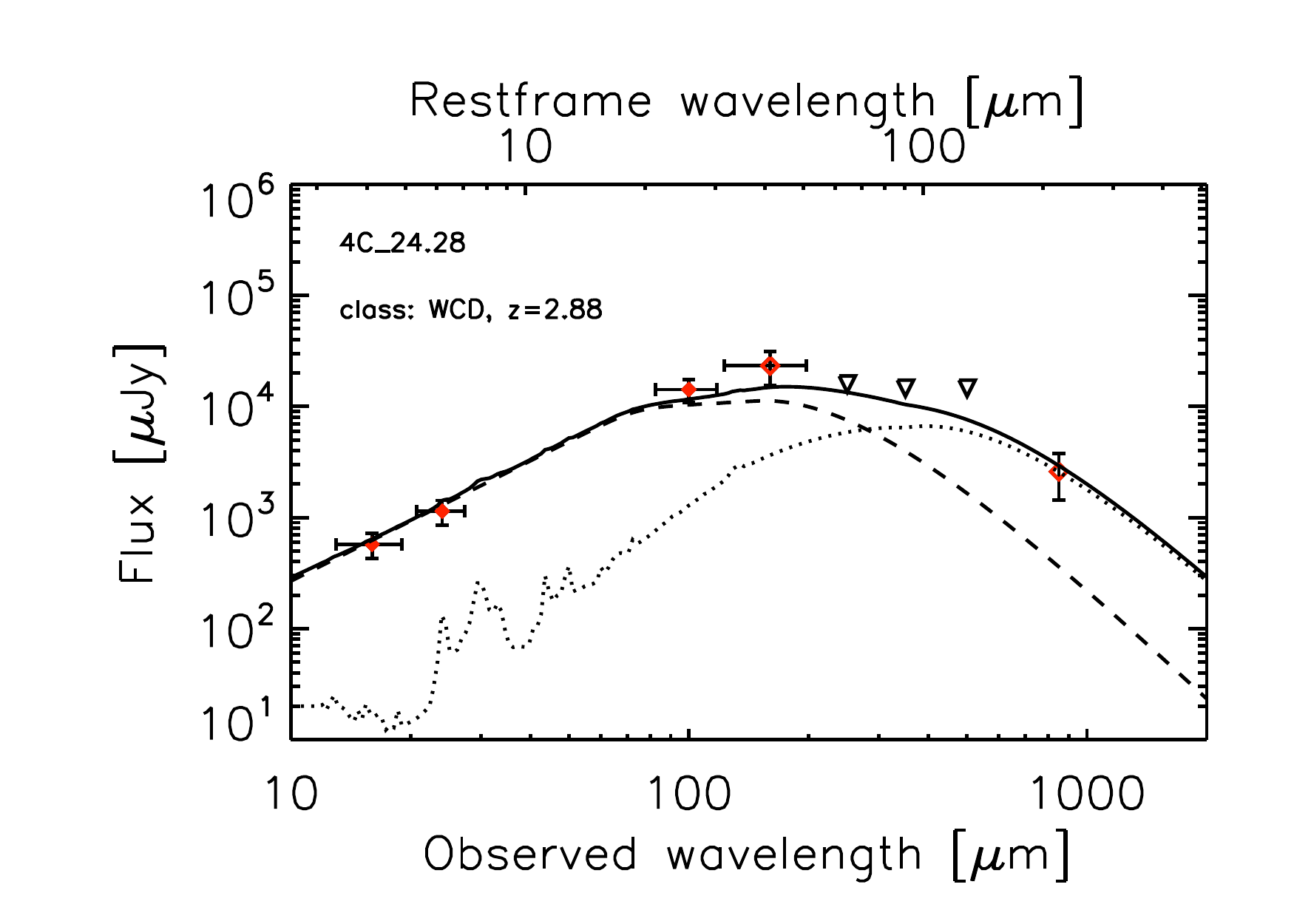}} \\
      24\mum & 100\mum & 160\mum & 250\mum & 350\mum & 500\mum \\
      \includegraphics[height=14mm,trim=70  -4 65 -10,clip=true]{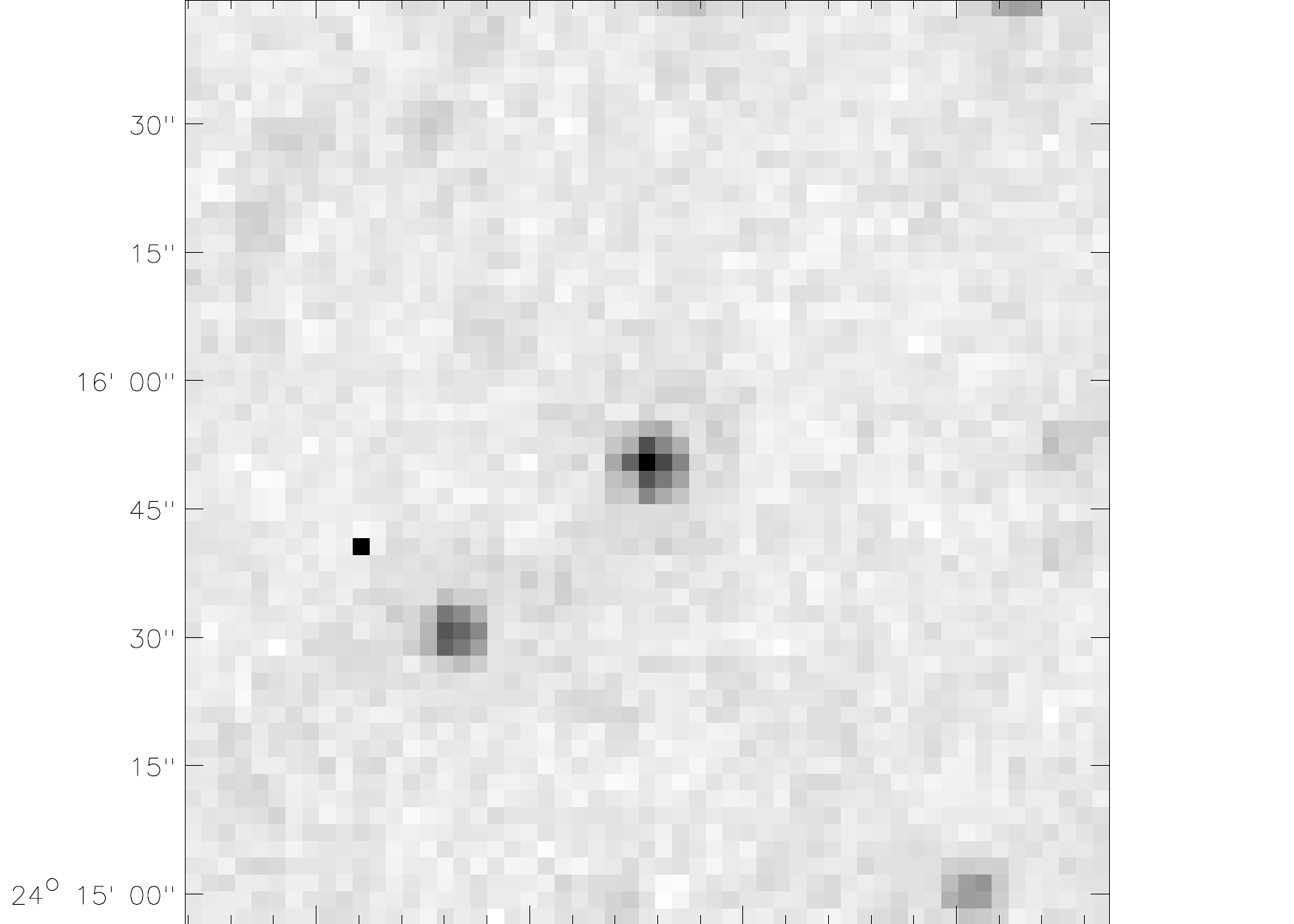} &
      \includegraphics[height=14mm,trim=70  -4 65 -10,clip=true]{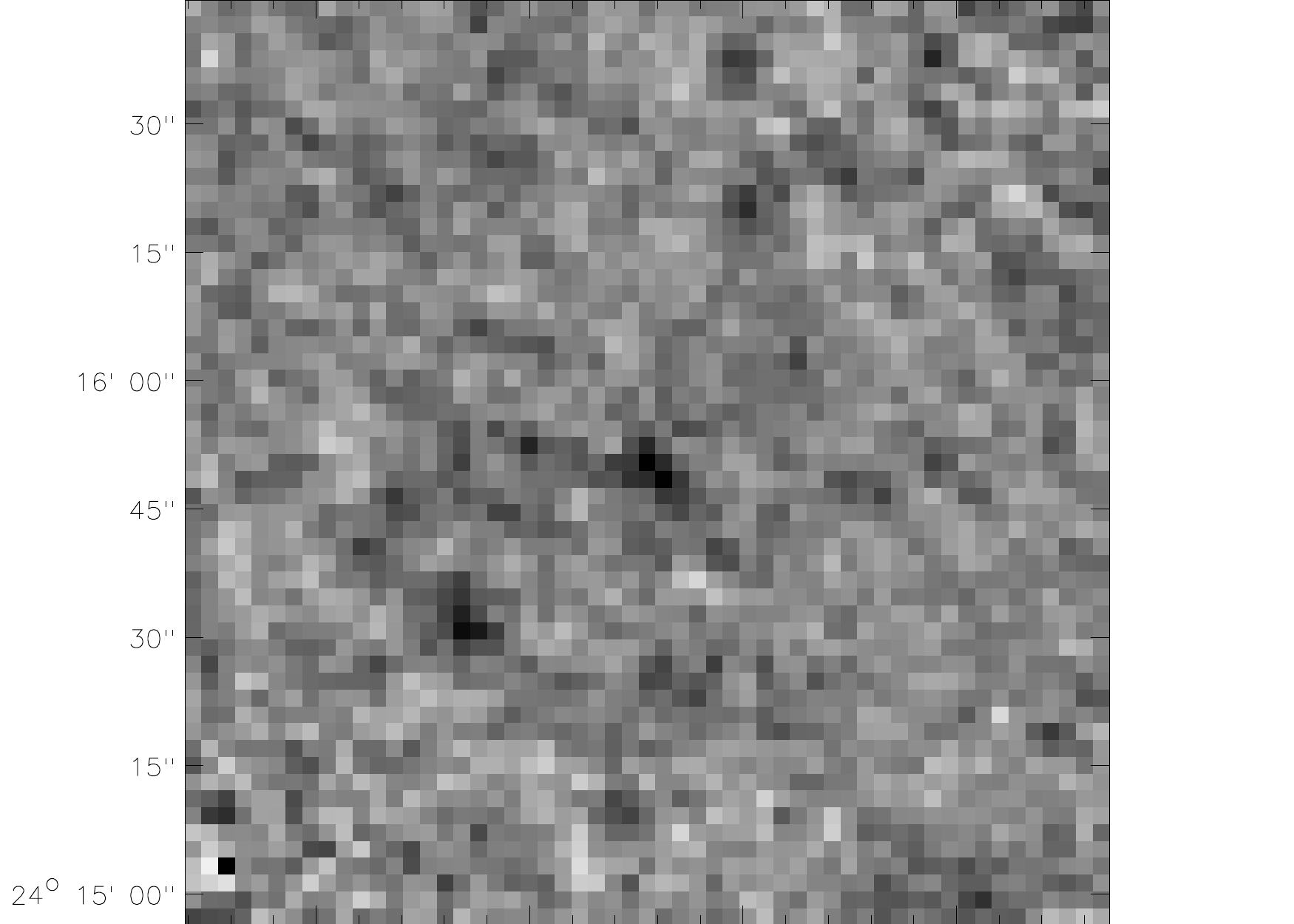} &
      \includegraphics[height=14mm,trim=70  -4 65 -10,clip=true]{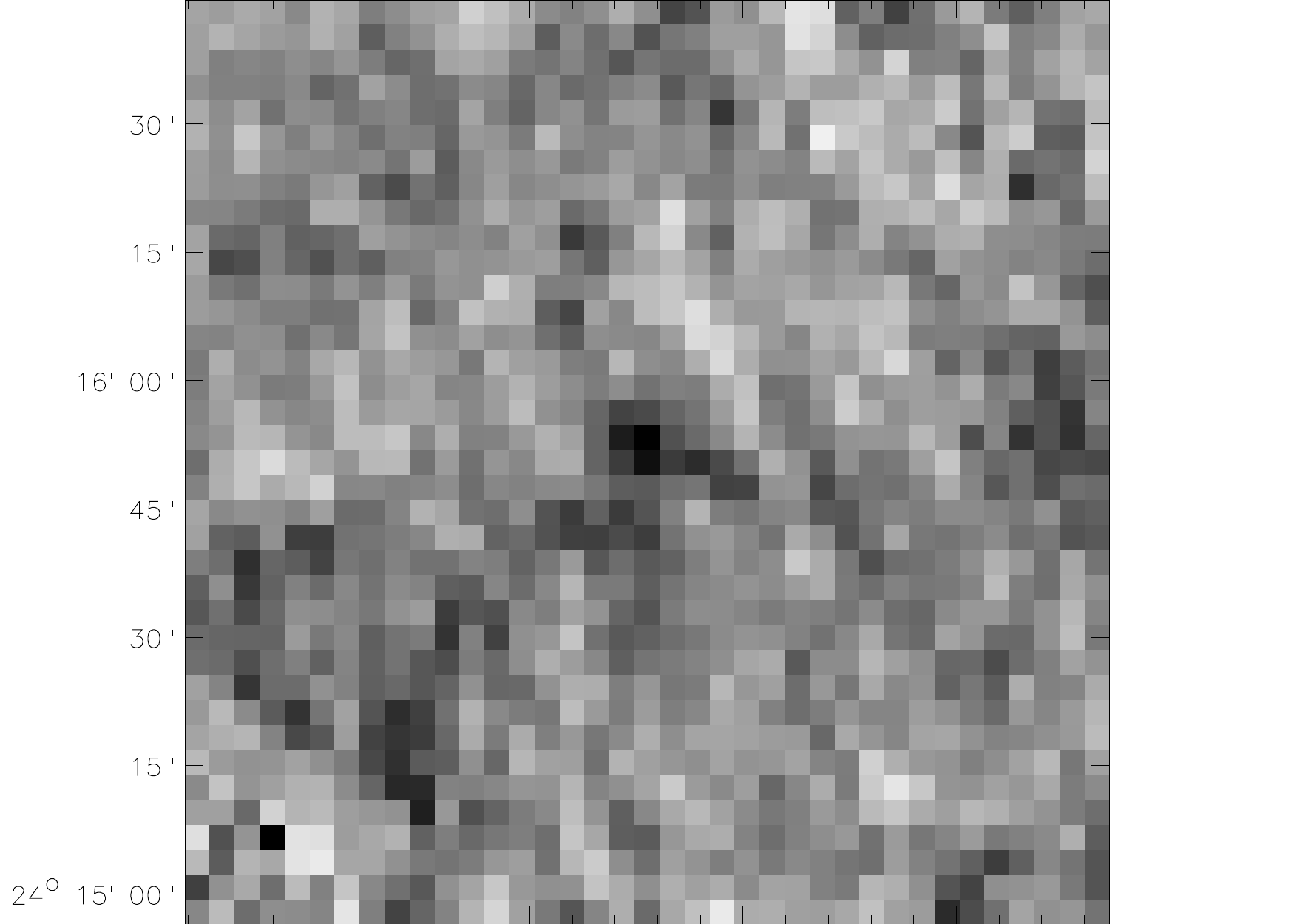} &
      \includegraphics[height=14mm,trim=70  -4 65 -10,clip=true]{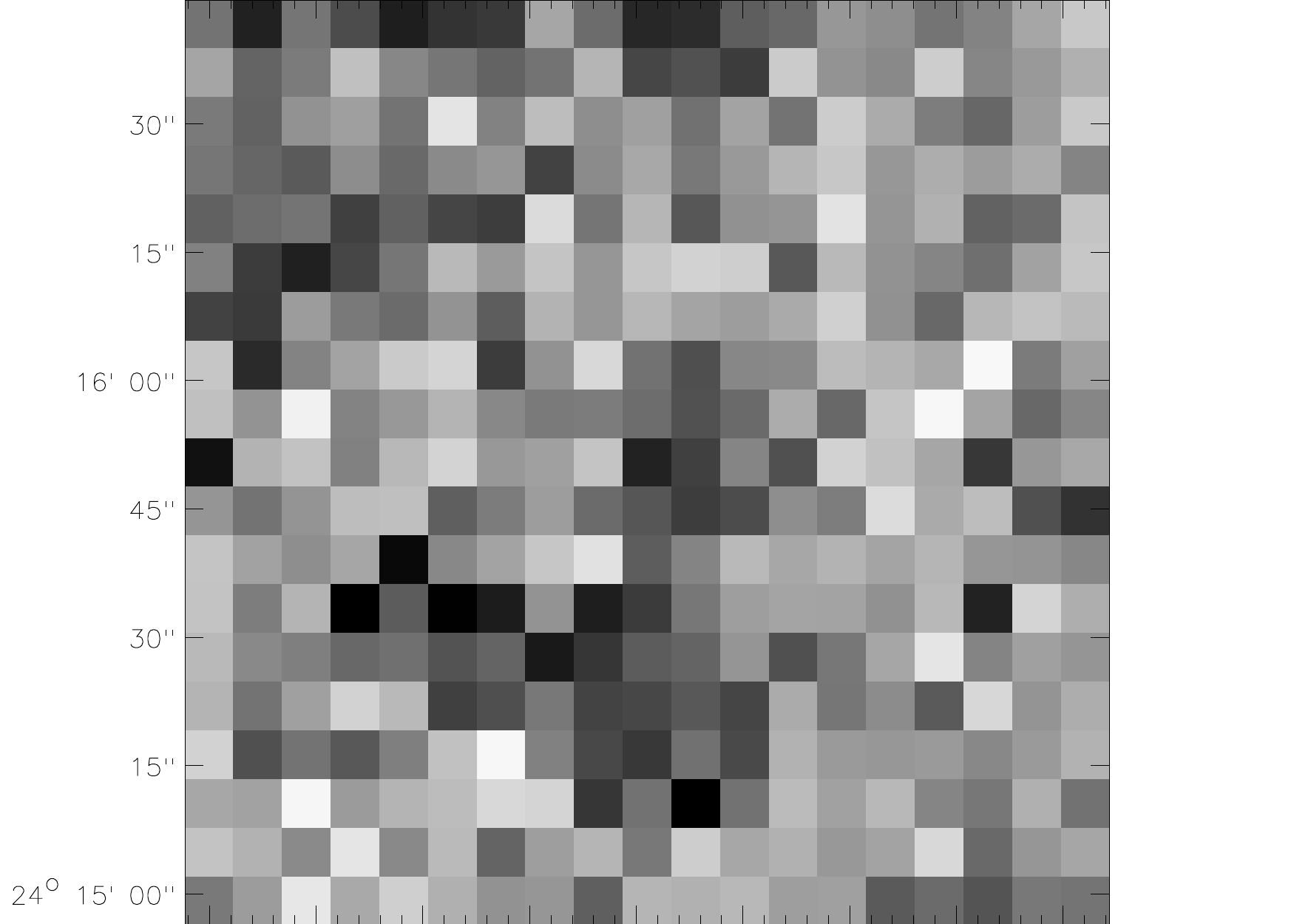}   &
      \includegraphics[height=14mm,trim=70  -4 65 -10,clip=true]{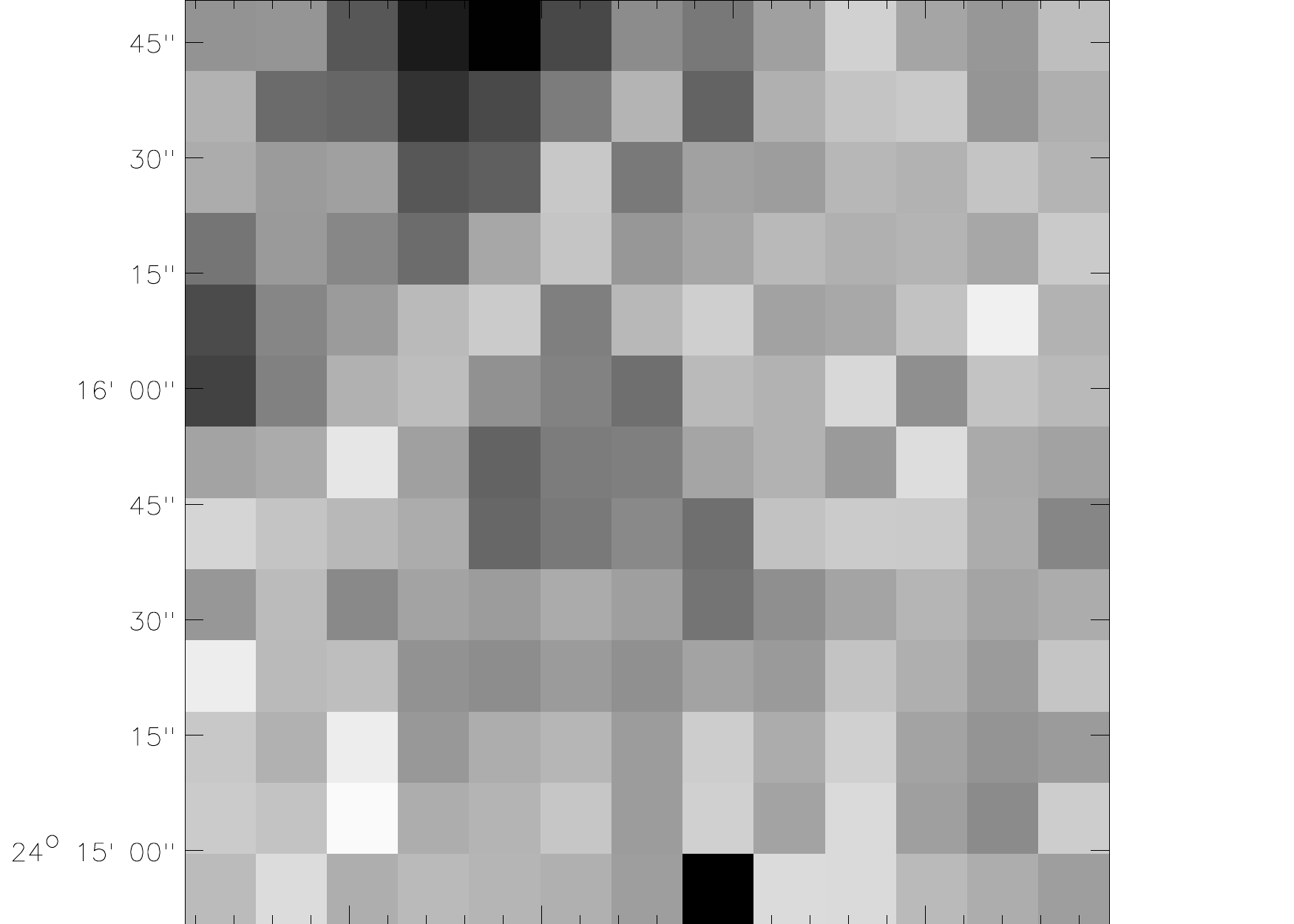}   &
      \includegraphics[height=14mm,trim=70  -4 65 -10,clip=true]{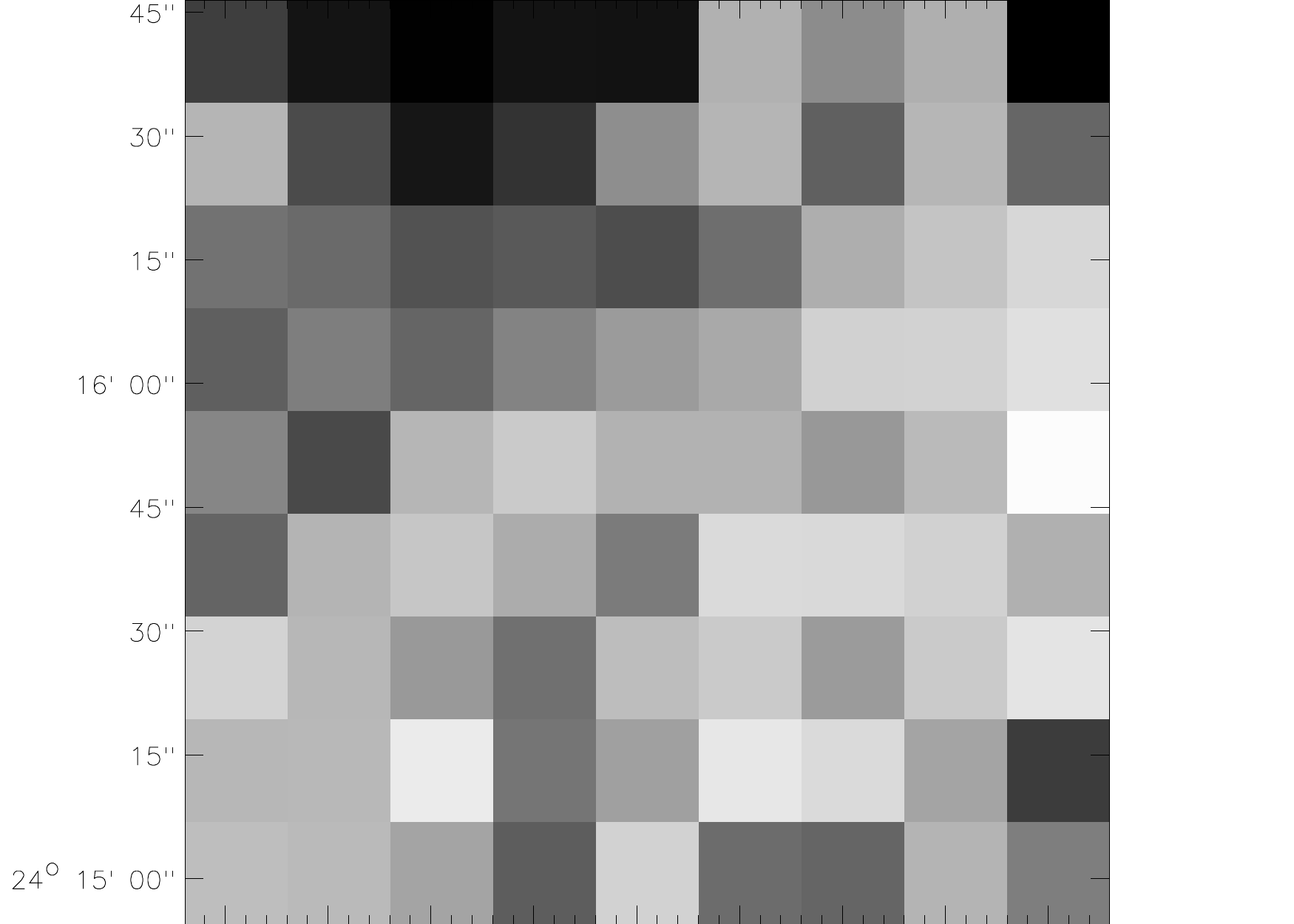} \\

      \multicolumn{6}{c}{\includegraphics[width=80mm,trim=30 0 20 0 ,clip=true]{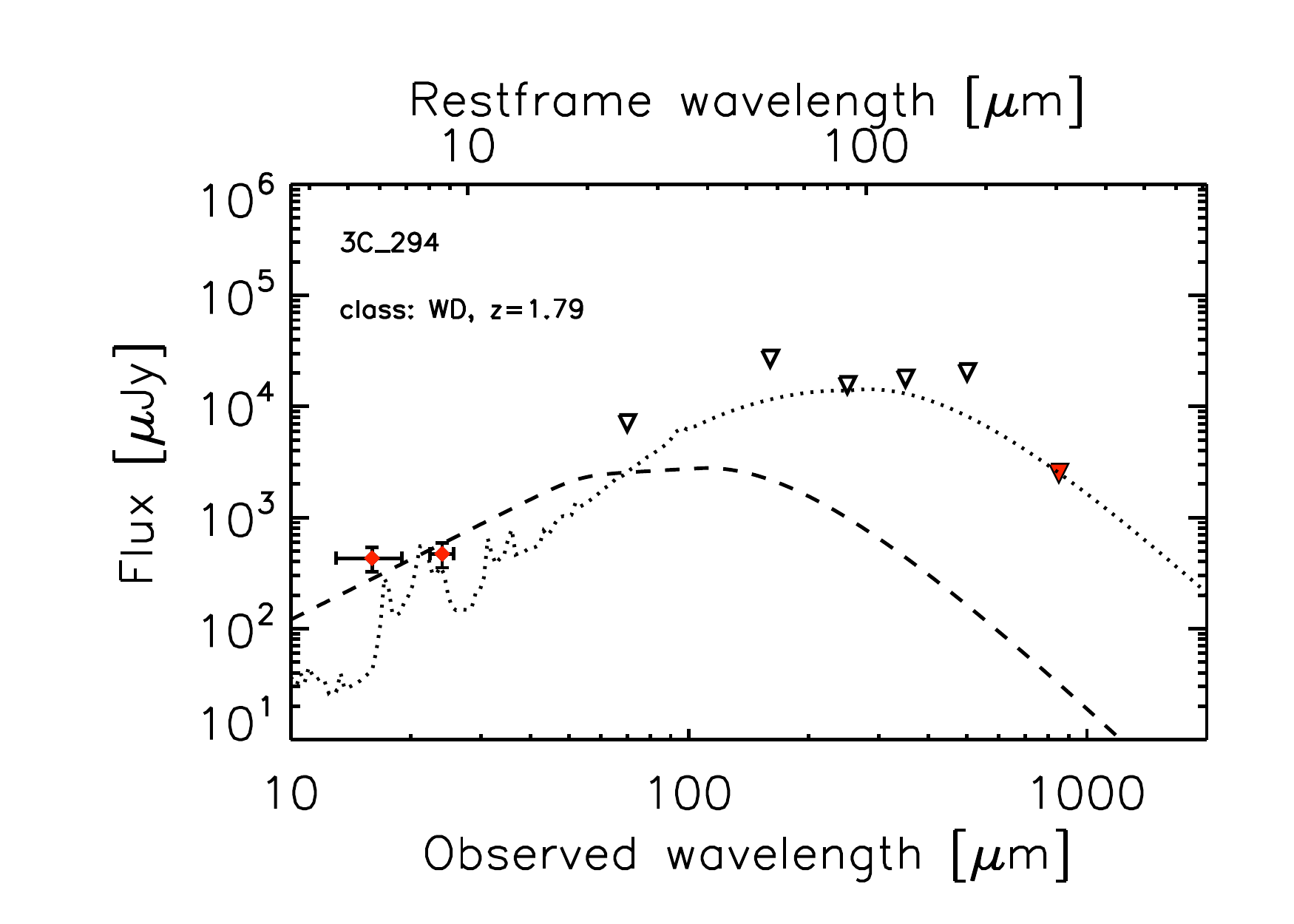}} \\
      24\mum & 70\mum & 160\mum & 250\mum & 350\mum & 500\mum \\
      \includegraphics[height=14mm,trim=70  -4 65 -10,clip=true]{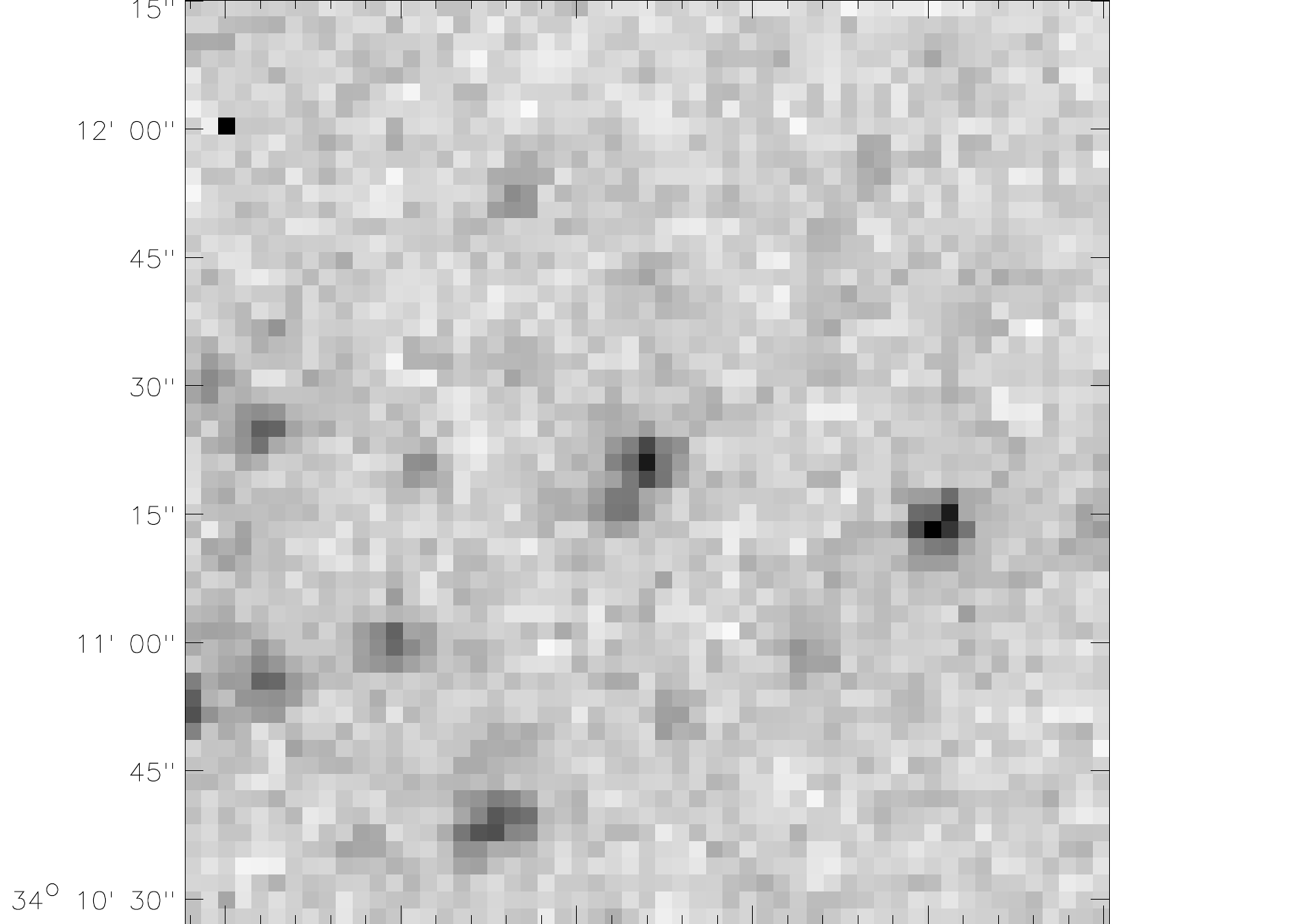} &
      \includegraphics[height=14mm,trim=70  -4 65 -10,clip=true]{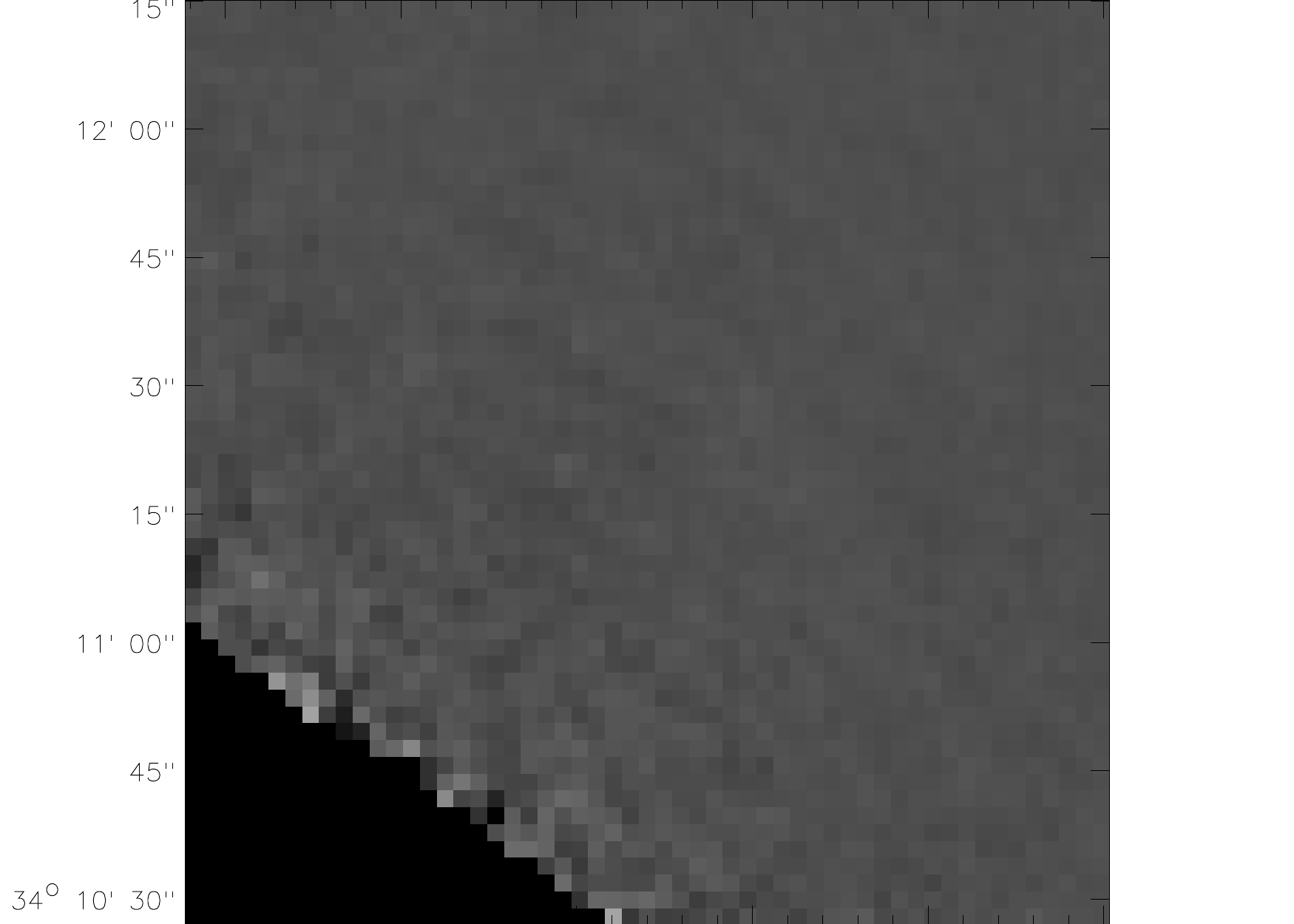} &
      \includegraphics[height=14mm,trim=70  -4 65 -10,clip=true]{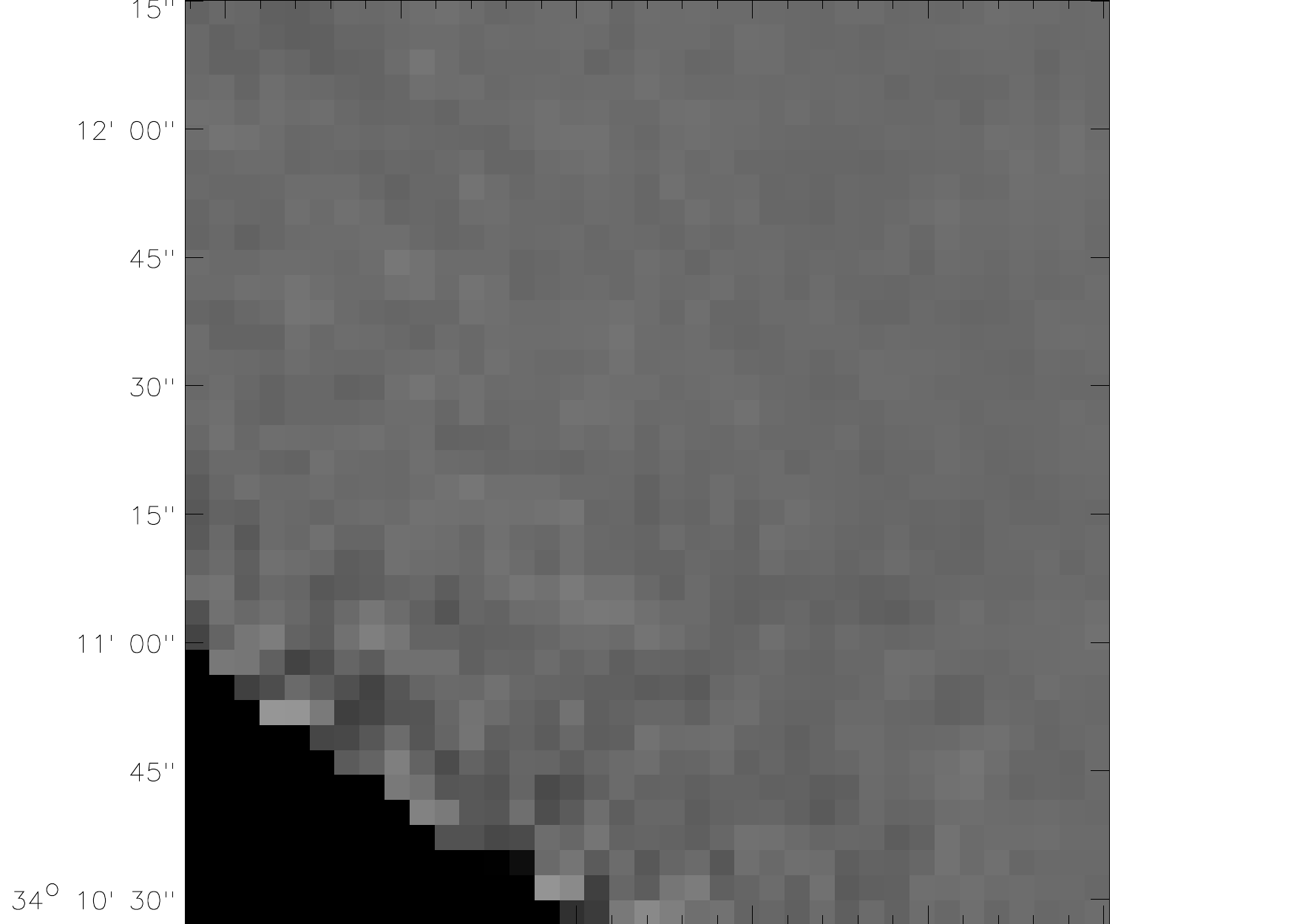} &
      \includegraphics[height=14mm,trim=70  -4 65 -10,clip=true]{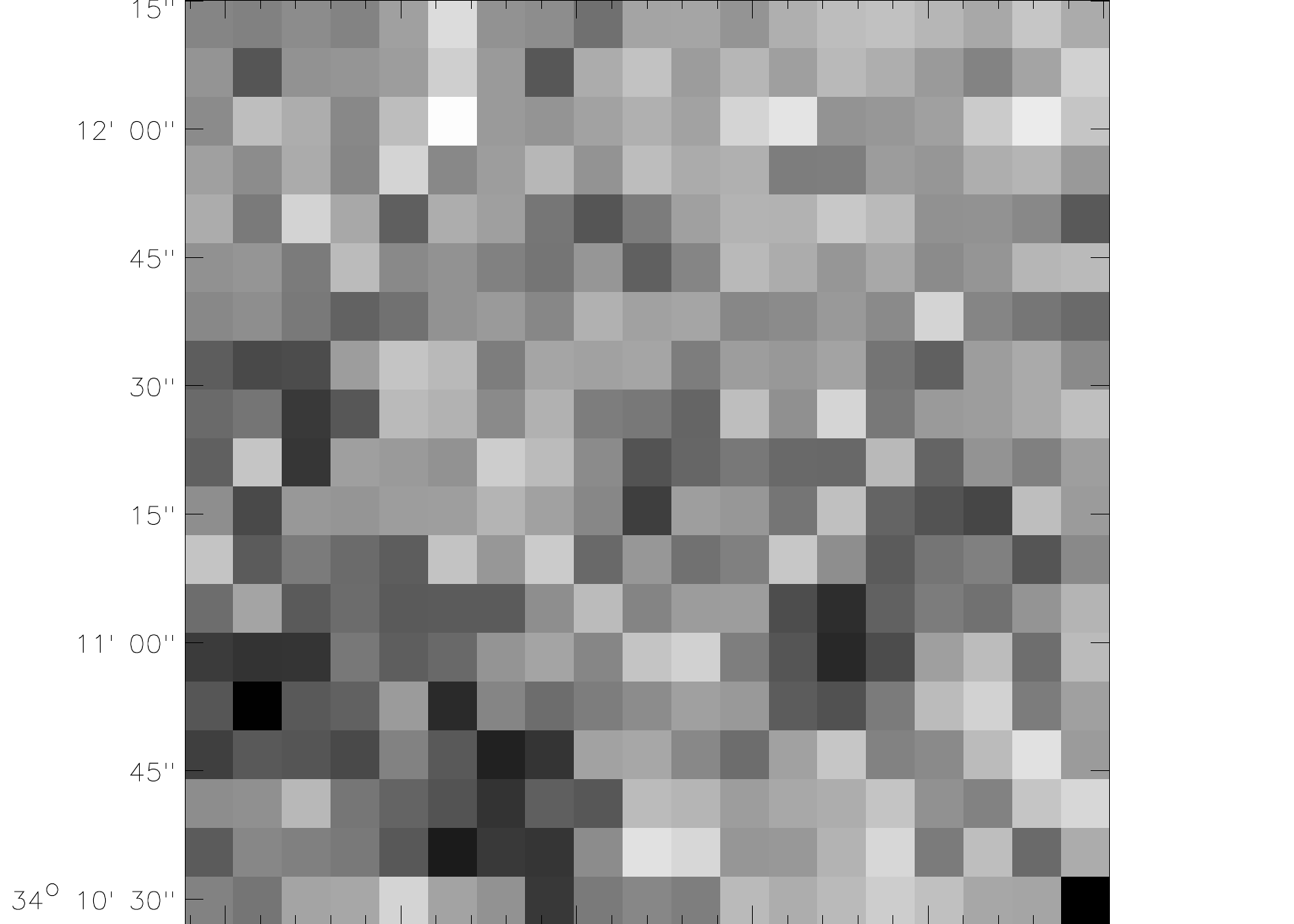}   &
      \includegraphics[height=14mm,trim=70  -4 65 -10,clip=true]{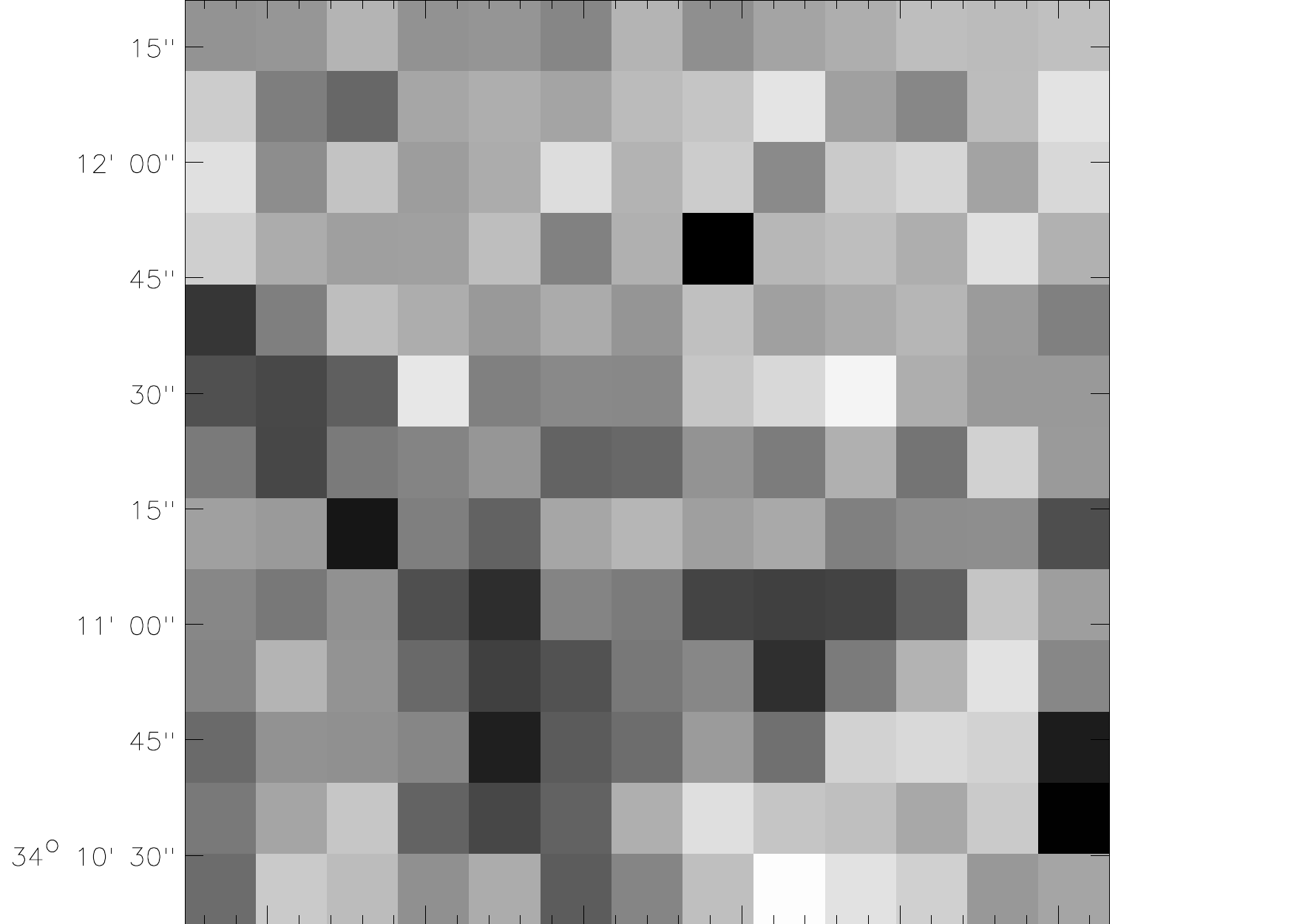}   &
      \includegraphics[height=14mm,trim=70  -4 65 -10,clip=true]{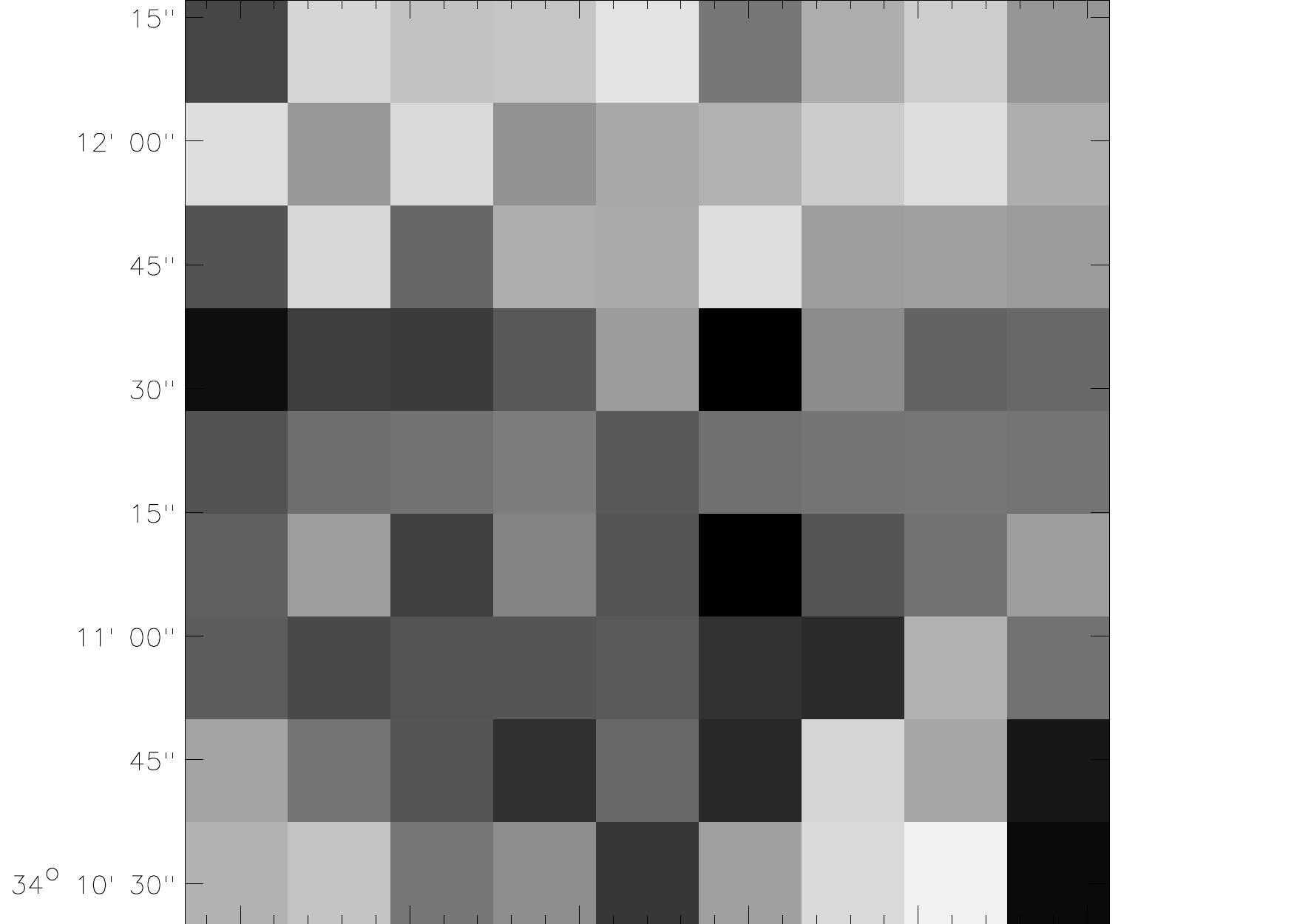} \\

      \multicolumn{6}{c}{\includegraphics[width=80mm,trim=30 0 20 0 ,clip=true]{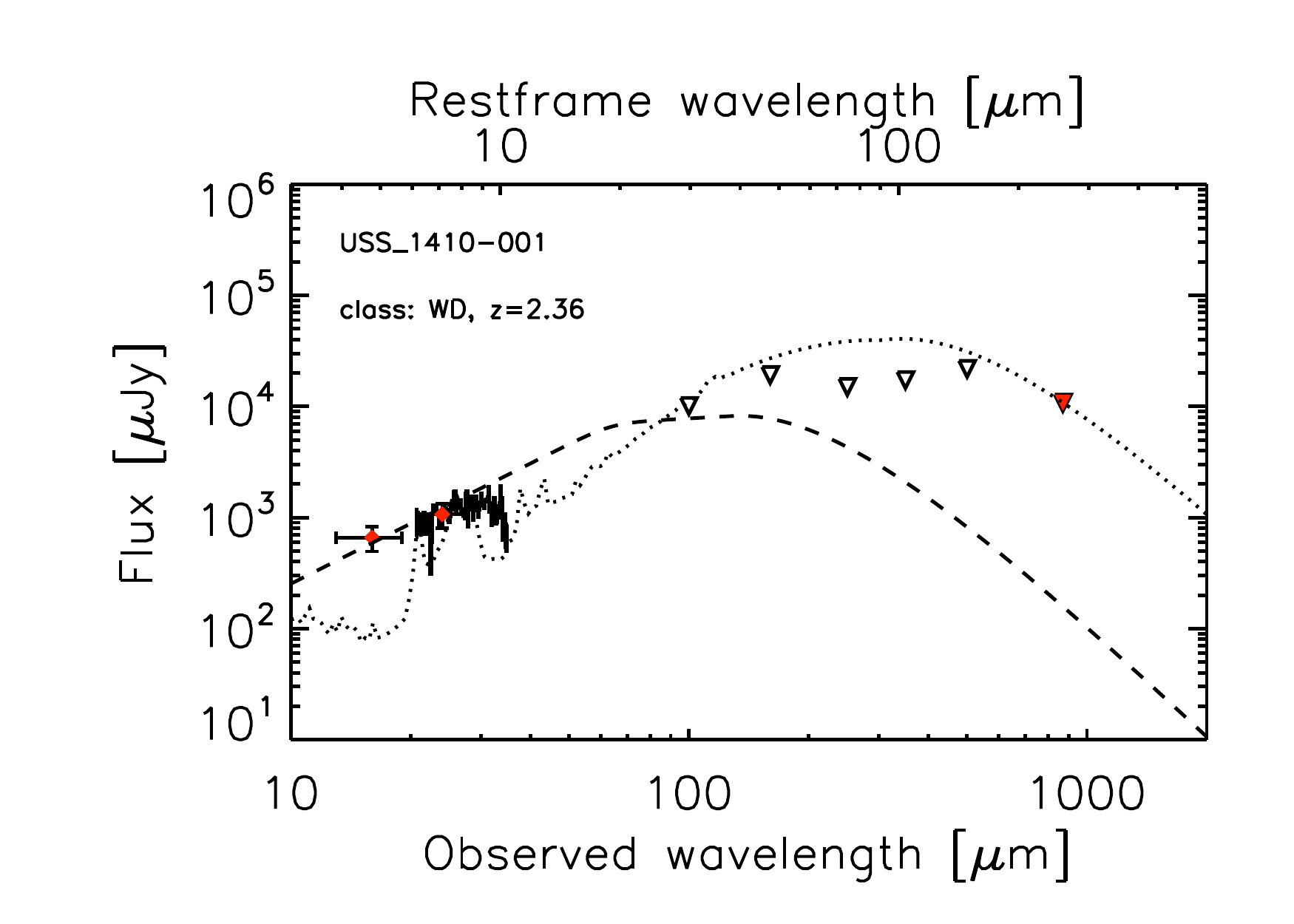}} \\
      24\mum & 100\mum & 160\mum & 250\mum & 350\mum & 500\mum \\
\end{tabular}
          \end{minipage}
  \end{centering}
\end{figure*}

\clearpage

\begin{figure*}[h]
  \begin{centering}
    \begin{minipage}[c]{0.46\linewidth}
    \begin{tabular}{c@{}c@{}c@{}c@{}c@{}c@{}}

      \multicolumn{6}{c}{\includegraphics[width=80mm,trim=30 0 20 0 ,clip=true]{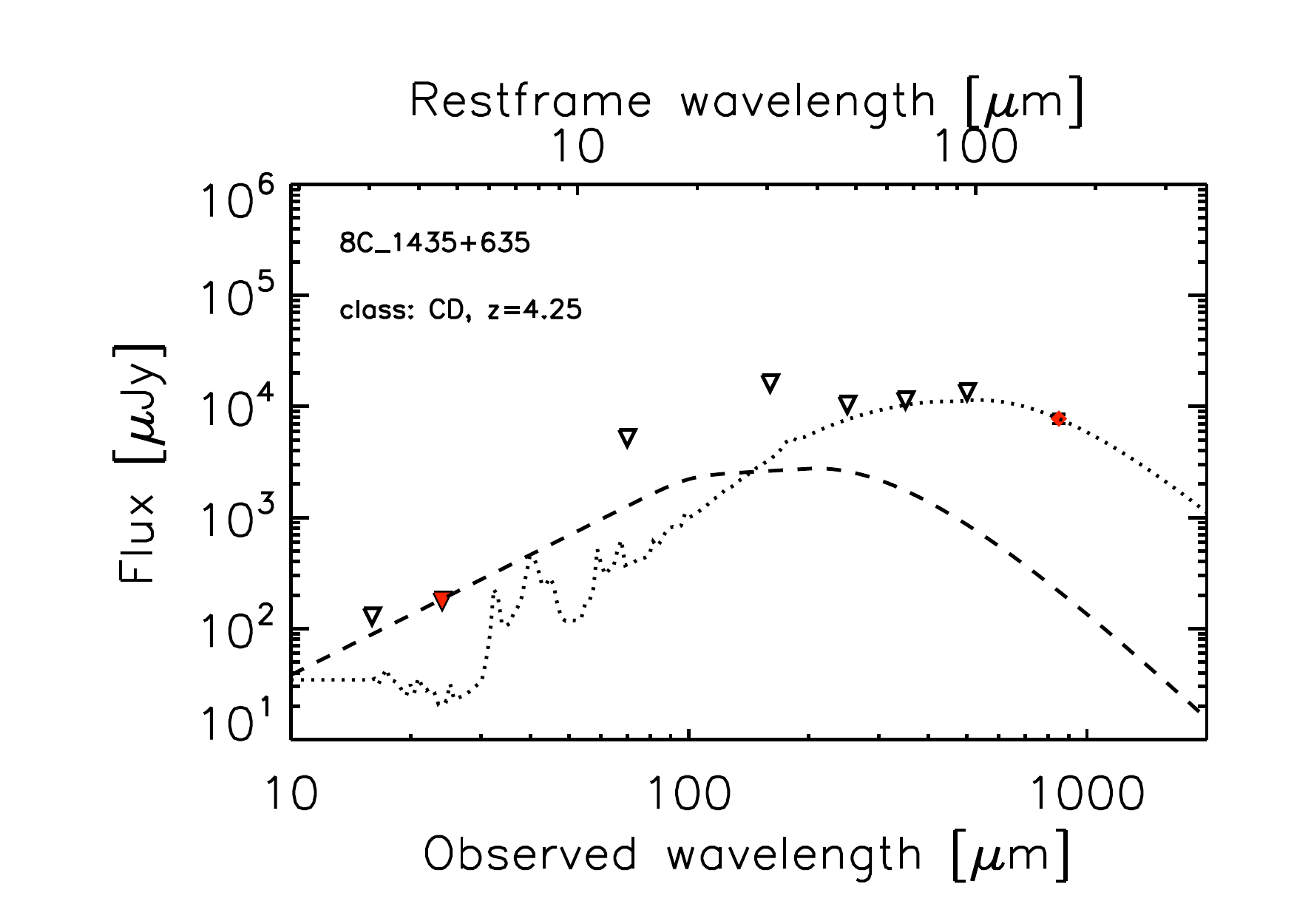}} \\
      24\mum & 70\mum & 160\mum & 250\mum & 350\mum & 500\mum \\
      \includegraphics[height=14mm,trim=70  -4 65 -10,clip=true]{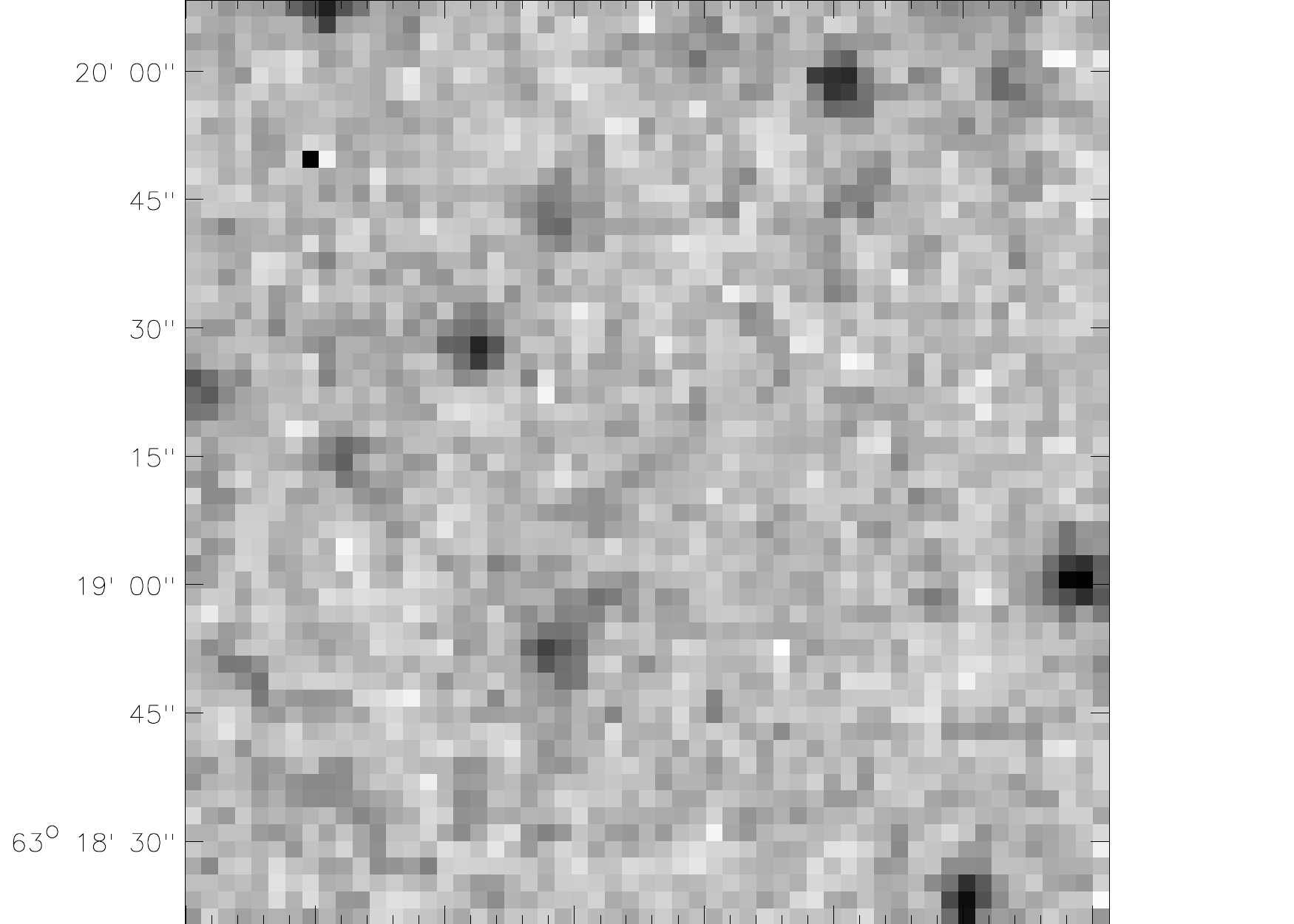} &
      \includegraphics[height=14mm,trim=70  -4 65 -10,clip=true]{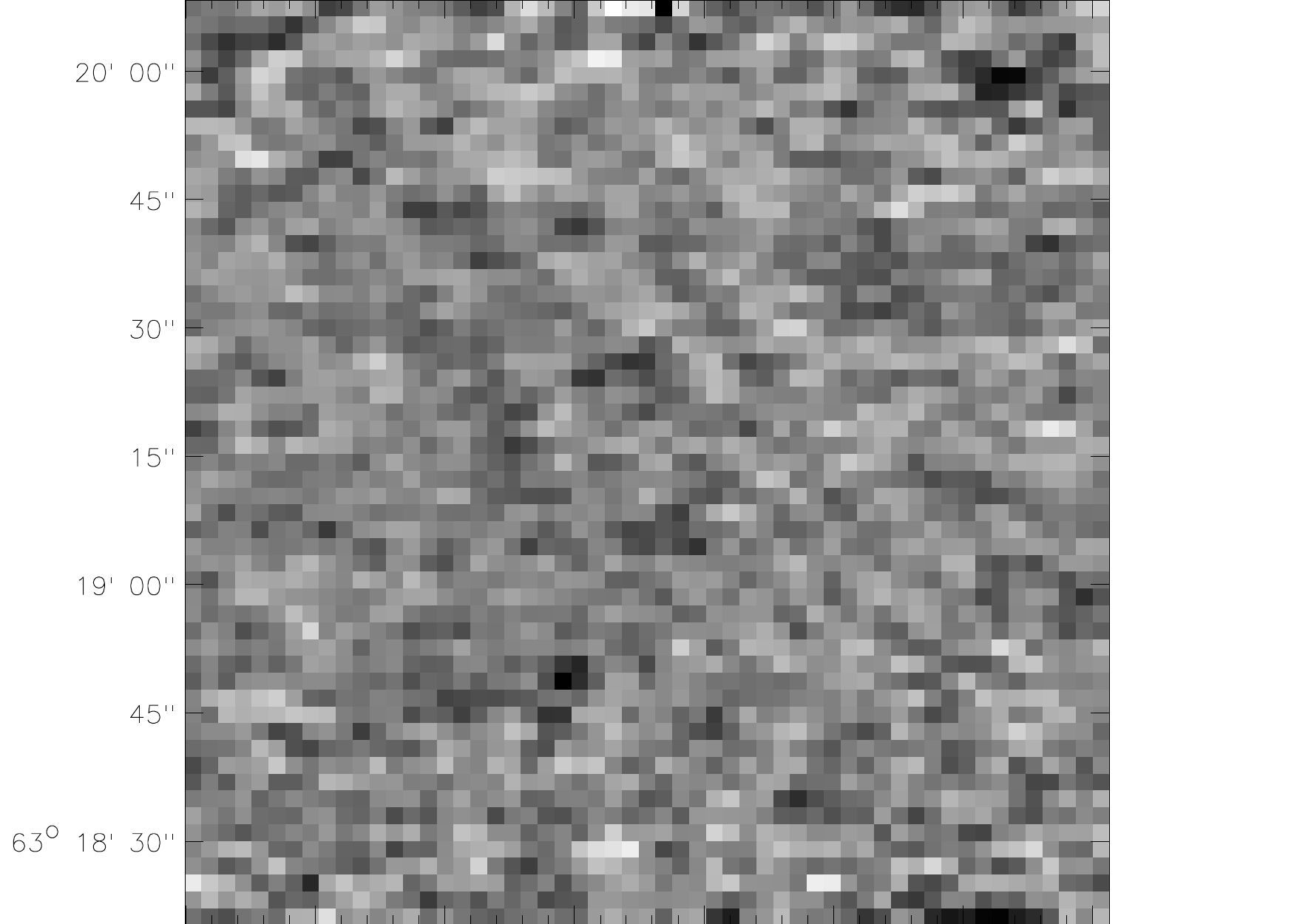} &
      \includegraphics[height=14mm,trim=70  -4 65 -10,clip=true]{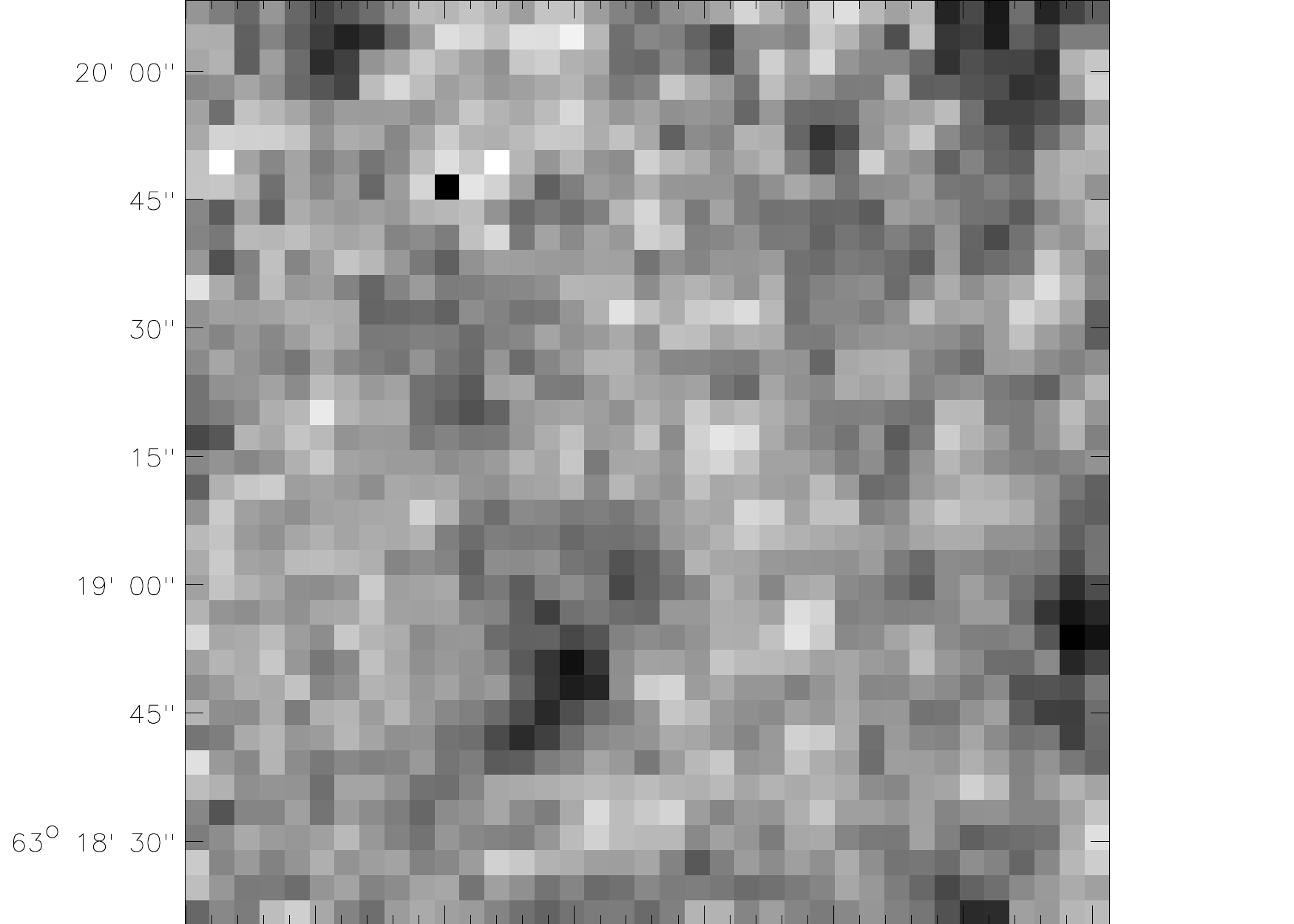} &
      \includegraphics[height=14mm,trim=70  -4 65 -10,clip=true]{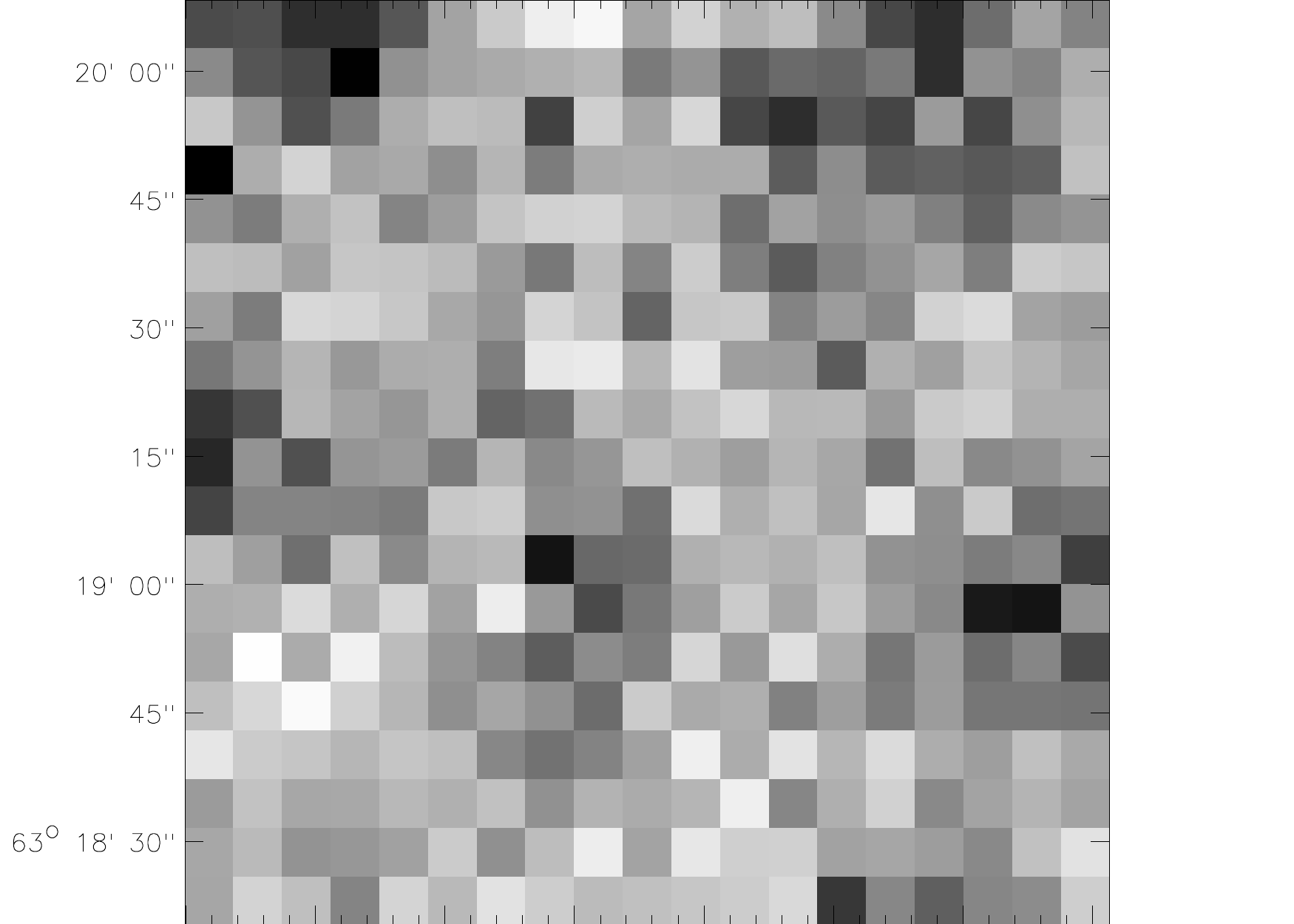}   &
      \includegraphics[height=14mm,trim=70  -4 65 -10,clip=true]{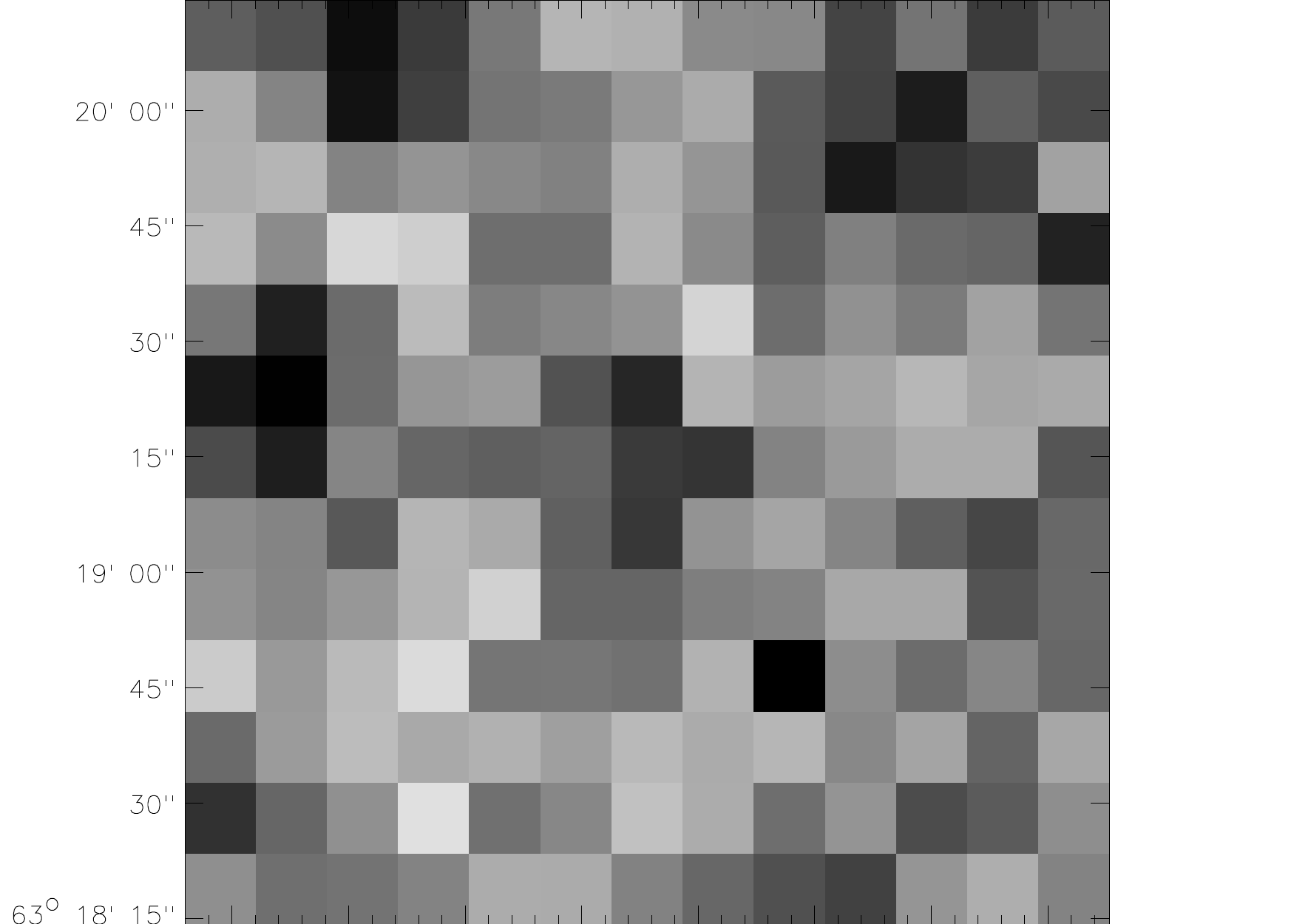}   &
      \includegraphics[height=14mm,trim=70  -4 65 -10,clip=true]{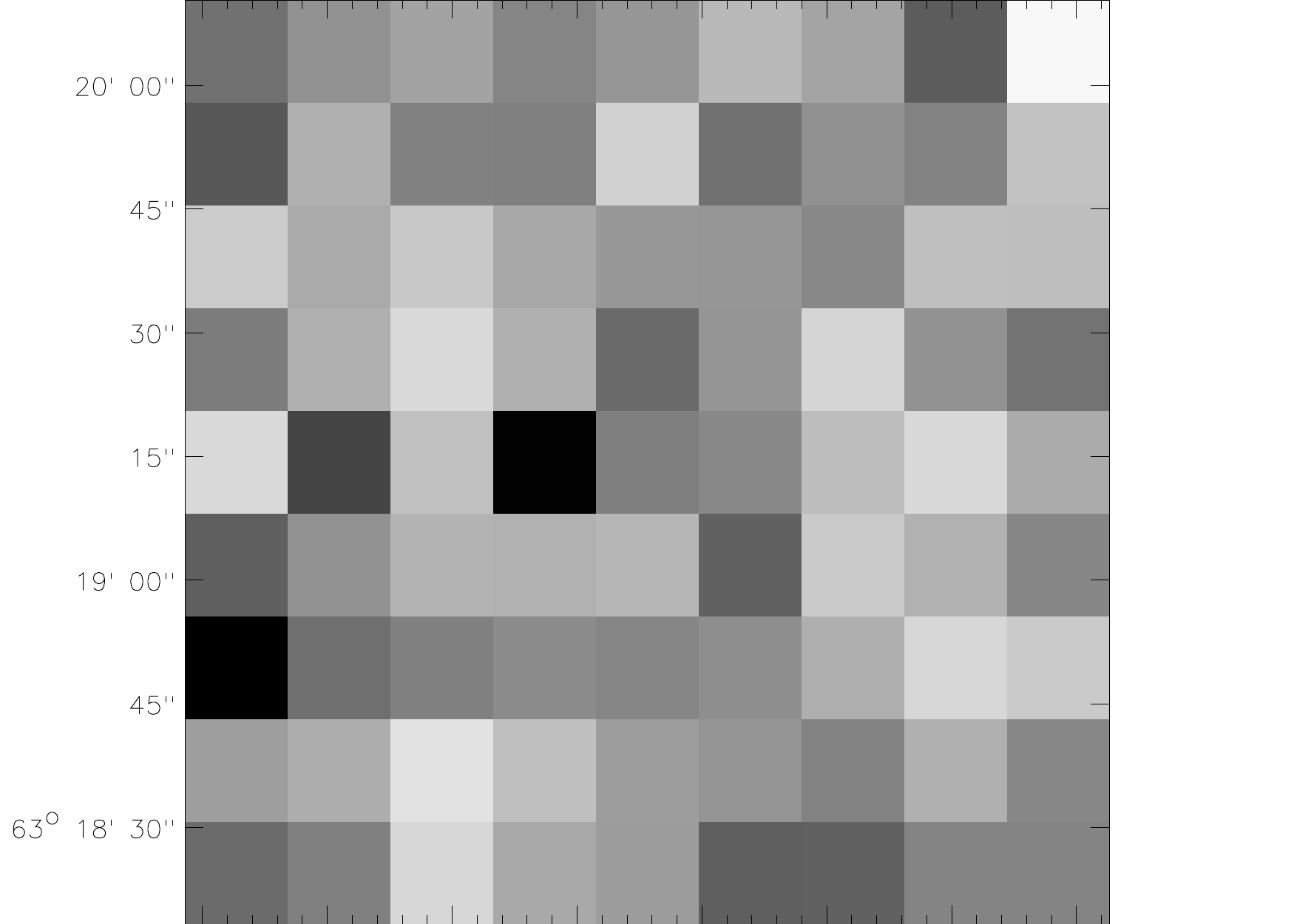} \\

      \multicolumn{6}{c}{\includegraphics[width=80mm,trim=30 0 20 0 ,clip=true]{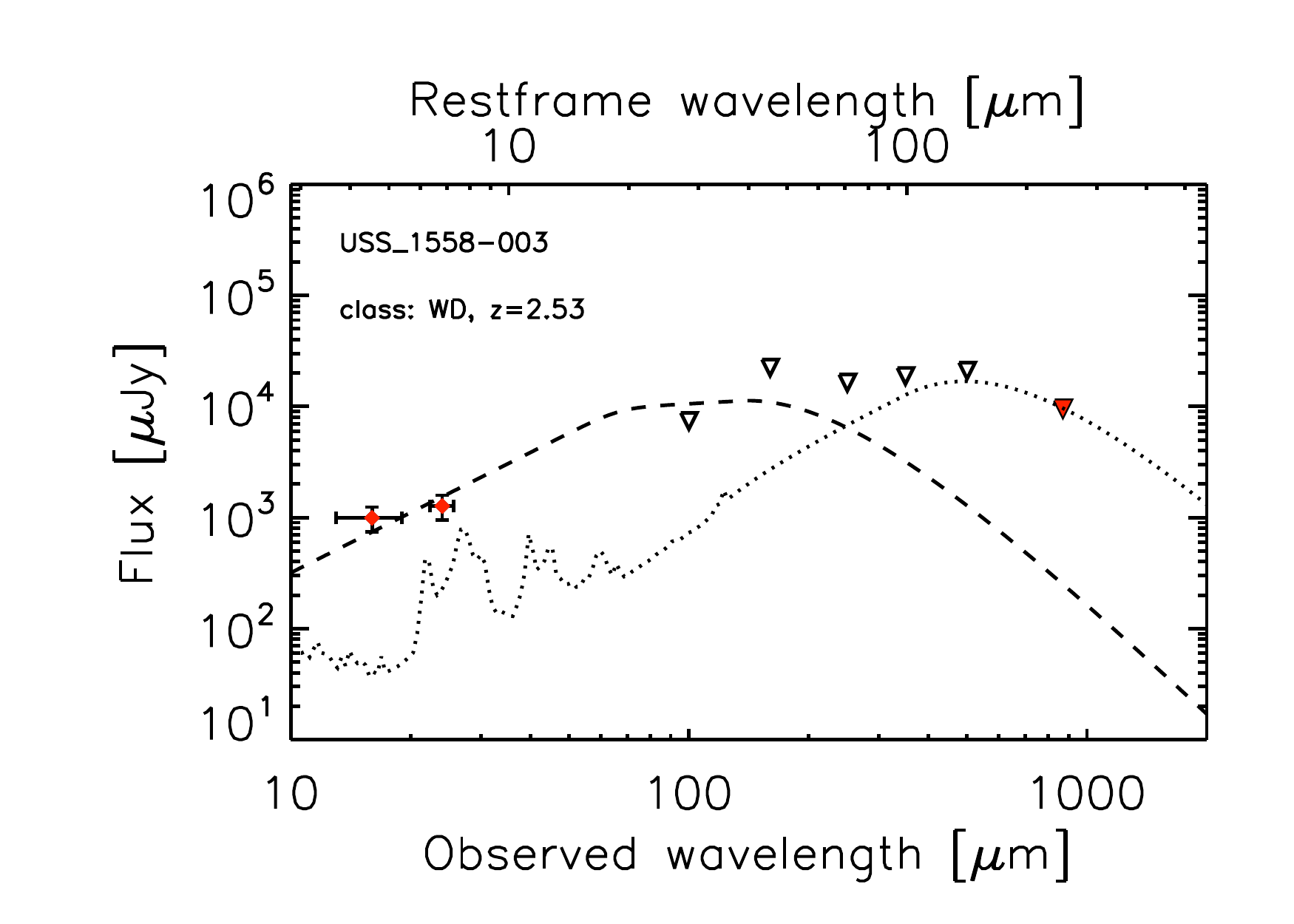}} \\
      24\mum & 100\mum & 160\mum & 250\mum & 350\mum & 500\mum \\
      \includegraphics[height=14mm,trim=70  -4 65 -10,clip=true]{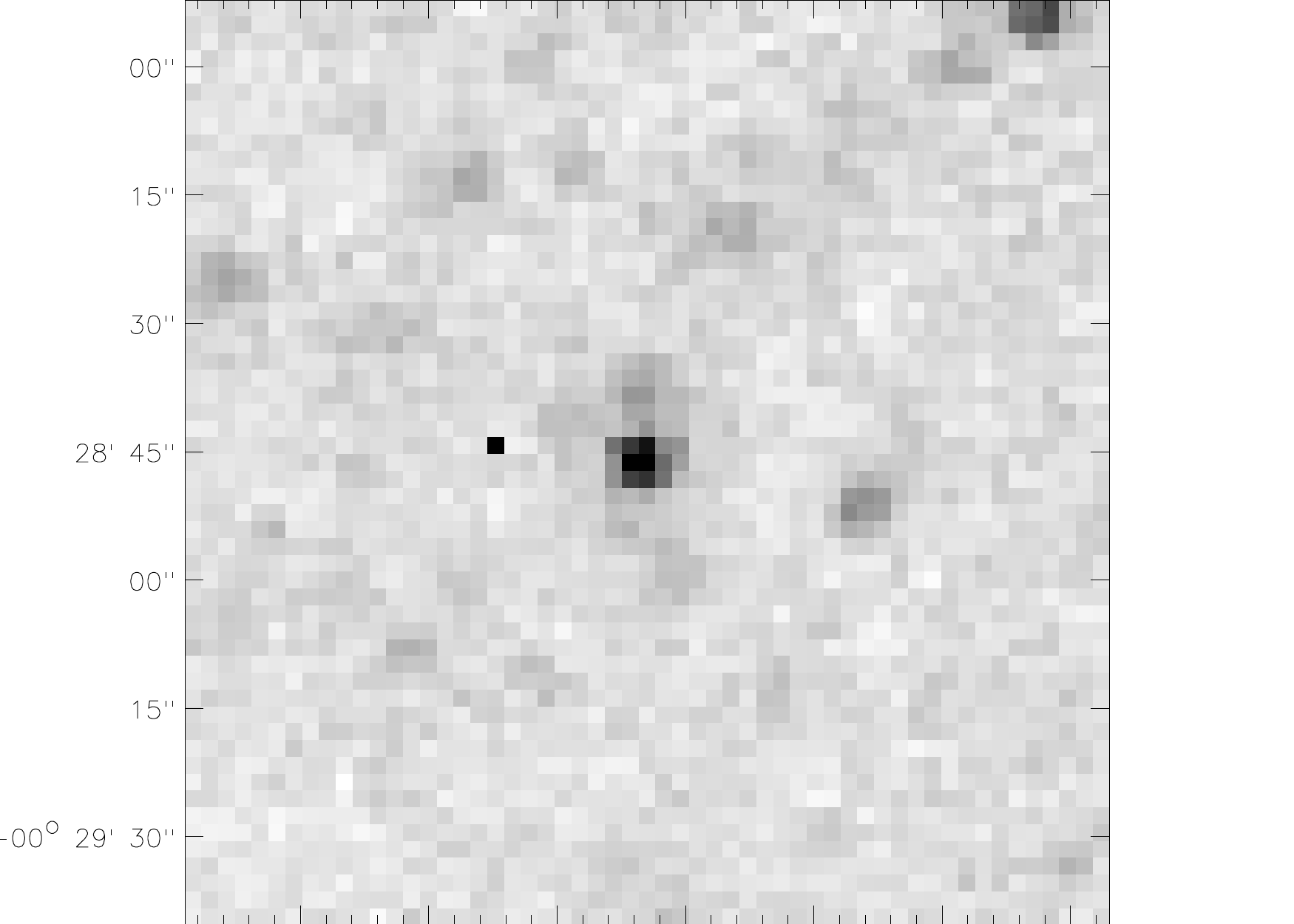} &
      \includegraphics[height=14mm,trim=70  -4 65 -10,clip=true]{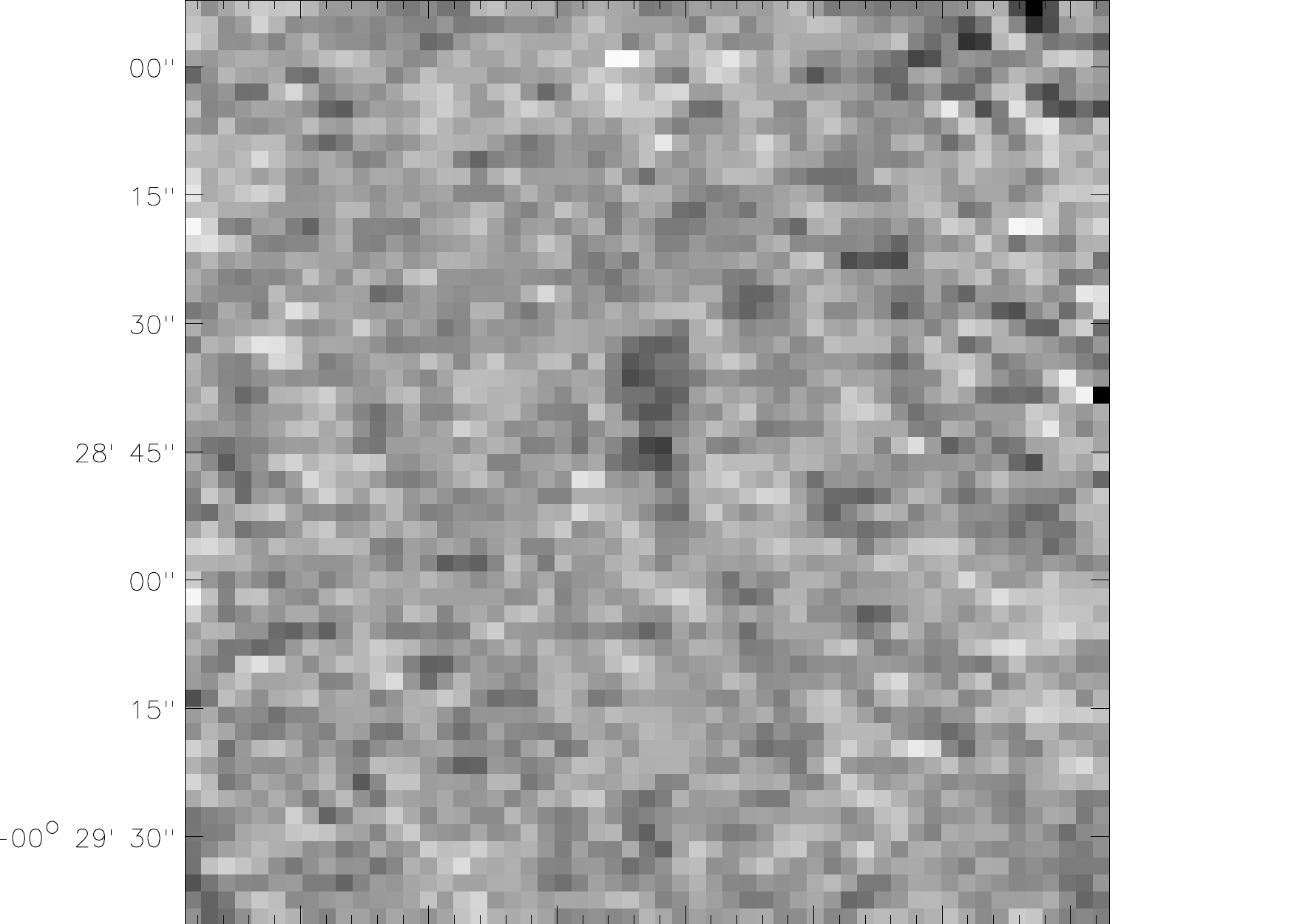} &
      \includegraphics[height=14mm,trim=70  -4 65 -10,clip=true]{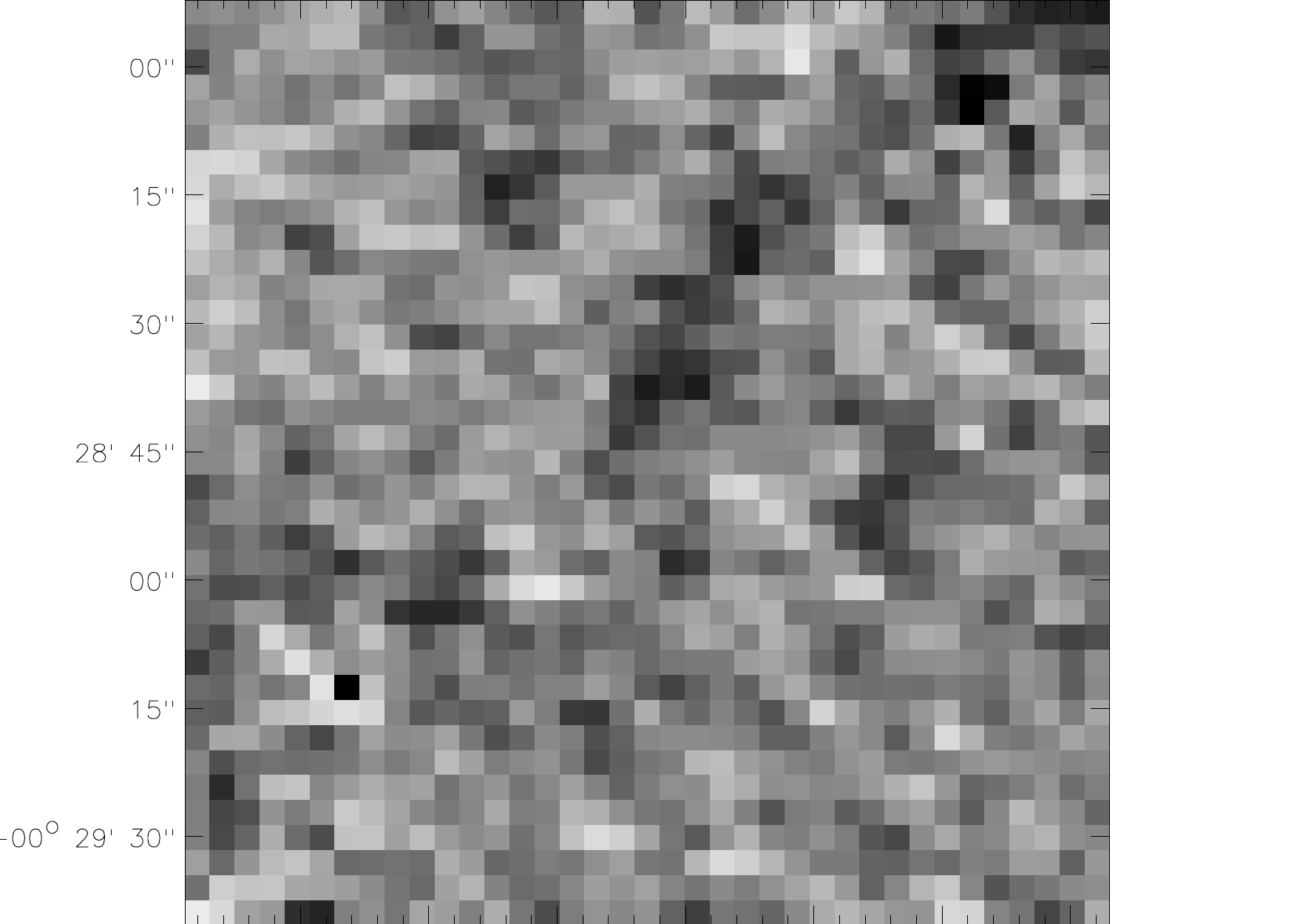} &
      \includegraphics[height=14mm,trim=70  -4 65 -10,clip=true]{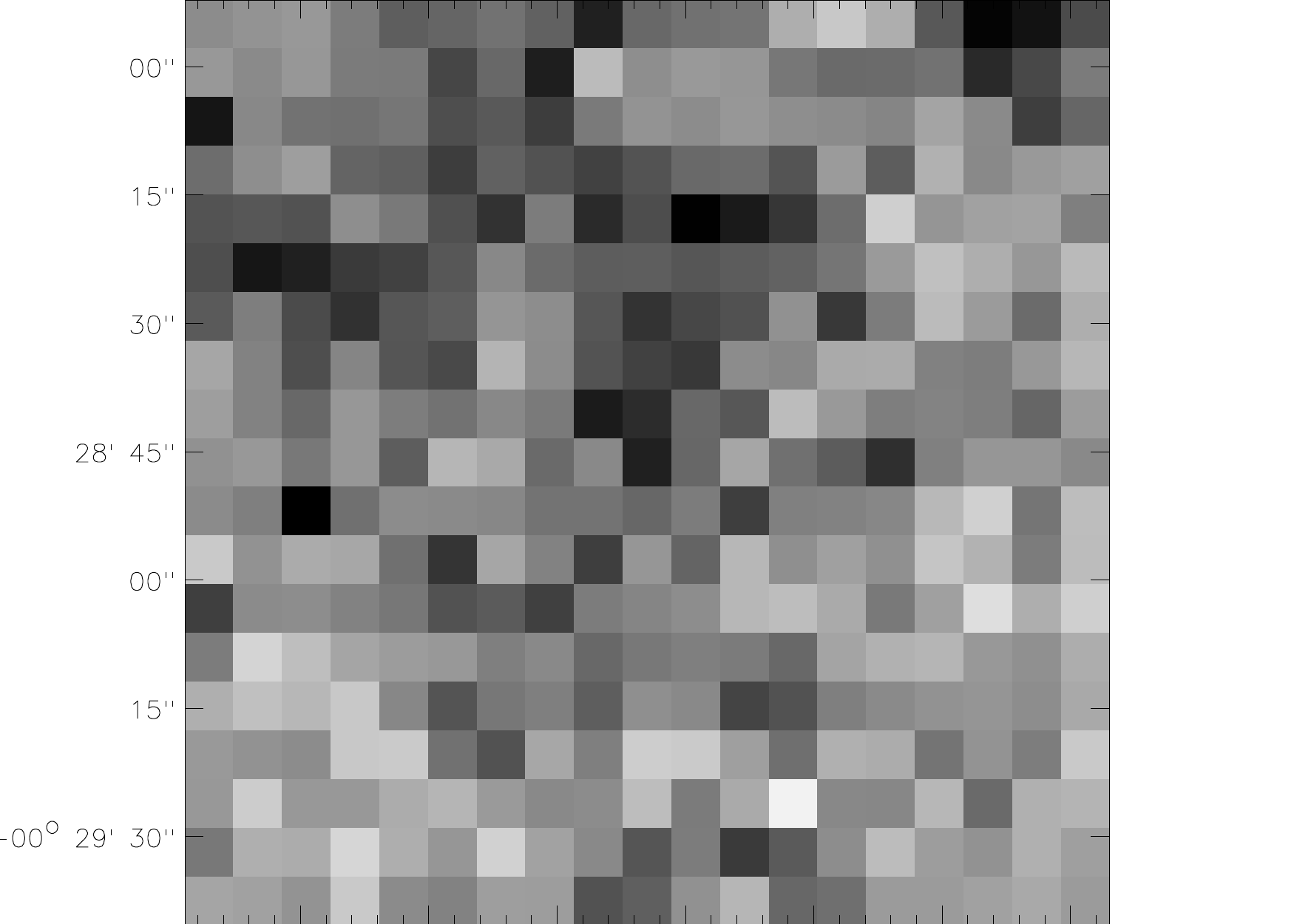}   &
      \includegraphics[height=14mm,trim=70  -4 65 -10,clip=true]{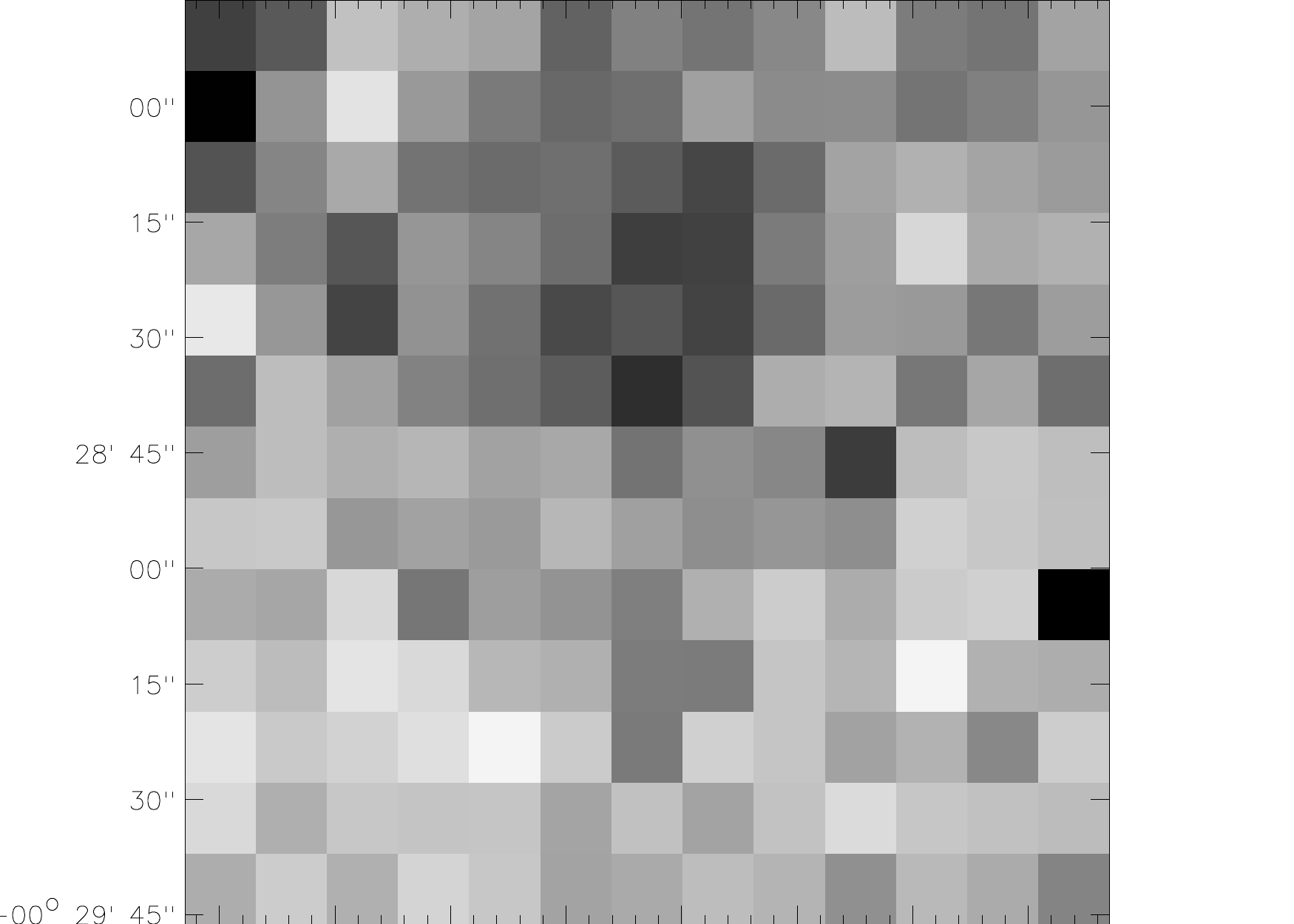}   &
      \includegraphics[height=14mm,trim=70  -4 65 -10,clip=true]{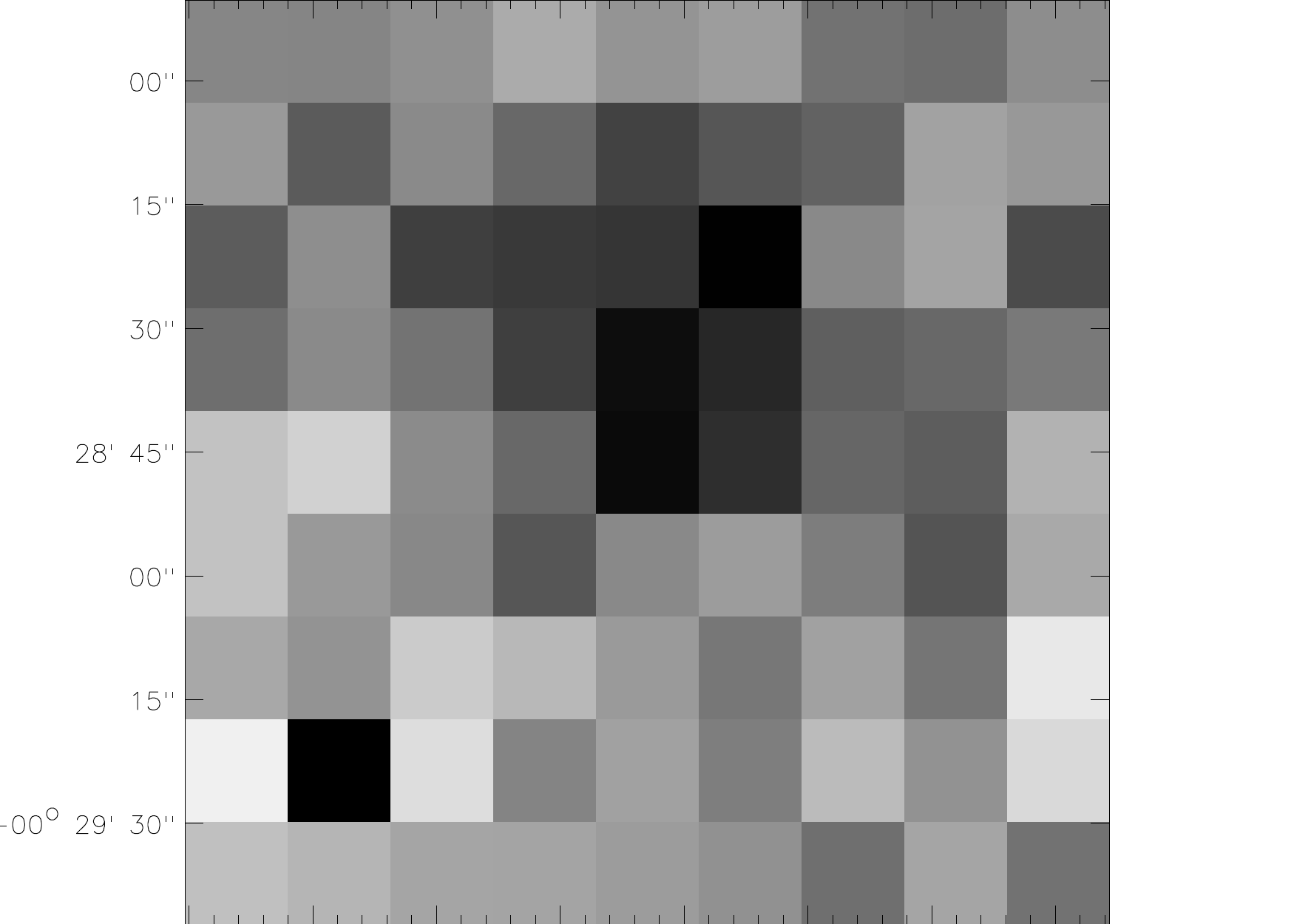} \\

      \multicolumn{6}{c}{\includegraphics[width=80mm,trim=30 0 20 0 ,clip=true]{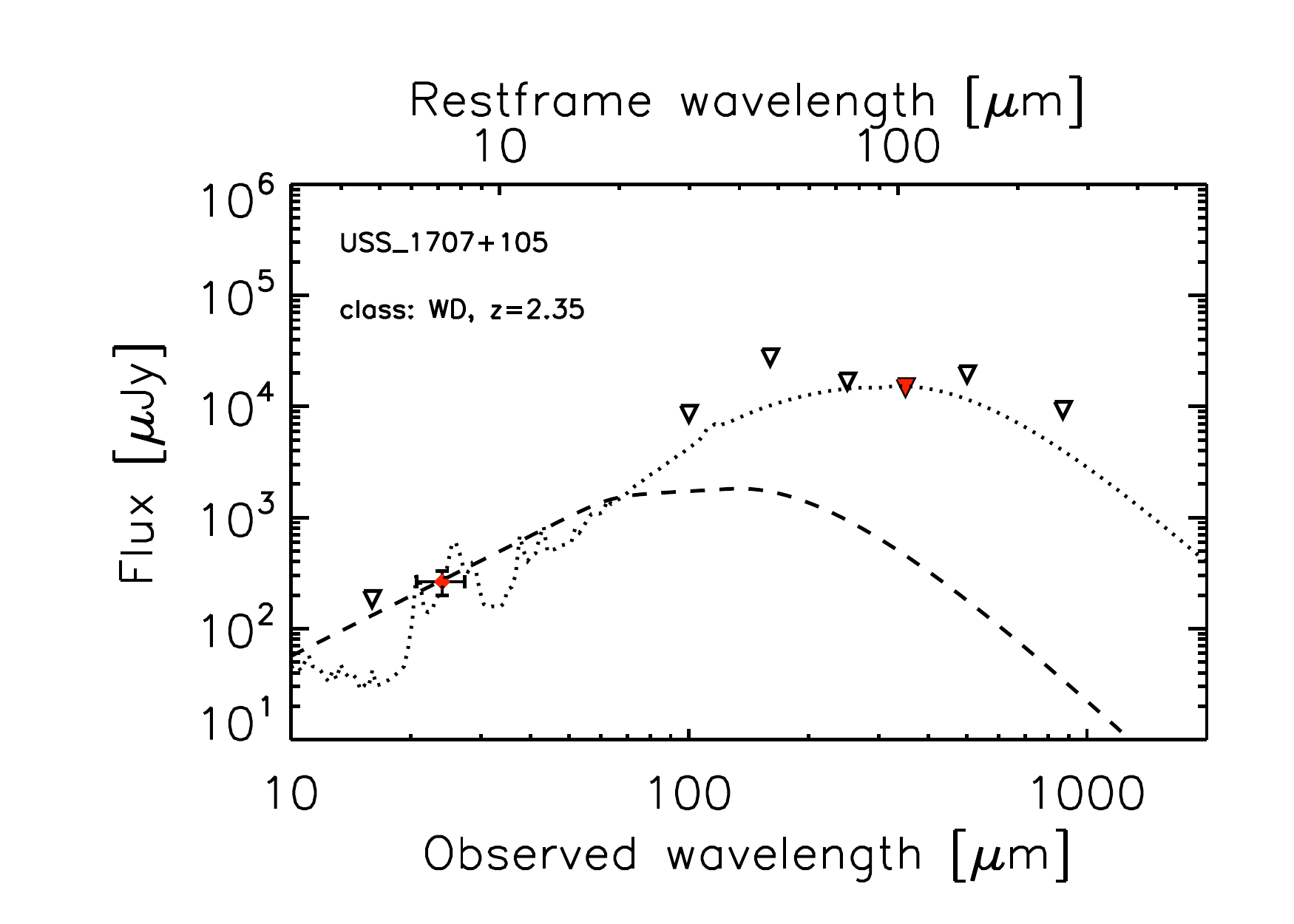}} \\
      24\mum & 100\mum & 160\mum & 250\mum & 350\mum & 500\mum \\
\end{tabular}
\end{minipage}
\begin{minipage}[c]{0.46\linewidth} 
\begin{tabular}{c@{}c@{}c@{}c@{}c@{}c@{}}

      \multicolumn{6}{c}{\includegraphics[width=80mm,trim=30 0 20 0 ,clip=true]{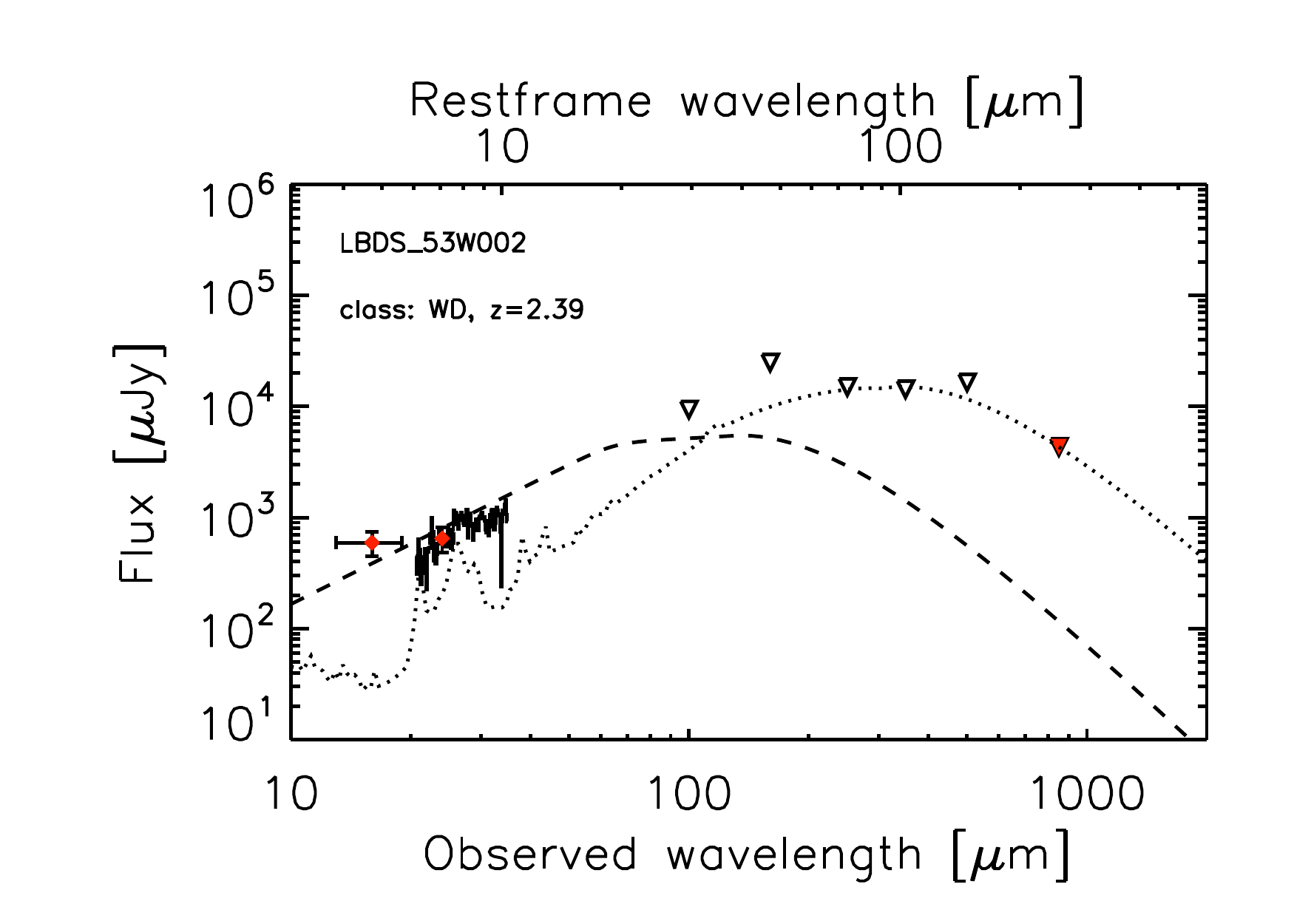}} \\
      24\mum & 100\mum & 160\mum & 250\mum & 350\mum & 500\mum \\
      \includegraphics[height=14mm,trim=70  -4 65 -10,clip=true]{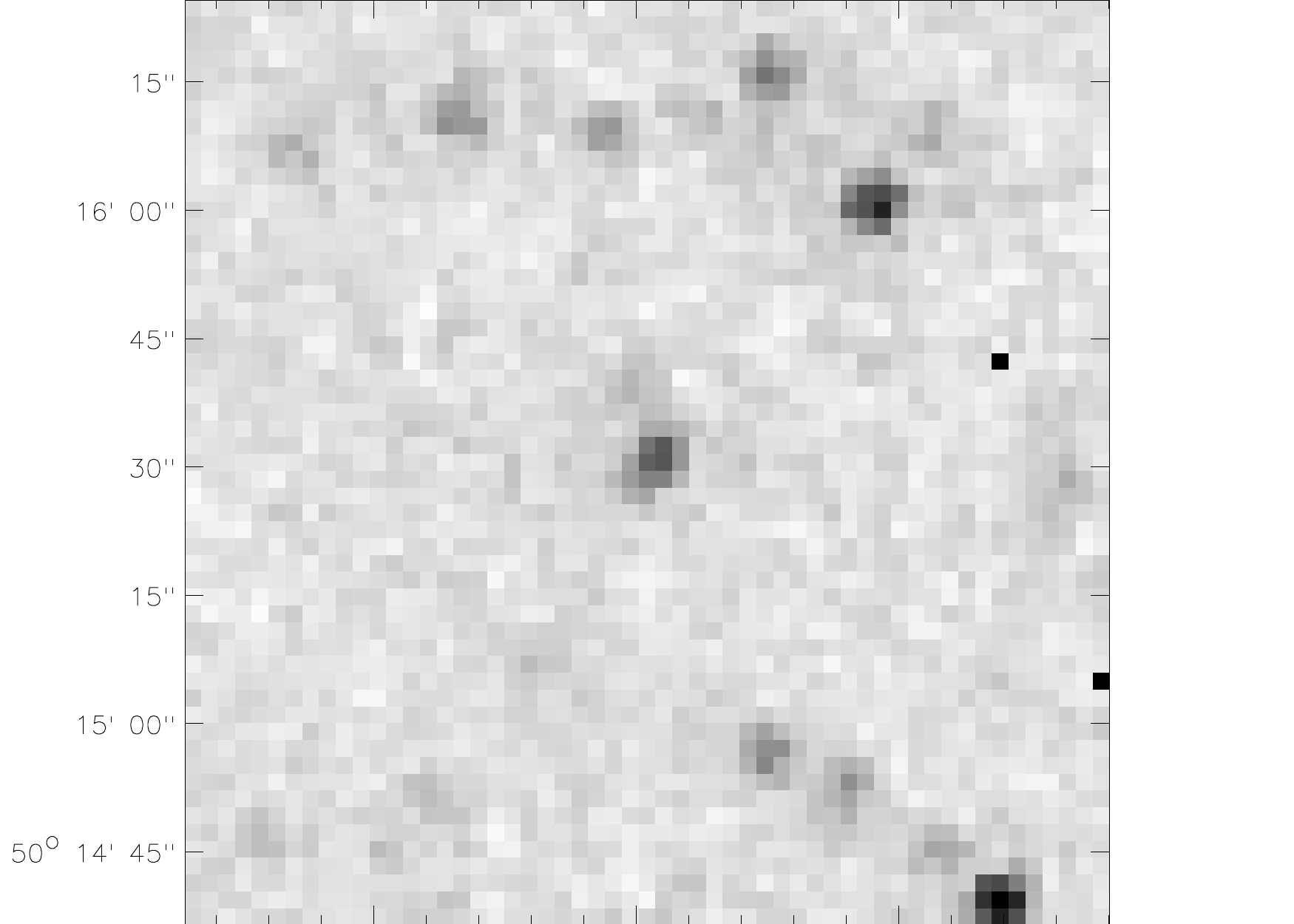} &
      \includegraphics[height=14mm,trim=70  -4 65 -10,clip=true]{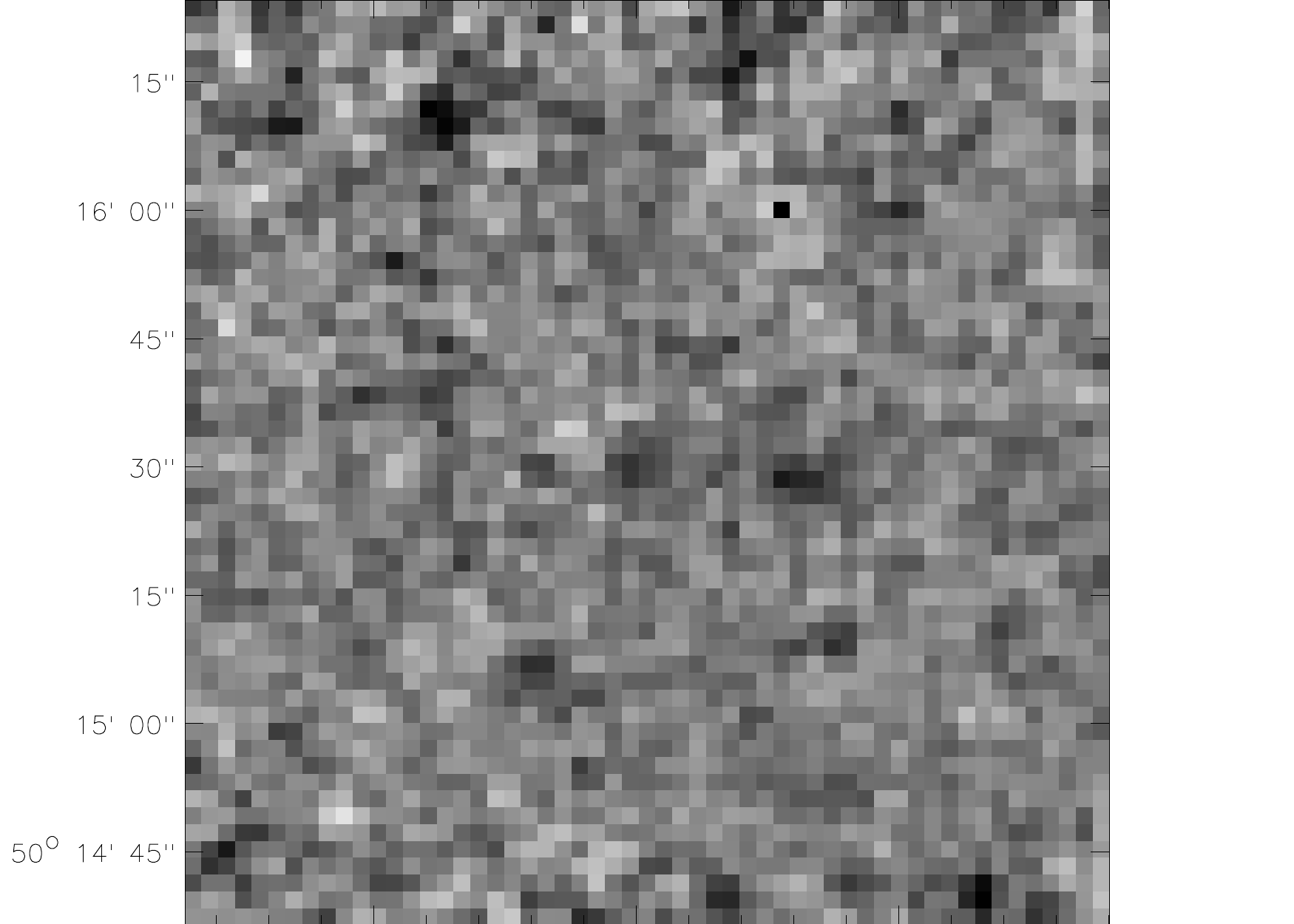} &
      \includegraphics[height=14mm,trim=70  -4 65 -10,clip=true]{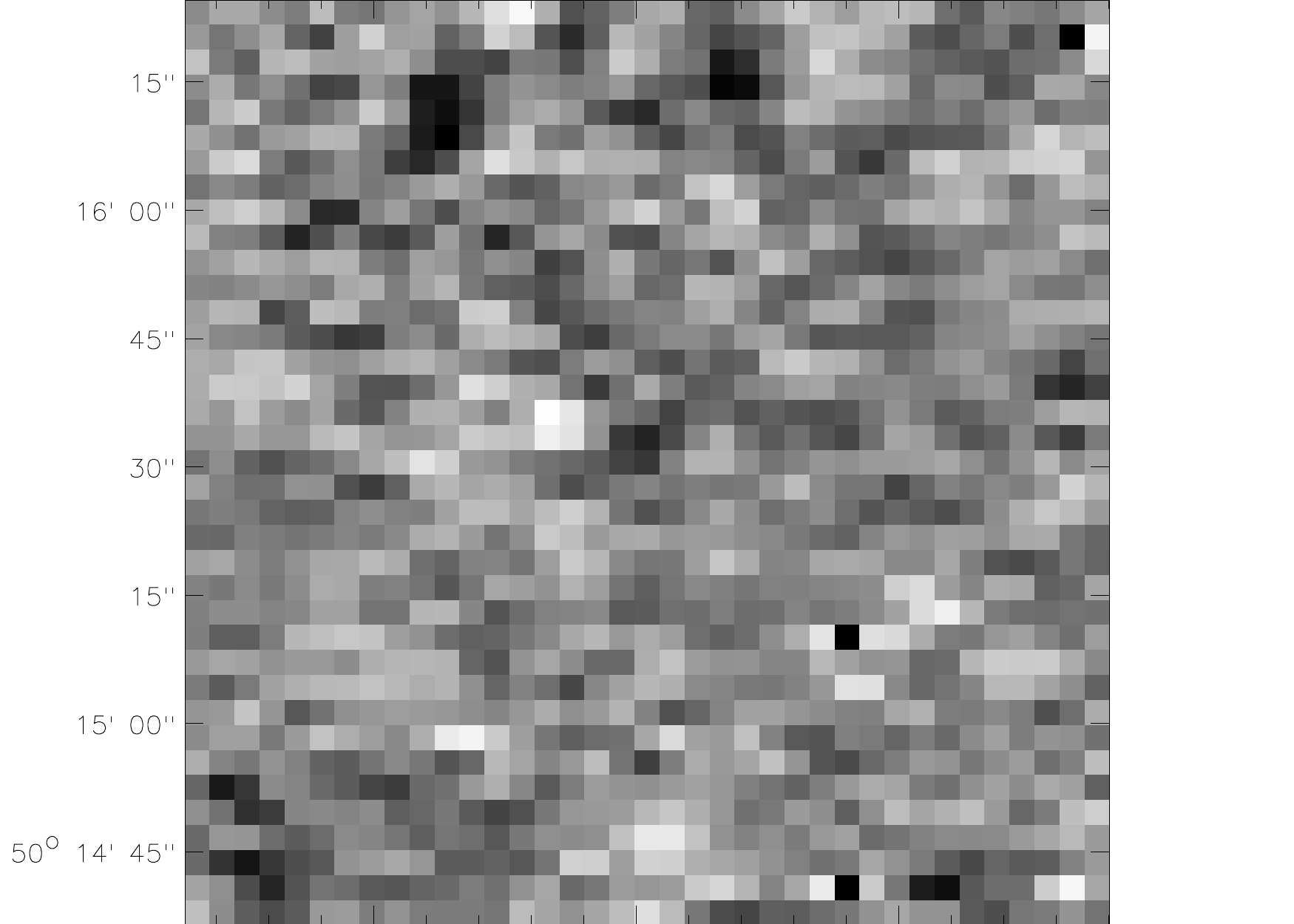} &
      \includegraphics[height=14mm,trim=70  -4 65 -10,clip=true]{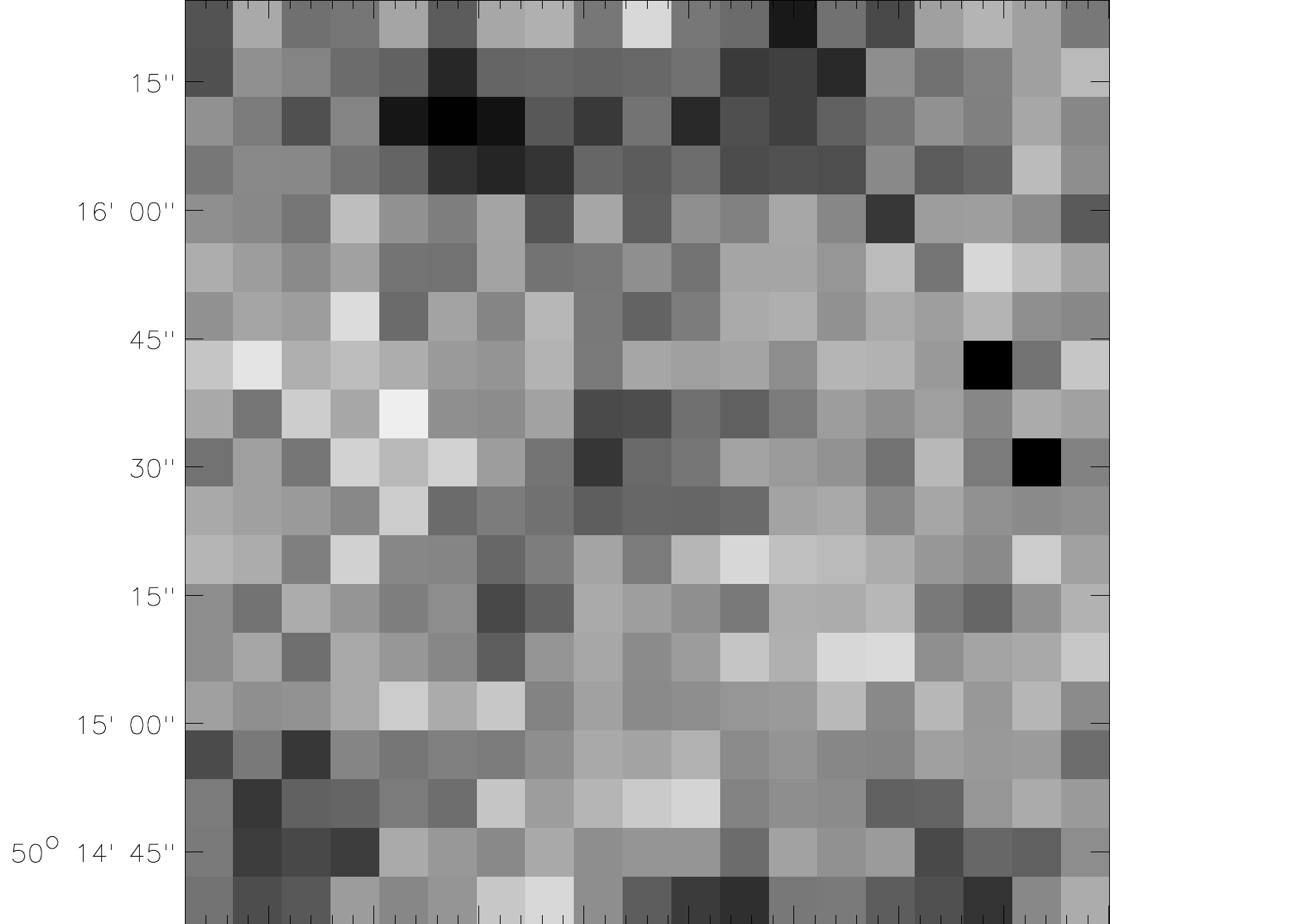}   &
      \includegraphics[height=14mm,trim=70  -4 65 -10,clip=true]{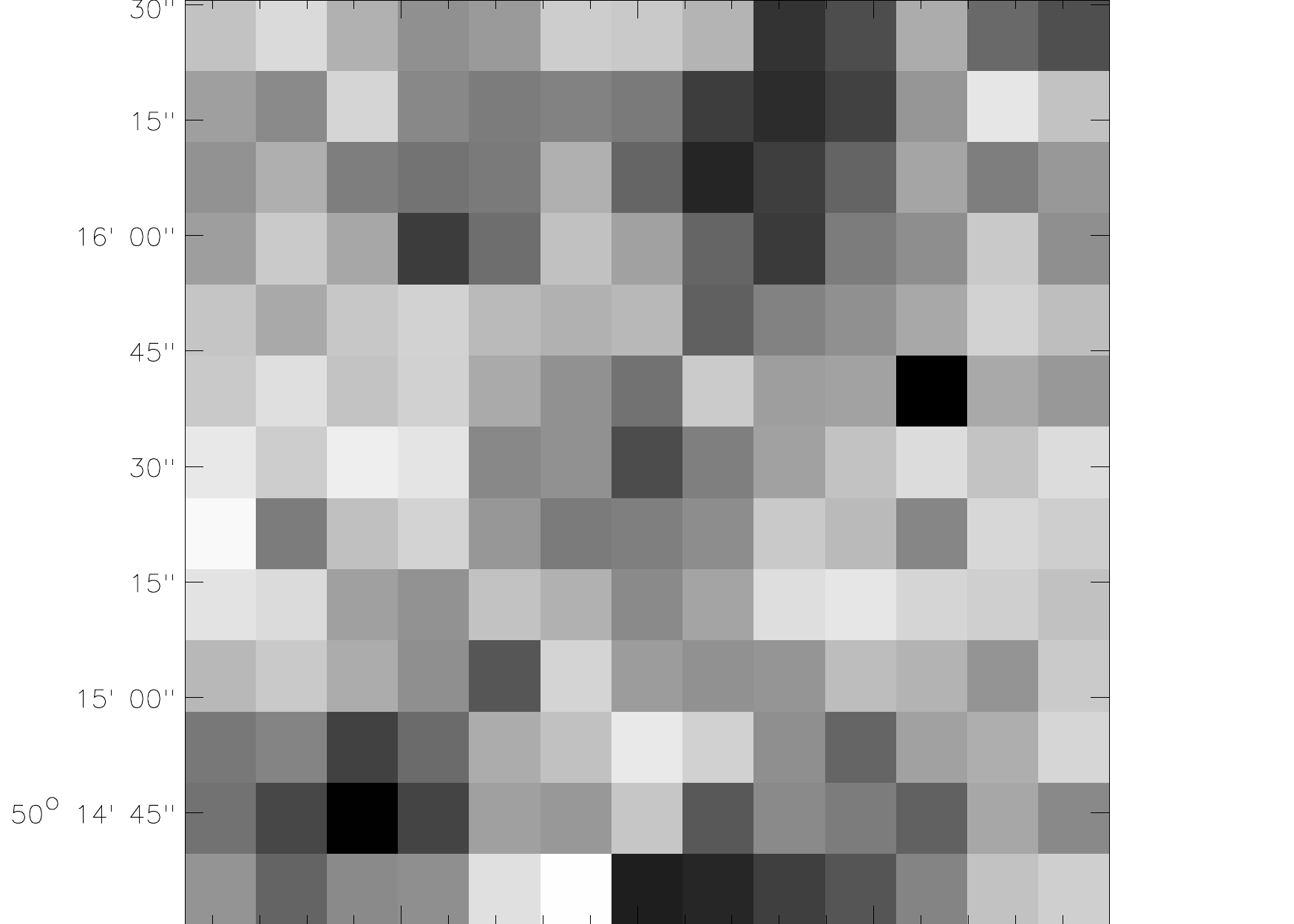}   &
      \includegraphics[height=14mm,trim=70  -4 65 -10,clip=true]{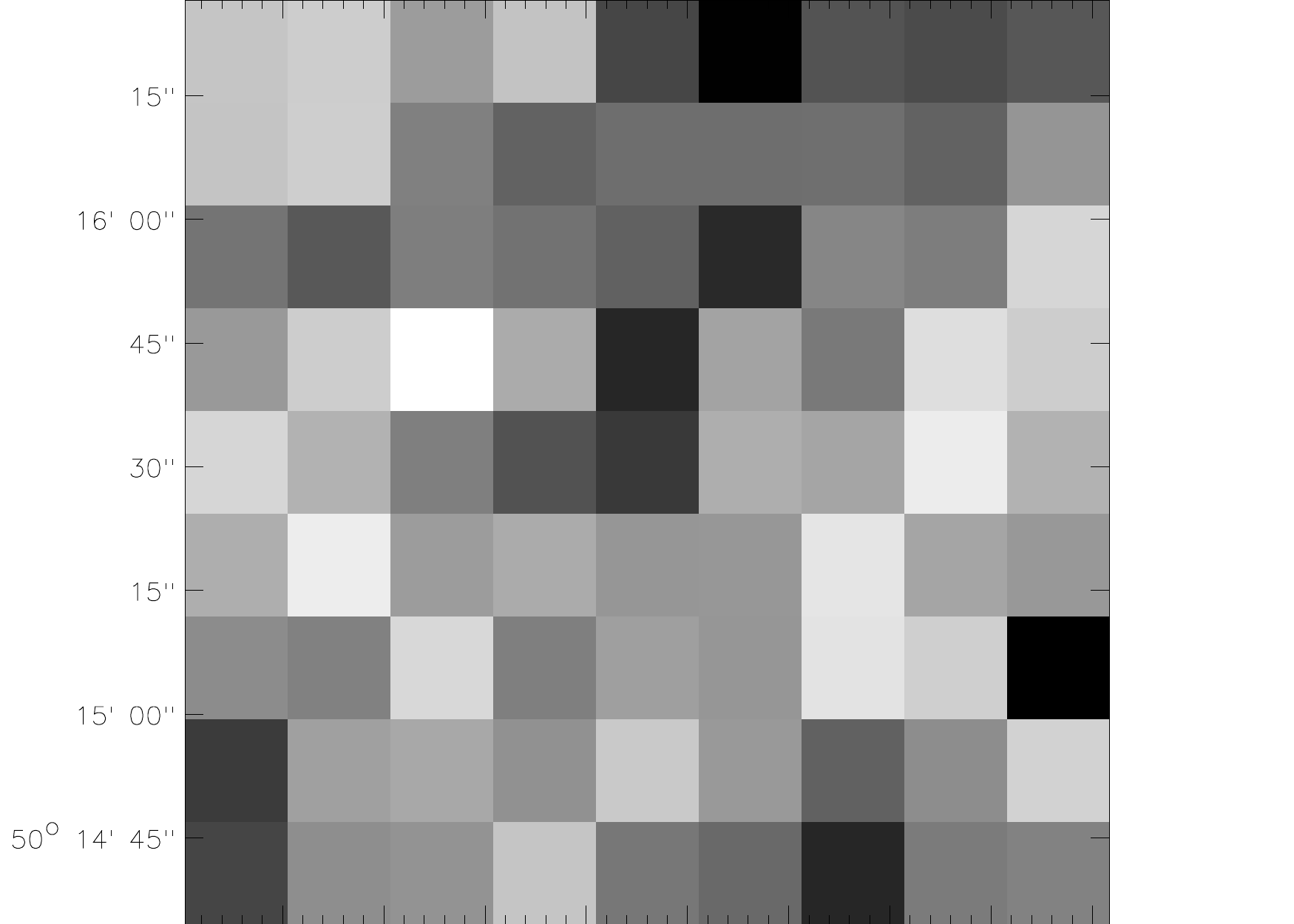} \\


      \multicolumn{6}{c}{\includegraphics[width=80mm,trim=30 0 20 0 ,clip=true]{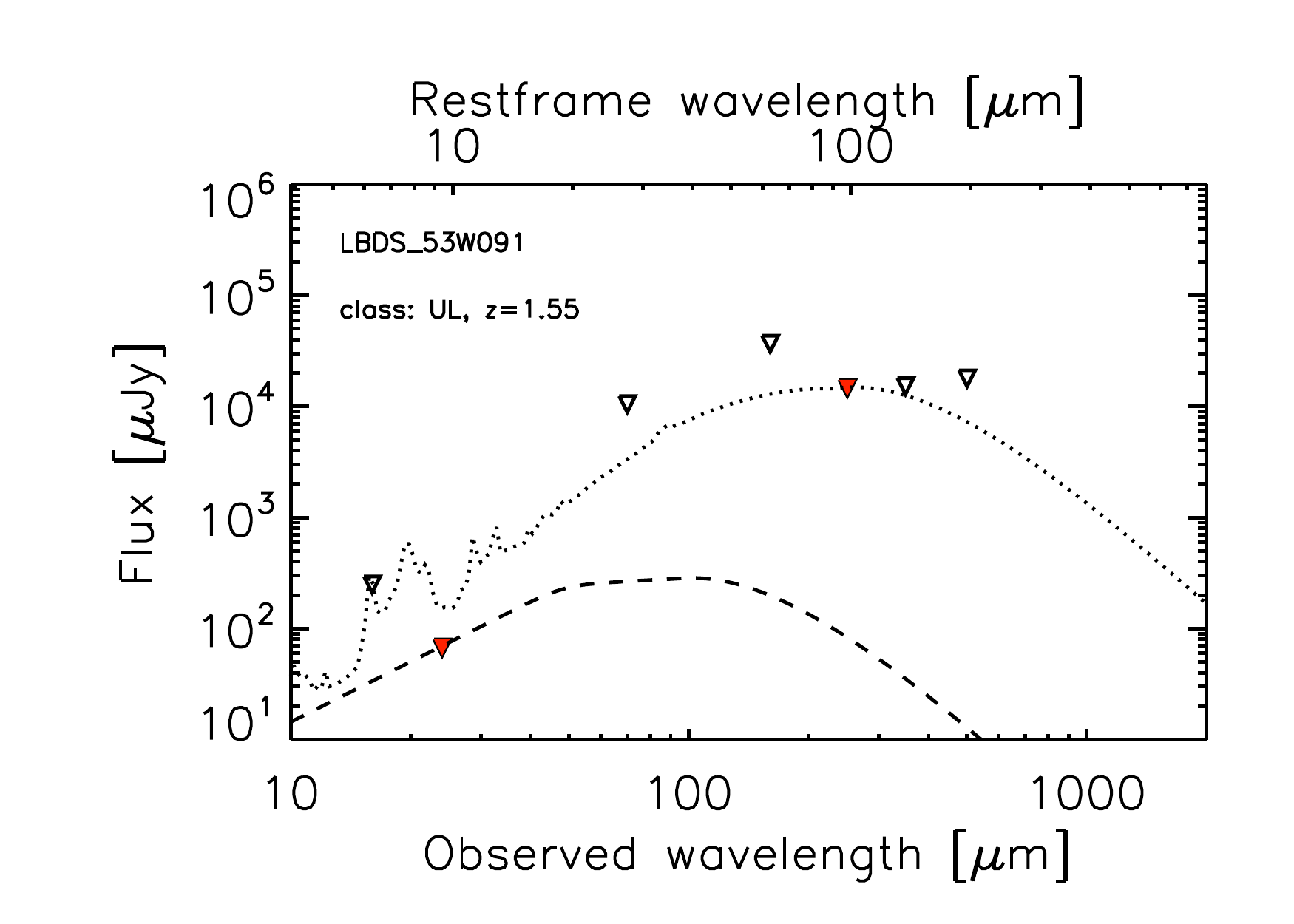}} \\
      24\mum & 70\mum & 160\mum & 250\mum & 350\mum & 500\mum \\
\end{tabular}
          \end{minipage}
  \end{centering}
\end{figure*}

\clearpage

\begin{figure*}[h]
  \begin{centering}
    \begin{minipage}[c]{0.46\linewidth}
    \begin{tabular}{c@{}c@{}c@{}c@{}c@{}c@{}}

      \multicolumn{6}{c}{\includegraphics[width=80mm,trim=30 0 20 0 ,clip=true]{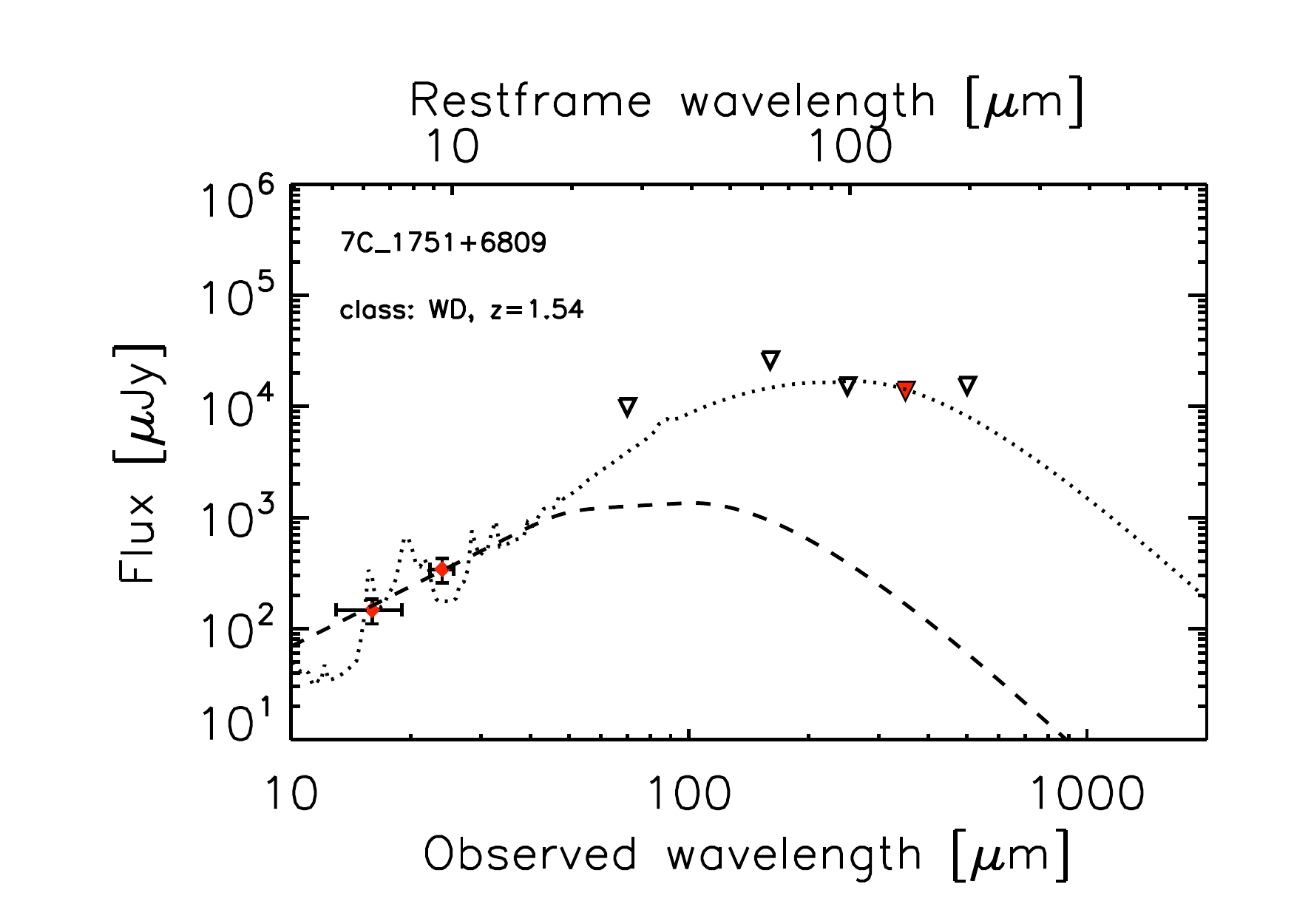}} \\
      24\mum & 70\mum & 160\mum & 250\mum & 350\mum & 500\mum \\
      \includegraphics[height=14mm,trim=70  -4 65 -10,clip=true]{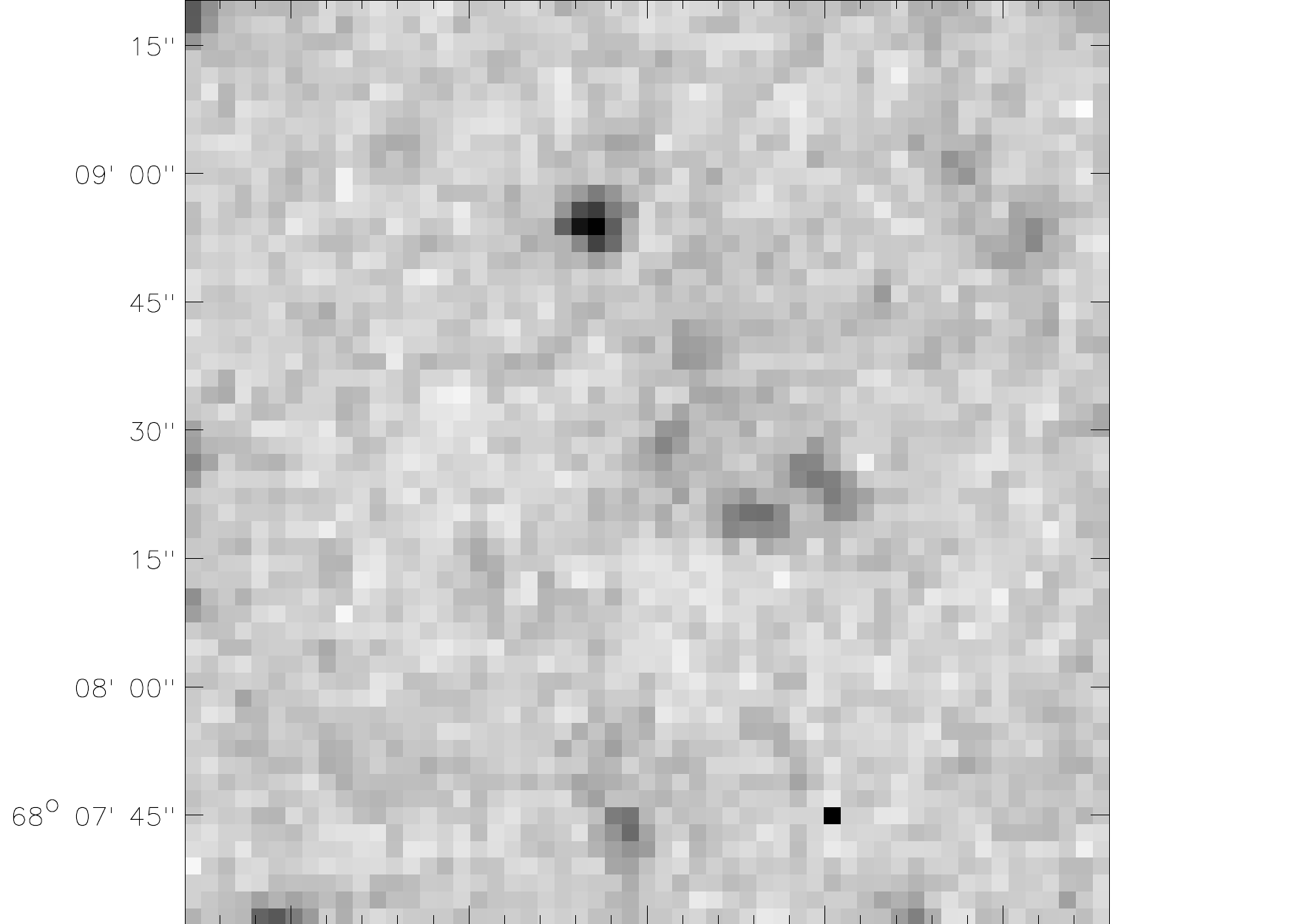} &
      \includegraphics[height=14mm,trim=70  -4 65 -10,clip=true]{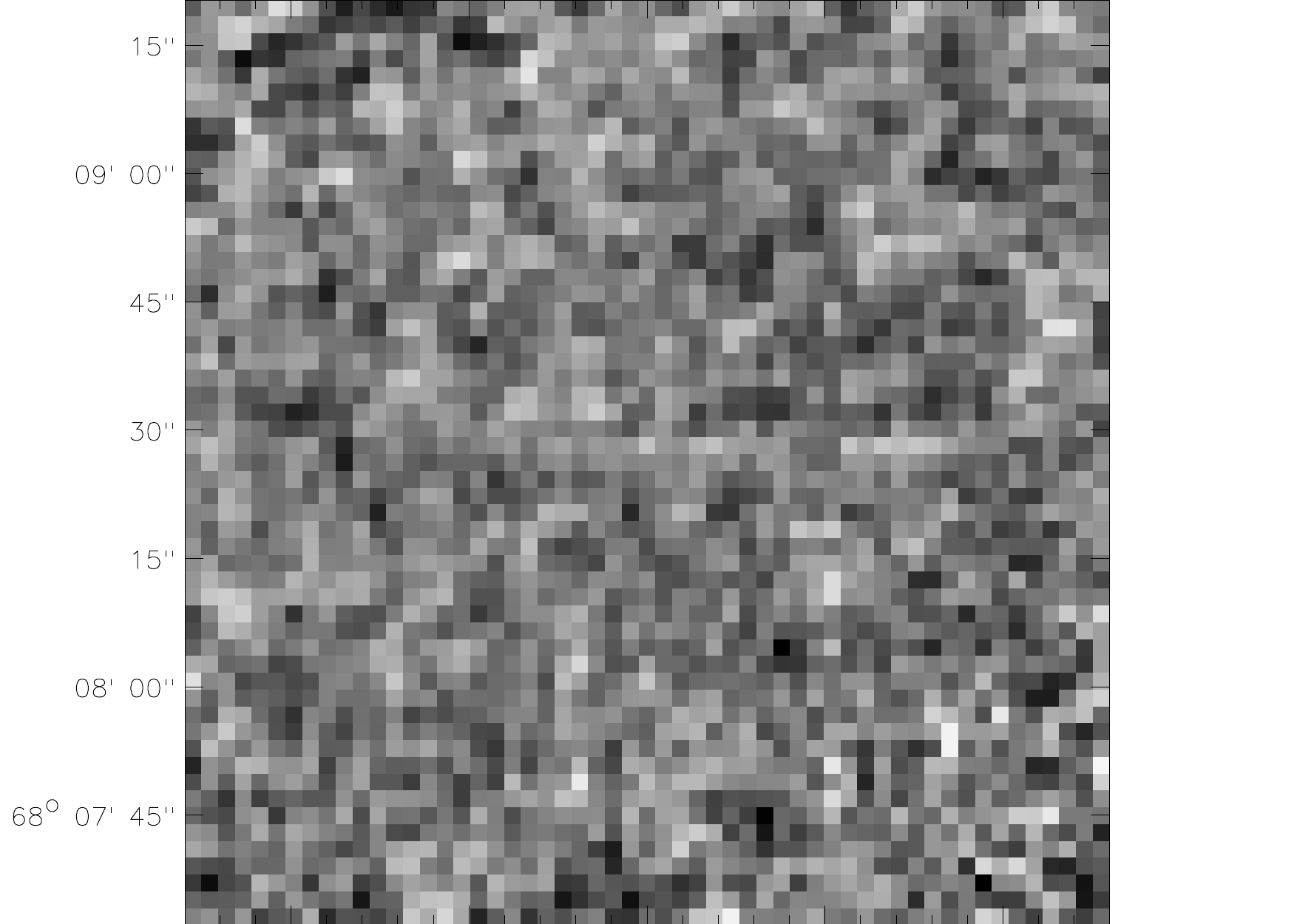} &
      \includegraphics[height=14mm,trim=70  -4 65 -10,clip=true]{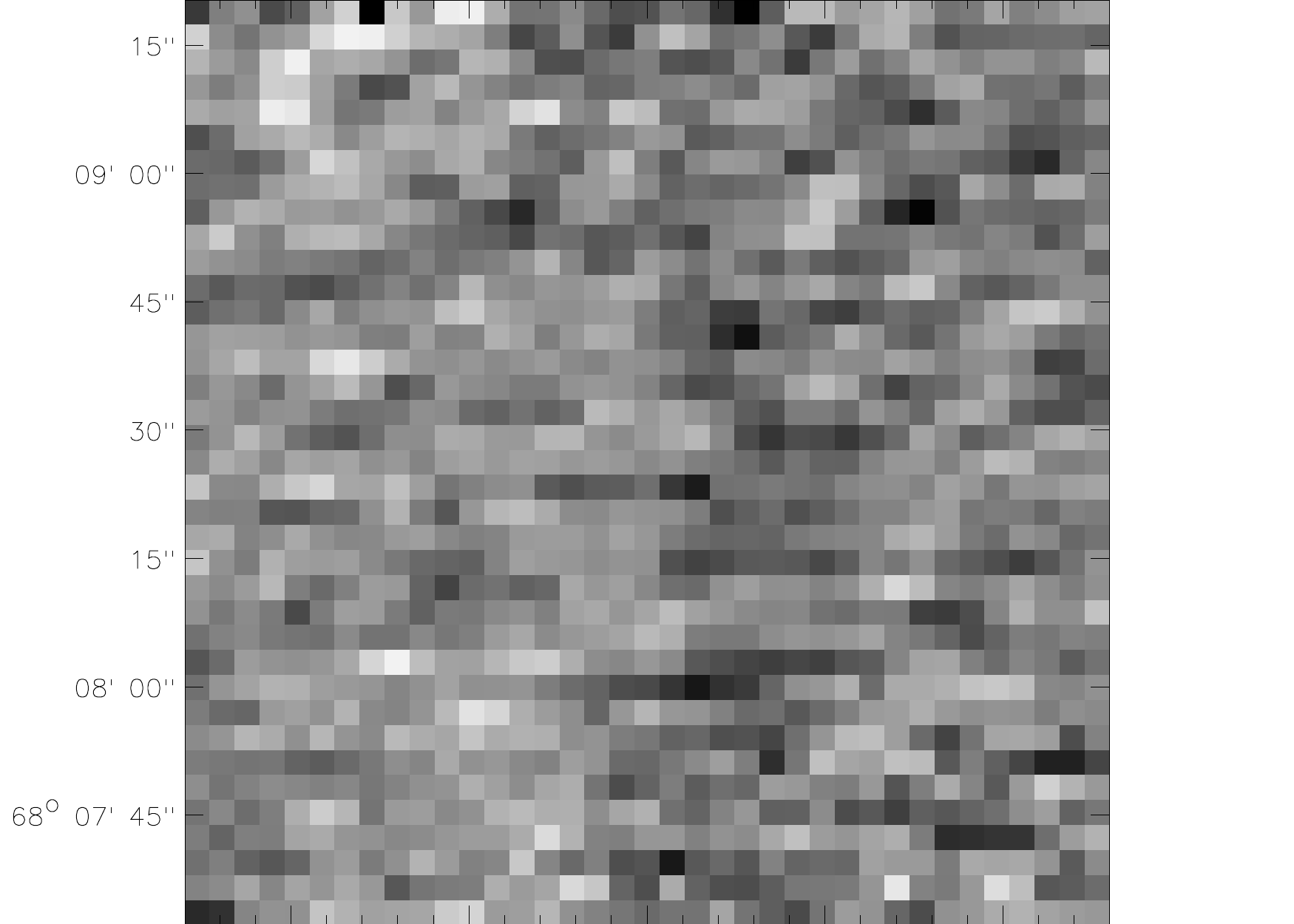} &
      \includegraphics[height=14mm,trim=70  -4 65 -10,clip=true]{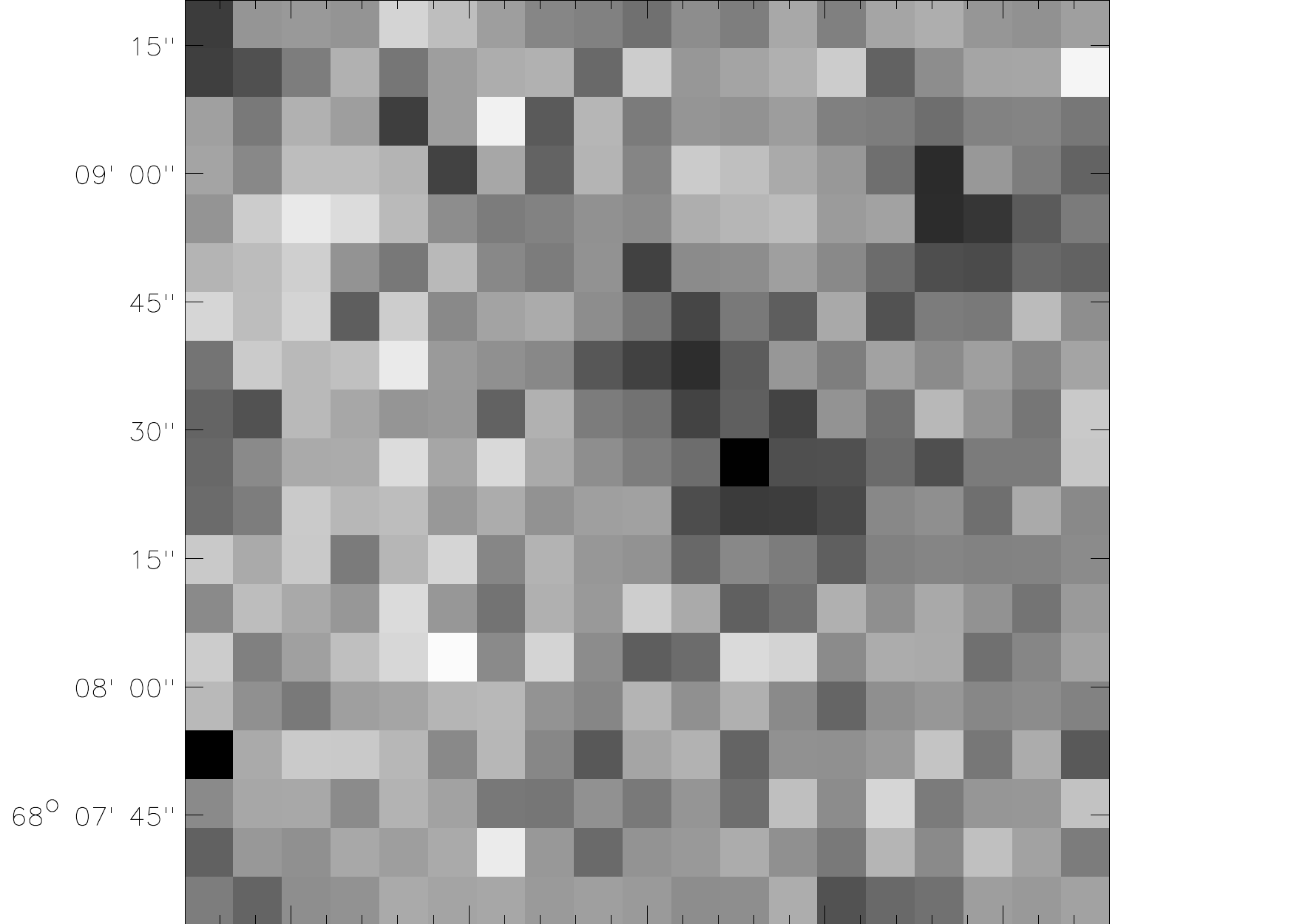}   &
      \includegraphics[height=14mm,trim=70  -4 65 -10,clip=true]{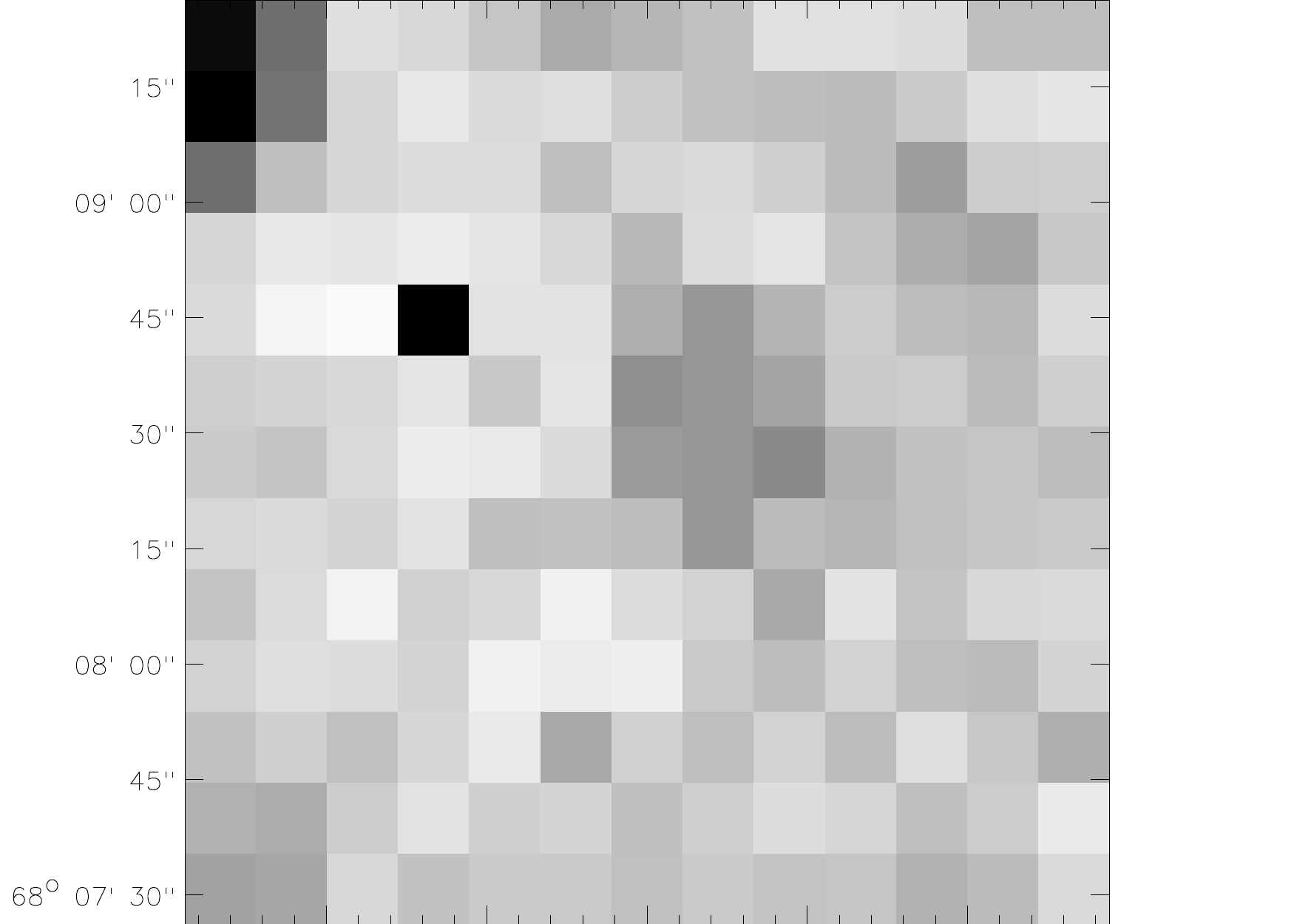}   &
      \includegraphics[height=14mm,trim=70  -4 65 -10,clip=true]{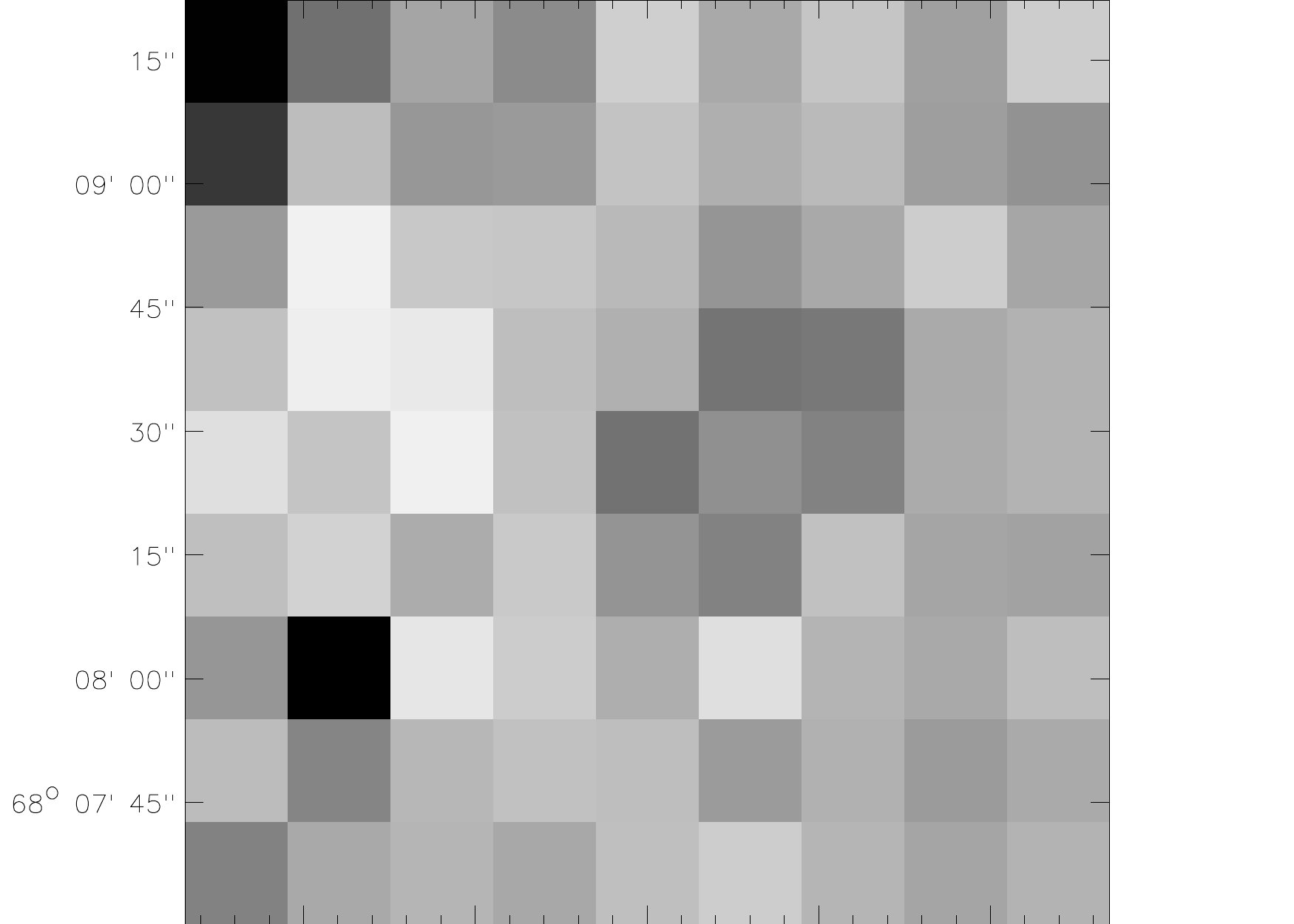} \\

      \multicolumn{6}{c}{\includegraphics[width=80mm,trim=30 0 20 0 ,clip=true]{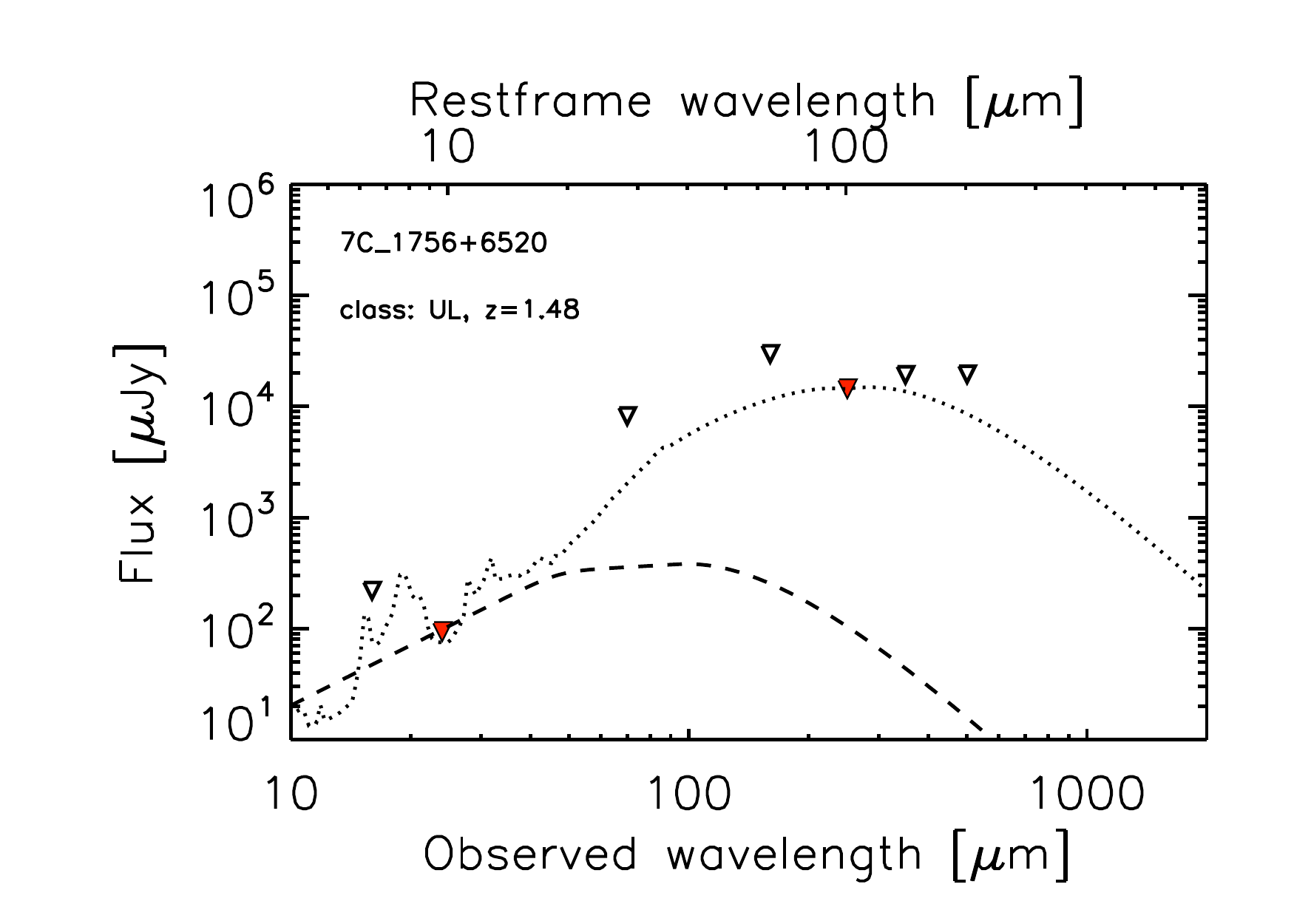}} \\
      24\mum & 70\mum & 160\mum & 250\mum & 350\mum & 500\mum \\
      \includegraphics[height=14mm,trim=70  -4 65 -10,clip=true]{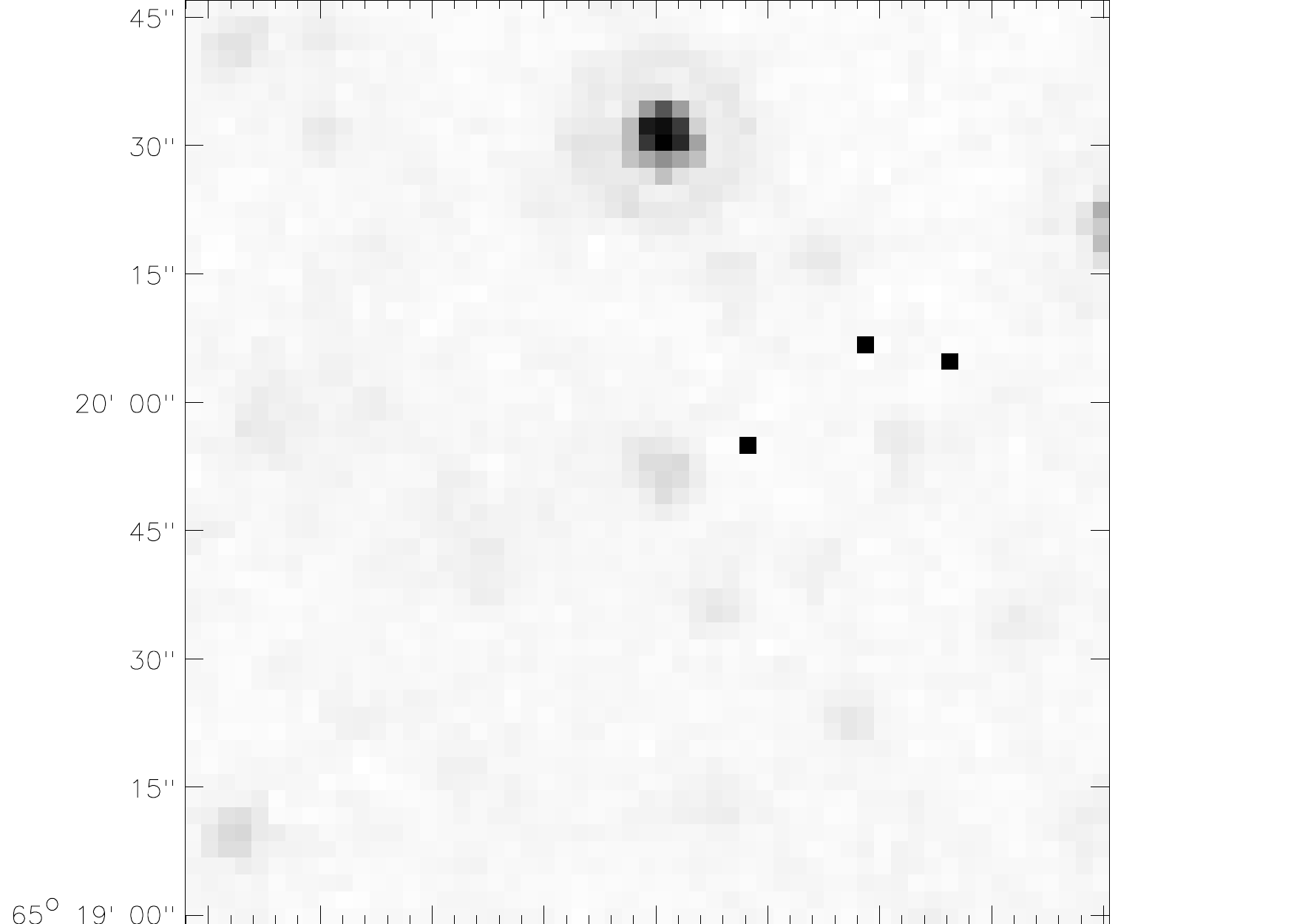} &
      \includegraphics[height=14mm,trim=70  -4 65 -10,clip=true]{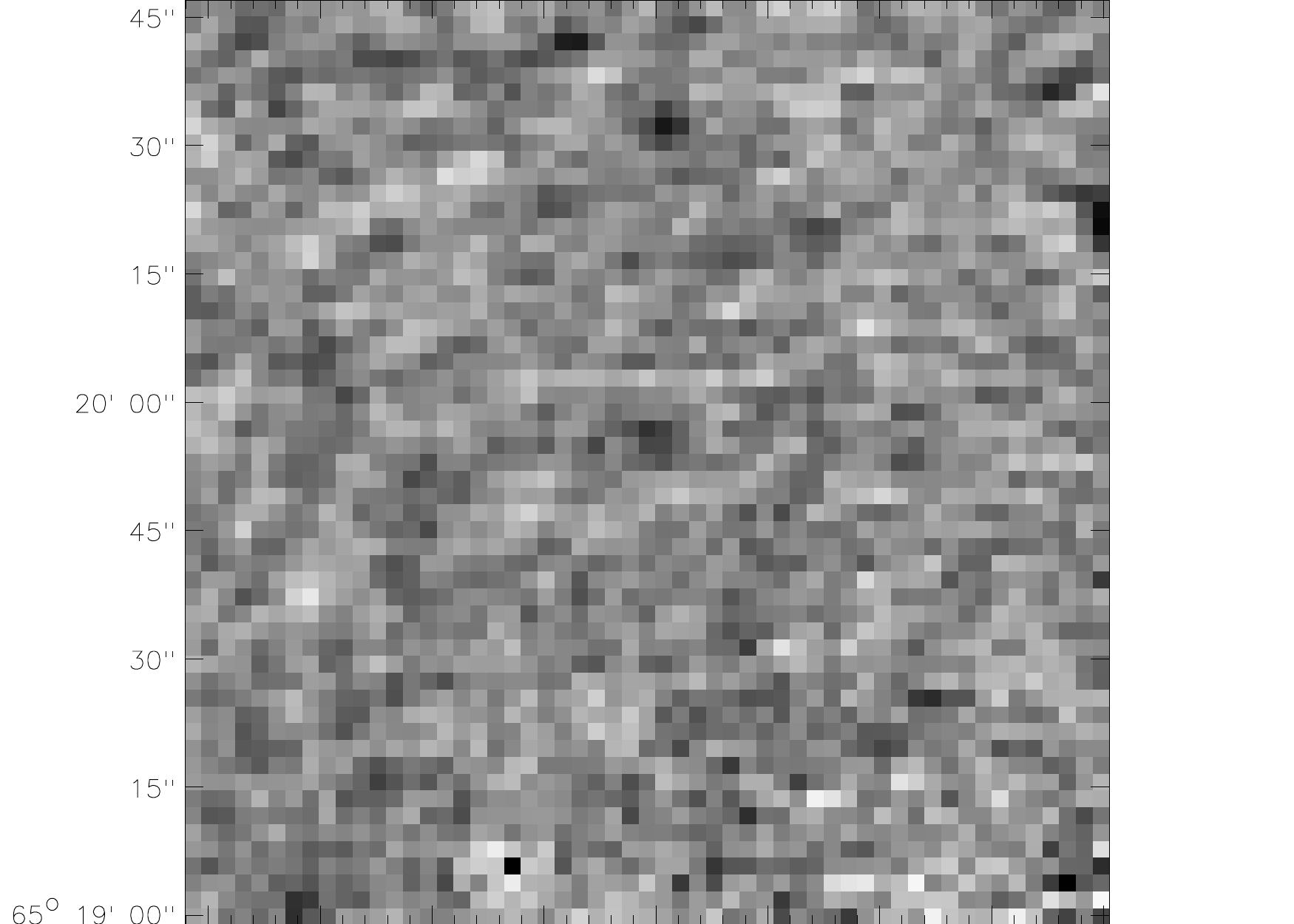} &
      \includegraphics[height=14mm,trim=70  -4 65 -10,clip=true]{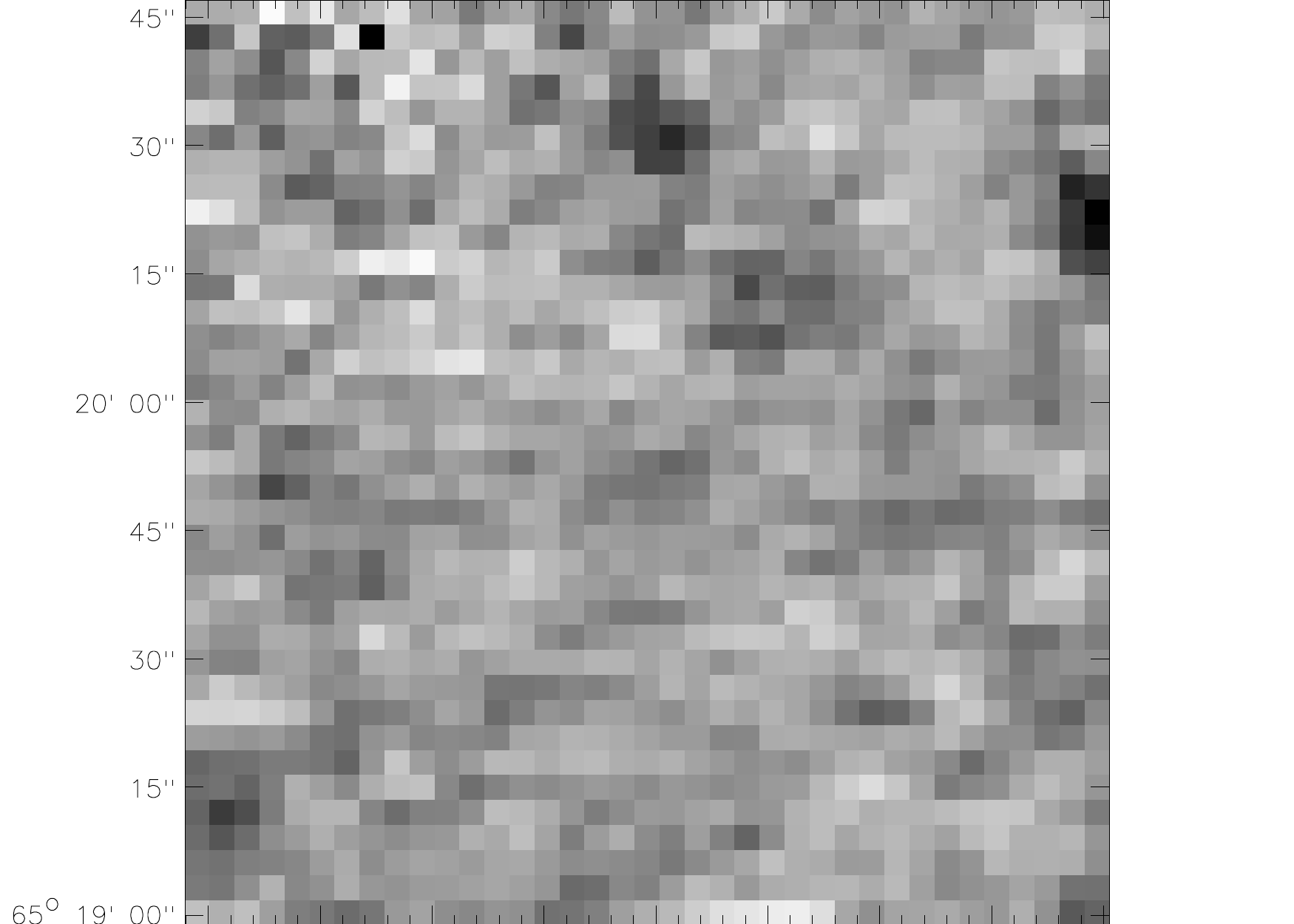} &
      \includegraphics[height=14mm,trim=70  -4 65 -10,clip=true]{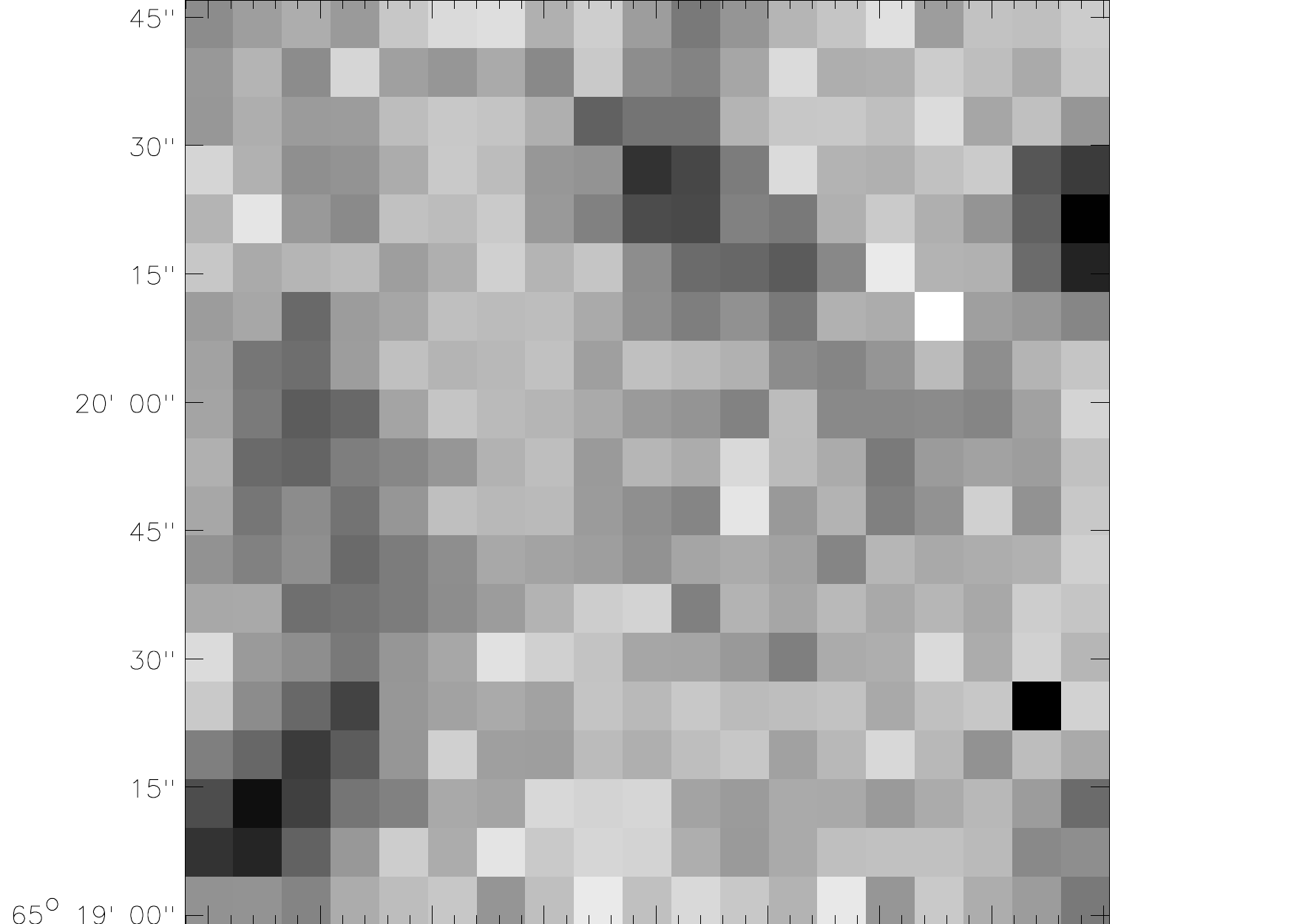}   &
      \includegraphics[height=14mm,trim=70  -4 65 -10,clip=true]{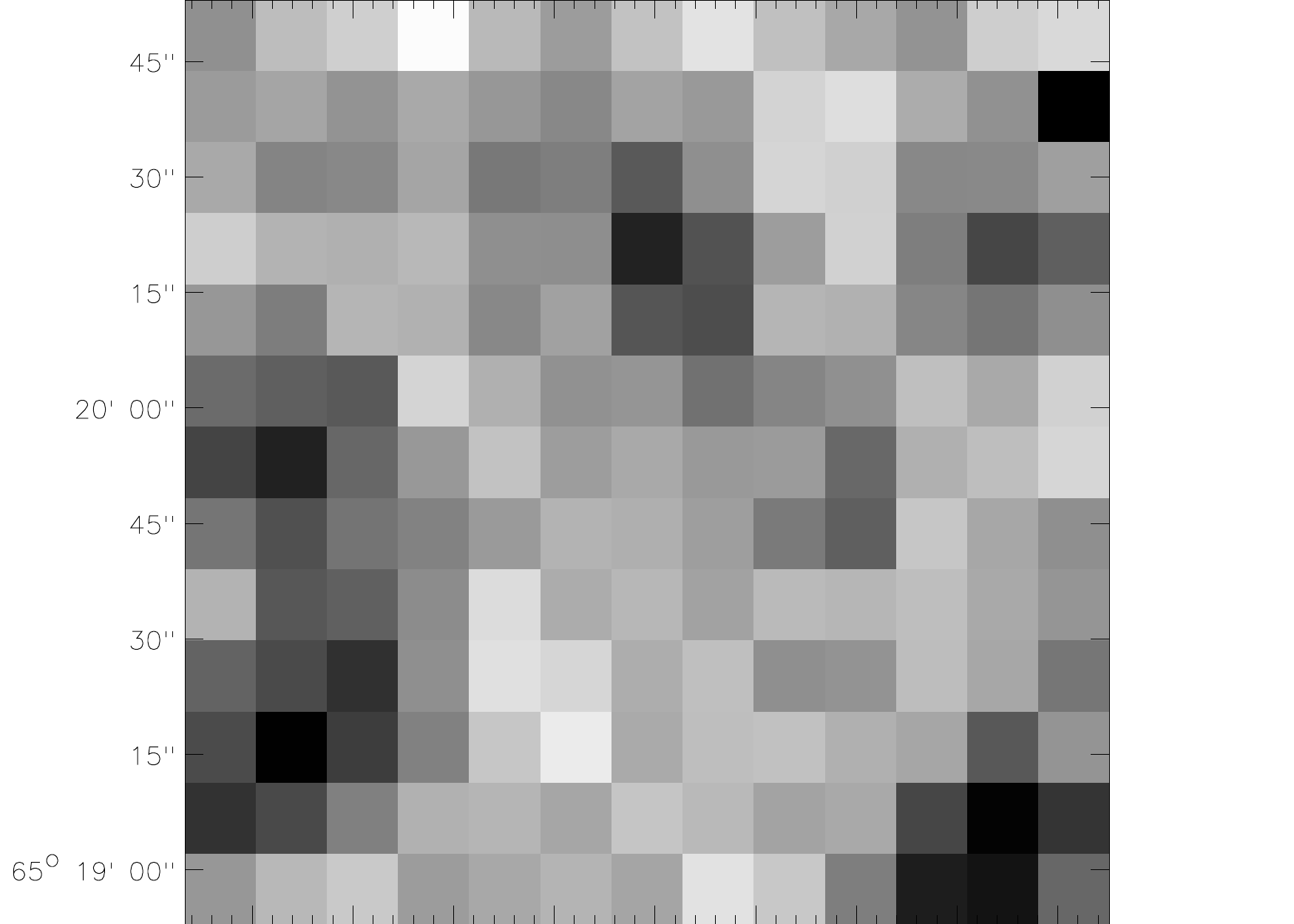}   &
      \includegraphics[height=14mm,trim=70  -4 65 -10,clip=true]{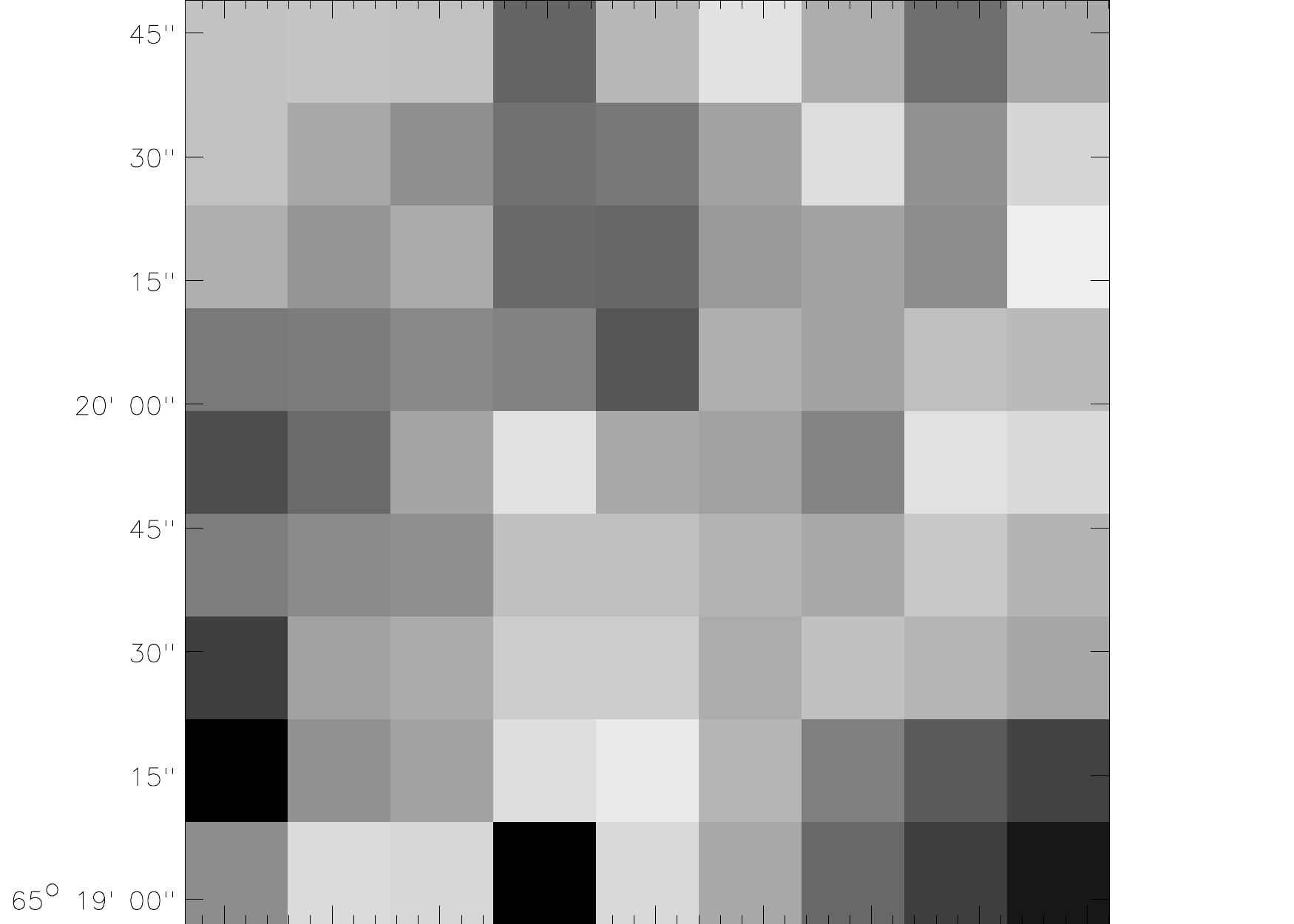} \\

      \multicolumn{6}{c}{\includegraphics[width=80mm,trim=30 0 20 0 ,clip=true]{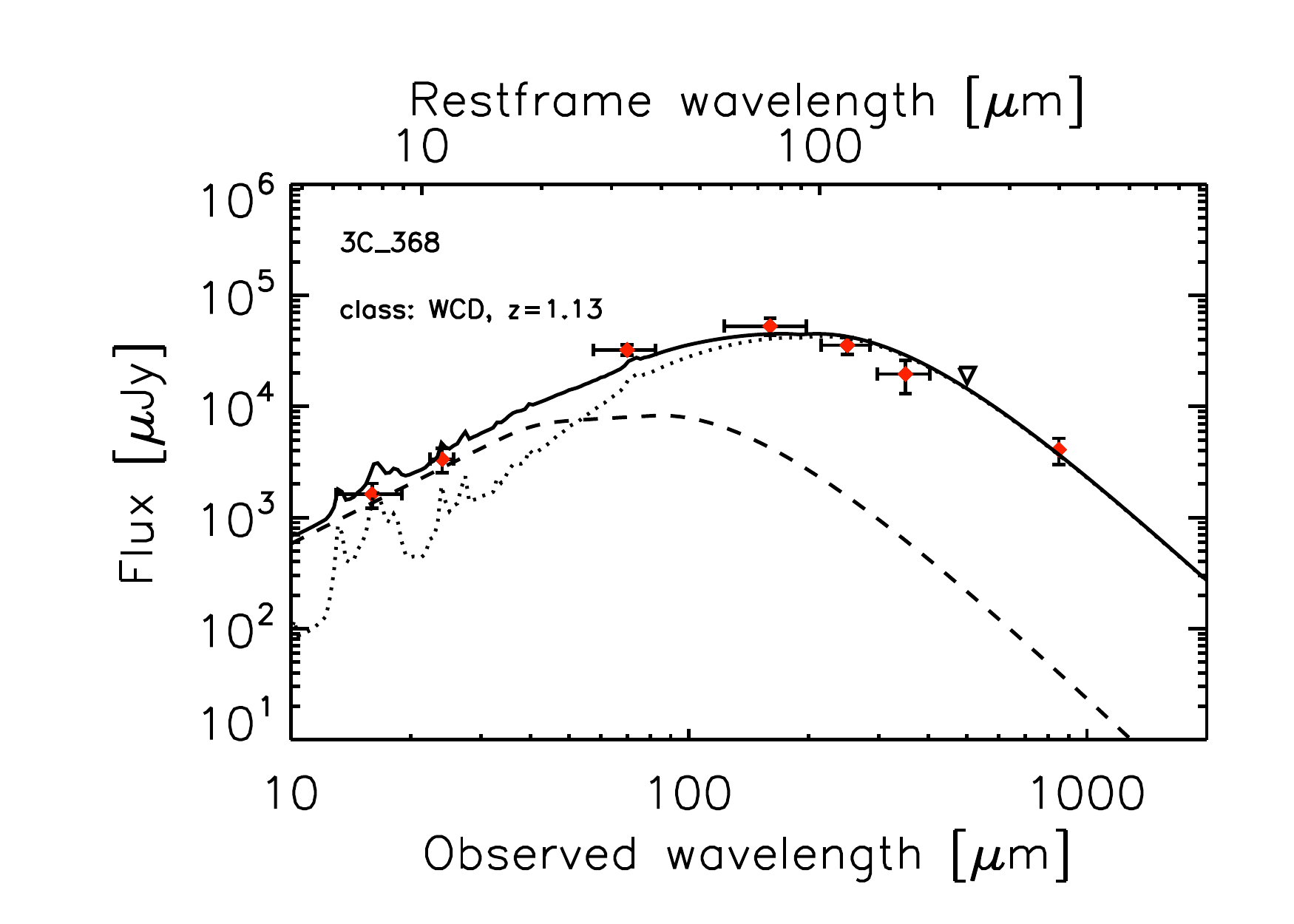}} \\
      24\mum & 70\mum & 160\mum & 250\mum & 350\mum & 500\mum \\
\end{tabular}
\end{minipage}
\begin{minipage}[c]{0.46\linewidth} 
\begin{tabular}{c@{}c@{}c@{}c@{}c@{}c@{}}

      \multicolumn{6}{c}{\includegraphics[width=80mm,trim=30 0 20 0 ,clip=true]{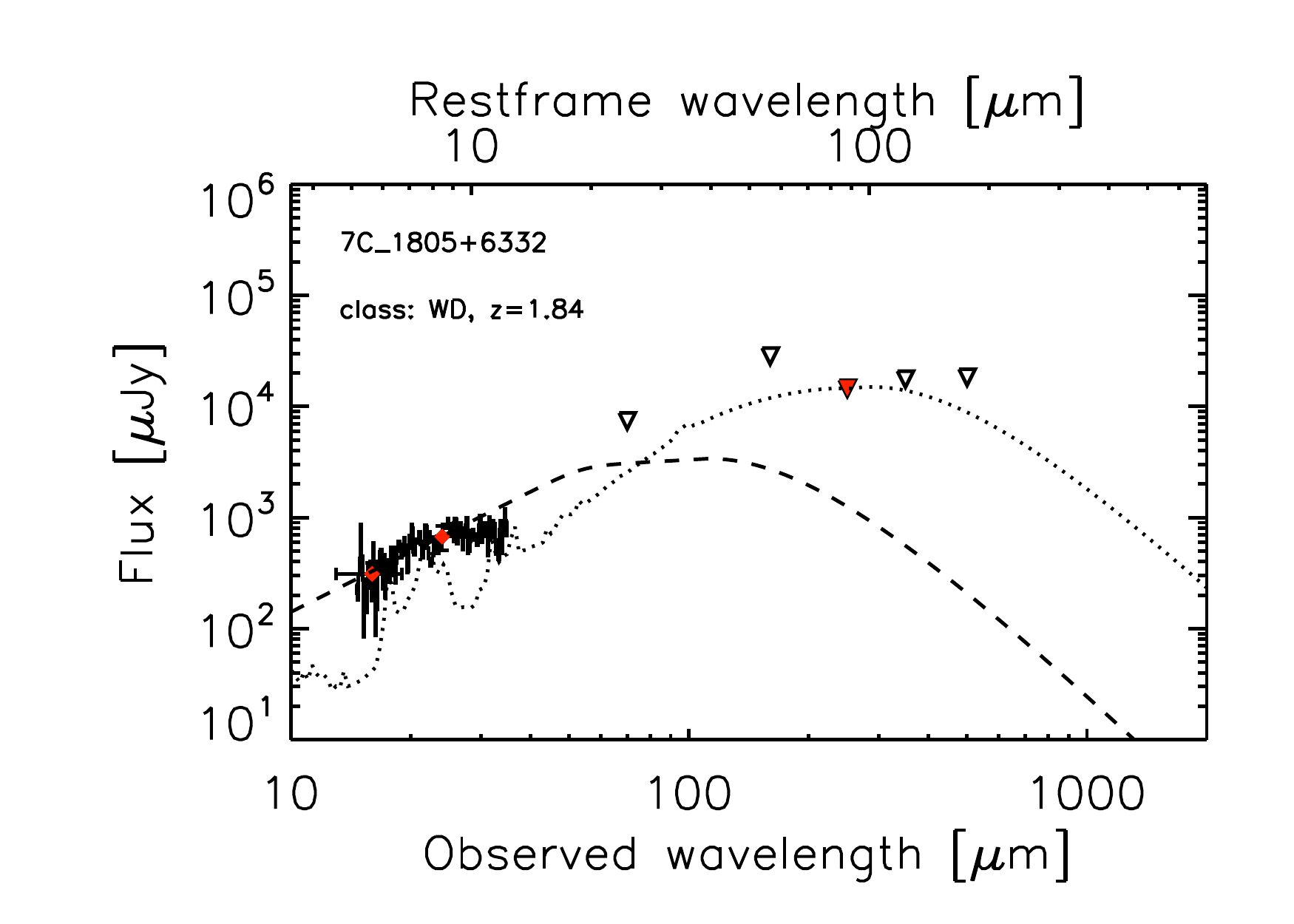}} \\
      24\mum & 100\mum & 160\mum & 250\mum & 350\mum & 500\mum \\
      \includegraphics[height=14mm,trim=70  -4 65 -10,clip=true]{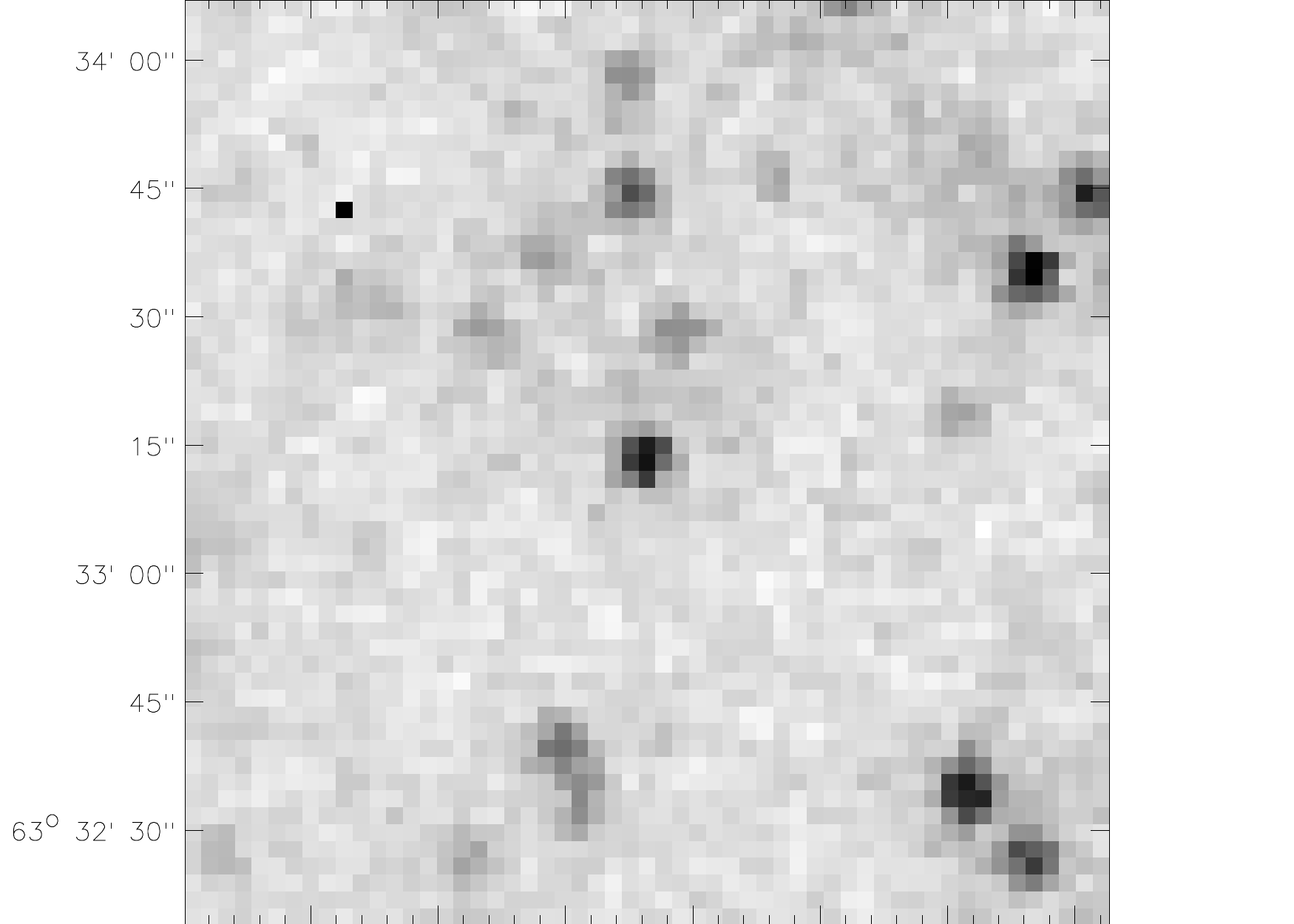} &
      \includegraphics[height=14mm,trim=70  -4 65 -10,clip=true]{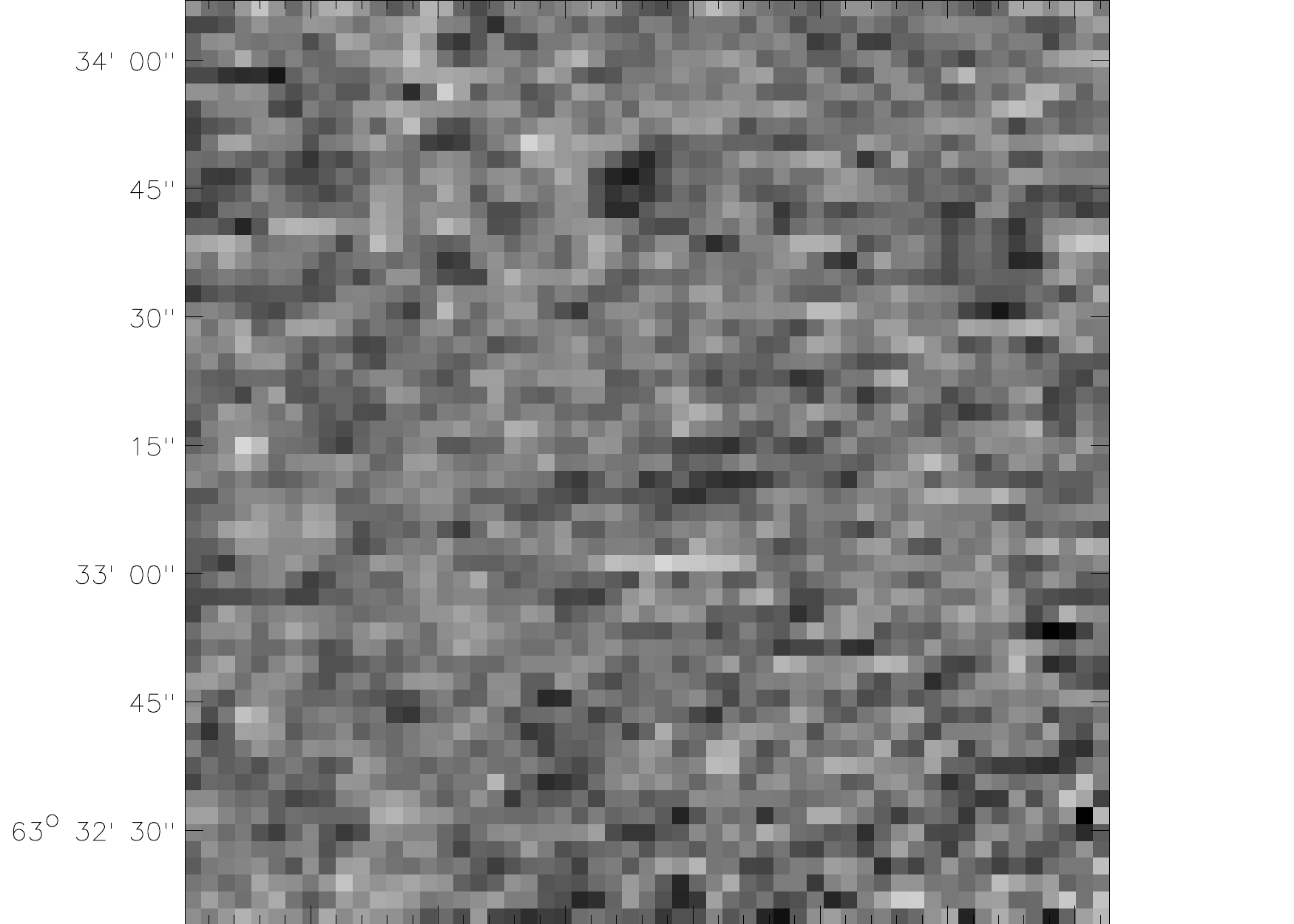} &
      \includegraphics[height=14mm,trim=70  -4 65 -10,clip=true]{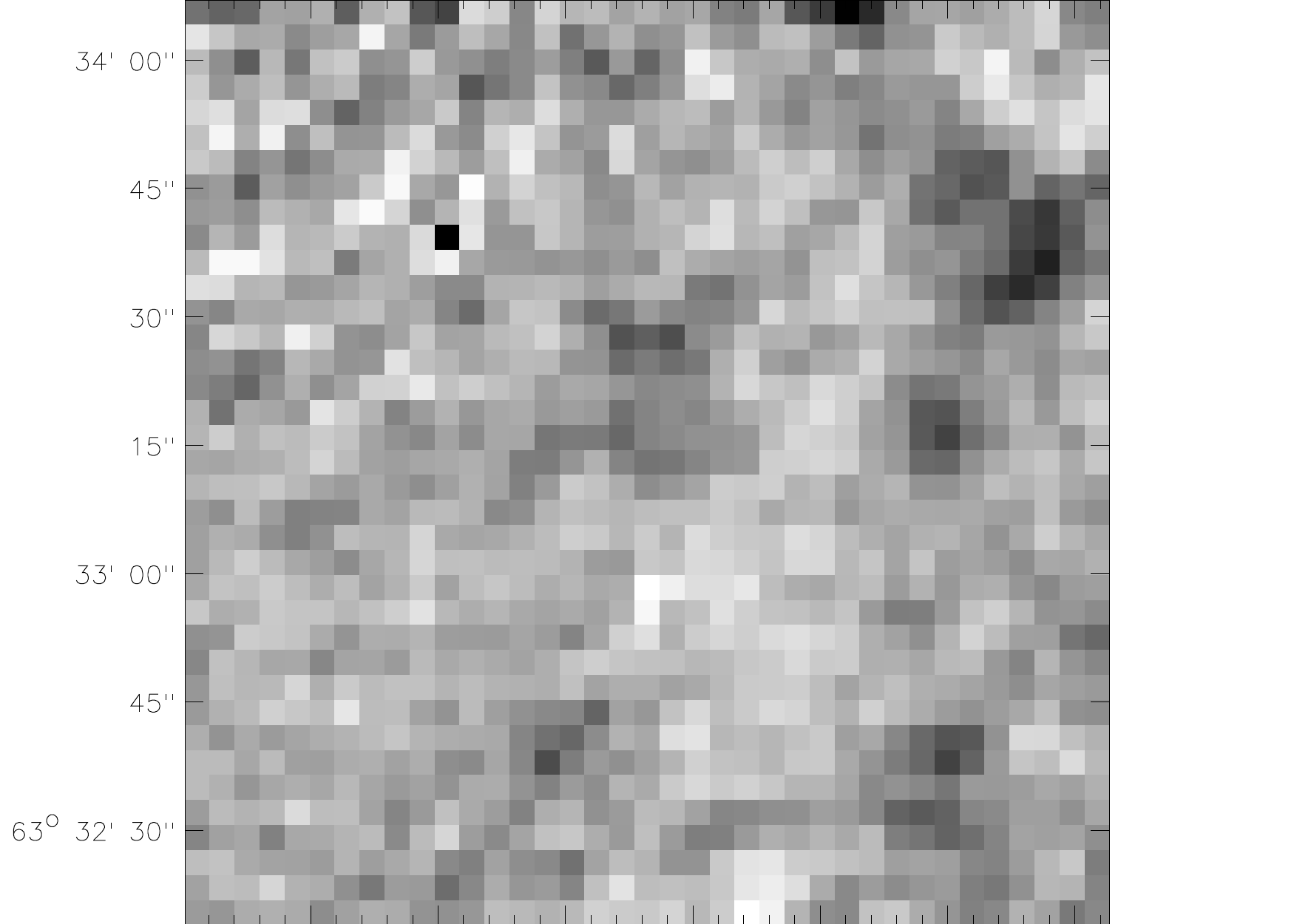} &
      \includegraphics[height=14mm,trim=70  -4 65 -10,clip=true]{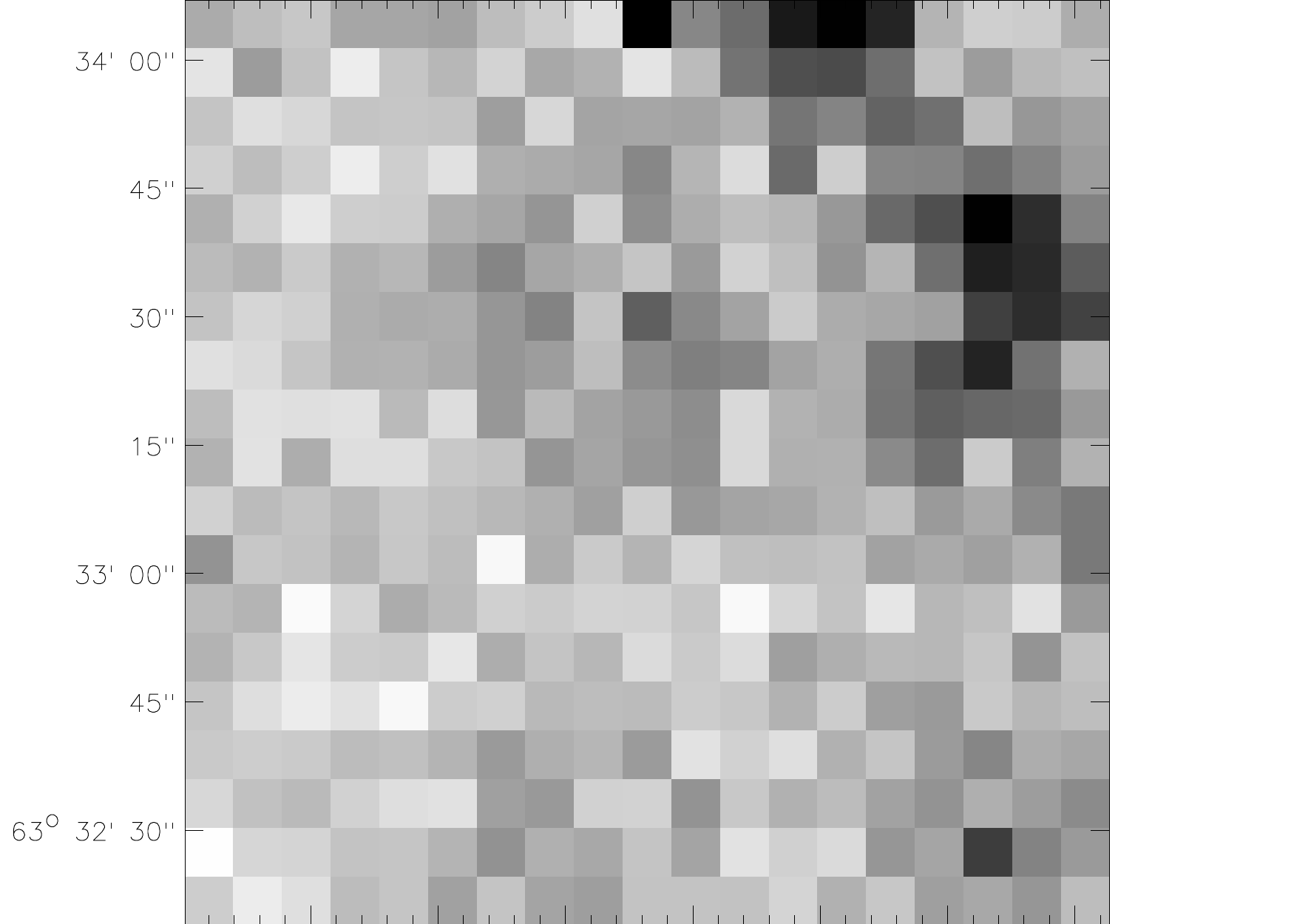}   &
      \includegraphics[height=14mm,trim=70  -4 65 -10,clip=true]{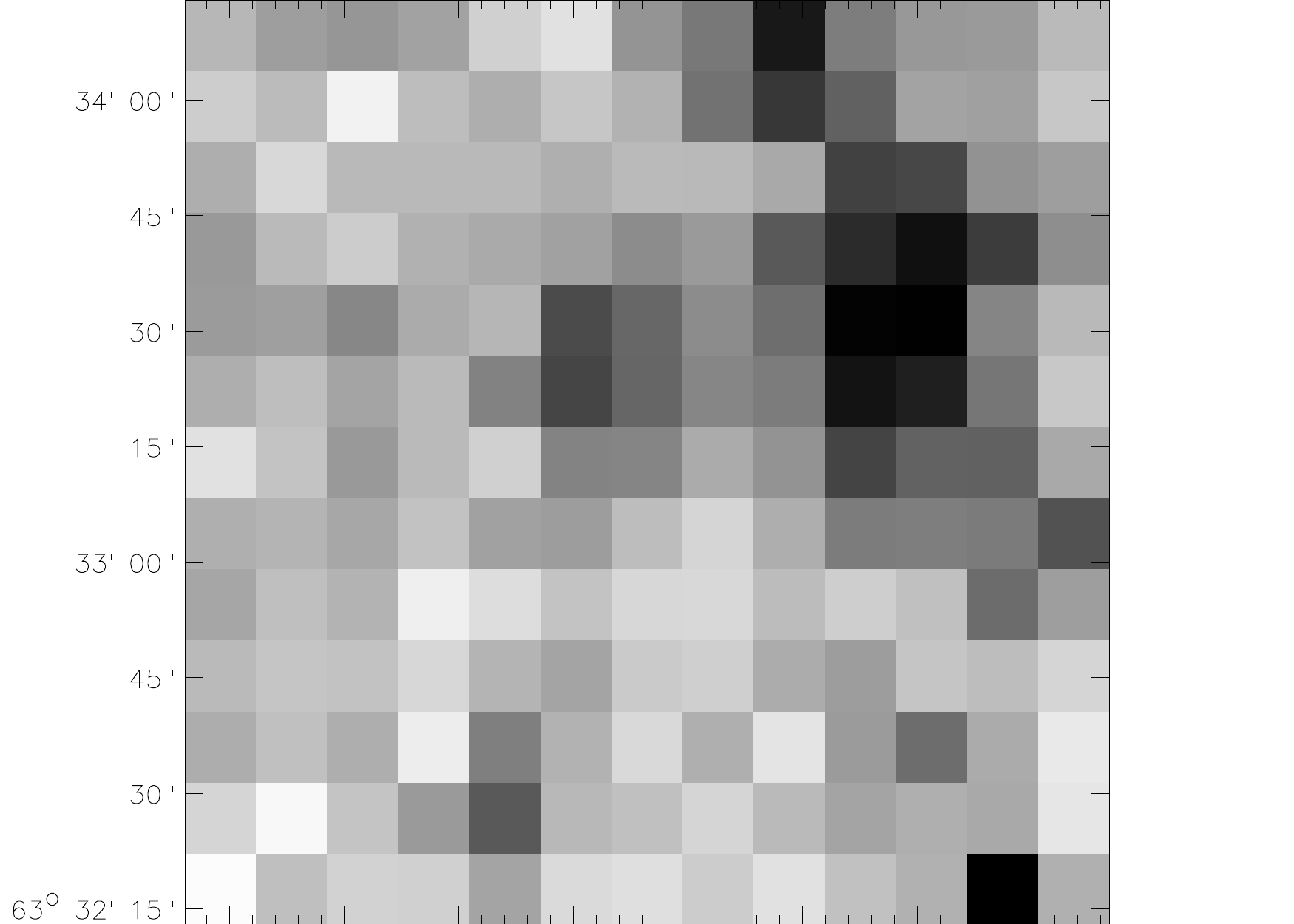}   &
      \includegraphics[height=14mm,trim=70  -4 65 -10,clip=true]{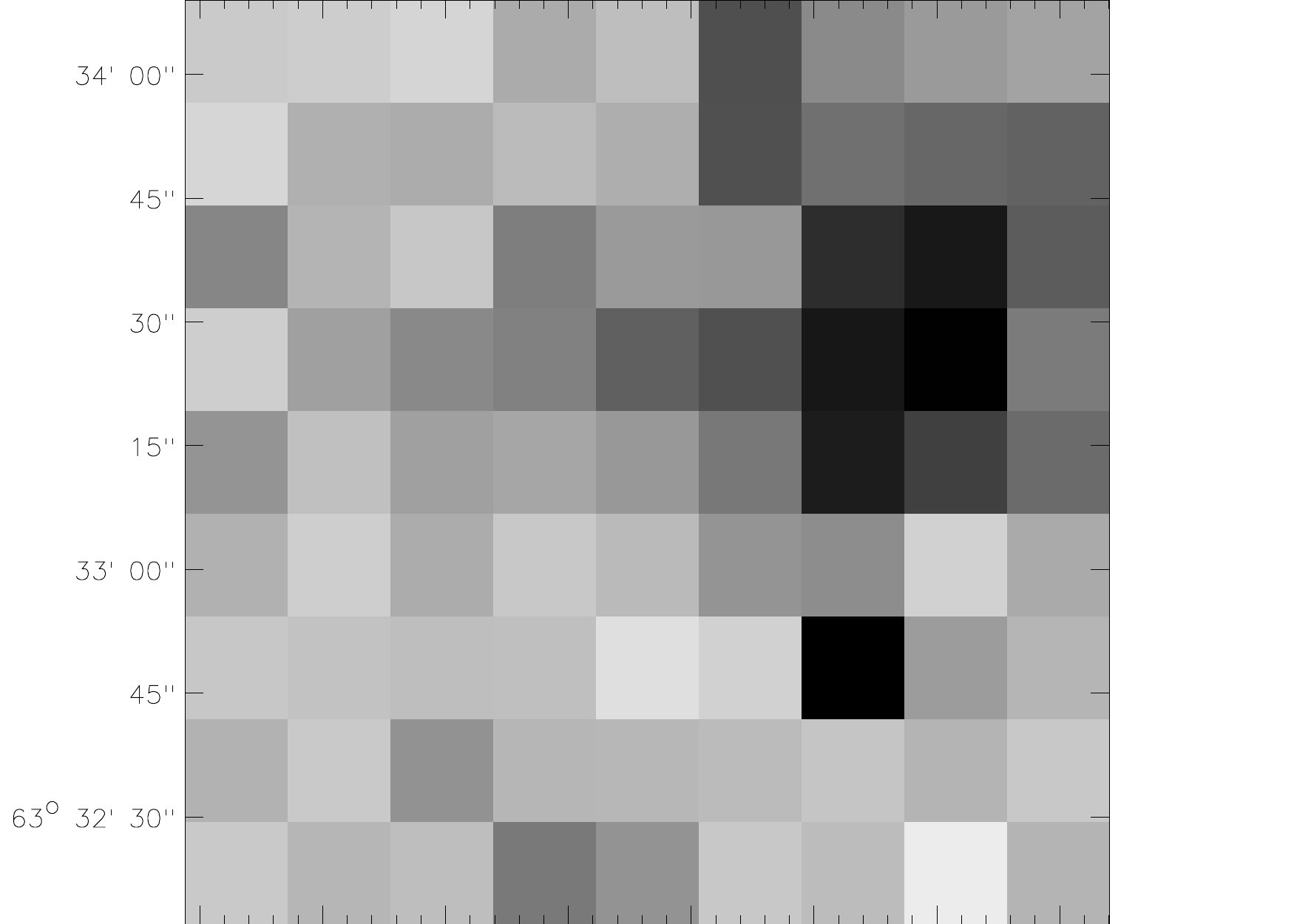} \\

      \multicolumn{6}{c}{\includegraphics[width=80mm,trim=30 0 20 0 ,clip=true]{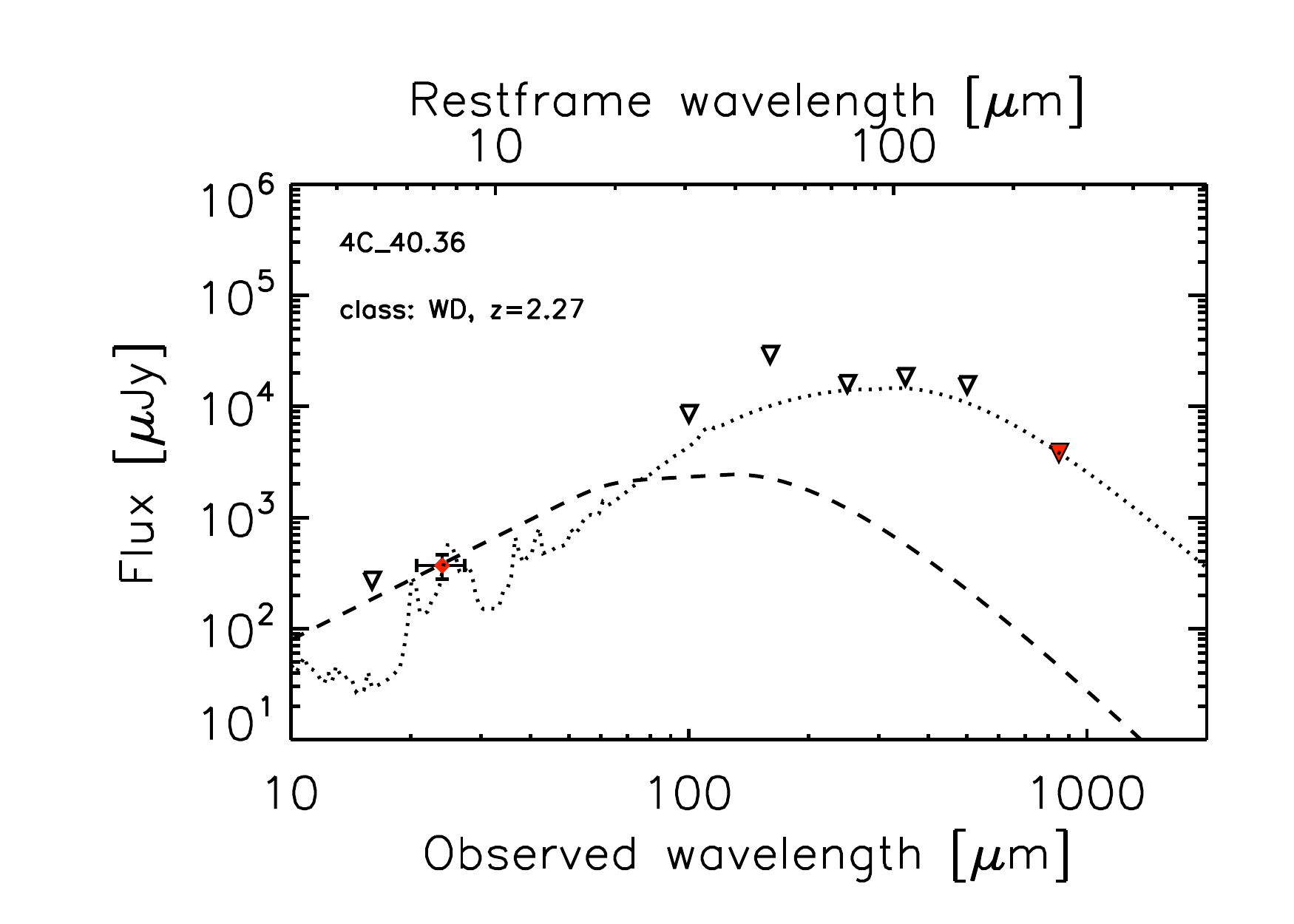}} \\
      24\mum & 100\mum & 160\mum & 250\mum & 350\mum & 500\mum \\
      \includegraphics[height=14mm,trim=70  -4 65 -10,clip=true]{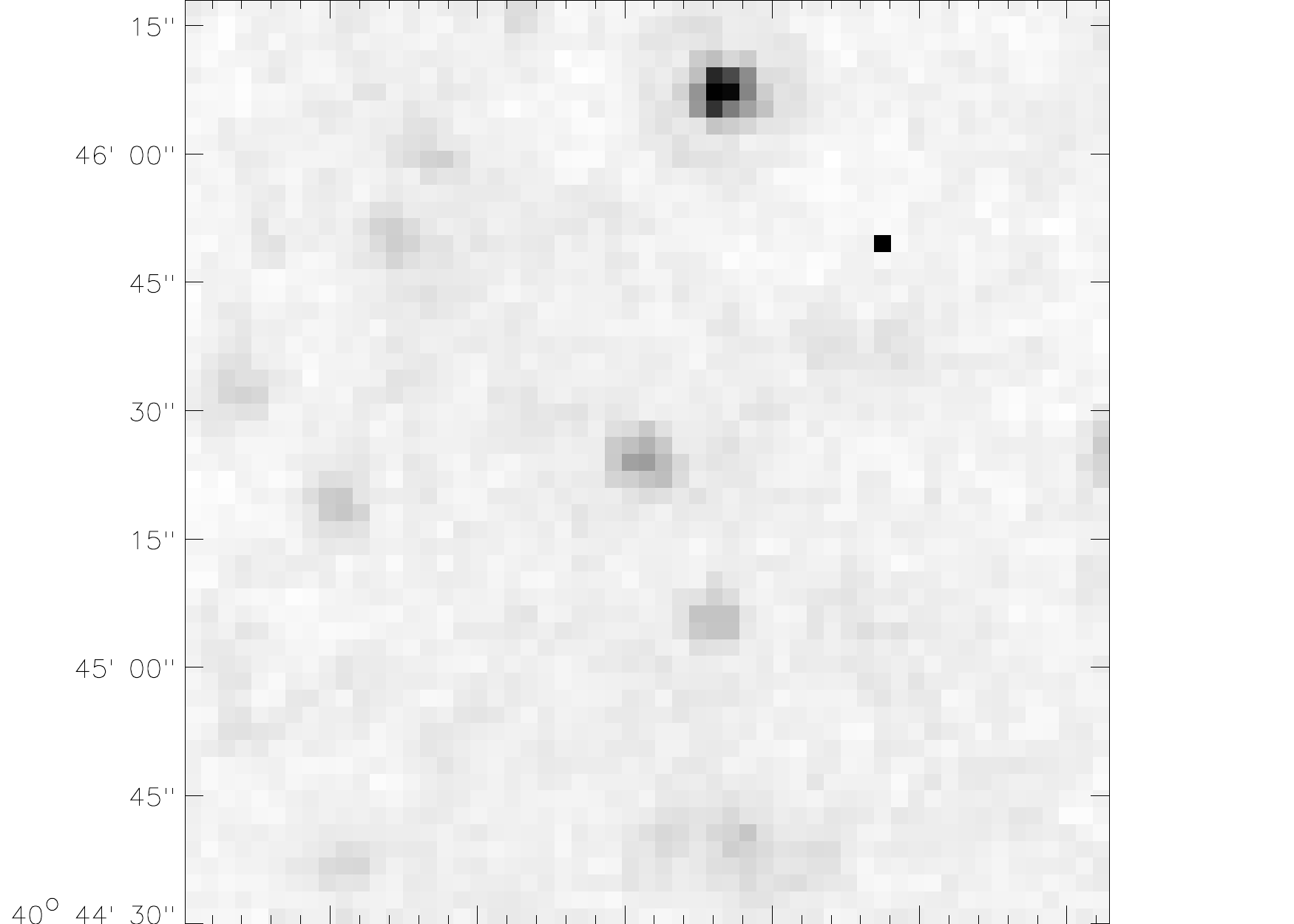} &
      \includegraphics[height=14mm,trim=70  -4 65 -10,clip=true]{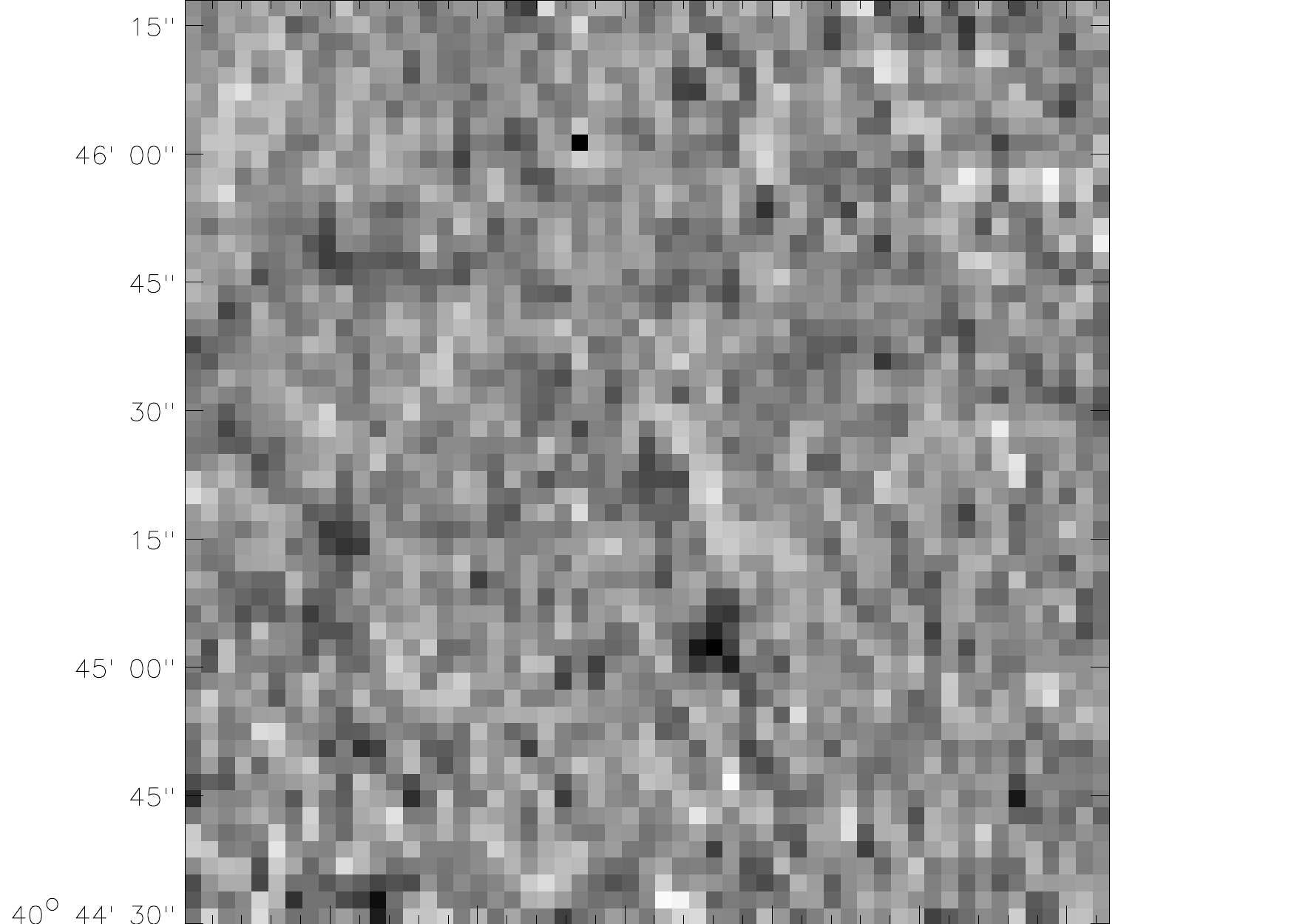} &
      \includegraphics[height=14mm,trim=70  -4 65 -10,clip=true]{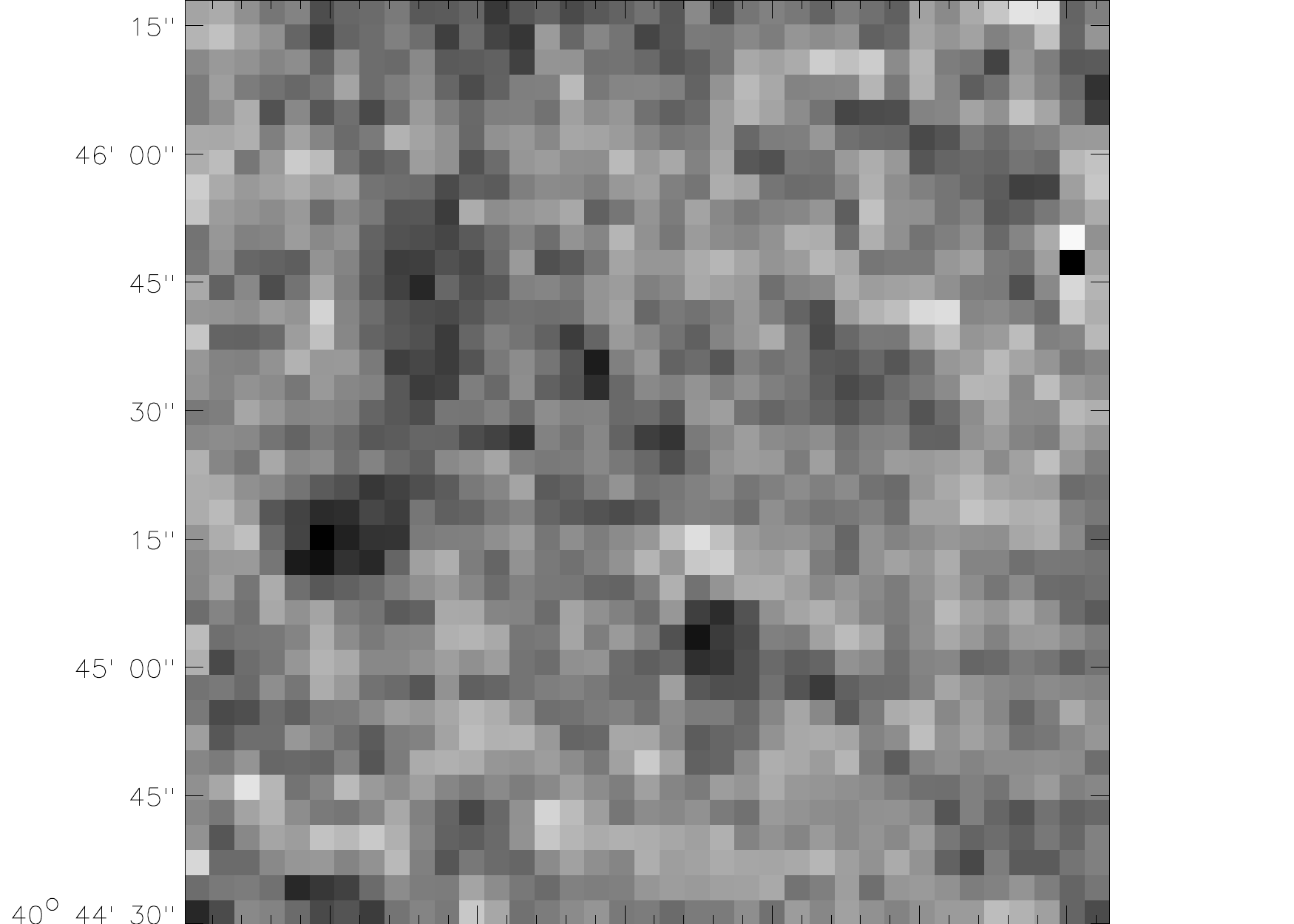} &
      \includegraphics[height=14mm,trim=70  -4 65 -10,clip=true]{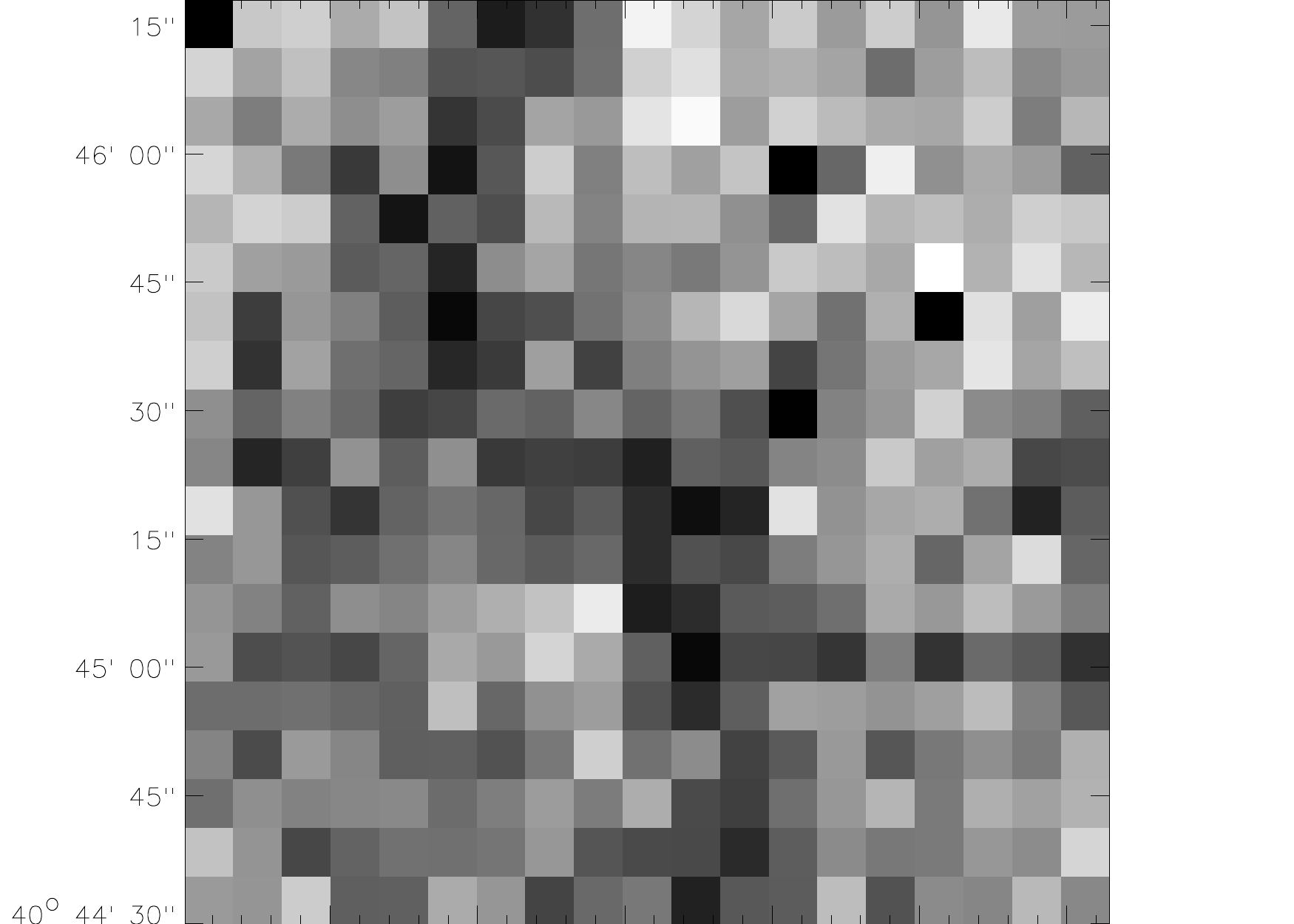}   &
      \includegraphics[height=14mm,trim=70  -4 65 -10,clip=true]{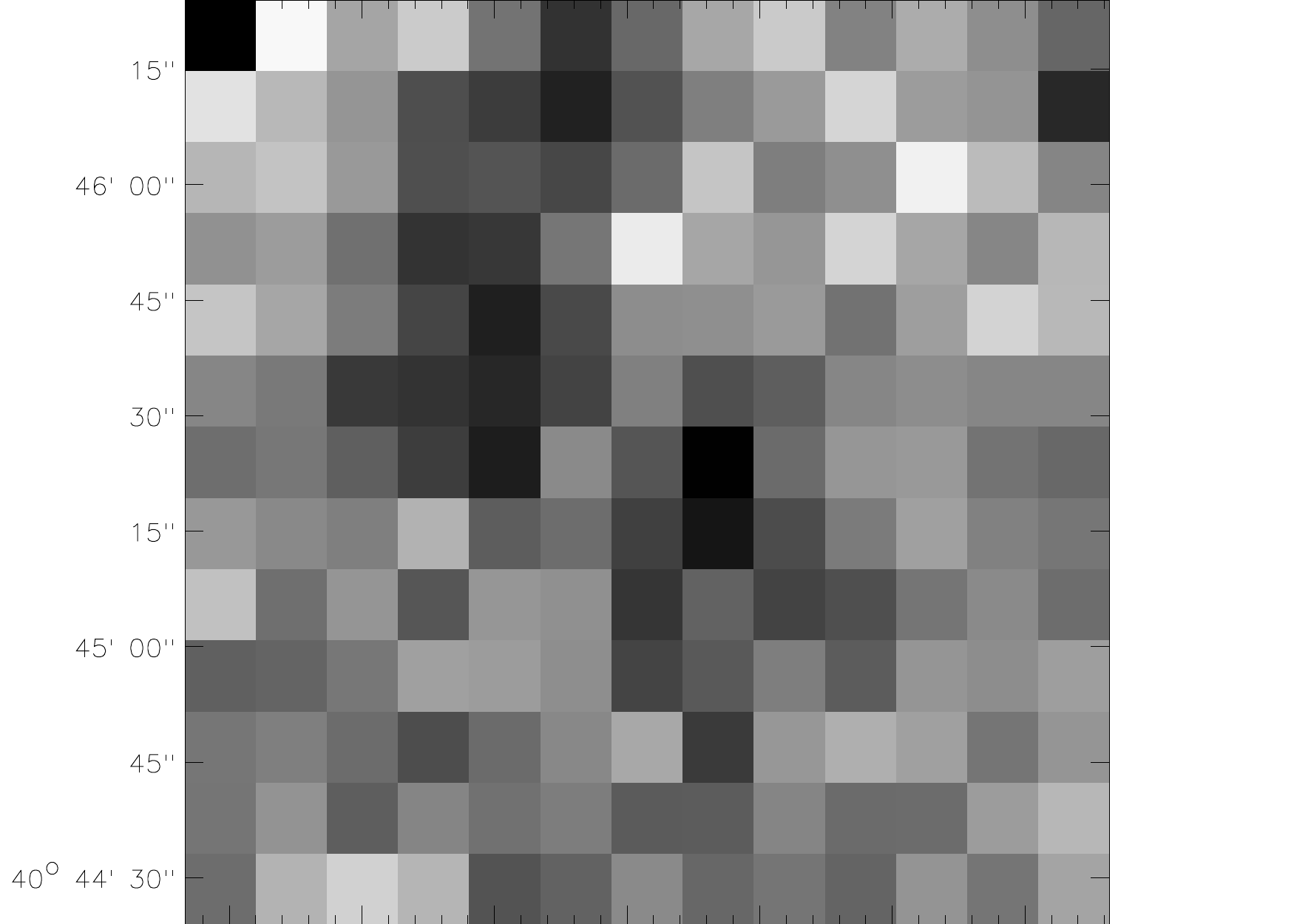}   &
      \includegraphics[height=14mm,trim=70  -4 65 -10,clip=true]{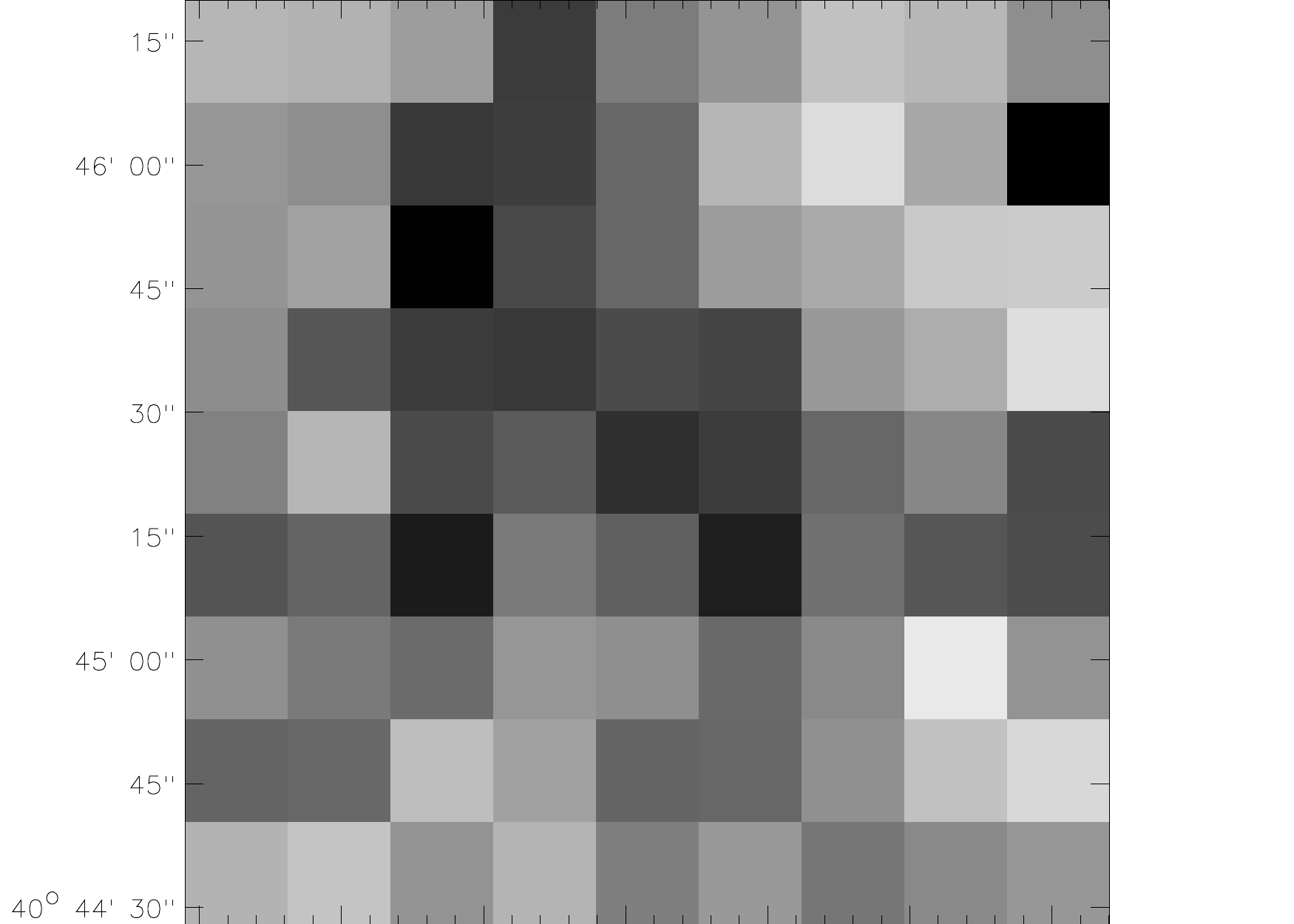} \\

      \multicolumn{6}{c}{\includegraphics[width=80mm,trim=30 0 20 0 ,clip=true]{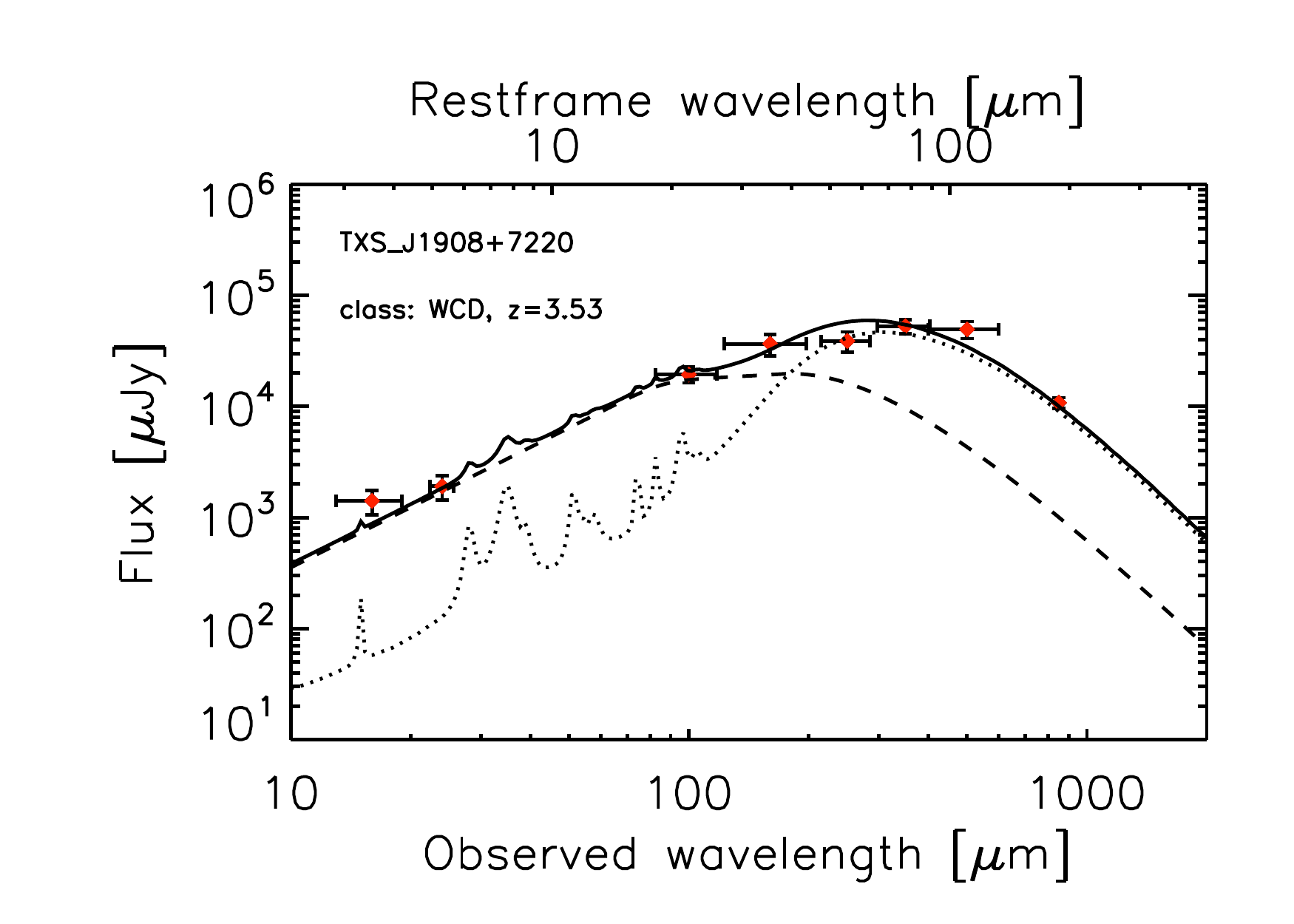}} \\
      24\mum & 100\mum & 160\mum & 250\mum & 350\mum & 500\mum \\
\end{tabular}
          \end{minipage}
  \end{centering}
\end{figure*}

\clearpage

\begin{figure*}[h]
  \begin{centering}
    \begin{minipage}[c]{0.46\linewidth}
    \begin{tabular}{c@{}c@{}c@{}c@{}c@{}c@{}}

      \multicolumn{6}{c}{\includegraphics[width=80mm,trim=30 0 20 0 ,clip=true]{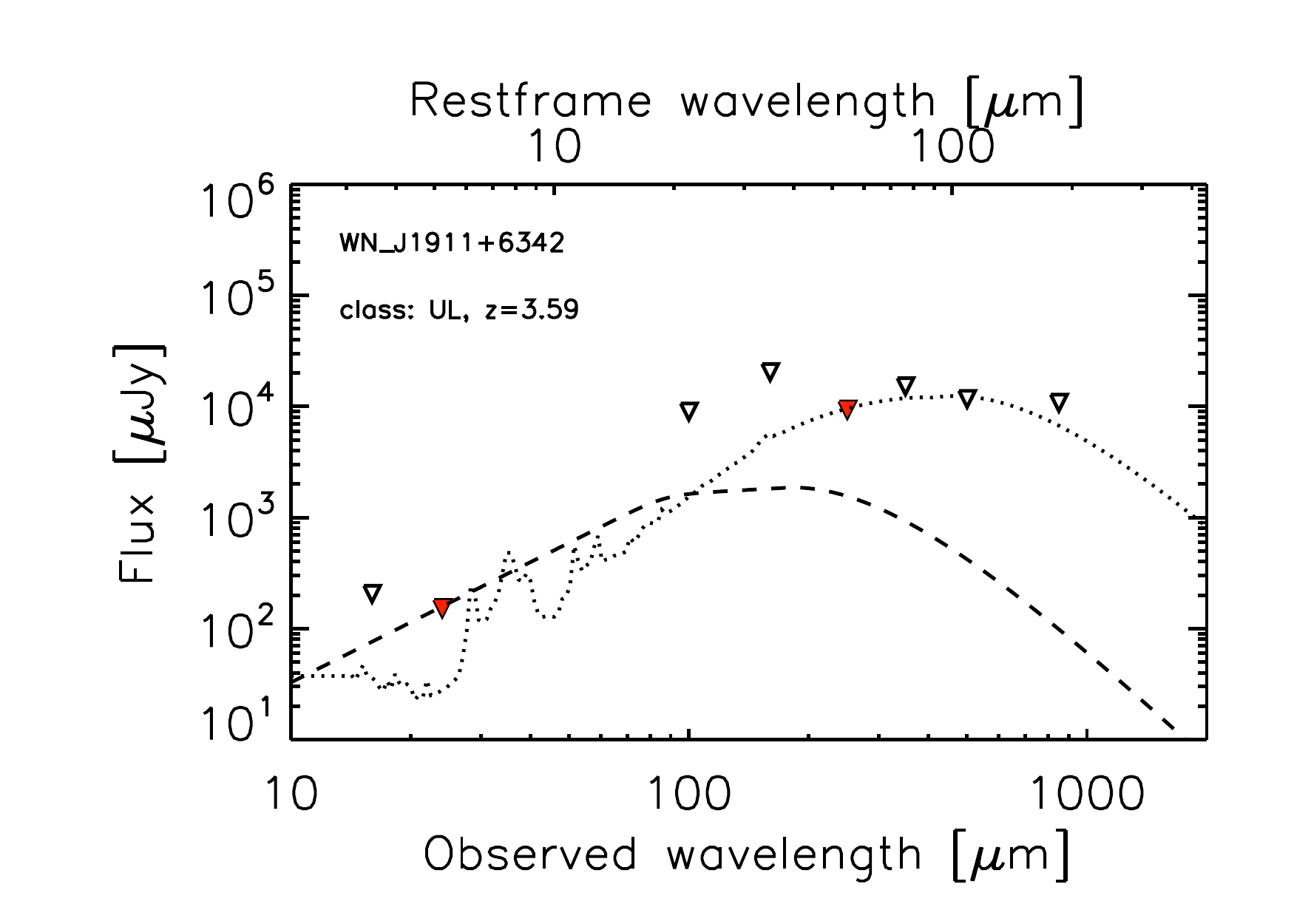}} \\
      24\mum & 100\mum & 160\mum & 250\mum & 350\mum & 500\mum \\
      \includegraphics[height=14mm,trim=70  -4 65 -10,clip=true]{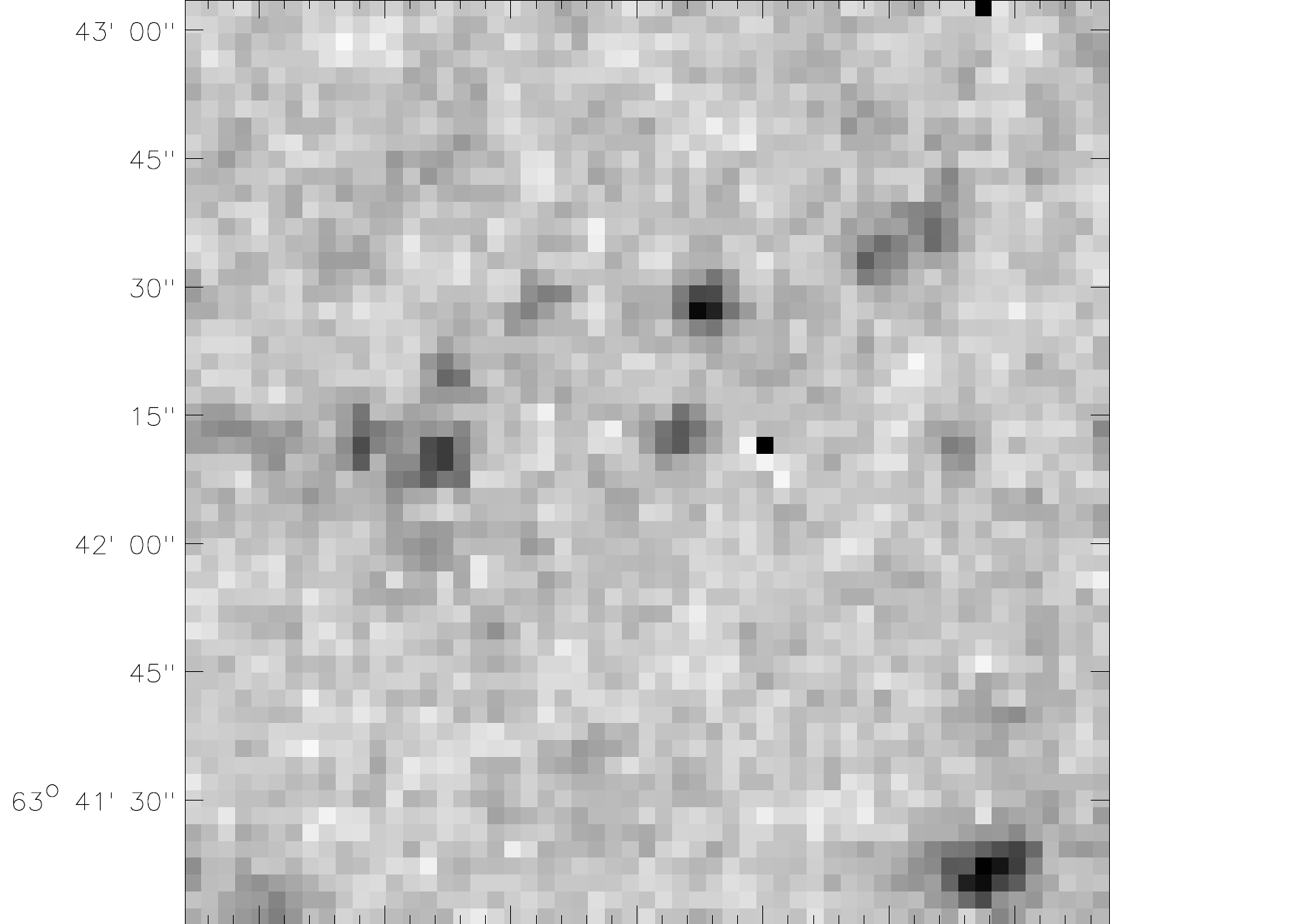} &
      \includegraphics[height=14mm,trim=70  -4 65 -10,clip=true]{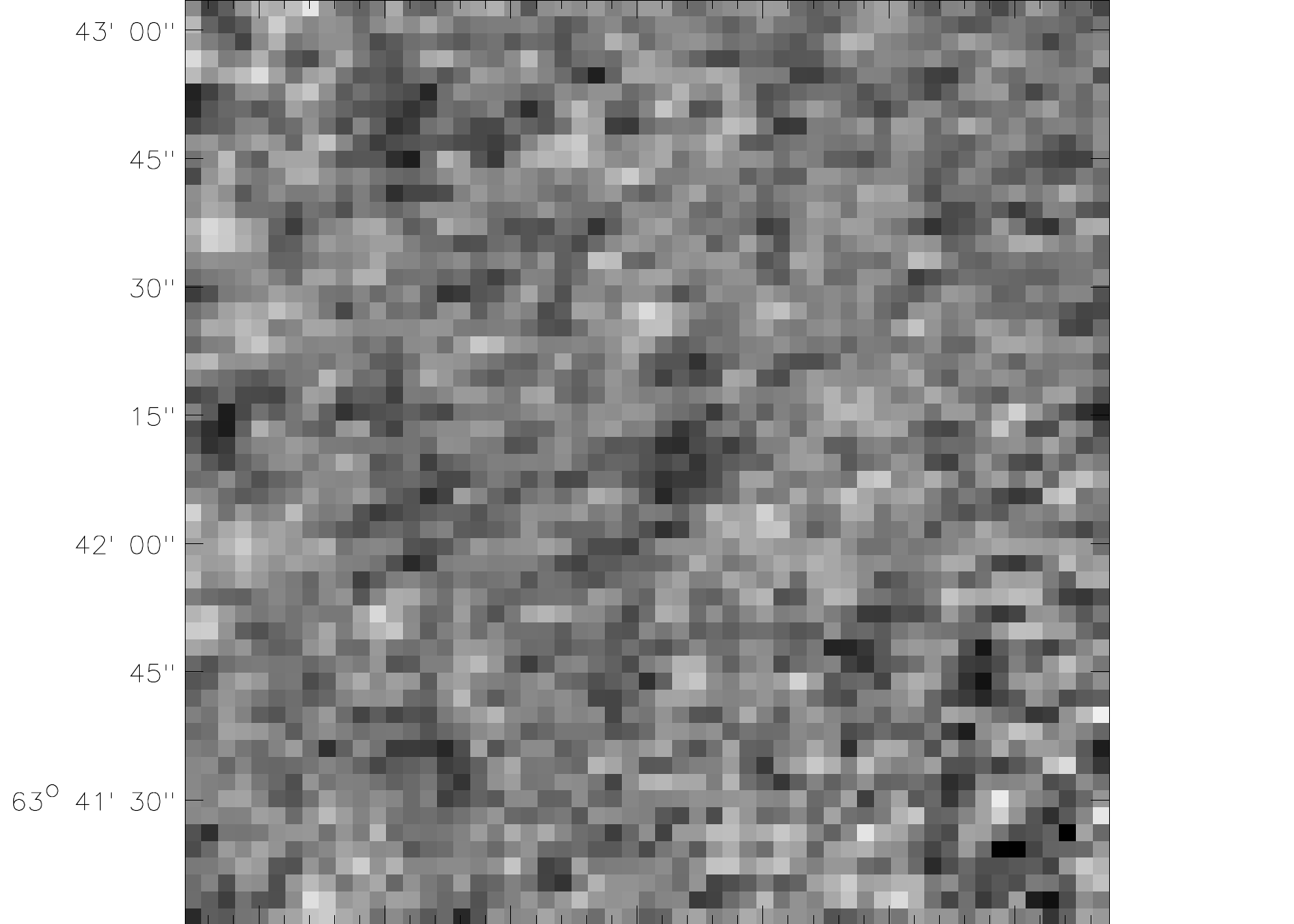} &
      \includegraphics[height=14mm,trim=70  -4 65 -10,clip=true]{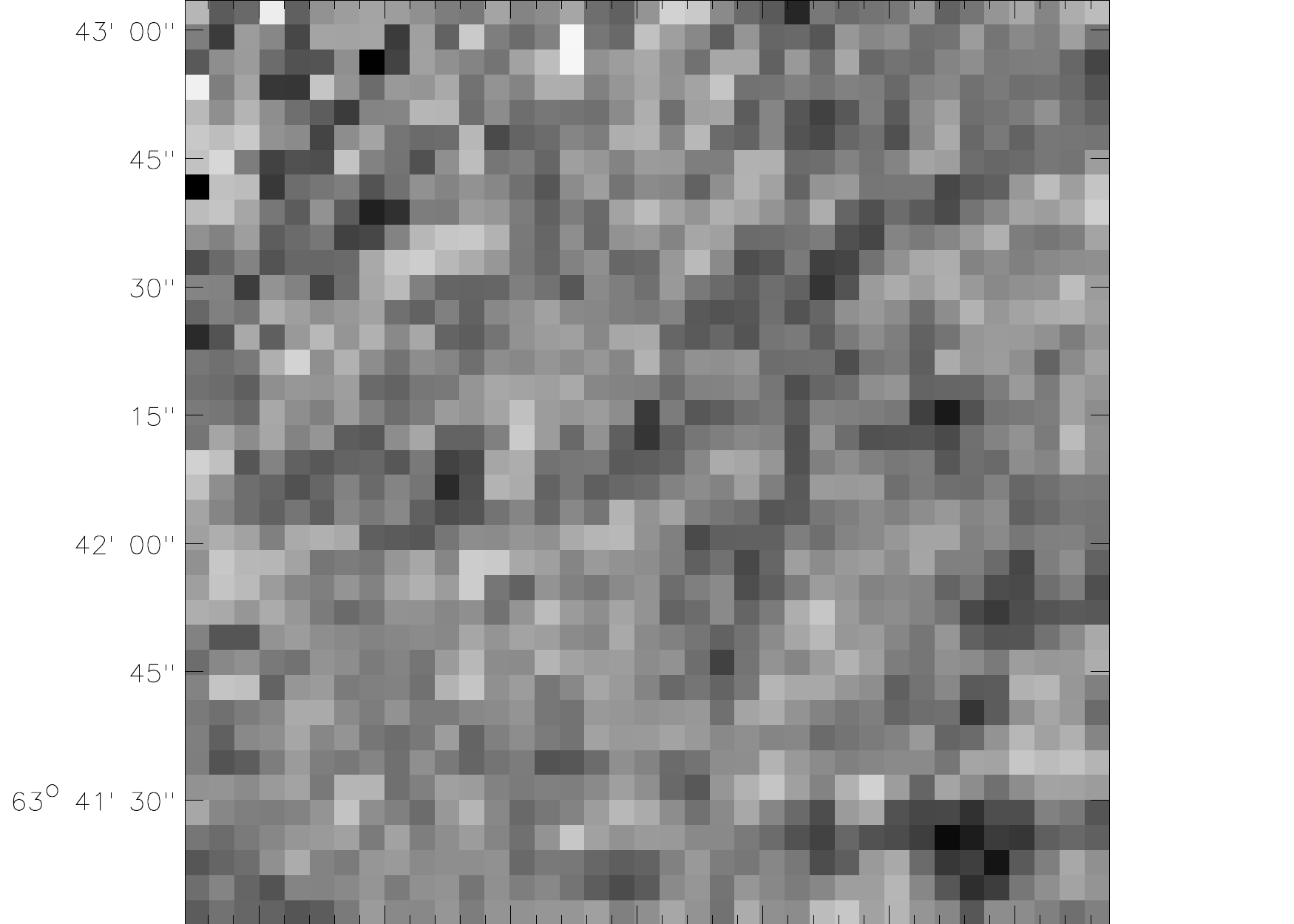} &
      \includegraphics[height=14mm,trim=70  -4 65 -10,clip=true]{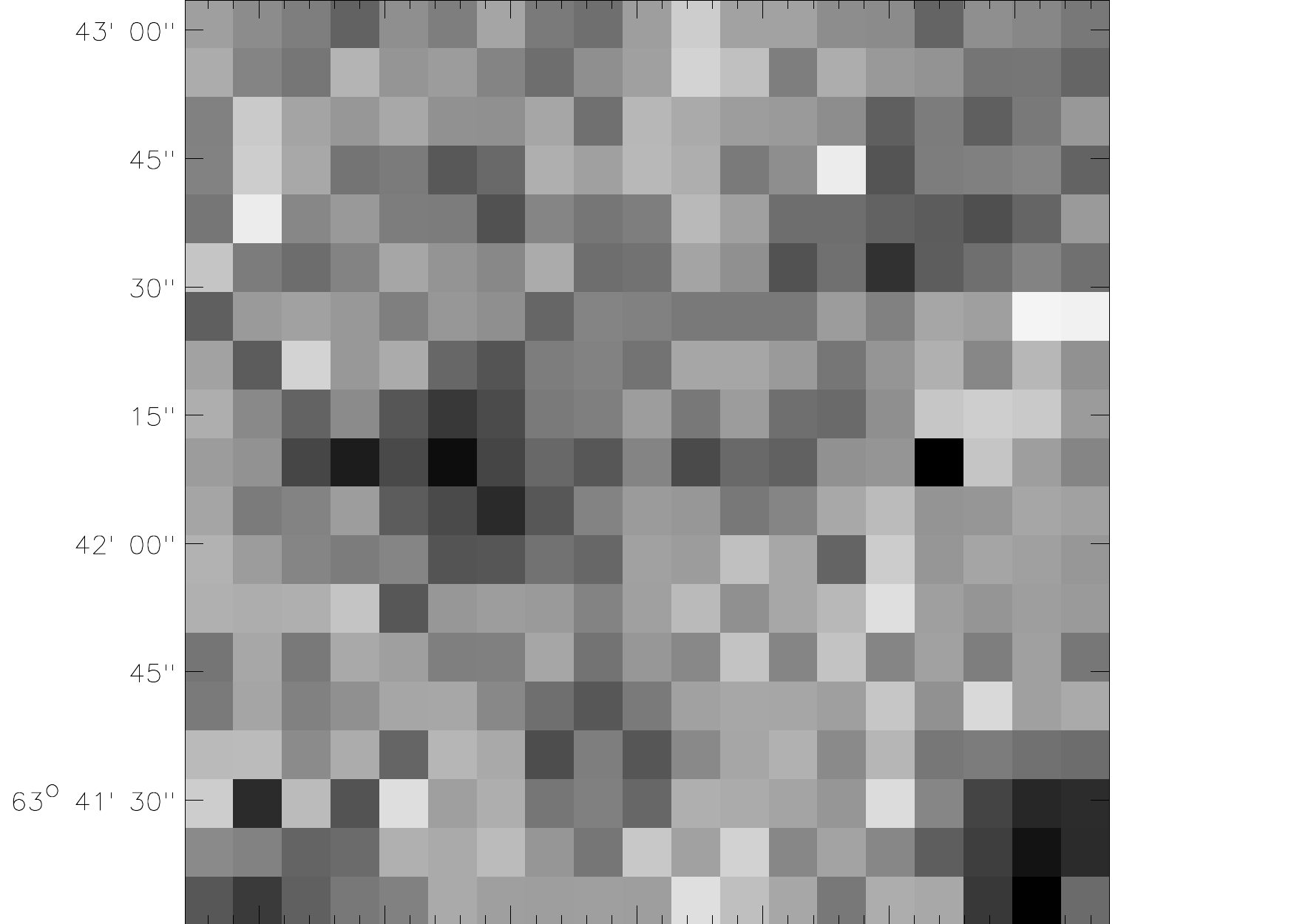}   &
      \includegraphics[height=14mm,trim=70  -4 65 -10,clip=true]{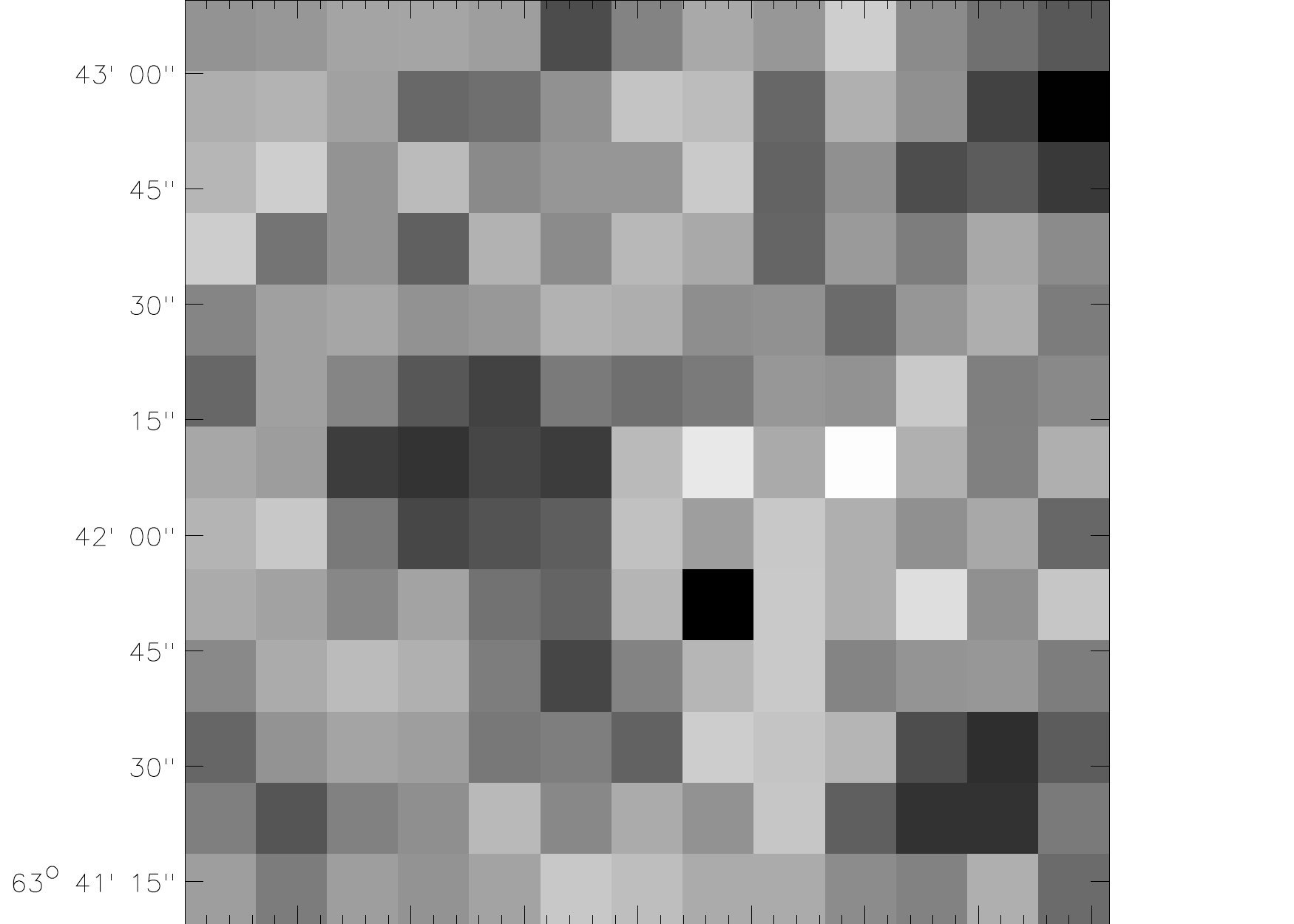}   &
      \includegraphics[height=14mm,trim=70  -4 65 -10,clip=true]{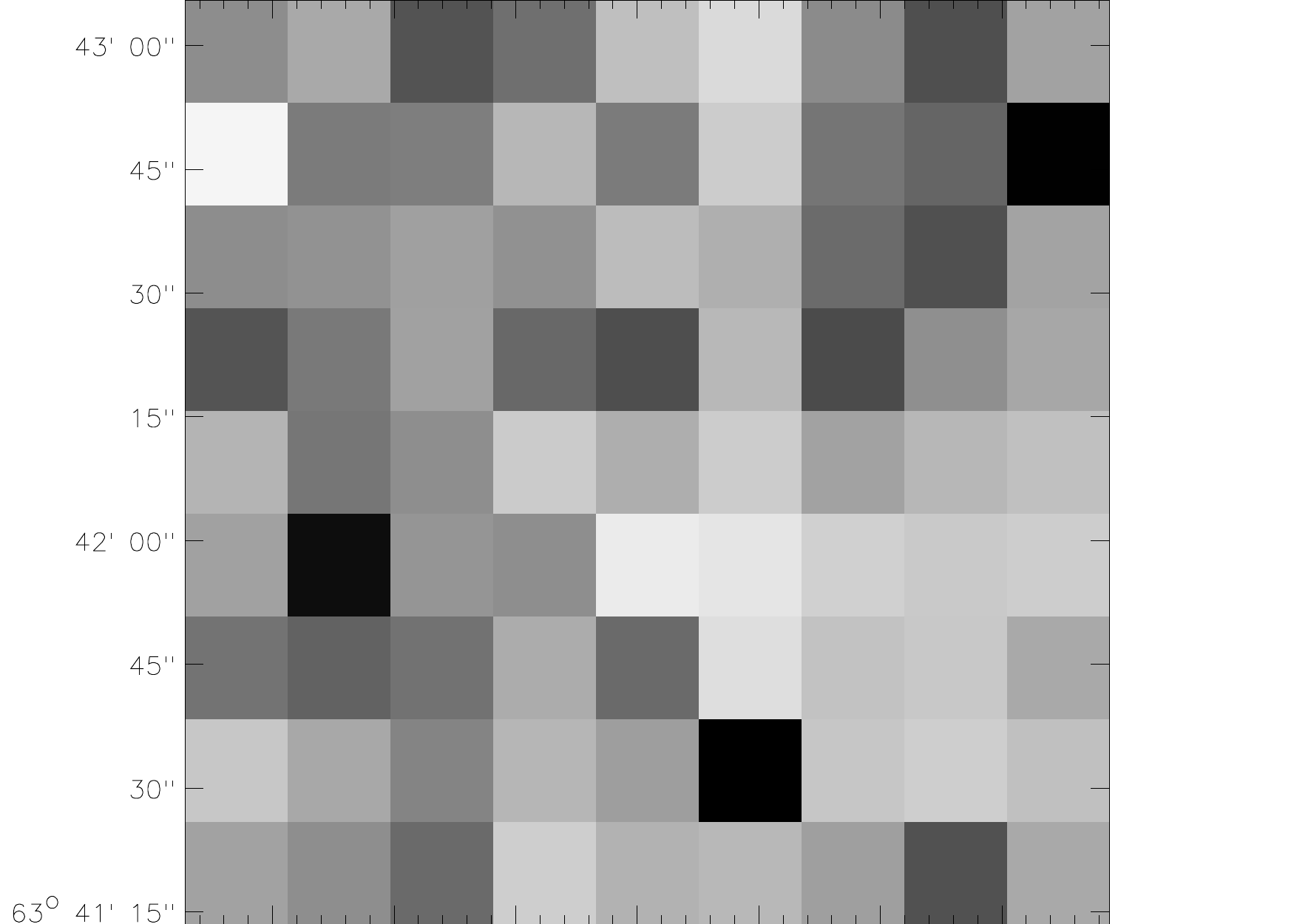} \\

      \multicolumn{6}{c}{\includegraphics[width=80mm,trim=30 0 20 0 ,clip=true]{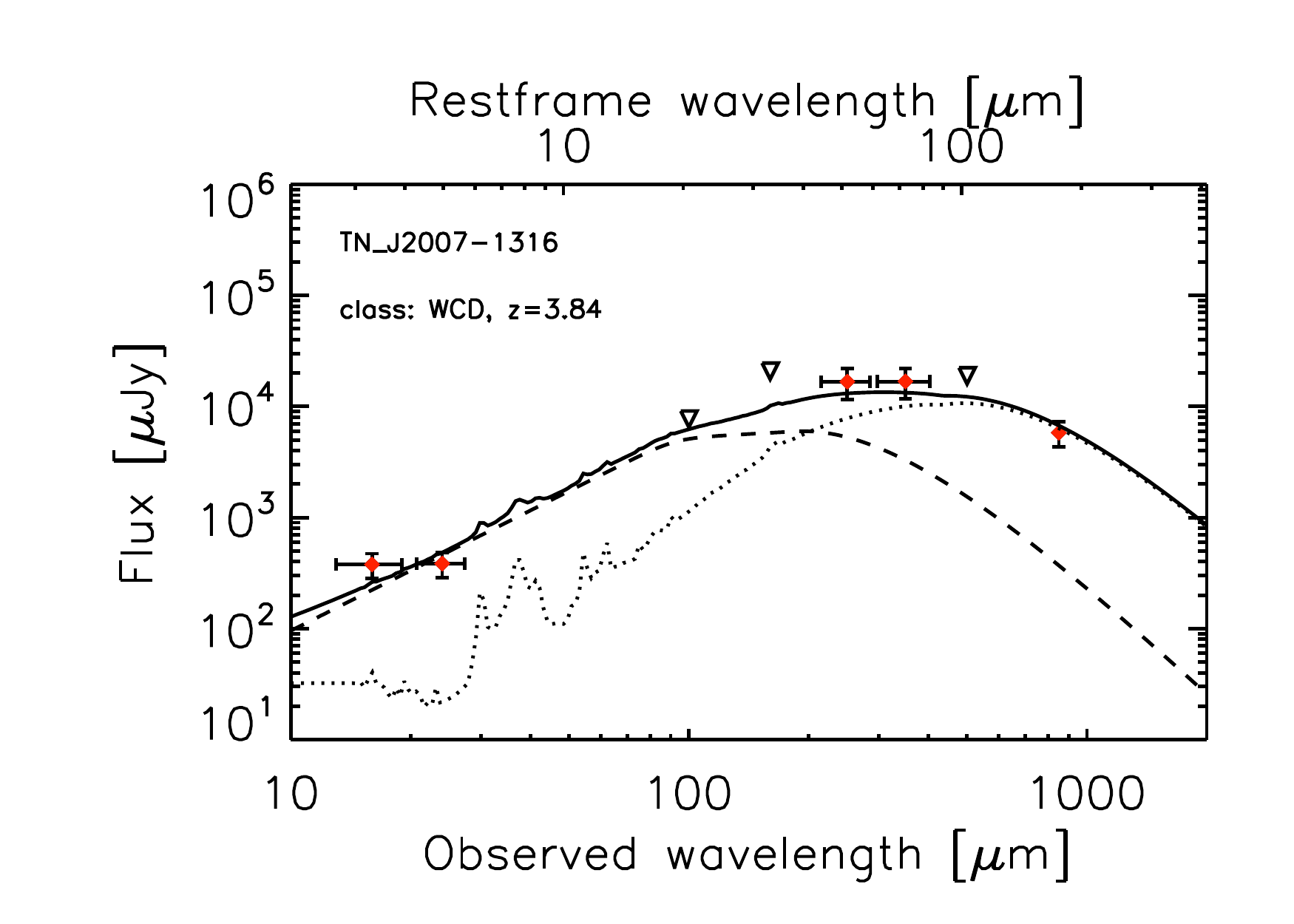}} \\
      24\mum & 100\mum & 160\mum & 250\mum & 350\mum & 500\mum \\
      \includegraphics[height=14mm,trim=70  -4 65 -10,clip=true]{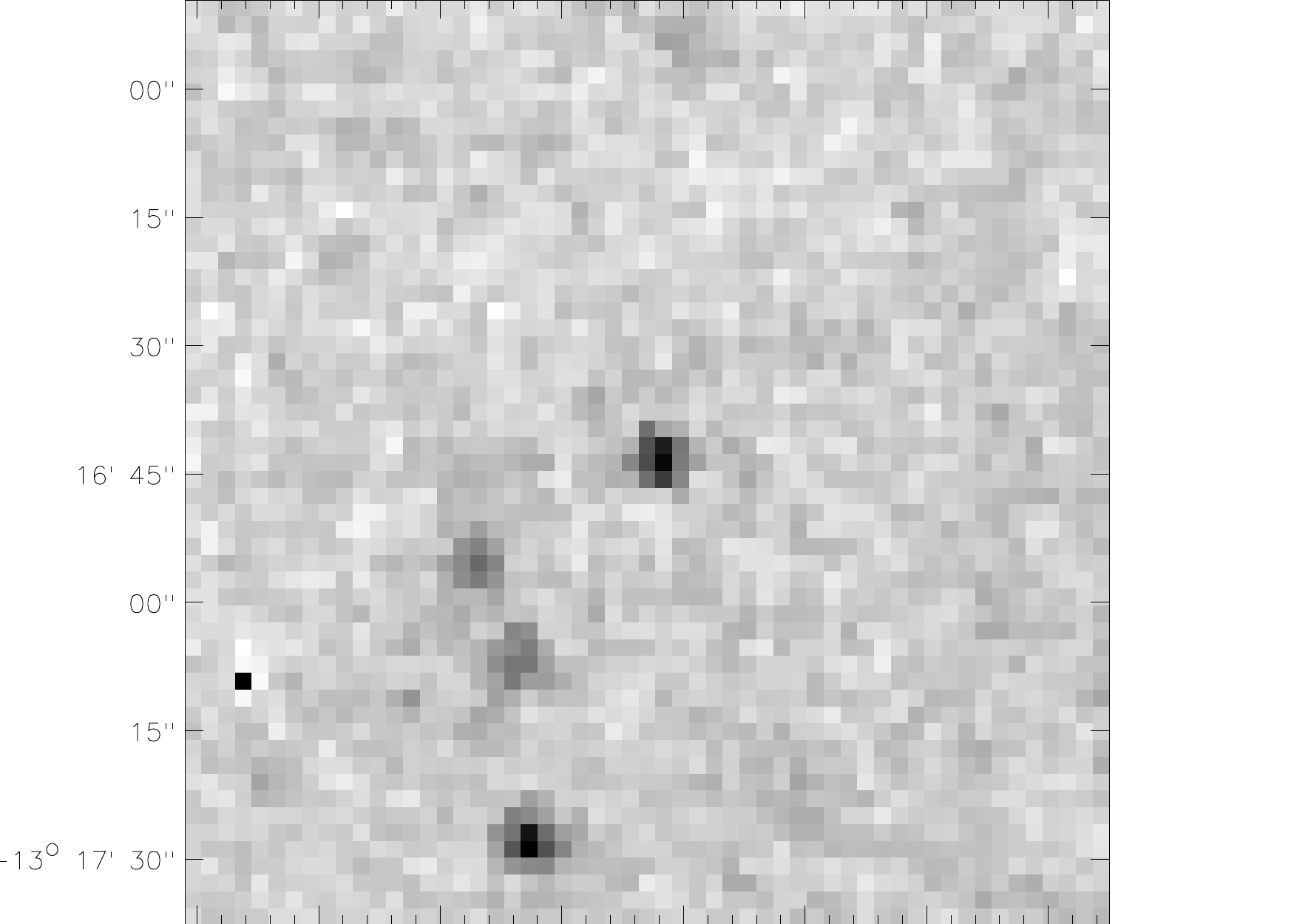} &
      \includegraphics[height=14mm,trim=70  -4 65 -10,clip=true]{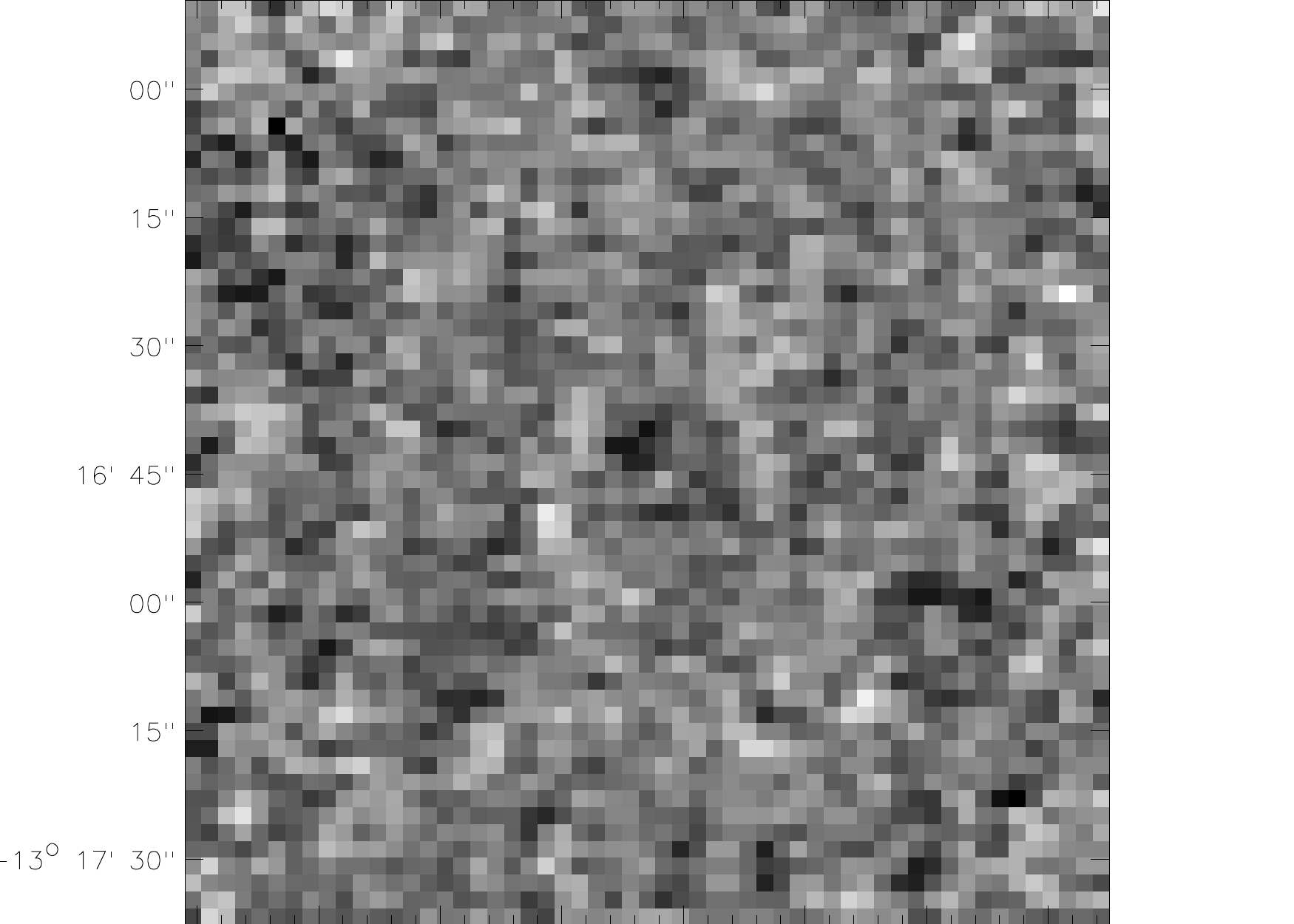} &
      \includegraphics[height=14mm,trim=70  -4 65 -10,clip=true]{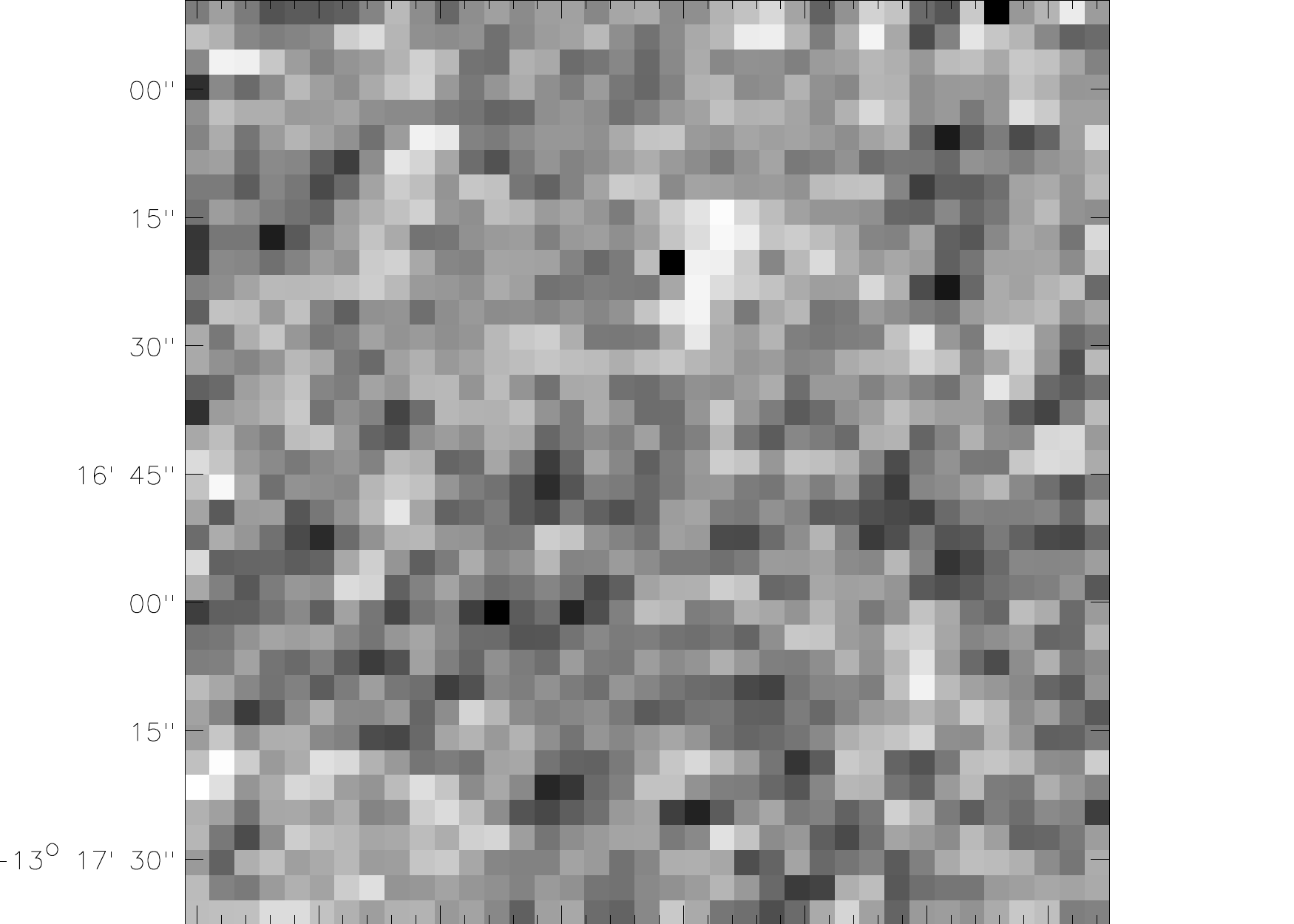} &
      \includegraphics[height=14mm,trim=70  -4 65 -10,clip=true]{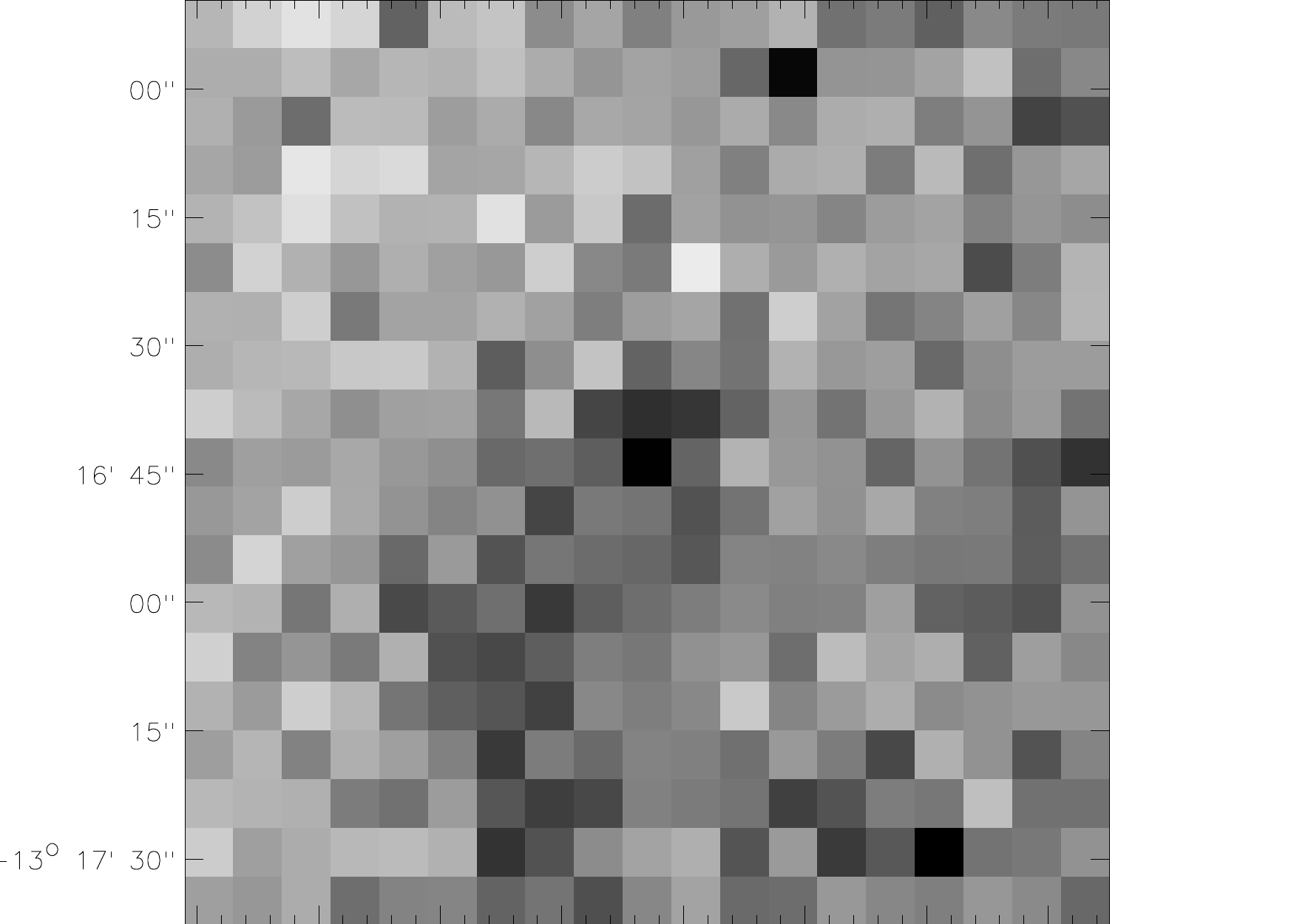}   &
      \includegraphics[height=14mm,trim=70  -4 65 -10,clip=true]{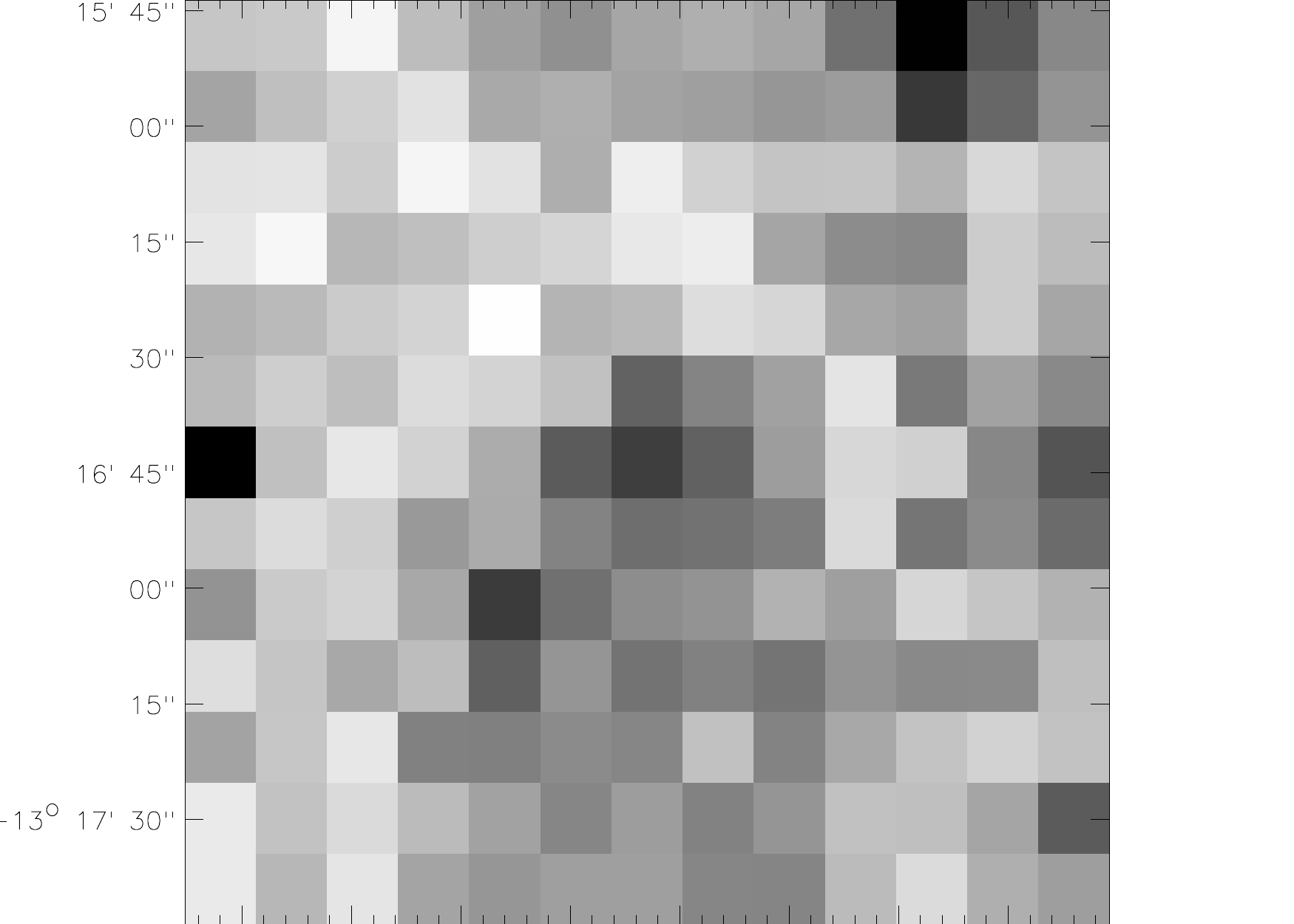}   &
      \includegraphics[height=14mm,trim=70  -4 65 -10,clip=true]{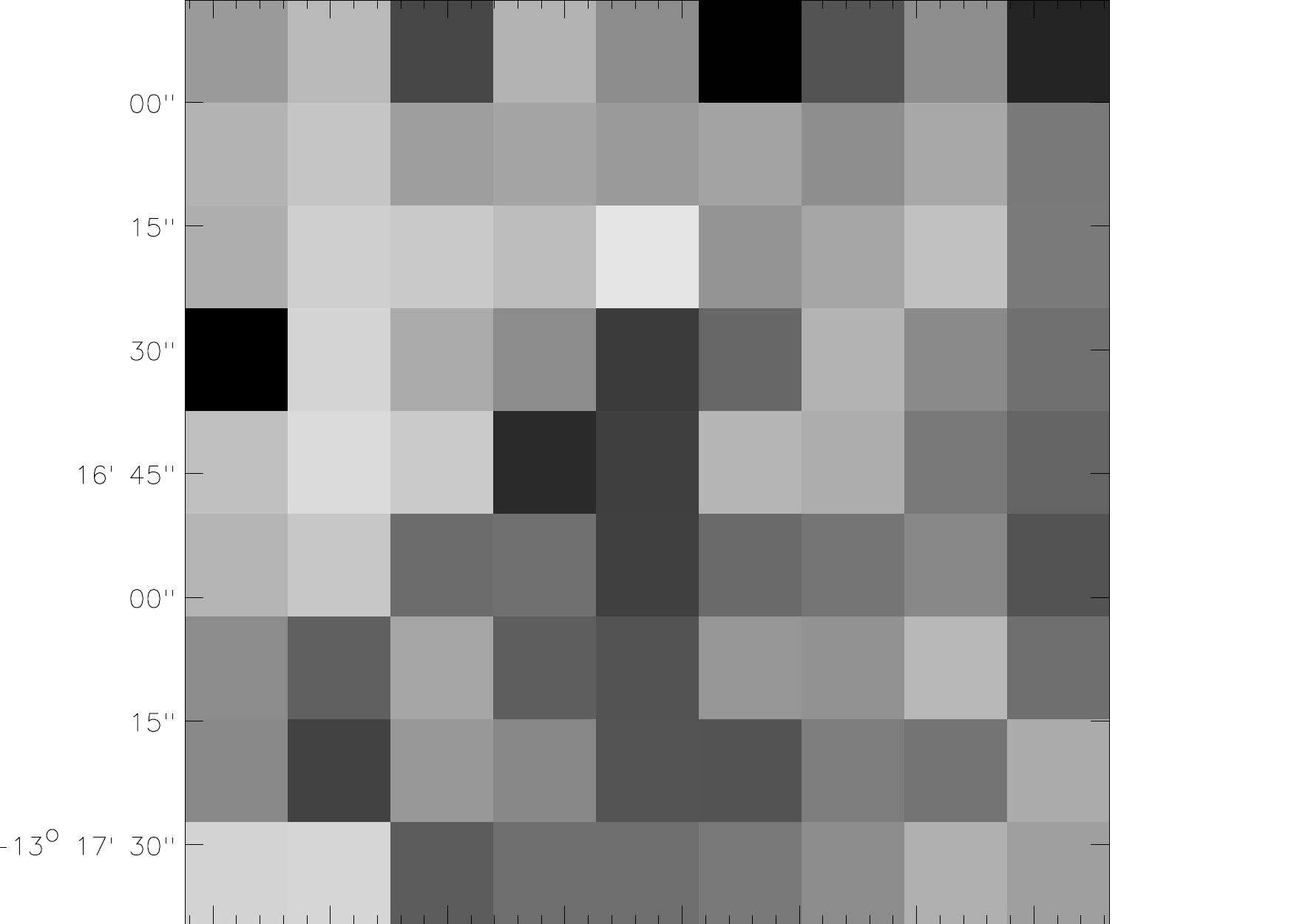} \\

      \multicolumn{6}{c}{\includegraphics[width=80mm,trim=30 0 20 0 ,clip=true]{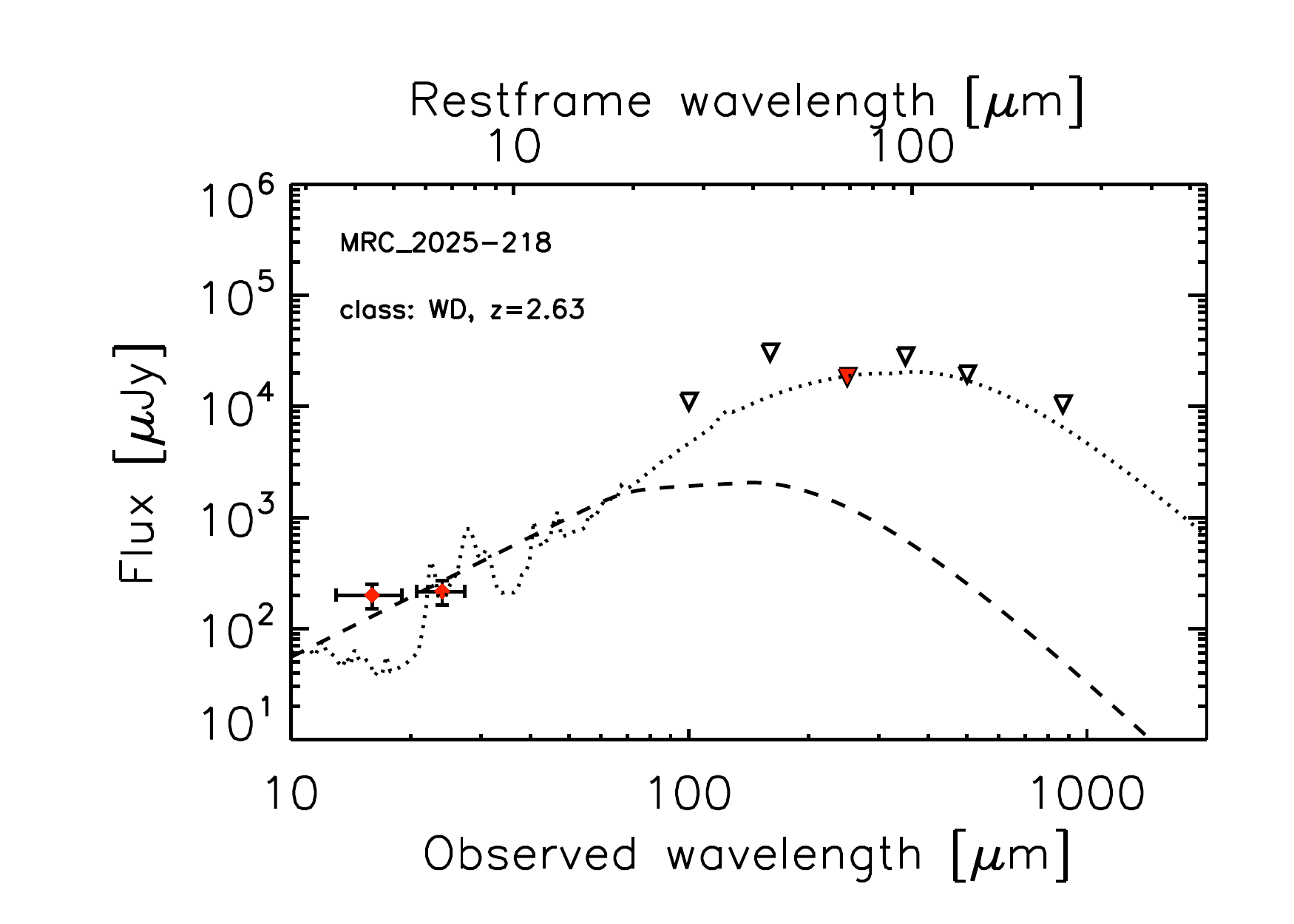}} \\
      24\mum & 100\mum & 160\mum & 250\mum & 350\mum & 500\mum \\
\end{tabular}
\end{minipage}
\begin{minipage}[c]{0.46\linewidth} 
\begin{tabular}{c@{}c@{}c@{}c@{}c@{}c@{}}

      \multicolumn{6}{c}{\includegraphics[width=80mm,trim=30 0 20 0 ,clip=true]{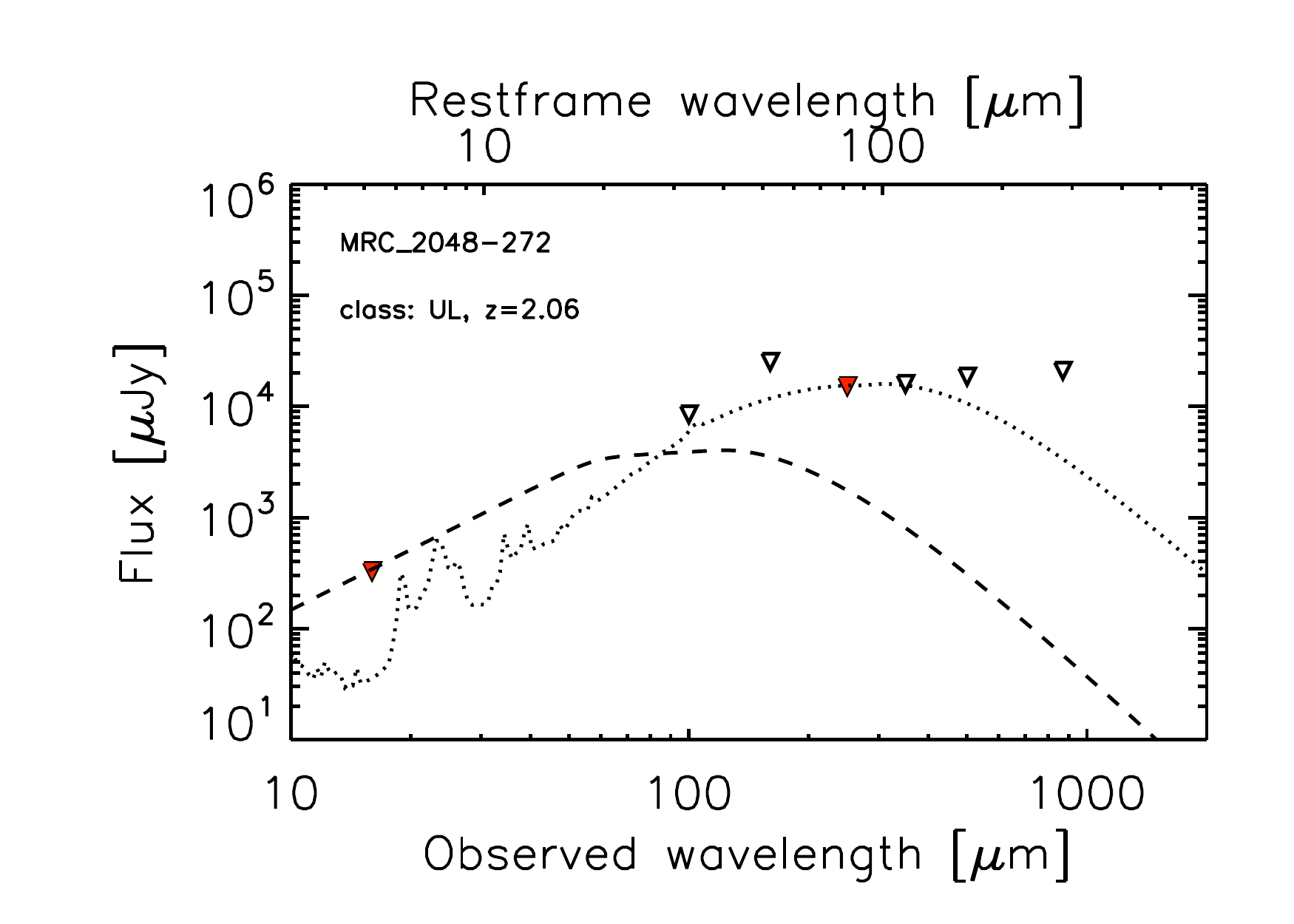}} \\
      24\mum & 100\mum & 160\mum & 250\mum & 350\mum & 500\mum \\
      \includegraphics[height=14mm,trim=70  -4 65 -10,clip=true]{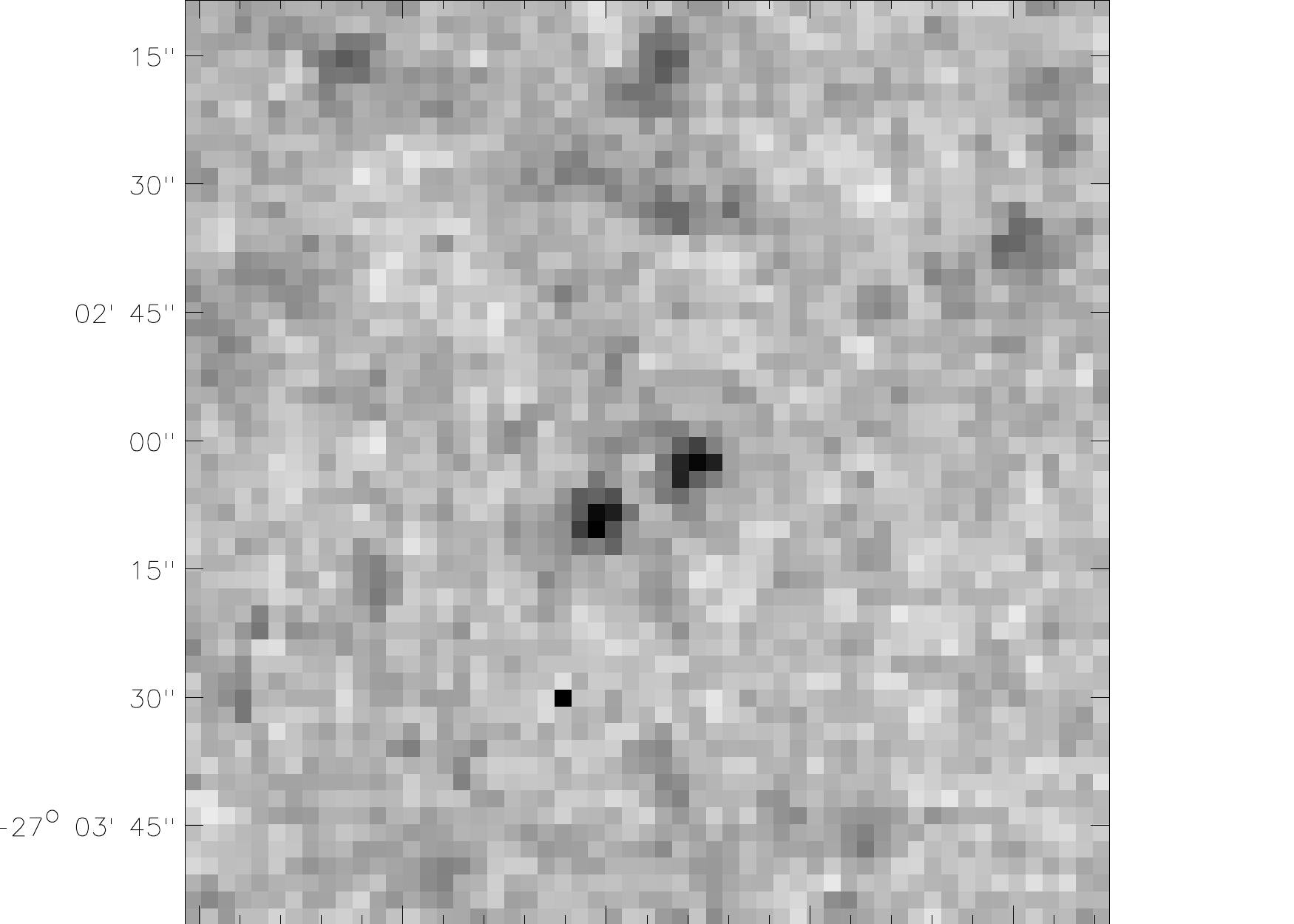} &
      \includegraphics[height=14mm,trim=70  -4 65 -10,clip=true]{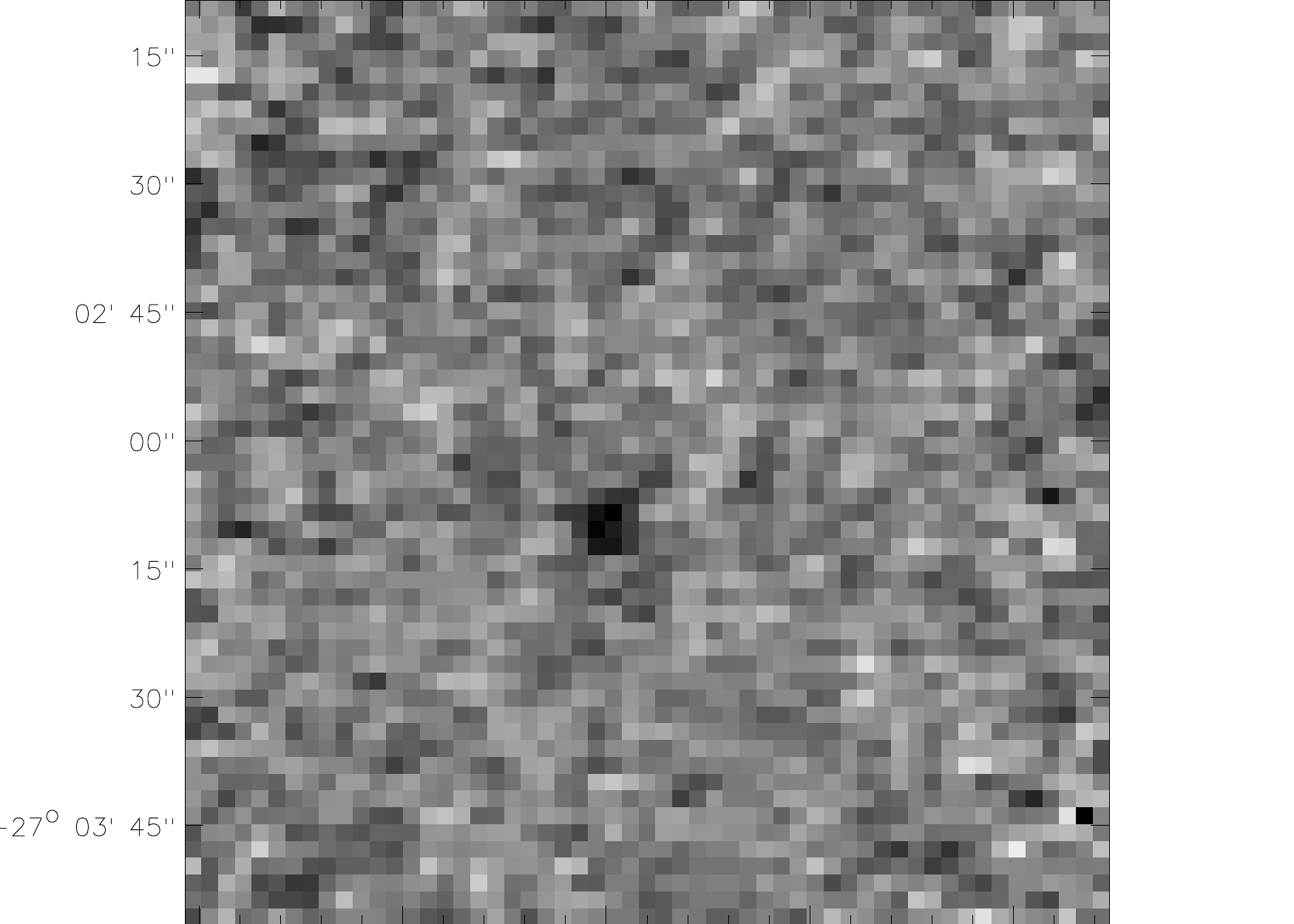} &
      \includegraphics[height=14mm,trim=70  -4 65 -10,clip=true]{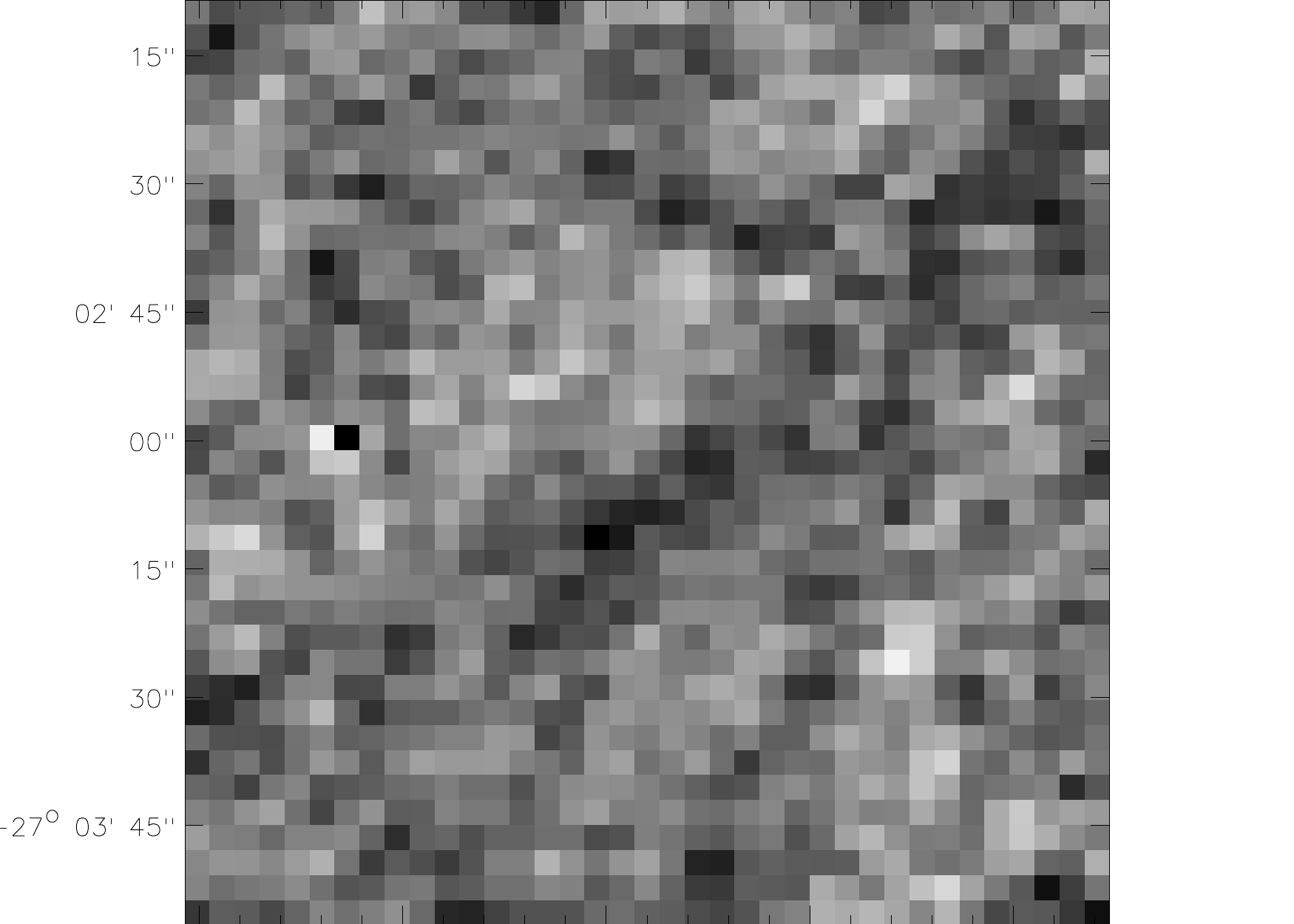} &
      \includegraphics[height=14mm,trim=70  -4 65 -10,clip=true]{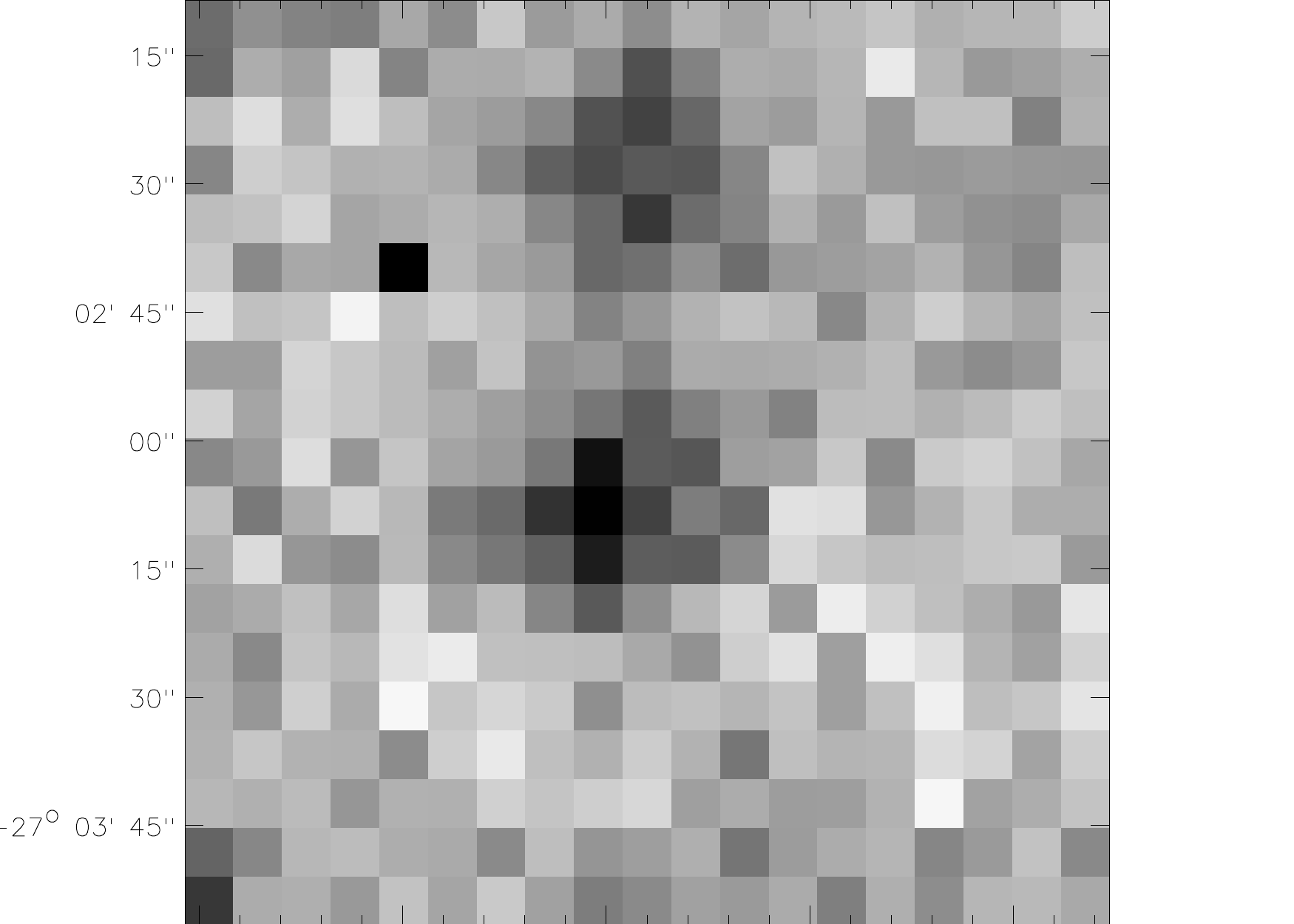}   &
      \includegraphics[height=14mm,trim=70  -4 65 -10,clip=true]{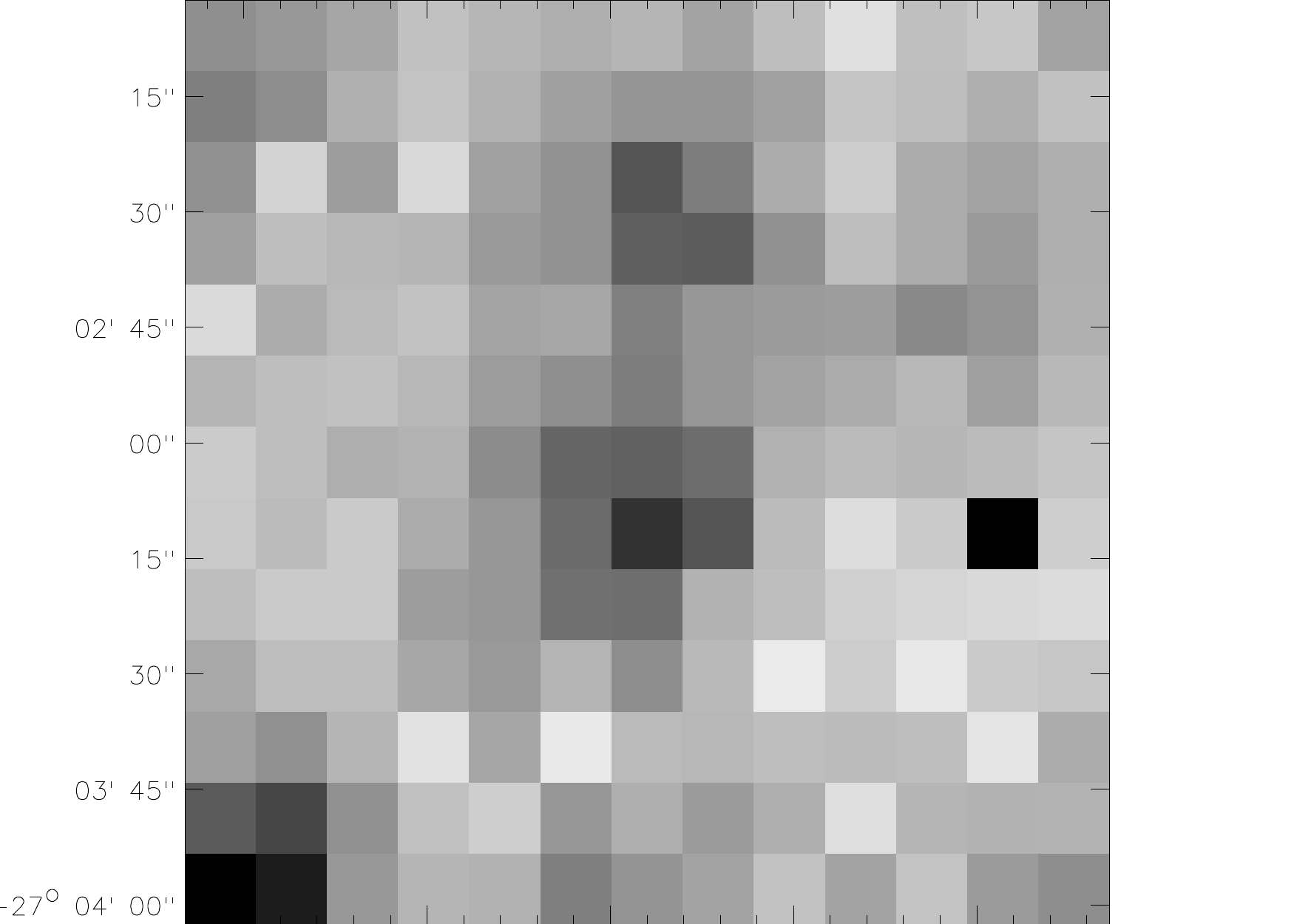}   &
      \includegraphics[height=14mm,trim=70  -4 65 -10,clip=true]{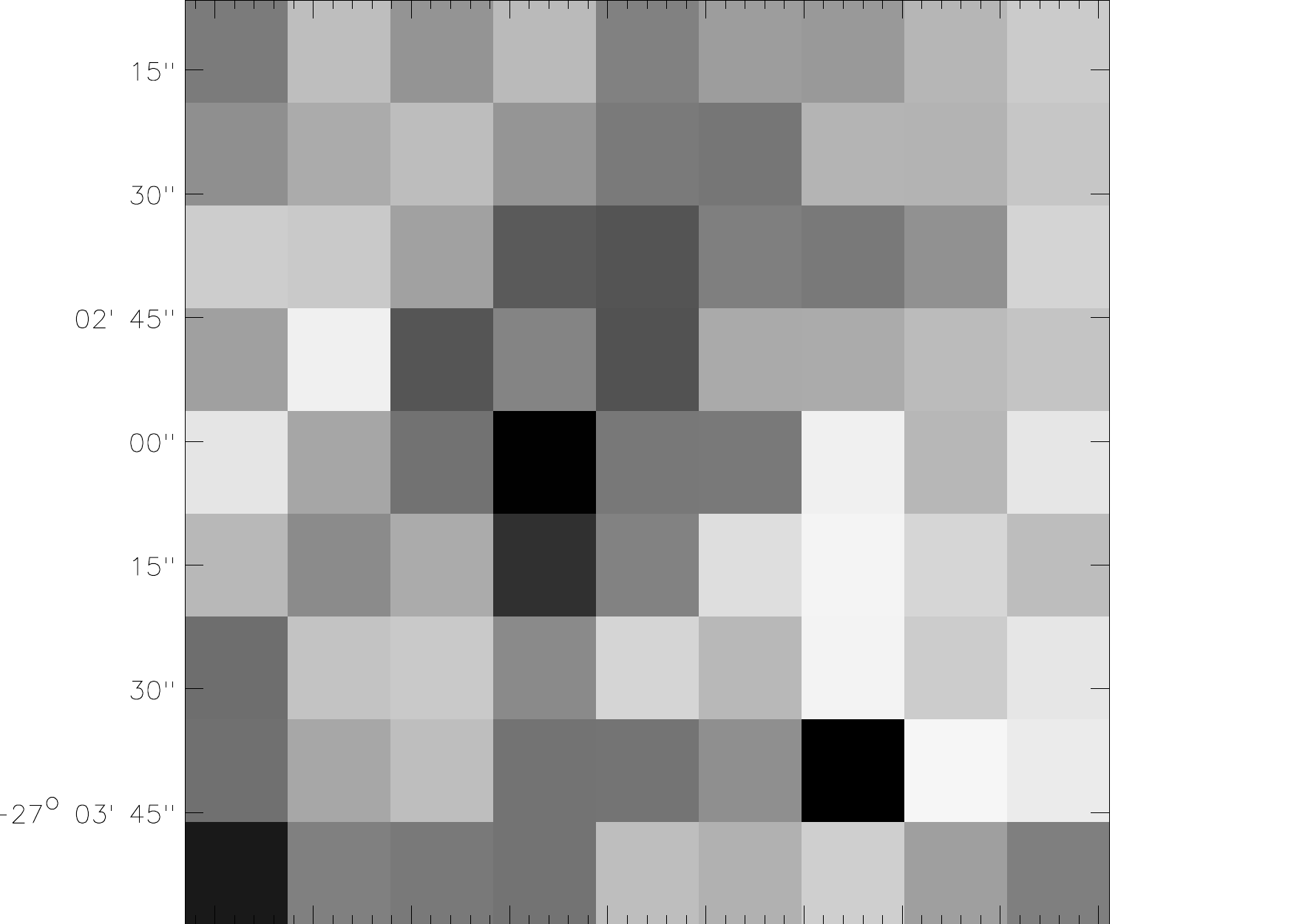} \\

      \multicolumn{6}{c}{\includegraphics[width=80mm,trim=30 0 20 0 ,clip=true]{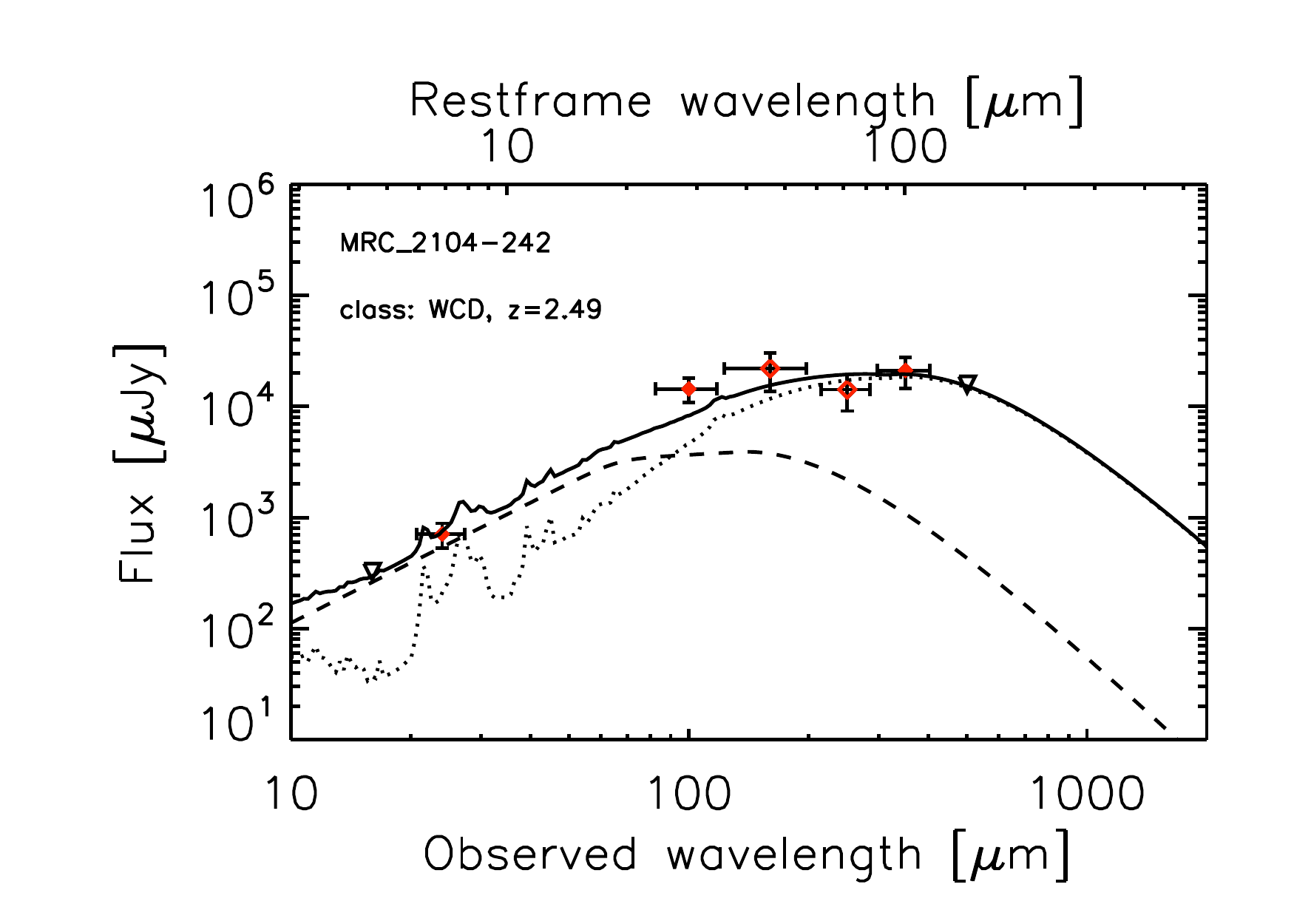}} \\
      24\mum & 100\mum & 160\mum & 250\mum & 350\mum & 500\mum \\
      \includegraphics[height=14mm,trim=70  -4 65 -10,clip=true]{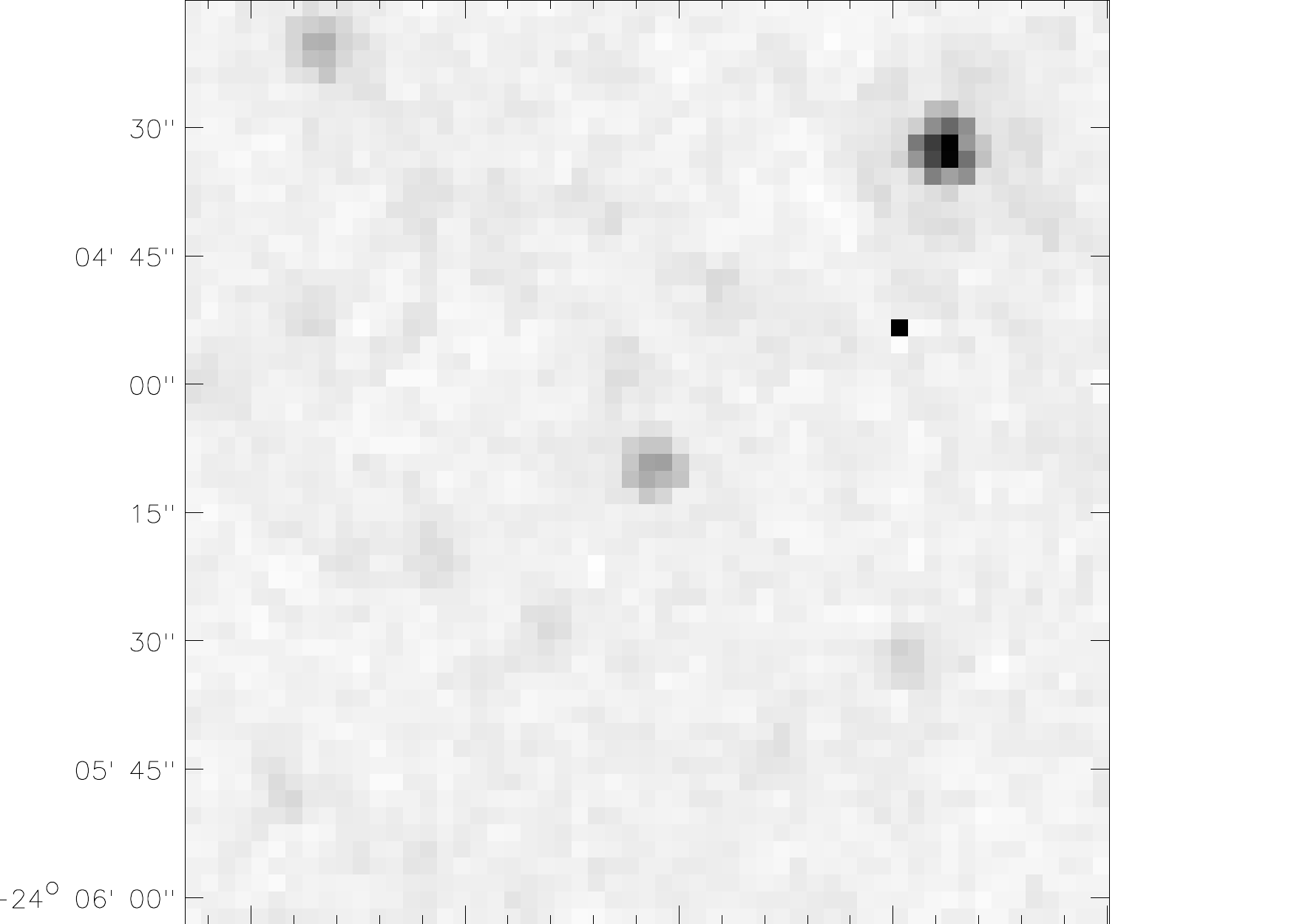} &
      \includegraphics[height=14mm,trim=70  -4 65 -10,clip=true]{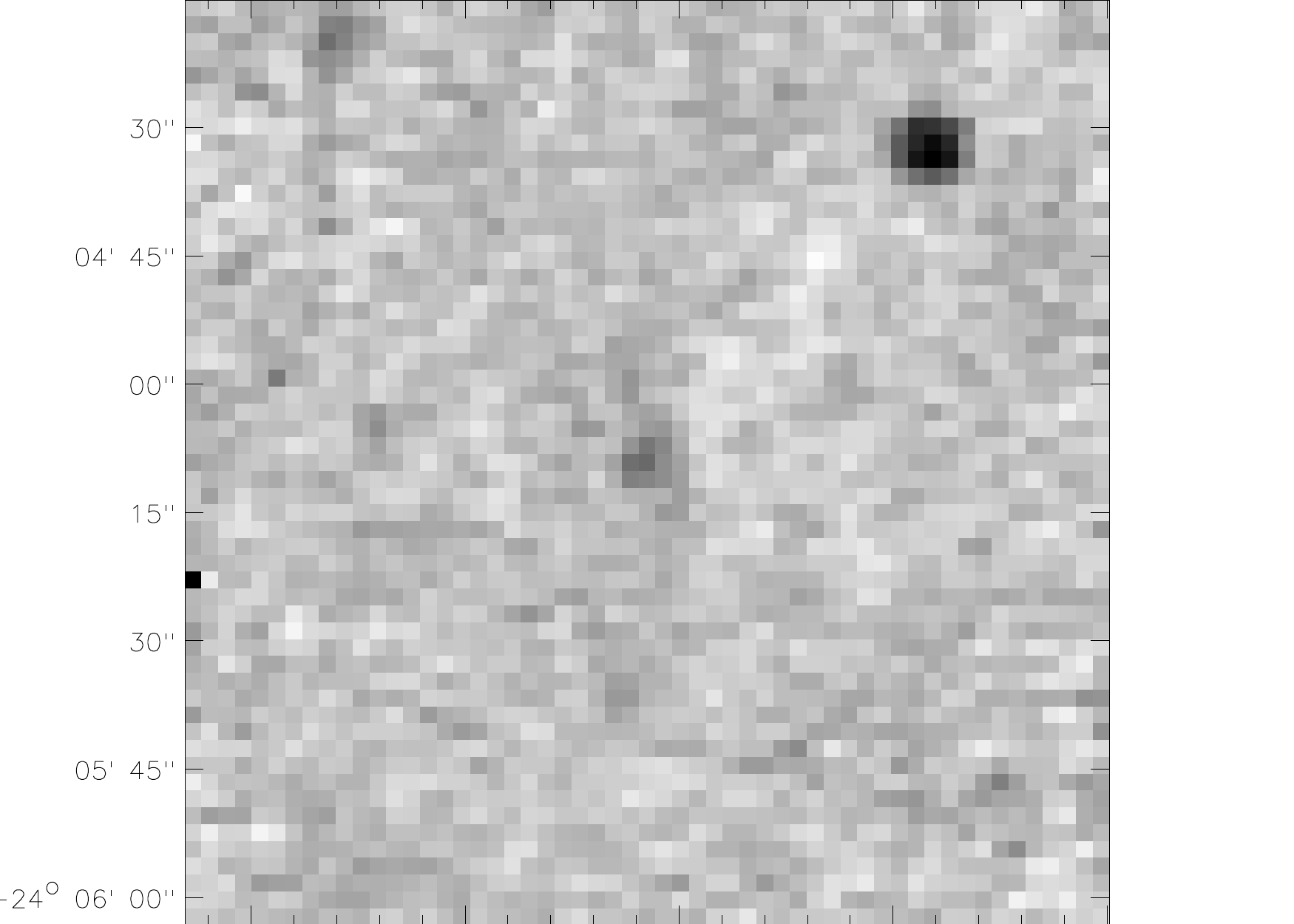} &
      \includegraphics[height=14mm,trim=70  -4 65 -10,clip=true]{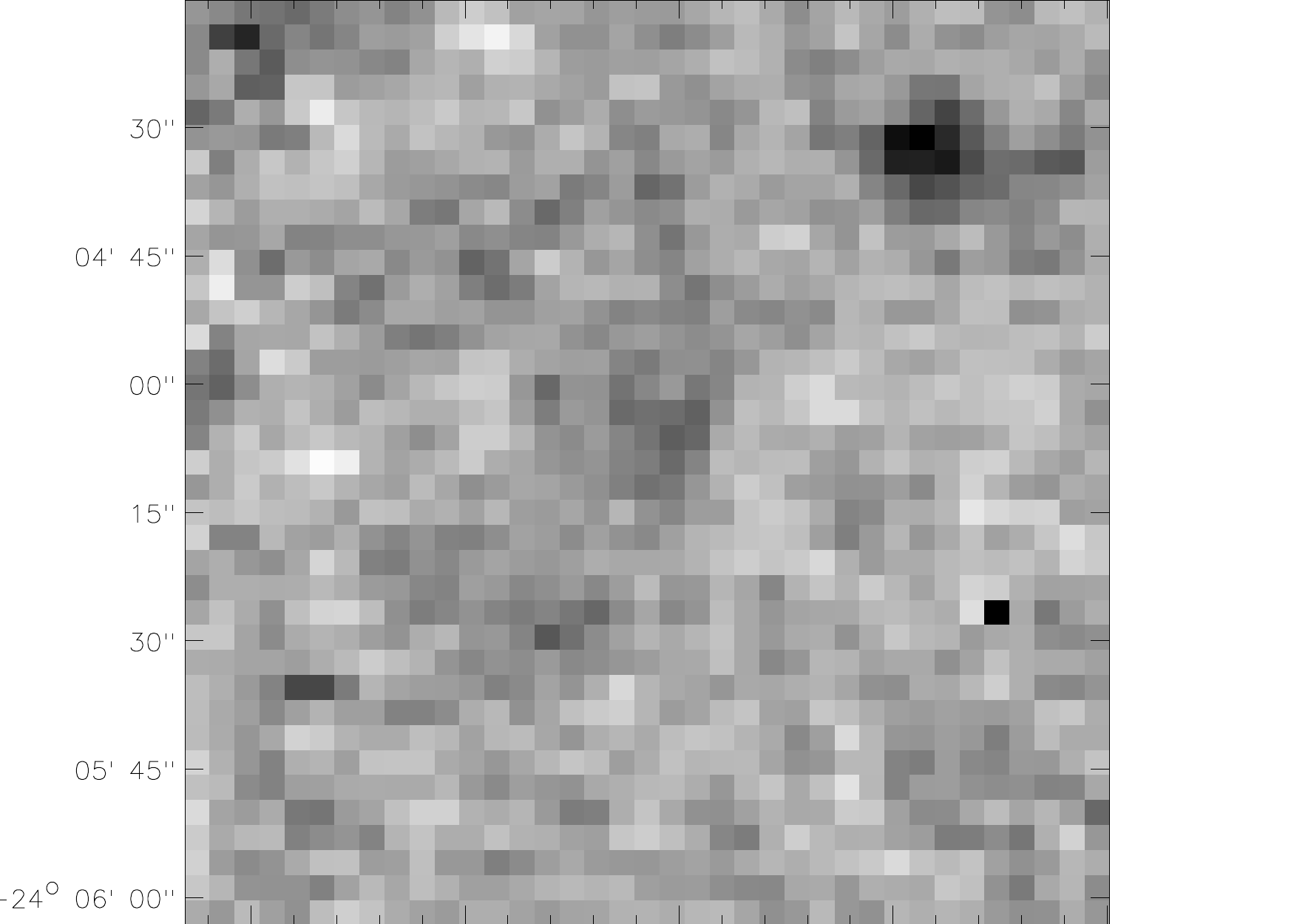} &
      \includegraphics[height=14mm,trim=70  -4 65 -10,clip=true]{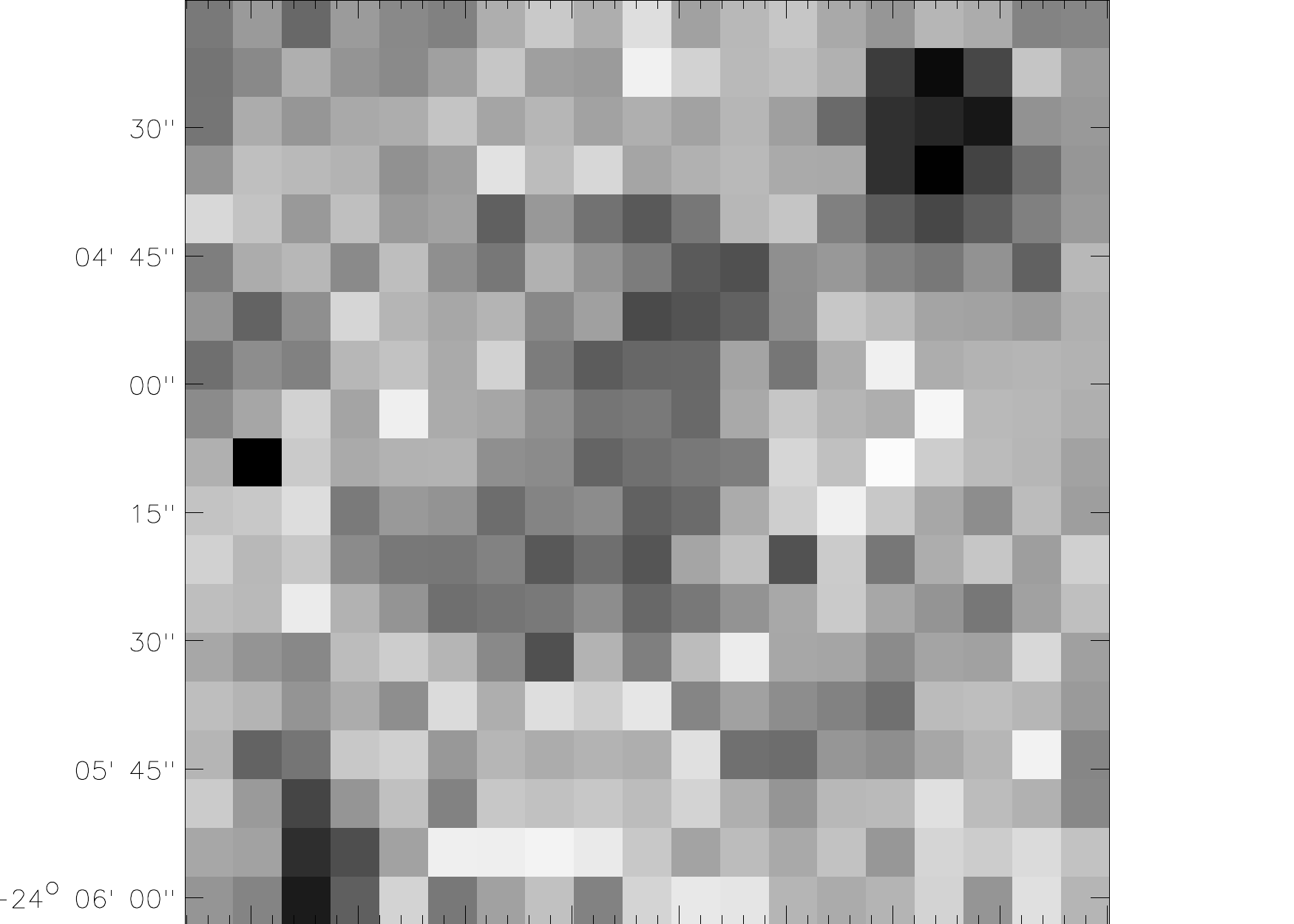}   &
      \includegraphics[height=14mm,trim=70  -4 65 -10,clip=true]{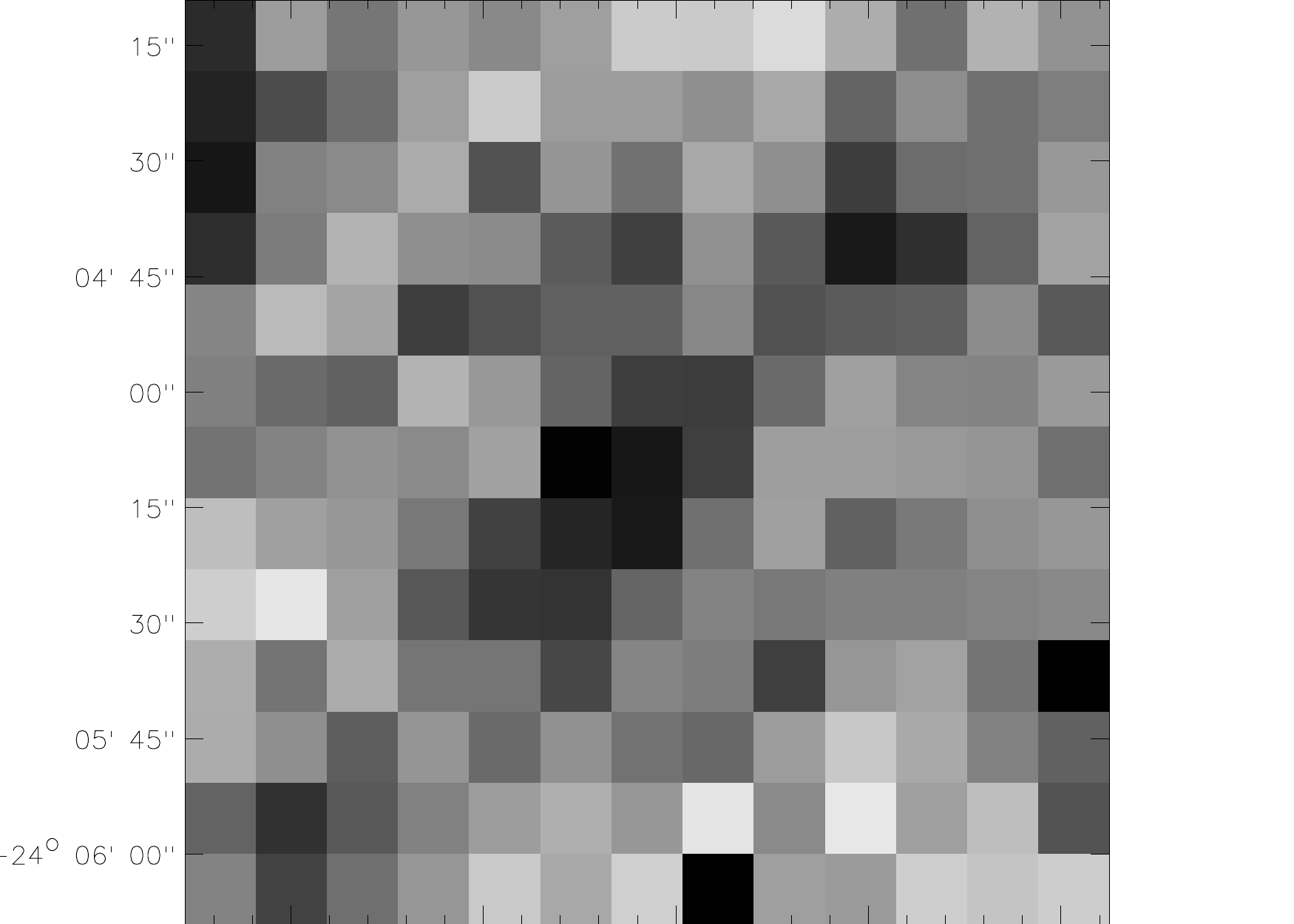}   &
      \includegraphics[height=14mm,trim=70  -4 65 -10,clip=true]{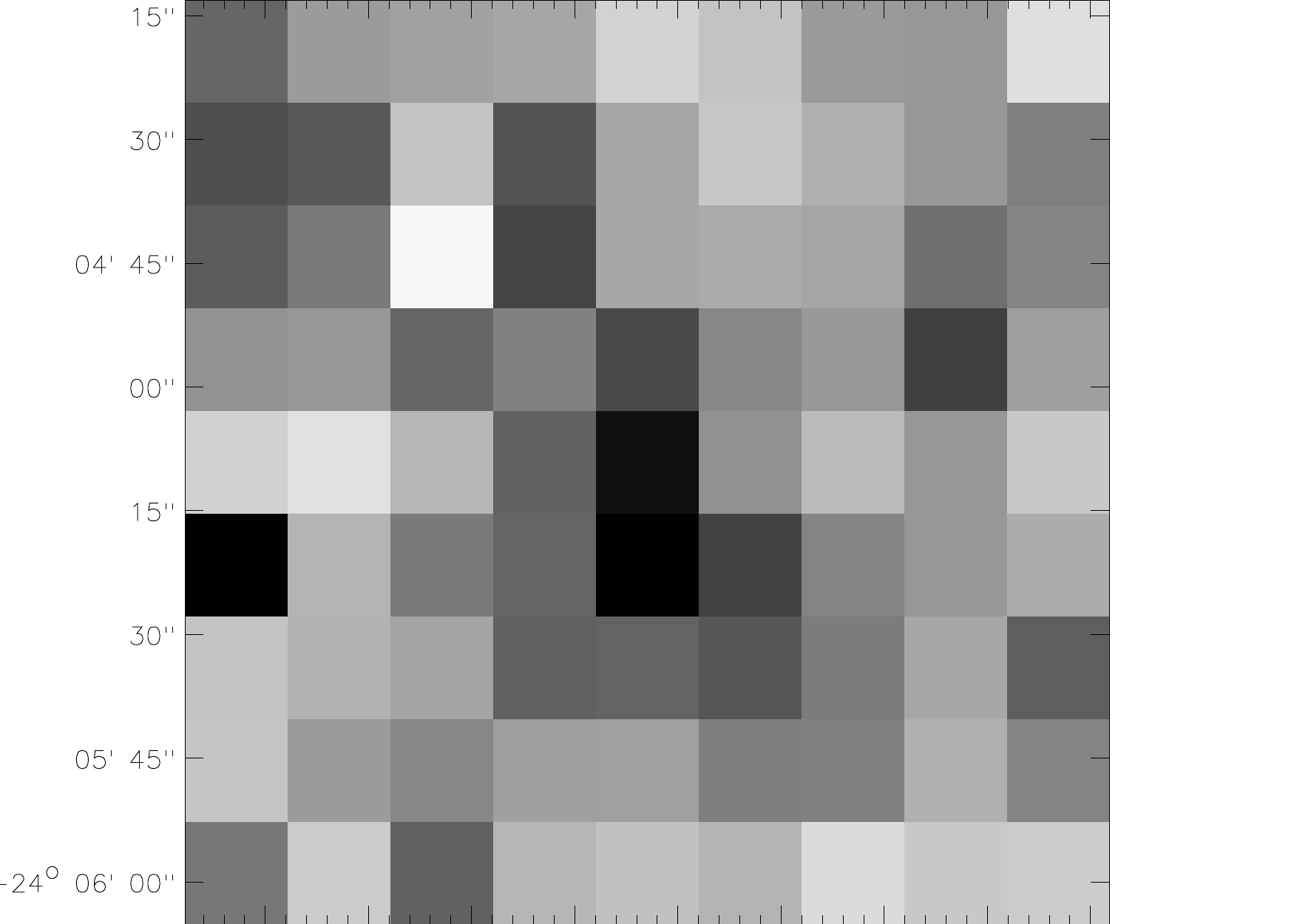} \\

      \multicolumn{6}{c}{\includegraphics[width=80mm,trim=30 0 20 0 ,clip=true]{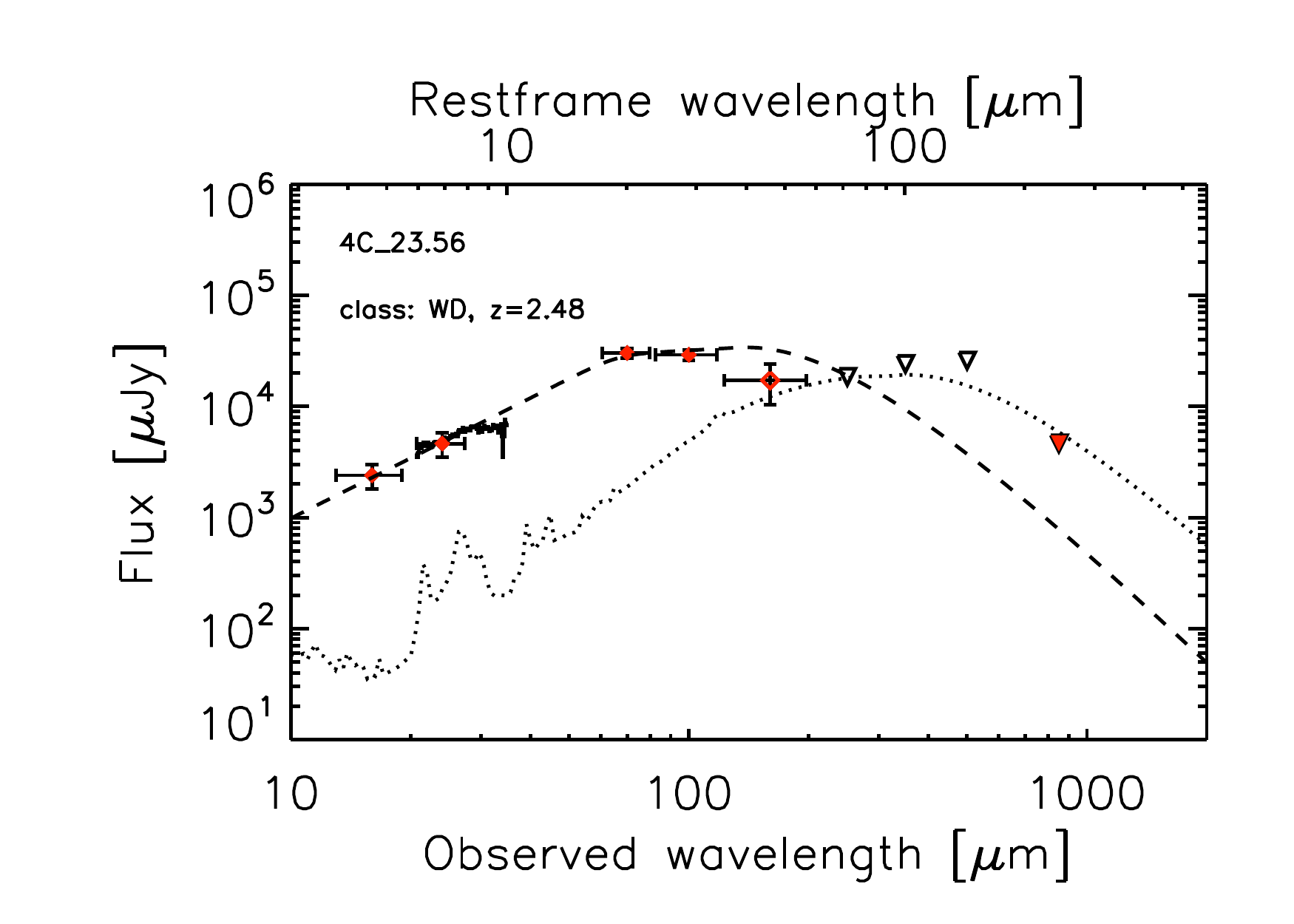}} \\
      24\mum & 100\mum & 160\mum & 250\mum & 350\mum & 500\mum \\
\end{tabular}
          \end{minipage}
  \end{centering}
\end{figure*}

\clearpage

\begin{figure*}[h]
  \begin{centering}
    \begin{minipage}[c]{0.46\linewidth}
    \begin{tabular}{c@{}c@{}c@{}c@{}c@{}c@{}}

      \multicolumn{6}{c}{\includegraphics[width=80mm,trim=30 0 20 0 ,clip=true]{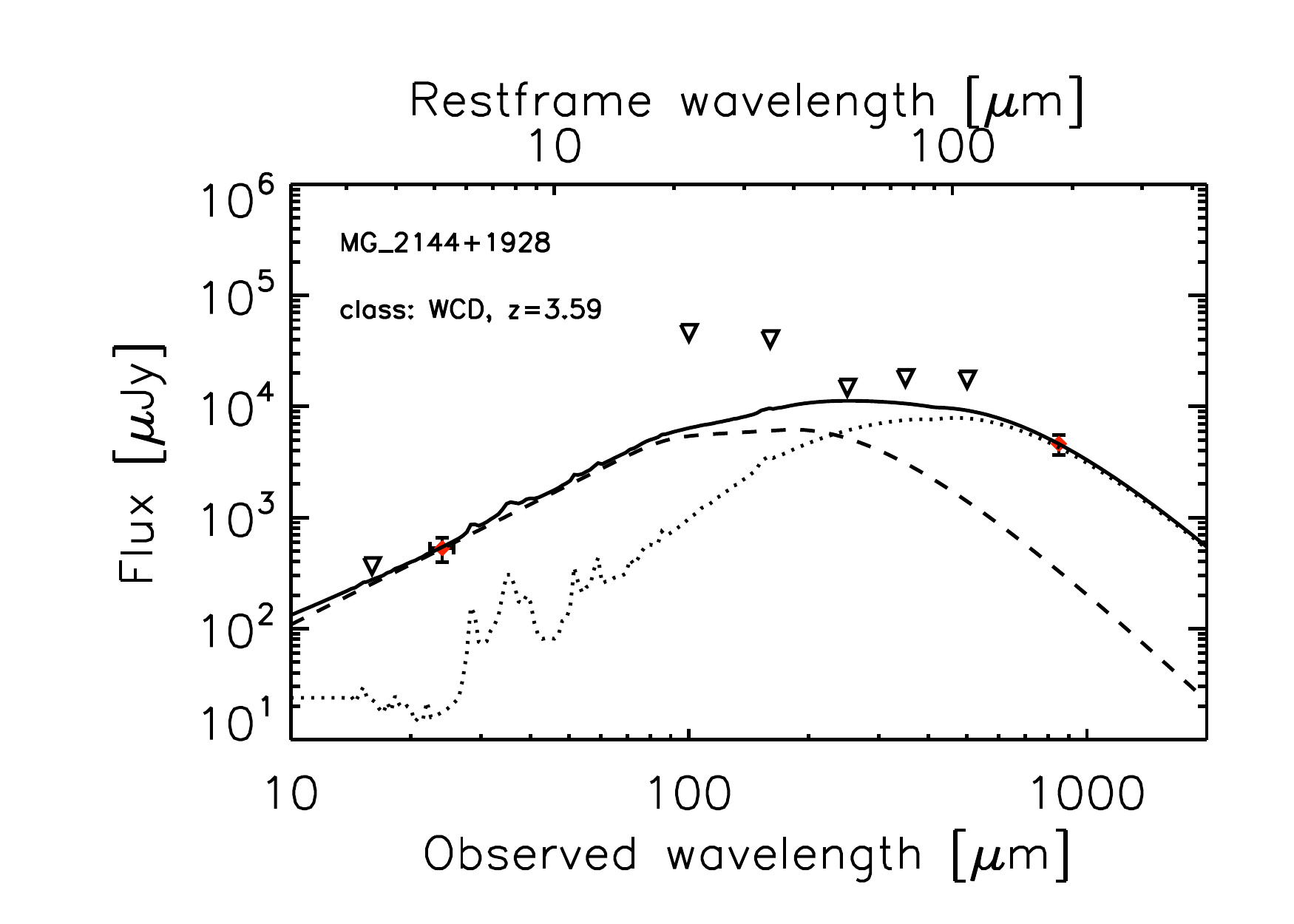}} \\
      24\mum & 100\mum & 160\mum & 250\mum & 350\mum & 500\mum \\
      \includegraphics[height=14mm,trim=70  -4 65 -10,clip=true]{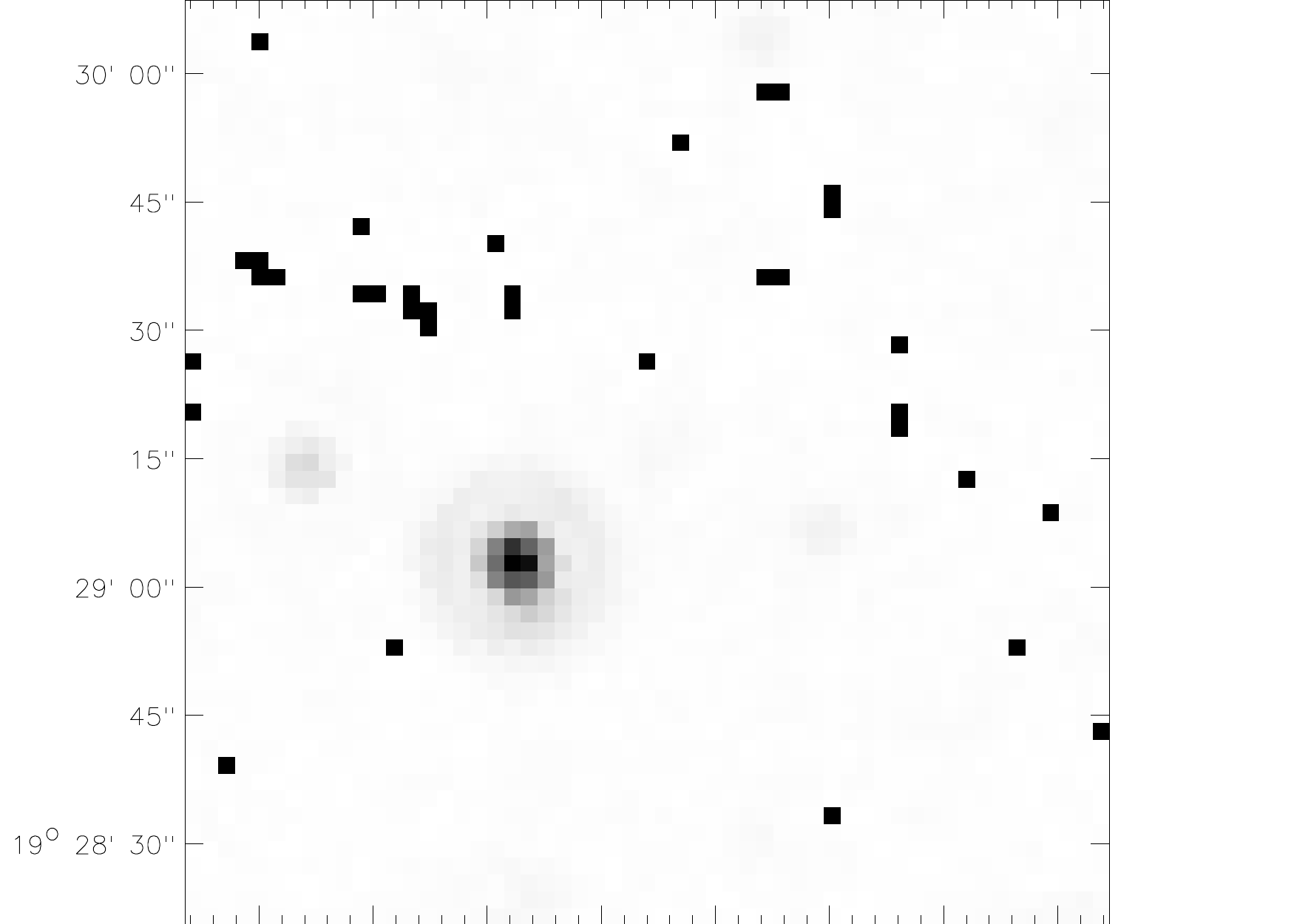} &
      \includegraphics[height=14mm,trim=70  -4 65 -10,clip=true]{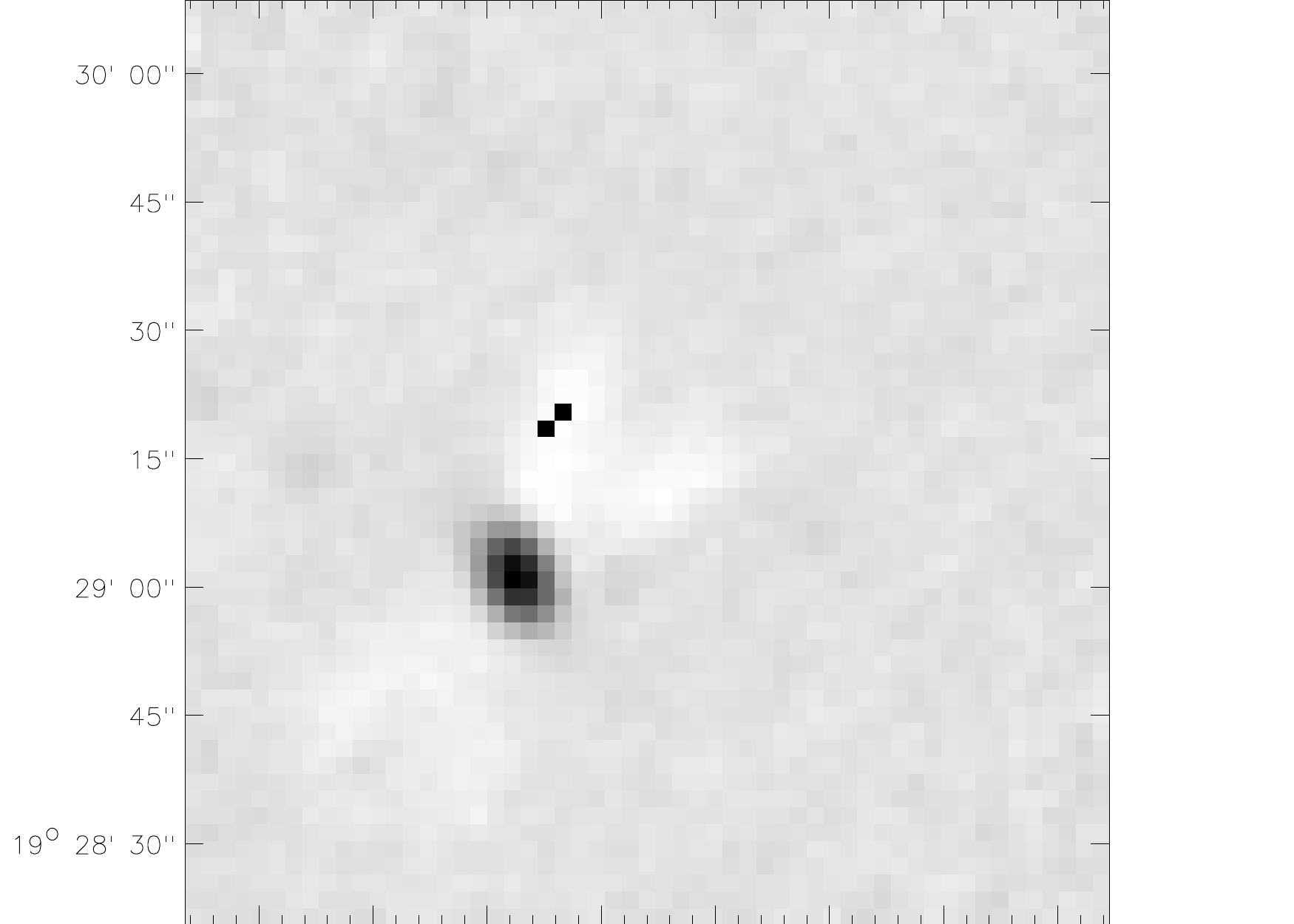} &
      \includegraphics[height=14mm,trim=70  -4 65 -10,clip=true]{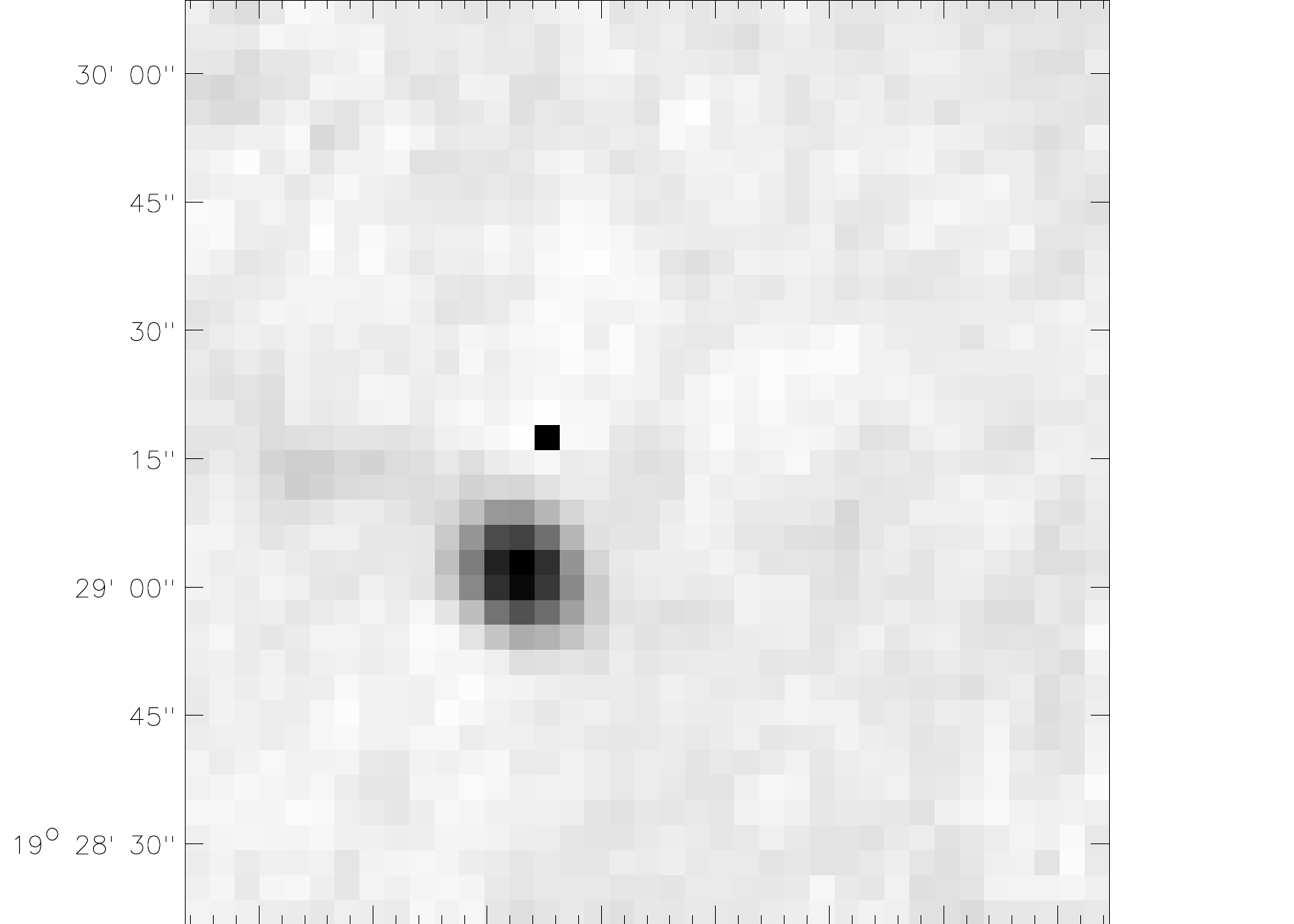} &
      \includegraphics[height=14mm,trim=70  -4 65 -10,clip=true]{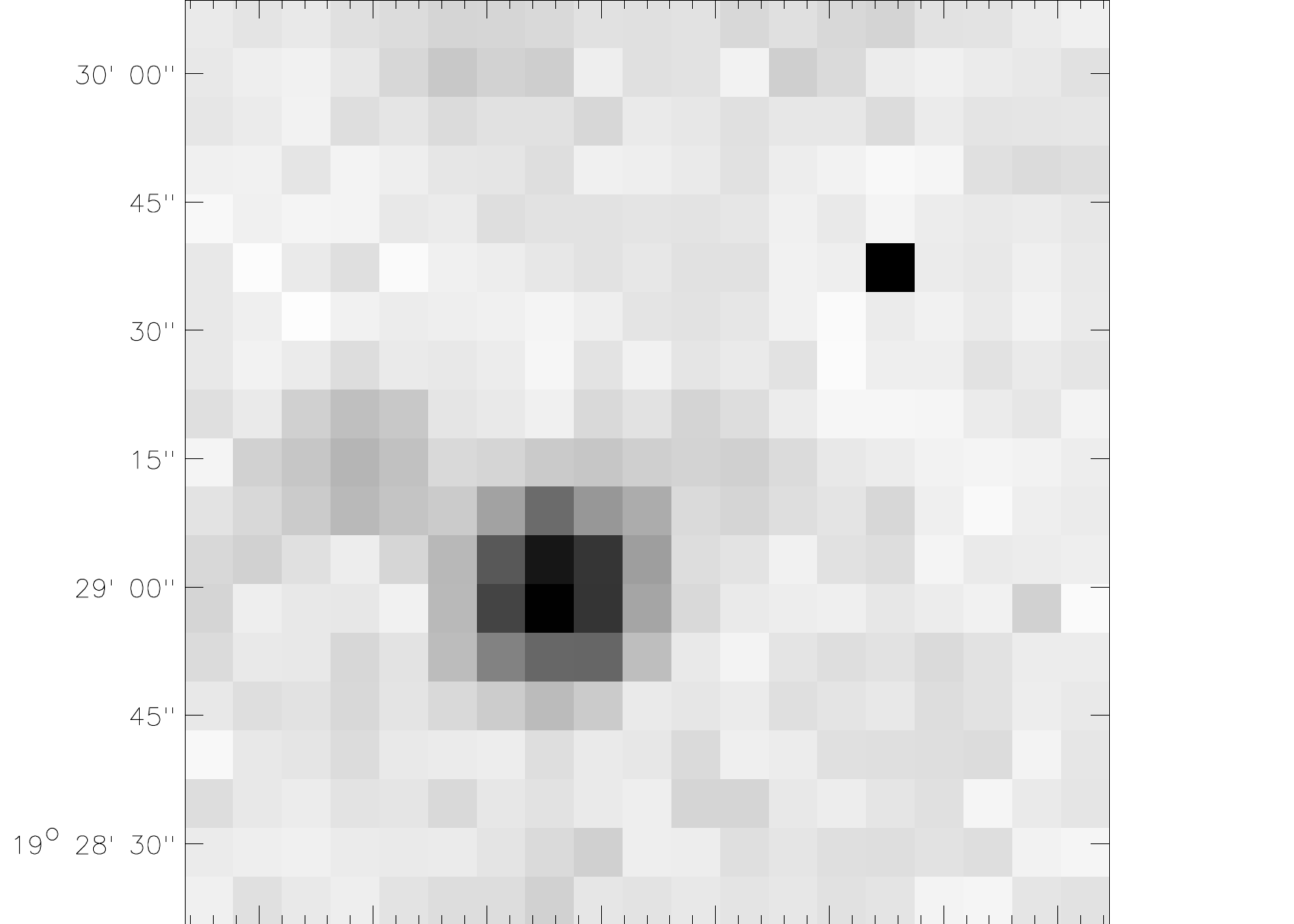}   &
      \includegraphics[height=14mm,trim=70  -4 65 -10,clip=true]{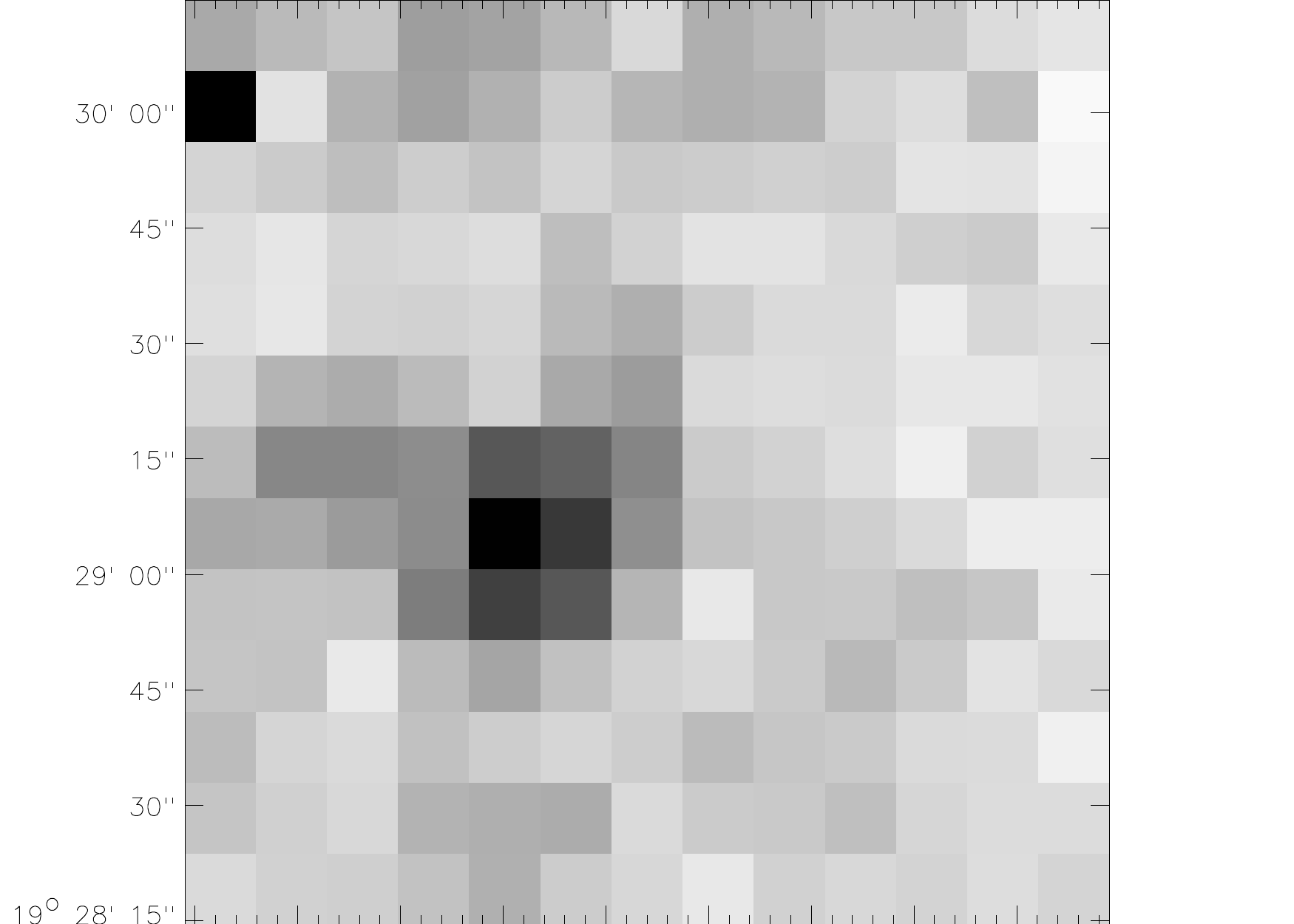}   &
      \includegraphics[height=14mm,trim=70  -4 65 -10,clip=true]{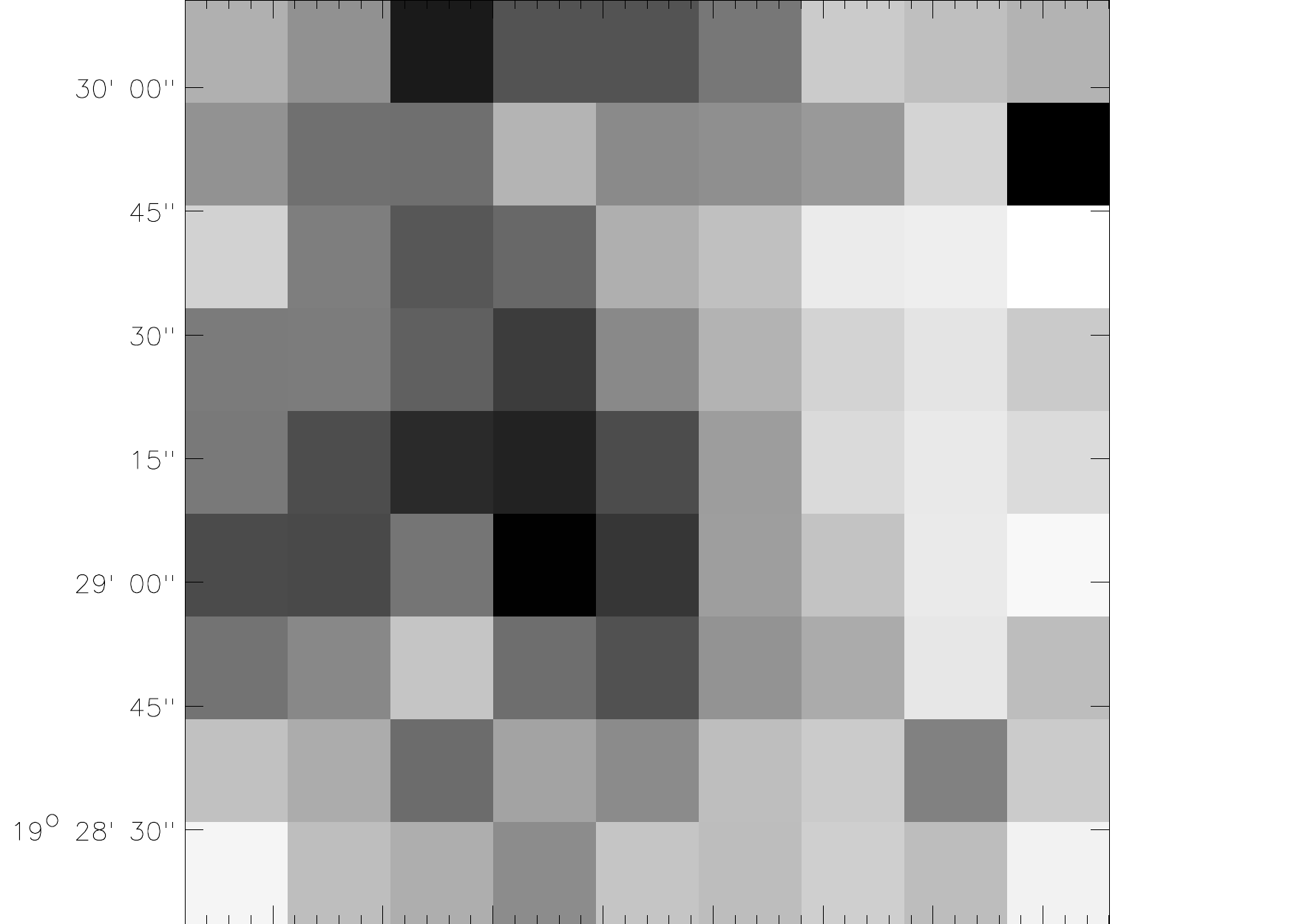} \\

      \multicolumn{6}{c}{\includegraphics[width=80mm,trim=30 0 20 0 ,clip=true]{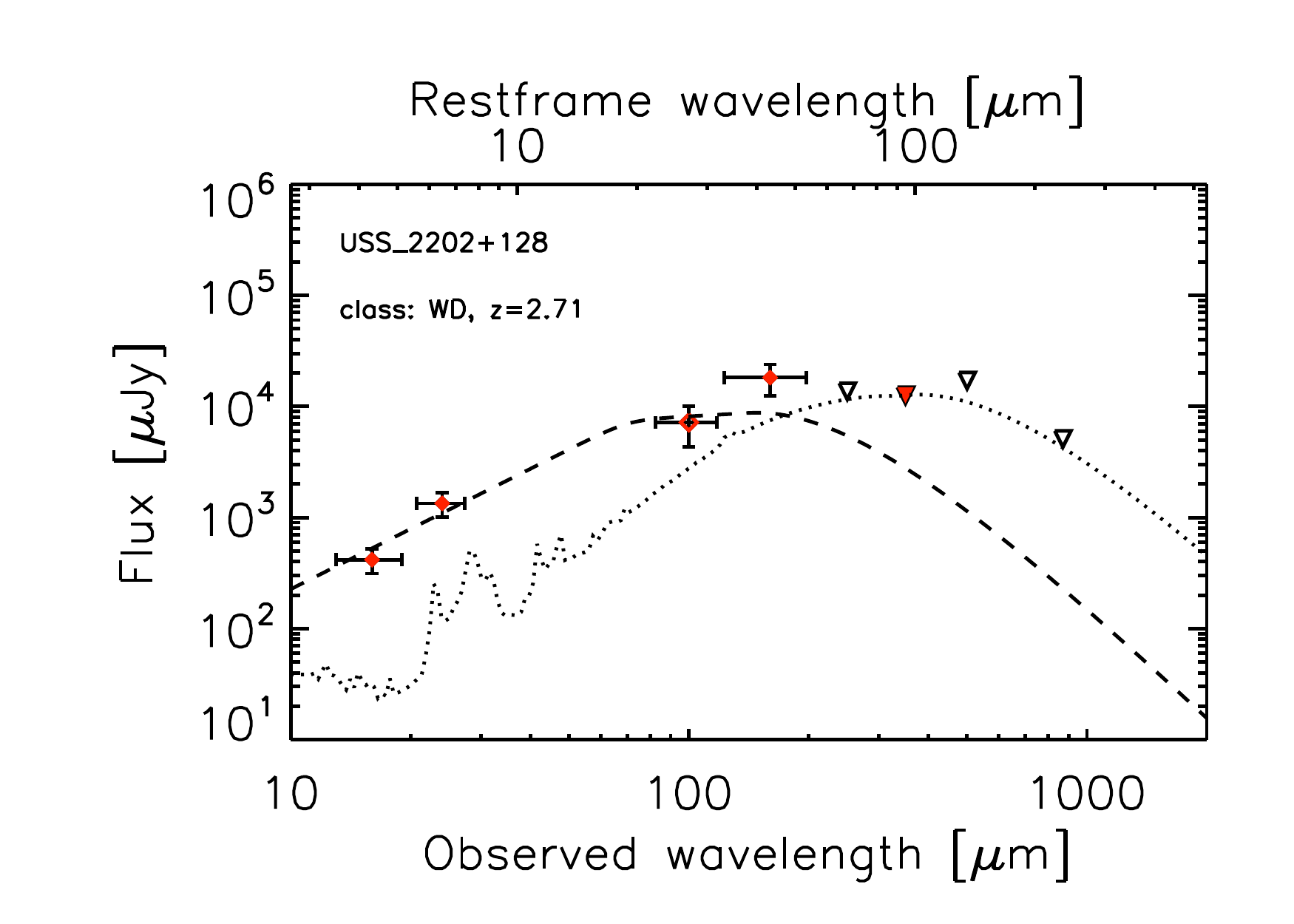}} \\
      24\mum & 100\mum & 160\mum & 250\mum & 350\mum & 500\mum \\
      \includegraphics[height=14mm,trim=70  -4 65 -10,clip=true]{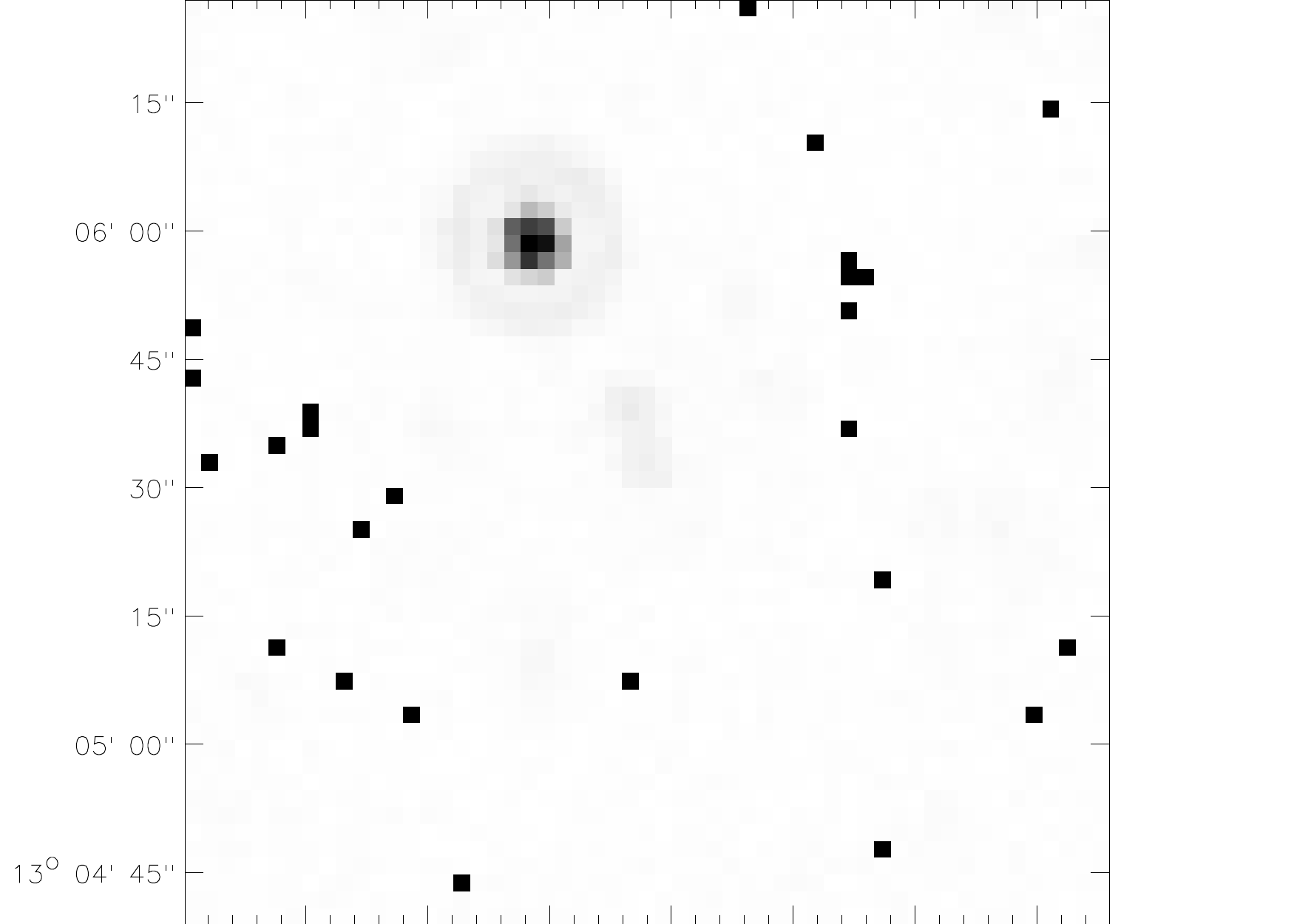} &
      \includegraphics[height=14mm,trim=70  -4 65 -10,clip=true]{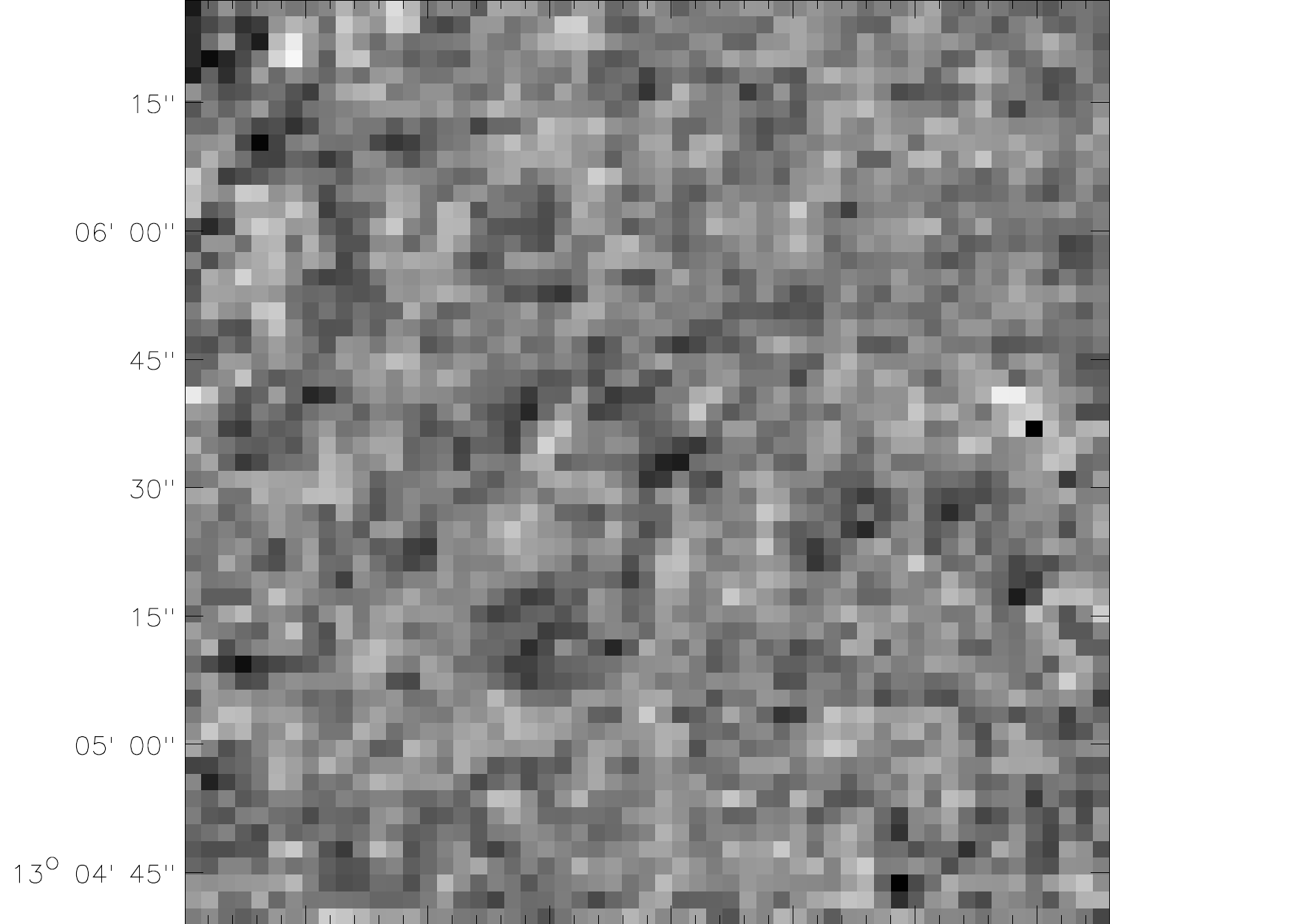} &
      \includegraphics[height=14mm,trim=70  -4 65 -10,clip=true]{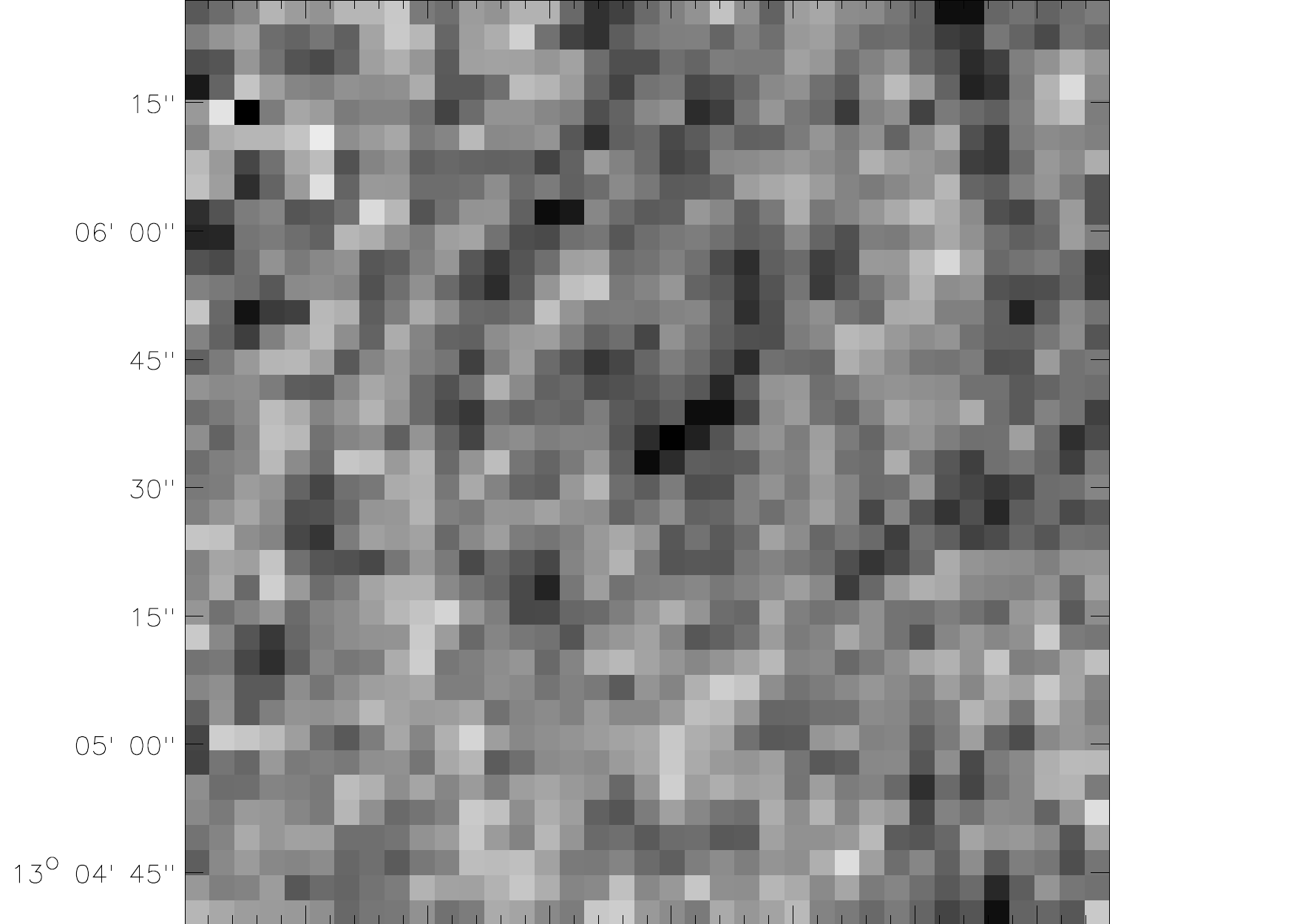} &
      \includegraphics[height=14mm,trim=70  -4 65 -10,clip=true]{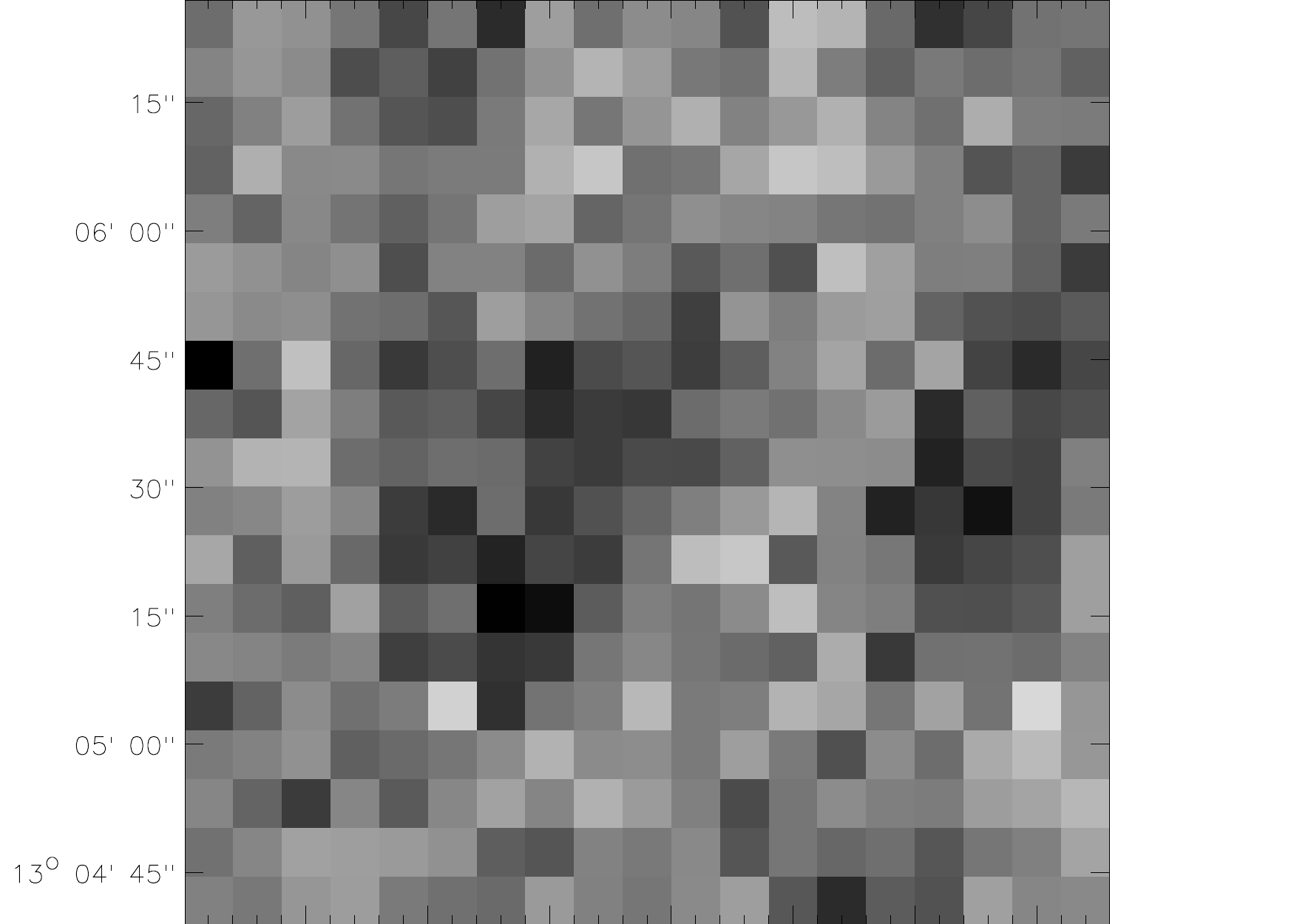}   &
      \includegraphics[height=14mm,trim=70  -4 65 -10,clip=true]{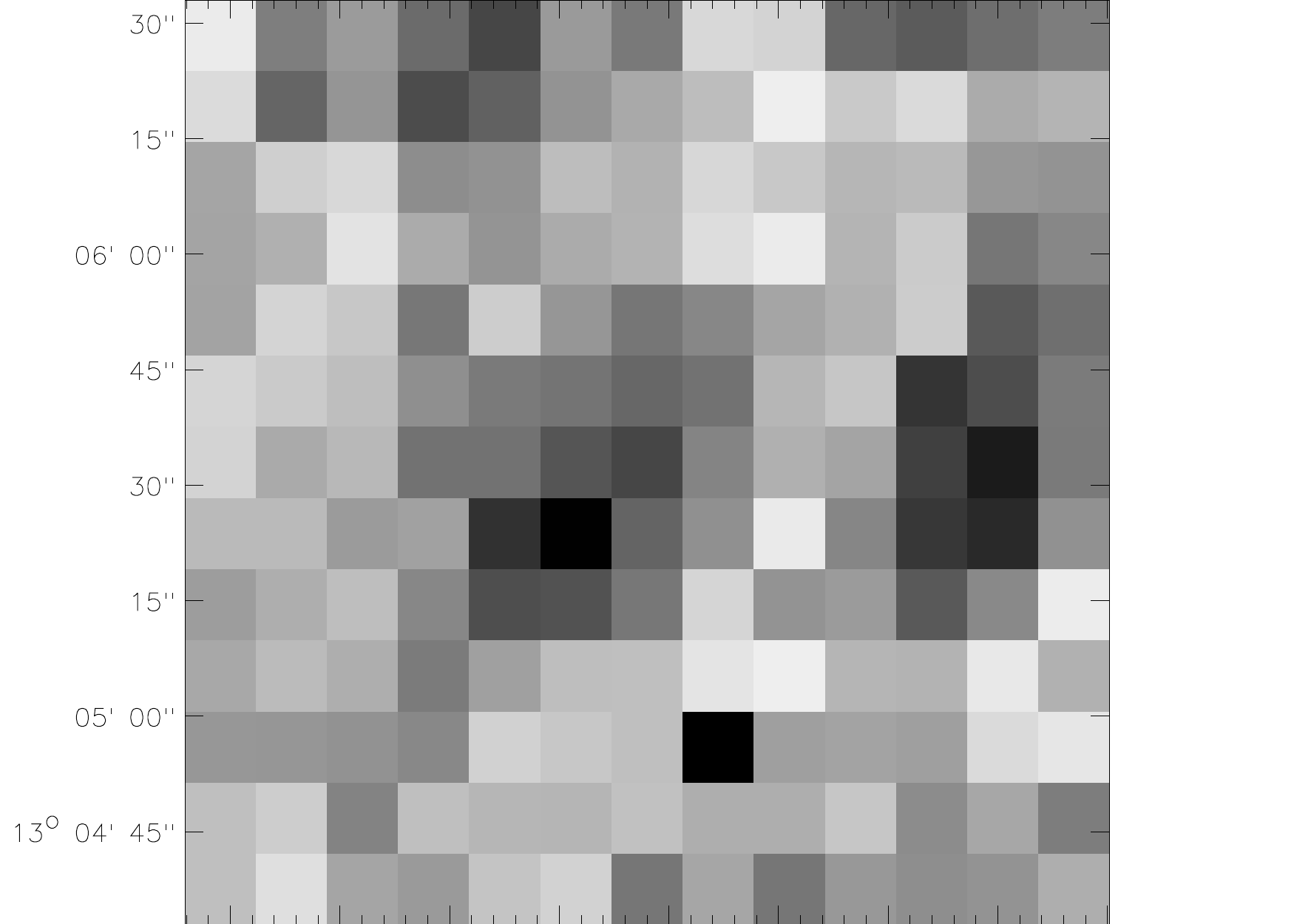}   &
      \includegraphics[height=14mm,trim=70  -4 65 -10,clip=true]{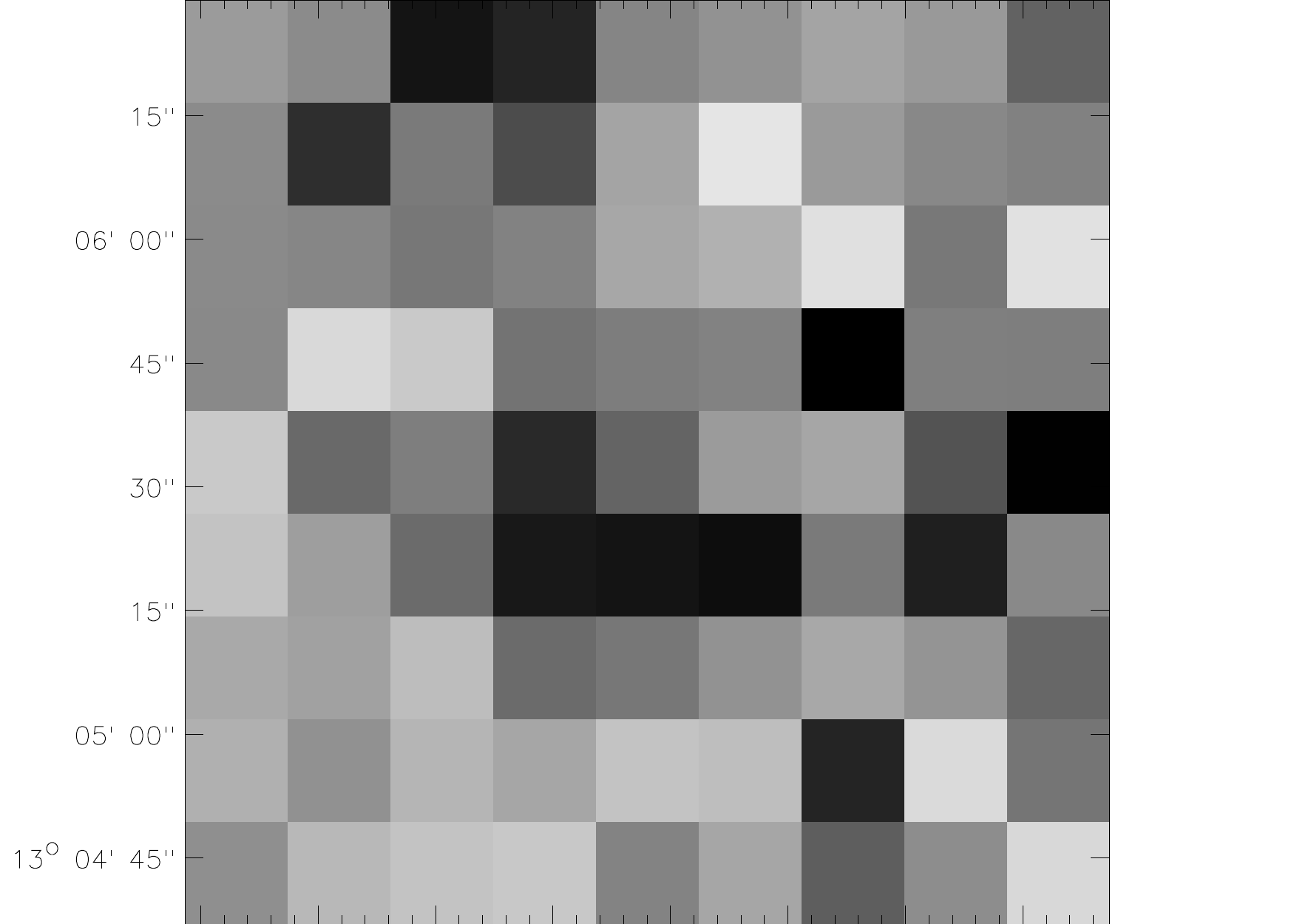} \\

      \multicolumn{6}{c}{\includegraphics[width=80mm,trim=30 0 20 0 ,clip=true]{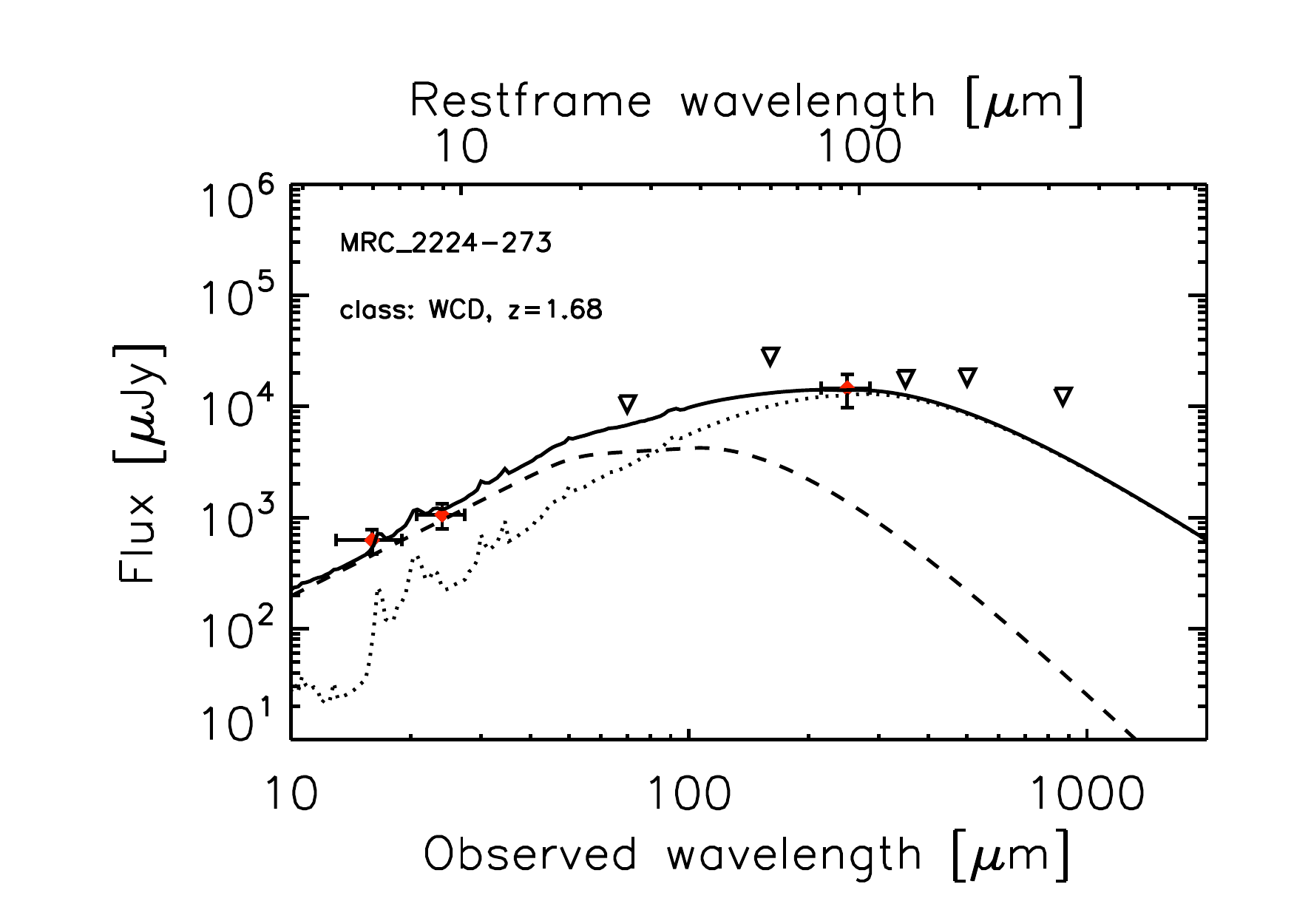}} \\
      24\mum & 70\mum & 160\mum & 250\mum & 350\mum & 500\mum \\
\end{tabular}
\end{minipage}
\begin{minipage}[c]{0.46\linewidth} 
\begin{tabular}{c@{}c@{}c@{}c@{}c@{}c@{}}

      \multicolumn{6}{c}{\includegraphics[width=80mm,trim=30 0 20 0 ,clip=true]{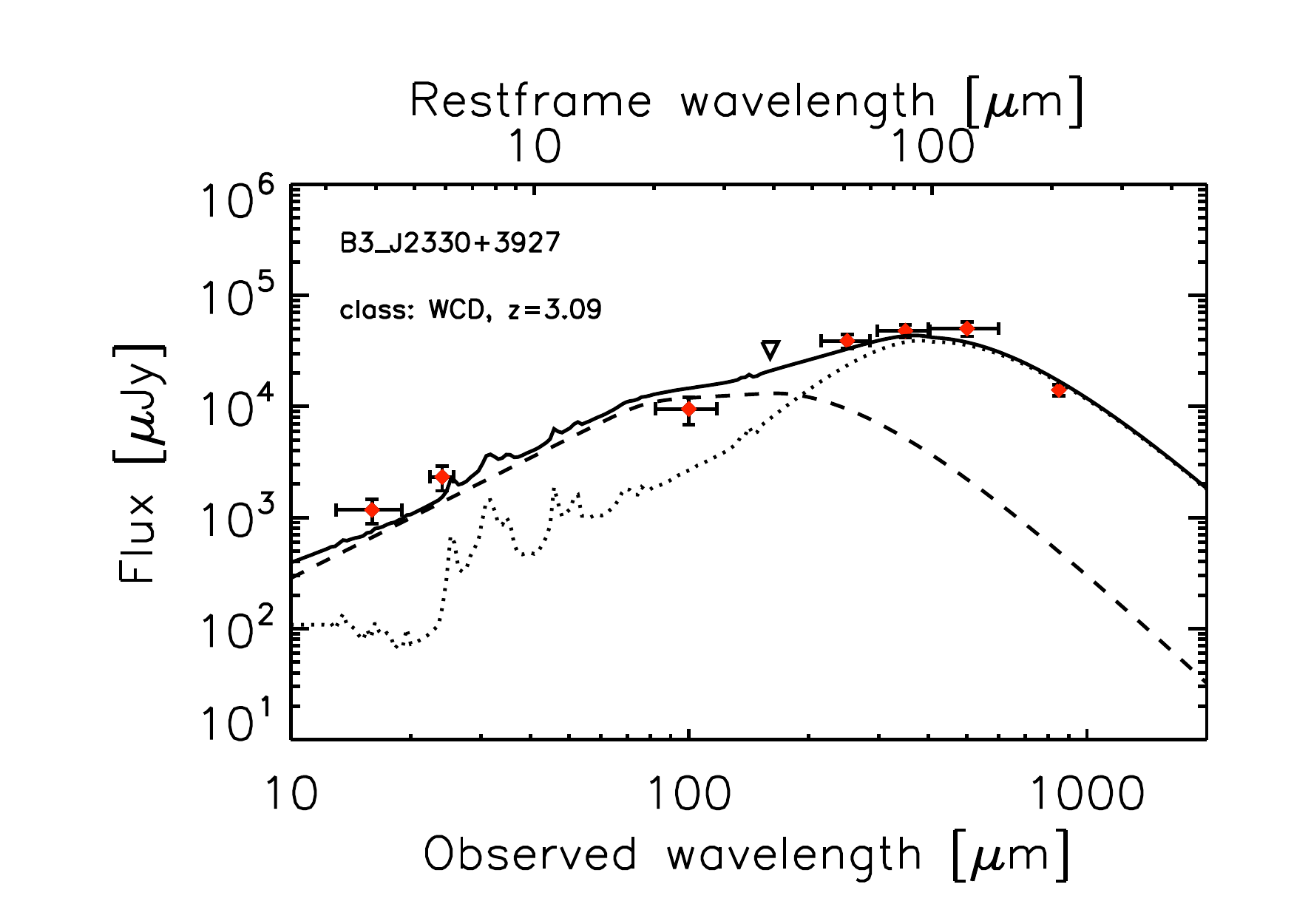}} \\
      24\mum & 100\mum & 160\mum & 250\mum & 350\mum & 500\mum \\
      \includegraphics[height=14mm,trim=70  -4 65 -10,clip=true]{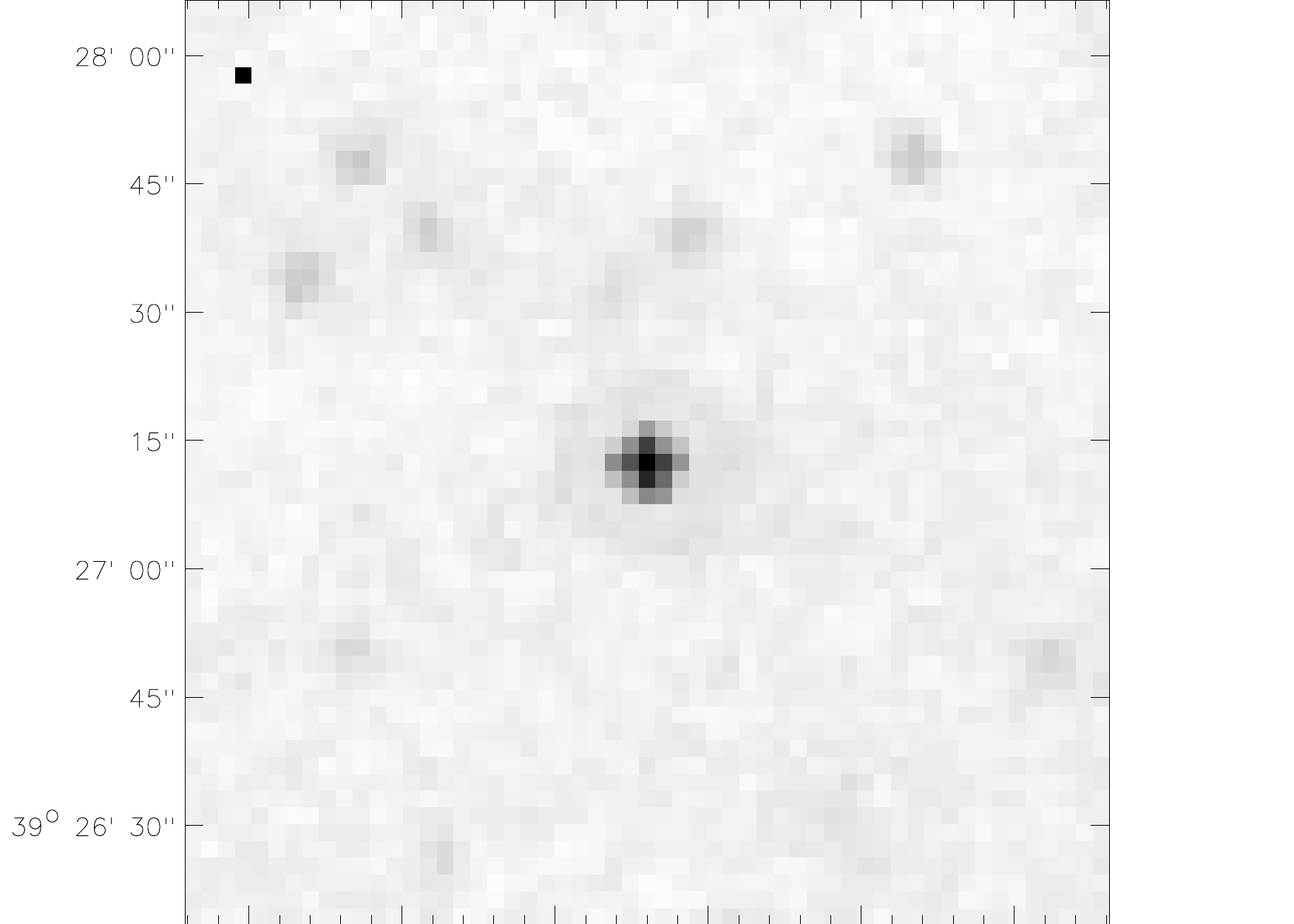} &
      \includegraphics[height=14mm,trim=70  -4 65 -10,clip=true]{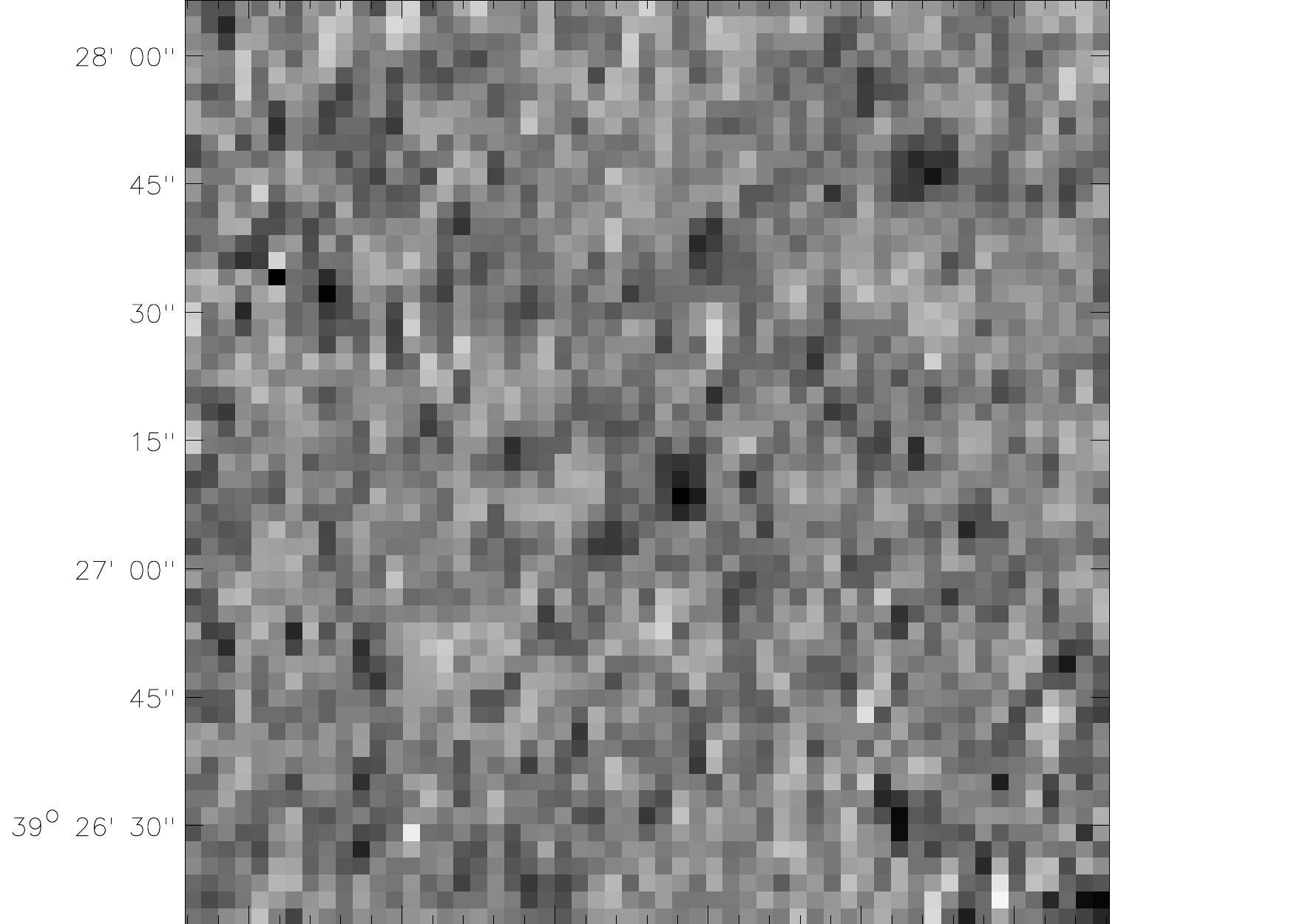} &
      \includegraphics[height=14mm,trim=70  -4 65 -10,clip=true]{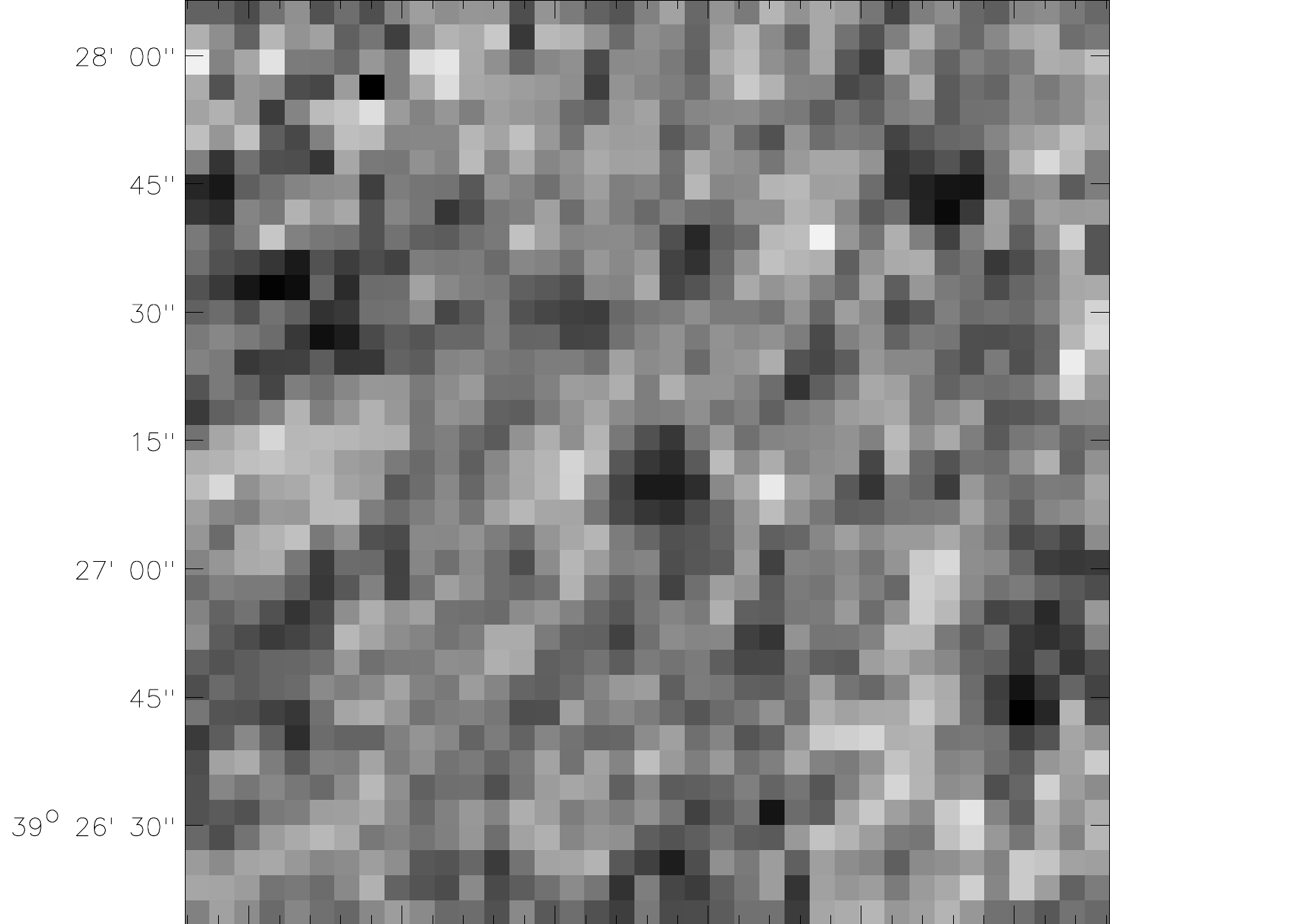} &
      \includegraphics[height=14mm,trim=70  -4 65 -10,clip=true]{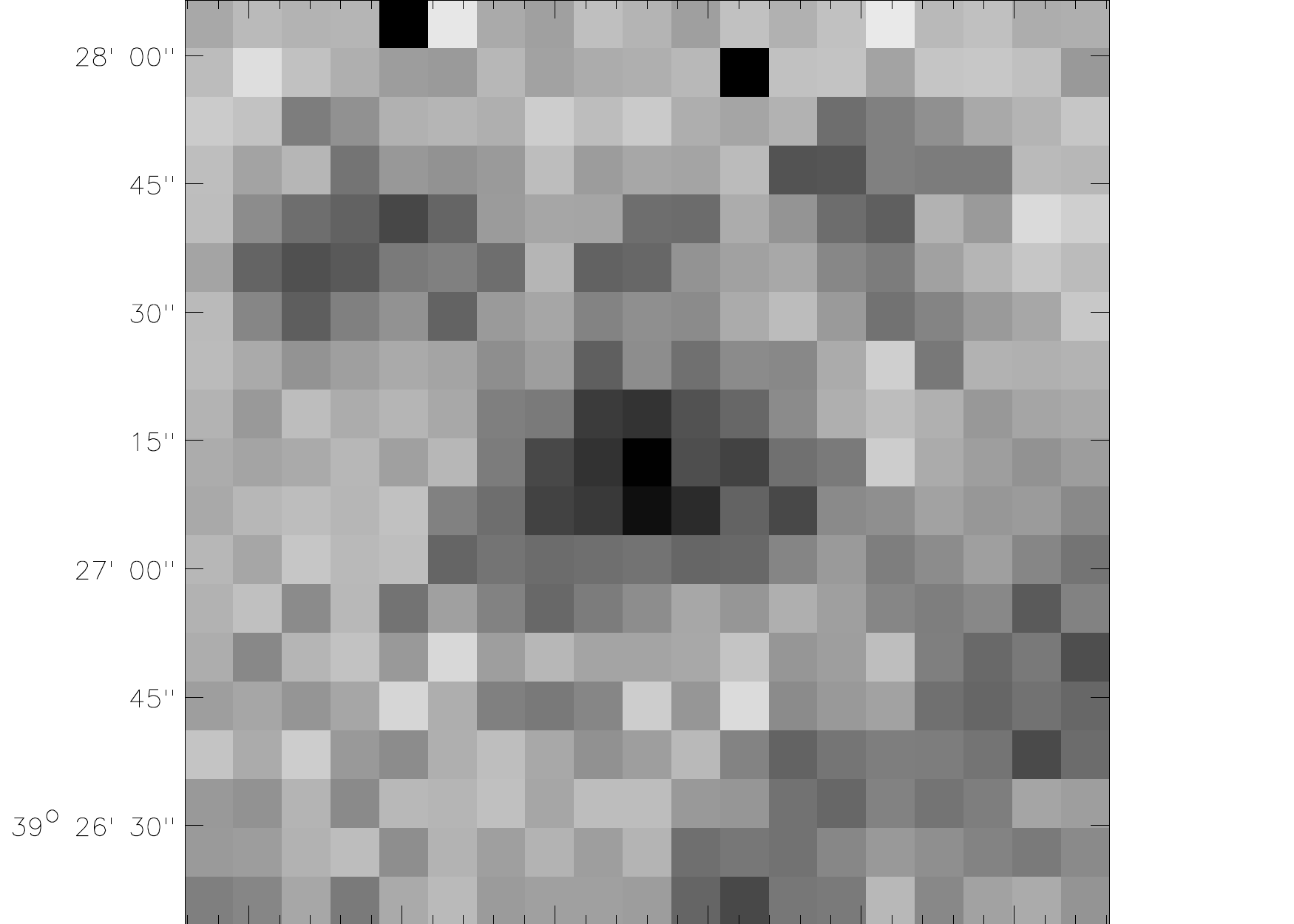}   &
      \includegraphics[height=14mm,trim=70  -4 65 -10,clip=true]{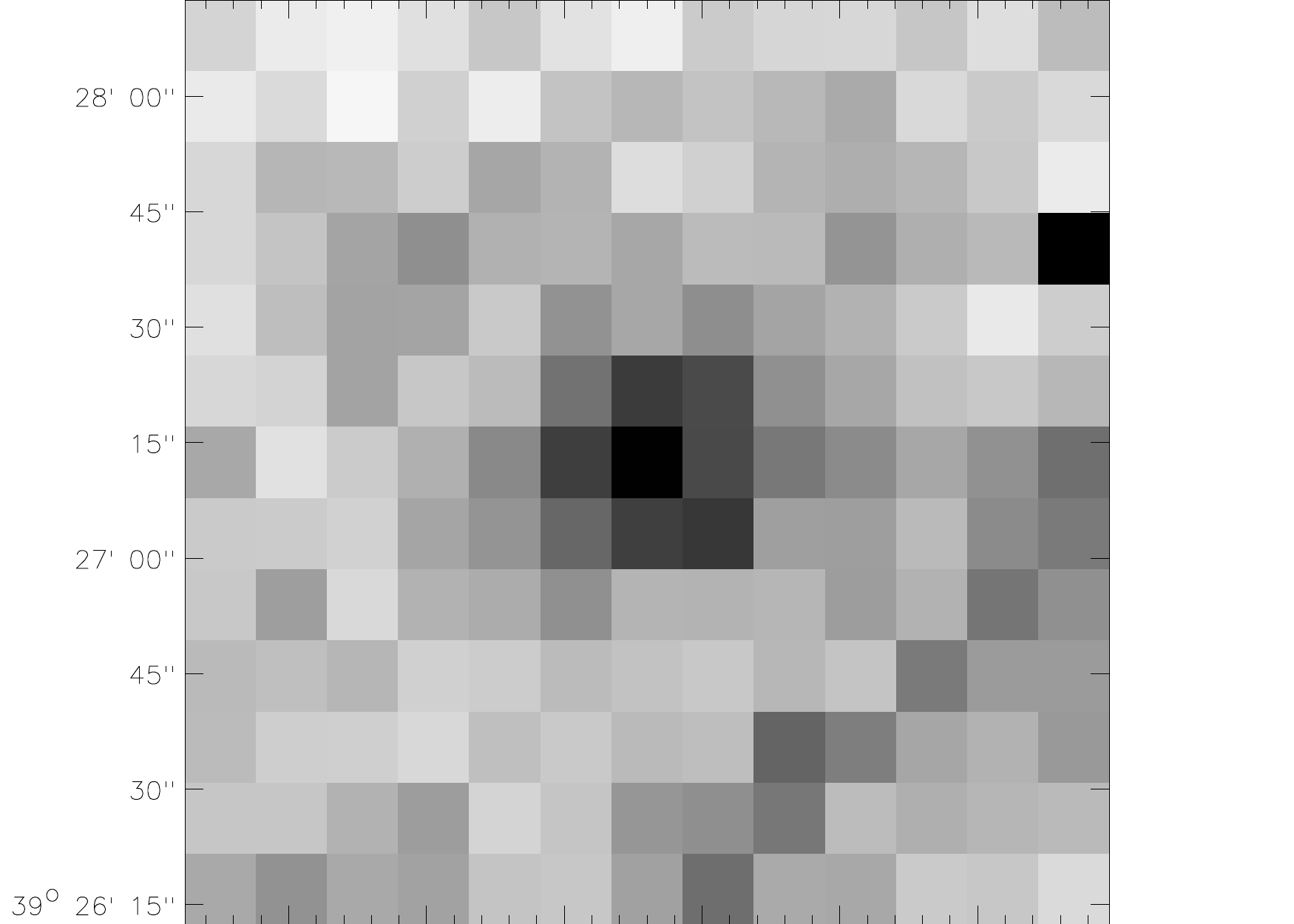}   &
      \includegraphics[height=14mm,trim=70  -4 65 -10,clip=true]{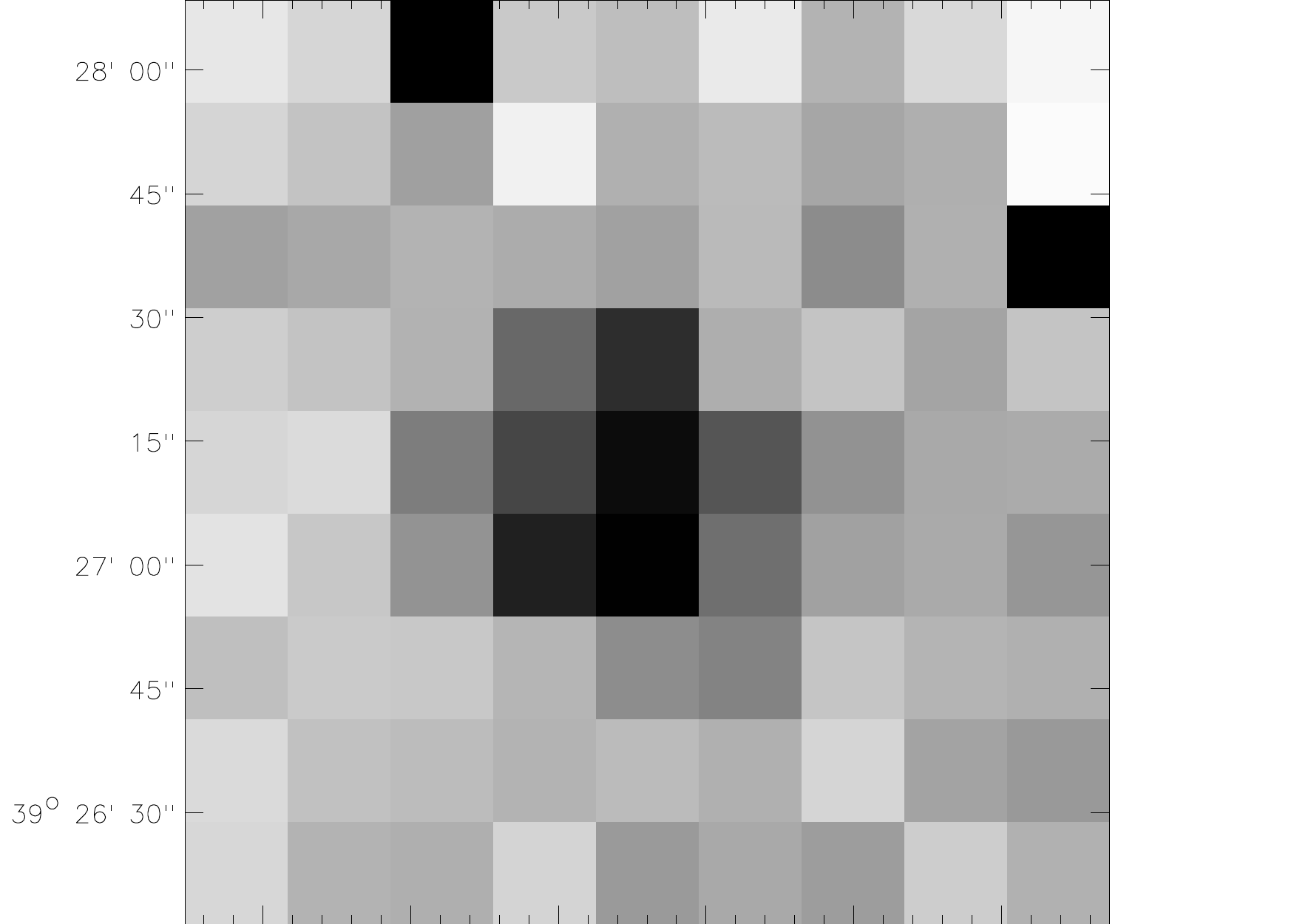} \\

      \multicolumn{6}{c}{\includegraphics[width=80mm,trim=30 0 20 0 ,clip=true]{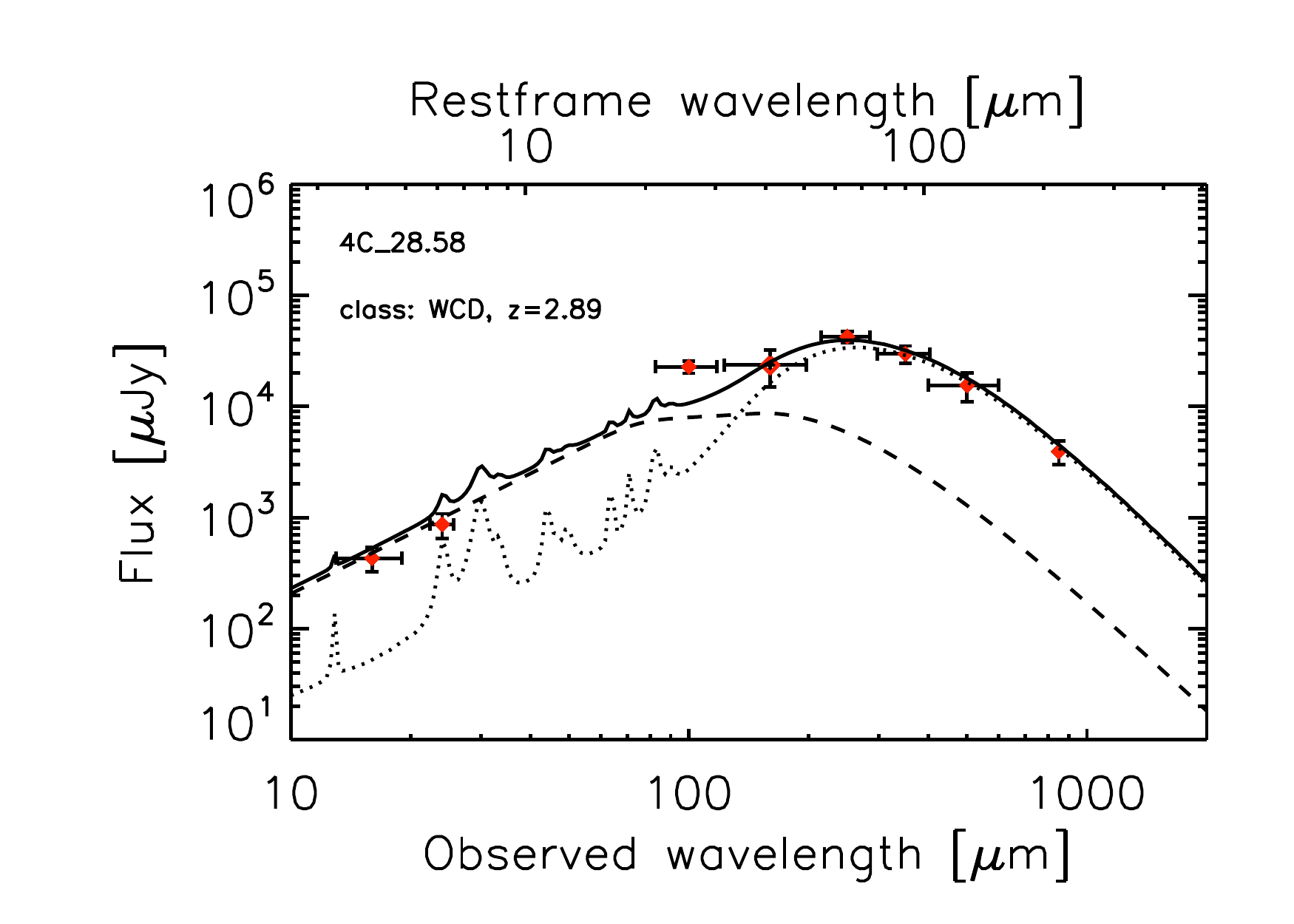}} \\
      24\mum & 100\mum & 160\mum & 250\mum & 350\mum & 500\mum \\
      \includegraphics[height=14mm,trim=70  -4 65 -10,clip=true]{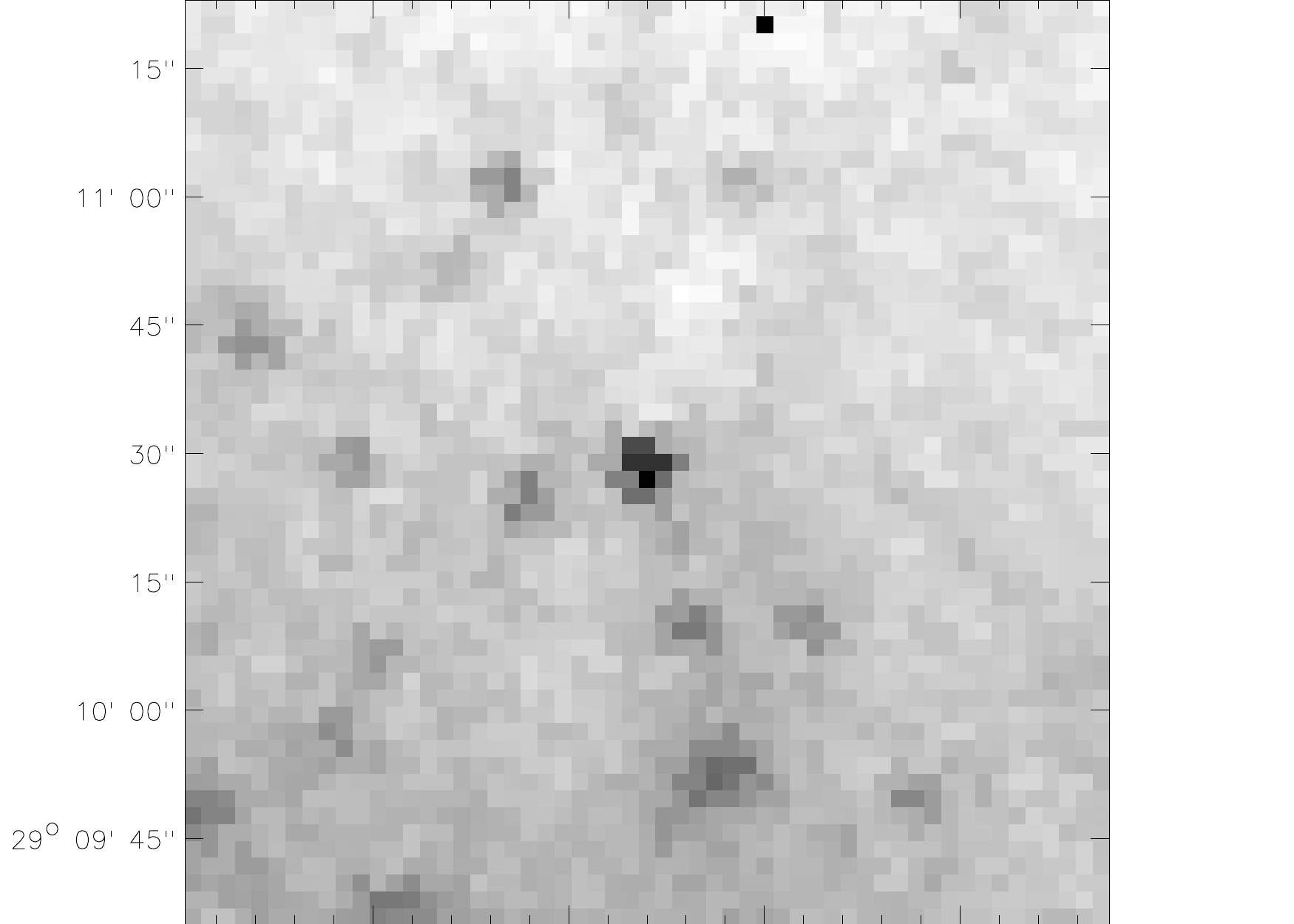} &
      \includegraphics[height=14mm,trim=70  -4 65 -10,clip=true]{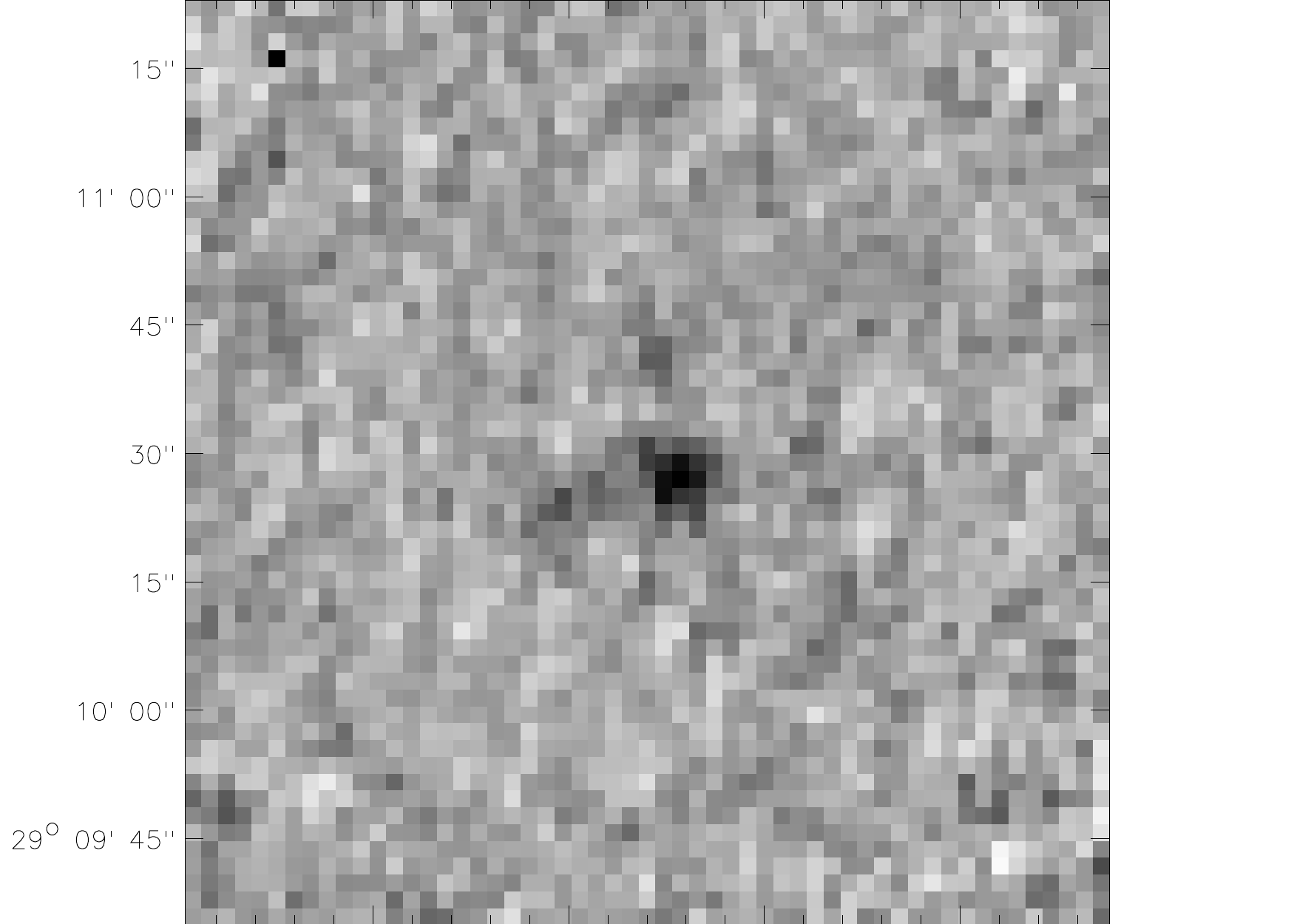} &
      \includegraphics[height=14mm,trim=70  -4 65 -10,clip=true]{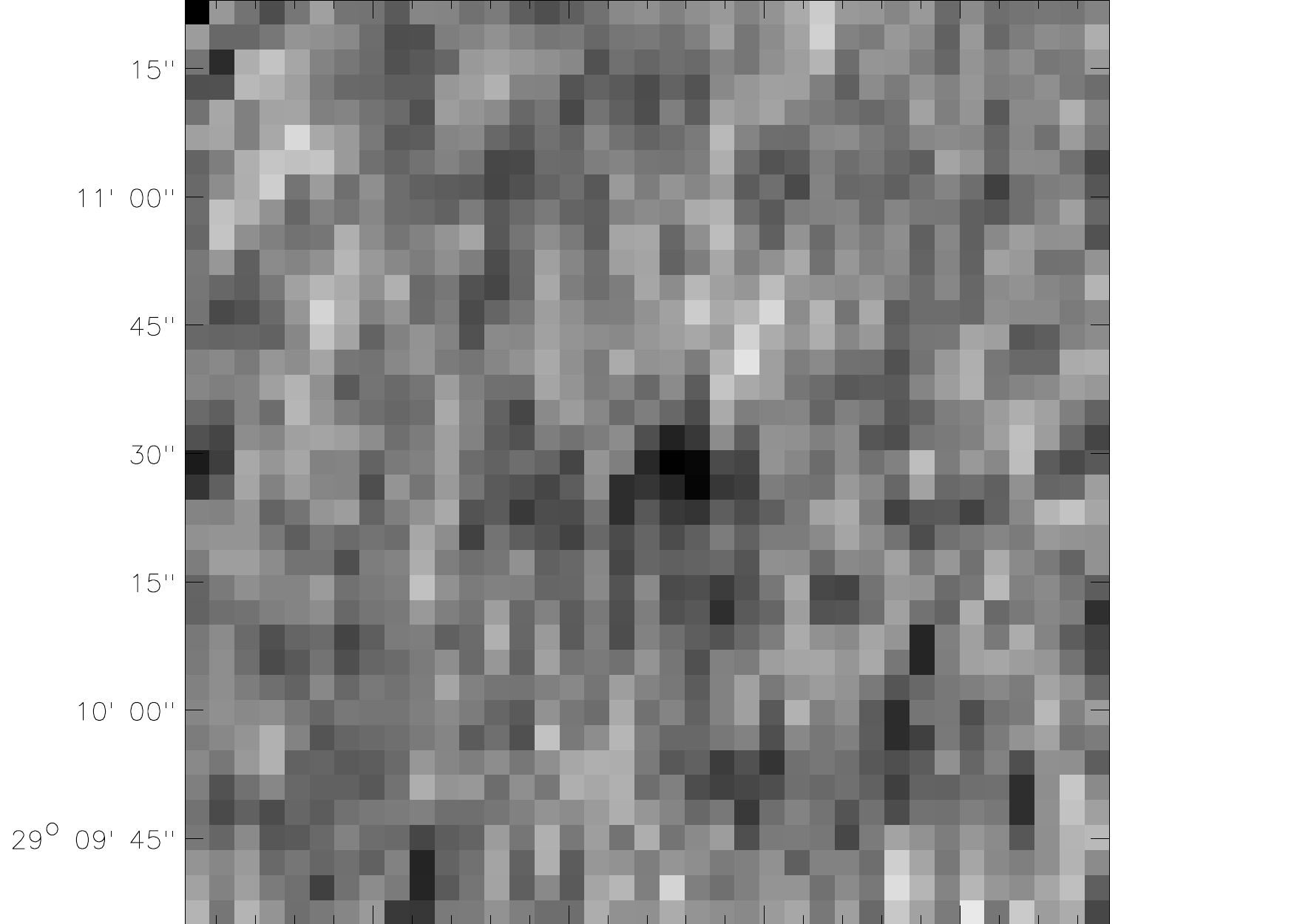} &
      \includegraphics[height=14mm,trim=70  -4 65 -10,clip=true]{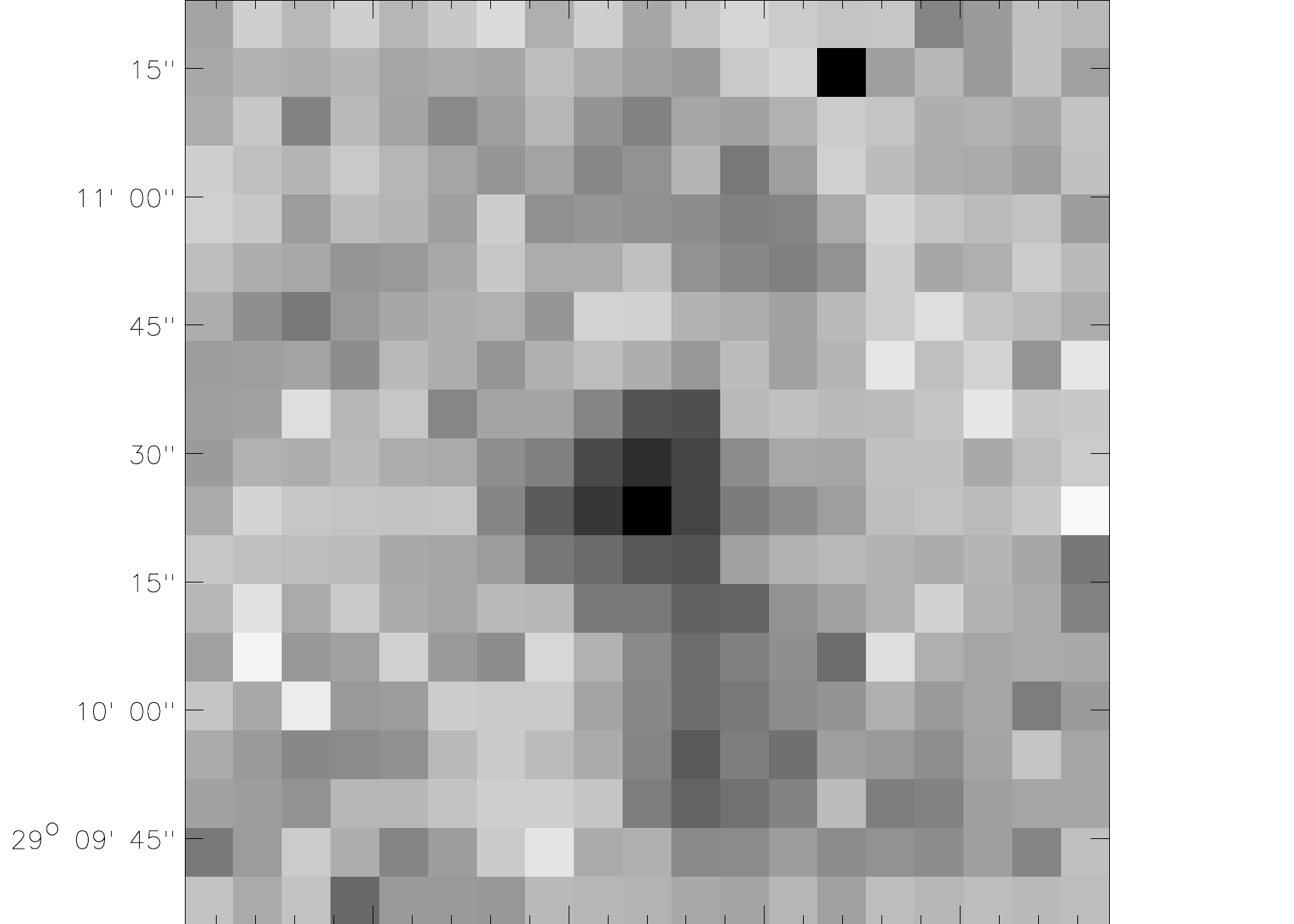}   &
      \includegraphics[height=14mm,trim=70  -4 65 -10,clip=true]{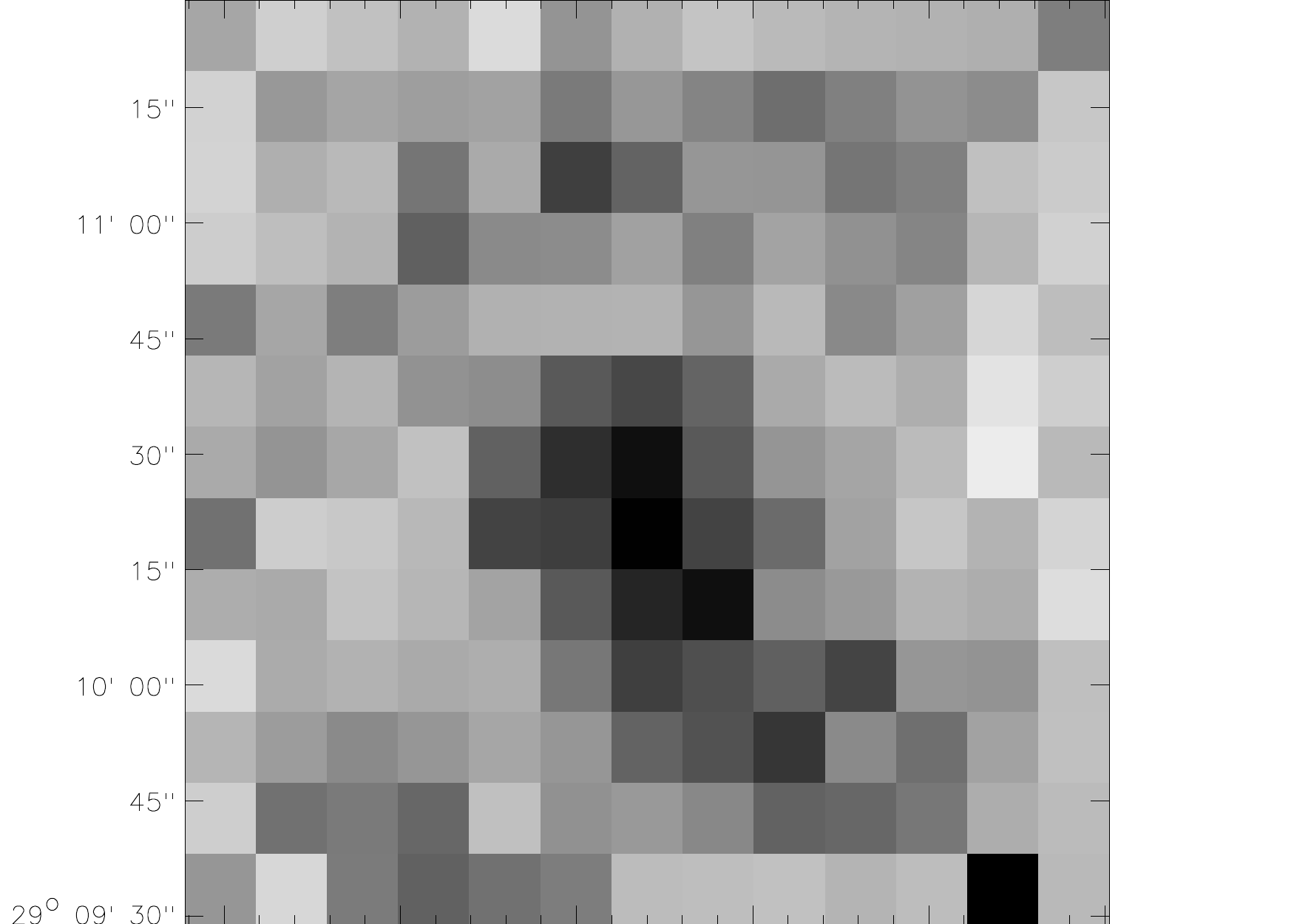}   &
      \includegraphics[height=14mm,trim=70  -4 65 -10,clip=true]{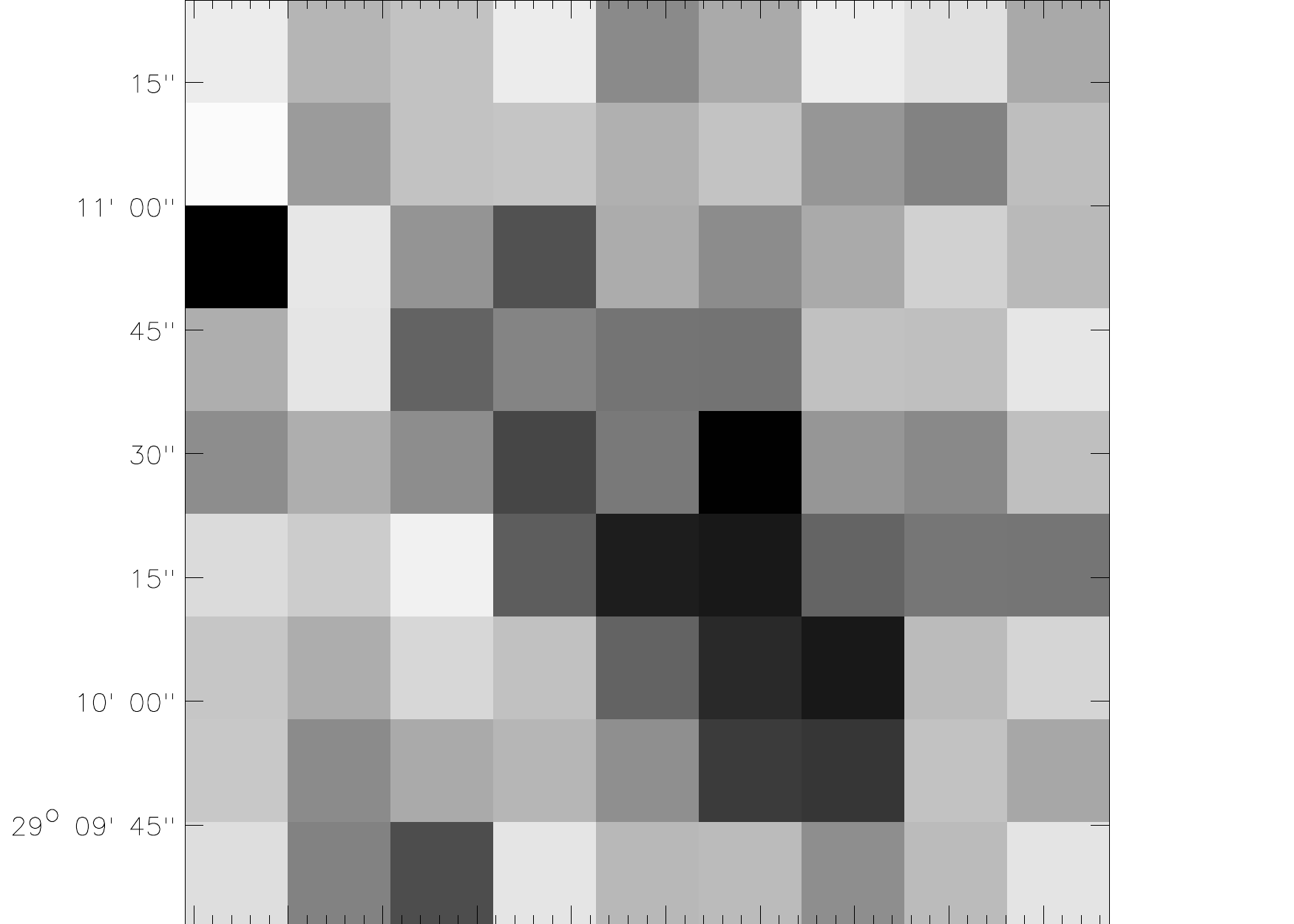} \\

      \multicolumn{6}{c}{\includegraphics[width=80mm,trim=30 0 20 0 ,clip=true]{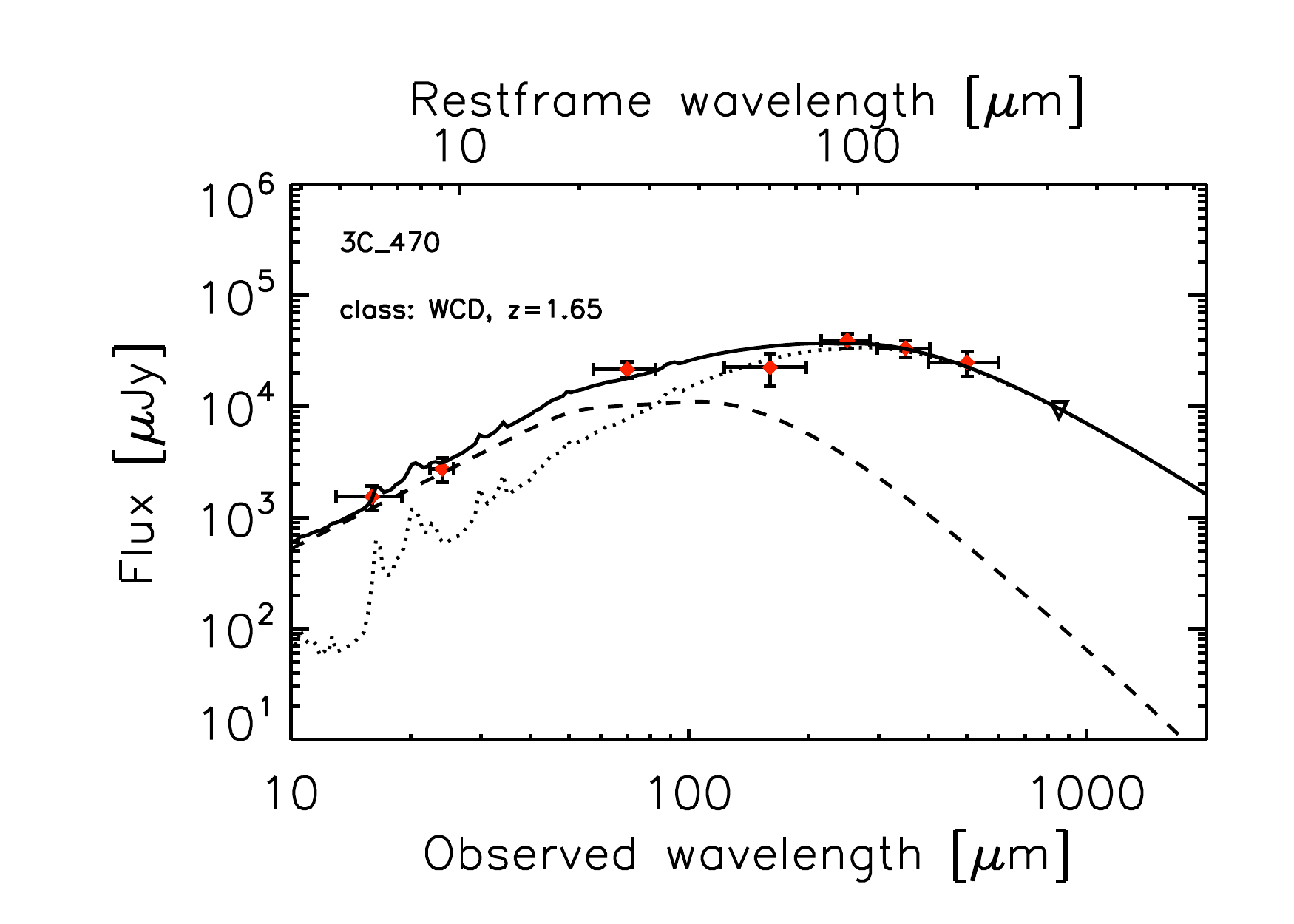}} \\
      24\mum & 70\mum & 160\mum & 250\mum & 350\mum & 500\mum \\
\end{tabular}
          \end{minipage}
  \end{centering}
\end{figure*}

\end{centering}



\end{document}